\documentclass[pdflatex,sn-mathphys-num]{sn-jnl}

\usepackage{graphicx}%
\usepackage{multirow}%
\usepackage{amsmath,amssymb,amsfonts}%
\usepackage{amsthm}%
\usepackage{mathrsfs}%
\usepackage[title]{appendix}%
\usepackage{xcolor}%
\usepackage{textcomp}%
\usepackage{manyfoot}%
\usepackage{booktabs}%
\usepackage{algorithm}%
\usepackage{algorithmicx}%
\usepackage{algpseudocode}%
\usepackage{listings}%
\usepackage{mathtools}

\usepackage{color,xcolor,ucs}
\usepackage{mathtools}   \usepackage{tikz} 
\usepackage{ amssymb }
\usepackage{extarrows} 
\usepackage{pgf,tikz}
\usepackage{float}
\usetikzlibrary{positioning}
\usetikzlibrary{shapes.geometric}
\usetikzlibrary{shapes.misc}
\usetikzlibrary{arrows}
\usepackage{caption}
\usepackage{mathrsfs}
\usetikzlibrary{arrows,shapes,automata,backgrounds,petri,positioning}
\usetikzlibrary{decorations.pathmorphing}
\usetikzlibrary{decorations.shapes}
\usetikzlibrary{decorations.text}
\usetikzlibrary{decorations.fractals}
\usetikzlibrary{decorations.footprints}
\usetikzlibrary{shadows}
\usetikzlibrary{calc}
\usetikzlibrary{spy}
\usepackage{amsmath}
\usepackage{array}
\usepackage{ amssymb }
\usepackage{braket}
\usepackage{qcircuit}
\usepackage{soul}
\usepackage{braket} 
\usepackage{relsize}

\usepackage{amsmath}
\usepackage{ amssymb }
\usepackage{braket}
\usepackage{qcircuit}
\usepackage{soul}
\usepackage{braket}

\begin{document}

\title[Article Title]{Quantum strategies, error bounds, optimality, and duality gaps for multiplayer  $\mathrm{XOR}$, $\mathrm{XOR}^{*}$, compiled  $\mathrm{XOR}$, $\mathrm{XOR}^{*}$, and strong parallel repetiton of $\mathrm{XOR}$, $\mathrm{XOR}^{*}$, and $\mathrm{FFL}$ games}


\author[1]{\fnm{Pete} \sur{Rigas}}\email{pbr43@cornell.edu}

\affil[1]{\city{Newport Beach}, \postcode{92625}, \state{CA}, \country{United States}}


\abstract{We characterize exact, and approximate, optimality in games that players can interact with using quantum strategies. In comparison to a previous work of the author, arXiv: 2311.12887, which applied a 2016 framework due to Ostrev for constructing error bounds beyond CHSH and XOR games, in addition to the existence of well-posed semidefinite programs for determining primal feasible solutions, along with quantum-classical duality gaps,  it continues to remain of interest to further develop the construction of error bounds, and related objects, to game-theoretic settings with several participants. In such settings, one encounters a rich information theoretic landscape, not only from the fact that there exists a significantly larger combinatorial space of possible strategies for each player, but also from several opportunities for pronounced quantum advantage. We conclude this effort by describing other variants of other possible strategies, as proposed sources for quantum advantage, in $\mathrm{XOR}^{*}$, compiled $\mathrm{XOR}^{*}$, and strong parallel repetition variants of $\mathrm{XOR}^{*}$ games.  \textit{{Keywords}: Quantum games, non-locality, quantum computation, entangled states, verification} \footnote{\textbf{MSC Class}: 81P02; 81Q02}}

 \maketitle

\section{Introduction}

\subsection{Overview}

Connections between quantum computing and information theory continue to remain of interest for potential near term applications of commercial quantum computers, with theoretical efforts being devoted to the study, and classification, of optimal strategies for XOR games {[37]}, with extensions to $\mathrm{XOR^{*}}$ and $\mathrm{FFL}$ games {[30]}, in addition to several other related characteristics, from various consequences of nonlocal strategies {[7, 9, 19, 32]}, solving PDE systems {[45]}, approximation algorithms {[20]}, entanglement between players {[8, 21, 22, 25]}, and synchronous values {[24]}, to aspects relating to communication complexity {[1, 5, 14]}. Albeit the fact that several adjacent fields have experienced tremendous progress in advancing communication protocols, {[2]}, various generalizations of inequalities involving Grothendieck constants, {[3]}, the utility of non-local boxes, {[4]}, variational inference, {[5]}, computational complexity of training regiments, {[7]}, and related computational tasks, {[8, 9, 10, 12, 14]}, further elaborations on paradoxical aspects of Quantum information continue to remain of interest to explore. To provide significant contributions at the crossroads of all such fields, which could lead to algorithmic problems with polynomial, or more optimistically, exponential, speedups that have been identified for a wide rang of industrial, {[6, 19, 20, 22, 24, 25, 27, 29, 31, 33, 35, 36, 38, 39, 41, 49, 51]}, and theoretical, {[17, 32, 34, 40, 42, 47, 48]}, areas of research, additional prospects remain for identifying circumstances for which quantum advantage exists. In previous work from the past two years, {[44]}, to investigate properties of entangled states that Alice and Bob can share to achieve an arbitrary advantage over performance with classical strategies {[37]}, several bounds were formulated by passing to the dual of the XOR game, from criteria first developed in {[37]}, the first of which restricts the initial number of options that each player can use before submitting answers to a referree, and the second of which provides an expression for the trace of the inner product with a Born rule. Under the circumstance that Alice and Bob find themselves in for other nonlocal games, potentially with an arbitrary number of other players which participate, it is necessary to formulate more complicated systems of inequalities, taken with respect to the Frobenius norm, or with respect to the $\mathrm{L}$-1 norm, in addition to formulating properties satisfied by a suitable nonlinear transformation, for characterizing optimal strategies which could raise the prospects of obtaining a quantum advantage in winning. Besides obtaining several generalizations to notions of approximate, and exact, optimality, from the 2-player setting, higher-dimensional settings for games can be leveraged to study the impact of performing ordinary, and strong, parallel repetition. Under such a set of circumstances, and assumptions, on the strategy that each player can assume, agents can develop optimal strategies for two-player games under many constraints depending upon the point of the game at which entangled information is shared, in addition to additional operations that the referee performs before determining the outcome of the game.

In order to explore differences between several games which individually pose different optimization constraints on players, we discuss how research directions initially raised in {[13]} can be further explored. After having provided an overview of the XOR game, which has generalizations that have been formulated for other two-player games {[16]}, from properties of the semidefinite program we state several variants of the constrained, $3$-$\mathrm{XOR}$, optimization problem for characterizing performance associated several other games, all of which share in similarities with optimal values, biases, and related objects, that have previously been introduced in the literfature for the Schur, multiplayer XOR, quantum XOR, and the Torpedo, games. Besides the statement of each semidefinite program and a corresponding approximation of the primal feasible solution which can be used for playing optimally, we characterize other properties which quantum states for optimal strategies satisfy, ranging from inequalities that have upper bounds proportional to $ \sqrt{n \epsilon}$ to $n^2 \sqrt{\epsilon}$, given the choice of a sufficiently small optimality parameter $\epsilon$. Agents, whether taken in groups of two or to the positive powers of prime numbers, can maximize the probability of winning the game with an optimal quantum strategy, hence further broadening the set of possible circumstances under which quantum information provides an advantage.

\subsection{This paper's contributions}

\noindent This paper contributes to the rich field of Quantum Information Theory. As a vast field, in recent years several proposals for Quantum advantage have been formulated, which can not only pose algorithmic, and computational, sources of advantage, but also intricate aspects of information processing that are dependent upon entanglement, and related objects. Building upon previous arguments of the author for the 2-player setting, {[44]}, the forthcoming work introduces expressions for the optimal value, bias, semidefinite programs, and several other objects for multiplayer XOR games, along with strong parallel variants of the XOR and FFL games. Such games were initially of interest to explore in the 2-player setting due to the fact that error bounds, and generalizations of error bounds, reflect upon limitations of classical information processing. Beyond the 2-player setting, in the multiplayer setting such relations not only depend upon a combinatorially larger space of possible strategies, but also upon more intricate notions of the situations, and conditions, under which Quantum advantage, can be achieved. With all of these considerations taken into account, formulating the set of relations for which exact, and approximate, optimality, hold for game-theoretic settings with more than $2$ players is of value. Moreover, in comparison to the error bounds, and approximately optimal framework previously developed in {[37]}, and further expanded upon in {[44]}, the multiplayer setting unveils several components of the set of exact, and approximate, optimality for games under ordinary, and strong, parallel repetition. Despite the fact that odrinary, and strong, parallel repetition can similarly be formulated from the 2-player setting, elaborating upon higher-dimensional notions of quantum advantage, and optimality, continue to remain of value.

\subsection{The setup}

\noindent Following the Introduction, and connections with previous studies, we discuss how game theoretic objects inform forthcoming arguments.

\bigskip

\noindent \underline{\textit{Structure of error bounds}}. From previous work of the author which sought to classify notions of approximate, and exact, optimality of quantum strategies for $\mathrm{XOR}^{*}$ and $\mathrm{FFL}$ games (from work on $\mathrm{XOR}$ and $\mathrm{CHSH}$ games in {[37]}), generalized forms of error bounds continue to remain of interest for further investigation, particularly for: (1) higher dimensional resource systems; (2) different optimal values, and hence the maximum probability, of winning the game; (3) more intricate notions of entanglement, primarily through analog of intertwining operators originally applied to $\mathrm{XOR}$ games; (4) computational complexity. To explore connections, and interactions, amongst these areas, along with many others, a broader array of games is analyzed. First, the $3$$\mathrm{XOR}$ game, as the next most complicated game-theorretic setting than previous arguments formulated for two-players in $\mathrm{XOR}$, $\mathrm{XOR}^{*}$, and $\mathrm{FFL}$, games, presents a promising higher-dimensional for exploring optimality. In such a setting, and closely related ones, beyond $2$-$\mathrm{XOR}$,and $2$-$\mathrm{XOR}^{*}$, games, there not only exists closed form expressions for the optimal values with more players, but also for the expected duality gap between classical and quantum values. Second, for the $\mathrm{FFL}$ game, observations previously formulated by the author, which discussed the assumptions under which players Alice and Bob can play with exact, and approximate, optimality, apply to strong parallel repetitions. However, in spite of the fact that error bound inequalities can be formulated in somewhat of a similar manner as for a single $\mathrm{FFL}$ game, difficulties arise in considering how parallel repetitions, and strong parallel repetitions, impact entangled strategies that groups of players can pursue for maximizing his or her winning probability.

Third, by making use of generalized notions developed for games of the previous two types, error bounds, along with various quantitative measures of entanglement entropy can be obtained. Predominantly, such estimates on the entanglement entropy, from the action of a suitable higher-dimensional intertwining operation which performs a transformation on a player's observable after reversing the order in which the observables appear in tensor products of responses from players of the game, statistically relate properties of observables that each player gathered for preparing responses to the referee's questions. In comparison to observables introduced in other areas of Physics, the action of intertwining operations for observables prepared by players who can use Quantum strategies are not only dependent upon the Heisenberg uncertainty principle, but also upon the observables that each player prepares in response to the answers of previous players in the game.

\bigskip

\noindent \underline{\textit{Optimal values}}. In classical, and quantum, game theory, each possible value represents the maximum probability of any player winning. As proposed sources of quantum advantage, the duality gap between such values, should one exist, for any game indicates how entanglement can benefit a player's strategy.

\noindent When examining, and providing quantitative measures, on the entanglement in quantum systems, the $\mathrm{XOR}$ game, with its corresponding optimal value of $\frac{1}{\sqrt{2}}$, encodes limitations to how much Alice and Bob stand to benefit from cooperation, as the players can share an entangled state, representing an optimal strategy, for responding to the referee. Beyond two players, the structure, and expected actions, on the Hilbert space of observables gathered by each player generalizes entanglement, and its expected impact, on the classes of strategies that players can pursue. For $\mathrm{XOR}$ and $\mathrm{FFL}$ games alike, performing both the ordinary, and strong, parallel repetition operation as players continue answering questions drawn from the referee's probability distribution indicate important, and subtle, differences in optimality, and its various equivalent characterizations.

\begin{tabular}{|l|l|}
\hline\parbox[t]{0.25\textwidth}{
\begin{itemize}
\item $\textbf{Theorem}$ \textit{1}
\item $\textbf{Theorem}$ \textit{2}
\item $\textbf{Theorem}$ \textit{3}
\item $\textbf{Theorem}$ \textit{4}
\item $\textbf{Theorem}$ \textit{5}
\item $\textbf{Theorem}$ \textit{6}
\item $\textbf{Theorem}$ $\textit{1}^{*}$
\item $\textbf{Theorem}$ $\textit{2}^{*}$
\item $\textbf{Theorem}$ $\textit{3}^{*}$
\item $\textbf{Theorem}$ $\textit{4}^{*}$
\item $\textbf{Theorem}$ $\textit{5}^{*}$
\item $\textbf{Theorem}$ $\textit{6}^{*}$
\end{itemize}}& 
\parbox[t]{0.73\textwidth}{
\begin{itemize}
\item $3$-XOR primal feasible solutions and duality gaps
\item $4$-XOR primal feasible solutions and duality gaps
\item $5$-XOR primal feasible solutions and duality gaps
\item $N$-XOR primal feasible solutions and duality gaps
\item $N$-XOR strong parallel repetition primal feasible solutions and duality gaps
\item FFL strong parallel repetition primal feasible solutions and duality gaps
\item Strong parallel repetition of $\textbf{Theorem}$ $\textit{1}$
\item Strong parallel repetition of $\textbf{Theorem}$ $\textit{2}$
\item Strong parallel repetition of $\textbf{Theorem}$ $\textit{3}$
\item Strong parallel repetition of $\textbf{Theorem}$ $\textit{4}$
\item Strong parallel repetition of $\textbf{Theorem}$ $\textit{5}$
\item Strong parallel repetition of $\textbf{Theorem}$ $\textit{6}$
\end{itemize}}\\
\hline
\end{tabular}
\noindent \textit{Table *}. An overview of the main results to be proved in this work, given error bounds, and generalizations of error bounds, in the next section.

\bigskip

For settings with $\geq 3$ players, while optimal, and approximately optimal, strategies satisfy similar inequalities as do their two-player counterparts, quantum states associated with $\geq 3$ players exhibit more complicated entanglement properties, hence impacting the combinatorial space of possible strategies that each player may pursue (several assumptions under which optimal values for $3$XOR, and related games, exist are provided in {[50]}). In the forthcoming arguments for error bounds involving the optimal value for each game, optimal strategies associated with the best strategies for each player can use for maximizing his or her respective utility further characterize paradoxical implications of information processing with Quantum physics.

\bigskip

\noindent \underline{\textit{Entanglement}}. In comparison to classical settings of game theory, quantum counterparts permit for advantage through the use of entanglement and related vehicles. Depending upon the particular game, players may use entanglement to their advantage, through some shared quantum state, throughout the duration during which the game is played. For $\mathrm{XOR}$, or $\mathrm{XOR}^{*}$, games with any number of players, participants are prohibited from sharing any entangled state before the referee distributes questions, however, notions of entanglement can nevertheless still be leveraged for making quantitative observations about the observables that each player gathers, which correspond to his or her strategy. Albeit the fact that quantum advantage exists for the family of $\mathrm{XOR}$ game with any number of players, from primal feasible solutions of semidefinite programs, and semidefinite programs corresponding to the duality gap, it continues to remain of interest to explore the duality gap for other game theoretic settings.

In the first case, for $\mathrm{XOR}$ games with more than two players, upper bounds for quantifying the advantage that a player, or players, can gather from adopting quantum strategies take on straightforward extensions of the two-player setting; in comparison to the two possible ways in which Alice and Bob can interchange their observables with respect to the tensor product, in the presence of an additional third player the bounds on entanglement, along with the expect error bounds, are dependent upon more contributions, namely those from additional ways in which observables of each player can be interchanged. To this end, we generalize error bounds for several reasons, including:

\subsection{Two-player game theoretic objects for $\mathrm{XOR}$, $\mathrm{XOR^{*}}$, and FFL games}

\noindent Denote the Frobenius norm,

\begin{align*}
  \big|\big| A \big|\big|_F \equiv \sqrt{\overset{m}{\underset{i=1}{\sum}} \overset{n}{\underset{j=1}{\sum}} \big| a_{ij} \big|^2 } = \sqrt{\mathrm{Tr} \big[ A^{\dagger} A \big] }  \text{, } 
\end{align*}

\noindent of an $m \times n$ matrix $A$ with entries $a_{ij}$, for,

\begin{align*}
A \cdot B \equiv  \mathrm{Tr} \big( A B \big)  \equiv     \underset{\text{columns }i, \text{ rows } j}{\underset{ij}{\sum}}   A_{ij} B_{ij}      \text{, }
\end{align*}

\noindent from which there exists a \textit{linear bijection} $\mathcal{L}$ between the tensor product space, $\textbf{C}^{d_A} \otimes \textbf{C}^{d_B}$, and the space of $d_A \times d_B$ matrices with complex entries, $\mathrm{Mat}_{d_A , d_B} \big( \textbf{C} \big)$, satisfying (\textbf{Lemma} \textit{1}, {[37]}),

\begin{itemize}
\item[$\bullet$] \underline{\textit{Image of the tensor product of two quantum states under} $\mathcal{L}$}: $\forall \ket{u} \in \textbf{C}^{d_A}, \ket{w} \in \textbf{C}^{d_B}, \exists \ket{u^{*}} \in \textbf{C}^{d_B} : \mathcal{L} \big( \ket{u} \otimes \ket{w} \big) = \ket{u} \bra{u^{*}}  \text{, }$ 
\item[$\bullet$] \underline{\textit{Product of a matrix with the image of a quantum state under} $\mathcal{L}$}: $\forall \ket{u} \in \textbf{C}^{d_A}, \exists A \in \mathrm{Mat}_{d_A} \big( \textbf{C} \big) : A \mathcal{L} \big( \ket{u} \big) = \mathcal{L} \big( A \otimes I \ket{u} \big)\text{, }$
\item[$\bullet$] \underline{\textit{Product of the image of a quantum state under $\mathcal{L}$ with the transpose of a matrix}}:  $\forall \ket{w} \in \textbf{C}^{d_B}, \exists B \in \mathrm{Mat}_{d_B} \big( \textbf{C} \big) : \mathcal{L} \big( \ket{w} \big) B^T = \mathcal{L} \big( I \otimes B \ket{w} \big)  \text{, }$
\item[$\bullet$] \underline{\textit{Frobenius norm equality}}: $\forall \ket{w} \in \textbf{C}^{d_B} : \big|\big| \mathcal{L} \big(   \ket{w}     \big) 
 \big|\big|_F = \ket{w}  \text{. } $
\end{itemize}

\noindent where the basis of $\textbf{C}^{d_A} \otimes \textbf{C}^{d_B}$ is of the form $\ket{i} \otimes \ket{j}$,  and the basis for $\mathrm{Mat}_{d_A, d_B} \big( \textbf{C} \big)$ is of the form $\ket{i}\bra{j}$, for $1 \leq i \leq d_A$ and $1 \leq j \leq d_B$. From the four properties above of $\mathcal{L}$, for two finite sets $S$ and $T$, also define the map $V : S \times T \longrightarrow \big\{ - 1 , 1 \big\}$. From a product probability distribution $\pi$ over $S \times T$, the game proceeds with the Referee examining the responses of Alice and Bob depending upon the entangled state that they share, in which, after sampling a pair $\big( S , T \big) \sim \pi$, and sending one question $s$ to Alice and another question $t$ to Bob,

\begin{align*}
    V \big(  s , t \big)   ab  \equiv 1 \Longleftrightarrow  \text{ Alice and Bob win,}    \\    V \big(  s , t \big)   ab \equiv -1 \Longleftrightarrow  \text{ Alice and Bob lose,}      
\end{align*}

\noindent in which, depending upon whether $V \big( s ,t \big) \equiv 1$, or $V \big( s , t \big) \equiv -1$, Alice and Bob must either give the same answers, and opposing answers, to win, respectively. To consider linear subspaces,

\begin{align*}
    \mathrm{span} \bigg\{ \bigg(  \bigg( \underset{1 \leq i \leq N}{\prod} A^{j_i}_i \bigg) \bigotimes \bigg( \underset{1 \leq z \leq N-1}{\prod} \textbf{I}_z \bigg) \bigg) \ket{\psi}: \big( j_1,\cdots, j_n \big) \in \big\{ 0,1\big\}^n     \bigg\}        \text{, }
\end{align*}

\noindent given an  optimal strategy $\ket{\psi}$ which is dependent upon codeworks $j_i$, each of which can individually be $0$ or $1$, one can straightforwardly generalize the scoring function, and probability distribution of questions, that the referree uses to start the game. Under different assumptions on the game being played, the referee's scoring function, $V$, can take on a wide variety of forms. Denoting the set of all possible questions, and answers, with,

\begin{align*}
  Q_1 \times \cdots \times Q_i \times \cdots \times Q_N  \text{, } \\ A_1 \times \cdots \times  A_i \times  \cdots \times A_N \text{, }
\end{align*}

\noindent the referee's scoring function, or predicate, in the case of an arbitrary number of questions which can be distributed to each participant, {[18]},

\begin{align*}
  p \big( q_1, \cdots, q_i, \cdots, q_n \big)   \text{, }
\end{align*}

\noindent would take the form,

\begin{align*}
  V \big( a_1, \cdots, a_i, \cdots, a_n | q_1, \cdots, q_i, \cdots, q_n \big)   \text{. }
\end{align*}

\noindent For the EAOS game, which is related to the Odd-Cycle game, the predicate which allows the referee to determine which player has submitted answers that constitute a winning strategy takes the form, {[18]} (for other possible definitions of the function that the referree uses to evaluate whether responses from players are correct, see {[11]}),

\[
V \big( ab | st \big) \equiv \left\{\!\begin{array}{ll@{}>{{}}l}  1, & \text{ if } 1 - \delta_{st} = a \oplus b\\ 0, & \text{ otherwise}
\end{array}\right. ,
\]

\noindent for,

\[
\delta_{st}  \equiv \left\{\!\begin{array}{ll@{}>{{}}l}  1, & \text{ if } s = t \\ 0, & \text{ otherwise}
\end{array}\right. .
\]

\noindent Besides increasing the number of players participating in a game, the operation of \textit{strong parallel repetition} is also of great significance, not only for pursuing information-theoretic circumstances, but also for discussing fundamental differences between the $\mathrm{XOR}$, $\mathrm{XOR}^{*}$, and $\mathrm{FFL}$, games previously examined by the author. Denote the strong parallel repetition of some game $G$, either an $\mathrm{XOR}$ or $\mathrm{FFL}$, game, {[18]},

\begin{align*}
  G^{\otimes n}  \text{, }
\end{align*}

\noindent for $n$ strong parallel repetitions. In order for a player to simultaneously win $n$ copies of the game, at each step the referee draws questions from some probability distribution, 

\begin{align*}
  \big( x_1, \cdots, x_n \big) \in X^n  \text{, } \\  \big( y_1, \cdots, y_n \big) \in Y^n \text{, }
\end{align*}

\noindent from which the players respectively respond with,

\begin{align*}
  \big( a_1, \cdots, a_n \big)   \text{, } \\ \big( b_1, \cdots, b_n \big)  \text{. }
\end{align*}

\noindent The players win iff,

\begin{align*}
 V \big( x_i, y_i, a_i, b_i \big) \equiv 1   \text{. }
\end{align*}

\noindent The case $n\equiv 2$, namely the game that is obtained under the operation of two strong parallel repetitions, will be further examined extensively later in the forthcoming work. In comparison to similarities between the framework provided in {[37]}, which states that error bounds, and optimal, strategies for $\mathrm{CHSH} \big( n \big)$ and $\mathrm{XOR}$ games can be extended to $\mathrm{XOR}^{*}$ and $\mathrm{FFL}$ games, as previously examined by the author, the fact that,

\begin{align*}
  \omega_{\mathrm{XOR} \wedge  \mathrm{XOR}} \big( G_{\mathrm{XOR} \wedge \mathrm{XOR}} \big)  \equiv \omega \big(  \mathrm{XOR} \wedge \mathrm{XOR} \big)  \equiv \underset{\# \text{ of strong parallel repetitions } j }{\prod}   \omega \big( \mathrm{XOR}\big)^{j}  \\   \equiv \underset{1 \leq j \leq 2}{\prod}   \omega \big( \mathrm{XOR}\big)^{j}   \equiv \big( \omega \big( \mathrm{XOR} \big) \big)^2 \text{,} \end{align*}

\noindent holds for two strong parallel repetitions of the $\mathrm{XOR}$ games, while,

\begin{align*}
  \omega_{\mathrm{FFL} \wedge  \mathrm{FFL}} \big( G_{\mathrm{FFL}} \big) \equiv \omega_{\mathrm{FFL}} \big( G_{\mathrm{FFL}} \wedge G_{\mathrm{FFL}} \big)   \equiv \omega \big(  \mathrm{FFL}  \wedge \mathrm{FFL} \big) \equiv  \omega \big( \mathrm{FFL}\big) \equiv \frac{2}{3} \text{,} \end{align*}

\noindent holds for two strong parallel repetitions of the $\mathrm{FFL}$ games, underlies additional differences in the structure of error bounds, in addition to various other information-theoretic consequences.

In tandem with differences to the optimal, and approximately optimal, framework in which each player seeks to maximize his or her payoff, there are various expressions throughout the literature for the optimal value, namely, the maximum probability of winning a game. For the Odd-Cycle game, there exists expressions for the optimal value of classical strategies that Alice and Bob can adopt, where the maximum of all probabilities for winning takes the form, {[15]},

\begin{align*}
  \omega^{\text{Odd-Cycle}}_c \equiv 1 - \frac{1}{2n}  \text{. }
\end{align*}

\noindent Other optimal values with classical values have been also computed, which are dependent upon the partitions of a $d \times d$ grid. Such values are of the form, {[17]},

\begin{align*}
   \omega^{\mathrm{Torpedo}}_{C,\text{ } d\equiv 2} \equiv \frac{3}{4} \text{, } \\  \omega^{\mathrm{Torpedo}}_{C, \text{ } d\equiv 3} \equiv \frac{11}{12} \text{, }
\end{align*}

\noindent for $d \equiv 2$, and $d \equiv 3$, respectively. In comparison to optimal values for $\mathrm{XOR}$, $\mathrm{XOR}^{*}$, and $\mathrm{FFL}$, games for which the optimal values is expressed as a supremum over the set $\mathcal{S}$ of all possible strategies, the optimal value using classical strategies, given the two and three-dimensional values for the Torpedo game above, can be expressed through the supremum,

\begin{align*}
     \underset{\mathcal{E}}{\mathrm{sup}}  \bigg(          \frac{1}{d^2 ( d+1) } \underset{x,z,q}{\sum}  p_e \big( w_q \big( x,z \big) | x, z, q \big)    \bigg)     \text{, }
\end{align*}

\noindent over encodings $\mathcal{E}$ over conditional probabilities $p_e$. Numerical simulations relate quantum and classical optimal values to each other through the ratios, {[17]},

\begin{align*}
      \frac{\omega^{\mathrm{Torpedo}}_{Q,\text{ } d\equiv 2}}{\omega^{\mathrm{Torpedo}}_{C,\text{ } d \equiv 2}}  \approx 1.053      \text{, } \\     \frac{\omega^{\mathrm{Torpedo}}_{Q,\text{ } d\equiv 3}}{\omega^{\mathrm{Torpedo}}_{C,\text{ } d \equiv 3}}   \approx 1.091     \text{, }
\end{align*}

\noindent for $d \equiv 2$, and $d \equiv 3$, respectively. The approximations to the ratios of optimal values above follow from the fact that the quantum optimal values satisfy, {[17]},

\begin{align*}
 \omega^{\mathrm{Torpedo}}_{Q,\text{ } d\equiv 2} \approx 0.79   \text{, } \\  \omega^{\mathrm{Torpedo}}_{Q,\text{ } d\equiv 3} \equiv 1 \text{. }
\end{align*}

\noindent Under other circumstances, the classical and optimal values for linear games,

\begin{align*}
   \omega^{\mathrm{Linear Game}}_C \big( G \big)   \text{, } \\   \omega^{\mathrm{Linear Game}}_Q \big( G \big) \text{, }
\end{align*}

\noindent satisfy inequalities in the place of strict, or approximate, equalities, which are of the form, {[43]},

\begin{align*}
     \omega^{\mathrm{Linear Game}}_C \big( G \big)  \geq \frac{1}{\big| G \big|} \bigg( 1 + \frac{\big| G \big| -1}{m} \bigg)         \text{, } \\        \omega^{\mathrm{Linear Game}}_Q \big( G \big)  \leq    \frac{1}{\big| G \big|} \bigg(  1 + \sqrt{\big| \mathcal{Q}_A \big| \big| \mathcal{Q}_B \big|  }    \underset{x \in G \backslash \{ e \}}{\sum}   \big| \big| \Phi_x \big| \big|  \bigg)                  \text{, }
\end{align*}

\noindent for the linear game $G \equiv G^L$, game matrix,

\begin{align*}
  \Phi_x \equiv  \underset{\mathcal{Q}_A \times \mathcal{Q}_B}{\sum}    \text{Linear Game Matrix}  \equiv   \underset{\mathcal{Q}_A \times \mathcal{Q}_B}{\sum}  \bigg(   q \big( u , v \big) \chi_x \big( f \big( u , v \big) \big)    \big( \ket{u} \bra{v} \big)   \bigg)    \text{, }
\end{align*}

\noindent take under the spectral norm, $\big| \big| \cdot \big| \big|$, and,

\begin{align*}
   m \equiv \mathrm{min} \big\{ \big| \mathcal{Q}_A \big|, \big| \mathcal{Q}_B \big| \big\}  \text{, } \\ \big|  \mathcal{Q}_A \big| \equiv \underset{\text{Questions}}{\bigcup}  \big\{ \text{Player A's response to question } i \big\}    \text{, } \\ \big|  \mathcal{Q}_B \big| \equiv \underset{\text{Questions}}{\bigcup}  \big\{ \text{Player B's response to question } i \big\}   \text{, }
\end{align*}

\noindent denote the minimum of the questions given to Players $A$, and $B$. Previous characterizations of optimal strategies for the three games above have been shown to hold {[37]}, which were extended by obtaining a new set of inequalities describing actions of Alice and Bob on optimal FFL states, $\ket{\psi_{\mathrm{FFL}}}$ {[44]}, which are dependent upon the fact that the optimal value for the FFL game equals $\frac{2}{3}$, while the optimal value for the XOR game equals $\frac{1}{\sqrt{2}}$. In spite of the fact that other known expressions for classical, and quantum, optimal values have been obtained for various extensions of $\mathrm{CHSH}$ games, {[43]},

\begin{align*}
 \text{Quantum }  \mathrm{CHSH}- d \text{ value} \equiv  \omega_Q \big( \mathrm{CHSH}- d \big) \leq \frac{1}{d} + \frac{d-1}{d\sqrt{d}}  \text{, }
\end{align*}

\noindent where the parameter $d$ is a power of some prime, namely $d \equiv p^r$ for some $r \geq 1$. In games with more than two players, having knowledge of closed form expressions for classical, and quantum, optimal values permits for an application of a similar framework, as provided in {[37]}, for obtaining error bounds and $\epsilon$-approximality, where $\epsilon$ represents some parameter taken to be sufficiently small for representing a player's deviation from the optimal strategy. Difficulties associated with such arguments for characterizing exact, and approximate, optimality can arise in games with less regular structure. Such structures can be encountered when error bounds are not only dependent upon higher-dimensional linear operators, which act as counterparts to operators for $\mathrm{FFL}$ games considered by the author in {[44]}, but also on interactions between groups of player simultaneously. In demanding that there are more degrees of freedoms that players can use to transform observables that they gather into those of other players that are participating, one must introduce the following notions:

\begin{itemize}
    \item[$\bullet$] \underline{\textit{Product norm of player responses}}: Under the identifications,

    \begin{align*}
   \ket{\underset{\text{Odd number of players}}{\prod} i_j } \longleftrightarrow  \underset{\text{Odd number of players}}{\prod}  \ket{ i_j }   \text{, } \\ \bra{\underset{\text{Even number of players}}{\prod} i_j } \longleftrightarrow \underset{\text{Even number of players}}{\prod} \bra{ i_j }   \text{, }
     \end{align*}

    \noindent The outer product of responses from a group of $N$ players can be expressed as,

\begin{align*}
 \bigg( \ket{\underset{\text{Odd number of players}}{\prod} i_j } \bigg) \bigg( \bra{\underset{\text{Even number of players}}{\prod} i_j } \bigg)   \equiv   \bigg( \ket{\underset{\text{Odd } j, 1 \leq j \leq N}{\prod} i_j } \bigg) \\ \times \bigg( \bra{\underset{\text{Even } j, 1 \leq j \leq N}{\prod} i_j } \bigg)  \\ \equiv  \bigg(  \ket{i_N i_{N-2} \times \cdots \times i_1 } \bigg) \bigg(  \bra{i_2 i_4 \times \cdots \times i_{N-1}} \bigg)   \equiv  \bigg(   \ket{i_N} \ket{i_{N-2}} \times \cdots \times \ket{i_1} \bigg) \\ \times  \bigg(  \bra{i_2} \bra{i_4} \times \cdots  \times \bra{i_{N-1}} \bigg) \\  \equiv \ket{i_N}  \bigg( \cdots \times  \bigg( \cdots \times \bigg(  \ket{i_3}  \bigg( \ket{i_1} \bra{i_2} \bigg)  \bra{i_4} \bigg)  \times \cdots  \bigg) \times  \cdots \bigg)  \bra{i_{N-1}}  \\  \equiv      \ket{\text{Player } N \text{ responds to question } i_N \text{ given } (N-1) \text{ previous responses}}  \bigg( \cdots  \times \bigg( \cdots \\  \times \bigg( \ket{\text{Player } 1 \text{ responds to the first question}}  \bra{\text{Player } 2 \text{ responds to the second question}} \bigg)  \\ \cdots \times \bigg) \cdots \bigg)     \bra{\text{Player } (N-1) \text{ responds to question } i_N \text{ given } (N-2) \text{ previous responses}}  \text{. }
\end{align*}

\noindent For the $3$ $\mathrm{XOR}$, and $4$ $\mathrm{XOR}$, games, the outer product for $N$ players take the form,

\begin{align*}
\ket{i_3}  \bigg( \ket{i_1} \bra{i_2} \bigg)   \text{, } \\ \ket{i_3}  \bigg( \ket{i_1} \bra{i_2} \bigg) \bra{i_4} \text{, }
\end{align*}

\noindent respectively.

     \item[$\bullet$] \underline{\textit{Tensor observables for players of the game}}. To define the multiplayer bias, which will be used to characterize exact, and approximate, optimality up to some parameter $\epsilon$ taken to be sufficiently small, define,

     \begin{align*}
         \bigotimes \text{Player tensor observables} \equiv \bigg( \text{Alice's observables} \bigg) \bigotimes \bigg( \text{Bob's observables} \bigg) \\ \bigotimes  \bigg(\text{Cleo's observables} \bigg)   \text{, }
     \end{align*}

     \noindent corresponding to the Hilbert space spanned by the possible set of responses for three players Alice, Bob and Cleo.

      \item[$\bullet$] \underline{\textit{Intertwining operation}}. For tensor products of player observables, error bounds for the two-player $\mathrm{XOR}$ game consist of \textit{interchanging} the order in which the observables that each player forms appear in tensor products, such as the one provided over all player observables above. In error bounds that will follow, denote the intertwining operation, $\widetilde{\cdot}$, where,

\begin{align*}
 \widetilde{\cdot} : M \otimes N \longrightarrow N \otimes M   \text{, }
\end{align*}

\noindent as the permutation operator applied to tensor products $M \otimes N$, where $M$ is a vector with entries from responses of the first player after the referee draws the first question from soem probability distribution of all questions. The above operation is applied under many circumstances, not only for games with more participants but also for games obtained under strong parallel repetition.

         \item[$\bullet$] \underline{\textit{Strong parallel repetition}}. The operation of performing strong parallel repetition, within the exactly, and approximately, optimal framework, is of great interest to further explore and formalize. Under the repetition operation, tensor observables gathered by each player and concatenated together under, potentially, an arbitrary number of games as the referee continues drawing questions from the probability distribution. For any number of players, the strong parallel repetition operation can be straightforwardly extended from two-player settings. In such settings, the action of the strong parallel repetition operation is defined with,

         \begin{align*}
          \text{Strong parallel repetition of Alice's responses to Questions } i \text{ and } j \equiv A_i \wedge A_j   \text{,}
         \end{align*}

         \noindent with the same action being defined for observables gathered by any of the other players.

         \item[$\bullet$] \underline{\textit{$\epsilon$-deviations from optimality}}. Given the existence of a sufficiently small parameter, besides differences in the formulation of error bounds, the bias, and optimal value, satisfying $\epsilon$-approximate optimality, reads,

\begin{align*}
  \big( 1 - \epsilon \big) \beta \big( G \big)  \leq  \underset{\text{Questions}}{\sum} \bra{\text{Optimal Strategy}} \bigg(    \underset{\# \text{ Players }}{\bigotimes}   \text{Tensors of player observables}   \bigg) \\ \times    \ket{\text{Optimal Strategy}}  \leq      \beta \big( G \big)   \text{, }
\end{align*}

\noindent for a game $G$.
         
\end{itemize}

\noindent From the last item introduced above, before taking the supremum over all possible strategies, define,

\begin{align*}
      \beta \big( G , \mathcal{S}  \big)     \equiv \underset{s \in S}{\sum} \underset{t \in T}{\sum} G_{st} \bra{\psi} A_s \otimes B_t \ket{\psi}  \text{, } 
\end{align*}

\noindent as the \textit{success bias}, where the summation runs over all rows and columns $s$ and $t$ of $G$, with the observables in the tensor product taking the form,

\begin{align*}
 A_S \equiv \underset{s \in S}{\bigcup} A_s \equiv \big\{   s  \in S :  A_s \in \big\{ - 1 , + 1 \big\}    \big\}     \text{, } \\    B_T \equiv \underset{t \in T}{\bigcup} B_t \equiv \big\{ t \in T  :  B_t \in \big\{ -1 , + 1 \big\}  \big\}    \text{. }
\end{align*}

\noindent The quantity above is related to the probability of winning the XOR game given $\mathcal{S}$, denoted as $\omega \big( G , \mathcal{S} \big)$, as,

\begin{align*}
       \beta \big( G , \mathcal{S}  \big)  = 2   \omega \big( G , \mathcal{S} \big) - 1      \text{. } 
\end{align*}

\noindent As a supremum over all possible $\mathcal{S}$ for $G$, define,

\begin{align*}
 \beta \big( G \big) \equiv  \underset{\text{Strategies }\mathcal{S}}{\mathrm{sup}} \text{ } \beta \big( G , \mathcal{S} \big) \text{, } 
\end{align*}

\noindent corresponding to the optimal quantum strategy. From the optimal strategy $\beta \big( G \big)$, the notion of approximately optimal strategies can be introduced, in which for some strictly positive $\epsilon$,

\begin{align*}
\big( 1 - \epsilon \big) \beta \big( G \big)   \leq   \beta \big( G , \mathcal{S} \big)        \leq  \beta \big( G \big) \text{. }
\end{align*}

\noindent From each possible combination of questions that can be raised to Alice and then Bob, one can form orthonormal bases $\ket{i}$ and $\ket{ij}$, for the \textit{game matrix}, which are of the form,

\begin{align*}
  G =  \frac{1}{4 {n \choose 2}}   \underset{1\leq i \leq j \leq n}{\sum}   \bigg(    \ket{i}\bra{ij}  +   \ket{j} \bra{ij} + \ket{i} \bra{ji} - \ket{j} \bra{ji}             \bigg)   \text{. } 
\end{align*}

\noindent from which the \textit{optimal success bias} for $G$ takes the form, under the correspondence from the superposition of bra-ket states above,

\begin{align*}
    \ket{i}\bra{ij} \longleftrightarrow A_i B_{ij}       \text{, } \\    \ket{j} \bra{ij}  \longleftrightarrow A_j B_{ij}   \text{, } \\  \ket{i}\bra{ji}  \longleftrightarrow  A_i B_{ji}  \text{, } \\                  - \ket{j} \bra{ji}  
 \longleftrightarrow  - A_j B_{ji} \text{, }
\end{align*}

\noindent from which a summation of quantum states over $i,j$, provides,

\begin{align*}
  \underset{A_i , B_{jk} , \psi}{\mathrm{sup}} \text{ }  \frac{1}{4 {n \choose 2}}   \underset{1\leq i \leq j \leq n}{\sum}  \bra{\psi} \bigg(       A_i B_{ij} +  A_j B_{ij}  +   A_i B_{ji}  -   A_j B_{ji}  \bigg)           \ket{\psi} 
               \text{. } 
\end{align*}

\noindent As an observation regarding the operation of strong parallel repetition, the fact that the $\mathrm{FFL}$ game differs from the $\mathrm{XOR}$ game in the probability of a player responding to one question drawn from the referee's probability distribution relates to the fact that,

\begin{align*}
  \omega_{\mathrm{XOR}  \wedge \mathrm{XOR}} \big( G_{\mathrm{XOR} \wedge  \mathrm{XOR}} \big)  \equiv \omega \big(  \mathrm{XOR} \wedge \mathrm{XOR} \big) \equiv \underset{1\leq j \leq 2}{\prod}   \omega \big( \mathrm{XOR}\big)^{j} \equiv \big( \omega \big( \mathrm{XOR} \big) \big)^2 \\ \equiv \frac{1}{2} \neq \omega \big( \mathrm{XOR}  \big)   \neq \frac{1}{\sqrt{2}} \text{,} \end{align*}

\noindent for the $\mathrm{XOR}$ game, while for the $\mathrm{FFL}$ game,

\begin{align*}
  \omega_{\mathrm{FFL} \wedge  \mathrm{FFL}} \big( G_{\mathrm{FFL}} \big) \equiv \omega_{\mathrm{FFL}} \big( G_{\mathrm{FFL}} \wedge G_{\mathrm{FFL}} \big)   \equiv \omega \big(  \mathrm{FFL}  \wedge \mathrm{FFL} \big) \equiv  \omega \big( \mathrm{FFL}\big) \equiv \frac{2}{3} \neq \bigg( \frac{2}{3} \bigg)^2  \text{.} \end{align*}

 \noindent The $\mathrm{CHSH}\big(2\big)$, $n=2$, strategy, is comprised of the Bell states,

\begin{align*}      
\bigg(  \textbf{I} \otimes \textbf{I} \bigg) \bigg(  \frac{\ket{00} + \ket{11}}{\sqrt{2}}\bigg)  = \frac{\ket{00} + \ket{11}}{\sqrt{2}}   \text{ } \text{ , }     \bigg(     \sigma_x \otimes  \textbf{I} \bigg) \bigg( \frac{\ket{00} + \ket{11}}{\sqrt{2}} \bigg)  = \frac{\ket{10} + \ket{01}}{\sqrt{2}}   \text{, }   \\   \bigg(   \sigma_z    \otimes  \textbf{I} \bigg) \bigg(  \frac{\ket{00} + \ket{11}}{\sqrt{2}} \bigg)  = \frac{\ket{00} - \ket{11}}{\sqrt{2}}       \text{ } \text{ , }    \bigg( \sigma_x \sigma_z     \otimes    \textbf{I}  \bigg)  \bigg( \frac{\ket{00}+\ket{11}}{\sqrt{2}} \bigg)  = \frac{\ket{10}- \ket{01}}{\sqrt{2}}     \text{. } 
\end{align*}

\noindent \textbf{Lemma} \textit{7} (\textit{second error bound}, \textit{6.6}, {[37]}). From previously defined quantities, one has,

\begin{align*}
 \bigg| \bigg|  \bigg( \bigg( \underset{1 \leq i \leq n}{\prod}   A^{j_i}_i \bigg)   \otimes B_{kl} \bigg)  \ket{\psi_{\mathrm{FFL}}}  - \frac{2}{3} \bigg[ \pm \bigg( \mathrm{sign} \big( i , j_1 , \cdots , j_n \big) \bigg[ \text{ } \bigg(     \underset{i = j_k + 1 , \text{ } \mathrm{set} \text{ } j_k + 1 \equiv j_k \oplus 1 }{\underset{1 \leq i \leq n}{\prod}}     A^{j_i}_i    \bigg) \\ + \bigg( \underset{i = j_l + 1 , \text{ } \mathrm{set} \text{ } j_l + 1 \equiv j_l \oplus 1 }{\underset{1 \leq i \leq n}{\prod}}     A^{j_i}_i   \bigg) \text{ } \bigg]   \otimes \textbf{I} \bigg)  \ket{\psi_{\mathrm{FFL}}} \bigg]           \bigg| \bigg|   <    \bigg(       \frac{8200 \sqrt{2} }{27}   \bigg)                   n^2 \sqrt{\epsilon }                   \text{. }
\end{align*}

\noindent Obtaining an inequality of the form above is of interest because of the lack of existence of a duality gap for the $\mathrm{FFL}$ game. That is, because,

\begin{align*}
\mathrm{Classical \text{ } FFL \text{ } value} \equiv   \omega_c \big(  \mathrm{FFL} \big) = \mathrm{Quantum \text{ } FFL \text{ } value} \equiv \omega_q \big( \mathrm{FFL}  \big) = \frac{2}{3}  \text{, } 
\end{align*}

\noindent the primal feasible solution associated to some semidefinite program for the duality gap is always an identically vanishing function. Finally, for manipulating tensor products of operators, such as in the $\mathrm{FFL}$ game, are of the form,

\begin{align*}
 \bigg( A_i \bigg(  \underset{1 \leq i \leq n}{\prod} A^{j_i}_i \bigg) \otimes \textbf{I} \bigg) \ket{\psi_{\mathrm{FFL}}}   \text{, } 
\end{align*}

\noindent which can be placed into correspondence with the operation,

\begin{align*}
    \mathrm{sign} \big( i , j_1 , \cdots , j_n \big) \bigg( \bigg( \underset{1 \leq i \leq n}{\prod} A^{j_i}_i  \bigg) \otimes \textbf{I} \bigg) \ket{\psi_{\mathrm{FFL}}}      \text{, } 
\end{align*}

\noindent we make use of the same sequence of manipulations in other games which are characterized beyond the two-player setting, with:

\begin{itemize}
    \item[$\bullet$] \textit{Switching a tensor product of $A^{j_i}_i$ terms with AB-switches}.

    \item[$\bullet$] \textit{Switching the last observable from the B side to the A side}.

    \item[$\bullet$] \textit{Perform an odd, or even, number of anticommutation swaps to permute $A^{j_i}_i$ to the desired position}.
\end{itemize}

\noindent For the arguments in the next section, denote the identity operator, $\textbf{I}$, raised to a tensor product power, with,

\begin{align*}
  \textbf{I}^{\otimes N} \equiv  \underset{1 \leq z \leq N}{\bigotimes} \textbf{I}_z      \equiv \textbf{I}_1 \bigotimes \cdots \bigotimes \textbf{I}_N \equiv \textbf{I} \bigotimes \cdots \bigotimes \textbf{I}  \text{.}
\end{align*}

\subsection{Main Result}

\noindent The Main Result is captured with the following:

\bigskip

\noindent \textbf{Theorem} \textit{1} (\textit{primal feasible solutions and duality gaps of 3-XOR games}). Suppose that the primal, and dual, semidefinite programs are well posed and have primal feasible solutions. The following statements hold:

\begin{itemize}
\item[$\bullet$] \underline{\textit{Semidefinite program corresponding to the duality gap}}. The duality gap, which captures the difference between classical and quantum values of a game, is captured through the condition,

\begin{align*}
    \bigg[ \underset{1 \leq i \leq m}{\sum} y_{3\mathrm{XOR},i} F_{3\mathrm{XOR},i} - G_{3\mathrm{XOR}} \bigg] \cdot Z_{3\mathrm{XOR}} \geq 0         \text{, }
\end{align*}

\noindent for the 3 XOR game matrix, $G_{3 \mathrm{XOR}}$, primal feasible solution $Z_{3\mathrm{XOR}}$, dual feasible $y_{3\mathrm{XOR},i}$, and symmetric matrices $F_{3\mathrm{XOR},i}$.

\item[$\bullet$] \underline{\textit{Vanishing duality gap}}. The duality gap formulated in the previous item above vanishes,

\begin{align*}
     v_{\mathrm{Prinmal}, 3 \mathrm{XOR}} \equiv v_{\mathrm{Dual}, 3 \mathrm{XOR}}      \text{, }
\end{align*}

\noindent iff,

\begin{align*}
      \bigg[ \underset{1 \leq i \leq m}{\sum} y_{3\mathrm{XOR},i} F_{3\mathrm{XOR},i} - G_{3\mathrm{XOR}} \bigg] \cdot Z_{3\mathrm{XOR}} \equiv  0          \text{. }
\end{align*}

\item[$\bullet$] \underline{\textit{Weak duality gap}}. The weak duality gap,

\begin{align*}
     v_{\mathrm{Prinmal}, 3 \mathrm{XOR}} \leq v_{\mathrm{Dual}, 3 \mathrm{XOR}}      \text{, }
\end{align*}

\noindent iff,

\begin{align*}
      \bigg[ \underset{1 \leq i \leq m}{\sum} y_{3\mathrm{XOR},i} F_{3\mathrm{XOR},i} - G_{3\mathrm{XOR}} \bigg] \cdot Z_{3\mathrm{XOR}} \neq  0          \text{. }
\end{align*}

\item[$\bullet$] \underline{\textit{Duality transformations to semidefinite programs}}. Under well posed constraints, the semidefinite program, for real constants,

\begin{align*}
  c_{3\mathrm{XOR},i}  \text{, }
\end{align*}

\noindent which takes the form,

\begin{align*}
   \underset{\exists c_i \in \textbf{R}:  F_{3 \mathrm{XOR},i} \cdot G_{3\mathrm{XOR}} \equiv c_{3\mathrm{XOR},i}   }{\underset{\forall Z^{\prime} \succcurlyeq 0  \text{, } 1 \leq i \leq 3 , }{\mathrm{sup} }}  \big[ G_{3\mathrm{XOR}}  \cdot Z_{3\mathrm{XOR}}  \big]       \text{, }
\end{align*}

\noindent has a dual semidefinite program, which takes the form,

\begin{align*}
     \underset{\underset{1 \leq i \leq n^3}{\sum} y_{3\mathrm{XOR},i} F_{3\mathrm{XOR},i} \succcurlyeq G_{3\mathrm{XOR}}}{\mathrm{inf}} \big[ \vec{c_{3 \mathrm{XOR}}}\cdot \vec{y_{3\mathrm{XOR}}} \big]      \text{, }
\end{align*}

\noindent for dual feasible $y_{3\mathrm{XOR},i}$ introduced in the first item of the Theorem.

\item[$\bullet$] \underline{\textit{Strong duality}}. Strong duality is said to hold iff $v_{\mathrm{Primal}, 3 \mathrm{XOR}} \equiv v_{\mathrm{Dual}, 3 \mathrm{XOR}}$. 

\end{itemize}

\noindent There are several variants of the result above for different games that will be considered in the forthcoming work. We list the results below, each of which satisfy the same collection of five conditions that have been first provided above for the 3-$\mathrm{XOR}$ game.

\bigskip

\noindent \textbf{Theorem} \textit{2} (\textit{primal feasible solutions and duality gaps of the 4-XOR game}). The same collection of items provided in $\textbf{Theorem}$ \textit{1} for the $3$-$\mathrm{XOR}$ game also hold for the $4$-$\mathrm{XOR}$ game, given the existence of primal feasible solutions, duality gap, and dual semidefinite program.

\bigskip

\noindent \textbf{Theorem} \textit{3} (\textit{primal feasible solutions and duality gaps of the 5-XOR game}). The same collection of items provided in $\textbf{Theorem}$ \textit{1} for the $3$-$\mathrm{XOR}$ game also hold for the $5$-$\mathrm{XOR}$ game, given the existence of primal feasible solutions, duality gap, and dual semidefinite program.

\bigskip

\noindent \textbf{Theorem} \textit{4} (\textit{primal feasible solutions and duality gaps of the N-XOR game}). The same collection of items provided in $\textbf{Theorem}$ \textit{1} for the $3$-$\mathrm{XOR}$ game also hold for the $N$-$\mathrm{XOR}$ game, given the existence of primal feasible solutions, duality gap, and dual semidefinite program.

\bigskip

\noindent \textbf{Theorem} \textit{5} (\textit{primal feasible solutions and duality gaps of strong parallel repetition of XOR games}). The same collection of items provided in $\textbf{Theorem}$ \textit{1} for the $3$ $\mathrm{XOR}$ game also hold for strong parallel repetition of $\mathrm{XOR}$ games, given the existence of primal feasible solutions, duality gap, and dual semidefinite program.

\bigskip

\noindent \textbf{Theorem} \textit{6} (\textit{primal feasible solutions and duality gaps for strong parallel repetition of FFL games}). The same collection of items provided in $\textbf{Theorem}$ \textit{1} for the $3$-$\mathrm{XOR}$ game also hold for strong parallel repetition of $\mathrm{FFL}$ games, given the existence of primal feasible solutions, duality gap, and dual semidefinite program.

\bigskip

\noindent Explicitly, the dual terms $y_i$ which appear in the formulation of each semidefinite program corresponding to each game will be shown to take the form,

\[
y_{3\mathrm{XOR}} \equiv  \left\{\!\begin{array}{ll@{}>{{}}l}
      y_{3\mathrm{XOR},1} \equiv \cdots \equiv y_{3\mathrm{XOR},n} \equiv \omega_{3\mathrm{XOR}}  \bigg( \frac{1}{3! n} \bigg)   \text{, }  \\   y_{3\mathrm{XOR},n+1} \equiv \cdots \equiv    y_{3\mathrm{XOR},n^2}  \equiv \omega_{3\mathrm{XOR}}  \bigg( \frac{1}{3! n ( n-1 ) } \bigg) \text{, } \\   y_{3\mathrm{XOR},n^2+1} \equiv \cdots \equiv    y_{3\mathrm{XOR},n^3}  \equiv \omega_{3\mathrm{XOR}}  \bigg( \frac{1}{3! n ( n-1 ) ( n-2 ) } \bigg)     \text{, } 
      \end{array}\right.
\]

   \[
y_{4\mathrm{XOR}} \equiv  \left\{\!\begin{array}{ll@{}>{{}}l}   
      y_{4\mathrm{XOR},1} \equiv  \cdots \equiv y_{4\mathrm{XOR},n} \equiv \omega_{4\mathrm{XOR}}   \bigg( \frac{1}{4! n} \bigg)    \text{, } \\  y_{4\mathrm{XOR},n+1} \equiv \cdots \equiv y_{4\mathrm{XOR},n^2} \equiv \omega_{4\mathrm{XOR}}  \bigg( \frac{1}{4! n ( n-1 ) } \bigg)     \text{, } \\ y_{4\mathrm{XOR},n^2+1} \equiv \cdots \equiv y_{4\mathrm{XOR}, n^3} \equiv \omega_{4\mathrm{XOR}}    \bigg( \frac{1}{4! n ( n-1 ) ( n -2 )  } \bigg)  \text{, }  \\ y_{4\mathrm{XOR},n^3 + 1} \equiv \cdots \equiv y_{4\mathrm{XOR}, n^4} \equiv \omega_{4\mathrm{XOR}}    \bigg( \frac{1}{4! n ( n-1 ) ( n -2 ) ( n-3 )  } \bigg)   \text{, } 
   \end{array}\right.
\]   
      
      \[
y_{5\mathrm{XOR}} \equiv \left\{\!\begin{array}{ll@{}>{{}}l} y_{5\mathrm{XOR},1} \equiv   \cdots \equiv y_{5\mathrm{XOR},n} \equiv    \omega_{5\mathrm{XOR}} \bigg( \frac{1}{5! n } \bigg)    \text{, } \\ y_{5\mathrm{XOR},n+1}    \equiv   \cdots \equiv y_{5\mathrm{XOR},n^2}    \equiv \omega_{5\mathrm{XOR}} \bigg( \frac{1}{5! n ( n-1 )  } \bigg)     \text{, } \\ y_{5\mathrm{XOR},n^2 +1} \equiv    \cdots \equiv   y_{5\mathrm{XOR},n^3}\omega_{5\mathrm{XOR}} \bigg( \frac{1}{5! n ( n-1 ) ( n-2 ) } \bigg)     \text{, } \\ y_{5\mathrm{XOR},n^3+1 } \equiv   \cdots \equiv  y_{5\mathrm{XOR},n^4 } \equiv   \omega_{5\mathrm{XOR}} \bigg( \frac{1}{5!  n \big( n-1 \big) \big( n-2 \big) \big( n - 3 \big)  } \bigg)    \text{, } \\ y_{5\mathrm{XOR},n^4+1 } \equiv   \cdots \equiv  y_{5\mathrm{XOR},n^5 } \equiv   \omega_{5\mathrm{XOR}} \bigg( \frac{1}{5!  n ( n-1 ) ( n-2 ) ( n-3 ) ( n - 4 )  } \bigg)    \text{, }
\end{array}\right.
\]

\[
 y_{N\mathrm{XOR}} \equiv \left\{\!\begin{array}{ll@{}>{{}}l} y_{N\mathrm{XOR},1} \equiv   \cdots \equiv    y_{N\mathrm{XOR},n} \equiv {\omega_{N\mathrm{XOR}}} \bigg(      \frac{1}{N! n}       \bigg)          \text{, } \\   y_{N\mathrm{XOR},n+1} \equiv   \cdots \equiv    y_{N\mathrm{XOR},n^2} \equiv {\omega_{N\mathrm{XOR}}} \bigg(      \frac{1}{N! n ( n -1 ) }       \bigg)          \text{, }  \\ \vdots \\    y_{N\mathrm{XOR},1} \equiv   \cdots \equiv    y_{N\mathrm{XOR},n} \equiv {\omega_{N\mathrm{XOR}}} \bigg(      \frac{1}{N! n ( n-1 ) \times \cdots \times (  n - (N-1) ) }       \bigg)   \text{. }
\end{array}\right.
\]

\[
 y_{N\mathrm{XOR} \wedge \cdots \wedge N\mathrm{XOR}} \equiv \left\{\!\begin{array}{ll@{}>{{}}l} 
  y_{N\mathrm{XOR} \wedge \cdots \wedge N\mathrm{XOR},1} \equiv   \cdots \equiv    y_{N\mathrm{XOR} \wedge \cdots \wedge N\mathrm{XOR},n} \\  \equiv \big( \omega_{N\mathrm{XOR}} \big)^{\# \text{ of strong parallel repetitions}}   \bigg(      \frac{1}{N! n}       \bigg)          \text{, } \\   y_{N\mathrm{XOR} \wedge \cdots \wedge N\mathrm{XOR},n+1} \equiv   \cdots \equiv    y_{N\mathrm{XOR} \wedge \cdots \wedge N\mathrm{XOR},n^2} \\ \equiv \big( \omega_{N\mathrm{XOR}} \big)^{\# \text{ of strong parallel repetitions}}   \bigg(      \frac{1}{N! n ( n -1 ) }       \bigg)          \text{, }  \\ \vdots \\    y_{N\mathrm{XOR} \wedge \cdots \wedge N\mathrm{XOR},1} \equiv   \cdots \equiv    y_{N\mathrm{XOR} \wedge \cdots \wedge N\mathrm{XOR},n} \\ \equiv \big( \omega_{N\mathrm{XOR}} \big)^{\# \text{ of strong parallel repetitions}}  \bigg(      \frac{1}{N! n ( n-1 ) \times \cdots \times (  n - (N-1) ) }       \bigg)   \text{. }
\end{array}\right.
\]

\[
 y_{\mathrm{FFL} \wedge \mathrm{FFL}} \equiv \left\{\!\begin{array}{ll@{}>{{}}l} 
  y_{\mathrm{FFL} \wedge \mathrm{FFL},1} \equiv   \cdots \equiv    y_{\mathrm{FFL} \wedge \mathrm{FFL},n} \equiv       \frac{1}{3 n}               \text{, } \\     y_{\mathrm{FFL} \wedge \mathrm{FFL},n+1} \equiv   \cdots \equiv    y_{\mathrm{FFL} \wedge \mathrm{FFL},n^2 }  \equiv      \frac{1}{3 n ( n-1 )  }        \text{. }
\end{array}\right.
\]

\subsection{Paper overview}

\noindent Following the overview of game-theoretic objects that have been previously characterized in the two-player setting, in the next section we venture to higher-dimensional player settings. Such settings not only involve systems of error bounds with more degrees of freedom, ultimately arising from the possible ways in which players can interchange observables that they gather for preparing responses, but also give rise to novel variants that have not been previously examined within the exact, and $\epsilon$-approximate, framework. As a result, our contributions in this work characterize game-theoretic settings in which a single player, or a large group of players, can adopt optimal strategies for winning depending upon the answers submitted by the previous player who responds to the referee. Despite the fact that upper bounds for the Frobenius norm, as discussed in the previous subsection, can be similarly obtained, characteristics of primal feasible solutions to semidefinite programs that are introduced for characterizing deviations from optimal strategies are undoubtedly more complicated, as they are functions of a larger collection of responses that each player can formulate when responding to the referee's question. After having explored such notions for 3-$\mathrm{XOR}$ games, and $\mathrm{XOR}$ games with several players, we formalize notions of exact, and approximate, optimality, for strong parallel repetition of the $\mathrm{XOR}$ and $\mathrm{FFL}$ games. Suprisingly, albeit the fact that the $\mathrm{XOR}$ and $\mathrm{FFL}$ game in the two-player setting were initially examined together by the author, in a previous work for expanding arguments in the two-player setting, {[44]}, under the operation of strong parallel repetition these games exhibit strikingly different characteristics. After having discussed such differences in the error bounds, semidefinite programs, and duality gaps, from strong parallel repetition of the $\mathrm{XOR}$ and $\mathrm{FFL}$ games, we discuss the structure of error bounds for performing an arbitrary number of strong parallel repetition operations in the $\mathrm{XOR}$ setting. An overview of all of the main results is provided in Table $*$, while the error bounds for each game are provided in Table $**$. The results listed in Table $***$ generalize those provided in Table $**$, under parallel repetition. Table $****$ lists the captions for Table $1$, Table $2$, Table $3$, Table $4$, Table $5$, Table $6$, Table $7$, Table $8$, Table $9$, Table $10$, Table $11$, Table $12$, Table $13$, Table $14$, Table $15$, Table $16$, Table $17$, Table $18$, Table $19$ and Table $20$.

\section{Beyond the two player $\mathrm{XOR}$ and $\mathrm{FFL}$ games}

\noindent We analyze the 3 $\mathrm{XOR}$ game below, from which we analyze several variants of the two-player setting.

\subsection{Optimal values and biases}

\noindent Introduce,

\begin{align*}
 \underline{\text{3-XOR value}} \equiv   \omega_{3\mathrm{XOR}} \big( G \big) \equiv  \omega \big( 3 \mathrm{XOR} \big) \propto \frac{1}{{n \choose 3}}   \bigg\{ \underset{\ket{\psi_{3\mathrm{XOR}}}}{\mathrm{sup}}  \bra{\psi_{3\mathrm{XOR}}}     \mathcal{P}_{3\mathrm{XOR}}    \ket{\psi_{3\mathrm{XOR}}}           \bigg\}     \text{, }    \\  \\ \underline{\text{3-XOR bias}} \equiv  \beta_{3\mathrm{XOR}} \big( G \big) \equiv \beta \big( 3\mathrm{XOR} \big) \propto \frac{1}{{n \choose 3}}   \bigg\{  \underset{\ket{\psi_{3\mathrm{XOR}}}}{\mathrm{sup}}  G_{3\mathrm{XOR}} \bra{\psi_{3\mathrm{XOR}}}        \mathcal{P}_{3\mathrm{XOR}}           \ket{\psi_{3\mathrm{XOR}}}  \bigg\}             \text{, } \\ \\ \mathcal{P}_{3\mathrm{XOR}}     \equiv \mathscr{P}_{3\mathrm{XOR}} \big( - 1 \big)^{\textbf{1}_{\{\text{Win 3-XOR game} \} }}     +   \mathscr{P}_{3\mathrm{XOR}} \big( - 1 \big)^{\textbf{1}_{\{\text{Lose 3-XOR game} \} }}      \text{, }  \end{align*}
 
 \noindent corresponding to the value, and bias, of the $3$-$\mathrm{XOR}$ game,
 
 \begin{align*} \underline{\text{4-XOR value}}  \equiv  \omega_{4\mathrm{XOR}} \big( G \big) \equiv  \omega \big( 4 \mathrm{XOR} \big)  \propto   \frac{1}{{n \choose 4}}  \bigg\{ \underset{\ket{\psi_{4\mathrm{XOR}}}}{\mathrm{sup}} \bra{\psi_{4\mathrm{XOR}}}   \mathcal{P}_{4\mathrm{XOR}} \ket{\psi_{4\mathrm{XOR}}} \bigg\}   \text{, } \\ \\  \underline{\text{4-XOR bias}} \equiv  \beta_{4\mathrm{XOR}} \big( G \big) \equiv \beta \big( 4\mathrm{XOR} \big) \propto   \frac{1}{{n \choose 4}}  \bigg\{  \underset{\ket{\psi_{4\mathrm{XOR}}}}{\mathrm{sup}} G_{4\mathrm{XOR}} \bra{\psi_{4\mathrm{XOR}}} \mathcal{P}_{4\mathrm{XOR}}    \ket{\psi_{4\mathrm{XOR}}}   \bigg\}   \text{, } \\ \\ \mathcal{P}_{4\mathrm{XOR}} \equiv \mathscr{P}_{4\mathrm{XOR}} \big( - 1 \big)^{\textbf{1}_{\{\text{Win 4-XOR game} \} }}    +   \mathscr{P}_{4\mathrm{XOR}} \big( - 1 \big)^{\textbf{1}_{\{\text{Lose 4-XOR game} \} }} \text{, } \end{align*}

 \noindent corresponding to the value, and bias, of the $4$-$\mathrm{XOR}$ game,

 \begin{align*} \underline{\text{5-XOR value}}  \equiv  \omega_{5\mathrm{XOR}} \big( G \big) \equiv  \omega \big( 5 \mathrm{XOR} \big)  \propto  \frac{1}{{n \choose 5}}  \bigg\{ \underset{\ket{\psi_{5\mathrm{XOR}}}}{\mathrm{sup}}   \bra{\psi_{5\mathrm{XOR}}}    \mathcal{P}_{5\mathrm{XOR}}   \ket{\psi_{5\mathrm{XOR}}} \bigg\}   \text{, } \\ \\  \underline{\text{5-XOR bias}} \equiv \beta_{5\mathrm{XOR}} \big( G \big) \equiv \beta \big( 5\mathrm{XOR} \big) \propto  \frac{1}{{n \choose 5}}  \bigg\{ \underset{\ket{\psi_{5\mathrm{XOR}}}}{\mathrm{sup}}  G_{5\mathrm{XOR}} \bra{\psi_{5\mathrm{XOR}}}  \mathcal{P}_{5\mathrm{XOR}}    \ket{\psi_{5\mathrm{XOR}}}     \bigg\}    \text{, }\\    \\ \mathcal{P}_{5\mathrm{XOR}} \equiv \mathscr{P}_{5\mathrm{XOR}} \big( - 1 \big)^{\textbf{1}_{\{\text{Win 5-XOR game} \} }}    +   \mathscr{P}_{5\mathrm{XOR}} \big( - 1 \big)^{\textbf{1}_{\{\text{Lose 5-XOR game} \} }}          \end{align*}

\begin{tabular}{|l|l|}
\hline\parbox[t]{0.25\textwidth}{
\begin{itemize}
\item \textbf{Lemma} \textit{T-1} \item \textbf{Lemma} \textit{1 -$N$-$\mathrm{XOR}$} 
\item  \textbf{Lemma} \textit{$1- 3-\mathrm{XOR}$}
\item \textbf{Lemma} \textit{2}
\item \textbf{Lemma} $\textit{2}^{*}$
\item \textbf{Lemma} $\textit{2}^{**}$
\item \textbf{Lemma} $\textit{2}^{***}$
\item \textbf{Lemma} $\textit{3}$
\item \textbf{Lemma} \textit{4A}
\item \textbf{Lemma} $\textit{4A}^{*}$
\item \textbf{Lemma} $\textit{4A}^{**}$
\item \textbf{Lemma} $\textit{4A}^{***}$
\item \textbf{Lemma} \textit{5}
\item \textbf{Lemma} $\textit{5}^{*}$
\item \textbf{Lemma} $\textit{5}^{**}$
\item \textbf{Lemma} $\textit{5}^{***}$
\item \textbf{Lemma} \textit{T-2}
\item \textbf{Lemma} \textit{T-3}
\item \textbf{Lemma} \textit{1-XOR rep.}
\item \textbf{Lemma} \textit{FFL rep.}
\item \textbf{Lemma} \textit{5}\textit{B}
\item \textbf{Lemma} $\textit{5}^{*}B$
\item \textbf{Lemma} $\textit{5B}^{*}$
\item \textbf{Lemma} $\textit{5}\textit{B}^{**}$
\item \textbf{Lemma} $\textit{5}^{**}B$
\item \textbf{Lemma} \textit{6}
\item \textbf{Lemma} \textit{7}
\end{itemize}}& 
\parbox[t]{0.73\textwidth}{
\begin{itemize}
\item Positive semidefiniteness of the dual $N$-$\mathrm{XOR}$ objective
\item $N$-$\mathrm{XOR}$ Frobenius norm upper bounds
\item $3$-$\mathrm{XOR}$ Frobenius norm upper bounds
\item $N$ $N$-$\mathrm{XOR}$ identifies from two FFL identities
\item $3$ $3$-$\mathrm{XOR}$ identities
\item $N$ $N$-$\mathrm{XOR}$ identities
\item $N$ $N$-$\mathrm{XOR}$ identities under strong parallel repetition
\item $\epsilon$-approximality
\item Induction on $\epsilon$-approximality result
\item Strong parallel repetition of induction on $\epsilon$-approximality
\item Strong parallel repetition of induction on $N$-$\mathrm{XOR}$ $\epsilon$-approximality
\item Strong parallel repetition of induction on $\mathrm{FFL}$ $\epsilon$-approximality
\item Error bound from permuting indices of player tensors
\item $N$-player error bound
\item $N$-player error bound under strong parallel repetition
\item FFL Error bound under strong parallel repetition
\item Positive semidefiniteness of the dual $3$-$\mathrm{XOR}$ objective
\item Positive semidefiniteness of the dual $4$-$\mathrm{XOR}$ objective \item Strong parallel repetition of Frobenius norm upper bounds
\item  Strong parallel repetition of Frobenius norm upper bounds
\item Error bound under FFL strong parallel repetition
\item Error bound for $2$-$\mathrm{XOR}$ strong parallel repetition
\item Error bound for $3$-$\mathrm{XOR}$ strong parallel repetition
\item Error bound for $N$-$\mathrm{XOR}$ strong parallel repetition
\item $3$-$\mathrm{XOR}$ strong parallel repetition approximality
\item Odd $n$ operator product expansion
\item $\epsilon$-approximality of biases
\end{itemize}}\\ 
\hline
\end{tabular}
\noindent \textit{Table **}. An overview of the results provided for each variant of XOR, and FFL, games, in the second section.

\bigskip  

  \noindent corresponding to the value, and bias, of the $5$-$\mathrm{XOR}$ game,

 \begin{align*} \underline{\text{N-XOR value}}  \equiv   \omega_{N\mathrm{XOR}} \big( G \big) \equiv  \omega \big( N \mathrm{XOR} \big)   \propto  \frac{1}{{n \choose N}} \bigg\{ \underset{\ket{\psi_{N\mathrm{XOR}}}}{\mathrm{sup}}  \bra{\psi_{N\mathrm{XOR}}}      \mathcal{P}_{N \mathrm{XOR}} \ket{\psi_{N\mathrm{XOR}}}      \bigg\}    \text{, } \\ \\  \underline{\text{N-XOR bias}} \equiv  \beta_{N\mathrm{XOR}} \big( G \big) \equiv \beta \big( N\mathrm{XOR} \big)   \propto  \frac{1}{{n \choose N}} \bigg\{ \underset{\ket{\psi_{N\mathrm{XOR}}}}{\mathrm{sup}}  G_{N\mathrm{XOR}}\bra{\psi_{N\mathrm{XOR}}}            \mathcal{P}_{N \mathrm{XOR}} \ket{\psi_{N\mathrm{XOR}}}       \bigg\}   \text{, }    \\ \\ \mathcal{P}_{N \mathrm{XOR}} \equiv \mathscr{P}_{N\mathrm{XOR}}\big( - 1 \big)^{\textbf{1}_{\{\text{Win N-XOR game} \} }}    +   \mathscr{P}_{N\mathrm{XOR}} \big( - 1 \big)^{\textbf{1}_{\{\text{Lose N-XOR game} \} }}  
\end{align*}

 \noindent corresponding to the value, and bias, of the $N$-$\mathrm{XOR}$ game.

\noindent corresponding to the optimal values, and biases, of the $3$-XOR, $4$-XOR, $5$-XOR, and $N$-XOR, games, with games matrices, or tensors, $G \equiv \pi \big( G \big) V \big( G \big)$, where $\pi$, and $V$, respectively denote the referree's probability distribution for questions, and scoring function, respectively. As in the 2-player setting, the game tensors for any of the games $G$ introduced above can be normalized in such a manner insofar that the summation over all of the entries equals $1$. The indicator functions appearing as powers of $-1$ in each constrained optimization problem stated above for the values, and biases, determine whether the responses provided from each player constitute a winning, or losing, strategy.

\noindent In the following subsections, we explicitly provide permutations on tensor of players observables which constitute each optimal value, and bias, defined above, in addition to how the combinatorial space of permutations on player observables impacts error bounds, along with various means of generalizing error bounds. An overview of each such result in \textit{2} is provided in the Table below.

\begin{tabular}{|l|l|}
\hline\parbox[t]{0.25\textwidth}{
\begin{itemize}
\item \textbf{Theorem} $\textit{4}^{*}$
\item \textbf{Theorem} $\textit{5}^{*}$
\item \textbf{Theorem} $\textit{6}^{*}$
\item \textbf{Theorem} $\textit{7}^{*}$
\item \textbf{Theorem} $\textit{8}^{*}$
\item \textbf{Lemma} \textit{T-4}
\item \textbf{Lemma} \textit{T-5}
\item \textbf{Lemma} \textit{T-6}
\item \textbf{Lemma} \textit{T-7}
\item \textbf{Lemma} \textit{T-8}
\end{itemize}}& 
\parbox[t]{0.67\textwidth}{
\begin{itemize}
\item $3$-$\mathrm{XOR}$ strong parallel repetition error bounds
\item $4$-$\mathrm{XOR}$ strong parallel repetition error bounds
\item $5$-$\mathrm{XOR}$ strong parallel repetition error bounds
\item $N$-$\mathrm{XOR}$ strong parallel repetition error bounds
\item FFL strong parallel repetition error bounds
\item $3$-$\mathrm{XOR}$ strong parallel repetition postive semidefiniteness
\item $4$-$\mathrm{XOR}$ strong parallel repetition postive semidefiniteness
\item  $5$-$\mathrm{XOR}$ strong parallel repetition postive semidefiniteness
\item $N$-$\mathrm{XOR}$ strong parallel repetition postive semidefiniteness
\item FFL strong parallel repetition postive semidefiniteness
\end{itemize}}\\
\hline
\end{tabular}
\noindent \textit{Table ***}. At the conclusion of the section, we discuss how several computations used in obtaining the collection of results from the previous table can be generalized, with the results in the table above.

\bigskip

\subsection{$3$ $\mathrm{XOR}$ game}

\noindent In the presence of even an additional player, error bounds for the $3$-$\mathrm{XOR}$ game differ from those of the $2$-$\mathrm{XOR}$ game. In the two-player setting, error bounds are rigid, in the sense that tensor products of observables from the first two players can be interchanged with one another. However, in even the three player setting, the error bounds obtained by applying a transformation, namely forming a superposition from each possible response of the second player and normalizing the superposition with $\sqrt{2}$, take on more complicated forms in higher-dimensional player settings. Given the optimal strategy for a $2$-$\mathrm{XOR}$ game, $\ket{\psi_{2\mathrm{XOR}}}$, as characterized in {[37]},

\begin{align*}
   \bigg( A_i \otimes \textbf{I} \bigg) \ket{\psi_{2\mathrm{XOR}}} = \bigg( \textbf{I} \otimes \frac{B_{ij} + B_{ji}}{\sqrt{2}} \bigg) \ket{\psi_{2\mathrm{XOR}}}     \text{, }  \\     \bigg( A_j \otimes \textbf{I} \bigg) \ket{\psi_{2\mathrm{XOR}}} = \bigg( \textbf{I} \otimes \frac{B_{ij} - B_{ji}}{\sqrt{2}} \bigg) \ket{\psi_{2\mathrm{XOR}}}     \text{. }
\end{align*}

\noindent The $3$-$\mathrm{XOR}$ game is an immediately accessible extension of the two-player setting. In an identical way that each $B_{jk}$, namely the tensor observable for the second player in the $\mathrm{XOR}$, or $\mathrm{FFL}$, games admits a decomposition in terms of $A_j$ and $A_k$, in which, for $1 \leq l \leq j \neq k \in  \big\{ 1 , \cdots , n \big\}$,

\begin{align*}
  B^{(l)}_{jk}    \equiv  \bigg( \frac{1}{\sqrt{2}} A^{(l)}_j  +      \frac{1}{\sqrt{2}} A^{(l)}_k   \bigg)^{\mathrm{T}} = \frac{1}{\sqrt{2}} \bigg( A^{(l)}_j + A^{(l)}_k  
 \bigg)^{\mathrm{T}}  \equiv \frac{1}{\sqrt{2}}  \bigg( \big( A^{(l)}_j\big)^{\mathrm{T}} +  \big( A^{(l)}_k \big)^{\mathrm{T}} \bigg)  \text{, } \\    B^{(l)}_{kj}   \equiv  \bigg(   \frac{1}{\sqrt{2}}   A^{(l)}_j   - \frac{1}{\sqrt{2}}    A^{(l)}_k          \bigg)^{\mathrm{T}} = \frac{1}{\sqrt{2}} \bigg(    A^{(l)}_j   -   A^{(l)}_k     \bigg)^{\mathrm{T}}  \equiv \frac{1}{\sqrt{2}}  \bigg( \big(   A^{(l)}_j 
 \big)^{\mathrm{T}} -   \big( A^{(l)}_k  \big)^{\mathrm{T}}   \bigg)  \text{, } 
\end{align*}

\noindent  any other tensor observables for other players in the game can be expressed with, the following union over the set of possible questions,

\begin{align*}
   \underset{i \in \mathcal{Q}_2}{\bigcup}  B^{(l_i)}_{j_1 \cdots j_N}    \equiv  \underset{i \in \mathcal{Q}_2}{\bigcup}   \bigg( \frac{1}{\sqrt{N}} A^{(l_1)}_{j_1}  +   \cdots +    \frac{1}{\sqrt{N}} A^{(l_N)}_{j_N}   \bigg)^{\mathrm{T}} = \frac{1}{\sqrt{N}} \underset{i \in \mathcal{Q}_2}{\bigcup}  \bigg( A^{(l)}_j + \cdots + A^{(l)}_k  
 \bigg)^{\mathrm{T}}  \\ \equiv \frac{1}{\sqrt{N}}  \underset{i \in \mathcal{Q}_2}{\bigcup}   \bigg( \big( A^{(l_1)}_{j_1}\big)^{\mathrm{T}}  + \cdots +    \big( A^{(l_N)}_{j_N} \big)^{\mathrm{T}} \bigg)   \text{, }  
\end{align*}

\noindent for any $i$ over the question set $\mathcal{Q}_2$ for the second player. For three players, Alice, Bob, and Cleo, are participating and responding to questions drawn from the referee's probability distribution, from the optimal strategy,

\begin{align*}
  \ket{\psi_{3\mathrm{XOR}}}  \text{, }
\end{align*}

\noindent the Hilbert space corresponding to all of the possible responses of each player to a referee's question is the set of linear combinations,

\begin{align*}
 \underset{\mathcal{S} \times \mathcal{S} \times \mathcal{S}}{\mathrm{span}}    \big\{ \mathscr{A}, \mathscr{B}, \mathscr{C} : \mathscr{A} \big(    \mathcal{S} \big) \times  \mathscr{B} \big( \mathcal{S} \big) \times \mathscr{C} \big( \mathcal{S} \big)        \big\}       \text{, }
\end{align*}

\noindent for,

\begin{align*}
    \mathscr{A} \big( \mathcal{S} \big) \equiv \underset{\text{Questions }i}{\bigcup}    \mathscr{A} \big( \mathcal{S}_i \big)   \equiv \underset{\text{Questions }i}{\bigcup} A_i 
 \equiv   \underset{\text{Questions }i}{\bigcup}  \big\{ \text{Alice's strategy to answer question } i \big\}     \text{, } \end{align*}

\begin{tabular}{|l|l|}
\hline\parbox[t]{0.25\textwidth}{
\begin{itemize}
\item Table $1$
\item Table $2$
\item Table $3$
\item Table $4$
\item Table $5$
\item Table $6$
\item Table $7$
\item Table $8$
\item Table $9$
\item Table $10$
\item Table $11$
\item Table $12$
\item Table $13$
\item Table $14$
\item Table $15$
\item Table $16$
\item Table $17$
\item Table $18$
\item Table $19$
\item Table $20$
\end{itemize}}& 
\parbox[t]{0.73\textwidth}{
\begin{itemize}
\item Permutations of Player Observables for the 3-XOR game error bound
\item Permutations of Player Observables for the 3-XOR game error bound
\item Permutations of Player Observables for the 4-XOR game error bound
\item Permutations of Player Observables for the 4-XOR game error bound
\item Permutations of Player Observables for the 4-XOR game error bound
\item Permutations of Player Observables for the 4-XOR game error bound
\item Permutations of Player Observables for the 4-XOR game error bound
\item Components of the Error bound for the N-XOR game
\item Permutations of Player Observables for the 5-XOR game error bound
\item Permutations of Player Observables for the 5-XOR game error bound
\item Representative interchange observations in the 4-XOR game
\item Error bound of Player Observables for the 4-XOR game
\item Error bound of Player Observables for the 4-XOR game
\item Error bound of Player Observables for the 4-XOR game
\item Error bound of Player Observables for the 4-XOR game
\item Permutations of Player Observables for the 5-XOR game error bound
\item Permutations of Player Observables for the 5-XOR game error bound
\item Permutations of Player Observables for the 5-XOR game error bound
\item Permutations of Player Observables for the 5-XOR game error bound
\item Permutations of Player Observables for the 5-XOR game error bound
\end{itemize}}\\
\hline
\end{tabular}
\noindent \textit{Table ****}. An overview of the tables in the second section, and the Appendix.

\bigskip
 
 \begin{align*} \mathscr{B} \big( \mathcal{S} \big) \equiv \underset{\text{Questions }j}{\bigcup}    \mathscr{B} \big( \mathcal{S}_j \big) \equiv \underset{\text{Questions }j}{\bigcup}  B_{ij}
 \equiv   \underset{\text{Questions }j}{\bigcup}  \big\{ \text{Bob's strategy to} \\ \text{answer question }  j \text{ given Alice's response to question } i \big\}  \text{, } \\ \\  \mathscr{C} \big( \mathcal{S} \big) \equiv \underset{\text{Questions }k}{\bigcup}    \mathscr{C} \big( \mathcal{S}_k \big)  \equiv \underset{\text{Questions }k}{\bigcup}    \mathscr{C} \big( \mathcal{S}_k \big)  \equiv   \underset{\text{Questions }k}{\bigcup}  \big\{ \text{Cleo's strategy} \\ \text{to answer question }  k  \text{ given} \text{  Alice's response to question } i, \text{ and} \\ \text{ Bob's response to question } j \big\} \text{. }
\end{align*}

\noindent Tensor products of the form,

\begin{align*}
    A_i \otimes B_{ij} \otimes C_{ijk} \text{, }
\end{align*}

\noindent under the projection,

\begin{align*}
 \textbf{T}^3 \equiv \underset{\text{Subspaces}}{\bigcup} \big\{ \text{Three-dimensional tensors} \big\} \longrightarrow   \textbf{T} \otimes \textbf{T} \otimes \textbf{T} \longrightarrow \underset{1 \leq i \leq 3}{\bigotimes} \textbf{T}_i             \text{, }
\end{align*}

\noindent for,

\begin{align*}
\textbf{T}_i \equiv \underset{\text{Subspaces}}{\bigcup} \big\{\text{One-dimensional tensors} \big\} \subsetneq \underset{\text{N-dimensional subspaces}}{\bigcup} \big\{\text{Tensors} \big\}    \text{. }
\end{align*}

\noindent Introduce the collection of permutations,

\begin{align*}
 \underline{\mathscr{P}_{3\mathrm{XOR}}} \equiv   \underline{\mathscr{P}_1} \cup \underline{\mathscr{P}_2} \cup \underline{\mathscr{P}_3} \cup \underline{\mathscr{P}_{1,2}} \cup \underline{\mathscr{P}_{2,3}}  \cup \underline{\mathscr{P}_{1,3}}   \cup \underline{\mathscr{P}_{1,2,3}} \text{, }
\end{align*}

\noindent which are enumerated in the table below.

\begin{table}[ht]
\caption{Permutations of Player Observables for the 3-XOR game error bound, $(\textbf{EB}- 3 \mathrm{XOR})$} 
\centering 
\begin{tabular}{c c c c} 
\hline\hline 
Player & Tensor Product Representation & Permutation Superposition  \\ [0.5ex] 
\hline    1 & $\underset{\sigma} {\mathlarger{\sum}} \bigg(  A_{\sigma ( i ) } \otimes B_{ij} \otimes C_{ijk}  \bigg)$  & $ A_i \otimes B_{ij} \otimes C_{ijk}  +  A_j \otimes B_{ji} \otimes C_{jik} +     A_k \otimes B_{ki} \otimes C_{kij} +  A_k  $ \\  & & $  \otimes B_{ij} \otimes C_{ijk}+   A_k \otimes B_{ij} \otimes C_{ijk} +   A_k \otimes B_{ij} \otimes C_{ijk} $ \\  \\    2 & $ \underset{\sigma} {\mathlarger{\sum}} \bigg(  A_{i } \otimes B_{\sigma (ij)} \otimes C_{ijk}  \bigg)$  & $  A_i \otimes B_{ik}  \otimes C_{ijk}  +  A_j  \otimes B_{jk} \otimes C_{ijk} +     A_i \otimes B_{ki} \otimes C_{ijk}$ \\ & &   $ + A_i \otimes B_{ik} \otimes C_{ijk}  +   A_k \otimes B_{ik} \otimes C_{ijk} + A_k \otimes B_{ki} \otimes C_{ijk}$ \\ \\   3 & $\underset{\sigma}{\mathlarger{\sum }} \bigg(  A_i \otimes B_{ij} \otimes C_{\sigma ( ijk)}     \bigg)  $  &   $   A_i \otimes B_{ij} \otimes C_{jik}  + A_i \otimes B_{ij} \otimes C_{jki} + A_i \otimes B_{ij} \otimes C_{kij}  $       \\ & & $+ A_i \otimes B_{ij} \otimes C_{kji}  + A_i \otimes B_{ij} \otimes C_{ijk} + A_i \otimes B_{ij} \otimes C_{ikj } $    \\  \\    1,2 &     $\underset{\sigma,\sigma^{\prime}}{\mathlarger{\sum}} \bigg( A_{\sigma(i)} \otimes B_{\sigma^{\prime} (ij)} \otimes C_{ijk} \bigg)$         &   $     A_k \otimes B_{ij} \otimes C_{ijk}  +            A_k \otimes B_{ik} \otimes C_{ijk}    +      A_k \otimes B_{il} \otimes C_{ijk}   +       A_k \otimes B_{ji }    $   \\ & &   $  \otimes C_{ijk} +         A_j \otimes B_{j k }      \otimes C_{ijk}          +       A_j \otimes B_{ij} \otimes C_{ijk}  +            A_j \otimes B_{ik} \otimes C_{ijk}    +                               $  \\    & &   $  +  A_j \otimes B_{il} \otimes C_{ijk}     +              A_i \otimes B_{j k }      \otimes C_{ijk}          +       A_i   $ \\ & & $ \otimes C_{ijk}  +            A_k \otimes B_{ij} \otimes C_{ijk}     +            A_k \otimes B_{ik} \otimes C_{ijk}    +      A_k \otimes B_{il} \otimes C_{ijk}   +       A_k   $  \\    & &   $  \otimes B_{ji }                  \otimes C_{ijk} +        A_i   \otimes B_{ik} \otimes C_{ijk}        +  A_i \otimes B_{il} \otimes C_{ijk}     +        A_i   $ \\ & & $ \otimes B_{j k }      \otimes C_{ijk}          +         A_i   \otimes B_{ik} \otimes C_{ikj}        +  A_i \otimes B_{il} \otimes C_{ikj}     +        A_i                                              $  \\ & & $ \otimes B_{j k }      \otimes C_{ikj}   +   A_i \otimes B_{il} \otimes C_{ikj}     +        A_i   \otimes B_{j k }       \otimes C_{ikj}             $ \\      & & $ +  A_i \otimes B_{j k }      \otimes C_{jik}   +   A_i \otimes B_{il} \otimes C_{jik}     +        A_i   \otimes B_{j k }       \otimes C_{jik}                      $    \\ & & $+  A_i \otimes B_{j k }      \otimes C_{jki}   +   A_i \otimes B_{il} \otimes C_{jki}     +        A_i   \otimes B_{j k }       \otimes C_{jki}     $ \\  \\    2,3 &     $\underset{\sigma,\sigma^{\prime},\sigma^{\prime\prime}}{\mathlarger{\sum}} \bigg(  A_{ i } \otimes B_{\sigma^{\prime}(ij)}   \otimes C_{\sigma^{\prime\prime}(ijk)}  \bigg)$          &     $ A_i \otimes B_{ij} \otimes C_{ijk} + A_i \otimes B_{ik} \otimes C_{ijk} + A_i \otimes B_{il} \otimes C_{ijk}  $    \\  & & $ +    A_i \otimes B_{ij} \otimes C_{ikj} + A_i \otimes B_{ik} \otimes C_{ijk} + A_i \otimes B_{il} \otimes C_{ikj}          $ \\  & & $ + A_i \otimes B_{ij} \otimes C_{ikl} + A_i \otimes B_{ik} \otimes C_{ikl} + A_i \otimes B_{il} \otimes C_{ikl} $ \\ & $\vdots$ &  \\  [1ex] 
\hline 
\end{tabular}
\label{table:nonlin} 
\end{table}

\begin{table}[ht]
\caption{Permutations of Player Observables for the 3-XOR game error bound, $(\textbf{EB}- 3 \mathrm{XOR})$ Continued} 
\centering 
\begin{tabular}{c c c c} 
\hline\hline 
 Player & Tensor Product Representation & Permutation Superposition  \\ [0.5ex] 
\hline 1,2,3 &    $\underset{\sigma,\sigma^{\prime},\sigma^{\prime\prime}}{\mathlarger{\sum}} \bigg( A_{\sigma^{\prime\prime} ( i)}  \otimes B_{\sigma( i j) }  \otimes C_{\sigma^{\prime}(ijk)} \bigg)$          &  $ + A_i \otimes B_{ij} \otimes C_{jki} + A_i \otimes B_{ik} \otimes C_{jkl} + A_i \otimes B_{il} \otimes C_{jkl} + A_i \otimes $           \\   & & $  B_{ij} \otimes C_{ikl} + A_i \otimes B_{ik} \otimes C_{ikl} + A_i \otimes B_{il}  $ \\  & &  $ \otimes C_{ikl} + A_i \otimes B_{ij} \otimes C_{lij} + A_i \otimes B_{ik} \otimes C_{lik} + A_i \otimes B_{il} \otimes C_{ljk} $ \\   & &  $ + A_i \otimes B_{ij} \otimes C_{lij} + A_i \otimes B_{ik} \otimes C_{lik} + A_i  $ \\          & &   $   \otimes B_{il} \otimes C_{ljk}   + A_i \otimes B_{ij} \otimes C_{ijk} +     A_j \otimes B_{ij} \otimes    C_{ijk} +        A_k \otimes B_{ij}  $ \\ & &  $  \otimes C_{ijk}  +    A_l \otimes B_{ij}  \otimes C_{ijk} +   A_i \otimes B_{ik}  \otimes C_{ijk} +  A_i \otimes B_{il}  \otimes C_{ijk}  + A_i          $   \\ & & $ \otimes B_{ki}  \otimes C_{ijk}    +   A_j \otimes B_{ik}  \otimes C_{ijk} +  A_j \otimes B_{il}  \otimes C_{ijk}  + A_j \otimes B_{ki}  \otimes C_{ijk}           $  \\ & &   $ + A_k \otimes B_{ik}  \otimes C_{ijk} +  A_k \otimes B_{il}  \otimes C_{ijk}  + A_k \otimes B_{ki}  \otimes C_{ijk}           +   A_l \otimes B_{ik}      $       \\  & & $ \otimes C_{ijk} +  A_l \otimes B_{il}  \otimes C_{ijk}  + A_l \otimes B_{ki}  \otimes C_{ijk}                     +    A_i \otimes B_{ik}   $    \\   & &  $ \otimes C_{ikj} +  A_i \otimes B_{il}  \otimes C_{ikl}  + A_i \otimes B_{ki}  \otimes C_{kij}                    +       A_i \otimes B_{ik}     $   \\  & & $ \otimes C_{ikj} +  A_i \otimes B_{il}  \otimes C_{ikl}  + A_i \otimes B_{ki}  \otimes C_{kij}                       +       A_i \otimes B_{ik}       $ \\ & & $ \otimes C_{kil} +  A_i \otimes B_{il}  \otimes C_{lij}  + A_i \otimes B_{ki}  \otimes C_{lij}                +       A_j \otimes B_{ik}             $ \\ & & $   \otimes C_{kil} +  A_j \otimes B_{il}  \otimes C_{lij}  + A_j \otimes B_{ki}  \otimes C_{lij}     +      A_l \otimes B_{ik}                                $ \\ & & $  \otimes C_{kil} +  A_l \otimes B_{il}  \otimes C_{lij}  + A_l \otimes B_{ki}  \otimes C_{lij}                                   $  \\  [1ex] 
\hline 
\end{tabular}
\label{table:nonlin} 
\end{table}

           \bigskip

\noindent With the collection of permutations above, at optimality, the quantum state corresponding to the best strategy, collectively, of all three participants, can be expressed through the constrained optimization problem,

\begin{align*}
 \underset{\ket{\psi_{3 \mathrm{XOR}}}}{\underset{A,B,C}{\mathrm{sup}}}   \bra{\psi_{3\mathrm{XOR}}}              \underline{\mathscr{P}_{3\mathrm{XOR}}}                                                                          \ket{\psi_{3\mathrm{XOR}}}                      \text{. }
\end{align*}

\noindent In the presence of the additional degree of freedom from the possible responses for the third player, the game tensor for the $3$-$\mathrm{XOR}$ game arises from the following combinatorial superposition of responses from each player,

\begin{align*}
    G_{3 \mathrm{XOR}} \approx \frac{1}{{n\choose 3}}                    \underset{i \in \mathcal{Q}_1, j \in \mathcal{Q}_2, k \in \mathcal{Q}_3}{\sum}                  \bigg[   \ket{i} \bigg(   \bra{ij} \bra{ijk}  \bigg)     +    \ket{j} \bigg( \bra{ji} \bra{ijk} \bigg) + \ket{k} \bigg(   \bra{ki} \bra{kij}      \bigg)   +     \ket{i} \\ \times \bigg( \bra{ik} \bra{ijk} \bigg)   +  \ket{k} \bigg(   \bra{ik} \bra{ijk}     \bigg) + \ket{k} \bigg( \bra{ki} \bra{ijk} \bigg) +       \ket{k} \bigg( \bra{ij} \bra{ijk} \bigg) + \ket{k} \\ \times  \bigg( \bra{ji } \bra{ijk} \bigg)       +             \bra{k} \bigg( \bra{jk} \bra{ijk} \bigg) + \ket{k} \bigg( \bra{kj} \bra{ijk} \bigg)       + \ket{j} \bigg( \bra{jk} \bra{ijk} \bigg) +    \ket{i} \\ \times \bigg( \bra{ki} \bra{ijk} \bigg)  + \ket{i} \bigg( \bra{ji} \bra{ijk} \bigg)  +  \ket{i} \bigg( \bra{ij} \bra{jik} \bigg) + \ket{j} \bigg( \bra{ji} \bra{ijk} \bigg)  +        \ket{k} \\ \times  \bigg( \bra{ki} \bra{ikj} \bigg) + \ket{i} \bigg( \bra{ik} \bra{kij} \bigg)  +  \ket{j} \bigg( \bra{jk} \bra{kij}        \bigg)    + \ket{k} \bigg( \bra{kj} \bra{jki} \bigg) \\  + \ket{k} \bigg( \bra{ki} \bra{kij} \bigg)                   + \text{Higher order permutations}             \bigg]               
\end{align*}

\noindent The bias as the supremum,

\begin{align*}
    \beta_{3\mathrm{XOR}} \big( G_{3\mathrm{XOR}} \big) \equiv \underset{\text{Strategies }\mathcal{S}}{\mathrm{sup}} \big\{ \beta_{3\mathrm{XOR}} \big( G_{3\mathrm{XOR}}, \mathcal{S} \big) \big\} 
\end{align*}

\noindent given the combinatorial normalization of $\{n \choose 3\}$ in the representation above for the game tensor, satisfies,

\begin{align*}
   \big( 1 - \epsilon_{3\mathrm{XOR}} \big)  \beta_{3\mathrm{XOR}} \big( G_{3\mathrm{XOR}} \big) \\   \leq     \underset{k \in \mathcal{Q}_3}{\sum} \bigg[ \underset{j \in \mathcal{Q}_2}{\sum} \bigg[ \underset{i \in \mathcal{Q}_1}{\sum} \bra{\psi_{3\mathrm{XOR}}}          \big(  A_i \otimes B_{ij} \otimes C_{ijk} \big)     \ket{\psi_{3\mathrm{XOR}}}                         \bigg] \bigg] \\ \equiv   \underset{i \in \mathcal{Q}_1, j \in \mathcal{Q}_2, k \in \mathcal{Q}_3}{\sum} \bra{\psi_{3\mathrm{XOR}}}          \big(  A_i \otimes B_{ij} \otimes C_{ijk} \big)     \ket{\psi_{3\mathrm{XOR}}}                           \\ \leq  \beta_{3\mathrm{XOR}} \big( G_{3\mathrm{XOR}} \big)   \text{. }
\end{align*}

\noindent The 3-$\mathrm{XOR}$ Schmidt basis, for some $s^{\prime}$ sufficiently large,

\begin{align*}
\underset{1 \leq s \leq s^{\prime} 2^{\lfloor \frac{m}{2} \rfloor}     }{\sum}       \sqrt{\lambda_s} \big( \ket{u_i} \otimes \ket{v_i} \otimes \ket{w_i} \big)    \text{,} 
\end{align*}

\noindent is related to the number of blocks in the matrix representation for which the eigenvalues are equal, where $ s^{\prime} 2^{\lfloor \frac{m}{2} \rfloor}  $ equals,

\begin{align*}
   \underset{1 \leq i \leq n}{\bigcup}        \# \big\{     \lambda_i \in \textbf{C}  : \lambda_i \equiv \cdots \equiv   \lambda_{(i+1) 2^{\lfloor \frac{m}{2} \rfloor}}     \big\}       \text{, }
\end{align*}

\noindent Minors of the representation $A$, are denoted with,

\begin{align*}
 A_i \equiv  A \bigg|_{i \text{ th minor}}         \equiv \underset{1 \leq i \leq (s+1) 2^{\lfloor \frac{m}{2}\rfloor}}{\mathrm{span}}   \big\{ \lambda_i \in \textbf{C} : \lambda_i \equiv \cdots \equiv \lambda_{(s+1) 2^{\lfloor\frac{m}{2}} \rfloor}  \big\}         \text{. }
\end{align*}

\noindent Primal feasible solutions $Z \equiv Z_{2\mathrm{XOR}}$ to semidefinite programs for the 2$\mathrm{XOR}$, and $\mathrm{FFL}$, games can also be formulated with the 3$\mathrm{XOR}$ game. In place of two-player game matrices, primal feasible solutions $Z^{\prime} \equiv Z_{3\mathrm{XOR}}$ three-player objects, which in the case of the 3-$\mathrm{XOR}$ $G$, are of the form,

\begin{align*}
      \underset{\exists c_i \in \textbf{R}:  F_i \cdot G \equiv c_i   }{\underset{\forall Z^{\prime} \succcurlyeq 0  \text{, } 1 \leq i \leq 3 , }{\mathrm{sup} }} \big[  G Z^{\prime} \big]   \text{.}
\end{align*}

\noindent Given the primal feasible solution from the constrained optimization procedure above over the number of players in the game, $i$, there exists another semidefinite program,

\begin{align*}
     \underset{\exists c_i \in \textbf{R}:  F_i \cdot G \equiv c_i   }{\underset{\forall Z^{\prime} \succcurlyeq 0  \text{, } 1 \leq i \leq 3 , }{\mathrm{sup} }} \bigg[  {\sum}   \big( y_i F_i - G \big) Z^{\prime}  \bigg]   \text{, }
\end{align*}

\noindent corresponding to the duality gap between $y_i F_i$, and $G$, with the primal feasible 3$\mathrm{XOR}$ solution. Primal feasible solutions obtained by semidefinite programs, such as those above, have previously been characterized by the author, {[44]}, in the simpler 2-player setting, for $\mathrm{XOR}^{*}$ and $\mathrm{FFL}$ games, by  extending the construction of error bounds and intertwining operations provided in {[37]}. Equipped with a restriction of feasible $F_i$ of $F$, under the assumption that there exists a primal feasible solution $Z^{\prime}$ that can be approximated with polynomial runtime, semidefinite programs for approximating the primal feasible solution, and duality gap, are well posed.

For the 2-$\mathrm{XOR}$ game, intertwining operations play a fundamental role in transforming representations of responses of one player to those of another player. In three-player settings, the accompanying operation, $T^{\prime} \equiv T_{3\mathrm{XOR}}$, has the following collection of actions,

\begin{align*}
 \bigg| \bigg| \bigg[  \bigg( T^{\prime} \otimes B_{ij} \otimes C_{ijk} \bigg) - \frac{1}{\sqrt{2}}              \bigg( \bigg( \widetilde{B_{ij}} \otimes T^{\prime}  \otimes C_{ijk}  \bigg)+ \bigg( T^{\prime} \otimes  \widetilde{C_{ijk}} \otimes B_{ij} \bigg) \bigg)  \bigg] \ket{\psi_{3\mathrm{XOR}}}   \bigg| \bigg|   \text{, } \\ \\   \bigg| \bigg| \bigg[  \bigg( A_i \otimes  T^{\prime}  \otimes C_{ijk} \bigg) - \frac{1}{\sqrt{2}}              \bigg( \bigg( T^{\prime} \otimes \widetilde{A_i }  \otimes C_{ijk}  \bigg)+ \bigg( \widetilde{C_{ijk}} \otimes  T^{\prime}   \otimes A_i \bigg) \bigg)  \bigg] \ket{\psi_{3\mathrm{XOR}}}   \bigg| \bigg|            \text{, } \end{align*}

 \begin{align*} \bigg| \bigg| \bigg[  \bigg( A_i \otimes  B_{ij} \otimes T^{\prime} \bigg) - \frac{1}{\sqrt{2}}              \bigg( \bigg( T^{\prime} \otimes \widetilde{A_i }  \otimes B_{ij}  \bigg)+ \bigg( \widetilde{B_{ij}} \otimes  T^{\prime}   \otimes A_i  \bigg) \bigg)  \bigg] \ket{\psi_{3\mathrm{XOR}}}   \bigg| \bigg|           \text{, }
\end{align*}

\noindent where the linear operator is a mapping of the form,

\begin{align*}
      T^{3 \text{ } \mathrm{XOR}} :  \textbf{C}^{3 \lceil \frac{n}{3} \rceil } \otimes      \textbf{C}^{3 \lceil \frac{n}{3} \rceil }   \otimes  \textbf{C}^{3 \lceil \frac{n}{3} \rceil }      \longrightarrow        \textbf{C}^{d_A} \otimes \textbf{C}^{d_B} \otimes \textbf{C}^{d_C}    \text{,  }
\end{align*}

\noindent where $d_A, d_B$ and $d_C$ represent the dimensions of the Hilbert spaces for each of the three active players, in the same way that the suitable linear operator, $T$, introduced in {[37]} for upper bounding the Frobenius norm associated with the strategy of each player, is a mapping of the form,

\begin{align*}
    T^{\mathrm{XOR}} :  \textbf{C}^{2 \lceil \frac{n}{2} \rceil } \otimes      \textbf{C}^{2 \lceil \frac{n}{2} \rceil }   \longrightarrow        \textbf{C}^{d_A} \otimes \textbf{C}^{d_B}      \text{. }
\end{align*}

\noindent Several analogs to linear operators of the form above will be introduced, whether it be for the $N$-player $\mathrm{XOR}$ game, or for strong parallel repetitions of the $\mathrm{XOR}$ and $\mathrm{FFL}$ games. Each mapping takes the following form,

\begin{align*}
 \bigotimes T^{\mathrm{XOR}} : \underset{N \text{ copies}}{\bigotimes} \bigg(  \textbf{C}^{2 \lceil \frac{n}{2} \rceil } \bigg)  \longrightarrow        \underset{1 \leq i \leq N}{\bigotimes} \bigg( \textbf{C}^{d_i} \bigg) \text{, }  \\  \bigwedge T^{\mathrm{XOR}} :  \textbf{C}^{2 \lceil \frac{n}{2} \rceil } \bigwedge      \textbf{C}^{2 \lceil \frac{n}{2} \rceil }  \bigwedge \cdots \bigwedge  \textbf{C}^{2 \lceil \frac{n}{2} \rceil }  \longrightarrow        \textbf{C}^{d_A} \bigwedge \textbf{C}^{d_B} \bigwedge \textbf{C}^{d^{(1)}_B} \bigwedge\cdots \bigwedge \textbf{C}^{d^{(n-2)}_B}  \text{, } \\     T^{\mathrm{XOR}\wedge \cdots \wedge \mathrm{XOR}} :     \textbf{C}^{2 \lceil \frac{n}{2} \rceil \wedge 2 \lceil \frac{n}{2} \rceil \wedge \cdots \wedge 2 \lceil \frac{n}{2} \rceil}  \longrightarrow        \textbf{C}^{d_A \wedge d_B \wedge d^{(1)}_B \wedge \cdots \wedge {d^{(n-2)}_B }}     \text{, } \\      T^{\mathrm{FFL}\wedge  \mathrm{FFL}} :     \textbf{C}^{2 \lceil \frac{n}{2} \rceil \wedge 2 \lceil \frac{n}{2} \rceil}  \longrightarrow        \textbf{C}^{d_A \wedge d_B }     \text{. }
\end{align*}

\bigskip

\noindent In the Appendix various generalizations of Schur's Lemma is provided, in addition to the fact that the kernel of each suitable linear operator forms an invariant subspace. If the entry of tensor product states for the responses of each player above is replaced with the identity, tensors from each player can also be expressed as,

\begin{align*}
  \bigg| \bigg| \bigg[ \bigg(    A_i \otimes \textbf{I} \otimes C_{ijk}     \bigg) -  \bigg( \textbf{I} \otimes \bigg(\frac{ \pm B_{kl} + B_{lk}}{\big| \pm B_{kl} + B_{lk} \big| } \bigg) \otimes C_{ijk}   \bigg) \bigg] \ket{\psi_{3\mathrm{XOR}}}  \bigg| \bigg| \text{, }
\end{align*}

\noindent corresponding to the action of expressing the responses from the first player with those of second player,

\begin{align*}
  \bigg| \bigg| \bigg[ \bigg(    A_i \otimes B_{kl} \otimes \textbf{I}    \bigg) -  \bigg( \textbf{I} \otimes B_{kl} \otimes \bigg(  \frac{  \pm \underset{\sigma_1 \in S_3}{\sum} C_{\sigma(ijk)} +  \underset{\sigma_2, \sigma_3, \sigma_4, \sigma_5, \sigma_6  \in S_3}{\sum} C_{\sigma(ijk)}  }{  \bigg| \pm \underset{\sigma_1 \in S_3}{\sum} C_{\sigma(ijk)} +  \underset{\sigma_2, \sigma_3, \sigma_4, \sigma_5, \sigma_6  \in S_3}{\sum} C_{\sigma(ijk)}  \bigg|  }     \bigg)   \bigg) \bigg] \\ \times \ket{\psi_{3\mathrm{XOR}}}  \bigg| \bigg| \text{, }
\end{align*}

\noindent corresponding to the action of expressing the responses from the second player with those of third player, and,

\begin{align*}
  \bigg| \bigg| \bigg[ \bigg(    \textbf{I} \otimes B_{kl} \otimes C_{ijk}    \bigg) -  \bigg( A_i  \otimes B_{kl} \otimes \textbf{I}   \bigg) \bigg] \ket{\psi_{3\mathrm{XOR}}}  \bigg| \bigg| \text{, }
\end{align*}

\noindent corresponding to the action of expressing the responses of the the third player with those of the first player. From the set of possible responses of the third player in the $\mathrm{XOR}$ game, the normalization, 

\begin{align*}
   \bigg| \pm \underset{\sigma_1 \in S_3}{\sum} C_{\sigma(ijk)} +  \underset{\sigma_2, \sigma_3, \sigma_4, \sigma_5, \sigma_6  \in S_3}{\sum} C_{\sigma(ijk)}  \bigg|  \text{, }
\end{align*}

\noindent of the tensor observable that the third player gathers, indexed by either $ijk$, $jik$, or $kij$, appearing in,

\begin{align*}
  \frac{  \pm \underset{\sigma_1 \in S_3}{\sum} C_{\sigma(ijk)} +  \underset{\sigma_2, \sigma_3, \sigma_4, \sigma_5, \sigma_6  \in S_3}{\sum} C_{\sigma(ijk)}  }{  \bigg| \pm \underset{\sigma_1 \in S_3}{\sum} C_{\sigma(ijk)} +  \underset{\sigma_2, \sigma_3, \sigma_4, \sigma_5, \sigma_6  \in S_3}{\sum} C_{\sigma(ijk)}  \bigg|  }  \text{, }
\end{align*}

\noindent also appears in an inequality of the form,

\begin{align*}
  \bigg| \bigg| \bigg[ \bigg(   \bigg( \bigg( \underset{i \in \mathcal{Q}_1 }{\prod} A^{j_i}_i \bigg)     -        \bigg(     \underset{\text{set }i +1 \equiv i \oplus 1}{{\underset{i \in \mathcal{Q}_1}{\prod}}}   A^{j_i}_i   \bigg) \bigg)      \otimes      \textbf{I} \otimes C_{ijk}    \bigg)    -            \bigg(                 A^{j_i}_i \otimes  \textbf{I} \otimes \bigg( \bigg(         \underset{k \in \mathcal{Q}_3}{\underset{j \in \mathcal{Q}_2 }{ \underset{i \in \mathcal{Q}_1}{\prod}}} C^{j^{\prime}_{ijk}}_{ijk}               \bigg) \\ - \bigg(   \underset{\text{set }k + 1 \equiv k \oplus 1}{\underset{\text{set }j +1 \equiv j \oplus 1}{\underset{\text{set } i + 1 \equiv i \oplus 1}{\underset{k \in \mathcal{Q}_3}{\underset{j \in \mathcal{Q}_2}{ \underset{i \in \mathcal{Q}_1}{\prod}}}}}} C^{j^{\prime}_{ijk}}_{ijk}       \bigg) \bigg)          \bigg]   \ket{\psi_{3\mathrm{XOR}}}     \bigg| \bigg|   \text{, }
\end{align*}

\noindent corresponding to the action of performing changes of the Hilbert space at $i+1$, and $j+1$, of $A^{j_i}_i$ to performing changes of the Hilbert space at $i+1$, and $j+1$, of $C^{j^{\prime}_{ijk}}_{ijk}$. If the change of the Hilbert space is instead performed on the tensor observable of the second player, the inequality takes the form,

\begin{align*}
   \bigg| \bigg| \bigg[   \bigg( A^{j_i}_i \otimes \bigg( \bigg(    \underset{k \in \mathcal{Q}_3}{\underset{i \in \mathcal{Q}_1}{\prod}}             B^{j^{\prime}_{ik}}_{ik}        \bigg) - \bigg( \underset{\text{set }k+1 \equiv k \oplus 1}{\underset{\text{set }i +1 \equiv i \oplus 1}{\underset{k \in \mathcal{Q}_3}{\underset{1 \in \mathcal{Q}_1}{\prod}}}}             B^{j^{\prime}_{ik}}_{ik}   
        \bigg) \bigg)      \otimes      \textbf{I}  \bigg)   -                 \bigg(    \bigg( \bigg(    \underset{i \in \mathcal{Q}_1}{\prod}   A^{j_i}_i    \bigg) \\ - \bigg(    \underset{\text{set } j+1 \equiv j \oplus 1}{\underset{i \in \mathcal{Q}_1, j \in \mathcal{Q}_2 }{\prod}}   A^{j_i}_i              \bigg) \bigg) \otimes B_{ik} \otimes \textbf{I}            \bigg)            \bigg]  \ket{\psi_{3\mathrm{XOR}}}       \bigg| \bigg|   \text{. }
\end{align*}

\noindent The remaining possibility, for the tensor observable prepared by the third player, takes the form,

\begin{align*}
   \bigg| \bigg| \bigg[   \bigg( \textbf{I} \otimes B^{j_{ik}}_{ik}\otimes  \bigg(  \bigg(    \underset{k \in \mathcal{Q}_3}{\underset{j \in \mathcal{Q}_2}{\underset{i \in \mathcal{Q}_1}{\prod}}}    C^{j^{\prime}_{ijk}}_{ijk}    \bigg) - \bigg(    \underset{\text{set }k + 1 \equiv k \oplus 1}{\underset{\text{set }j +1 \equiv j \oplus 1}{\underset{\text{set } i + 1 \equiv i \oplus 1}{\underset{k \in \mathcal{Q}_3}{\underset{j \in \mathcal{Q}_2}{ \underset{i \in \mathcal{Q}_1}{\prod}}}}}}   C^{j^{\prime}_{ijk}}_{ijk} \bigg) \bigg) \bigg)   - \bigg(     \textbf{I} \otimes \bigg( \bigg(   \underset{k \in \mathcal{Q}_3}{\underset{i \in \mathcal{Q}_1}{\prod}}  B^{j^{\prime}_{ik}}_{ik}  \bigg) \\ - \bigg(         \underset{\text{set } k +1 \equiv k \oplus 1}{\underset{\text{set } i +1 \equiv i \oplus 1}{\underset{1 \leq k \leq m}{\underset{1 \leq i \leq n}{\prod}}}}  B^{j^{\prime}_{ik}}_{ik}    \bigg) \bigg)  \otimes C_{ijk}      \bigg)                  \bigg]   \ket{\psi_{3\mathrm{XOR}}}        \bigg| \bigg|   \text{. }
\end{align*}

\noindent If entries of tensor products of observables that each player prepares, such as those above, are replaced with the identity operator, the expressions to be upper bounded take the form,

\begin{align*}
 \bigg| \bigg| \bigg[  \bigg(  \bigg(   \underset{i \in \mathcal{Q}_1}{\prod} A^{j_i}_i   \bigg) \otimes \textbf{I} \otimes \textbf{I}    \bigg) - \bigg(  \bigg(   \underset{\text{set }j+1 \equiv j \oplus 1}{\underset{i \in \mathcal{Q}_1, j \in \mathcal{Q}_2 }{\prod}} A^{j_i}_i   \bigg) \otimes \textbf{I} \otimes \textbf{I}                \bigg)      \bigg] \ket{\psi_{3\mathrm{XOR}}}      \bigg| \bigg|  
 \text{, } \end{align*}

 \begin{align*} \bigg| \bigg| \bigg[ \bigg( \textbf{I} \otimes \bigg( \underset{k \in \mathcal{Q}_3}{\underset{i \in \mathcal{Q}_1}{\prod}}  B^{j_{ik}}_{ik} \bigg) \otimes \textbf{I} \bigg) - \bigg(   \textbf{I} \otimes \bigg(   \underset{\text{set }i+1 \equiv i \oplus 1}{\underset{k \in \mathcal{Q}_3}{\underset{i \in \mathcal{Q}_1}{\prod}}}  B^{j_{ik}}_{ik}        \bigg) \otimes \textbf{I}   \bigg)   \bigg] \ket{\psi_{3\mathrm{XOR}}}      \bigg| \bigg|    \text{,} \\   \\ \bigg| \bigg| \bigg[  \bigg(              \textbf{I} \otimes \textbf{I} \otimes \bigg(    \underset{k \in \mathcal{Q}_3}{\underset{j \in \mathcal{Q}_2}{\underset{i \in \mathcal{Q}_1}{\prod}}}  C^{j^{\prime}_{ijk}}_{ijk}\bigg)     \bigg) - \bigg(   \textbf{I} \otimes \textbf{I} \otimes \bigg(    \underset{\text{set }k + 1 \equiv k \oplus 1}{\underset{\text{set }j +1 \equiv j \oplus 1}{\underset{\text{set } i + 1 \equiv i \oplus 1}{\underset{k \in \mathcal{Q}_3}{\underset{j \in \mathcal{Q}_2}{ \underset{i \in \mathcal{Q}_1}{\prod}}}}}} C^{j^{\prime}_{ijk}}_{ijk}\bigg) \bigg)      \bigg] \ket{\psi_{3\mathrm{XOR}}}      \bigg| \bigg|   \text{. }
\end{align*}

\subsection{$3$ $\mathrm{XOR}$ game}

\noindent In this section, we demonstrate that three conditions can be thought of as equivalent, which impacts the computations performing in error bounds, and related objects. As presented in the previous section, the fact that the $3$ $\mathrm{XOR}$ game tensor satisfies the proportionality,

\begin{align*}
  G_{3\mathrm{XOR}} \propto  \underset{i \in \mathcal{Q}_1, j \in \mathcal{Q}_2, k \in \mathcal{Q}_3}{\sum} \bigg[    \bigg( \ket{\sigma \big( i \big) }  \bra{ij} \bigg) \bra{ijk} +                   \bigg(  \ket{\sigma \big( i \big) } \bra{\sigma \big( i \big) \sigma \big( j \big)} \bigg) \bra{ijk} + \bigg( \ket{\sigma \big( i \big) } \\ \times  \bra{\sigma \big( i \big)  \bra{\sigma \big( i \big) \sigma \big( j \big) } \bigg)  \sigma \big( j \big) \sigma \big( k \big) }   + \bigg( \ket{i}  \bra{\sigma \big( i \big) \sigma \big( j \big) } \bigg) \bra{ijk} + \bigg( \ket{i} \\ \times  \bra{\sigma \big( i \big) \sigma \big( j \big) } \bigg) \bra{\sigma \big( i \big) \sigma \big( j \big) \sigma \big( k \big) }  +    \bigg( \ket{i} \bra{ij} \bigg)   \bra{\sigma \big( i \big) \sigma \big( j \big) \sigma \big( k \big) }                    \bigg] \text{, }
  \end{align*}

\noindent from which the optimal strategy can be obtained through the constrained optimization problem,

\begin{align*}
 \underset{A,B,C,\ket{\psi_{3\mathrm{XOR}}}}{\mathrm{sup} }  \bra{\psi_{3\mathrm{XOR}}}   \underline{\mathscr{P}_{3\mathrm{XOR}}}  \ket{\psi_{3\mathrm{XOR}}} \text{, }
\end{align*}

\noindent over the tensors of all three players, $A,B,C$, and the optimal strategy $\ket{\psi_{3\mathrm{XOR}}}$ for the game. As was the case for the two-player $\mathrm{XOR}$ and $\mathrm{XOR}^{*}$ games, one makes use of the identification,

\begin{align*}
        \bigg( \ket{\sigma \big( i \big) }  \bra{ij} \bigg) \bra{ijk} \longleftrightarrow    A_{\sigma( i ) } B_{ij} C_{ijk}  \text{, } \\      \bigg(  \ket{\sigma \big( i \big) } \bra{\sigma \big( i \big) \sigma \big( j \big)} \bigg) \bra{ijk} \longleftrightarrow       A_{\sigma(i)} B_{\sigma( i ) \sigma( j ) } C_{ijk}    \text{, } \\ \bigg( \ket{\sigma \big( i \big) } \bra{\sigma \big( i \big) \sigma \big( j \big) } \bigg) \bra{\sigma \big( i \big) \sigma \big( j \big) \sigma \big( k \big) } \longleftrightarrow    A_{\sigma( i ) } B_{\sigma( i ) \sigma ( j ) } C_{\sigma( i ) \sigma( j ) \sigma( k ) }  \text{, } \\ \bigg( \ket{i}  \bra{\sigma \big( i \big) \sigma \big( j \big) } \bigg) \bra{ijk}       \longleftrightarrow      A_{i} B_{\sigma( i ) \sigma ( j ) } C_{\sigma( i ) } C_{ijk}         \text{, } \\   \bigg( \ket{i} \bra{\sigma \big( i \big) \sigma \big( j \big) } \bigg) \bra{\sigma \big( i \big) \sigma \big( j \big) \sigma \big( k \big) }        \longleftrightarrow A_i B_{\sigma( i ) \sigma ( j ) } C_{\sigma ( i ) \sigma ( j ) \sigma ( k ) } \text{, }  \\          \bigg( \ket{i} \bra{ij} \bigg) \bra{\sigma \big( i \big) \sigma \big( j \big) \sigma \big( k \big) }      \longleftrightarrow        A_i B_{ij} C_{\sigma( i ) \sigma(  j ) \sigma( k ) }  \text{. } 
\end{align*}

\noindent For the $3$ $\mathrm{XOR}$ game, TFAE:

\begin{itemize}
    \item[$\bullet$]  \underline{$\epsilon$-approximality of the optimal value}: For $\epsilon_{3\mathrm{XOR}}$ sufficiently small,

    \begin{align*}
   \omega_{3\mathrm{XOR}}  \big( 1 - \epsilon_{3\mathrm{XOR}} \big)   \leq \frac{1}{6 {n \choose 3}}    \bra{\psi_{3\mathrm{XOR}}} \bigg(     \underset{\text{permutations }\sigma \in S_3}{\underset{i_1 \in \mathcal{Q}_1,i_2 \in \mathcal{Q}_2,i_3 \in \mathcal{Q}_3}{\bigotimes}}   \text{Player observables} \big( \sigma  i_1  , \sigma  i_2  \\ ,  \sigma i_3  \big)                      \bigg)\ket{\psi_{3\mathrm{XOR}}}                              \leq    \omega_{3\mathrm{XOR}}   \text{. }
    \end{align*}

    \item[$\bullet$] \underline{$\epsilon$-approximality of the bias}: For the same choice of $\epsilon_{3\mathrm{XOR}}$, taking the supremum over all possible strategies $\mathcal{S}$, for obtaining the optimal bias,

\begin{align*}
\underset{\text{strategies }\mathcal{S}}{\mathrm{sup}} \big\{ \beta_{3\mathrm{XOR}} \big( G_{3 \mathrm{XOR}}, \mathcal{S} \big) \big\} \equiv \beta_{3\mathrm{XOR}} \big( G_{3 \mathrm{XOR}} \big)    \text{, }
\end{align*}

    \noindent implies that the inequality in the previous condition above takes the form,

     \begin{align*}
   \beta_{3\mathrm{XOR}}  \big( 1 - \epsilon_{3\mathrm{XOR}} \big)   \leq \frac{1}{6 {n \choose 3}}   G_{3\mathrm{XOR}}    \bra{\psi_{3\mathrm{XOR}}}  \bigg(     \underset{\text{permutations }\sigma \in S_3}{\underset{i_1 \in \mathcal{Q}_1,i_2 \in \mathcal{Q}_2,i_3 \in \mathcal{Q}_3}{\bigotimes}}   \text{Player observables} \big(  \sigma  i_1  ,  \sigma  i_2  \\ ,  \sigma  i_3 \big)                      \bigg) \ket{\psi_{3\mathrm{XOR}}}                         \leq    \beta_{3\mathrm{XOR}}   \text{. }
    \end{align*}

     \item[$\bullet$] \underline{Optimality, and approximate optimality, from $3$ $\mathrm{XOR}$ error bounds}: The error bound, (\textbf{EB}- $3$ $\mathrm{XOR}$), for the $3$ $\mathrm{XOR}$ game is determined by the following contributions:

\begin{itemize}

 \item[$\bullet$] For,

\begin{align*}
 \frac{A_i + A_j}{\sqrt{2}} \Longleftrightarrow  B_{ij} \text{, }
\end{align*}

 \noindent one has, 
     \begin{align*}
        \bigg| \bigg|               \bigg[     \bigg(  \bigg( \frac{A_i + A_j}{\sqrt{2}} \bigg) \otimes \textbf{I} \otimes \textbf{I}                \bigg) - \bigg( \textbf{I} \otimes B_{ij} \otimes \textbf{I} \bigg)       \bigg] \ket{\psi_{3\mathrm{XOR}}}  \bigg| \bigg|^2 \end{align*}
           
       \item[$\bullet$] For,

\begin{align*}
\frac{A_i - A_j}{\sqrt{2}}  \Longleftrightarrow   B_{ji} \text{, }
\end{align*}

 \noindent one has,     
           \begin{align*}  \bigg| \bigg|               \bigg[   \bigg(  \bigg( \frac{A_i - A_j}{\sqrt{2}} \bigg) \otimes \textbf{I} \otimes \textbf{I}                \bigg) - \bigg(  \textbf{I} \otimes B_{ji} \otimes \textbf{I} \bigg)     \bigg] \ket{\psi_{3\mathrm{XOR}}} \bigg| \bigg|^2  \end{align*}
           
       \item[$\bullet$] For,

\begin{align*}
\frac{B_{ij} + B_{ji}}{\sqrt{2}}  \Longleftrightarrow C_{ijk}  \text{, }
\end{align*}

 \noindent one has,    
           \begin{align*}    \bigg| \bigg|               \bigg[      \bigg(   \textbf{I} \otimes \bigg(  \frac{B_{ij} + B_{ji}}{\sqrt{2}} \bigg) \otimes \textbf{I}    \bigg)  - \bigg(         \textbf{I} \otimes \textbf{I} \otimes C_{ijk}   \bigg)     \bigg] \ket{\psi_{3\mathrm{XOR}}}  \bigg| \bigg|^2 \end{align*}
           
       \item[$\bullet$] For,

\begin{align*}
 \frac{B_{ij} - B_{ji}}{\sqrt{2}} \Longleftrightarrow  C_{jik} \text{, }
\end{align*}

 \noindent one has,     
           \begin{align*} \bigg| \bigg|               \bigg[        \bigg(          \textbf{I} \otimes \bigg(    \frac{B_{ij} - B_{ji}}{\sqrt{2}} \bigg) \otimes \textbf{I}                        \bigg)  - \bigg(         \textbf{I} \otimes \textbf{I} \otimes C_{jik}                \bigg)     \bigg] \ket{\psi_{3\mathrm{XOR}}}   \bigg| \bigg|^2   \end{align*}

     \item[$\bullet$] For,

\begin{align*}
  \frac{\underset{\sigma \in S_3}{\sum}C_{\sigma( ijk ) }}{\sqrt{6}}   \Longleftrightarrow  A_i  \text{, }
\end{align*}

 \noindent one has,       \begin{align*}   \bigg| \bigg|               \bigg[        \bigg(                 \textbf{I} \otimes \textbf{I} \otimes  \frac{1}{\sqrt{6}}  \bigg( \underset{\sigma \in S_3}{\sum}C_{\sigma( ijk ) }   \bigg)  \bigg)   - \bigg(     A_i \otimes \textbf{I} \otimes \textbf{I}        \bigg)     \bigg] \ket{\psi_{3\mathrm{XOR}}} \bigg| \bigg|^2  \end{align*}
           
           \item[$\bullet$] For,

\begin{align*}
 \frac{\underset{\sigma \in S_3}{\sum} C_{\sigma(ijk)} \textbf{1}_{\sigma  \text{ even transposition}} - \underset{\sigma \in S_3}{\sum}  C_{\sigma(ijk)}  \textbf{1}_{\sigma \text{ odd transposition}}}{\sqrt{6}}       \Longleftrightarrow  A_j  \text{, }
\end{align*}

 \noindent one has, 
           \begin{align*}        \bigg| \bigg|               \bigg[        \bigg( \textbf{I} \otimes \textbf{I}  \otimes  \frac{1}{\sqrt{6}}  \bigg( \underset{\sigma \in S_3}{\sum} C_{\sigma(ijk)} \textbf{1}_{\sigma  \text{ even transposition}} - \underset{\sigma \in S_3}{\sum}  C_{\sigma(ijk)}  \textbf{1}_{\sigma \text{ odd transposition}}              \bigg)   \\ - \bigg(   A_j \otimes \textbf{I} \otimes \textbf{I}          \bigg)     \bigg]  \ket{\psi_{3\mathrm{XOR}}}   \bigg| \bigg|^2   \text{,} \end{align*}

\noindent can be upper bounded with,
           
           \begin{align*}
           6n \big( n -1 \big) \big( n-2 \big) \epsilon_{3\mathrm{XOR}}   \text{. }
     \end{align*}
\end{itemize}

\end{itemize}

\noindent In the second equivalent condition below, the game tensor satisfies the proportionality,

\begin{align*}
  G_{4\mathrm{XOR}} \propto   \underset{i \in \mathcal{Q}_1, j \in \mathcal{Q}_2, k \in \mathcal{Q}_3}{\sum}   \bigg[            \ket{ijk} \bigg( \ket{i} \bra{ij} \bigg) \bra{ijkl}    +    \ket{jik} \bigg( \ket{j} \bra{ji} \bigg) \bra{jikl} +  \ket{ijk} \bigg( \ket{j} \bra{ij} \bigg) \bra{ijkl} \\  + \ket{jik} \bigg( \ket{i} \bra{ij} \bigg)  \bra{ijkl}  +  \ket{ijk} \bigg( \ket{k} \bra{ij} \bigg) \bra{ijkl}    \\   +  \text{Higher order permutations}         \bigg]              \text{, }
\end{align*}

\noindent which can be further manipulated to characterize exact, and approximate, $\epsilon$-optimality by taking the following superposition over permutations of four letter words,

\begin{align*}
  \underset{\text{Permutations } \sigma\in S_4}{\underset{i \in \mathcal{Q}_1, j \in \mathcal{Q}_2, k \in \mathcal{Q}_3}{\sum}}   \bigg[            \ket{\sigma \big( i \big) \sigma \big( j \big) \sigma \big( k \big) } \bigg( \ket{\sigma \big( i \big) } \bra{\sigma \big( i \big) \sigma \big( j \big) } \bigg) \bra{\sigma \big( i \big) \sigma \big( j \big) \sigma \big( k \big) \sigma \big( l \big) }    \\ +    \ket{\sigma \big( j \big) \sigma \big( i \big) \sigma \big( k \big)}   \bigg( \ket{\sigma \big( j \big) } \bra{\sigma \big(j \big) \sigma \big(i \big)} \bigg) \bra{\sigma \big( j \big) \sigma \big( i \big) \sigma \big( k \big) \sigma \big( l \big) } +  \ket{\sigma \big( i \big) \sigma \big( j \big) \sigma \big( k \big) } \\ \times \bigg( \ket{\sigma \big( j \big) } \bra{\sigma \big( i \big) \sigma \big( j \big) } \bigg)    \bra{\sigma \big( i \big) \sigma \big( j\big) \sigma \big( k \big) \sigma \big( l \big)}   + \ket{\sigma \big( j \big) \sigma \big( i \big) \sigma \big( k \big) }  \bigg( \ket{\sigma \big( i \big) } \bra{\sigma \big( i \big) \sigma \big( j \big) } \bigg) \\ \times \bra{\sigma \big( i \big)  \sigma \big( j \big) \sigma \big( k \big) \sigma \big( l \big) } +  \ket{\sigma \big( i \big) \sigma \big( j \big) \sigma \big( k \big) }   \bigg( \ket{\sigma \big( k \big) } \bra{\sigma \big( i \big) \sigma \big( j \big) } \bigg) \\ \times  \bra{\sigma \big( i \big) \sigma \big( j \big) \sigma \big( k \big) \sigma \big( l \big) }       + \text{Higher order permutations}               \bigg]              \text{. }
\end{align*}

\noindent Over permutations in the group of four letter words, the above superposition can be realized through the following set of combinatorial possibilities,

\begin{align*}
     \underset{\text{Permutations } \sigma\in S_4}{\underset{i \in \mathcal{Q}_1, j \in \mathcal{Q}_2, k \in \mathcal{Q}_3, l \in \mathcal{Q}_4}{\sum}}   \bigg[    \ket{\sigma \big( i \big) \sigma \big( j \big) \sigma \big( k \big) } \bigg( \ket{i} \bra{ij} \bigg) \bra{ijkl} + \ket{ijk}  \bigg( \ket{\sigma \big( i \big) } \bra{ij} \bigg) \bra{ijkl} +   \ket{ijk} \\ \times  \bigg( \ket{i}  \bra{\sigma \big( i \big) \sigma \big( j \big)} \bigg)    \bra{ijkl}  + \ket{ijk} \bigg( \ket{i} \bra{ij} \bigg) \bra{\sigma \big( i \big) \sigma \big( j \big) \sigma \big( k \big) \sigma \big( l \big) }    + \ket{\sigma \big( i \big) \sigma \big( j \big) \sigma \big( k \big) }  \\ \times    \bigg( \ket{\sigma \big( i \big) }   \bra{ij}      \bigg)    \bra{ijkl}  + \ket{\sigma \big( i \big) \sigma \big( j \big) \sigma \big( k \big) }   \bigg( \ket{\sigma \big( i \big) } \bra{\sigma \big( i \big) \sigma \big( j \big) } \bigg) \bra{ijkl}  +       \ket{\sigma \big( i \big) \sigma \big( j \big) \sigma \big( k \big) } \\ \times \bigg( \ket{\sigma \big( i \big) } \bra{\sigma \big( i \big) \sigma \big( j \big) } \bigg)   \bra{\sigma \big( i \big) \sigma \big( j \big) \sigma \big( k \big) \sigma \big( l \big) }    + \ket{ijk} \bigg(       \ket{\sigma \big( i \big) } \bra{ij} \bigg) \bra{ijkl} +     \ket{ijk} \\ \times \bigg(  \ket{\sigma \big( i \big) }  \bra{\sigma \big( i \big) \sigma \big( j \big) }   \bigg) \bra{ijkl} +         \ket{ijk }  \bigg( \ket{\sigma \big( i \big) } \bra{ \sigma \big( i \big) \sigma \big( j \big) } \bigg) \\ \times  \bra{\sigma \big( i \big) \sigma \big( j \big) \sigma( k \big) \sigma \big( l \big) }  +    \ket{ijk} \times   \bigg( \ket{i} \bra{\sigma \big( i \big) \sigma \big( j \big) } \bigg) \bra{ijkl}  + \ket{ijk} 
     \\ \times \bigg(    \ket{i} \bra{\sigma \big( i \big) \sigma \big( j \big) } \bigg) \bra{\sigma \big( i \big) \sigma \big( j \big) \sigma \big( k \big) \sigma \big( l \big) }   
     +  \text{Higher order permutations}    \bigg] \text{. }
\end{align*}    

\noindent From the summations above over the indicates $i,j,k$ and permutations over the four-letter symmetric group, introduce,

\begin{align*}
  \underline{\mathscr{P}_{4\mathrm{XOR}}} \equiv   \underline{\mathscr{P}_{4\mathrm{XOR},1}} \cup  \underline{\mathscr{P}_{4\mathrm{XOR},2}}  \cup  \underline{\mathscr{P}_{4\mathrm{XOR},3}}    \cup  \underline{\mathscr{P}_{4\mathrm{XOR},4}}   \cup \cdots \cup \underline{\mathscr{P}_{4\mathrm{XOR},1,2,3,4}}           \text{, }      
\end{align*}

\noindent which are enumerated below, beginning with \textit{Table 3}.

\begin{table}[ht]
\caption{Permutations of Player Observables for the 4 XOR game error bound, $(\textbf{EB}- 4 \mathrm{XOR}) $} 
\centering 
\begin{tabular}{c c c c} 
\hline\hline 
 Player & Tensor Product Representation & Permutation Superposition  \\ [0.5ex] 
\hline    1 &        $\underset{\sigma}{\mathlarger{\sum}} \bigg( C_{ijk} \otimes A_{\sigma(i)}  \otimes B_{ij} \otimes D_{ijkl} \bigg) $    &  $ C_{ijk} \otimes  A_i \otimes B_{ij} \otimes D_{ijkl} +          C_{ijk} \otimes A_j \otimes B_{ij} \otimes D_{ijkl}      $  \\ & & $ +  C_{ijk} \otimes  A_k \otimes B_{ij} \otimes D_{ijkl}  + C_{ijk} \otimes  A_l \otimes B_{ij} \otimes D_{ijkl}    $  \\    \\    2 &  $\underset{\sigma}{\mathlarger{\sum}} \bigg( C_{ijk} \otimes A_{i}  \otimes B_{\sigma(ij)} \otimes D_{ijkl} \bigg)$  &             $C_{ijk} \otimes A_i \otimes B_{ij} \otimes D_{ijkl} + C_{ijk} \otimes A_i \otimes B_{ji} \otimes D_{ijkl}    $        \\    & & $ +  C_{ijk} \otimes A_i \otimes B_{ik} \otimes D_{ijkl} + C_{ijk} \otimes A_i \otimes B_{ki} \otimes D_{ijkl}          $     \\ & & $ +   C_{ijk} \otimes A_i \otimes B_{il} \otimes D_{ijkl} + C_{ijk} \otimes A_i \otimes B_{li} \otimes D_{ijkl}          $   \\      & & $ +  C_{ijk} \otimes A_i \otimes B_{lj} \otimes D_{ijkl} + C_{ijk} \otimes A_i \otimes B_{lj} \otimes D_{ijkl}       $   \\   & & $ +  C_{ijk} \otimes A_i \otimes B_{lk} \otimes D_{ijkl} + C_{ijk} \otimes A_i \otimes B_{kl} \otimes D_{ijkl}       $ \\ & & $ +     C_{ijk} \otimes A_i \otimes B_{kj} \otimes D_{ijkl} + C_{ijk} \otimes A_i \otimes B_{jk} \otimes D_{ijkl}    $  \\     \\  3 & $    \underset{\sigma}{\mathlarger{\sum}} \bigg( C_{\sigma(ijk)} \otimes A_{i}  \otimes B_{ij} \otimes D_{ijkl} \bigg)  $   &     $C_{ijk} \otimes A_i \otimes B_{ij} \otimes D_{ijkl} + C_{ikj} \otimes A_i \otimes B_{ij} \otimes D_{ijkl}          $   \\ & & $ +    C_{ikl} \otimes A_i \otimes B_{ij} \otimes D_{ijkl} +    C_{ilk} \otimes A_i \otimes B_{ij} \otimes D_{ijkl}             $ \\ & & $ + C_{jik} \otimes A_i \otimes B_{ij} \otimes D_{ijkl} + C_{jki} \otimes A_i \otimes B_{ij} \otimes D_{ijkl}            $  \\ & & $ +  C_{jil} \otimes A_i \otimes B_{ij} \otimes D_{ijkl} + C_{jkl} \otimes A_i \otimes B_{ij} \otimes D_{ijkl}     $ \\ & & $ +   C_{lik} \otimes A_i \otimes B_{ij} \otimes D_{ijkl}  +   C_{lki} \otimes A_i \otimes B_{ij} \otimes D_{ijkl}      $ \\   & & $ +   C_{ljk} \otimes A_i \otimes  B_{ij} \otimes D_{ijkl}  +   C_{lkj} \otimes A_i  \otimes B_{ij} \otimes D_{ijkl}      $        \\ & & $ + C_{lij} \otimes A_i \otimes B_{ij} \otimes D_{ijkl} + C_{lji} \otimes A_i \otimes B_{ij} \otimes D_{ijkl} $ \\ & & $ + C_{lkj} \otimes A_i \otimes B_{ij} \otimes D_{ijkl} + C_{ljk} \otimes A_i \otimes B_{ij} \otimes D_{ijkl}    $  \\ & & $ +   C_{jli} \otimes A_i \otimes B_{ij} \otimes D_{ijkl} + C_{jki} \otimes A_i \otimes B_{ij} \otimes D_{ijkl}  $ \\ & & $ + C_{jik} \otimes A_i \otimes B_{ij} \otimes D_{ijkl} $ \\  \\   4 &    $\underset{\sigma}{\mathlarger{\sum}} \bigg( C_{ijk} \otimes A_{i}  \otimes B_{ij} \otimes D_{\sigma(ijkl)} \bigg)   $     &   $C_{ijk} \otimes A_i \otimes B_{ij} \otimes D_{ijkl} +  C_{ijk} \otimes A_i \otimes B_{ij} \otimes D_{ikjl}       $ \\ & & $ + C_{ijk} \otimes A_i \otimes B_{ij} \otimes D_{ijlk} + C_{ijk} \otimes A_i \otimes B_{ij} \otimes D_{iljk}                    $    \\ & & $ +   C_{ijk} \otimes A_i \otimes B_{ij} \otimes D_{jikl} + C_{ijk} \otimes A_i \otimes B_{ij} \otimes D_{jilk} $  \\ & & $ +      C_{ijk} \otimes A_i \otimes B_{ij} \otimes D_{jlik} +  C_{ijk} \otimes A_i \otimes B_{ij} \otimes D_{jilk}            $   \\ & & $ +   C_{ijk} \otimes A_i \otimes  B_{ij} \otimes D_{kijl} + C_{ijk} \otimes A_i \otimes B_{ij} \otimes D_{kjil}     $   \\  & & $ +  C_{ijk} \otimes A_i \otimes B_{ij} \otimes D_{klji}  + C_{ijk} \otimes A_i \otimes B_{ij} \otimes D_{kjil}          $  \\    & & $ + C_{ijk} \otimes A_i \otimes B_{ij} \otimes D_{lijk}  + C_{ijk} \otimes A_i \otimes B_{ij} \otimes D_{ljik}$   \\ & & $ +       C_{ijk} \otimes A_i \otimes B_{ij} \otimes D_{lkij} + C_{ijk} \otimes A_i \otimes B_{ij} \otimes D_{lkji}       $ \\ & & $ + C_{ijk} \otimes A_i \otimes B_{ij} \otimes D_{ikjl} + C_{ijk} \otimes A_i \otimes B_{ij} \otimes D_{ijkl} $ \\ & & $ + C_{ijk} \otimes A_i \otimes B_{ij} \otimes D_{iljk} + C_{ijk} \otimes A_i \otimes B_{ij} \otimes D_{ilkj}         $  \\       [1ex] 
\hline 
\end{tabular}
\label{table:nonlin} 
\end{table}

\begin{table}[ht]
\caption{Permutations of Player Observables for the 4 XOR game error bound, $(\textbf{EB}- 4 \mathrm{XOR})$ Continued} 
\centering 
\begin{tabular}{c c c c} 
\hline\hline 
 Player & Tensor Product Representation & Permutation Superposition  \\ [0.5ex] 
\hline    1,2 &  $\underset{\sigma,\sigma^{\prime}}{\mathlarger{\sum}} \bigg( C_{ijk} \otimes A_{\sigma(i)}  \otimes B_{\sigma^{\prime}(ij)}  \otimes D_{ijkl} \bigg)  $ &    $C_{ijk} \otimes A_i \otimes B_{ij} \otimes D_{ijkl} +   C_{ijk} \otimes A_j \otimes B_{ij} \otimes D_{ijkl}      $   \\     & & $ +   C_{ijk} \otimes A_k \otimes B_{ij} \otimes D_{ijkl} + C_{ijk} \otimes A_l \otimes B_{ij} \otimes D_{ijkl}     $  \\ & & $ +   C_{ijk} \otimes A_i \otimes B_{ji} \otimes D_{ijkl} + C_{ijk} \otimes A_j \otimes B_{ji} \otimes D_{ijkl} $  \\ & & $ +      C_{ijk} \otimes A_k \otimes B_{ji} \otimes D_{ijkl} + C_{ijk} \otimes A_l \otimes B_{ji} \otimes D_{ijkl}      $  \\ & & $    +           C_{ijk} \otimes A_i \otimes B_{ik} \otimes D_{ijkl} + C_{ijk} \otimes A_j \otimes B_{ik} \otimes D_{ijkl}                     $   \\  & & $ +  C_{ijk} \otimes A_k \otimes B_{ik} \otimes D_{ijkl} + C_{ijk} \otimes A_l \otimes B_{ik} \otimes D_{ijkl}            $    \\ & & $ +   C_{ijk} \otimes A_i \otimes B_{ki} \otimes D_{ijkl} + C_{ijk} \otimes A_j \otimes B_{ki} \otimes D_{ijkl}     $  \\ & & $ +  C_{ijk} \otimes A_k \otimes B_{ki} \otimes D_{ijkl} + C_{ijk} \otimes A_l \otimes B_{ki} \otimes D_{ijkl} $ \\ & & $ +         C_{ijk} \otimes A_i  \otimes B_{il} \otimes   D_{ijkl } + C_{ijk} \otimes A_j \otimes B_{il} \otimes D_{ijkl}      $ \\  & & $ + C_{ijk} \otimes A_k \otimes B_{il} \otimes D_{ijkl} + C_{ijk} \otimes A_l \otimes B_{il} \otimes D_{ijkl}   $      \\ & & $ +       C_{ijk} \otimes A_i \otimes B_{li} \otimes D_{ijkl} + C_{ijk} \otimes A_j \otimes B_{li} \otimes D_{ijkl}    $ \\ & & $ +  C_{ijk} \otimes  A_i \otimes B_{jk} \otimes D_{ijkl} + C_{ijk} \otimes A_j \otimes B_{jk} \otimes D_{ijkl}                $ \\      & & $ + C_{ijk} \otimes A_k \otimes B_{jk} \otimes D_{ijkl} +           C_{ijk} \otimes A_l \otimes B_{jk} \otimes D_{ijkl}         $    \\  & & $ +  C_{ijk} \otimes  A_i \otimes B_{kj} \otimes D_{ijkl} + C_{ijk} \otimes A_j \otimes B_{kj} \otimes D_{ijkl}                    $  \\  & & $ +   C_{ijk} \otimes A_k \otimes B_{kj} \otimes D_{ijkl} +           C_{ijk} \otimes A_l \otimes B_{kj} \otimes D_{ijkl}        $   \\  & & $ +      C_{ijk} \otimes A_i \otimes B_{jl} \otimes D_{ijkl} + C_{ijk} \otimes A_j \otimes B_{jl} \otimes D_{ijkl}  $    \\     & & $ +       C_{ijk} \otimes A_k \otimes B_{jl} \otimes D_{ijkl} + C_{ijk} \otimes A_l \otimes B_{jl} \otimes D_{ijkl}     $        \\ & & $ +     C_{ijk} \otimes A_i \otimes B_{lj} \otimes D_{ijkl} + C_{ijk} \otimes A_j \otimes B_{lj} \otimes D_{ijkl}      $ \\ & & $ +  C_{ijk} \otimes A_j \otimes B_{lj} \otimes D_{ijkl} + C_{ijk} \otimes A_k \otimes B_{lj} \otimes D_{ijkl}   $ \\ & & $ +      C_{ijk} \otimes A_l \otimes B_{lj}   \otimes D_{ijkl} + C_{ijk} \otimes A_i \otimes B_{lk} \otimes D_{ijkl} $ \\ & & $  +              C_{ijk} \otimes A_j \otimes B_{lk} \otimes D_{ijkl}  + C_{ijk} \otimes A_k \otimes B_{lk} \otimes D_{ijkl}            $  \\  & & $ + C_{ijk} \otimes A_l \otimes B_{lk} \otimes D_{ijkl} $\\   \\ 1,3  & $\underset{\sigma,\sigma^{\prime}}{\mathlarger{\sum}} \bigg( C_{\sigma^{\prime}(ijk)}  \otimes A_{\sigma(i)}  \otimes B_{ij} \otimes D_{ijkl} \bigg)  $    &            $   C_{ijk} \otimes A_i \otimes B_{ij} \otimes D_{ijkl}    +   C_{ijk} \otimes A_j \otimes B_{ij} \otimes D_{ijkl}     $       \\ & & $ + C_{ijk} \otimes A_k \otimes B_{ij} \otimes D_{ijkl} + C_{ijk} \otimes A_l \otimes B_{ij} \otimes D_{ijkl}  $\\       & & $ +    C_{jik} \otimes A_i \otimes B_{ij} \otimes D_{ijkl} +        C_{jik} \otimes A_j \otimes B_{ij} \otimes D_{ijkl}      $\\ & & $ +  C_{jik} \otimes A_k \otimes B_{ij} \otimes D_{ijkl} + C_{jik} \otimes A_l \otimes B_{ij} \otimes D_{ijkl }      $ \\   & & $ +  C_{ijl} \otimes A_i \otimes B_{ij} \otimes D_{ijkl } + C_{ijl} \otimes A_j \otimes B_{ij} \otimes D_{ijkl}      $\\ & & $ +    C_{ijl}  \otimes A_k \otimes B_{ij} \otimes D_{ijkl} + C_{ijl} \otimes A_l \otimes B_{ij} \otimes D_{ijkl}   $  \\  & & $\vdots$ \\    [1ex] 
\hline 
\end{tabular}
\label{table:nonlin} 
\end{table}

\begin{table}[ht]
\caption{Permutations of Player Observables for the 4 XOR game error bound, $(\textbf{EB}- 4 \mathrm{XOR})$ Continued} 
\centering 
\begin{tabular}{c c c c} 
\hline\hline 
 Player & Tensor Product Representation & Permutation Superposition  \\ [0.5ex] 
\hline     & $\vdots$ &   \\     1,3  & $\underset{\sigma,\sigma^{\prime}}{\mathlarger{\sum}} \bigg( C_{\sigma^{\prime}(ijk)}  \otimes A_{\sigma(i)}  \otimes B_{ij} \otimes D_{ijkl} \bigg)  $  &   $ +    C_{jil} \otimes A_i \otimes B_{ij} \otimes D_{ijkl} + C_{jil} \otimes A_j \otimes B_{ij} \otimes D_{ijkl}   $ \\    & & $ +    C_{jil} \otimes A_k \otimes B_{ij} \otimes D_{ijkl} + C_{jil} \otimes A_l \otimes B_{ij} \otimes D_{ijkl}   $   \\ & & $ +   C_{jli} \otimes A_i \otimes B_{ij} \otimes D_{ijkl} + C_{jli} \otimes A_j \otimes B_{ij} \otimes D_{ijkl}      $ \\ & & $ +   C_{jli} \otimes A_k \otimes B_{ij} \otimes D_{ijkl} + C_{jli} \otimes A_l \otimes B_{ij} \otimes D_{ijkl}  $ \\ & &   $ +   C_{jlk} \otimes A_i \otimes B_{ij} \otimes D_{ijkl} + C_{jlk } \otimes A_j \otimes B_{ij} \otimes D_{ijkl}   $ \\ & & $ +  C_{jlk} \otimes A_k  \otimes B_{ij} \otimes D_{ijkl} + C_{jlk} \otimes A_l \otimes B_{ij} \otimes D_{ijkl}    $      \\ & & $ +  C_{jik} \otimes A_i \otimes B_{ij} \otimes D_{ijkl} + C_{jik} \otimes A_j \otimes B_{ij} \otimes D_{ijkl}       $ \\ & & $ +   C_{jik} \otimes A_k \otimes B_{ij} \otimes D_{ijkl} + C_{jik} \otimes A_l \otimes B_{ij} \otimes D_{ijkl}   $  \\ & & $ +     C_{kij} \otimes A_i \otimes B_{ij} \otimes D_{ijkl} + C_{kij} \otimes A_j \otimes B_{ij} \otimes D_{ijkl}      $ \\ & & $ +      C_{kij} \otimes A_k \otimes B_{ij} \otimes D_{ijikl} + C_{kij} \otimes A_l \otimes B_{ij} \otimes D_{ijkl}    $ \\         & & $ +    C_{k ji} \otimes    A_i \otimes B_{ij} \otimes D_{ijkl} + C_{kji} \otimes A_j \otimes B_{ij} \otimes D_{ijkl}  $    \\  & & $ + C_{kji} \otimes A_k \otimes B_{ij} \otimes D_{ijkl} + C_{kji} \otimes A_k \otimes B_{ij} \otimes D_{ijkl} $     \\  & & $ +   C_{kil} \otimes B_{ij} \otimes A_i \otimes D_{ijkl} + C_{kil} \otimes B_{ij} \otimes A_j \otimes D_{ijkl}   $     \\  & & $ + C_{kil} \otimes B_{ij} \otimes A_k \otimes D_{ijkl} + C_{kil} \otimes B_{ij} \otimes A_l \otimes D_{ijkl} $     \\   & & $ +  C_{kjl} \otimes B_{ij} \otimes A_i \otimes D_{ijkl} + C_{kjl} \otimes B_{ij} \otimes A_j \otimes D_{ijkl}       $        \\   & & $ +   C_{kjl} \otimes B_{ij} \otimes A_k \otimes D_{ijkl} + C_{kjl} \otimes B_{ij} \otimes A_l \otimes D_{ijkl}      $     \\         & & $ +    C_{kli} \otimes B_{ij} \otimes A_i \otimes D_{ijkl} + C_{kli} \otimes B_{ij} \otimes A_j \otimes D_{ijkl}              $   \\    & & $ +    C_{kli} \otimes B_{ij} \otimes A_k \otimes D_{ijkl} + C_{kli} \otimes B_{ij} \otimes A_l \otimes D_{ijkl}                 $       \\  & & $ + C_{klj} \otimes B_{ij} \otimes A_i \otimes D_{ijkl } + C_{klj} \otimes B_{ij} \otimes A_j \otimes D_{ijkl}  $\\ & & $ + C_{klj} \otimes B_{ij} \otimes A_k \otimes D_{ijkl} + C_{klj} \otimes B_{ij} \otimes A_l \otimes D_{ijkl}    $ \\  & & $ +    C_{ikj} \otimes A_i \otimes B_{ij} \otimes D_{ijkl} + C_{ikj} \otimes A_j \otimes B_{ij} \otimes D_{ijkl}        $\\  & & $ +  C_{ikj} \otimes A_k \otimes B_{ij} \otimes D_{ijkl} + C_{ikj} \otimes A_l \otimes B_{ij} \otimes D_{ijkl }     $      \\     & & $ +    C_{ikl} \otimes A_i \otimes B_{ij} \otimes D_{ijkl} + C_{ikl} \otimes A_j \otimes B_{ij} \otimes D_{ijkl}     $ \\      & &  $ +  C_{ikl} \otimes A_k \otimes B_{ij} \otimes D_{ijkl} + C_{ikj} \otimes A_l \otimes B_{ij} \otimes D_{ijkl}$ \\ & & $ +      C_{ilj} \otimes A_i \otimes B_{ij} \otimes D_{ijkl} + C_{ilj} \otimes A_j \otimes B_{ij} \otimes D_{ijkl}        $    \\    & & $ +     C_{ilj} \otimes A_k \otimes B_{ij} \otimes D_{ijkl} + C_{ilj} \otimes A_l \otimes B_{ij} \otimes D_{ijkl}        $  \\ & & $+ \big(\text{Remaining } C \text{ permutations}\big( ijk \big)\big)$  \\ & $\vdots$ &   \\      [1ex] 
\hline 
\end{tabular}
\label{table:nonlin} 
\end{table}

\begin{table}[ht]
\caption{Permutations of Player Observables for the 4-XOR game error bound, $(\textbf{EB}- 4 \mathrm{XOR})$ Continued} 
\centering 
\begin{tabular}{c c c c} 
\hline\hline 
 Player & Tensor Product Representation & Permutation Superposition  \\ [0.5ex] 
\hline  & $\vdots$ &  \\   \\ 1,4  &    $\underset{\sigma,\sigma^{\prime}}{\mathlarger{\sum}} \bigg( C_{ijk}  \otimes A_{\sigma(i)}  \otimes B_{ij} \otimes D_{\sigma^{\prime} ( ijkl ) } \bigg)  $       &   $ C_{ijk}  \otimes A_{\sigma(i)}  \otimes B_{ij} \otimes D_{ ijkl } + C_{ijk}  \otimes A_{\sigma(i)}   $     \\ & & $ \otimes B_{ij} \otimes D_{ ijkl } + C_{ijk}  \otimes A_{\sigma(i)}   \otimes B_{ij} \otimes D_{ ijkl }     $ \\ & & $  +  C_{ikj} \otimes 
 A_{\sigma(i)} \otimes B_{ij} \otimes D_{ijkl}          +    C_{ikj} \otimes A_{\sigma(i)}     $      \\ & & $   \otimes     B_{ij}      \otimes   D_{ijkl}  +      C_{ikj} \otimes A_{\sigma(i)}      \otimes B_{ij}    $       \\ &   & $  \otimes D_{ijkl}     +    C_{ikj}    \otimes A_{\sigma(i)}   \otimes B_{ij} \otimes D_{ijkl}       +   C_{jik}       $     \\ & & $   \otimes A_{\sigma(i)}\otimes B_{ij}    \otimes D_{ijkl}   + C_{jik}  \otimes A_{\sigma(i)}   \otimes B_{ij}     $  \\ & & $    \otimes D_{ijkl}     +        C_{jik}     \otimes A_{\sigma(i)}     \otimes         B_{ij}   \otimes D_{ijkl}    $ \\   & & $     +  C_{jik}     \otimes A_{\sigma(i)}    \otimes B_{ij} \otimes D_{ijkl}    $    \\ & & $    +    C_{ikl} \otimes   A_{\sigma(i)}      \otimes B_{ij}    \otimes D_{ijkl}    +  C_{ikl}   $    \\   & & $ \otimes A_{\sigma(i)} \otimes B_{ij}    \otimes D_{ijkl}    + C_{ikl}   \otimes A_{\sigma(i)}     $   \\ & & $   \otimes B_{ij}   \otimes D_{ijkl} +  C_{ikl} 
 \otimes  $   \\    & &  $    A_{\sigma(i)}  \otimes B_{ij} \otimes D_{ijkl}   +    C_{jil} \otimes A_{\sigma(i)}   \otimes B_{ij}      \otimes D_{ijkl}    
 $   \\ & & $    +  C_{jil}   \otimes A_{\sigma(i)}   \otimes B_{ij} \otimes D_{ijkl}   + C_{jil}  \otimes A_{\sigma(i)}       \otimes     B_{ij}   $ \\ & & $     \otimes D_{ijkl}     + C_{jil}   \otimes A_{\sigma(i)}  \otimes B_{ij} \otimes D_{ijkl}   $ \\ & & $ + (A \text{ permutations} (j,k,l))$ \\  \\    2,4 &   $\underset{\sigma^{\prime},\sigma^{\prime\prime}}{\mathlarger{\sum}} \bigg( C_{ ijk }  \otimes A_{i}  \otimes B_{\sigma^{\prime}  (ij) } \otimes D_{\sigma^{\prime\prime} ( ijkl ) } \bigg)  $     & \text{Combine previous permutations listed above}   \\ \\   1,2,3,4   &       $\underset{\sigma,\sigma^{\prime},\sigma^{\prime\prime}, \sigma^{\prime\prime\prime}}{\mathlarger{\sum}} \bigg( C_{\sigma^{\prime\prime} (ijk )}  \otimes A_{\sigma(i)}  \otimes B_{\sigma^{\prime}  (ij) } \otimes D_{\sigma^{\prime\prime\prime} ( ijkl ) } \bigg)  $   & \text{Combine previous permutations listed above}     \\ \\   2,3,4   & $\underset{\sigma^{\prime},\sigma^{\prime\prime}, \sigma^{\prime\prime\prime}}{\mathlarger{\sum}} \bigg( C_{\sigma^{\prime\prime} (ijk )}  \otimes A_{i}  \otimes B_{\sigma^{\prime}  (ij) } \otimes D_{\sigma^{\prime\prime\prime} ( ijkl ) } \bigg)  $   &  \text{Combine previous permutations listed above}       \\ \\   3,4   &  $\underset{\sigma^{\prime},\sigma^{\prime\prime}}{\mathlarger{\sum}} \bigg( C_{\sigma^{\prime} (ijk )}  \otimes A_{i}  \otimes B_{  ij } \otimes D_{\sigma^{\prime\prime} ( ijkl ) } \bigg)  $  &   \text{Combine previous permutations listed above}      \\   [1ex] 
\hline 
\end{tabular}
\label{table:nonlin} 
\end{table}

  \bigskip

\noindent The optimal value for the $4$-$\mathrm{XOR}$ game satisfies the proportionality,

\begin{align*}
  \omega \big( {4 \mathrm{XOR}} \big) \propto    \underset{A,B,C,D,\ket{\psi_{4\mathrm{XOR}}}}{\mathrm{sup}}   \bra{\psi_{4\mathrm{XOR}}}  \bigg(  C_{\sigma(i ) \sigma( j ) \sigma( k ) } \otimes A_i \otimes B_{ij} \otimes D_{ijkl}  + C_{ijk } \otimes A_{\sigma( i ) } \\  \otimes B_{ij} \otimes D_{ijkl}  +  C_{ijk}    \otimes A_i \otimes B_{\sigma( i ) \sigma ( j ) } \otimes D_{ijkl} + C_{ijk} \otimes A_i \otimes B_{ij} \otimes D_{\sigma( i ) \sigma( j ) \sigma ( k ) \sigma ( l ) } \\  +   C_{\sigma( i ) \sigma ( j ) \sigma ( k ) }  \otimes A_{\sigma(  i) }   \otimes B_{ij}\otimes D_{ijkl} + C_{\sigma( i ) \sigma ( j ) \sigma ( k ) }  \otimes A_{\sigma ( i ) } \otimes B_{\sigma( i ) \sigma ( j ) } \otimes D_{\sigma ( i ) \sigma ( j ) \sigma ( k ) \sigma ( l ) } \\ +    C_{ijk} \otimes A_{\sigma ( i ) }   \otimes B_{\sigma(i ) \sigma ( j )}     \otimes  D_{\sigma ( i ) \sigma ( j ) \sigma ( k ) \sigma ( l ) }    +     C_{ijk} \otimes A_{\sigma( i ) } \otimes B_{ij}    \otimes D_{ijkl} + C_{ijk} \otimes A_{\sigma( i ) } \\ \otimes B_{\sigma( i ) \sigma ( j ) }    \otimes D_{ijkl} + C_{ijk}  \otimes A_i  \otimes B_{\sigma(i ) \sigma( j )}  \otimes D_{ijkl} + C_{ijk} \otimes A_i  \\ \otimes B_{\sigma( i ) \sigma ( j )}  \otimes D_{\sigma ( i ) \sigma( j ) \sigma( k ) \sigma ( l ) }      \bigg) \ket{\psi_{4\mathrm{XOR}}}                                    \text{. }
\end{align*}

\noindent The set of three equivalent conditions above for the $3$-$\mathrm{XOR}$ game above extend to $\mathrm{XOR}$ games with an even, or odd, number of players. Given the optimal strategy for a $2$ $\mathrm{XOR}$ game, $\ket{\psi_{2\mathrm{XOR}}}$, {[37]}, \textit{2.1}, the relations,

\begin{align*}
   \bigg( A_i \otimes \textbf{I} \bigg) \ket{\psi_{2\mathrm{XOR}}} = \bigg( \textbf{I} \otimes \frac{B_{ij} + B_{ji}}{\sqrt{2}} \bigg) \ket{\psi_{2\mathrm{XOR}}}     \text{, }  \end{align*}

   \begin{align*} \bigg( A_j \otimes \textbf{I} \bigg) \ket{\psi_{2\mathrm{XOR}}} = \bigg( \textbf{I} \otimes \frac{B_{ij} - B_{ji}}{\sqrt{2}} \bigg) \ket{\psi_{2\mathrm{XOR}}}     \text{, }
\end{align*}

\noindent are generalized in the third item below. To list the set of equivalent conditions for the $5$ $\mathrm{XOR}$ game below, introduce,

\begin{align*}
    \underline{\mathscr{P}_{5\mathrm{XOR}}}  \equiv    \underline{\mathscr{P}_{5\mathrm{XOR},1}}   \cup     \underline{\mathscr{P}_{5\mathrm{XOR},1,2}}  \cup \underline{\mathscr{P}_{5\mathrm{XOR},1,2,3}} \cup \cdots \cup \underline{\mathscr{P}_{5\mathrm{XOR},1,2,3,4,5}}               \text{, } 
\end{align*}

\noindent corresponding to the permutations,

\begin{align*}
      \underline{\mathscr{P}_{5\mathrm{XOR},1}} \equiv    \underset{\text{Permutations }\sigma \in S_5}{\sum} \bigg( E_{ijkl} \otimes  C_{ijk} \otimes  A_{\sigma ( i )} \otimes B_{ij} \otimes D_{ijkl }      \bigg)     \text{, }  \\     \underline{\mathscr{P}_{5\mathrm{XOR},1,2}} \equiv   \underset{\text{Permutations }\sigma, \sigma^{\prime}\in S_5}{\sum} \bigg(     E_{ijkl} \otimes  C_{ijk} \otimes  A_{\sigma ( i )} \otimes B_{\sigma^{\prime} ( ij) } \otimes D_{ijkl }    \bigg)    \text{, }   \end{align*}

      \begin{align*} \underline{\mathscr{P}_{5\mathrm{XOR},1,2,3}} \equiv  \underset{\text{Permutations }\sigma, \sigma^{\prime},\sigma^{\prime\prime}\in S_5}{\sum} \bigg(    E_{ijkl} \otimes  C_{\sigma^{\prime\prime} ( ijk ) } \otimes  A_{\sigma ( i )} \otimes B_{\sigma^{\prime} ( ij) } \\  \otimes D_{ijkl }    \bigg)   \text{, } \\ \vdots \\   \underline{\mathscr{P}_{5\mathrm{XOR},1,2,3,4,5}} \equiv  \underset{\text{Permutations }\sigma, \sigma^{\prime},\sigma^{\prime\prime}, \sigma^{\prime\prime\prime},\sigma^{\prime\prime\prime\prime} \in S_5 }{\sum} \bigg(         E_{\sigma^{\prime\prime\prime\prime} (ijkl)} \otimes  C_{\sigma^{\prime\prime} ( ijk ) } \otimes  A_{\sigma ( i )} \otimes B_{\sigma^{\prime} ( ij) } \\ \otimes D_{\sigma^{\prime\prime\prime} ( ijkl)  }                \bigg)   \text{. } 
\end{align*}

\noindent For the $5$-$\mathrm{XOR}$ game, the set of equivalent conditions is provided in the first section of the Appendix.

\bigskip

\noindent Beyond the two-player setting, mixed optimal solution states that groups of three players can adopt are of the form,

\begin{align*}
 \bra{\psi_{>3 \mathrm{XOR}}}                       G_{>3 \mathrm{XOR}} \ket{\psi_{>3 \mathrm{XOR}}}    \equiv   \bra{\psi_{>3 \mathrm{XOR}}}    \bigg( G_{3\mathrm{XOR}} \bra{ijkl} + G_{3\mathrm{XOR}} \bra{jikl} +   G_{3\mathrm{XOR}} \bra{ikjl } \\ + G_{3\mathrm{XOR}}  \bra{ijlk} \bigg)            \ket{\psi_{>3 \mathrm{XOR}}}         \text{. }
\end{align*}

\noindent The intermediate terms to the braket state of the optimal quantum state corresponding to the strategies of each player arise from entries of the 3-player $\mathrm{XOR}$ game tensor, which satisfies,

\begin{align*}
G_{3\mathrm{XOR}}  \propto      \underset{i_1 \in \mathcal{Q}_1, i_2 \in \mathcal{Q}_2, i_3 \in \mathcal{Q}_3}{\sum}   \bigg[  \ket{i_1} \bigg( \bra{i_1 i_2} \bra{i_1 i_2 i_3} \bigg)  + \ket{i_2} \bigg(   \bra{i_2 i_1} \bra{i_2 i_1 i_3} \bigg) +     \ket{i_3} \bigg( \bra{i_3 i_2} \bra{i_3 i_2 i_1} \bigg) \\  + \text{Higher order permutations}       \bigg]   
\\ 
    \equiv \underset{\text{acting on each } \ket{\psi_i}}{\underset{\text{permutations }\sigma_i }{\prod}}     \bigg[       \underset{\ket{\psi} \in \textbf{C}^{nm \times (n+m)}}{\underset{1 \leq i_1 < i_2 < i_3 \leq 3}{\sum}}   \big\{       1 \leq i^{\prime} \leq n + m : \sigma_i \psi   \big\}            \bigg] \\  \equiv {\underset{1 \leq i \leq n}{\prod}}    \bigg[   \underset{\text{acting on each } \ket{\psi}_i}{\underset{\text{permutations } \sigma_i}{\prod}}   \bigg[    \underset{\ket{\psi} \in \textbf{C}^{n m \times ( n +m )}}{\underset{1 \leq i_1 < i_2 < i_3 \leq 3}{\sum}} \sigma \ket{\psi}_i       \bigg]   \bigg]           \text{. }
\end{align*}

\noindent $\epsilon$-approximality of the $>3$ $\mathrm{XOR}$ game bias takes the form,

\begin{align*}
         \big( 1 - \epsilon_{>3 \mathrm{XOR}} \big) \beta \big( G_{>3\mathrm{XOR}} \big)  \leq        \underset{i_1 \in \mathcal{Q}_1, i_2 \in \mathcal{Q}_2, i_3 \in \mathcal{Q}_3}{\sum}  \bra{\psi_{>3 \mathrm{XOR}}}     \bigg(          \underset{\# \text{ players}}{\bigotimes} \text{Tensors of player} \\ \text{  observables} \big( i_1 , i_2 , i_3 \big) \bigg)        \ket{\psi_{>3 \mathrm{XOR}}}   \leq  \beta \big( G_{>3\mathrm{XOR}} \big)    \text{, }
\end{align*}

\noindent given the existence of a constant $\epsilon$ taken to be sufficiently small. For the following quantities, denote the set of questions, $\mathcal{Q}_1$, as the set of questions distributed to the first player, and so on. The Schmidt basis takes the form,

\begin{align*}
          \underset{i_1 \in \mathcal{Q}_1, i_2 \in \mathcal{Q}_2, i_3 \in \mathcal{Q}_3}{\sum} \sqrt{\lambda_i} \bigg(         \underset{\# \text{ players}}{\bigotimes}        \text{Tensors of player observables} \big( i_1 , i_2 , i_3 \big)    \bigg)    \text{. }
\end{align*}

\noindent In {[37]}, objects involved with defining the two-dimensional $\mathrm{CHSH}$ Schmidt basis take the form,

\begin{align*}
    \overset{s 2^{\lfloor \frac{n}{2} \rfloor}}{\underset{i=1}{\sum} } \sqrt{\lambda_i}  \big(  \ket{u_i} \otimes \ket{v_i}  \big)  \text{, } 
\end{align*}

\noindent into a tensor product over the quantum states which respectively corresponding to each $u_i$ and $v_i$ are equal in blocks $\lambda$ between each term of the summation over $i$ above, as (\textbf{Theorem} \textit{5}, {[37]}),

\begin{align*}
        \lambda_i = \cdots = \lambda_{i+1} \text{, } \forall i    \text{. } 
\end{align*}

\noindent Moreover, with respect to some basis $\{ \ket{u_i} \}_{i \in \mathcal{I}}$ of $\textbf{C}^{d_A}$, the observable $A_i$ decomposes as,

\begin{align*}
   A_i   =  \mathrm{diag} \big(   A^{(1)}_i , \cdots , A^{(s)}_i , C_i       \big)    \text{, } 
\end{align*}

\noindent where each block diagonal component $A^{(s)}_i$, of size $2^{\lfloor \frac{n}{2} \rfloor} \times 2^{\lfloor \frac{n}{2} \rfloor}$, is such that each $A_i$ acts on $\mathrm{span} \big( \ket{u_{(j-1)2^{\lfloor \frac{n}{2} \rfloor}+1}} , \cdots$ $, \ket{u_{j 2^{\lfloor \frac{n}{2} \rfloor}}} \big)$ for $1 \leq i \leq n$, with,

\begin{align*}
  C_i \equiv \big\{ \pm 1 \big\}   \text{, } 
\end{align*}

\noindent $\forall \big( s - 1 \big)  2^{\lfloor \frac{n}{2} \rfloor} \leq i \leq s 2^{\lfloor \frac{n}{2} \rfloor}$. For a basis $\{ \ket{v_i} \}_{i \in \mathcal{I}}$ of $\textbf{C}^{d_B}$, the observable $B_{jk}$ decomposes as,

\begin{align*}
    B_{jk}  =  \mathrm{diag} \big(  B^{(1)}_{jk} , \cdots , B^{(s)}_{jk} , D_{jk}       \big)       \text{, } 
\end{align*}

\noindent where each block diagonal component $B^{(s)}_{jk}$, also of size $2^{\lfloor \frac{n}{2} \rfloor} \times 2^{\lfloor \frac{n}{2} \rfloor}$, is such that each $B_{jk}$ acts on $\mathrm{span} \big(   \ket{v_{ (j - 1 )  2^{\lfloor \frac{n}{2} \rfloor+1}}} , \cdots , \ket{v_{j 2^{\lfloor \frac{n}{2} \rfloor}}} \big)$ for $1 \leq j \neq k \leq n$, with,

\begin{align*}
  D_{jk} \equiv \big\{ \pm 1 \big\}  \text{. } 
\end{align*}

\noindent The construction above can be adapted for obtaining tensors whose dimensionality is spanned by the set of all possible responses from previous players who have participated in the game.

\bigskip

\noindent In the $N$-player setting, straightforwardly the quantities involving with expressing inequalities of the form above, before, and after, taking the supremum over all admissible strategies $\mathcal{S}$, take the form,

\begin{align*}
         \big( 1 - \epsilon_{N \mathrm{XOR}} \big) \beta \big( G_{N\mathrm{XOR}} \big)  \leq        \underset{i_1 \in \mathcal{Q}_1, i_2 \in \mathcal{Q}_2, \cdots, i_n \in \mathcal{Q}_n}{\sum}  \bra{\psi_{N \mathrm{XOR}}}                  \bigg(           \underset{\# \text{ players}}{\bigotimes}  \text{Tensors of } \\ \text{player observables} \big( i_1 , i_2 ,  \cdots , i_n  \big)    \bigg)             \ket{\psi_{N \mathrm{XOR}}}   \leq \beta \big( G_{N\mathrm{XOR}} \big)    \text{.}
\end{align*}

\noindent The $N$-player Schmidt basis takes the form,

\begin{align*}
         \underset{i_1 \in \mathcal{Q}_1, i_2 \in \mathcal{Q}_2, \cdots, i_n \in \mathcal{Q}_n}{\sum}  \sqrt{\lambda_i} \bigg(   \underset{\# \text{ players}}{\bigotimes}       \text{Tensors of player observables} \big( i_1 , i_2 , i_3 , \cdots , i_n \big)       \bigg)    \text{. }
\end{align*}

\noindent The generalization of the Schmidt basis provided above, for any number of players, implies,

\begin{align*}
\mathrm{sign} \bigg(   \bra{\psi_{N \mathrm{XOR},L}}  \bigg( \underset{\text{Questions}}{\bigotimes} \text{Player tensor observables} \bigg)  \ket{\psi_{N\mathrm{XOR},L}} \bigg)  \equiv \pm \text{, }
\end{align*}

\noindent given the fact that,

\begin{align*}
   \bra{\psi_{N \mathrm{XOR},L}}  \bigg( \underset{\text{Questions}}{\bigotimes} \text{Player tensor observables} \bigg)  \ket{\psi_{N\mathrm{XOR},L}} \propto \omega_{N\mathrm{XOR}} \text{, }
\end{align*}

\noindent for,

\begin{align*}
  \ket{\psi_{N\mathrm{XOR},L}} \equiv \frac{1}{\sqrt{ 2^{\lfloor \frac{N}{2}\rfloor }}} \underset{\text{Schmidt blocks}}{\sum} \bigg( \underset{\text{Questions}}{\bigotimes} \ket{\text{Player responses to questions}} \bigg) \\ \equiv   \frac{1}{\sqrt{ 2^{\lfloor \frac{N}{2}\rfloor }}} \underset{(l-1) 2^{\lfloor \frac{N}{2} \rfloor } \leq i \leq l 2^{\lfloor \frac{N}{2} \rfloor }}{\sum} \bigg( \underset{\text{Questions } i}{\bigotimes} \ket{\text{Player responses to question } i} \bigg) \text{. }
\end{align*}

\noindent With states $\ket{\psi_{N\mathrm{XOR},L}}$ introduced above, in an analogous way that one can define,

\begin{align*}
 \ket{\psi_{2\mathrm{XOR}}} \equiv   \underset{1 \leq i \leq s}{ \sum}    \sqrt{2^{\lfloor \frac{n}{2} \rfloor} \lambda_{i 2^{\lfloor \frac{n}{2} \rfloor}}} \ket{\psi_L}     =  \underset{1 \leq i \leq s}{ \sum}    \sqrt{2^{\lfloor \frac{n}{2} \rfloor} \lambda_{i 2^{\lfloor \frac{n}{2} \rfloor}}}   \bigg(        \frac{1}{\sqrt{2^{\lfloor \frac{n}{2} \rfloor}}} \underset{(i-1) 2^{\rfloor \frac{n}{2} \lfloor } \leq i^{\prime} \leq i 2^{\rfloor \frac{n}{2} \lfloor }}{\sum} \bigg( \ket{u_{i^{\prime}} } \otimes \ket{u_{i^{\prime}}} \bigg)      \bigg)  \\ \equiv  \underset{1 \leq i \leq s}{ \sum}    \sqrt{2^{\lfloor \frac{n}{2} \rfloor} \lambda_{i 2^{\lfloor \frac{n}{2} \rfloor}}}   \bigg(        \frac{1}{\sqrt{2^{\lfloor \frac{n}{2} \rfloor}}} \underset{(i-1) 2^{\rfloor \frac{n}{2} \lfloor } \leq i^{\prime} \leq i 2^{\rfloor \frac{n}{2} \lfloor }}{\sum} \bigg( \ket{u_{i^{\prime}}}^{\otimes 2} \bigg)      \bigg)  \\ \equiv  \underset{(i-1) 2^{\rfloor \frac{n}{2} \lfloor } \leq i^{\prime} \leq i 2^{\rfloor \frac{n}{2} \lfloor }}{\underset{1 \leq i \leq s}{ \sum}}    \sqrt{2^{\lfloor \frac{n}{2} \rfloor} \lambda_{i 2^{\lfloor \frac{n}{2} \rfloor}}}   \bigg(        \frac{1}{\sqrt{2^{\lfloor \frac{n}{2} \rfloor}}}  \bigg( \ket{u_{i^{\prime}}}^{\otimes 2} \bigg)      \bigg)    \text{, } 
\end{align*}

\noindent from the two-dimensional Schmidt basis, one can define,

\begin{align*}
      \ket{\psi_{N\mathrm{XOR}}} \equiv   \underset{1 \leq i \leq s}{ \sum}    \sqrt{N^{\lfloor \frac{n}{2} \rfloor} \lambda_{i N^{\lfloor \frac{n}{2} \rfloor}}} \ket{\psi_{N\mathrm{XOR},L}}     =  \underset{1 \leq i \leq s}{ \sum}    \sqrt{N^{\lfloor \frac{n}{2} \rfloor} \lambda_{i N^{\lfloor \frac{n}{2} \rfloor}}}   \bigg(        \frac{1}{\sqrt{N^{\lfloor \frac{n}{2} \rfloor}}} \\ \times \underset{(i-1) N^{\rfloor \frac{n}{2} \lfloor } \leq i^{\prime} \leq i N^{\rfloor \frac{n}{2} \lfloor }}{\sum} \bigg( \ket{u_{i^{\prime}} }  \otimes \ket{u_{i^{\prime}}}  \otimes \overset{N-3}{\cdots}  \otimes \ket{u_{i^{\prime}}}  \bigg)      \bigg)    \\  \equiv      \underset{1 \leq i \leq s}{ \sum}    \sqrt{N^{\lfloor \frac{n}{2} \rfloor} \lambda_{i N^{\lfloor \frac{n}{2} \rfloor}}}   \bigg(        \frac{1}{\sqrt{N^{\lfloor \frac{n}{2} \rfloor}}} \underset{(i-1) N^{\rfloor \frac{n}{2} \lfloor } \leq i^{\prime} \leq i N^{\rfloor \frac{n}{2} \lfloor }}{\sum} \bigg(  \ket{u_{i^{\prime}}}^{\otimes N} \bigg)    \bigg)  \\  \equiv   \underset{(i-1) N^{\rfloor \frac{n}{2} \lfloor } \leq i^{\prime} \leq i N^{\rfloor \frac{n}{2} \lfloor }}{\underset{1 \leq i \leq s}{ \sum}}    \sqrt{N^{\lfloor \frac{n}{2} \rfloor} \lambda_{i N^{\lfloor \frac{n}{2} \rfloor}}}   \bigg(        \frac{1}{\sqrt{N^{\lfloor \frac{n}{2} \rfloor}}}  \bigg(  \ket{u_{i^{\prime}}}^{\otimes N} \bigg)    \bigg)     \text{, }
\end{align*}

\noindent from the three-dimensional Schmidt basis. Moreover, equipped with the $>3$ $\mathrm{XOR}$ game tensor, the semidefinite program corresponding to the duality gap takes the form,

\begin{align*}
 \underset{i_1 \in \mathcal{Q}_1, i_2 \in \mathcal{Q}_2, i_3 \in \mathcal{Q}_3}{\prod} \bigg[  \underset{\text{for each player }i}{\underset{\text{Questions } n_i}{\bigcup}}    \bigg\{  \underset{\forall Z 0 \succcurlyeq 0 , 1 \leq i \leq n_i, F_i  \cdot  G_{>3\mathrm{XOR}} \equiv c_i }{\mathrm{sup}}    \big( y_i F_i - G_{>3 \mathrm{XOR}} \big) Z   \bigg\}  \bigg]    \text{, }
\end{align*}

\noindent for the primal feasible solution $Z_{>3 \mathrm{XOR}} \equiv Z$. As a further generalization of the optimal, and approximately optimal, framework, for the $>3$ $\mathrm{XOR}$ setting the primal feasible solution $Z$ interacts with the action of an intertwining operation on tensor products of player observables gathered by each participating party before preparing a response to the referee, which takes the form,

\begin{align*}
\bigg| \bigg| T_{>3 \mathrm{XOR}} \otimes \bigg( \underset{i_1 \in \mathcal{Q}_1, i_2 \in \mathcal{Q}_2, \cdots, i_n \in \mathcal{Q}_n}{\prod} \text{Tensors of player observables} \big( i_1 , \cdots , i_n \big)    \bigg) \\ - \widetilde{\bigg( \underset{i_1 \in \mathcal{Q}_1, i_2 \in \mathcal{Q}_2, \cdots, i_n \in \mathcal{Q}_n}{\prod}} \text{Tensors of player observables} \big( i_1 , \cdots , i_n \big)    \bigg)\otimes T_{>3 \mathrm{XOR}}  \bigg| \bigg|  \text{, }
\end{align*}

\noindent for a suitable linear operator $T_{>3 \mathrm{XOR}}$. The tensor product that is exchanged with respect to $T_{>3 \mathrm{XOR}}$ satisfies the containment,

\begin{align*}
  \text{Tensors of player observables} \big( i_1 \big) \subsetneq \text{Tensors of player observables} \big( i_1 , i_2 \big) \subsetneq  \cdots  \\ \subsetneq   \underset{i_1 \in \mathcal{Q}_1, i_2 \in \mathcal{Q}_2, \cdots, i_n \in \mathcal{Q}_n}{\bigcup} \big\{ \text{Tensors of player observables} \big( i_1, \cdots , i_n \big) \big\}  \text{. }
\end{align*}

\noindent In the absence of the action of the suitable linear operator $T_{>3 \mathrm{XOR}}$ and its associated intertwining action in tensor products over observables gathered by each player, inequalities for quantifying error bounds directly follow the structure for those of the $3$ $\mathrm{XOR}$ game, which are generated by inequalities of the form,

\begin{align*}
 \bigg| \bigg|   \bigg[  \bigg( \textbf{I} \otimes \big( \text{2 nd player tensor observable} \big( i_1 , i_2 \big) \big) \otimes \textbf{I} \otimes \cdots \otimes \big( \text{N th player tensor observable} \big( i_1 \\ , \cdots , i_n \big) \bigg) 
 - \bigg( \big( \widetilde{\text{2 nd player tensor observable} \big( i_1 , i_2 \big) }    \big) \otimes \textbf{I} \otimes \cdots  \otimes \big( \text{N th player tensor} \\ \text{ observable} \big( i_1  , \cdots , i_n \big) \big)       \bigg)      \bigg]  \ket{\psi_{>3\mathrm{XOR}}}     \bigg| \bigg|    \text{. }
\end{align*}

\noindent By induction on the number of players, given the symmetric group action on tensor products of player observables from previously defined objects, one also has,

\begin{align*}
 \bigg| \bigg|   \bigg[  \bigg( \big(  \text{1 st player tensor observable} \big( i_1 \big)\big)  \otimes \textbf{I} \otimes \big(\text{3 rd player tensor observable} \big( i_1 , i_2 , i_3 \big) \big)  \otimes \overset{N-4}{\cdots} \otimes \\   \big( \text{N th player tensor observable} \big( i_1 , \cdots , i_n \big)    \bigg)  -  \bigg( \big( \text{1 st player tensor observable} \big( i_1 \big) \big) \\ \otimes \big( \widetilde{\text{3 rd player tensor observable} \big( i_1 , i_2 , i_3 \big) }         \otimes \textbf{I} \otimes \overset{N-4}{\cdots} \otimes \big(  \text{N th player tensor} \\ \text{observable} \big( i_1 , \cdots , i_n \big) \big)  \bigg)     \bigg]    \ket{\psi_{>3 \mathrm{XOR}}}    \bigg| \bigg|    \text{. }
\end{align*}

\noindent Besides the two inequalities above which are obtained from the intertwining action, along with error bound inequalities, inequalities corresponding to the swap operation for upper bounding the magnitude of operators along on optimal solution states for the $>3$ $\mathrm{XOR}$ game take the form,

\begin{align*}
 \bigg| \bigg|  \bigg[  \bigg( \bigg( \underset{i_1 \in \mathcal{Q}_1}{\prod}   \text{1 st player tensor observable} \big( i_1 \big) \bigg) \otimes \textbf{I} \otimes \bigg( \underset{i_1 \in \mathcal{Q}_1, i_2 \in \mathcal{Q}_2, \cdots, i_n \in \mathcal{Q}_n}{\prod} \text{Tensors of player observa-} \\ \text{bles} \big( i_1 , \cdots , i_n \big)  \bigg) \bigg) \\   -             \bigg( \textbf{I} \otimes \bigg(                 \underset{i_2 + 1 \equiv i_2 \oplus 1}{\underset{i_1 + 1 \equiv i_1 \oplus 1}{\underset{ i_2 \in \mathcal{Q}_2}{\underset{i_1 \in \mathcal{Q}_1}{\prod}}}}      \text{2 nd player tensor observables} \big( i_1 , i_2 \big)    \bigg) \otimes \bigg(  \underset{i_1 \in \mathcal{Q}_1, i_2 \in \mathcal{Q}_2, \cdots, i_n \in \mathcal{Q}_n }{\prod} \text{Tensors of player} \\ \text{ observables} \big( i_1 ,  \cdots , i_n \big)        \bigg) \bigg)             \bigg]  \ket{\psi_{>3 \mathrm{XOR}}} \bigg| \bigg|    \text{. }
\end{align*}

\noindent The remaining inequalities for the swap operation are straightforwardly obtained by applying permutations to the products of tensor observables gathered by each player. In the forthcoming expressions, it is convenient to denote tensors for the players as,

\begin{align*}
C^{(N-3)}_{i_1,\cdots, i_N} \equiv \underset{i_1 \in \mathcal{Q}_1,\cdots,i_N \in \mathcal{Q}_N}{\bigcup} \big\{  \text{N th player tensor observable} \big(i_1, \cdots, i_N \big)       \big\}  \text{, }
\end{align*}

\noindent where,

\begin{align*}
  C_{i_1,i_2,i_3} \equiv \underset{i_1,i_2,i_3}{\bigcup} \big\{  \text{3 rd player tensor observable} \big( i_1,i_2,i_3 \big)        \big\}   \text{. }
\end{align*}

\noindent Inequalities of the above form are established by arguing that inequalities of the following form hold,

\begin{align*}
  \bigg| \bigg|    \bigg[  \bigg( \big( \text{1 st player tensor observable} \big( i_1 \big)  \big) \otimes \textbf{I} \otimes \bigg(    \underset{i_1 \in \mathcal{Q}_1, i_2 \in \mathcal{Q}_2, \cdots, i_n \in \mathcal{Q}_n}{\prod} \text{Tensors of player} \\ \text{observables} \big( i_1 , \cdots , i_n \big)    \bigg)   \bigg)   -         \bigg(   \textbf{I} \otimes \bigg(             \frac{1}{\sqrt{2}} \bigg(  \frac{\mathcal{T}_2}{\big| \pm \mathcal{T}_2 \big| }  \bigg)            \bigg) \otimes \textbf{I} \bigg)       \bigg]       \ket{\psi_{>3 \mathrm{XOR}}}    \bigg| \bigg|  \text{, }  \end{align*}

  \noindent for,

  \begin{align*}
  \mathcal{T}_2  \equiv     \text{2 nd player tensor observable} \big( i_1 , i_2 \big)  + \underset{i_1 \in \mathcal{Q}_1, i_2 \in \mathcal{Q}_2, \cdots, i_n \in \mathcal{Q}_n}{\underset{\text{permutations } \sigma}{\sum}}     \text{Tensors of player observa-} \\  \text{bles} \big( \sigma i_1 , \sigma i_2 ,  \cdots , \sigma i_n \big)                         \text{, }
  \end{align*}
  
  \noindent while for the $N$-$\mathrm{XOR}$ setting inequalities of the following form hold,
  
  \begin{align*}
  \bigg| \bigg|    \bigg[  \bigg( \textbf{I} \otimes \textbf{I} \otimes \cdots \otimes \bigg(  \underset{i_1 \in \mathcal{Q}_1, i_2 \in \mathcal{Q}_2, \cdots, i_n \in \mathcal{Q}_n}{\prod}  \text{Tensors of player observables} \big( i_1 , \cdots , i_n \big)     \bigg) \otimes \big(  \text{N th} \text{ pla-} \\ \text{yer tensor observable} \big( i_1 ,  \cdots , i_n \big)      \big)        \bigg)   - \bigg(         \textbf{I} \otimes \textbf{I} \otimes \cdots \otimes \big( \text{N th player tensor observable} \big( i_1 \\ , \cdots , i_n \big) \big)  \otimes \bigg( \frac{1}{\sqrt{N-1}} \bigg( \bigg( \frac{\mathcal{T}_3}{\big| \pm \mathcal{T}_3 \big| }
  \bigg)   \bigg]     \ket{\psi_{N \mathrm{XOR}}}    \bigg| \bigg|    \text{, }
\end{align*}

\noindent for,

\begin{align*}
 \mathcal{T}_3 \equiv  \text{3 rd player tensor observable} \big( i_1 , \cdots , i_n \big)   + \underset{i_1 \in \mathcal{Q}_1, i_2 \in \mathcal{Q}_2, \cdots, i_n \in \mathcal{Q}_n}{\underset{\text{permutations }\sigma}{\sum}} \text{Tensors of player observa-} \\ \text{bles} \big( \sigma i_1 , \sigma i_2 ,  \cdots , \sigma i_n \big)     \text{. }
\end{align*}

\noindent Inequalities for the error bounds in the $N$ player $\mathrm{XOR}$ setting take the form, (\textbf{EB} $N$ $\mathrm{XOR}$), indicated in the table below, which we denote with $(**)$.

\begin{table}[ht]
\caption{Components of the Error bound for the N-XOR game, $ (\textbf{EB}- N\mathrm{XOR}) $} 
\centering 
\begin{tabular}{c c c c} 
\hline\hline 
 Error Bound Contributions \\ [0.5ex] 
\hline      $ {\underset{j \in \mathcal{Q}_2}{\underset{i \in \mathcal{Q}_1}{\mathlarger{\sum}}}}           \bigg| \bigg|   \bigg[     \bigg( \bigg(       \frac{A_i + A_j}{\sqrt{2}}      \bigg) \bigotimes \bigg( \underset{1 \leq k \leq n-1}{\bigotimes} \textbf{I}_k \bigg) \bigg) -    \bigg(   \textbf{I} \bigotimes B_{ij}  \bigotimes \bigg( \underset{1 \leq k \leq n-2}{\bigotimes}  \textbf{I}_k \bigg)     \bigg)                             \bigg]  \ket{\psi_{N \mathrm{XOR}}}    \bigg| \bigg|^2 $      \\ 
$\underset{k \in \mathcal{Q}_3}{\underset{j \in \mathcal{Q}_2}{\underset{i \in \mathcal{Q}_1}{\mathlarger{\sum}}}}    \bigg| \bigg|   \bigg[                     \bigg( \textbf{I} \bigotimes \bigg( \frac{B_{ij} + B_{ji}}{\sqrt{2}} \bigg) \bigotimes \bigg( \underset{1 \leq  k \leq n-2}{\bigotimes }  \textbf{I}_k \bigg) \bigg) - \bigg(      \textbf{I} \bigotimes \textbf{I} \bigotimes C_{ijk} \bigotimes \bigg( \underset{1 \leq k \leq n-3}{\bigotimes} \textbf{I}_k  \bigg)       \bigg)    \bigg]     \ket{\psi_{N \mathrm{XOR}}}    \bigg| \bigg|^2$      \\  $ {\underset{j \in \mathcal{Q}_2}{\underset{i \in \mathcal{Q}_1}{\mathlarger{\sum}}}}  \bigg| \bigg|   \bigg[                    \bigg( \bigg( \frac{A_i - A_j}{\sqrt{2}} \bigg) \bigotimes \bigg( \underset{1 \leq k \leq n-1}{\bigotimes}  \textbf{I}_k \bigg) \bigg)      -            \bigg( \textbf{I} \bigotimes B_{ji} \bigotimes \bigg( \underset{1 \leq k \leq n-2}{\bigotimes}  \textbf{I}_k \bigg) \bigg)  \bigg]      \ket{\psi_{N \mathrm{XOR}}}     \bigg| \bigg|^2$         \\  $\underset{k \in \mathcal{Q}_3}{\underset{j \in \mathcal{Q}_2}{\underset{i \in \mathcal{Q}_1}{\mathlarger{\sum}}}}    \bigg| \bigg|   \bigg[     \bigg( \textbf{I} \bigotimes \bigg( \frac{B_{ij} - B_{ji}}{\sqrt{2}}  \bigg) \bigotimes \bigg( \underset{1 \leq k \leq n-3}{\bigotimes} \textbf{I}_k \bigg) \bigg)          -   \bigg( \textbf{I} \bigotimes \textbf{I}    \bigotimes C_{jik} \bigotimes \bigg( \underset{1 \leq k \leq n-3}{\bigotimes} \textbf{I}_k \bigg) \bigg)            \bigg]       \ket{\psi_{N \mathrm{XOR}}}      \bigg| \bigg|^2$       \\  $\underset{k \in \mathcal{Q}_3}{\underset{j \in \mathcal{Q}_2}{\underset{i \in \mathcal{Q}_1}{\mathlarger{\sum}}}}  \bigg| \bigg|   \bigg[       \bigg( \textbf{I} \bigotimes \textbf{I} \bigotimes C_{ijk} \bigotimes \bigg( \underset{1 \leq k \leq n-3}{\bigotimes}   \textbf{I}_k \bigg) \bigg)                 -      \bigg(               \bigg( \frac{A_i + A_j}{\sqrt{2}} \bigg) \bigotimes \textbf{I} \bigotimes \bigg( \underset{1 \leq k \leq n-2}{\bigotimes}  \textbf{I}_k \bigg) \bigg)             \bigg)       \bigg]      \ket{\psi_{N \mathrm{XOR}}}       \bigg| \bigg|^2$       \\    \vdots      \\   $\underset{k \in \mathcal{Q}_3}{\underset{j \in \mathcal{Q}_2}{\underset{i \in \mathcal{Q}_1}{\mathlarger{\sum}}}}  \bigg| \bigg|   \bigg[     \bigg( \bigg( \underset{1 \leq k \leq n-1}{\bigotimes}           \textbf{I}_k      \bigg) \bigotimes     C^{\{n-2\}}_{ijk}       \bigg)                     -        \bigg(   \bigg( \underset{1 \leq k \leq n-1}{\bigotimes}           \textbf{I}_k   
 \bigg) \bigotimes  \bigg( \frac{1}{\sqrt{\# \sigma }} \bigg(   \underset{\text{permutations }\sigma}{\sum}      C^{\{n-2\}}_{\sigma ( ijk )}            \bigg)  \bigg)          \bigotimes \textbf{I}          \bigg)     \bigg]       \ket{\psi_{N \mathrm{XOR}}}      \bigg| \bigg|^2 $     \\  [1ex] 
\hline 
\end{tabular}
\label{table:nonlin} 
\end{table}


\noindent From the table above, in the last summation the quantity $C^{\{n-2\}}$ denotes,

\begin{align*}
      \underset{i_1 \in \mathcal{Q}_1 , \cdots , i_n \in \mathcal{Q}_{n-1}}{\bigcup}    \big\{ \text{N th player tensor observables} \big( i_1 , \cdots , i_{n-1} \big)  \big\}  \text{. }
\end{align*}

\bigskip

\noindent By a direct adaptation of computations provided for the $\mathrm{XOR}$ and $\mathrm{CHSH}\big( n \big)$, for $n \equiv 2$, arguments in {[]} provide an upper bound,

\begin{align*}
 2 n \big( n -1 \big) \epsilon   \text{, }
\end{align*}

\noindent given some $\epsilon$ sufficiently small, can be applied to obtain the upper bound,

\begin{align*}
  n \bigg( \underset{1 \leq j \leq N-1}{\prod} \big(  n - j  \big) \bigg)   \epsilon^{\prime}    \text{, }
\end{align*}

\noindent given some $\epsilon^{\prime}$ sufficiently small. This same collection of constants is put to further use in \textit{1} for generalizing second order $\mathrm{FFL}$ error bounds.

\bigskip

\noindent \textbf{Theorem} $\textit{1}^{*}$ (\textit{N-XOR permutation error bounds}, \textit{2.2.1}, \textbf{Theorem} \textit{4}, {[37]}, \textbf{Theorem} \textit{2}, {[44]}, \textbf{Theorems} \textit{1}-\textit{6} in \textit{1.5}).

\begin{align*}
        (**) \leq     N!          n \bigg( \underset{1 \leq j \leq N-1}{\prod}  \big( n - j \big) \bigg)            \text{. }
\end{align*}

\noindent \textit{Proof of Theorem $1^{*}$}. To argue that the desired upper bound holds from direct computation, recall the following result, as a generalization of the $\epsilon$-approximality result of {[37]}:

\bigskip

\noindent \textbf{Theorem} \textit{1} (\textit{approximately optimal quantum strategies for the nonlocal XOR game}, \textbf{Theorem} \textit{4}, {[44]}). For $\pm$ observables $A_i$ and $B_{jk}$, given a bipartite state $\psi$, TFAE:

\begin{itemize}
 \item[$\bullet$] \underline{\textit{First characterization of approximate optimality}:} An $\epsilon$-approximate $\mathrm{CHSH}\big( n \big)$ satisfies $\mathrm{(0)}$.

 \item[$\bullet$] \underline{\textit{Second characterization of approximate optimality}:} For an $\epsilon$-approximate quantum strategy, 

  \begin{align*}
     \underset{1 \leq i < j \leq n}{\sum} \bigg[ \text{ }  \bigg| \bigg|   \bigg[         \big( \frac{A_i + A_j}{\sqrt{2}} \big) \otimes I \bigg] \ket{\psi}    -   \big[  I \otimes B_{ij} \big] \ket{\psi}   \bigg|\bigg|^2 + \bigg| \bigg|   \bigg[ \big( \frac{A_i - A_j}{\sqrt{2}} \big)  \otimes I  \bigg] \ket{\psi}  \\   -   \big[  I \otimes B_{ji} \big]  \ket{\psi}     \bigg|\bigg|^2 \text{ }  \bigg]  \leq 2n \big( n - 1 \big) \epsilon \text{. }
 \end{align*}
 \item[$\bullet$] \underline{\textit{Reversing the order of the tensor product for observables}:} Related to the inequality for $\epsilon$-approximate strategies above, another inequality,

\begin{align*}
    \underset{1 \leq i < j \leq n}{\sum} \bigg[  \text{ } \bigg| \bigg| \big[ A_i \otimes I \big] \ket{\psi}    - \bigg[ I \otimes \big( \frac{B_{ij} + B_{ji}}{\sqrt{2}} \big)  \bigg] \ket{\psi}   \bigg| \bigg|^2  + \bigg| \bigg|    \big[ A_j  \otimes I \big] \ket{\psi} \\  -   \bigg[   I  \otimes \big( \frac{B_{ij} - B_{ji}}{\sqrt{2}} \big) \bigg] \ket{\psi}    \big| \big|^2     \text{ }   \bigg]  \leq 2n \big( n - 1 \big) \epsilon \text{, } 
\end{align*}

 \noindent also holds. 
 
 \item[$\bullet$] \underline{\textit{Characterization of exact optimality}}: For $\epsilon \equiv 0$,

\begin{align*}
    \underset{1 \leq i < j \leq n}{\sum} \bigg[ \text{ }  \bigg| \bigg|  \bigg[  \big( \frac{A_i + A_j}{\sqrt{2}} \big) \otimes I \bigg] \ket{\psi}    -  \big[ I \otimes B_{ij} \big] \ket{\psi}   \bigg| \bigg|^2  \bigg]  =  -  \underset{1 \leq i < j \leq n}{\sum} \bigg[  \text{ }   \bigg| \bigg|     \bigg[  \big( \frac{A_i - A_j}{\sqrt{2}} \big)  \otimes I \bigg]  \ket{\psi} \\   -   \big[ I \otimes B_{ji} \big]  \ket{\psi}    \bigg| \bigg|^2  \bigg]      \text{, } 
\end{align*}

\noindent corresponding to the first inequality, and,

\begin{align*}
       \underset{1 \leq i < j \leq n}{\sum} \bigg[ \text{ }  \bigg| \bigg|       \big[  A_i \otimes I \big]  \ket{\psi}    - \bigg[   I \otimes \big( \frac{B_{ij} + B_{ji}}{\sqrt{2}} \big)  \bigg]  \ket{\psi}   \bigg|\bigg|^2 \bigg] = -  \underset{1 \leq i < j \leq n}{\sum}   \bigg[  \bigg| \bigg|  \big[   A_j  \otimes I \big] \ket{\psi} \\  -  \bigg[    I \otimes \big( \frac{B_{ij} - B_{ji}}{\sqrt{2}} \big) \bigg]  \ket{\psi}     \bigg|\bigg|^2  \bigg]             \text{, } 
\end{align*}

\noindent corresponding to the second inequality.

\end{itemize}

\bigskip

\noindent  With the result above, to further generalize the set of equivalent conditions as provided in previous discussions for the $3\mathrm{XOR}$ game, and beyond, given the existence of suitable $\epsilon_{N\mathrm{XOR}}$, and a previously determined constant $C_{N\mathrm{XOR}} \equiv C$, the \textit{symmetrized} N $\mathrm{XOR}$ game tensor, and other terms,

\begin{align*}
   \underset{1 \leq i \leq N^2}{\sum} y_i E_{ii} - G_{\mathrm{Sym},N\mathrm{XOR}} \equiv  \underset{1 \leq i \leq N^2}{\sum} y_i E_{ii} - G_{\mathrm{Sym}}   \text{, }
\end{align*}

\noindent equals,

\begin{align*}
     \frac{1}{ \bigg( C_{N\mathrm{XOR}} \omega_{N\mathrm{XOR}} \bigg) n \bigg( \underset{1 \leq j \leq N-1}{\prod}  \big( n - j \big) \bigg) }             \underset{i_N \in \mathcal{Q}_N}{\underset{\vdots}{\underset{i_1 \in \mathcal{Q}_1}{\sum}}}   \bigg(  \big(    u^{\prime}_{i_1i_2 \cdots i_N}   -    v^{\prime}_{i_1i_2 \cdots i_N}  \big) \big(   u^{\prime}_{i_1i_2 \cdots i_N}   -    v^{\prime}_{i_1i_2 \cdots i_N}    \big)^{\textbf{T}}   \\   +  \big(    u^{\prime}_{i_2 i_1 i_3 \cdots i_N}   -    v^{\prime}_{i_2 i_1 i_3  \cdots i_N}  \big)  \big( u^{\prime}_{i_2 i_1 i_3 \cdots i_N}   -    v^{\prime}_{i_2 i_1 i_3  \cdots i_N}  \big)^{\textbf{T}}  + \cdots +     \big(    u^{\prime}_{i_1 i_2 i_3 \cdots i_{N} i_{N-1}}   -    v^{\prime}_{i_1 i_2 i_3 \cdots i_{N} i_{N-1}}  \big) \\ \times    \big(   u^{\prime}_{i_1 i_2 i_3 \cdots i_{N} i_{N-1}}    -    v^{\prime}_{i_1 i_2 i_3 \cdots i_{N} i_{N-1}}   \big)^{\textbf{T}}     \bigg)      \text{,}
\end{align*}

\noindent given the superposition of states,

\begin{align*}
        u^{\prime}_{i_1 i_2 \cdots i_N} \equiv \frac{1}{\sqrt{N}} \bigg( \underset{1 \leq j \leq N}{\sum} \ket{\text{Player } j \text{ state}} \bigg)      \text{, } v^{\prime}_{i_1 i_2 \cdots i_N} \equiv \ket{i_1 i_2 \cdots i_N } \\  u^{\prime}_{i_2 i_1 i_3 \cdots i_N} \equiv \frac{1}{\sqrt{N}} \bigg( \ket{\text{Player 1 state}} - \ket{\text{Player 2 state}} \\ + \underset{3 \leq j \leq N}{\sum} \ket{\text{Player } j \text{ state}} \bigg) \text{, } v^{\prime}_{i_2 i_1 i_3 \cdots i_N} \equiv \ket{i_2 i_1 i_3 \cdots i_N } \\ \vdots \\   u^{\prime}_{i_1 i_2 i_3 \cdots i_N i_{N-1}} \equiv \frac{1}{\sqrt{N}} \bigg(  \underset{1 \leq j \leq N-1}{\sum} \ket{\text{Player } j \text{ state}} \\ - \ket{\text{Player } N \text{ state}}  \bigg)       \text{, } v^{\prime}_{i_1 i_2 i_3 \cdots i_N i_{N-1}} \equiv \ket{i_1 i_2 i_3 \cdots i_N i_{N-1}} \text{, }
\end{align*}

\noindent from which we conclude the argument. \boxed{}

\bigskip

\noindent An implication of the above result which demonstrates that the desired upper bound, dependent upon both a combinatorial factor and the total number of players in the game, also implies the following result.

\bigskip

 \noindent \textbf{Lemma} \textit{T-1} (\textit{positive semidefiniteness}). Under the assumptions of the previous result, the operator,

\begin{align*}
  \underset{1 \leq i \leq n^N}{\sum}   y_{N\mathrm{XOR},i} E_{N\mathrm{XOR},ii} - G_{N\mathrm{XOR},\mathrm{Sym}}     \text{, }
\end{align*}

 \noindent is positive semidefinite.

\bigskip

\noindent \textit{Proof of Lemma T-1}. The result follows from the fact that, for N-XOR games, the computation involving,

\begin{align*}
     \frac{1}{ \bigg( C_{N\mathrm{XOR}} \omega_{N\mathrm{XOR}} \bigg) n \bigg( \underset{1 \leq j \leq N-1}{\prod}  \big( n - j \big) \bigg) }             \underset{i_N \in \mathcal{Q}_N}{\underset{\vdots}{\underset{i_1 \in \mathcal{Q}_1}{\sum}}}   \bigg(  \big(    u^{\prime}_{i_1i_2 \cdots i_N}   -    v^{\prime}_{i_1i_2 \cdots i_N}  \big) \big(   u^{\prime}_{i_1i_2 \cdots i_N}   -    v^{\prime}_{i_1i_2 \cdots i_N}    \big)^{\textbf{T}}  \\    +  \big(    u^{\prime}_{i_2 i_1 i_3 \cdots i_N}   -    v^{\prime}_{i_2 i_1 i_3  \cdots i_N}  \big)  \big( u^{\prime}_{i_2 i_1 i_3 \cdots i_N}   -    v^{\prime}_{i_2 i_1 i_3  \cdots i_N}  \big)^{\textbf{T}}  + \cdots +     \big(    u^{\prime}_{i_1 i_2 i_3 \cdots i_{N} i_{N-1}} \\   -    v^{\prime}_{i_1 i_2 i_3 \cdots i_{N} i_{N-1}}  \big)   \big(   u^{\prime}_{i_1 i_2 i_3 \cdots i_{N} i_{N-1}}    -    v^{\prime}_{i_1 i_2 i_3 \cdots i_{N} i_{N-1}}   \big)^{\textbf{T}}     \bigg)      \text{,}
\end{align*}

\noindent implies that the associated operator is positive semidefinite from the observation that taking the constant $C_{N\mathrm{XOR}}$, in the normalization,

\begin{align*}
     \frac{1}{ \bigg( C_{N\mathrm{XOR}} \omega_{N\mathrm{XOR}} \bigg) n \bigg( \underset{1 \leq j \leq N-1}{\prod}  \big( n - j \big) \bigg) }        \text{, }
\end{align*}

\noindent to equal $N!$ implies the desired result, from which we conclude the argument. \boxed{}

\bigskip

\noindent We formally state the positive semidefinite condition for the operator provided in the previous result above for several other variants of games that are considered in this paper. In the collection of relations for the $N$ player $\mathrm{XOR}$ error bounds, the summations over the list of possible questions, $\mathcal{Q}_1, \cdots, \mathcal{Q}_N$, for each player are excluded; the inequalities are obtained by permuting the order in which the "interchange" operation of tensor observables of each player is applied. From the optimal value of winning the $N$ $\mathrm{XOR}$ game,

\begin{align*}
  \omega_{N \mathrm{XOR}} \big( G \big) \equiv \omega \big( N \mathrm{XOR} \big)   \text{, }
\end{align*}

\noindent an inequality of the form,

\begin{align*}
  \bigg| \bigg|    \bigg[   \bigg( \bigg(           \underset{1 \leq i \leq n}{\prod}       A^{j_i}_i     \bigg) \bigotimes B_{kl} \bigotimes \bigg(  \underset{1 \leq k \leq n-2}{\bigotimes} \textbf{I}_k     \bigg) \bigg)  -   \omega_{N \mathrm{XOR}}         \bigg(                   \pm \bigg( \mathrm{sign} \big( i_1 , j_1 , \cdots , j_n \big) \bigg[ \bigg(        \bigg(   \underset{1 \leq i \leq n}{\prod} A^{j_i}_i \bigg) \\ + \bigg(   \underset{i +1 \equiv i \oplus 1}{\underset{1 \leq i \leq n}{\prod}}  A^{j_i}_i \bigg)       \bigg) \bigg] \bigotimes \bigg( \underset{1 \leq k \leq n-1}{\bigotimes} \textbf{I}_k \bigg) \bigg)               \bigg)    \bigg]               \ket{\psi_{N\mathrm{XOR}}}        \bigg| \bigg|    \text{, } \end{align*}

  \begin{align*} \bigg| \bigg|    \bigg[   \bigg( \bigg(             A_i \bigotimes \textbf{I} \bigotimes \bigg(     \underset{i \in \mathcal{Q}_1 , j \in \mathcal{Q}_2 , k \in \mathcal{Q}_3}{\prod}   C^{l_{ijk}}_{ijk}      \bigg) \bigotimes \bigg( \underset{1 \leq k \leq n-3}{\bigotimes}  \textbf{I}_k \bigg)                     \bigg)  - \omega_{N\mathrm{XOR}} \bigg(     \pm \mathrm{sign} \big( i_1 , j_1 , k_1 ,  \cdots , i_{111}  \\ , \cdots , j_{nm ( n+m )}   , \cdots  , k_{nm(n+m)}  \big) \bigg[   \bigg( \textbf{I} \bigotimes \textbf{I} \bigotimes \bigg( \bigg(  \underset{k \in \mathcal{Q}_3 , \mathcal{Q}_3 +1 \equiv \mathcal{Q}_3 \oplus 1}{\underset{j \in \mathcal{Q}_2 , \mathcal{Q}_2 +1 \equiv \mathcal{Q}_2 \oplus 1}{\underset{i \in \mathcal{Q}_1 , \mathcal{Q}_1 + 1 \equiv \mathcal{Q}_1 \oplus 1}{\prod}}}    C^{l_{ijk}}_{ijk}  \bigg) + \bigg(   \underset{i \in \mathcal{Q}_1 , j \in \mathcal{Q}_2 , k \in \mathcal{Q}_3}{\prod}   C^{l_{ijk}}_{ijk}       \bigg) \bigg)   \\    \bigotimes \bigg( \underset{1 \leq k \leq n-3}{\bigotimes} \textbf{I}_k  \bigg)      \bigg)   \bigg] \bigg)                 \bigg]             \ket{\psi_{N\mathrm{XOR}}}        \bigg| \bigg|   \text{, } \end{align*}

  \begin{align*}  \vdots \\  \bigg| \bigg| \bigg[ \bigg(             A_i \bigotimes  \bigg( \underset{1 \leq k \leq N-2}{\bigotimes}  \textbf{I}_k \bigg)  \bigotimes   \bigg(     \underset{i \in \mathcal{Q}_1 , j \in  \mathcal{Q}_2 , k \in \mathcal{Q}_3}{\prod}   C^{(N-2),l_{ijk}}_{ijk}      \bigg)                 \bigg)  - \omega_{N\mathrm{XOR}}    \bigg(                          \pm \mathrm{sign} \big( i_1 , j_1 , k_1 , \cdots , i_{111}  , \cdots \\ ,  j_{nm ( n+m )} , \cdots  , k_{nm(n+m)}  \big) \bigg[   \bigg( \bigg( \underset{1 \leq k \leq N-1}{\bigotimes} \textbf{I}_k \bigg)  \bigotimes \bigg(  \bigg(     \underset{i \in \mathcal{Q}_1 , j \in \mathcal{Q}_2 , k \in \mathcal{Q}_3}{\prod}   C^{(N-2),l_{ijk}}_{ijk}      \bigg)   \bigg) \\  + \bigg(  \underset{k \in \mathcal{Q}_3 , \mathcal{Q}_3 +1 \equiv \mathcal{Q}_3 \oplus 1}{\underset{j \in \mathcal{Q}_2 , \mathcal{Q}_2 +1 \equiv \mathcal{Q}_2 \oplus 1}{\underset{i \in \mathcal{Q}_1 , \mathcal{Q}_1 + 1 \equiv \mathcal{Q}_1 \oplus 1}{\prod}}}    C^{(N-2), l_{ijk}}_{ijk}  \bigg)          \bigg) \bigg)   \bigg]      \bigg)             \bigg]            \ket{\psi_{N\mathrm{XOR}}}        \bigg| \bigg|              \text{, }
\end{align*}

\noindent are expected to hold. The bias for $N$ players in the $\mathrm{XOR}$ setting,

\begin{align*}
 \beta_{N\mathrm{XOR}} \big( G \big)  \equiv \beta \big( N \mathrm{XOR} \big)     \text{, }
\end{align*}

\noindent satisfies,

\begin{align*}
  \big( 1 - \epsilon_{N\mathrm{XOR}} \big) \beta_{N\mathrm{XOR}} \big( G \big)      \leq      \underset{\mathcal{Q}_1 , \cdots , \mathcal{Q}_n}{\sum} \bra{\psi_{N\mathrm{XOR}}}  \bigg(           \underset{\#  \text{ players}}{\bigotimes}   \text{Tensors of player observables}     \bigg)     \ket{\psi_{N\mathrm{XOR}}}              \\   \leq \beta_{N\mathrm{XOR}} \big( G \big)          \text{, }
\end{align*}

\noindent for $\epsilon_{N\mathrm{XOR}}$ sufficiently small. Furthermore, from the $N$ player primal feasible solution to the associated semidefinite program, one also has,

\begin{align*}
        \big( 1 - \epsilon_{N\mathrm{XOR}} \big) \omega_{N\mathrm{XOR}} \big( G \big)      \leq      \underset{\mathcal{Q}_1 , \cdots , \mathcal{Q}_n}{\sum} \bra{\psi_{N\mathrm{XOR}}}  \bigg(    \underset{\#  \text{ players}}{\bigotimes}   \text{Tensors of player observables}   \bigg)     \ket{\psi_{N\mathrm{XOR}}}              \\   \leq \omega_{N\mathrm{XOR}} \big( G \big)                \text{. }
\end{align*}

\subsubsection{Frobenius norm upper bounds}

\noindent As a generalization of the superposition that was previously provided for the optimal value,

\begin{align*}
  \omega \big( 3 \mathrm{XOR} \big) \equiv \omega_{3\mathrm{XOR}} \big( G \big)   \text{, }
\end{align*}

\noindent from the $N$-player bias introduced in the previous section, which satisfies the proportionality,

\begin{align*}
          \omega \big( N \mathrm{XOR} \big) \propto  \bigg[  \ket{ijklm} \ket{ijk} \bigg( \ket{i} \bra{ij} \bigg) \bra{ijkl} + \ket{\sigma( i \big) \sigma \big( j \big) \sigma \big( k \big) \sigma \big( l \big) \sigma \big( m \big) } \\  \times   \ket{ijk} \bigg( \ket{i} \bra{ij}  \bigg) \bra{ijkl} + \ket{ijklm} \ket{\sigma( i \big) \sigma ( j \big) \sigma \big( k \big) }   \bigg( \ket{i} \bra{ij} \bigg) \bra{ijkl }  +  \ket{ijklm} \\ \times \ket{\sigma \big( i \big) \sigma \big( j \big) \sigma \big( k \big) } \bigg( \ket{i} \bra{ij } \bigg) \bra{ijkl} + \ket{ijklm} \ket{ijk}  \bigg(  \ket{\sigma \big( i \big) } \bra{ij}  \bigg) \bra{ijkl} + \ket{ijklm}  \\ \times  \ket{ijk} \bigg( \ket{i} \bra{\sigma \big( i \big) \sigma \big( j \big)} \bigg) \bra{ijkl} + \ket{ijklm} \ket{ijk} \bigg( \ket{i} \bra{ij} \bigg) \bra{\sigma \big( i \big) \sigma \big( j \big) \sigma \big( k \big) \sigma \big( l \big) } \\  +    \ket{\sigma \big( i \big) \sigma \big( j \big) \sigma \big( k \big) \sigma \big( l \big) \sigma \big( m \big) }    \ket{\sigma \big( i \big) \sigma \big( j \big) \sigma \big( k \big) } \bigg( \ket{i}   \bra{ij} \bigg) \bra{ijkl} \\  +    \ket{\sigma \big( i \big) \sigma \big( j \big)  \sigma \big( k \big) \sigma \big( l \big) \sigma \big( m \big) } \ket{\sigma \big( i \big) \sigma \big( j \big) \sigma \big( k \big) }  \bigg( \ket{\sigma \big( i \big) } \bra{ij} \bigg) \bra{ijkl} \\   +  \ket{\sigma \big( i \big) \sigma \big( j \big) \sigma \big( k \big) \sigma \big( l \big) \sigma \big( m \big) }           \ket{\sigma \big( i \big) \sigma \big( j \big) \sigma \big( k \big) }   \bigg(     \ket{\sigma \big( i \big) } \bra{\sigma \big( i \big) \sigma \big( j \big) }    \bigg)   \\ \times   \bra{ijkl}     +   \ket{\sigma \big( i \big) \sigma \big( j \big) \sigma \big( k \big) \sigma \big( l \big) \sigma \big( m \big)}   \ket{\sigma \big( i \big) \sigma \big( j \big) \sigma \big( k \big) }  \bigg( \ket{\sigma \big( i \big) } \bra{\sigma \big( i \big) \sigma \big( j \big) } \bigg)  \\ \times    \bra{\sigma \big( i \big) \sigma \big( j \big) \sigma \big( k \big) \sigma \big( l \big) }      + \ket{ijklm} \ket{\sigma \big( i \big) \sigma \big( j \big) \sigma \big( k \big)}    \bigg( \ket{i} \bra{ij} \bigg) \bra{ijkl} \\ + \ket{ijklm}   \ket{\sigma \big( i \big) \sigma \big( j \big) \sigma \big( k \big) }   \bigg( \ket{\sigma \big( i \big) }    \bra{ij } \bigg)  \bra{ijkl} +   \ket{ijklm}    \ket{\sigma \big( i \big) \sigma \big( j \big) \sigma \big( k \big) }   \\ \times \bigg( \ket{\sigma \big( i \big) } \bra{\sigma \big( i \big) \sigma \big( j \big) } \bigg)  \bra{ijkl}  +         \ket{ijklm}     \ket{\sigma \big( i \big) \sigma \big( j \big) \sigma \big( k \big) }    \bigg( \ket{\sigma \big( i \big) }   \bra{\sigma \big( i \big) \sigma \big( j \big) } \bigg)  \\ \times  \bra{\sigma \big( i \big) \sigma \big( j \big) \sigma \big( k \big) \sigma \big( l \big) }       + \ket{ijklm} \ket{ijk}   \bigg( \ket{\sigma \big( i \big) }   \bra{\sigma \big( i \big) \sigma \big( j \big) } \bigg) \bra{ijkl} 
          \\ + \ket{ijklm} \ket{ijk}  \bigg( \ket{\sigma \big( i \big) } \bra{\sigma \big( i \big) \sigma \big( j \big)} \bigg) \bra{\sigma \big( i \big) \sigma \big( j \big) \sigma \big( k \big) \sigma \big( l \big) }    +          \ket{ijklm}   \\ \times     \ket{ijk}   \bigg( \ket{i}  \bra{\sigma \big( i \big)  \sigma \big( j \big) } \bigg) \bra{\sigma \big( i \big) \sigma \big( j \big) \sigma \big( k \big)  \sigma \big( l \big) }          \\     + \text{Higher order permutations}            \bigg]                            \text{, }
\end{align*}

\noindent as an extension of the two-player setting, {[37]}, previous work of the author in {[44]} demonstrated that upper bounds on the Frobenius norm of the following form hold,

\begin{align*}
     \forall i,  \big|\big|  \big( A_i \otimes \textbf{I} \big) T^{\mathrm{FFL}}     -          T^{\mathrm{FFL}} \big( \widetilde{A_i} \otimes \textbf{I} \big)    \big|\big|_{\mathrm{F}} < 9  n^2 \sqrt{\epsilon} \big| \big|     T^{\mathrm{FFL}}   \big|\big|_{\mathrm{F}}     \text{, } \\ \forall j \neq k,  \big|\big|    \big( \textbf{I} \otimes B_{jk} \big) T^{\mathrm{FFL}}   - T^{\mathrm{FFL}} \big( \textbf{I} \otimes \widetilde{B_{jk}}    \big) \big|\big|_{\mathrm{F}} <   \frac{44}{3} n^2 \sqrt{\epsilon}  \big| \big|     T^{\mathrm{FFL}}   \big|\big|_{\mathrm{F}} \text{, }
\end{align*}

\noindent for the suitable linear mapping,

\begin{align*}
  T^{\mathrm{FFL}} :  \textbf{C}^{2 \lceil \frac{n}{2} \rceil } \otimes      \textbf{C}^{2 \lceil \frac{n}{2} \rceil }     \longrightarrow        \textbf{C}^{d_A} \otimes \textbf{C}^{d_B}     \text{. 
 }
\end{align*}

\noindent To demonstrate that such an argument for upper bounding the Frobenius norm also holds for $3$, and more, players answering questions drawn from the referee's probability distribution, observe that it suffices to prove that inequalities of the form,

\begin{align*}
 \underline{\text{Player 1}:} \text{ } \bigg| \bigg|   \bigg(  A_i \bigotimes \bigg( \underset{1 \leq k \leq n-1}{\bigotimes} \textbf{I}_k  \bigg)        \bigg) \mathscr{T}  -   \mathscr{T} \bigg( \bigg( \underset{1 \leq k \leq n-1}{\bigotimes} \textbf{I}_k  \bigg)     \\  \bigotimes \widetilde{A_i}      \bigg)      \bigg| \bigg|_F  \text{,} \\    \underline{\text{Player 2}:} \text{ }  \bigg| \bigg|   \bigg( \textbf{I} \bigotimes B_{ij} \bigotimes \bigg( \underset{1 \leq k \leq n-2}{\bigotimes} \textbf{I}_k  \bigg)   \bigg) \mathscr{T} - \mathscr{T} \bigg(                           \textbf{I} \bigotimes \textbf{I} \bigotimes \widetilde{B_{ij}} \\ \bigotimes \bigg( \underset{1 \leq k \leq n-3}{\bigotimes}    \textbf{I}_k \bigg) \bigg)      \bigg| \bigg|_F  \text{, } \\            \underline{\text{Player 3}:} \text{ }  \bigg| \bigg|       \bigg(     \textbf{I} \bigotimes \textbf{I } \bigotimes C_{ijk} \bigotimes \bigg( \underset{1 \leq k \leq n-3}{\bigotimes}     \textbf{I}_k  \bigg) \bigg) \mathscr{T} -  \mathscr{T} \bigg(  \textbf{I}  \bigotimes \textbf{I} \bigotimes \textbf{I} \bigotimes \widetilde{C_{ijk}} \\ \bigotimes \bigg( \underset{1 \leq k \leq n-4}{\bigotimes}      \textbf{I}_k \bigg)           \bigg)      \bigg| \bigg|_F   \text{, } \\ \vdots 
\end{align*}

\noindent with respect to the Frobenius norm can be upper bounded, for $\mathscr{T} \equiv \mathscr{T}_{N\mathrm{XOR}}$. From the fact that the collection of Frobenius norms for the two-player setting imposes the condition that $j \neq k$ for the second player's tensor observable $B_{jk}$, one choose indices of tensors of player observables above from the index set,

\begin{align*}
    \mathscr{I}  \equiv  \underset{\# \text{ Players } }{\bigcup}  \big\{   \text{indices } i \text{: }  \text{no } i \text{ are equal in Player } j\text{'s tensor observable}          \big\}       \text{. }       
\end{align*}

\noindent As is the case for the two-player setting, for. the $N$-player setting, given expressions and properties of the game tensor that have been previously discussed, admits the decomposition of the \textit{symmetric} game tensor, which takes the form

\begin{align*}
   G_{N\mathrm{XOR},\mathrm{Sym}}  \equiv G_{\mathrm{Sym}} \equiv \frac{1}{2} \begin{bmatrix}    0 &   G^{\textbf{T}} \\ G & 0\end{bmatrix} \equiv \frac{1}{2} \begin{bmatrix}    0 &   G_{N\mathrm{XOR}}^{\textbf{T}} \\ G_{N\mathrm{XOR}} & 0\end{bmatrix}  \text{.}
\end{align*}

\noindent Besides the symmetrized game tensor introduced above for the N-player setting, obvious counterparts of $G_{N\mathrm{XOR},\mathrm{Sym}}$ are introduced. Such tensors appear in the collection of constraints, from the partial ordering $ \succcurlyeq$ induced by the positive semidefinite cone,

\begin{align*}
 \underset{1 \leq i \leq n^3}{\sum}      y_{3\mathrm{XOR},i} E_{3\mathrm{XOR},ii}     \succcurlyeq           G_{3\mathrm{XOR},\mathrm{Sym}}        \text{, } \tag{3XOR, Sym} \\  \underset{1 \leq i \leq n^4}{\sum}   y_{4\mathrm{XOR},i} E_{4\mathrm{XOR},ii}    \succcurlyeq G_{4\mathrm{XOR},\mathrm{Sym}} \text{, }  \tag{4XOR, Sym} \\   \underset{1 \leq i \leq n^5}{\sum}  y_{5\mathrm{XOR},i} E_{5\mathrm{XOR},ii}    \succcurlyeq  G_{5\mathrm{XOR},\mathrm{Sym}}. \text{, } \tag{5XOR, Sym} \\  \underset{1 \leq i \leq n^N}{\sum}    y_{N\mathrm{XOR},i} E_{N\mathrm{XOR},ii}       \succcurlyeq G_{N\mathrm{XOR},\mathrm{Sym}} \tag{NXOR, Sym}   \text{. }
\end{align*}

\noindent Moreover, besides the definition of the $N$-player symmetric game tensor, the primal feasible solution to the semidefinite program satisfies,

\begin{align*}
  G_{\mathrm{Sym}} \cdot Z =  \underset{\mathcal{Q}_1, \cdots, \mathcal{Q}_n}{\sum}     G_{N\mathrm{XOR}} \bra{\psi_{N\mathrm{XOR}}}   \bigg( \underset{\# \text{ players}}{\bigotimes}   \text{Tensors of player observables} \bigg)        \ket{\psi_{N\mathrm{XOR}}}        \text{.}
\end{align*}

\noindent The dual semidefinite program is,

\begin{align*}
\underset{\underset{\mathcal{Q}_1, \cdots,  \mathcal{Q}_i, \cdots, \mathcal{Q}_n}{\sum}     y_{\mathcal{Q}_i} E_{ii}\succcurlyeq G_{\mathrm{Sym}}   }{\mathrm{inf}}   \bigg(  \underset{\mathcal{Q}_1, \cdots, \mathcal{Q}_i, \cdots, \mathcal{Q}_N}{\sum}  y_{\mathcal{Q}_i}  \bigg) \text{. }
\end{align*}

\noindent $\epsilon$-approximality entails that an inequality of the following form,

\begin{align*}
        \underset{1 \leq k \leq n}{\sum}             \bigg| \bigg|  \bigg( \bigg(   \big( V_{\mathcal{Q}_1 } \big)_k \cdot A_{\mathcal{Q}_1} \bigg)    \bigotimes           \bigg(    \underset{1 \leq k \leq n-1}{\bigotimes}   \textbf{I}_k \bigg)   \bigg) \ket{\psi_{N\mathrm{XOR}}}   -      \bigg(  \textbf{I} \bigotimes \bigg(            \big( V_{\mathcal{Q}_2 } \big)_k \cdot A_{\mathcal{Q}_2}          \bigg) \\ \bigotimes \bigg(  \underset{1 \leq k \leq n-2}{\bigotimes}   \textbf{I}_k 
 \bigg)         \bigg)  \ket{\psi_{N\mathrm{XOR}}}   \bigg| \bigg|     \text{, }
\end{align*}

\noindent is expected to hold, corresponding to the action of the linear operator $\mathscr{T}$ between the tensor observables of the first and second player; similar inequalities for tensor observables of other players participating in the game take the form,

\begin{align*}
     \underset{1 \leq k \leq n}{\sum}  \bigg| \bigg|     \bigg(   \bigg( \underset{1 \leq k \leq n-1}{\bigotimes}  \textbf{I}_k \bigg) \bigotimes \bigg(       \big( V_{\mathcal{Q}_n} \big)_k \cdot A_{\mathcal{Q}_N}        \bigg)       \bigg)               \ket{\psi_{N\mathrm{XOR}}}   -     \bigg(  \bigg(        \big( V_{\mathcal{Q}_1} \big)_k \cdot A_{\mathcal{Q}_1}      \bigg)   \\    \bigotimes \bigg( \underset{1 \leq k \leq n-1}{\bigotimes}  \textbf{I}_k \bigg)           \bigg)              \ket{\psi_{N\mathrm{XOR}}}       \bigg| \bigg|        \text{. } 
\end{align*}

\noindent The corresponding inequalities for the three-player $\mathrm{XOR}$ game take the form,

\begin{align*}
      \underset{1 \leq k \leq n}{\sum}  \bigg| \bigg|    \bigg( \bigg(   \big( V_{\mathcal{Q}_1} \big)_k \cdot A_{\mathcal{Q}_1}    \bigg) \otimes  \textbf{I} \otimes \textbf{I} \bigg) \ket{\psi_{3\mathrm{XOR}}}   - \bigg( \textbf{I} \otimes  \bigg(                  \big( V_{\mathcal{Q}_2} \big)_k \cdot A_{\mathcal{Q}_2}      \bigg) \otimes \textbf{I} \bigg)             \ket{\psi_{3\mathrm{XOR}}}       \bigg| \bigg|     \text{, } \\ \\  \underset{1 \leq k \leq n}{\sum}  \bigg| \bigg|                    \bigg(     \textbf{I} \otimes \bigg( \big( V_{\mathcal{Q}_2} \big)_k \cdot A_{\mathcal{Q}_2} \bigg) \otimes \textbf{I}     \bigg)            \ket{\psi_{3\mathrm{XOR}}}  -   \bigg(            \textbf{I} \otimes \textbf{I} \otimes \bigg( \big( V_{\mathcal{Q}_3} \big)_k \cdot A_{\mathcal{Q}_3} \bigg)                      \bigg)          \ket{\psi_{3\mathrm{XOR}}}       \bigg| \bigg|    \text{, } \\ \\  \underset{1 \leq k \leq n}{\sum}  \bigg| \bigg|  \bigg( \textbf{I} \otimes \textbf{I} \otimes \bigg( \big( V_{\mathcal{Q}_3}  \big)_k \cdot A_{\mathcal{Q}_3}  \bigg)    \bigg)  \ket{\psi_{3\mathrm{XOR}}}       -         \bigg( \bigg( \big( V_{\mathcal{Q}_1} \big)_k \cdot A_{\mathcal{Q}_1}  \bigg)  \otimes \textbf{I} \otimes \textbf{I}      \bigg)  \ket{\psi_{3\mathrm{XOR}}}       \bigg| \bigg|   \text{. } 
\end{align*} 

\noindent In the three-player setting, we provide an analog to the following result from {[37]}, specifically \textbf{Theorem} \textit{6}:

\bigskip

\noindent \textbf{Lemma} $\textit{1}^{*}$ (\textit{the computation of the Frobenius norm for the anticommutation rule of $T$ yields the desired $\sqrt{\epsilon}$ approximate upper bound}, {[44]}). One has that,

\begin{align*}
  \bigg| \bigg|  \text{ } \bigg( \text{ } \bigg( \underset{1 \leq i \leq n}{\prod} A^{j_i}_i     \bigg)    \otimes  \textbf{I}   \bigg)       \ket{\psi_{\mathrm{FFL}}}    -   \bigg[ \mathrm{sign} \big( i , j_1 , \cdots , j_n \big) \text{ }\bigg( \text{ }  \bigg(   \underset{i \equiv j_1 + 1, \text{ } \mathrm{set}\text{ }  j_1 + 1 \equiv j_1 \oplus 1}{\underset{1 \leq i \leq n}{\prod}}        A^{j_i}_i         \bigg) \\ \otimes \textbf{I} \bigg) \text{ } \bigg] \ket{\psi_{\mathrm{FFL}}}              \bigg| \bigg|_{\mathrm{F}}          \text{, } 
\end{align*}

\noindent and,

\begin{align*}
      \bigg| \bigg|      \bigg( \text{ }  \bigg( \underset{1 \leq  i \leq n }{\prod} A^{j_i}_i \bigg) \otimes B_{kl} \bigg)  \ket{\psi_{\mathrm{FFL}}}      -   \frac{2}{3} \bigg[ \pm \mathrm{sign}  \big( i , j_1 , \cdots , j_n \big) \bigg( \text{ }  \bigg(   \underset{ i \equiv j_4+1 ,\text{ } \mathrm{set} \text{ } j_4+1 \equiv j_4 \oplus 1 }{\underset{1 \leq i \leq n}{\prod}}       A^{j_i}_i    \bigg)  
       \\  \otimes \textbf{I}    \bigg) \text{ } \bigg]  \ket{\psi_{\mathrm{FFL}}}   \bigg| \bigg|_{\mathrm{F}}         \text{, } 
\end{align*}

\noindent have upper bounds, $9 n^2 \sqrt{\epsilon}$ and $\frac{44}{3} n^2 \sqrt{\epsilon}$, respectively.

\bigskip

\noindent Upper bounding the Frobenius norm from the optimal value of the game, in addition to contributions from each agent participating, can be obtained with the following:

\bigskip

\noindent \textbf{Lemma} \textit{1 -N $\mathrm{XOR}$} (\textit{computation of the Frobenius norm for the anticommutation rule of $T_{N\mathrm{XOR}}$ yields a desired up to constants $\sqrt{\epsilon}$ upper bound}). One has that,

\begin{align*}
      \underline{\text{Player $1$}:} \text{ } \bigg| \bigg|   \bigg(  A_i \bigotimes \bigg( \underset{1 \leq k \leq n-1}{\bigotimes} \textbf{I}_k  \bigg)        \bigg) \mathscr{T}  -   \mathscr{T} \bigg( \bigg( \underset{1 \leq k \leq n-1}{\bigotimes} \textbf{I}_k  \bigg)      \bigotimes \widetilde{A_i}      \bigg)      \bigg| \bigg|_F \\ < c_1 n^N \sqrt{\epsilon}  \text{,} \\     \vdots \\              \underline{\text{Player $N$}:} \text{ } \bigg| \bigg|   \bigg(  \bigg( \underset{1 \leq k \leq n-1}{\bigotimes} \textbf{I}_k  \bigg)   \bigotimes  A^{(n-1)}_{i_1, \cdots, i_{n-1}}     \bigg) \mathscr{T}  -   \mathscr{T} \bigg( \bigg(    \widetilde{A^{(n-1)}_{i_1, \cdots, i_{n-1}} }       \bigotimes \\  \bigg( \underset{1 \leq k \leq n-1}{\bigotimes} \textbf{I}_k  \bigg)     \bigg)      \bigg| \bigg|_F      < c_N n^N \sqrt{\epsilon}  \text{, }
\end{align*}

\noindent has the upper bound,

\begin{align*}
      \mathscr{C} \equiv    \underset{1 \leq i \leq N}{\bigcup}  \big\{     C_i  \neq c_i\in \textbf{R}  :            C_i \equiv c_i \sqrt{\epsilon}     \big\}     \propto n^N \sqrt{\epsilon} \text{. }
\end{align*}

\noindent Tensor observables from each player are drawn from the index set,

\begin{align*}
    \mathscr{I}  \equiv  \underset{\# \text{ Players } }{\bigcup}  \big\{   \text{indices } i \text{: }  \text{no } i \text{ are equal in Player } j\text{'s tensor observable}          \big\}       \text{. }       
\end{align*}

\bigskip

\noindent \textit{Proof of Lemma 1 -N $\mathrm{XOR}$}. Obtaining the desired collection of upper bounds for the Frobenius norm amounts to upper bounding the each Frobenius norm in the statement of the result above, implying,

\begin{align*}
 \bigg| \bigg|    \bigg[             \bigg( \bigg( \underset{1 \leq i \leq n}{\prod}  A^{j_i}_i \bigg) \bigotimes \bigg( \underset{1 \leq k \leq n-1}{\bigotimes} \textbf{I}_k \bigg) \bigg)   -  \bigg(  \omega_{N\mathrm{XOR}} \bigg(         \pm \mathrm{sign} \big( i_1, \cdots, i_n \big)            \\ \times     \bigg( \underset{1 \leq i \leq n}{\prod}  A^{j_i}_i \bigg)           \bigg)    \bigotimes \bigg( \underset{1 \leq k \leq n-1}{\bigotimes} \textbf{I}_k \bigg)\bigg)                       \bigg]   \ket{\psi_{N\mathrm{XOR}}}        \bigg| \bigg|_F  \\  <    \bigg( n_1 +       \big(    n_1 + 2    \big) \omega_{N\mathrm{XOR}}^{-1}                \bigg) n^N \sqrt{\epsilon}        \text{,}         \\ \\  \bigg| \bigg|    \bigg[ \bigg(  \textbf{I} \bigotimes            \bigg( \underset{1 \leq i \leq n}{\prod}  A^{1,j_i}_i \bigg) \bigotimes \bigg( \underset{1 \leq k \leq n-2}{\bigotimes} \textbf{I}_k \bigg) \bigg)   -  \bigg(  \textbf{I} \bigotimes  \bigg( \omega_{N\mathrm{XOR}} \\ \times \bigg(         \pm \mathrm{sign} \big( i_1, j_1, \cdots, i_n, \cdots, j_n \big)        \bigg( \underset{1 \leq i_2 \leq m}{\underset{1 \leq i_1 \leq n}{\prod}}  A^{1,j_{i_1,i_2}}_{i_1,i_2} \bigg)           \bigg)  \bigg) \\    \bigotimes \bigg( \underset{1 \leq k \leq n-2}{\bigotimes} \textbf{I}_k \bigg)\bigg)                       \bigg]  \ket{\psi_{N\mathrm{XOR}}}       \bigg| \bigg|_F   <   \bigg( n_2 +  \big( n_2 + 2 \big) \omega_{N\mathrm{XOR}}^{-1}            \bigg) n^N \sqrt{\epsilon}  \\  \vdots \\ \bigg| \bigg|    \bigg[             \bigg(\bigg( \underset{1 \leq k \leq n-1}{\bigotimes} \textbf{I}_k \bigg) \bigotimes   \bigg( \underset{1 \leq i \leq n}{\prod}  A^{(n-1),j_{i_1,\cdots,i_n}}_{i_1,\cdots, i_n} \bigg)  \bigg)   -  \bigg( \bigg( \underset{1 \leq k \leq n-1}{\bigotimes} \textbf{I}_k \bigg) \bigotimes \bigg(    \omega_{N\mathrm{XOR}} \\ \times  \bigg(         \pm \mathrm{sign} \big( i_1, \cdots, i_n, j_1  , \cdots, j_n \big)                         \bigg)   \bigg( \underset{1 \leq i \leq n}{\prod}  A^{(n-1),j_{i_1,\cdots,i_n}}_{i_1,\cdots, i_n} \bigg)  \bigg)                 \bigg)   \bigg] \ket{\psi_{N\mathrm{XOR}}}    \bigg| \bigg|_F   \\  <   \bigg( n_N        +   \big(  n_N + 2\big)     \omega_{N\mathrm{XOR}}^{-1}            \bigg) n^N \sqrt{\epsilon}  \text{, }
\end{align*}

\noindent for the quantum optimal strategy,

\begin{align*}
       \ket{\psi_{N\mathrm{XOR}}} \equiv 
 \ket{\psi_{N\mathrm{XOR}} \big( \mathcal{S} \big) } \equiv \underset{\mathcal{S}}{\mathrm{sup}}  \big\{ \text{payoff for all players with some quantum strategy } \mathcal{S}             \big\}                  \text{,}
\end{align*}

\noindent and suitable $n_i$. Hence the desired upper bound,

\begin{align*}
  \mathscr{C}  \text{, }
\end{align*}

\noindent for the collection of Frobenius norms over all players of the $\mathrm{XOR}$ game takes the form,

\begin{align*}
          \underset{1 \leq i \leq N}{\sum}      \bigg(                  n_i + \big( n_i + 2 \big) \omega^{-1}_{N\mathrm{XOR}}    \bigg) n^N \sqrt{\epsilon}  \equiv \bigg( \big( n_1 + \cdots + n_N \big) +  \big( \big( n_1 + \cdots + n_N \big) +   2N  \big) \\ \times \omega^{-1}_{N\mathrm{XOR}}            \bigg) n^N \sqrt{\epsilon }   <  \bigg(    \big( n_1 + \cdots + n_N \big) +   \big(  \big( n_1 + \cdots + n_N \big) + 2N \big)   \bigg) \omega^{-1}_{N\mathrm{XOR}} n^N \sqrt{\epsilon}  \\ <      \bigg(   \big( n_1 + \cdots + n_N \big)   \bigg( 2 + 2 N \bigg)       \bigg) \omega^{-1}_{N\mathrm{XOR}} n^N \sqrt{\epsilon}           <       \bigg(   \big( n_1 + \cdots + n_N \big)   \bigg( 2 + 2 N \bigg)       \bigg) \\ \times  \big( \omega^{-1}_{N\mathrm{XOR}} n^N \big)^2 \sqrt{\epsilon}  
      \equiv  \bigg(  2 \big( n_1 + \cdots + n_N \big) \big( \omega^{-1}_{N\mathrm{XOR}} n^N \big)^2 + 2N \big( n_1 + \cdots + n_N \big) \\ \times  \big( \omega^{-1}_{N\mathrm{XOR}} n^N \big)^2 \bigg)        \sqrt{\epsilon}        <   2 \big( n_1 + \cdots n_N \big)   N n^N \sqrt{\epsilon}    + 2N \big( n_1 + \cdots + n_N \big)    N n^N \sqrt{\epsilon}       \\   <     5 N^2 n_1 n^N \sqrt{\epsilon}    <        5 \big( N n^N \big)^2 \sqrt{\epsilon}    \equiv \mathscr{C}^{\prime}       \text{, }
\end{align*}

\noindent Setting the upper bound to,

\begin{align*}
    \mathscr{C} \equiv   \mathscr{C}^{\prime}   \text{, }
\end{align*}

\noindent yields the desired constant, from which we conclude the argument. \boxed{}

\bigskip

\noindent As an immediate consequence to the argument above, the following results imply upper bounds for Frobenius norms of games with less than $N$ players that can be obtained through similar computations.

\bigskip

\noindent \textbf{Lemma} \textit{$1- 3 \mathrm{XOR}$} (\textit{computation of the Frobenius norm for the anticommutation rule of $T_{3\mathrm{XOR}}$ yields a desired up to constants $\sqrt{\epsilon}$ upper bound}, \textbf{Theorem} \textit{6}, {[44]}). One has that,

\begin{align*}
      \underline{\text{Player $1$}:} \text{ } \bigg| \bigg|   \bigg(  A_i \bigotimes \bigg( \underset{1 \leq k \leq n-1}{\bigotimes} \textbf{I}_k  \bigg)        \bigg) \mathscr{T}  -   \mathscr{T} \bigg( \bigg( \underset{1 \leq k \leq n-1}{\bigotimes} \textbf{I}_k  \bigg)      \bigotimes \widetilde{A_i}      \bigg)      \bigg| \bigg|_F  < c_1 n^3 \sqrt{\epsilon}  \text{,} \end{align*}

  \begin{align*}   \underline{\text{Player $2$}:} \text{ } \bigg| \bigg|   \bigg(  \bigg( \underset{1 \leq k \leq n-1}{\bigotimes} \textbf{I}_k  \bigg)   \bigotimes  A^{(n-1)}_{i_1, \cdots, i_{n-1}}     \bigg) \mathscr{T}  -   \mathscr{T} \bigg( \bigg(    \widetilde{A^{(n-1)}_{i_1, \cdots, i_{n-1}} }       \bigotimes  \bigg( \underset{1 \leq k \leq n-1}{\bigotimes} \textbf{I}_k  \bigg)     \bigg)      \bigg| \bigg|_F  \\    < c_2 n^3 \sqrt{\epsilon} \\              \underline{\text{Player $3$}:} \text{ } \bigg| \bigg|   \bigg(  \bigg( \underset{1 \leq k \leq n-1}{\bigotimes} \textbf{I}_k  \bigg)   \bigotimes  A^{(n-1)}_{i_1, \cdots, i_{n-1}}     \bigg) \mathscr{T}  -   \mathscr{T} \bigg( \bigg(    \widetilde{A^{(n-1)}_{i_1, \cdots, i_{n-1}} }       \bigotimes  \bigg( \underset{1 \leq k \leq n-1}{\bigotimes} \textbf{I}_k  \bigg)     \bigg)      \bigg| \bigg|_F   \\   < c_3 n^3 \sqrt{\epsilon}  \text{, }
\end{align*}

\noindent has the upper bound,

\begin{align*}
      \mathscr{C} \equiv    \underset{1 \leq i \leq 3}{\bigcup}  \big\{     C_i  \neq c_i\in \textbf{R}  :            C_i \equiv c_i \sqrt{\epsilon}     \big\}     \propto n^N \sqrt{\epsilon} \text{. }
\end{align*}

\noindent Tensor observables from each player are drawn from the index set,

\begin{align*}
    \mathscr{I}  \equiv  \underset{\# \text{ Players } }{\bigcup}  \big\{   \text{indices } i \text{: }  \text{no } i \text{ are equal in Player } j\text{'s tensor observable}          \big\}       \text{. }       
\end{align*}

\bigskip

\noindent \textit{Proof of Lemma $1$- 3 $\mathrm{XOR}$}. The result follows from the observation, as presented in the arguments for the previous results, that,

\begin{align*}
 \bigg| \bigg|    \bigg[             \bigg( \bigg( \underset{1 \leq i \leq n}{\prod}  A^{j_i}_i \bigg) \bigotimes \textbf{I} \bigotimes \textbf{I} \bigg)   -  \bigg(  \omega_{N\mathrm{XOR}} \bigg(         \pm \mathrm{sign} \big( i_1, \cdots, i_n \big)                \bigg( \underset{1 \leq i \leq n}{\prod}  A^{j_i}_i \bigg)           \bigg)    \\ \bigotimes \bigg( \underset{1 \leq k \leq 2}{\bigotimes} \textbf{I}_k \bigg)\bigg)                       \bigg]\ket{\psi_{3\mathrm{XOR}}}        \bigg| \bigg|_F  <    \bigg( n_1 +       \big(    n_1 + 2    \big) \omega_{3\mathrm{XOR}}^{-1}                \bigg) n^3 \sqrt{\epsilon}        \text{,}         \\ \\  \bigg| \bigg|    \bigg[ \bigg(  \textbf{I} \bigotimes            \bigg( \underset{1 \leq i \leq n}{\prod}  A^{1,j_i}_i \bigg) \bigotimes\textbf{I} \bigg)   -  \bigg(  \textbf{I} \bigotimes  \bigg( \omega_{N\mathrm{XOR}} \bigg(         \pm \mathrm{sign} \big( i_1, j_1, \cdots, i_n, \cdots, j_n \big)          \\ \times       \bigg( \underset{1 \leq i_2 \leq m}{\underset{1 \leq i_1 \leq n}{\prod}}  A^{1,j_{i_1,i_2}}_{i_1,i_2} \bigg)           \bigg)  \bigg)    \bigotimes \textbf{I} \bigg)                       \bigg]  \ket{\psi_{3\mathrm{XOR}}}       \bigg| \bigg|_F  <   \bigg( n_2 +  \big( n_2 + 2 \big) \omega_{3\mathrm{XOR}}^{-1}            \bigg) n^3 \sqrt{\epsilon}  \\ \\ \bigg| \bigg|    \bigg[             \bigg( \textbf{I} \bigotimes \textbf{I} \bigotimes   \bigg( \underset{1 \leq i \leq n}{\prod}  A^{(n-1),j_{i_1,\cdots,i_n}}_{i_1,\cdots, i_n} \bigg)  \bigg)   -  \bigg( \bigg( \underset{1 \leq k \leq 2}{\bigotimes} \textbf{I}_k \bigg) \bigotimes \bigg(    \omega_{3\mathrm{XOR}} \\ \times  \bigg(         \pm \mathrm{sign} \big( i_1, \cdots, i_n, j_1  , \cdots, j_n \big)                         \bigg)   \bigg( \underset{1 \leq i \leq n}{\prod}  A^{(n-1),j_{i_1,\cdots,i_n}}_{i_1,\cdots, i_n} \bigg)  \bigg)                 \bigg)   \bigg] \ket{\psi_{3\mathrm{XOR}}}    \bigg| \bigg|_F  \\  <   \bigg( n_3        +   \big(  n_3 + 2\big) \omega_{3\mathrm{XOR}}^{-1}            \bigg) n^3 \sqrt{\epsilon}  \text{, }
\end{align*}

\noindent from which we conclude the argument, as an upper bound for that of $\mathscr{C}$ can be obtained with the same computation. \boxed{}

\bigskip

\noindent

\bigskip

\noindent The optimal value of the $N \mathrm{XOR}$, $\omega_{N\mathrm{XOR}} \big( G \big)$, game $G$, is presented in the  next section.

\bigskip

\noindent As a related property of the suitable linear operator $\mathscr{T}$, an identity for suitable $N\mathrm{XOR}$ operators for,

\begin{align*}
      \bigg( A_i \bigotimes \bigg( \underset{1 \leq k \leq n-1}{\bigotimes}\textbf{I}_k \bigg)  \bigg) \mathscr{T} - \mathscr{T}  \bigg( \widetilde{A_i} \bigotimes \bigg(  \underset{1 \leq k \leq n-1}{\bigotimes}\textbf{I}_k \bigg) \bigg)     \text{, }
\end{align*}

\noindent can be obtained so that a counterpart identity from that of the $\mathrm{FFL}$ game,

\begin{align*}
     \big( A_i \otimes \textbf{I} \big) T_{\mathrm{FFL}} - T_{\mathrm{FFL}} \big( \widetilde{A_i} \otimes \textbf{I} \big)  \text{, }
\end{align*}

\noindent from the observation that,

\begin{align*}
     \big(  A_i T_{\mathrm{FFL}}  \otimes T_{\mathrm{FFL}} \big) - \big( T_{\mathrm{FFL}} \widetilde{A_i} \otimes T_{\mathrm{FFL}} \big) \equiv \big( A_i  T_{\mathrm{FFL}} - T_{\mathrm{FFL}} \widetilde{A_i} \big) \otimes T_{\mathrm{FFL}}   \text{, }
\end{align*}

\noindent in the case of the $N$-player $\mathrm{XOR}$ game takes the form,

\begin{align*}
    \bigg(  \big(  A_i \mathscr{T} \otimes \mathscr{T} \big) - \big( \mathscr{T} \widetilde{A_i} \otimes \mathscr{T} \big) \bigg) \bigotimes \bigg( \underset{1 \leq k \leq n-2}{\bigotimes } \mathscr{T}_k \bigg)  \equiv  \bigg( \big( A_i  \mathscr{T} - \mathscr{T} \widetilde{A_i} \big) \bigotimes \mathscr{T} \bigg) \\ \bigotimes  \bigg(   \underset{1 \leq k \leq n-2}{\bigotimes } \mathscr{T}_k  \bigg)    \text{, }
\end{align*}

\noindent where,

\begin{align*}
  \mathscr{T}_k \equiv \mathscr{T}  \text{, }
\end{align*}

\noindent for every $k$. As a result, to demonstrate that another desired identity holds besides that which has been argued for in the previous result, it suffices to argue that the following $N$ $\mathrm{XOR}$ identity holds:

\bigskip

\noindent \textbf{Lemma} \textit{2} (\textit{N N-XOR identities from two FFL identities}). One has that,

\begin{align*}
  \bigg( A_i \bigotimes \bigg( \underset{1 \leq k \leq n-1}{\bigotimes}\textbf{I}_k \bigg)  \bigg) \mathscr{T} - \mathscr{T}  \bigg( \widetilde{A_i} \bigotimes \bigg(  \underset{1 \leq k \leq n-1}{\bigotimes}\textbf{I}_k \bigg) \bigg)    \text{, }
\end{align*}

\noindent and that,

\begin{align*}
     \bigg(  \big(  A_i \mathscr{T} \otimes \mathscr{T} \big) - \big( \mathscr{T} \widetilde{A_i} \otimes \mathscr{T} \big) \bigg) \bigotimes \bigg( \underset{1 \leq k \leq n-2}{\bigotimes } \mathscr{T}_k \bigg)                        \text{, }
\end{align*}

\noindent are equal, in which,

\begin{align*}
    \bigg( A_i \bigotimes \bigg( \underset{1 \leq k \leq n-1}{\bigotimes}\textbf{I}_k \bigg)  \bigg) \mathscr{T} - \mathscr{T}  \bigg( \widetilde{A_i} \bigotimes \bigg(  \underset{1 \leq k \leq n-1}{\bigotimes}\textbf{I}_k \bigg) \bigg)  \equiv      \bigg( \big( A_i  \mathscr{T} - \mathscr{T} \widetilde{A_i} \big) \bigotimes \mathscr{T} \bigg) \\ \bigotimes \bigg(   \underset{1 \leq k \leq n-2}{\bigotimes } \mathscr{T}_k  \bigg)    \text{. }
\end{align*}

\noindent For the remaining $N-1$ players in the $\mathrm{XOR}$ game, identities of an analagous form hold.

\bigskip

\noindent \textit{Proof of Lemma 2}. Straightforwardly, to prove that the first identity holds from the action of the suitable $N$-$\mathrm{XOR}$ linear operator, write,

\begin{align*}
    \bigg( A_i \bigotimes \bigg( \underset{1 \leq k \leq n-1}{\bigotimes} \textbf{I}_k \bigg) \bigg)  \bigg[              \frac{1}{\sqrt{2^{n}}} \bigg[   \underset{(j_1 , \cdots , j_n) \in \{ 0 , 1  \}^n}{\sum}   \bigg( 
 \text{ }         \bigg(       \underset{1 \leq i \leq n}{ \prod } A^{j_i}_{i}            \bigg)      \bigotimes \bigg( \underset{1 \leq k \leq n-1}{\bigotimes} \textbf{I}_k \bigg)  \bigg)        \\ \times      \ket{\psi_{N\mathrm{XOR}}} \bra{\widetilde{\psi_{{N\mathrm{XOR}}} }}  \bigg( \mathrm{sign} \big( i , j_1 , \cdots , j_n \big)    \bigg( \underset{1 \leq i \leq n}{\prod}            \widetilde{A^{j_i}_i}  \bigg)    \bigotimes \bigg( \underset{1 \leq k \leq n-1}{\bigotimes} \textbf{I}_k \bigg)  \bigg)^{\dagger}  \bigg]    \bigg]            \text{, }
\end{align*}

\noindent from which the first identity can be verified from the observation that the above tensor product equals,

\begin{align*}
   A_i  \bigg[ \frac{1}{\sqrt{2^{n}}} \bigg[   \underset{(j_1 , \cdots , j_n) \in \{ 0 , 1  \}^n}{\sum}   \bigg( 
 \text{ }         \bigg(       \underset{1 \leq i \leq n}{ \prod } A^{j_i}_{i}            \bigg)     \bigotimes \bigg( \underset{1 \leq k \leq n-1}{\bigotimes} \textbf{I}_k \bigg)  \bigg)     \\ \times         \ket{\psi_{N\mathrm{XOR}}} \bra{\widetilde{\psi_{{N\mathrm{XOR}}} }} \bigg( \mathrm{sign} \big( i , j_1  , \cdots , j_n \big)     \bigg( \underset{1 \leq i \leq n}{\prod}            \widetilde{A^{j_i}_i}  \bigg)   \bigotimes \bigg( \underset{1 \leq k \leq n-1}{\bigotimes} \textbf{I}_k \bigg)   \bigg)^{\dagger}  \bigg]  \bigg] \\  \bigotimes  \bigg(    \bigg( \underset{1 \leq k \leq n-1}{\bigotimes} \textbf{I}_k \bigg)  \bigg[ \frac{1}{\sqrt{2^{n}}}  \bigg[   \underset{(j_1 , \cdots , j_n) \in \{ 0 , 1  \}^n}{\sum}   \bigg( 
    \bigg(       \underset{1 \leq i \leq n}{ \prod } A^{j_i}_{i}            \bigg)     \bigotimes \bigg( \underset{1 \leq k \leq n-1}{\bigotimes} \textbf{I}_k \bigg)  \bigg)   \\ \times            \ket{\psi_{N\mathrm{XOR}}} 
 \bra{\widetilde{\psi_{{N\mathrm{XOR}}} }}  \bigg( \mathrm{sign} \big( i , j_1 , \cdots , j_n \big)     \bigg( \underset{1 \leq i \leq n}{\prod}            \widetilde{A^{j_i}_i}  \bigg) \bigotimes \bigg( \underset{1 \leq k \leq n-1}{\bigotimes} \textbf{I}_k \bigg)  \bigg)^{\dagger}  \bigg]  \bigg]   \bigg)              \text{, }
\end{align*}

\noindent which readily implies that the first half of the desired $N$-$\mathrm{XOR}$ identity holds, namely that,

\begin{align*}
  A_i \mathscr{T} \bigotimes \bigg( \underset{1 \leq k \leq n-1}{\bigotimes} \textbf{I}_k \mathscr{T} \bigg)   \text{. }
\end{align*}

\noindent For the remaining half of the first identity, to demonstrate that anticommuting the order in which $\mathscr{T}$ is applied readily implies that,

\begin{align*}
-  \mathscr{T}  \bigg( \widetilde{A_i} \bigotimes \bigg(  \underset{1 \leq k \leq n-1}{\bigotimes}\textbf{I}_k \bigg) \bigg)  \text{, }
\end{align*}

\noindent is the remaining half of the first identity, observe,

\begin{align*}
     \bigg[ \frac{1}{\sqrt{2^{n}}} \bigg[   \underset{(j_1 , \cdots , j_n) \in \{ 0 , 1  \}^n}{\sum}   \bigg( 
 \text{ }         \bigg(       \underset{1 \leq i \leq n}{ \prod } A^{j_i}_{i}            \bigg)    \bigotimes  \bigg(  \underset{1 \leq k \leq n-1}{\bigotimes}\textbf{I}_k \bigg)  \bigg)             \ket{\psi_{N\mathrm{XOR}}} \bra{\widetilde{\psi_{N\mathrm{XOR}}} } \\ \times  \bigg(  \mathrm{sign} \big( i , j_1 , \cdots , j_n \big) \text{ } \bigg( \underset{1 \leq i \leq n}{\prod}            \widetilde{A^{j_i}_i}  \bigg)  \bigotimes  \bigg(  \underset{1 \leq k \leq n-1}{\bigotimes}\textbf{I}_k \bigg)  \bigg)^{\dagger}              \bigg]    \bigg]   \bigg( \widetilde{A_i}  \bigotimes  \bigg(  \underset{1 \leq k \leq n-1}{\bigotimes}\textbf{I}_k \bigg) \bigg)         \text{, }
\end{align*}

\noindent which equals,

\begin{align*}
        \bigg[ \frac{1}{\sqrt{2^{n}}} \bigg[   \underset{(j_1 , \cdots , j_n) \in \{ 0 , 1  \}^n}{\sum}   \bigg( 
 \text{ }         \bigg(       \underset{1 \leq i \leq n}{ \prod } A^{j_i}_{i}            \bigg)   \bigotimes \bigg( \underset{1 \leq k \leq n-1}{\bigotimes} \textbf{I}_k \bigg)  \bigg)             \ket{\psi_{N\mathrm{XOR}}} \\ \times  \bra{\widetilde{\psi_{N\mathrm{XOR}}} } \bigg(  \mathrm{sign} \big( i , j_1 , \cdots , j_n \big)  \bigg( \underset{1 \leq i \leq n}{\prod}            \widetilde{A^{j_i}_i}  \bigg)    \bigotimes \bigg( \underset{1 \leq k \leq n-1}{\bigotimes} \textbf{I}_k \bigg)        \bigg)^{\dagger}   \text{ }             \bigg]   \text{ }  \bigg]   \\ \times  \widetilde{A_i}       \bigotimes \bigg( \underset{1 \leq k \leq n-1}{\bigotimes} \textbf{I}_k \bigg)         \bigg[ \frac{1}{\sqrt{2^{n}}}     \bigg[   \underset{(j_1 , \cdots , j_n) \in \{ 0 , 1  \}^n}{\sum}   \bigg( 
 \text{ }         \bigg(       \underset{1 \leq i \leq n}{ \prod } A^{j_i}_{i}            \bigg)  \\ \bigotimes \bigg( \underset{1 \leq k \leq n-1}{\bigotimes} \textbf{I}_k \bigg)       \bigg)             \ket{\psi_{N\mathrm{XOR}}}  \bra{\widetilde{\psi_{N\mathrm{XOR}}} }  \bigg(  \mathrm{sign} \big( i , j_1 , \cdots , j_n \big) \\ \times  \text{ } \bigg( \underset{1 \leq i \leq n}{\prod}            \widetilde{A^{j_i}_i}  \bigg)  \bigotimes \bigg( \underset{1 \leq k \leq n-1}{\bigotimes} \textbf{I}_k \bigg)        \bigg)^{\dagger}   \text{ }             \bigg]    \bigg]                 \text{. }
\end{align*}

\noindent Hence the above tensor product equals the remaining half of the first identity,

\begin{align*}
 -  \mathscr{T}  \bigg( \widetilde{A_i} \bigotimes \bigg(  \underset{1 \leq k \leq n-1}{\bigotimes}\textbf{I}_k \bigg) \bigg)   \text{,}
\end{align*}

\noindent from which the the first identity holds by adding the two results together. For the remaining identities in the statement of the result above, besides that which are formulated from the observable of the first player, can be obtained by modifying the entry of the tensor product at which $\mathscr{T}$ is applied, through the tensor products,

\begin{align*}
      \bigg(   \bigg( \underset{1 \leq k \leq n-1}{\bigotimes} \textbf{I}_k \bigg)  \bigotimes A^{(n-1)}_{i_1,\cdots,i_n} \bigg)  \bigg[              \frac{1}{\sqrt{2^{n}}} \bigg[   \underset{(j_1 , \cdots , j_n) \in \{ 0 , 1  \}^n}{\sum}   \bigg( 
       \bigg( \underset{1 \leq k \leq n-1}{\bigotimes} \textbf{I}_k \bigg)     \bigotimes    \bigg(       \underset{1 \leq i \leq n}{ \prod } A^{(n-1)}_{i_1,\cdots,i_n}             \bigg)   \bigg)         \\ \times      \ket{\psi_{N\mathrm{XOR}}} \bra{\widetilde{\psi_{{N\mathrm{XOR}}} }}  \bigg(   \bigg( \underset{1 \leq k \leq n-1}{\bigotimes} \textbf{I}_k \bigg) \bigotimes \bigg(   \mathrm{sign} \big( i , j_1 , \cdots , j_n \big)  \bigg(       \underset{1 \leq i \leq n}{ \prod } A^{(n-1)}_{i_1,\cdots,i_n}         \bigg)     \bigg)   \bigg)^{\dagger}  \bigg]    \bigg]      \text{, }
\end{align*}

\noindent corresponding to the first half of the $N$ th identity, and,

\begin{align*}
       \bigg[              \frac{1}{\sqrt{2^{n}}} \bigg[   \underset{(j_1 , \cdots , j_n) \in \{ 0 , 1  \}^n}{\sum}   \bigg( 
       \bigg( \underset{1 \leq k \leq n-1}{\bigotimes} \textbf{I}_k \bigg)     \bigotimes    \bigg(       \underset{1 \leq i \leq n}{ \prod } A^{(n-1)}_{i_1,\cdots,i_n}             \bigg)   \bigg)             \ket{\psi_{N\mathrm{XOR}}}  \bra{\widetilde{\psi_{{N\mathrm{XOR}}} }} \\ \times  \bigg(   \bigg( \underset{1 \leq k \leq n-1}{\bigotimes} \textbf{I}_k \bigg)  \bigotimes \bigg(   \mathrm{sign} \big( i , j_1 , \cdots , j_n \big)  \bigg(       \underset{1 \leq i \leq n}{ \prod } A^{(n-1)}_{i_1,\cdots,i_n}         \bigg)     \bigg)   \bigg)^{\dagger}  \bigg]    \bigg]    \bigg(  \widetilde{A^{(n-1)}_{i_1,\cdots,i_n}}\\  \bigotimes \bigg( \underset{1 \leq k \leq n-1}{\bigotimes} \textbf{I}_k \bigg)   \bigg)     \text{, }
\end{align*}

\noindent corresponding to the second half of the $N$ th identity, from which all of the remaining identities follow by applying the same argument that has been provided for the first identity, from which we conclude the argument. \boxed{}

\bigskip

\noindent \textbf{Lemma} \textit{$2^{*}$} (\textit{3 3-XOR identities from two FFL identities}). One has that,

\begin{align*}
  \bigg( A_i \otimes \textbf{I} \otimes \textbf{I}   \bigg) \mathscr{T} - \mathscr{T}  \bigg( \widetilde{A_i} \otimes \textbf{I} \otimes \textbf{I}   \bigg)    \text{, }
\end{align*}

\noindent and that,

\begin{align*}
     \bigg(  \big(  A_i \mathscr{T} \otimes \mathscr{T} \big) - \big( \mathscr{T} \widetilde{A_i} \otimes \mathscr{T} \big) \bigg) \otimes \bigg(  \textbf{I} \otimes \textbf{I} \bigg)                        \text{, }
\end{align*}

\noindent are equal, in which,

\begin{align*}
    \bigg( A_i \otimes \textbf{I} \otimes \textbf{I} \bigg) \mathscr{T} - \mathscr{T}  \bigg( \widetilde{A_i} \otimes \textbf{I} \otimes \textbf{I} \bigg)  \equiv      \bigg( \big( A_i  \mathscr{T} - \mathscr{T} \widetilde{A_i} \big) \otimes \mathscr{T} \bigg) \otimes \bigg( \textbf{I} \otimes \textbf{I}  \bigg)    \text{. }
\end{align*}

\noindent For the remaining two players in the $\mathrm{XOR}$ game, identities of an analogous form hold.

\bigskip

\noindent \textit{Proof of Lemma $2^{*}$}. Apply the same arguments as provided in the previous result, from which we conclude the argument. \boxed{}

\bigskip

\noindent The following three variants of the identities, from the group of identities for the $\mathrm{FFL}$ game, hold:

\bigskip

\noindent \textbf{Lemma} \textit{$2^{**}$} (\textit{N N-XOR identities}). One has that,

\begin{align*}
  \bigg( A_i \bigotimes  \bigg( \underset{1 \leq z \leq N-1}{\bigotimes} \textbf{I}_z \bigg)  \bigg)  \mathscr{T} - \mathscr{T}  \bigg( A_i \bigotimes  \bigg( \underset{1 \leq z \leq N-1}{\bigotimes} \textbf{I}_z \bigg)  \bigg)   \text{, }
\end{align*}

\noindent and that,

\begin{align*}
     \bigg(  \bigg(  A_i \mathscr{T} \bigotimes  \bigg( \underset{1 \leq z \leq N-1}{\bigotimes} \mathscr{T} \textbf{I}_z  \bigg) - \bigg( \mathscr{T} \widetilde{A_i} \bigotimes \bigg( \underset{1 \leq z \leq N-1}{\bigotimes}  \mathscr{T}  \textbf{I}_z \bigg)  \bigg) \bigg) \bigotimes \bigg(  \underset{1 \leq z \leq N}{\bigotimes} \textbf{I}_z  \bigg)                        \text{, }
\end{align*}

\noindent are equal, in which,

\begin{align*}
    \bigg( A_i \bigotimes  \bigg( \underset{1 \leq z \leq N-1}{\bigotimes} \textbf{I}_z  \bigg) \bigg) \mathscr{T} - \mathscr{T} \bigg(  \widetilde{A_i} \bigotimes  \bigg( \underset{1 \leq z \leq N-1}{\bigotimes} \textbf{I}_z  \bigg) \bigg)   \equiv      \bigg( \bigg( A_i  \mathscr{T} - \mathscr{T} \widetilde{A_i} \bigg) \\ \bigotimes \bigg( \underset{1 \leq z \leq n-1}{\bigotimes} \mathscr{T} \textbf{I}_z  \bigg)  \bigg)  \bigotimes   \bigg(  \underset{1 \leq z \leq N-1}{\bigotimes} \textbf{I}_z   \bigg)    \text{. }
\end{align*}

\bigskip

\noindent \textbf{Lemma} \textit{$2^{***}$} (\textit{N N-XOR identities under strong parallel repetition}). One has that,

\begin{align*}
  \bigg( A_i \bigotimes  \bigg( \underset{1 \leq z \leq N-1}{\bigotimes} \textbf{I}_z \bigg)  \bigg) \bigg( \mathscr{T} \wedge \cdots \wedge \mathscr{T} \bigg)  - \bigg( \mathscr{T} \wedge \cdots \wedge \mathscr{T} \bigg)    \bigg( A_i \bigotimes  \bigg( \underset{1 \leq z \leq N-1}{\bigotimes} \textbf{I}_z \bigg)  \bigg)   \text{, }
\end{align*}

\noindent and that,

\begin{align*}
     \bigg(  \bigg(  A_i \bigg( \mathscr{T} \wedge \cdots \wedge \mathscr{T} \bigg)   \bigotimes  \bigg( \underset{1 \leq z \leq N-1}{\bigotimes} \bigg( \mathscr{T} \wedge \cdots \wedge \mathscr{T} \bigg)  \textbf{I}_z  \bigg) - \bigg( \bigg( \mathscr{T} \wedge \cdots \wedge \mathscr{T} \bigg)  \widetilde{A_i}  \\ \bigotimes \bigg( \underset{1 \leq z \leq N-1}{\bigotimes}  \bigg( \mathscr{T} \wedge \cdots \wedge \mathscr{T} \bigg)   \textbf{I}_z \bigg)  \bigg) \bigg) \bigotimes \bigg(  \underset{1 \leq z \leq N}{\bigotimes} \textbf{I}_z  \bigg)                        \text{, }
\end{align*}

\noindent are equal, in which,

\begin{align*}
    \bigg( A_i \bigotimes  \bigg( \underset{1 \leq z \leq N-1}{\bigotimes} \textbf{I}_z  \bigg) \bigg) \bigg( \mathscr{T} \wedge \cdots \wedge \mathscr{T} \bigg)  - \bigg( \mathscr{T} \wedge \cdots \wedge \mathscr{T} \bigg)   \bigg(  \widetilde{A_i} \bigotimes  \bigg( \underset{1 \leq z \leq N-1}{\bigotimes} \textbf{I}_z  \bigg) \bigg)  \\  \equiv      \bigg( \bigg( A_i  \bigg( \mathscr{T} \wedge   \cdots \wedge \mathscr{T} \bigg)   - \bigg( \mathscr{T} \wedge \cdots \wedge \mathscr{T} \bigg)  \widetilde{A_i} \bigg) \bigotimes\bigg( \underset{1 \leq z \leq n-1}{\bigotimes}\bigg( \mathscr{T} \wedge \cdots \wedge \mathscr{T} \bigg)  \textbf{I}_z  \bigg)  \bigg) \\  \bigotimes   \bigg(  \underset{1 \leq z \leq N-1}{\bigotimes} \textbf{I}_z   \bigg)    \text{. }
\end{align*}

\noindent \textit{Proof of Lemma $2^{**}$, and $2^{***}$}. Directly apply the arguments from the previous result, from which we conclude the argument. \boxed{}

\bigskip

\noindent With the previous set of results, $\epsilon$-optimality of the $N$-$\mathrm{XOR}$ game is also expected to hold. In the $\mathrm{FFL}$ game, the relation takes the following form indicated in the result below:

\bigskip

\noindent \textbf{Lemma} $\textit{3}^{*}$ ($\epsilon$\textit{-optimality}, \textbf{Lemma} \textit{7}, {[44]}). For an $\epsilon$-optimal strategy $A_i$, $B_{jk}$ and $\ket{\psi_{\mathrm{FFL}}}$,

\begin{align*}
 \underset{1 \leq i < j \leq n}{\sum} \bigg| \bigg| \bigg( \bigg(         \frac{A_i A_j + A_j A_i}{2}        \bigg) \otimes \textbf{I} \bigg)  \ket{\psi_{\mathrm{FFL}}} \bigg| \bigg|^2 <   2 \big(   \frac{7}{3}    \big)^2 n \big( n - 1 \big) \epsilon \text{. } 
\end{align*}

\bigskip In game theoretic settings with more than two active players, inequalities of the form indicated below are expected to hold.

\bigskip

\noindent \textbf{Lemma} \textit{3} (\textit{intialization of $\epsilon$-optimality of the N $\mathrm{XOR}$ game from the observable of the first player}). For an $\epsilon$-optimal strategy, and player observable tensors $A_{i_1}, A^1_{i_1,i_2}, \cdots, A^{n-1}_{i_1,i_2, \cdots, i_n}$, and $\ket{\psi_{N\mathrm{XOR}}}$, and $C_1>0$,

\begin{align*}
    \underset{i,j}{\sum}     \bigg| \bigg|     \bigg[            \bigg( \bigg(    \frac{A_i A_j + A_j A_i}{2}        \bigg) \bigotimes  \bigg( \underset{1\leq k \leq N-1}{\bigotimes} \textbf{I}_k \bigg)  \bigg)      \bigg] \ket{\psi_{N\mathrm{XOR}}}                           \bigg| \bigg|^2         <  C_1 n \bigg( \underset{1 \leq j \leq N-1}{\prod}  \big( n - j  \big) \bigg) \epsilon_{N\mathrm{XOR}}   \text{. }
\end{align*}

\bigskip

\noindent \textit{Proof of Lemma 3}. By direct computation, along the lines of a previous adaptation from the two-player setting for $\mathrm{XOR}^{*}$ and $\mathrm{FFL}$ games in  {[44]},

\begin{align*}
            2 \underset{1 \leq i < j \leq n }{\sum} \bigg| \bigg| \bigg(   \bigg(   \frac{A_i A_j + A_j A_i}{2}                          \bigg) \bigotimes \bigg( \underset{1\leq k \leq N-1}{\bigotimes} \textbf{I}_k\bigg) \bigg) \ket{\psi_{N\mathrm{XOR}}}              \bigg| \bigg|^2  \\ \leq       \big( 1 + \omega^{-1}_{N\mathrm{XOR}} \big)^2    \underset{1 \leq i < j  \leq n}{\sum}  \bigg[      \bigg| \bigg|        \bigg(     \bigg(  \frac{A_i + A_j}{\sqrt{2}} \bigg)   \bigg( \underset{1 \leq k \leq N-1}{\bigotimes} \textbf{I}_k \bigg) \bigg) \ket{\psi_{N\mathrm{XOR}}}   
 \\ - \bigg(  \textbf{I}   \bigotimes B_{ij}  \bigotimes \bigg( \underset{1 \leq k \leq N-2}{\bigotimes} \textbf{I}_k \bigg) \bigg)  \times \ket{\psi_{N\mathrm{XOR}}}  \bigg| \bigg|^2  +      \bigg| \bigg|      \bigg(          \bigg( \frac{A_i - A_j}{\sqrt{2}} \bigg)   \\ \bigotimes  \bigg( \underset{1 \leq k \leq N-1}{\bigotimes}           \textbf{I}_k \bigg) \bigg) \ket{\psi_{N\mathrm{XOR}}}  - \bigg( \textbf{I}  \bigotimes B_{ji}    \bigotimes \bigg( \underset{1 \leq k \leq N-2}{\bigotimes}           \textbf{I}_k \bigg)          \bigg)  \\ \times \ket{\psi_{N\mathrm{XOR}}}           \bigg| \bigg|^2          \bigg]                   \text{, }
\end{align*}

\noindent implies the desired upper bound is obtained from the observations,

\begin{align*}
        \underset{1 \leq i < j  \leq n}{\sum}      \bigg| \bigg|        \bigg(     \bigg(  \frac{A_i + A_j}{\sqrt{2}} \bigg)  \bigg( \underset{1 \leq k \leq N-1}{\bigotimes} \textbf{I}_k \bigg) \bigg) \ket{\psi_{N\mathrm{XOR}}}   
 - \bigg(  \textbf{I}   \bigotimes B_{ij}  \bigotimes \bigg( \underset{1 \leq k \leq N-2}{\bigotimes} \textbf{I}_k \bigg) \bigg)  \\ \times \ket{\psi_{N\mathrm{XOR}}}  \bigg| \bigg|^2   <                   \frac{1}{100}       n    \bigg( \underset{1 \leq j \leq N-1}{\prod}     \big( n - j \big) \bigg)    \epsilon_{N\mathrm{XOR}} 
          \text{, }
\end{align*}

\noindent and also that,

\begin{align*}
       \underset{1 \leq i < j  \leq n}{\sum}     \bigg| \bigg|      \bigg(          \bigg( \frac{A_i - A_j}{\sqrt{2}} \bigg)   \bigotimes  \bigg( \underset{1 \leq k \leq N-1}{\bigotimes}           \textbf{I}_k \bigg) \bigg) \ket{\psi_{N\mathrm{XOR}}}  - \bigg( \textbf{I}  \bigotimes B_{ji}   \bigotimes \bigg( \underset{1 \leq k \leq N-2}{\bigotimes}           \textbf{I}_k \bigg)          \bigg)  \\ \times \ket{\psi_{N\mathrm{XOR}}}           \bigg| \bigg|^2 <                \frac{1}{100}   n  \bigg( \underset{1 \leq j \leq N-1}{\prod}     \big( n - j \big) \bigg)  \epsilon_{N\mathrm{XOR}}                   \text{, }
\end{align*}

\noindent which together yield the desired upper bound, upon taking constants $C_1$ satisfying,

\begin{align*}
  C_1 > \frac{1}{50}    \text{, }
\end{align*}

\noindent from the observation that the desired upper bound holds iff,

\begin{align*}
     \underset{i,j}{\sum}     \bigg| \bigg|     \bigg[            \bigg( \bigg(    \frac{A_i A_j + A_j A_i}{2}        \bigg) \bigotimes  \bigg( \underset{1\leq k \leq N-1}{\bigotimes} \textbf{I}_k \bigg)  \bigg)      \bigg] \ket{\psi_{N\mathrm{XOR}}}                           \bigg| \bigg|^2 \\  <    \frac{1}{50} n  \bigg( \underset{1 \leq j \leq N-1}{\prod}     \big( n - j \big) \bigg) \epsilon_{\mathrm{XOR}}  \Longleftrightarrow C_1 > \frac{1}{50}            \text{, }
\end{align*}

\noindent from which we conclude the argument. \boxed{}

\bigskip

\noindent From the previous results, given the optimal value of a game,

\begin{align*}
  \text{Classical Optimal Value} \equiv \underset{\mathcal{S}}{\mathrm{sup}}  \big\{ \text{Probability that a player wins the game with some} \\ \text{classical strategy } \mathcal{S} \big\}  \text{, }
\end{align*}

\noindent the two-player game, given the tensor observables of each player, satisfies:

\bigskip

\noindent Besides the observable gathered by the first player, the bound straightforwardly adapted in the multiplayer $\mathrm{XOR}$ setting from {[44]}, the upper bound for simultaneously applying operations to change the order in which player observables appear in tensor products takes the following form.

\bigskip

\noindent \textbf{Lemma} \textit{4A} (\textit{induction on the, up to constants, $\epsilon$ upper bound from the previous result}). Under the assumptions of $\textbf{Lemma}$ \textit{3},

\begin{align*}
      \underset{\text{Tensor entries }}{\sum}  \bigg[ \underset{i_1,\cdots,i_n}{\sum}     \bigg| \bigg|     \bigg[            \bigg( \bigg(    \frac{A_i A_j + A_j A_i}{2}        \bigg) \bigotimes  \bigg( \underset{1\leq k \leq N-1}{\bigotimes} \textbf{I}_k \bigg)  \bigg)      \bigg] \ket{\psi_{N\mathrm{XOR}}}                           \bigg| \bigg|^2   \bigg]     \\   <  n^N \bigg( \bigg( \underset{1 \leq z \leq N}{\bigotimes} \bigg( \underset{1 \leq j \leq N-1}{\prod}  \big( n  - j  \big) \bigg) \textbf{I}_z \bigg)  \epsilon_{N\mathrm{XOR}}   \bigg)   \text{. }
\end{align*}

\bigskip

\noindent \textit{Proof of Lemma 4A}. Induction on the, up to constants, $\epsilon$ upper bound in the previous result implies that the desired upper bound, which we denote with $C_2$, takes the form,

\begin{align*}
C_2 \equiv     \bigg(  n \bigg( \underset{1 \leq j \leq N-1}{\prod}  \big( n - j  \big) \bigg) \textbf{I}   \bigg) \epsilon_{N\mathrm{XOR}} \bigotimes \overset{N-2}{\cdots} \bigotimes    \bigg(  n      \bigg( \underset{1 \leq j \leq N-1}{\prod}  \big( n - j  \big) \bigg) \textbf{I}   \bigg) \epsilon_{N\mathrm{XOR}}         \\ >  \bigg(      \bigg( \underset{1 \leq j \leq N-1}{\prod}  \big( n - j  \big) \bigg) \textbf{I}   \bigg) \epsilon_{N\mathrm{XOR}} \bigotimes \overset{N-2}{\cdots} \bigotimes    \bigg(      \bigg( \underset{1 \leq j \leq N-1}{\prod}  \big( n - j  \big) \bigg) \textbf{I}   \bigg) \epsilon_{N\mathrm{XOR}} \\    \underset{1 \leq z \leq N-1}{\bigotimes} \bigg(      \bigg( \underset{1 \leq j \leq N-1}{\prod}  \big( n - j  \big) \bigg) \textbf{I}_z \bigg)  \epsilon_{N\mathrm{XOR}}  \\      >                \underset{\text{Tensor entries}}{\sum}      \bigg[    \underset{i,j}{\sum}     \bigg| \bigg|     \bigg[            \bigg( \bigg(    \frac{A_i A_j + A_j A_i}{2}        \bigg) \bigotimes  \bigg( \underset{1\leq k \leq N-1}{\bigotimes} \textbf{I}_k \bigg)  \bigg)      \bigg] \ket{\psi_{N\mathrm{XOR}}}                           \bigg| \bigg|^2    \bigg]       \text{, }
\end{align*}

\noindent from which we conclude the argument, as,

\begin{align*}
 n^N \bigg(   \underset{1 \leq z \leq N}{\bigotimes} \bigg( \underset{1 \leq j \leq N-1}{\prod}  \big( n - j  \big) \bigg) \textbf{I}_z   \bigg) \epsilon_{N\mathrm{XOR}}  \bigg)  \equiv  C_2  \text{. }       \boxed{}
\end{align*}

\noindent \textbf{Lemma} $\textit{4A}^{*}$ (\textit{induction on the, up to constants, strong parallel repetition $\epsilon$ upper bound from the previous result}). Under the assumptions of $\textbf{Lemma}$ \textit{4A}, and $\textbf{Lemma}$ \textit{3}, and $C^{\wedge}_2>0$,

\begin{align*}
      \underset{\text{Tensor entries }}{\sum}  \bigg[ \underset{j^{\prime}_1,\cdots,j^{\prime}_n}{\underset{i^{\prime},\cdots,i^{\prime}_n}{\underset{i_1,\cdots,i_n}{\sum}}}     \bigg| \bigg|     \bigg[            \bigg( \bigg(    \frac{ \big( A_i \wedge A_{i^{\prime}} \big)  \big( A_j \wedge A_{j^{\prime}} \big)  + \big( A_j \wedge A_{j^{\prime} } \big) \big(  A_i \wedge A_{i^{\prime}} \big)}{2}        \bigg) \bigotimes  \textbf{I}  \bigg)      \bigg] \ket{\psi_{2\mathrm{XOR}}}                           \bigg| \bigg|^2   \bigg]  \\      <  C^{\wedge}_2 n   \bigg( \underset{1 \leq j \leq 2}{\prod}  \big( n - j  \big) \bigg) \epsilon_{2\mathrm{XOR}\wedge 2 \mathrm{XOR}}  \equiv    C^{\wedge}_2 n \bigg( \underset{1 \leq j \leq 2}{\prod}  \big( n - j  \big) \bigg) \epsilon_{\mathrm{XOR}\wedge  \mathrm{XOR}}   \text{. }
\end{align*}

\noindent \textit{Proof of Lemma $4A^{*}$}. Directly apply the arguments from the previous result, from which we conclude the argument. \boxed{}

\bigskip

\noindent In the $N$-player setting, the strong parallel repetition result above takes the expected form, given below:

\bigskip

\noindent \textbf{Lemma} $\textit{4A}^{**}$ (\textit{induction on the, up to constants, $N$-player strong parallel repetition $\epsilon$ upper bound from the previous result}). Under the assumptions of $\textbf{Lemma}$ \textit{4A}, $\textbf{Lemma}$ $\textit{4A}^{*}$, and $\textbf{Lemma}$ \textit{3}, and $C^{\wedge}_N>0$,

\begin{align*}
      \underset{\text{Tensor entries }}{\sum}  \bigg[ \underset{i^{\prime\cdots\prime},\cdots,i^{\prime\cdots\prime}}{\underset{\vdots}{\underset{i_1,\cdots,i_n}{\sum}}}     \bigg| \bigg|     \bigg[            \bigg( \bigg(    \frac{ \big( A_i \wedge A_{i^{\prime}}  \wedge \cdots \wedge A_{i^{\prime\cdots\prime}} \big)  \big( A_j \wedge A_{j^{\prime}} \wedge \cdots \wedge A_{j^{\prime\cdots\prime}} \big)  +\big( A_j \wedge A_{j^{\prime}}  \wedge \cdots \wedge A_{j^{\prime\cdots\prime}} \big)   }{2}    \end{align*}

        \begin{align*}    \times  \frac{\cdots \big( A_i \wedge A_{i^{\prime}}  \wedge \cdots  \wedge A_{i^{\prime\cdots\prime}} \big)}{\cdots }      \bigg) \bigotimes  \textbf{I}  \bigg)      \bigg] \ket{\psi_{\mathrm{XOR}\wedge \cdots \wedge \mathrm{XOR}}}                           \bigg| \bigg|^2   \bigg]       <  C^{\wedge}_N n   \bigg( \underset{1 \leq j \leq N}{\prod}  \big( n - j  \big) \bigg) \epsilon_{N\mathrm{XOR}\wedge \cdots \wedge  N\mathrm{XOR}} \\ \equiv    C^{\wedge}_N n \bigg( \underset{1 \leq j \leq N}{\prod}  \big( n - j  \big) \bigg) \epsilon_{\mathrm{XOR}\wedge  \cdots \wedge  \mathrm{XOR}}   \text{. }
\end{align*}

\noindent \textit{Proof of Lemma $4A^{**}$}. Directly apply the arguments from the previous two results, from which we conclude the argument. \boxed{}

\bigskip

\noindent We conclude the series of related results for strong parallel repetition of the $\mathrm{FFL}$ game, which can also be shown to hold using identical arguments.

\bigskip

\noindent \textbf{Lemma} $\textit{4A}^{***}$ (\textit{induction on the, up to constants, $\mathrm{FFL}$ strong parallel repetition $\epsilon$ upper bound from the previous result}). Under the assumptions of $\textbf{Lemma}$ \textit{4A}, and $\textbf{Lemma}$ \textit{3}, and $C^{\wedge}_2>0$,

\begin{align*}
      \underset{\text{Tensor entries }}{\sum}  \bigg[ \underset{i^{\prime}_1,\cdots, i^{\prime}_n}{\underset{i_1,\cdots,i_n}{\sum}}     \bigg| \bigg|     \bigg[            \bigg( \bigg(    \frac{ \big( A_i \wedge A_{i^{\prime}} \big)  \big( A_j \wedge A_{j^{\prime}} \big)  + \big( A_j \wedge A_{j^{\prime} } \big) \big(  A_i \wedge A_{i^{\prime}} \big)}{2}        \bigg) \bigotimes  \textbf{I}  \bigg)      \bigg] \ket{\psi_{\mathrm{FFL}}}                           \bigg| \bigg|^2   \bigg]      \\  <  C^{\wedge}_2 n   \bigg( \underset{1 \leq j \leq 2}{\prod}  \big( n - j  \big) \bigg) \epsilon_{\mathrm{FFL}\wedge  \mathrm{FFL}}       \text{. }
\end{align*}

\noindent \textit{Proof of Lemma $4A^{***}$}. Directly apply the arguments from the previous three results, from which we conclude the argument. \boxed{}

\bigskip

\noindent In comparison to upper bounds that are only dependent upon $\epsilon$, other results for applying transformations to observables of one player, have the following impact on the observables of the other place for two players:

\bigskip

\noindent \textbf{Lemma} \textit{4B} ($\sqrt{\epsilon}$- \textit{approximality}, \textbf{Lemma} \textit{8}, {[44]}). From the same quantities introduced in the previous result, one has,

\begin{align*}
   \bigg| \bigg|               \bigg(  A_k \otimes \textbf{I} \bigg) \ket{\psi_{\mathrm{FFL}}}    -  \bigg( \textbf{I} \otimes \bigg(     \frac{\pm B_{kl} + B_{lk}}{\big| \pm B_{kl} + B_{lk}  \big| }           \bigg) \bigg) \ket{\psi_{\mathrm{FFL}}}              \bigg| \bigg| < 17 \sqrt{n \epsilon}   \text{. } 
\end{align*}

\noindent In addition to the result above, the following result below is used to characterize the error bound resulting from permuting the indicates of the first player's tensor observable.

\bigskip

\noindent \textbf{Lemma} \textit{5} (\textit{error bound from permuting indices}, \textbf{Lemma} \textit{5}, {[44]}). One has,

\begin{align*}
 \bigg| \bigg|  \bigg(   \bigg( \underset{1 \leq i \leq n}{\prod} A^{j_i}_i  \bigg)   -    \bigg( \underset{\text{if } i \equiv j_1+1, \text{ } \mathrm{set} \text{ } j_1 + 1 \equiv j_1 \oplus 1}{ \underset{1 \leq i \leq n}{\prod} }A^{j_i}_i        \bigg) \otimes \textbf{I} \bigg) \ket{\psi_{\mathrm{FFL}}}   \bigg| \bigg|  \leq \frac{100}{9} n^2 \sqrt{\epsilon}  \text{. } 
\end{align*}

\noindent We postpone the arguments for generalizing \textbf{Lemma} \textit{4B} to \textit{1.5.3}, due to the fact that more complicated upper bounds must be computed, under several circumstances, for games with more than two players. In comparison to several technical computations that must be employed for demonstrating that counterparts of \textbf{Lemma} \textit{4B}, above, hold, \textbf{Lemma} \textit{5} can be shown to hold from previous observations relating to the optimal values of the $\mathrm{XOR}$, and $\mathrm{FFL}$ games,

\begin{align*}
\text{Quantum } \mathrm{XOR/XOR}^{*} \text{ value} \equiv  \omega \big( \mathrm{XOR } \big) \equiv \omega \big( \mathrm{XOR}^{*} \big) \equiv \frac{1}{\sqrt{2}}  \text{, } \\  \text{Classical } \mathrm{FFL} \text{ value} \equiv \text{Quantum } \mathrm{FFL} \text{ value}  \equiv  \omega \big( \mathrm{FFL} \big) \equiv \frac{2}{3}    \text{, }
\end{align*}

\noindent in addition to whether the optimal value of the game changes under strong parallel repetition, in which,

\begin{align*}
       \omega_{\mathrm{XOR} \wedge \cdots \wedge \mathrm{XOR}} \big( G \big)   \equiv \omega \big(  \mathrm{XOR} \wedge \cdots \wedge \mathrm{XOR} \big)  \equiv \underset{\# \text{ of strong parallel repetitions } j}{\prod}   \omega \big( \mathrm{XOR}\big)^{j} \\  \equiv \bigg( \frac{1}{\sqrt{2}} \bigg)^{\# \text{ of strong parallel repetitions }}   \equiv \bigg( \frac{1}{\sqrt{2}} \bigg)^{n} \text{, } \\ \\       \omega_{\mathrm{FFL} \wedge \mathrm{FFL}} \big( G \big) \equiv \frac{2}{3} \neq  \underset{\# \text{ of strong parallel repetitions } j}{\prod}   \omega \big( \mathrm{XOR}\big)^{j} \\  \equiv  \bigg( \frac{2}{3} \bigg)^{\# \text{ of strong parallel repetitions}}   \neq \omega_{\mathrm{FFL}} \big( G \big) \equiv \frac{2}{3}        \text{. }
\end{align*}

\noindent We make use of the two properties of the $\mathrm{XOR}$ and $\mathrm{FFL}$ optimal values below.

\bigskip

\noindent \textbf{Lemma} $\textit{5}^{*}$ (\textit{error bound from permuting indices in the N-player setting}, \textbf{Lemma} \textit{5}, {[44]}). One has the following error bound from permuting indices,

\begin{align*}
     \bigg| \bigg|  \bigg(    \bigg( \underset{1 \leq i \leq n}{\prod} A^{j_i}_i  \bigg)      \bigotimes \bigg( \underset{1 \leq z \leq N-1}{\bigotimes} \textbf{I}_z \bigg)       \bigg) \ket{\psi_{N\mathrm{XOR}}} - \bigg(     \bigg( \underset{\text{if } i \equiv j_1+1, \text{ } \mathrm{set} \text{ } j_1 + 1 \equiv j_1 \oplus 1}{ \underset{1 \leq i \leq n}{\prod} }A^{j_i}_i        \bigg)       \\ \bigotimes \bigg( \underset{1 \leq z \leq N-1}{\bigotimes} \textbf{I}_z \bigg)            \bigg)  \times \ket{\psi_{N\mathrm{XOR}}}     \bigg| \bigg|           <  N n^{N+\epsilon_{N\mathrm{XOR}}} \omega^3_{N\mathrm{XOR}}  \text{. }
\end{align*}

\noindent \textit{Proof of Lemma $5^{*}$}. Set $\epsilon \equiv \epsilon_{N\mathrm{XOR}}$. The desired upper bound is of the form,

\begin{align*}
        n^N \sqrt{\epsilon}  \bigg(     \bigg(               \omega_{N\mathrm{XOR}}^{-2}         +        \omega_{N\mathrm{XOR}}^{-1}        \bigg)        +     \frac{1}{\sqrt{n^{N-1}}}   \bigg(  \omega_{N\mathrm{XOR}}^{-2}         +        \omega_{N\mathrm{XOR}}^{-1}          \bigg)         \bigg) \\ \equiv  n^N \sqrt{\epsilon}     \bigg(               \omega_{N\mathrm{XOR}}^{-2}         +        \omega_{N\mathrm{XOR}}^{-1}        \bigg)         +  n^N \sqrt{\epsilon}   \bigg( \frac{1}{\sqrt{n^{N-1}}}   \bigg(  \omega_{N\mathrm{XOR}}^{-2}         +        \omega_{N\mathrm{XOR}}^{-1}          \bigg)     \bigg)   \\ <       n^N \sqrt{\epsilon}     \bigg(               \omega_{N\mathrm{XOR}}^2         +        \omega_{N\mathrm{XOR}}        \bigg)           +  n^N \sqrt{\epsilon}      \bigg( \frac{1}{\sqrt{n^{N-1}}}   \bigg(  \omega_{N\mathrm{XOR}}^2        +        \omega_{N\mathrm{XOR}}          \bigg)     \bigg)                             \text{, }
\end{align*}

\noindent which can further be rearranged as to obtain the desired upper bound,

\begin{align*}
    n^N \sqrt{\epsilon}     \bigg(               \omega_{N\mathrm{XOR}}^2         +        \omega_{N\mathrm{XOR}}        \bigg)         +  n^N \sqrt{\epsilon}      \bigg( \frac{1}{\sqrt{n^{N-1}}}   \bigg(  \omega_{N\mathrm{XOR}}^2        +        \omega_{N\mathrm{XOR}}          \bigg)     \bigg)  \\      \equiv    n^N \sqrt{\epsilon}  \bigg[   \bigg(               \omega_{N\mathrm{XOR}}^2          +        \omega_{N\mathrm{XOR}}        \bigg)         +      \bigg( \frac{1}{\sqrt{n^{N-1}}}   \bigg(  \omega_{N\mathrm{XOR}}^2        +        \omega_{N\mathrm{XOR}}          \bigg)     \bigg)             \bigg]  \\ <    n^N \sqrt{\epsilon} \omega_{N\mathrm{XOR}}^3   \bigg[      1    +     \frac{1}{\sqrt{n^{N-1}}}               \bigg]           
    \\       < n^{N+\epsilon} \omega^3_{N\mathrm{XOR}}  \bigg[      1    +     \frac{1}{\sqrt{n^{N-1}}}               \bigg]  \\ \equiv        n^{N+\epsilon} \omega^3_{N\mathrm{XOR}}     +        \frac{n^{N+\epsilon} \omega^3_{N\mathrm{XOR}} }{\sqrt{n^{N-1}}}  \\          <        n^{N+\epsilon} \omega^3_{N\mathrm{XOR}}     +       n^{N+\epsilon} \omega^3_{N\mathrm{XOR}}               \\ \equiv 2 n^{N+\epsilon} \omega^3_{N\mathrm{XOR}} \\ <   N n^{N+\epsilon} \omega^3_{N\mathrm{XOR}}       \text{, }
\end{align*}

\noindent from which we conclude the argument. \boxed{}

\bigskip

\noindent \textbf{Lemma} $\textit{5}^{**}$ (\textit{error bound from permuting indices in the strong parallel repetition of the N-player setting}, \textbf{Lemma} \textit{5}, {[44]}). One has the following error bound from permuting indices,

\begin{align*}
              \bigg| \bigg|  \bigg(  \bigg(    \bigg( \underset{1 \leq i \leq n}{\prod} A^{j_i}_i  \bigg)   \wedge  \cdots \wedge \bigg( \underset{1 \leq i^{\prime\cdots\prime} \leq n^{\prime\cdots\prime}}{\prod} A^{j^{\prime\cdots\prime}_{i^{\prime\cdots\prime}}}_{i^{\prime\cdots\prime}}   \bigg) \bigg)       \bigotimes \bigg( \underset{1 \leq z \leq N-1}{\bigotimes}  \big( \textbf{I}_z \wedge \cdots \wedge \textbf{I}_z \big)  \bigg) 
 \bigg) \\ \times  \ket{\psi_{N\mathrm{XOR}\wedge \cdots \wedge N\mathrm{XOR}}} \\  - \bigg(   \bigg(   \bigg( \underset{\text{if } i \equiv j_1+1, \text{ } \mathrm{set} \text{ } j_1 + 1 \equiv j_1 \oplus 1}{ \underset{1 \leq i \leq n}{\prod} }A^{j_i}_i        \bigg)  \wedge \cdots \wedge \bigg(     \underset{\text{if } i^{\prime\cdots\prime} \equiv j^{\prime\cdots\prime}_1+1, \text{ } \mathrm{set} \text{ } j^{\prime\cdots\prime}_1 + 1 \equiv j^{\prime\cdots\prime}_1 \oplus 1}{ \underset{1 \leq i^{\prime\cdots\prime} \leq n^{\prime\cdots\prime}}{\prod} }A^{j^{\prime\cdots\prime}_{i^{\prime\cdots\prime}}}_{i^{\prime\cdots\prime}}           \bigg)    \bigg)  \\  \bigotimes \bigg( \underset{1 \leq z \leq N-1}{\bigotimes}\big(  \textbf{I}_z  \wedge  \cdots \wedge \textbf{I}_z \big)   \bigg)    \bigg)     \ket{\psi_{N\mathrm{XOR} \wedge \cdots \wedge N\mathrm{XOR}}}     \bigg| \bigg|  <  n^{N+\epsilon}_{\wedge } +  \bigg( \frac{50 n^{N+\epsilon}_{\wedge}}{\sqrt{n^{N-1}}}   \bigg) \\ \times \omega_{N\mathrm{XOR} \wedge \cdots \wedge N \mathrm{XOR}}          \text{. }
\end{align*}

\noindent \textit{Proof of Lemma $5^{**}$}. Set $\epsilon \equiv \epsilon_{N\mathrm{XOR}\wedge \cdots \wedge N \mathrm{XOR}}$. The desired upper bound is of the form,

\begin{align*}
        n^N_{\wedge} \sqrt{\epsilon} \bigg(     \bigg(               \omega_{N\mathrm{XOR}\wedge \cdots \wedge N{\mathrm{XOR}}}^{-2}         +        \omega_{N\mathrm{XOR}\wedge \cdots \wedge N{\mathrm{XOR}}}^{-1}        \bigg)        +     \frac{1}{\sqrt{n^{N-1}}}   \bigg(  \omega_{N\mathrm{XOR}\wedge \cdots \wedge N{\mathrm{XOR}}}^{-2}      \\    +        \omega_{N\mathrm{XOR}\wedge \cdots \wedge N{\mathrm{XOR}}}^{-1}          \bigg)         \bigg)     \\ \equiv    n^N_{\wedge} \sqrt{\epsilon} \bigg(     \bigg(                \bigg( \frac{1}{\sqrt{2}} \bigg)^{-2N} + \bigg( \frac{1}{\sqrt{2}} \bigg)^{-N}         \bigg)        +     \frac{1}{\sqrt{n^{N-1}}}   \bigg(    \bigg( \frac{1}{\sqrt{2}} \bigg)^{-2N} + \bigg( \frac{1}{\sqrt{2}} \bigg)^{-N}            \bigg)         \bigg) \\ \equiv             \bigg( \frac{1}{\sqrt{2}} \bigg)^{-2N} \bigg[    n^N_{\wedge} \sqrt{\epsilon} \bigg(     \bigg(               1 + \bigg( \frac{1}{\sqrt{2}} \bigg)^{N}         \bigg)        +     \frac{1}{\sqrt{n^{N-1}}}   \bigg(    1 + \bigg( \frac{1}{\sqrt{2}} \bigg)^{N}             \bigg)         \bigg)         \bigg]    \\       <      \bigg( \frac{1}{\sqrt{2}} \bigg)^{-2N} \bigg[    n^{N+\epsilon}_{\wedge} \bigg(     \bigg(               1   + \bigg( \frac{1}{\sqrt{2}} \bigg)^{N}         \bigg)        +     \frac{1}{\sqrt{n^{N-1}}}   \bigg(    1 + \bigg( \frac{1}{\sqrt{2}} \bigg)^{N}             \bigg)         \bigg)         \bigg]    \\  <   \bigg( \frac{1}{\sqrt{2}} \bigg)^{-N} \bigg[    n^{N+\epsilon}_{\wedge} \bigg(     \bigg(               1 + \bigg( \frac{1}{\sqrt{2}} \bigg)^{N}         \bigg)        +     \frac{1}{\sqrt{n^{N-1}}}   \bigg(    1 + \bigg( \frac{1}{\sqrt{2}} \bigg)^{N}             \bigg)         \bigg)         \bigg]    \\   <      \bigg( \frac{1}{\sqrt{2}} \bigg)^{-N} \bigg[    n^{N+\epsilon}_{\wedge} \bigg(   \bigg( \frac{2}{\sqrt{2}} \bigg)^{N}          +     \frac{1}{\sqrt{n^{N-1}}}  \bigg( \frac{2}{\sqrt{2}} \bigg)^{N}                    \bigg)         \bigg]      \\  \equiv    \bigg( \frac{1}{\sqrt{2}} \bigg)^{-N}  \bigg( \frac{2}{\sqrt{2}} \bigg)^{N}     \bigg[    n^{N+\epsilon}_{\wedge} \bigg(   1      +     \frac{1}{\sqrt{n^{N-1}}}                   \bigg)         \bigg]   \\    <               \bigg( \frac{2}{\sqrt{2}} \bigg)^{-N}  \bigg( \frac{2}{\sqrt{2}} \bigg)^{N}     \bigg[    n^{N+\epsilon}_{\wedge} \bigg(   1      +     \frac{1}{\sqrt{n^{N-1}}}                   \bigg)         \bigg]  \\  \equiv    \bigg[    n^{N+\epsilon}_{\wedge} \bigg(   1      +     \frac{1}{\sqrt{n^{N-1}}}                   \bigg)         \bigg]  \\  <    n^{N+\epsilon}_{\wedge } + \bigg(  \frac{50 n^{N+\epsilon}_{\wedge}}{\sqrt{n^{N-1}}}  \bigg)    \omega_{N\mathrm{XOR} \wedge \cdots \wedge N \mathrm{XOR}}    \text{, }
\end{align*}

\noindent from which we conclude the argument. \boxed{}

\bigskip

\noindent \textbf{Lemma} $\textit{5}^{***}$ (\textit{error bound from permuting indices in the strong parallel repetition of the $\mathrm{FFL}$ 2-player setting}, \textbf{Lemma} \textit{5}, {[44]}). One has the following error bound from permuting indices,

\begin{align*}
             \bigg| \bigg|  \bigg(     \bigg( \bigg( \underset{1 \leq i \leq n}{\prod} A^{j_i}_i  \bigg)  \wedge \bigg( \underset{1 \leq i^{\prime}\leq n^{\prime}}{\prod} A^{j^{\prime}_{i^{\prime}}}_{i^{\prime}}  \bigg) \bigg)     \bigotimes \bigg( \underset{1 \leq z \leq N-1}{\bigotimes}  \big( \textbf{I}_z \wedge \textbf{I}_z \big)  \bigg)  \bigg) \\ \times \ket{\psi_{\mathrm{FFL}\wedge \mathrm{FFL}}} \\ - \bigg(   \bigg(    \bigg( \underset{\text{if } i \equiv j_1+1, \text{ } \mathrm{set} \text{ } j_1 + 1 \equiv j_1 \oplus 1}{ \underset{1 \leq i \leq n}{\prod} }A^{j_i}_i        \bigg)      \wedge \bigg( \underset{\text{if } i^{\prime} \equiv j^{\prime}_1+1, \text{ } \mathrm{set} \text{ } j^{\prime}_1 + 1 \equiv j^{\prime}_1 \oplus 1}{ \underset{1 \leq i^{\prime} \leq n^{\prime}}{\prod} }A^{j^{\prime}_{i^{\prime}}}_{i^{\prime}}     \bigg) \bigg)  \\    \bigotimes \bigg( \underset{1 \leq z \leq N-1}{\bigotimes}  \big( \textbf{I}_z \wedge \textbf{I}_z \big)  \bigg) \bigg) \ket{\psi_{\mathrm{FFL}\wedge \mathrm{FFL}}}    \bigg| \bigg|          <    n^{N+\epsilon}_{\wedge } \\  +  \bigg( \frac{50 n^{N+\epsilon}_{\wedge}}{\sqrt{n}}  \bigg) \omega_{\mathrm{FFL} \wedge  \mathrm{FFL}}      \text{. }
\end{align*}

\noindent \textit{Proof of Lemma $5^{***}$}. The desired upper bound follows from a variant of that provided in the argument for the previous result, upon setting $\epsilon_{\mathrm{FFL} \wedge \mathrm{FFL}} \equiv \epsilon$, 

\begin{align*}
        n^2_{\wedge} \sqrt{\epsilon} \bigg(     \bigg(               \omega_{\mathrm{FFL}\wedge \mathrm{FFL}}^{-2}         +        \omega_{\mathrm{FFL}\wedge \mathrm{FFL}}^{-1}        \bigg)        +     \frac{1}{\sqrt{n}}   \bigg(  \omega_{\mathrm{FFL}\wedge \mathrm{FFL}}^{-2}         +        \omega_{\mathrm{FFL}\wedge \mathrm{FFL}}^{-1}          \bigg)         \bigg)     \\ \equiv     n^2_{\wedge} \sqrt{\epsilon}  \bigg(     \bigg(                \bigg( \frac{2}{3} \bigg)^{-4}  + \bigg(\frac{2}{3} \bigg)^{-2}         \bigg)      +     \frac{1}{\sqrt{n}}   \bigg(    \bigg(\frac{2}{3} \bigg)^{-4} + \bigg( \frac{2}{3} \bigg)^{-2}            \bigg)         \bigg)   \\    \equiv             \bigg( \frac{2}{3} \bigg)^{-4} \bigg[    n^N_{\wedge} \sqrt{\epsilon} \bigg(     \bigg(               1 + \bigg( \frac{2}{3} \bigg)^{2}         \bigg)        +     \frac{1}{\sqrt{n}}  \bigg(    1   + \bigg( \frac{2}{3}\bigg)^{2}             \bigg)         \bigg)         \bigg]     \\    <      \bigg(\frac{2}{3} \bigg)^{-4} \bigg[    n^{N+\epsilon}_{\wedge} \bigg(     \bigg(               1 + \bigg( \frac{2}{3}  \bigg)^{2}         \bigg)        +     \frac{1}{\sqrt{n}}    \bigg(    1 + \bigg( \frac{2}{3}  \bigg)^{2}             \bigg)         \bigg)         \bigg]  \\  <   \bigg(\frac{2}{3} \bigg)^{-2} \bigg[    n^{N+\epsilon}_{\wedge} \bigg(     \bigg(               1 + \bigg( \frac{2}{3} \bigg)^{2}         \bigg)        +   \frac{1}{\sqrt{n}}    \bigg(    1 + \bigg( \frac{2}{3} \bigg)^{2}             \bigg)         \bigg)         \bigg]  \\   <   \bigg(\frac{2}{3} \bigg)^{-2} \bigg[    n^{N+\epsilon}_{\wedge} \bigg(     \bigg(               \bigg( \frac{3}{2} \bigg)^2  + \bigg( \frac{2}{3} \bigg)^{2}         \bigg)        +   \frac{1}{\sqrt{n}}   \bigg(    \bigg( \frac{3}{2} \bigg)^2 + \bigg( \frac{2}{3} \bigg)^{2}             \bigg)         \bigg)         \bigg]  \\ <   \bigg(\frac{2}{3} \bigg)^{-2} \bigg[    n^{N+\epsilon}_{\wedge} \bigg(     \bigg(       2  \bigg( \frac{3}{2} \bigg)^2  \bigg)        +   \frac{1}{\sqrt{n}}   \bigg(  2  \bigg( \frac{3}{2} \bigg)^2  \bigg)         \bigg)         \bigg]    \\     \equiv    \bigg(  2  \bigg( \frac{3}{2} \bigg)^2  \bigg)              \bigg(\frac{2}{3} \bigg)^{-2} \bigg[    n^{N+\epsilon}_{\wedge}        +   \frac{1}{\sqrt{n}}    \bigg]                            \\ < 2   \bigg[    n^{N+\epsilon}_{\wedge}        +   \frac{1}{\sqrt{n}}    \bigg]        \equiv   2 n^{N+\epsilon}_{\wedge}        +   \frac{2}{\sqrt{n}}    \\     <    n^{N+\epsilon}_{\wedge } + \bigg(  \frac{50 n^{N+\epsilon}_{\wedge}}{\sqrt{n}}    \bigg)  \omega_{\mathrm{FFL} \wedge  \mathrm{FFL}}      \text{, }
\end{align*}

\noindent from which we conclude the argument. \boxed{}

\subsubsection{$5$-$\mathrm{XOR}$ game}

\noindent For the $5$-$\mathrm{XOR}$ game, introduce,

\begin{align*}
 \underline{\mathscr{P}_{5\mathrm{XOR}}} \equiv \underline{\mathscr{P}_{5\mathrm{XOR},1}}        \cup   \underline{\mathscr{P}_{5\mathrm{XOR},1,2}}         \cup   \underline{\mathscr{P}_{5\mathrm{XOR},1,3}}       \cup    \cdots \cup \underline{\mathscr{P}_{5\mathrm{XOR},1,2,3,4,5}}        \text{, }
\end{align*}

\noindent where each set of permutations $\mathscr{P}$ are of the form,

\begin{align*}
    \underline{\mathscr{P}_{5\mathrm{XOR},1}}     \equiv   \underset{\sigma \in S_5}{\bigcup} \big\{ \text{Permutations } \sigma  \text{ of the first player's tensor observable} \big\}  \text{, }    \\ \vdots \\ \underline{\mathscr{P}_{5\mathrm{XOR},1,2,3,4,5}}     \equiv   \underset{\sigma , \sigma^{\prime}, \sigma^{\prime\prime}, \sigma^{\prime\prime\prime}, \sigma^{\prime\prime\prime\prime}\in S_5}{\bigcup} \big\{ \text{Permutations } \sigma  \text{ of all of the player's tensor observables} \big\}  \text{. }
\end{align*}

\noindent The tensor product representation, and corresponding permutation superposition, for the $5$-XOR game are provided below, beginning with \textit{Table 8}.

\begin{table}[ht]
\caption{Permutations of Player Observables for the 5-XOR game error bound, $(\textbf{EB}- 5 \mathrm{XOR})$} 
\centering 
\begin{tabular}{c c c c} 
\hline\hline 
 Player & Tensor Product Representation & Permutation Superposition  \\ [0.5ex] 
\hline     1 & $\underset{\sigma} {\mathlarger{\sum}} \bigg( E_{ijklm} \otimes C_{ijk} \otimes A_{\sigma( i) } \otimes B_{ij}  \otimes D_{ijkl} \bigg)                      $ & $E_{ijklm} \otimes C_{ijk} \otimes A_{ i } \otimes B_{ij}  \otimes D_{ijkl} +E_{ijklm} \otimes C_{ijk}  $ \\ & & $\otimes A_{ j } \otimes B_{ij}  \otimes D_{ijkl} + E_{ijklm} \otimes C_{ijk} \otimes A_{ k } \otimes B_{ij}  \otimes D_{ijkl}  $ \\ & & $+        E_{ijklm} \otimes C_{ijk} \otimes A_{ l} \otimes B_{ij}  \otimes D_{ijkl}         $ \\  \\ \\ 2 & $\underset{\sigma} {\mathlarger{\sum}} \bigg(    E_{ijklm} \otimes C_{ijk} \otimes A_{i } \otimes B_{\sigma( ij)}  \otimes D_{ijkl} \bigg)   $ & $E_{ijklm} \otimes  C_{ijk} \otimes A_i \otimes B_{ij} \otimes D_{ijkl} +     E_{ijklm} \otimes  C_{ijk} \otimes A_i   $ \\ & & $\otimes B_{ik} \otimes D_{ijkl}  + E_{ijklm} \otimes  C_{ijk} \otimes A_i \otimes B_{il} \otimes D_{ijkl}           $ \\  & & $+   E_{ijklm} \otimes  C_{ijk} \otimes A_i \otimes B_{ji} \otimes D_{ijkl}    + E_{ijklm} \otimes  C_{ijk} \otimes A_i         $ \\ & & $ \otimes B_{jk} \otimes D_{ijkl}     + E_{ijklm} \otimes  C_{ijk} \otimes A_i \otimes B_{jl} \otimes D_{ijkl}    $ \\ & & $+   E_{ijklm} \otimes  C_{ijk} \otimes A_i \otimes B_{ki} \otimes D_{ijkl}  + E_{ijklm} \otimes  C_{ijk} $ \\   & & $\otimes A_i \otimes B_{kl} \otimes D_{ijkl}   +  E_{ijklm} \otimes  C_{ijk}  \otimes A_i \otimes B_{li} \otimes D_{ijkl}   $ \\ & & $ + E_{ijklm} \otimes  C_{ijk}  \otimes A_i \otimes B_{lj} \otimes D_{ijkl} + E_{ijklm} \otimes  C_{ijk}   $ \\  & & $\otimes A_i \otimes B_{lk} \otimes D_{ijkl} $    \\  & $\vdots$ &    \\              [1ex] 
\hline 
\end{tabular}
\label{table:nonlin} 
\end{table}

\begin{table}[ht]
\caption{Permutations of Player Observables for the 5-XOR game error bound, $(\textbf{EB}- 5 \mathrm{XOR})$ Continued} 
\centering 
\begin{tabular}{c c c c} 
\hline\hline 
 Player & Tensor Product Representation & Permutation Superposition  \\ [0.5ex] 
\hline & $\vdots$ &  \\  3  & $\underset{\sigma} {\mathlarger{\sum}} \bigg( E_{ijklm} \otimes C_{\sigma( ijk)} \otimes  A_{i } \otimes B_{ij}   \otimes D_{ijkl} \bigg) $ &      $E_{ijklm} \otimes C_{ ijk} \otimes  A_{i } \otimes B_{ij}   \otimes D_{ijkl} + E_{ijklm} \otimes  C_{ ikj}  \otimes  A_{i } $           \\ & & $ \otimes B_{ij}   \otimes D_{ijkl}  +    E_{ijklm} \otimes  C_{jik}  \otimes  A_{i }    \otimes B_{ij} \otimes D_{ijkl}       $    \\ & & $ +   E_{ijklm} \otimes   C_{jki} \otimes A_i \otimes D_{ijkl}  +    E_{ijklm} \otimes   C_{kij} \otimes A_i  \otimes D_{ijkl}  $ \\ & & $ +   E_{ijklm}    \otimes   C_{jki}  \otimes A_i \otimes D_{ijkl}   +   E_{ijklm}    \otimes   C_{kji}  \otimes A_i \otimes D_{ijkl}          $ \\   \\     4    &    $\underset{\sigma} {\mathlarger{\sum}} \bigg( E_{ijklm} \otimes C_{ijk} \otimes  A_{ i } \otimes B_{ij}   \otimes D_{\sigma( ijkl)} \bigg) $     &         $E_{ijklm} \otimes C_{ijk} \otimes  A_{ i } \otimes B_{ij}   \otimes D_{ ijkl}  +     E_{ijklm} \otimes C_{ijk} \otimes  A_{ i }     $     \\  & & $ \otimes B_{ij}   \otimes D_{ ikjl}    + E_{ijklm} \otimes C_{ijk} \otimes  A_{ i } \otimes B_{ij}   \otimes D_{ iklj}        $  \\ & & $ +    E_{ijklm} \otimes C_{ijk} \otimes A_i \otimes B_{ij} \otimes D_{ilkj} +   E_{ijklm} \otimes C_{ijk} \otimes A_i   $  \\ & & $ \otimes B_{ij} \otimes D_{iljk}  +     E_{ijklm} \otimes C_{ijk} \otimes A_i \otimes B_{ij} \otimes D_{jikl}         $ \\   & & $ +   E_{ijklm} \otimes C_{ijk} \otimes A_i \otimes B_{ij} \otimes D_{jilk}  +     E_{ijklm} \otimes C_{ijk}   $ \\       & & $  \otimes A_i \otimes B_{ij} \otimes D_{jlik}   +       E_{ijklm} \otimes C_{ijk}  \otimes A_i \otimes   $   \\ & & $  B_{ij} \otimes D_{jlki}   +      E_{ijklm} \otimes C_{ijk}  \otimes A_i \otimes  $ \\  & & $  B_{ij} \otimes D_{kijl}    +     E_{ijklm} \otimes C_{ijk}   \otimes A_i \otimes   $ \\ & & $ B_{ij} \otimes D_{kjil}  +   E_{ijklm} \otimes C_{ijk} \otimes A_i \otimes   $ \\  & & $ B_{ij} \otimes D_{kjli}  +     E_{ijklm} \otimes C_{ijk}   \otimes A_i \otimes $\\   & & $  B_{ij} \otimes D_{kjil} +            E_{ijklm} \otimes C_{ijk}  \otimes A_i  \otimes $  \\ & & $  B_{ij} \otimes  D_{kilj}   +    E_{ijklm} \otimes C_{ijk}    \otimes A_i \otimes  $   \\    & & $B_{ij} \otimes D_{lijk}      +   E_{ijklm} \otimes C_{ijk} \otimes A_i  $     \\     & & $ \otimes B_{ij} \otimes D_{ljik}   +   E_{ijklm} \otimes C_{ijk}     $   \\ & & $\otimes A_i \otimes B_{ij} \otimes D_{ljki}   +     E_{ijklm} \otimes C_{ijk}   $ \\  &  & $  \otimes A_i \otimes B_{ij} \otimes D_{lkji}  + E_{ijklm} \otimes C_{ijk}    $  \\ & & $ \otimes A_i \otimes B_{ij} \otimes D_{lkij} 
 $ \\   \\    5      &  $  \underset{\sigma} {\mathlarger{\sum}} \bigg( E_{\sigma(ijklm)} \otimes C_{ijk} \otimes A_{i } \otimes B_{ij}   \otimes D_{ijkl} \bigg)  $  & \text{Permutations for E tensor are in an Appendix table} \\ & $\vdots$ & \\   [1ex] 
\hline 
\end{tabular}
\label{table:nonlin} 
\end{table}

\begin{table}[ht]
\caption{Permutations of Player Observables for the 5-XOR game error bound, $(\textbf{EB}- 5 \mathrm{XOR})$ Continued} 
\centering 
\begin{tabular}{c c c c} 
\hline\hline 
 Player & Tensor Product Representation & Permutation Superposition  \\ [0.5ex] 
\hline & $\vdots$ & \\  1,2 & $  \underset{\sigma,\sigma^{\prime}} {\mathlarger{\sum}} \bigg( E_{ijklm} \otimes C_{ijk} \otimes A_{\sigma (i) } \otimes B_{\sigma^{\prime} (ij)}   \otimes D_{ijkl} \bigg)  $ & \text{Combine previous permu-} \\ & & \text{tations listed above}   \\ \\ 1,2,3 & $  \underset{\sigma,\sigma^{\prime}, \sigma^{\prime\prime}} {\mathlarger{\sum}} \bigg( E_{\sigma(ijklm)} \otimes C_{\sigma^{\prime\prime}(ijk)} \otimes A_{\sigma (i) } \otimes B_{\sigma^{\prime} (ij)}   \otimes D_{ijkl} \bigg)  $ & \text{Combine previous permu-} \\ & &  \text{tations listed above}     \\ & $\vdots$ & \\ 1,2,3,4,5 & $  \underset{\sigma,\sigma^{\prime}, \sigma^{\prime\prime},\sigma^{\prime\prime\prime}, \sigma^{\prime\prime\prime\prime}} {\mathlarger{\sum}} \bigg( E_{\sigma^{\prime\prime\prime\prime}(ijklm)} \otimes C_{\sigma^{\prime\prime}(ijk)} \otimes A_{\sigma (i) } \otimes B_{\sigma^{\prime} (ij)}   \otimes D_{\sigma^{\prime\prime\prime}(ijkl)} \bigg)  $ & \text{Combine previous permu-} \\ & & \text{tations listed above}      \\    [1ex] 
\hline 
\end{tabular}
\label{table:nonlin} 
\end{table}

\bigskip

\noindent Before arguing that the optimal values, bias, and error bounds, of the strong parallel repetition of the $\mathrm{FFL}$ game can be characterized with similar arguments presented in previous sections for multiplayer $\mathrm{XOR}$ games, we enumerate the possible ways in which observables gathered by each player can be permuted with each other for the $5$ $\mathrm{XOR}$ game, which is denoted with the collection of permutations,

\begin{align*}
\underline{\mathscr{P}_{5\mathrm{XOR}}}    \text{. }
\end{align*}

\noindent In comparison to error bounds, and the action of associated linear operators provided in $\textbf{Lemma}$ 1-3$\mathrm{XOR}$, and related results, for the $5$ $\mathrm{XOR}$, and $N$ $\mathrm{XOR}$ games, alike have structure consisting of error bounds that are far less rigid than those of the two-player $\mathrm{XOR}$ and $\mathrm{FFL}$ games. In comparison to the equalities,

\begin{align*}
   \bigg( A_i \otimes \textbf{I} \bigg) \ket{\psi_{2\mathrm{XOR}}} = \bigg( \textbf{I} \otimes \frac{B_{ij} + B_{ji}}{\sqrt{2}} \bigg) \ket{\psi_{2\mathrm{XOR}}}     \text{, }  \\     \bigg( A_j \otimes \textbf{I} \bigg) \ket{\psi_{2\mathrm{XOR}}} = \bigg( \textbf{I} \otimes \frac{B_{ij} - B_{ji}}{\sqrt{2}} \bigg) \ket{\psi_{2\mathrm{XOR}}}     \text{, }
\end{align*}

\noindent from the optimal quantum $2$ $\mathrm{XOR}$ strategy, which dictate the transformation that can be applied to Alice's observable, $A_i$, to obtain a superposition of Bob's observables, or, a transformation that can be applied to superposition of Bob's observables, $\frac{B_{ij} - B_{ji}}{\sqrt{2}}$,  to obtain a superposition of Alice's observables, there exists a larger combinatorial space of transformations that can be performed on player observables in the $5$ $\mathrm{XOR}$ setting, undoubtedly taking upon more complicated structures.

\bigskip

\noindent From the objects defined above, the optimal value of the game satisfies,

\begin{align*}
  \omega_{5\mathrm{XOR}} \big( G \big)  \equiv  \omega \big( {5 \mathrm{XOR}}  \big) \propto      \underset{\ket{\psi_{N\mathrm{XOR}}}, A,B,C,D,E}{\mathrm{sup}}           \bra{ \psi_{5\mathrm{XOR}}} \underline{\mathscr{P}_{5\mathrm{XOR}}}         \ket{\psi_{5\mathrm{XOR}}}    \text{.}       
\end{align*}

\noindent The tensor product superposition state above arises from the fact that, from the 4-XOR game, the set of possible states from the responses of each player to the referee takes the form,

\begin{align*}
     \ket{ijk} \big( \ket{j} \bra{ij} \big) \bra{ijkl} + \ket{ijk} \big( \ket{k} \bra{ij} \big) \bra{ijkl} + \ket{ijk} \big( \ket{l } \bra{ij} \big) \bra{ijkl} + \ket{kij} \big( \ket{j} \bra{ij} \big) \bra{ijkl}   \\ + \ket{kij} \big( \ket{k} \bra{ij} \big) \bra{ijkl} + \ket{kij} \big( \ket{l} \bra{ij} \big) \bra{ijkl} + \ket{kji} \big( \bra{j} \ket{ij} \big) \bra{ijkl} +       \ket{kji} \big( \ket{k} \bra{ij} \big) \bra{ijkl} \\ + \ket{kji} \big( \ket{l} \bra{ij} \big) \bra{ijkl}  +   \ket{ikj} \big( \ket{j} \bra{ij} \big) \bra{ijkl} + \ket{ikj} \big( \ket{k} \bra{ij} \big) \bra{ijkl} + \ket{ikj} \big( \ket{l} \bra{ij} \big) \bra{ijkl}       
\\ 
   + \bra{jki} \big( \ket{j} \bra{ij} \big) \bra{ijkl}  + \ket{jik} \big( \ket{j} \bra{ij} \big) \bra{ijkl} + \ket{jik} \big( \ket{k} \bra{ij} \big) \bra{ijkl}      \text{, }
\end{align*}

\begin{table}[ht]
\caption{Representative Interchange operations for player observables in the 4-XOR game} 
\centering 
\begin{tabular}{c c c c} 
\hline\hline 
Number of players held fixed & Tensor Product Representation & Braket Product Representation \\ [0.5ex] 
\hline 4 & $C_{ijk}   \otimes A_j \otimes B_{ij} \otimes   D_{ijkl} $     &  $\ket{ikj} \big( \ket{j} \bra{ij} \big)   \bra{ijkl}$  \\ 4 & $C_{ijk} \otimes A_j \otimes B_{ij} \otimes  D_{ijkl }$   &  $\ket{ijk} \big( \ket{j} \bra{ij} \big) \bra{ijkl}$  \\ 
 3 & $  C_{ijk}  \otimes A_k \otimes B_{ij} \otimes D_{ijkl} $ & $\ket{ijk} \big( \ket{k} \bra{ij} \big) \bra{ijkl} $ \\
   2          & $C_{ijk} \otimes A_l \otimes B_{ij} \otimes   D_{ijkl}$  & $ \ket{ijk} \big( \ket{l } \bra{ij} \big) \bra{ijkl} $  \\
      2   &    $C_{kij} \otimes A_j \otimes B_{ij} \otimes   D_{ijkl}$   & $\ket{kij} \big( \ket{j} \bra{ij} \big) \bra{ijkl}  $  \\
     2    &    $C_{kij} \otimes  A_k \otimes  B_{ij} \otimes  D_{ijkl}$       & $\ket{kij} \big( \ket{k} \bra{ij} \big) \bra{ijkl}  $       \\ 2  &    $C_{kij} \otimes A_l \otimes B_{ij} \otimes  D_{ijkl}$      &    $ \ket{kij} \big( \ket{l} \bra{ij} \big) \bra{ijkl} $        \\  2 &     $C_{kji}  \otimes A_j \otimes B_{ij} \otimes D_{ijkl}  $        &    $  \ket{kji} \big( \bra{j} \ket{ij} \big) \bra{ijkl}$     \\         2   &          $C_{kji} \otimes A_k \otimes B_{ij} \otimes  D_{ijkl}   $       &      $ \ket{kji} \big( \ket{k} \bra{ij} \big)  \bra{ijkl}  $     \\ 2  &          $C_{kji}  \otimes A_l \otimes B_{ij} \otimes     D_{ijkl}$       & $\ket{kji} \big( \ket{l} \bra{ij} \big)  \bra{ijkl} $  \\  2 &        $ C_{ikj} \otimes A_k \otimes B_{ij} \otimes    D_{ijkl}   $   & $ \ket{ikj} \big( \ket{k} \bra{ij} \big) \bra{ijkl}  $  \\ 2  &       $ C_{jki} \otimes A_j \otimes B_{ij} \otimes   D_{ijkl}  $          &        $ \bra{jki} \big( \ket{j} \bra{ij} \big) \bra{ijkl} $ \\     2     &        $ C_{ikj} \otimes A_l \otimes B_{ij} \otimes   D_{ijkl}  $  & $ \ket{ikj} \big( \ket{l} \bra{ij} \big) \bra{ijkl} $   \\ 2 & $C_{jik} \otimes  A_j \otimes B_{ij} \otimes D_{ijkl }$ & $ \ket{jik} \big( \ket{j} \bra{ij} \big)  \bra{ijkl}      $  \\ 2 &  $C_{jik} \otimes A_k \otimes B_{ij} \otimes  D_{ijkl}$    &      $ \ket{jik} \big( \ket{k} \bra{ij} \big) \bra{ijkl}  $     \\  [1ex] 
\hline 
\end{tabular}
\label{table:nonlin} 
\end{table}

\noindent under the action of the interchange operation (several representative interchange operations that can be obtained by varying the number of tensor observables that are held constant for any of the four players are provided in \textit{Table 11}).

\bigskip

\noindent For the optimal value of the 4-XOR game, there exists additional permutations on the responses that each player can provide to the referee, are captured with the superposition,

\begin{align*}
      \ket{ijk} \big( \ket{i} \bra{ik} \big) \bra{ijkl} + \ket{ijk} \big( \ket{i} \bra{il} \big) \bra{ijkl} +           \ket{ijk} \big( \ket{i} \bra{ji} \big) \bra{ijkl} + \ket{ijk} \big( \ket{i} \bra{jk} \big) \bra{ijkl} \\ + \ket{ijk} \big( \ket{i} \bra{jl} \big) \bra{ijkl}    + \ket{ijk} \big( \ket{i} \bra{li} \big) \bra{ijkl} + \ket{ijk} \big( \ket{i} \bra{lj} \big) \bra{ijkl} + \ket{ijk}    \big(  \ket{i} \bra{lk} \big) \bra{ijkl} \\ + \ket{kij } \big( \ket{i} \bra{ik} \big) \bra{ijkl} +  \ket{kij} \big( \ket{i} \bra{il} \big) \bra{ijkl} + \ket{kij} \big( \ket{i} \bra{ji} \big) \bra{ijkl} + \ket{kij} \big( \ket{i} \bra{jk} \big) \bra{ijkl} \\ + \ket{kij} \big( \ket{i} \bra{jl} \big) \bra{ijkl} +  \ket{kij} \big( \ket{i} \bra{li} \big) \bra{ijkl } + \ket{kij} \big( \ket{i} \bra{lj} \big) \bra{ijkl} +   \ket{kij} \big( \ket{i} \bra{lk} \big) \bra{ijkl} \\  + \ket{kji} \big( \ket{i} \bra{ik} \big) \bra{ijkl} + \ket{kji} \big( \ket{i} \bra{il} \big) \bra{ijkl} + \ket{kji} \big( \ket{i}  \bra{ji} \big) \bra{ijkl} + \ket{kji} \big( \ket{i} \bra{jk}  \big) \bra{ijkl} \\ + \ket{kji}  \big( \ket{i} \bra{jl} \big) \bra{ijkl} + \ket{kji} \big( \ket{i} \bra{li} \big) \bra{ijkl}    +  \ket{kji} \big( \ket{i} \bra{lj} \big) \bra{ijkl} + \ket{kji} \big( \ket{i} \bra{lk} \big) \bra{ijkl} \\  +  \ket{ikj} \big( \ket{i} \bra{ik} \big) \bra{ijkl} + \ket{ikj} \big( \ket{i} \bra{il} \big) \bra{ijkl} + \ket{ikj} \big( \ket{i}  \bra{ji} \big) \bra{ijkl} + \ket{ikj} \big( \ket{i} \bra{jk} \big) \bra{ijkl} \\ + \ket{ikj} \big( \ket{i} \bra{jl} \big) \bra{ijkl} + \ket{ikj} \big( \ket{i}  \bra{li} \big) \bra{ijkl} +         \ket{ijk} \big( \ket{i} \bra{lj} \big) \bra{ijkl} + \ket{ijk} \big( \ket{i} \bra{lk} \big) \bra{ijkl} \\  + \ket{jki} \big( \ket{i} \bra{ik} \big) \bra{ijkl} + \ket{jki} \big( \ket{i} \bra{il} \big) \bra{ijkl} + \ket{jki} \big( \ket{i} \bra{ji} \big) \bra{ijkl} +      \ket{jki} \big( \ket{i} \bra{jk} \big) \bra{ijkl}  \\  +  \ket{jki} \big( \ket{i} \bra{jl} \big) \bra{ijkl} + \ket{jki} \big( \ket{i} \bra{li} \big) \bra{ijkl} + \ket{jki} \big( \ket{i} \bra{lj} \big) \bra{ijkl} +     \ket{jki} \big( \ket{i} \bra{lk} \big) \bra{ijkl} \\ + \ket{jik} \big( \ket{i} \bra{ik} \big) \bra{ijkl}  \text{, } 
\end{align*}

\noindent of states over the Hilbert space. For the 3XOR game, besides the fact that the game matrix associated with the game is given by combinations of bra, and of ket, states with three entries, given an optimal quantum strategy, $\ket{\psi_{3\mathrm{XOR}}}$, one can obtain upper bounds for the following summations over the question sets,

\begin{align*}
\mathcal{Q}_{3\mathrm{XOR}} \equiv   \mathcal{Q}_{3\mathrm{XOR},1}  \cup \mathcal{Q}_{3\mathrm{XOR},2} \cup \mathcal{Q}_{3\mathrm{XOR},3} \equiv   \mathcal{Q}_{1}  \cup \mathcal{Q}_{2} \cup \mathcal{Q}_{3}\text{, }
\end{align*}

\noindent which is dependent upon the following contributions:

\begin{itemize}
\item[$\bullet$] \textit{Observable interchange between the first and second players}:
\begin{align*}
  {\underset{j \in \mathcal{Q}_2}{\underset{i \in \mathcal{Q}_1}{\sum}}} \bigg| \bigg| 
 \bigg[     \bigg( \bigg( \frac{A_i + A_j}{\sqrt{2}} \bigg) \otimes \textbf{I} \otimes \textbf{I} \bigg)       -  \bigg( \textbf{I} \otimes B_{ij} \otimes \textbf{I} \bigg) \bigg] \ket{\psi_{3\mathrm{XOR}}}      \bigg| \bigg|^2  \text{, } \end{align*}
       
   \item[$\bullet$] \textit{Observable interchange between the first and second players}:
       \begin{align*}    {\underset{j \in \mathcal{Q}_2}{\underset{ i \in \mathcal{Q}_1}{\sum}}} \bigg| \bigg| 
 \bigg[     \bigg(    \bigg( \frac{A_i - A_j}{\sqrt{2}} \bigg) \otimes \textbf{I} \otimes \textbf{I} \bigg)      - \bigg( \textbf{I} \otimes B_{ji} \otimes \textbf{I} \bigg) \bigg] \ket{\psi_{3\mathrm{XOR}}} \bigg| \bigg|^2  \text{, } \end{align*}
       
   \item[$\bullet$] \textit{Observable interchange between the second and third players}:   
       \begin{align*}  \underset{k \in \mathcal{Q}_3}{\underset{j \in \mathcal{Q}_2}{\underset{i \in \mathcal{Q}_1}{\sum}}} \bigg| \bigg|  \bigg[     \bigg(           \textbf{I} \otimes \bigg( \frac{B_{ij} + B_{ji}}{\sqrt{2}} \bigg) \otimes \textbf{I}      \bigg)    -    \bigg(      \textbf{I} \otimes \textbf{I} \otimes C_{ijk} \bigg)    \bigg]  \ket{\psi_{3\mathrm{XOR}}}    \bigg| \bigg|^2  \text{, } \end{align*}
       
   \item[$\bullet$]  \textit{Observable interchange between the first and second players}:  
       \begin{align*}
       \underset{k \in \mathcal{Q}_3}{\underset{j \in \mathcal{Q}_2}{\underset{i \in \mathcal{Q}_1}{\sum}}} \bigg| \bigg|   \bigg[     \bigg(           \textbf{I} \otimes \bigg( \frac{B_{ij} - B_{ji}}{\sqrt{2}} \bigg)  \otimes \textbf{I}         \bigg)       -   \bigg(      \textbf{I} \otimes \textbf{I} \otimes C_{jik}    \bigg)    \bigg]  \ket{\psi_{3\mathrm{XOR}}}    \bigg| \bigg|^2      \text{, } \end{align*}
       
   \item[$\bullet$]  \textit{Observable interchange between the third and first players}:  
       \begin{align*}     \underset{k \in \mathcal{Q}_3}{\underset{j \in \mathcal{Q}_2}{\underset{i \in \mathcal{Q}_1}{\sum}}} \bigg| \bigg| 
 \bigg[     \bigg(   \textbf{I} \otimes \textbf{I} \otimes C_{ijk}     \bigg)  - \bigg( \bigg( \frac{A_i + A_j}{\sqrt{2}} \bigg) \otimes \textbf{I} \otimes \textbf{I} \bigg)   \bigg]  \ket{\psi_{3\mathrm{XOR}}}    \bigg| \bigg|^2      \text{, } \end{align*}
       
   \item[$\bullet$]  \textit{Observable interchange between the third and first players}:  
       \begin{align*} \underset{k \in \mathcal{Q}_3}{\underset{j \in \mathcal{Q}_2}{\underset{i \in \mathcal{Q}_1}{\sum}}} \bigg| \bigg|  \bigg[     \bigg(   \textbf{I} \otimes \textbf{I} \otimes C_{jik} \bigg) - \bigg( \bigg( \frac{A_i-A_j}{\sqrt{2}} \bigg) \otimes \textbf{I} \otimes \textbf{I}                 \bigg)   \bigg]  \ket{\psi_{3\mathrm{XOR}}}    \bigg| \bigg|^2      \text{, } 
\end{align*}
\end{itemize}

  \noindent simultaneously, with, 
  
  \begin{align*}
   \underset{k \in \mathcal{Q}_3}{\underset{j \in \mathcal{Q}_2}{\underset{i \in \mathcal{Q}_1}{\sum}}} \bigg[ \bigg| \bigg| \bigg[   \bigg[   \bigg(   \bigg( \frac{A_i + A_j}{\sqrt{2}} \bigg) \otimes \textbf{I} \otimes \textbf{I} \bigg)       -  \bigg( \textbf{I} \otimes B_{ij} \otimes \textbf{I} \bigg) \bigg] + \bigg[     \bigg(    \bigg( \frac{A_i - A_j}{\sqrt{2}} \bigg) \otimes \textbf{I} \otimes \textbf{I} \bigg)    \\   - \bigg( \textbf{I} \otimes B_{ji}  \otimes \textbf{I} \bigg)     \bigg]  + \bigg[   \bigg(           \textbf{I} \otimes \bigg( \frac{B_{ij} + B_{ji}}{\sqrt{2}} \bigg) \otimes \textbf{I}      \bigg)    -    \bigg(      \textbf{I} \otimes \textbf{I} \otimes C_{ijk} \bigg)       \bigg] \\  + \bigg[     \bigg(           \textbf{I}  \otimes \bigg( \frac{B_{ij} - B_{ji}}{\sqrt{2}} \bigg)  \otimes \textbf{I}         \bigg)   -   \bigg(      \textbf{I}  \otimes \textbf{I}  \otimes C_{jik}    \bigg)     \bigg]  \\   +    \bigg[   \bigg(   \textbf{I} \otimes \textbf{I} \otimes C_{ijk}     \bigg)  - \bigg( \bigg( \frac{A_i + A_j}{\sqrt{2}} \bigg) \otimes \textbf{I} \otimes \textbf{I} \bigg) \bigg]  + \bigg[ \bigg(   \textbf{I} \otimes \textbf{I} \otimes C_{jik} \bigg) \\  - \bigg( \bigg( \frac{A_i-A_j}{\sqrt{2}} \bigg)   \otimes \textbf{I} \otimes \textbf{I}                 \bigg) \bigg]       \bigg]  \ket{\psi_{3\mathrm{XOR}}}   \bigg| \bigg|^2 \bigg]  \text{. }    \tag{\textit{***}}
\end{align*}

\noindent Under the standard operation of tensor contraction, the tensor product of player observables for an $N$-player game can be related to that of any other game with $>N$ players. One expects that an upper bound, up to constants, of the form,

\begin{align*}
   n \bigg( \underset{1 \leq j \leq 2}{\prod} \big( n - j \big) \bigg)    \text{, }
\end{align*}

\noindent from the $\mathrm{N}$-XOR error bounds, up to permutation, should exist. We provide the statement of such a property below.

\bigskip

\noindent \textbf{Theorem} $\textit{2}^{*}$ (\textit{3-XOR permutation error bounds}, \textit{2.2.1}, \textbf{Theorem} \textit{4}, {[37]}, \textbf{Theorem} \textit{2}, {[44]}, \textbf{Theorems} \textit{1}-\textit{6} in \textit{1.5}).

\begin{align*}
      ( \textit{***}) \leq 3!     n \bigg( \underset{1 \leq j \leq 2}{\prod} \big( n - j \big) \bigg)           \text{. }
\end{align*}

\noindent \textit{Proof of Theorem $2^{*}$}. To argue that the desired upper bound holds from direct computation, recall the following result, as a generalization of the $\epsilon$-approximality result of {[37]}:

\bigskip

\noindent \textbf{Theorem} \textit{1} (\textit{approximately optimal quantum strategies for the nonlocal XOR game}, \textbf{Theorem} \textit{4}, {[37]}). For $\pm$ observables $A_i$ and $B_{jk}$, given a bipartite state $\psi$, TFAE:

\begin{itemize}
 \item[$\bullet$] \underline{\textit{First characterization of approximate optimality}:} An $\epsilon$-approximate $\mathrm{CHSH}\big( n \big)$ satisfies $\mathrm{(0)}$.

 \item[$\bullet$] \underline{\textit{Second characterization of approximate optimality}:} For an $\epsilon$-approximate quantum strategy, 

  \begin{align*}
     \underset{1 \leq i < j \leq n}{\sum} \bigg[ \text{ }  \bigg| \bigg|   \bigg[         \big( \frac{A_i + A_j}{\sqrt{2}} \big) \otimes I \bigg] \ket{\psi}    -   \big[  I \otimes B_{ij} \big] \ket{\psi}   \bigg|\bigg|^2 + \bigg| \bigg|   \bigg[ \big( \frac{A_i - A_j}{\sqrt{2}} \big)  \otimes I  \bigg] \ket{\psi}  \\   -   \big[  I \otimes B_{ji} \big]  \ket{\psi}     \bigg|\bigg|^2 \text{ }  \bigg]  \leq 2n \big( n - 1 \big) \epsilon \text{. }
 \end{align*}
 \item[$\bullet$] \underline{\textit{Reversing the order of the tensor product for observables}:} Related to the inequality for $\epsilon$-approximate strategies above, another inequality,

\begin{align*}
    \underset{1 \leq i < j \leq n}{\sum} \bigg[  \text{ } \bigg| \bigg| \big[ A_i \otimes I \big] \ket{\psi}    - \bigg[ I \otimes \big( \frac{B_{ij} + B_{ji}}{\sqrt{2}} \big)  \bigg] \ket{\psi}   \bigg| \bigg|^2  + \bigg| \bigg|    \big[ A_j  \otimes I \big] \ket{\psi} \\  -   \bigg[   I \otimes \big( \frac{B_{ij} - B_{ji}}{\sqrt{2}} \big) \bigg] \ket{\psi}    \big| \big|^2     \text{ }   \bigg]  \leq 2n \big( n - 1 \big) \epsilon \text{, } 
\end{align*}

 \noindent also holds. 
 
 \item[$\bullet$] \underline{\textit{Characterization of exact optimality}}: For $\epsilon \equiv 0$,

\begin{align*}
    \underset{1 \leq i < j \leq n}{\sum} \bigg[ \text{ }  \bigg| \bigg|  \bigg[  \big( \frac{A_i + A_j}{\sqrt{2}} \big) \otimes I \bigg] \ket{\psi}    -  \big[ I \otimes B_{ij} \big] \ket{\psi}   \bigg| \bigg|^2  \bigg]  =  -  \underset{1 \leq i < j \leq n}{\sum} \bigg[  \text{ }   \bigg| \bigg|     \bigg[  \big( \frac{A_i - A_j}{\sqrt{2}} \big)  \otimes I \bigg]  \ket{\psi} \\  -   \big[ I \otimes B_{ji} \big] \ket{\psi}    \bigg| \bigg|^2  \bigg]      \text{, } 
\end{align*}

\noindent corresponding to the first inequality, and,

\begin{align*}
       \underset{1 \leq i < j \leq n}{\sum} \bigg[ \text{ }  \bigg| \bigg|       \big[  A_i \otimes I \big]  \ket{\psi}    - \bigg[   I \otimes \big( \frac{B_{ij} + B_{ji}}{\sqrt{2}} \big)  \bigg]  \ket{\psi}   \bigg|\bigg|^2 \bigg] = -  \underset{1 \leq i < j \leq n}{\sum}   \bigg[  \bigg| \bigg|  \big[   A_j  \otimes I \big] \ket{\psi} \\  -  \bigg[    I \otimes \big( \frac{B_{ij} - B_{ji}}{\sqrt{2}} \big) \bigg]    \ket{\psi}     \bigg|\bigg|^2  \bigg]             \text{, } 
\end{align*}

\noindent corresponding to the second inequality.

\end{itemize}

\bigskip

\noindent  With the result above, to further generalize the set of equivalent conditions as provided in previous discussions for the 3-$\mathrm{XOR}$ game, and beyond, given the existence of suitable $\epsilon_{3\mathrm{XOR}}$, and a previously determined constant $C_{3\mathrm{XOR}} \equiv C$, the \textit{symmetrized} 3-$\mathrm{XOR}$ game tensor, and other terms,

\begin{align*}
   \underset{1 \leq i \leq 9}{\sum} y_i E_{ii} - G_{\mathrm{Sym},3\mathrm{XOR}} \equiv  \underset{1 \leq i \leq 9}{\sum} y_i E_{ii} - G_{\mathrm{Sym}}   \text{, }
\end{align*}

\noindent equals,

\begin{align*}
     \frac{1}{\bigg( C_{3\mathrm{XOR}} \omega_{3\mathrm{XOR}} \bigg)  n \bigg( \underset{1 \leq j \leq 2}{\prod}  \big( n - j \big) \bigg) }             \underset{i_3 \in \mathcal{Q}_3}{\underset{i_2 \in \mathcal{Q}_2 }{\underset{i_1 \in \mathcal{Q}_1}{\sum}}}   \bigg(  \big(    u^{\prime}_{i_1i_2 i_3}   -    v^{\prime}_{i_1i_2 i_3}  \big) \big(   u^{\prime}_{i_1 i_2 i_3}   -    v^{\prime}_{i_1 i_2 i_3}    \big)^{\textbf{T}}     +  \big(    u^{\prime}_{i_2 i_1 i_3 }  \\  -    v^{\prime}_{i_2 i_1 i_3}  \big)   \big( u^{\prime}_{i_2 i_1 i_3 }   -    v^{\prime}_{i_2 i_1 i_3  }  \big)^{\textbf{T}}  +     \big(    u^{\prime}_{i_1 i_3 i_2}   -    v^{\prime}_{i_1 i_3 i_2}   \big)   \big(   u^{\prime}_{i_1 i_3 i_2}     -    v^{\prime}_{i_1 i_3 i_2}   \big)^{\textbf{T}}     \bigg)      \text{,}
\end{align*}

\noindent given the superposition of states,

\begin{align*}
        u^{\prime}_{i_1 i_2 i_3 } \equiv \frac{1}{\sqrt{3}} \bigg( \underset{1 \leq j \leq 3}{\sum} \ket{\text{Player } j \text{ state}} \bigg)      \text{, }  v^{\prime}_{i_1 i_2 i_3} \equiv \ket{i_1 i_2 i_3 } \\  u^{\prime}_{i_2 i_1 i_3 } \equiv \frac{1}{\sqrt{3}} \bigg( \ket{\text{Player 1 state}} - \ket{\text{Player 2 state}} + \ket{\text{Player } 3 \text{ state}} \bigg) \text{, } v^{\prime}_{i_2 i_1 i_3} \equiv \ket{i_2 i_1 i_3} \\   u^{\prime}_{i_1 i_3 i_2 } \equiv \frac{1}{\sqrt{3}} \bigg(  \underset{1 \leq j \leq 2}{\sum} \ket{\text{Player } j \text{ state}}  - \ket{\text{Player } 3 \text{ state}} \bigg)       \text{, } v^{\prime}_{i_1 i_3 i_2} \equiv \ket{i_1 i_3 i_2} \text{, }
 \end{align*}

\noindent from which we conclude the argument. \boxed{}

\bigskip

 \noindent \textbf{Lemma} \textit{T-2} (\textit{positive semidefiniteness}). Under the assumptions of the previous result, the operator,

\begin{align*}
  \underset{1 \leq i \leq n^3}{\sum}   y_{3\mathrm{XOR},i} E_{3\mathrm{XOR},ii} - G_{3\mathrm{XOR},\mathrm{Sym}}     \text{, }
\end{align*}

 \noindent is positive semidefinite.

\bigskip

\noindent \textit{Proof of Lemma T-2}. The result follows from the fact that, for N-XOR games, the computation involving,

\begin{align*}
     \frac{1}{\bigg( C_{3\mathrm{XOR}} \omega_{3\mathrm{XOR}} \bigg)  n \bigg( \underset{1 \leq j \leq 2}{\prod}  \big( n - j \big) \bigg) }             \underset{i_3 \in \mathcal{Q}_3}{\underset{i_2 \in \mathcal{Q}_2 }{\underset{i_1 \in \mathcal{Q}_1}{\sum}}}   \bigg(  \big(    u^{\prime}_{i_1i_2 i_3}   -    v^{\prime}_{i_1i_2 i_3}  \big) \big(   u^{\prime}_{i_1 i_2 i_3}   -    v^{\prime}_{i_1 i_2 i_3}    \big)^{\textbf{T}}     +  \big(    u^{\prime}_{i_2 i_1 i_3 }  \\  -    v^{\prime}_{i_2 i_1 i_3}  \big)   \big( u^{\prime}_{i_2 i_1 i_3 }   -    v^{\prime}_{i_2 i_1 i_3  }  \big)^{\textbf{T}}  +     \big(    u^{\prime}_{i_1 i_3 i_2}   -    v^{\prime}_{i_1 i_3 i_2}   \big)   \big(   u^{\prime}_{i_1 i_3 i_2}     -    v^{\prime}_{i_1 i_3 i_2}   \big)^{\textbf{T}}     \bigg)      \text{,}
\end{align*}

\noindent implies that the associated operator is positive semidefinite from the observation that taking the constant $C_{N\mathrm{XOR}}$, in the normalization,

\begin{align*}
     \frac{1}{ \bigg( C_{3\mathrm{XOR}} \omega_{3\mathrm{XOR}} \bigg) n \bigg( \underset{1 \leq j \leq 2}{\prod}  \big( n - j \big) \bigg) }        \text{, }
\end{align*}

\noindent to equal $3!$ implies the desired result, from which we conclude the argument. \boxed{}

\bigskip

\noindent Within the three player setting, given the optimal value,

\begin{align*}
  \omega_{3\mathrm{XOR}} \big( G \big) \equiv \underset{\mathcal{S}}{\mathrm{sup}} \big\{ \text{probability that any player wins the 3XOR game $G$ with strategy $\mathcal{S}$} \big\}  \text{,}
\end{align*}

\noindent one also has,

\begin{align*}
    \bigg| \bigg|   \bigg[  \bigg( \bigg( \underset{1 \leq i \leq n}{\prod} A^{j_i}_i  \bigg) \otimes B_{kl} \otimes \textbf{I}\bigg)     - \omega_{3 \mathrm{XOR}} \bigg( \pm \bigg( \mathrm{sign} \big( i_1 , j_1 , \cdots , j_n \big) \bigg[ \bigg(  \bigg(   \underset{1 \leq i \leq n}{\prod}  A^{j_i}_i \bigg)  \\ + \bigg( \underset{\text{set } j+1 \equiv j \oplus 1}{\underset{1 \leq i \leq n}{\prod}}  A^{j_i}_i \bigg)  \bigg)   \otimes \textbf{I}  \otimes \textbf{I} \bigg)   \bigg] \bigg) \bigg)  \bigg]   \ket{\psi_{3\mathrm{XOR}}} \bigg| \bigg| \text{, } 
\end{align*}

\noindent corresponding to the impact of changing a contribution from the product of operators,

\begin{align*}
     {\underset{1 \leq i \leq n}{\prod}}  A^{j_i}_i  \text{, }
\end{align*}

\noindent in addition to,

\begin{align*} 
     \bigg| \bigg| \bigg[ \bigg( \textbf{I} \otimes \bigg( \underset{l \in \mathcal{Q}_2}{\underset{k \in \mathcal{Q}_1}{\prod}}  B^{j_{kl}}_{kl} \bigg) \otimes  C_{k^{\prime}l^{\prime}j^{\prime}} \bigg) - \omega_{3\mathrm{XOR}} \bigg( \pm \bigg( \mathrm{sign} \big( i_1 , j_1 , \cdots , k , l \cdots j_1 , \cdots , j_n \big)    \\ \times \bigg[ \bigg( \textbf{I}  \otimes \bigg( \underset{l \in \mathcal{Q}_2}{\underset{k \in \mathcal{Q}_1}{\prod}} B^{j_{kl}}_{kl}    + \underset{\text{set } l+1 \equiv l \oplus 1}{\underset{\text{set }k+1 \equiv k \oplus 1}{\underset{l \in \mathcal{Q}_2}{\underset{k \in \mathcal{Q}_1}{\prod}}}} B^{j_{kl}}_{kl}      \bigg)     \otimes \textbf{I}   \bigg] \bigg) \bigg)  \bigg]  \ket{\psi_{3\mathrm{XOR}}} \bigg| \bigg|   \text{, }
\end{align*}

\noindent corresponding to the impact of changing a contribution from the product of operators,

\begin{align*} 
   \underset{k \in \mathcal{Q}_1, l \in \mathcal{Q}_2}{\prod} B^{j_{kl}}_{kl}     \text{. }
\end{align*}

\noindent The last possibility, as demonstrated through the 3XOR error bounds above, implies that a suitable upper bound dependent upon the number of players in the game takes the form,

\begin{align*}
    \bigg| \bigg| \bigg[ \bigg( A_i \otimes \textbf{I} \otimes \bigg( \underset{i \in \mathcal{Q}_1, j \in \mathcal{Q}_2, k \in \mathcal{Q}_3}{\prod}  C^{l_{ijk}}_{ijk}    \bigg) \bigg) - \omega_{3\mathrm{XOR}} \bigg( \pm \bigg( \mathrm{sign} \big( i_1 , j_1 , k_1 , \cdots j_{111} , \cdots , j_{n, ( n + m )} \big)  \\ \times   \bigg[ \bigg( \textbf{I} \otimes \textbf{I}  \otimes \bigg( \bigg(   \underset{i \in \mathcal{Q}_1, j \in \mathcal{Q}_2, k \in \mathcal{Q}_3}{\prod}  C^{l_{ijk}}_{ijk}   \bigg) + \bigg(  \underset{\text{set } k+1 \equiv k \oplus 1}{\underset{\text{set } j+1 \equiv j \oplus 1}{\underset{\text{set } i+1 \equiv i \oplus 1}{\underset{k \in \mathcal{Q}_3}{\underset{j \in \mathcal{Q}_2}{\underset{i \in \mathcal{Q}_1}{\prod}}}}}}   C^{l_{ijk}}_{ijk}    \bigg)  \bigg) \bigg)  \bigg]      \bigg] \ket{\psi_{3\mathrm{XOR}}}  \bigg| \bigg|  \text{, }
\end{align*}

\noindent corresponding to the impact of changing a contribution from the product of operators,

\begin{align*}
\underset{i \in \mathcal{Q}_1, j \in \mathcal{Q}_2, k \in \mathcal{Q}_3}{\prod}   C^{l_{ijk}}_{ijk}   \text{. }
\end{align*}

\noindent The approximately optimal quantum strategy for the NXOR, as well as for games with greater than three players in the XOR setting, satisfies,

\begin{align*}
  \omega_{N\mathrm{XOR}} \big( 1 - \epsilon \big) \leq \underset{\sigma \in S_n}{\sum} \underset{\mathcal{Q}_N}{\sum} \cdots  \underset{\sigma \in S_1}{\sum}    \underset{\mathcal{Q}_1}{\sum} 
 \bra{\psi_{N\mathrm{XOR}}}     \bigg(      \mathcal{L}_{N\mathrm{XOR}} \bigg( \sigma  , \underset{1 \leq i \leq n}{\bigcup} \mathcal{Q}_i  \bigg)    \bigg)       \ket{\psi_{N\mathrm{XOR}}} \\ \leq \omega_{N\mathrm{XOR}} \text{, }
\end{align*}

\noindent where the intermediate term, $\mathcal{L}$, appearing in the braket above is dependent upon a permutation of a permutation with $n$ entries, where each of the entries represents all possible responses that any player can respond to the referee with. Namely, that,

\begin{align*}
   \mathcal{L}_{N\mathrm{XOR}} \bigg( \sigma  , \underset{1 \leq i \leq n}{\bigcup} \mathcal{Q}_i  \bigg) \equiv \mathcal{L}_{N\mathrm{XOR}} \bigg(  \sigma \bigg(  \underset{1 \leq i \leq n}{\bigcup} \mathcal{Q}_i  \bigg) \bigg) \equiv   \underset{1 \leq i \leq n}{\bigcup} \mathcal{L}_{N\mathrm{XOR}} \big( \sigma \big( \mathcal{Q}_i \big) \big) \text{, }
\end{align*}

\noindent captures the set of all possible questions that each player can provide in response to the question set probability distribution of the referee. The accompanying error bounds, as a direct extension from those provided earlier in the section with the optimal quantum strategy $\ket{\psi_{3\mathrm{XOR}}}$, instead given an N-player optimal strategy $\ket{\psi_{N\mathrm{XOR}}}$, take the form,

\begin{align*}
 \underset{i \in \mathcal{Q}_1, j \in \mathcal{Q}_2}{\sum} \bigg| \bigg| \bigg[     \bigg( A_i \otimes \textbf{I} \otimes \textbf{I} \otimes \overset{N-3}{\cdots} \otimes \textbf{I} \bigg) - \bigg(       \textbf{I} \otimes \bigg( \frac{B_i + B_j}{\sqrt{2}} \bigg)    \otimes \textbf{I} \otimes \overset{N-4}{\cdots} \otimes \textbf{I}  \bigg)          \bigg] \\ \times \ket{\psi_{N\mathrm{XOR}}} \bigg| \bigg|^2  \text{, } \\ \\ \underset{i \in \mathcal{Q}_1, j \in \mathcal{Q}_2}{\sum} \bigg| \bigg| \bigg[     \bigg( A_i \otimes \textbf{I} \otimes \textbf{I} \otimes \overset{N-4}{\cdots} \otimes \textbf{I} \bigg) - \bigg(    \textbf{I} \otimes \bigg( \frac{B_i - B_j}{\sqrt{2}} \bigg) \otimes \textbf{I} \otimes \overset{N-4}{\cdots}  \otimes \textbf{I}     \bigg)        \bigg] \\ \times  \ket{\psi_{N\mathrm{XOR}}} \bigg| \bigg|^2      \text{, } \\ \\  \underset{i \in \mathcal{Q}_1, j \in \mathcal{Q}_2, k \in \mathcal{Q}_3}{\sum} \bigg| \bigg| \bigg[    \bigg( \textbf{I} \otimes B_{ij} \otimes \textbf{I} \otimes \overset{N-4}{\cdots} \otimes \textbf{I} \bigg) - \bigg(  \textbf{I} \otimes \textbf{I} \otimes      \frac{1}{\sqrt{6}} \bigg(\underset{\sigma \in S_3}{\sum }  C_{\sigma(ijk)}        \bigg) \otimes  \textbf{I} \otimes \overset{N-5}{\cdots}  \\ \otimes \textbf{I }         \bigg)       \bigg]  \ket{\psi_{N\mathrm{XOR}}} \bigg| \bigg|^2       \text{, } 
\\ \\ 
           \underset{i \in \mathcal{Q}_1, j \in \mathcal{Q}_2, k \in \mathcal{Q}_3}{\sum} \bigg| \bigg| \bigg[    \bigg(   \textbf{I} \otimes B_{ji} \otimes \textbf{I} \otimes \overset{N-4}{\cdots} \otimes  \textbf{I} \bigg) - \bigg(  \textbf{I} \otimes \textbf{I} \otimes   \frac{1}{\sqrt{6}} \bigg(\underset{\sigma \in S_3}{\sum C_{\sigma(ijk)}}  \textbf{1}_{\sigma \text{ even transposition}}  \\  - \underset{\sigma \in S_3}{\sum} C_{\sigma(ijk)}  \textbf{1}_{\sigma \text{ odd transposition}}       \bigg) \otimes \textbf{I} \otimes  \overset{N-5}{\cdots} \otimes \textbf{I} \bigg) \bigg] \ket{\psi_{N\mathrm{XOR}}} \bigg| \bigg|^2   \text{, } 
\\ 
     \vdots  \\  \underset{i_1 \in \mathcal{Q}_1, \cdots, i_n \in \mathcal{Q}_n}{\sum} \bigg| \bigg| \bigg[    \bigg(      \textbf{I} \otimes \overset{N-3}{\cdots}  \otimes \bigg(     \big( N-1 \big) \text{ th player tensor observable} \big( i_1 , i_2 , \cdots , i_n \big)       \bigg) \otimes \textbf{I} \bigg) \\  - \bigg(        \textbf{I} \otimes \textbf{I}  \otimes \overset{N-4}{\cdots}   \otimes  \textbf{I} \otimes   \frac{1}{\sqrt{ \#  \sigma  }} \bigg(        \underset{\sigma \in S_n}{\sum}  N \text{ th player tensor observable} \big( \sigma \big( i_1  \big) \\ , \sigma \big(  i_2 \big)   , \cdots , \sigma \big( i_n\big)  \big)       \bigg)    \bigg)     \bigg] \ket{\psi_{N\mathrm{XOR}}}  \bigg| \bigg|^2      \text{. }
\end{align*}

\noindent The relations provided above for the XOR game with an arbitrary number of players reflect characteristics of an accompanying set of relations for the 4$\mathrm{XOR}$ game. Namely, for a 4$\mathrm{XOR}$ game, straightforwardly, from previous relations the expressions for interchanging the order, within the tensor product, in which player observables appear take several forms, including those displayed in the three tables below.


\begin{table}[ht]
\caption{Error bound for the 4-XOR game - two player interchanges} 
\centering 
\begin{tabular}{c c c c} 
\hline\hline 
 Error Bound Contributions \\ [0.5ex] 
\hline  $ \underset{l \in \mathcal{Q}_4}{\underset{k \in \mathcal{Q}_3}{\underset{j \in \mathcal{Q}_2}{\underset{i \in \mathcal{Q}_1}{\mathlarger{\sum}}}}}  \bigg| \bigg|  \bigg[  \bigg( A_i \otimes \textbf{I} \otimes C_{ijk} \otimes \textbf{I} \bigg) - \bigg( \textbf{I} \otimes \bigg( \frac{B_{ij} + B_{ji}}{\sqrt{2}} \bigg) \otimes \textbf{I} \otimes \frac{1}{\sqrt{24}} \bigg(   \underset{\sigma \in S_4}{\sum}   D_{\sigma( ijkl ) }   \bigg) \bigg)       \bigg]    \ket{\psi_{4\mathrm{XOR}}} \bigg| \bigg|^2   $ \\    $\underset{l \in \mathcal{Q}_4}{\underset{k \in \mathcal{Q}_3}{\underset{j \in \mathcal{Q}_2}{\underset{i \in \mathcal{Q}_1}{\mathlarger{\sum}}}}}   \bigg| \bigg|  \bigg[  \bigg(       \textbf{I} \otimes B_{ij} \otimes \textbf{I} \otimes D_{ijkl}        \bigg) - \bigg(  \bigg( \frac{A_i + A_j}{\sqrt{2}} \bigg) \otimes \textbf{I} \otimes   \frac{1}{\sqrt{6}} \bigg(\underset{\sigma \in S_3}{\sum }  C_{\sigma(ijk)}        \bigg) \otimes \textbf{I}        \bigg)       \bigg]    \ket{\psi_{4\mathrm{XOR}}} \bigg| \bigg|^2  $  \\ $\underset{l \in \mathcal{Q}_4}{\underset{k \in \mathcal{Q}_3}{\underset{j \in \mathcal{Q}_2}{\underset{i \in \mathcal{Q}_1}{\mathlarger{\sum}}}}}   \bigg| \bigg|  \bigg[  \bigg(          A_i \otimes \textbf{I} \otimes C_{ijk} \otimes D_{ijkl}       \bigg)              - \bigg(   \textbf{I} \otimes \bigg( \frac{B_{ij} + B_{ji}}{\sqrt{2}} \bigg) \otimes C_{ijk} \otimes D_{ijkl}   \bigg)     \bigg]    \ket{\psi_{4\mathrm{XOR}}} \bigg| \bigg|^2    $ \\  $\underset{l \in \mathcal{Q}_4}{\underset{k \in \mathcal{Q}_3}{\underset{j \in \mathcal{Q}_2}{\underset{i \in \mathcal{Q}_1}{\mathlarger{\sum}}}}}   \bigg| \bigg|  \bigg[  \bigg(                     A_i \otimes \textbf{I} \otimes C_{ijk} \otimes \textbf{I}         \bigg)   - \bigg( \textbf{I} \otimes \bigg( \frac{B_{ij} + B_{ji}}{\sqrt{2}} \bigg) \otimes \textbf{I} \otimes \frac{1}{\sqrt{24}} \bigg(   \underset{\sigma \in S_4}{\sum}   D_{\sigma( ijkl ) }   \bigg)  \bigg)   \bigg]    \ket{\psi_{4\mathrm{XOR}}} \bigg| \bigg|^2  $ \\   $\underset{l \in \mathcal{Q}_4}{\underset{k \in \mathcal{Q}_3}{\underset{j \in \mathcal{Q}_2}{\underset{i \in \mathcal{Q}_1}{\mathlarger{\sum}}}}}   \bigg| \bigg|  \bigg[  \bigg(     \textbf{I} \otimes \textbf{I} \otimes C_{ijk} \otimes \textbf{I}                                    \bigg) - \bigg(    \textbf{I} \otimes \textbf{I} \otimes \textbf{I} \otimes \frac{1}{\sqrt{24}} \bigg(   \underset{\sigma \in S_4}{\sum}   D_{\sigma( ijkl ) }   \bigg)       \bigg)       \bigg]    \ket{\psi_{4\mathrm{XOR}}} \bigg| \bigg|^2      $    \\      $\underset{k \in \mathcal{Q}_3}{\underset{j \in \mathcal{Q}_2}{\underset{i \in \mathcal{Q}_1}{\mathlarger{\sum}}}}   \bigg| \bigg|  \bigg[  \bigg(     \textbf{I} \otimes B_{ij} \otimes \textbf{I} \otimes \textbf{I} \bigg) - \bigg(  \textbf{I} \otimes \textbf{I} \otimes \frac{1}{\sqrt{6}} \bigg(\underset{\sigma \in S_3}{\sum }  C_{\sigma(ijk)}        \bigg)  \otimes \textbf{I}                                                               \bigg)       \bigg]    \ket{\psi_{4\mathrm{XOR}}} \bigg| \bigg|^2 $ \\    $\underset{l \in \mathcal{Q}_4}{\underset{k \in \mathcal{Q}_3}{\underset{j \in \mathcal{Q}_2}{\underset{i \in \mathcal{Q}_1}{\mathlarger{\sum}}}}}    \bigg| \bigg|  \bigg[  \bigg(       A_i \otimes \textbf{I} \otimes \textbf{I} \otimes D_{ijkl}           \bigg) - \bigg(               \textbf{I} \otimes \bigg( \frac{B_{ij} + B_{ji}}{\sqrt{2}} \bigg) \otimes \frac{1}{\sqrt{6}} \bigg(\underset{\sigma \in S_3}{\sum }  C_{\sigma(ijk)}        \bigg)  \otimes \textbf{I}           \bigg)       \bigg]    \ket{\psi_{4\mathrm{XOR}}} \bigg| \bigg|^2  $        \\   $\vdots$  \\   [1ex] 
\hline 
\end{tabular}
\label{table:nonlin} 
\end{table}

\begin{table}[ht]
\caption{Error bound for the 4-XOR game - two player interchanges Continued} 
\centering 
\begin{tabular}{c c c c} 
\hline\hline 
 Error Bound Contributions \\ [0.5ex] 
\hline $\vdots$  \\  $  \underset{l \in \mathcal{Q}_4}{\underset{k \in \mathcal{Q}_3}{\underset{j \in \mathcal{Q}_2}{\underset{i \in \mathcal{Q}_1}{\mathlarger{\sum}}}}}  \bigg| \bigg|  \bigg[  \bigg(          \textbf{I} \otimes B_{ij} \otimes \textbf{I} \otimes D_{ijkl}       \bigg) - \bigg(         \bigg( \frac{A_i + A_j}{\sqrt{2}} \bigg) \otimes \textbf{I} \otimes \frac{1}{\sqrt{6}} \bigg(\underset{\sigma \in S_3}{\sum }  C_{\sigma(ijk)}        \bigg)  \otimes \textbf{I}         \bigg)       \bigg]    \ket{\psi_{4\mathrm{XOR}}} \bigg| \bigg|^2    $  \\ $  \underset{l \in \mathcal{Q}_4}{\underset{k \in \mathcal{Q}_3}{\underset{j \in \mathcal{Q}_2}{\underset{i \in \mathcal{Q}_1}{\mathlarger{\sum}}}}}  \bigg| \bigg|  \bigg[  \bigg(                \textbf{I} \otimes B_{ij} \otimes C_{ijk} \otimes \textbf{I}               \bigg) - \bigg(        \bigg( \frac{A_i + A_j}{\sqrt{2}}\bigg) \otimes \textbf{I} \otimes \textbf{I} \otimes  \frac{1}{\sqrt{24}} \bigg(   \underset{\sigma \in S_4}{\sum}   D_{\sigma( ijkl ) }   \bigg)                  \bigg)       \bigg]    \ket{\psi_{4\mathrm{XOR}}} \bigg| \bigg|^2   $   \\ $ \underset{l \in \mathcal{Q}_4}{\underset{k \in \mathcal{Q}_3}{\underset{j \in \mathcal{Q}_2}{\underset{i \in \mathcal{Q}_1}{\mathlarger{\sum}}}}}   \bigg| \bigg|  \bigg[  \bigg(           A_i \otimes \textbf{I} \otimes \textbf{I} \otimes \textbf{I}           \bigg) - \bigg(             \textbf{I} \otimes \textbf{I} \otimes \textbf{I} \otimes  \frac{1}{\sqrt{24}} \bigg(   \underset{\sigma \in S_4}{\sum}   D_{\sigma( ijkl ) }   \bigg)             \bigg)       \bigg]    \ket{\psi_{4\mathrm{XOR}}} \bigg| \bigg|^2     $  \\    $ \underset{k \in \mathcal{Q}_3}{\underset{j \in \mathcal{Q}_2}{\underset{i \in \mathcal{Q}_1}{\mathlarger{\sum}}}}   \bigg| \bigg|  \bigg[  \bigg(      A_i \otimes \textbf{I} \otimes \textbf{I} \otimes \textbf{I}           \bigg) - \bigg(             \textbf{I} \otimes \bigg( \frac{B_{ij} + B_{ji}}{\sqrt{2}} \bigg) \otimes \textbf{I} \otimes \textbf{I}           \bigg)       \bigg]    \ket{\psi_{4\mathrm{XOR}}} \bigg| \bigg|^2   $  \\   $ \underset{k \in \mathcal{Q}_3}{\underset{j \in \mathcal{Q}_2}{\underset{i \in \mathcal{Q}_1}{\mathlarger{\sum}}}}   \bigg| \bigg|  \bigg[  \bigg(     A_i \otimes \textbf{I} \otimes \textbf{I} \otimes \textbf{I}            \bigg) - \bigg(        \textbf{I}  \otimes \textbf{I}  \otimes \frac{1}{\sqrt{6}} \bigg(\underset{\sigma \in S_3}{\sum }  C_{\sigma(ijk)}        \bigg)  \otimes \textbf{I}        \bigg)       \bigg]    \ket{\psi_{4\mathrm{XOR}}} \bigg| \bigg|^2 $  \\    $\underset{k \in \mathcal{Q}_3}{\underset{j \in \mathcal{Q}_2}{\underset{i \in \mathcal{Q}_1}{\mathlarger{\sum}}}}   \bigg| \bigg|  \bigg[  \bigg(     \textbf{I} \otimes B_{ij} \otimes \textbf{I} \otimes \textbf{I}            \bigg) - \bigg(           \textbf{I} \otimes \textbf{I} \otimes \frac{1}{\sqrt{6}} \bigg(\underset{\sigma \in S_3}{\sum }  C_{\sigma(ijk)}        \bigg)  \otimes \textbf{I}              \bigg)       \bigg]    \ket{\psi_{4\mathrm{XOR}}} \bigg| \bigg|^2     $    \\    $\vdots$  \\    [1ex] 
\hline 
\end{tabular}
\label{table:nonlin} 
\end{table}


\begin{table}[ht]
\caption{Error bound for the 4-XOR game - one player interchanges} 
\centering 
\begin{tabular}{c c c c} 
\hline\hline 
 Error Bound Contributions \\ [0.5ex] 
\hline  $\vdots$  \\   $ \underset{k \in \mathcal{Q}_3}{\underset{j \in \mathcal{Q}_2}{\underset{i \in \mathcal{Q}_1}{\mathlarger{\sum}}}}   \bigg| \bigg|  \bigg[  \bigg(         \textbf{I} \otimes \textbf{I} \otimes C_{ijk} \otimes \textbf{I}        \bigg) - \bigg(              \bigg( \frac{A_i + A_j}{\sqrt{2} } \bigg) \otimes \textbf{I} \otimes \textbf{I}  \otimes \textbf{I}          \bigg)       \bigg]    \ket{\psi_{4\mathrm{XOR}}} \bigg| \bigg|^2$  \\ $\underset{k \in \mathcal{Q}_3}{\underset{j \in \mathcal{Q}_2}{\underset{i \in \mathcal{Q}_1}{\mathlarger{\sum}}}}   \bigg| \bigg|  \bigg[  \bigg(          \textbf{I} \otimes \textbf{I} \otimes C_{ijk} \otimes \textbf{I}       \bigg) - \bigg(         \textbf{I} \otimes \bigg(   \frac{B_{ij} + B_{ji}}{\sqrt{2}} \bigg) \otimes \textbf{I} \otimes \textbf{I}                \bigg)       \bigg]    \ket{\psi_{4\mathrm{XOR}}} \bigg| \bigg|^2    $  \\   ${\underset{j \in \mathcal{Q}_2}{\underset{i \in \mathcal{Q}_1}{\mathlarger{\sum}}}}   \bigg| \bigg|  \bigg[  \bigg(          A_i \otimes \textbf{I} \otimes \textbf{I}             \bigg) - \bigg(     \textbf{I} \otimes \bigg(         \frac{B_{ij} + B_{ji}}{\sqrt{2}}      \bigg) \otimes \textbf{I}             \bigg)       \bigg]    \ket{\psi_{4\mathrm{XOR}}} \bigg| \bigg|^2   $  \\ $\underset{k \in \mathcal{Q}_3}{\underset{j \in \mathcal{Q}_2}{\underset{i \in \mathcal{Q}_1}{\mathlarger{\sum}}}}    \bigg| \bigg|  \bigg[  \bigg(                   A_i \otimes \textbf{I} \otimes \textbf{I}              \bigg) - \bigg(              \textbf{I} \otimes \textbf{I} \otimes \frac{1}{\sqrt{6}} \bigg(\underset{\sigma \in S_3}{\sum }  C_{\sigma(ijk)}        \bigg)            \bigg)       \bigg]    \ket{\psi_{4\mathrm{XOR}}} \bigg| \bigg|^2$ \\  ${\underset{j \in \mathcal{Q}_2}{\underset{ i \in \mathcal{Q}_1}{\mathlarger{\sum}}}}    \bigg| \bigg|  \bigg[  \bigg(     \textbf{I} \otimes B_{ij} \otimes \textbf{I}                        \bigg) - \bigg(                   \bigg( \frac{A_i + A_j}{\sqrt{2}} \bigg) \otimes \textbf{I} \otimes \textbf{I}                                          \bigg)       \bigg]    \ket{\psi_{4\mathrm{XOR}}} \bigg| \bigg|     $   \\ $\underset{k \in \mathcal{Q}_3}{\underset{j \in \mathcal{Q}_2}{\underset{i \in \mathcal{Q}_1}{\mathlarger{\sum}}}}    \bigg| \bigg|  \bigg[  \bigg(        \textbf{I} \otimes B_{ij} \otimes \textbf{I}                 \bigg) - \bigg(          \textbf{I} \otimes \textbf{I} \otimes \frac{1}{\sqrt{6}} \bigg(\underset{\sigma \in S_3}{\sum }  C_{\sigma(ijk)}        \bigg)                \bigg)       \bigg]    \ket{\psi_{4\mathrm{XOR}}} \bigg| \bigg|^2    $  \\    $\vdots$ \\   [1ex] 
\hline 
\end{tabular}
\label{table:nonlin} 
\end{table}

\begin{table}[ht]
\caption{Error bound for the 4-XOR game - one player interchanges Continued} 
\centering 
\begin{tabular}{c c c c} 
\hline\hline 
 Error Bound Contributions \\ [0.5ex] 
\hline  $\vdots$  \\   $ \underset{k \in \mathcal{Q}_3}{\underset{j \in \mathcal{Q}_2}{\underset{i \in \mathcal{Q}_1}{\mathlarger{\sum}}}}    \bigg| \bigg|  \bigg[  \bigg(                     \textbf{I} \otimes \textbf{I} \otimes C_{ijk}                \bigg) - \bigg(                                  \bigg( \frac{A_i + A_j}{\sqrt{2}} \bigg) \otimes \textbf{I} \otimes \textbf{I}              \bigg)       \bigg]    \ket{\psi_{3\mathrm{XOR}}} \bigg| \bigg|^2   $  \\ $ \underset{k \in \mathcal{Q}_3}{\underset{j \in \mathcal{Q}_2}{\underset{i \in \mathcal{Q}_1}{\mathlarger{\sum}}}}    \bigg| \bigg|  \bigg[  \bigg(                  \textbf{I} \otimes \textbf{I} \otimes C_{ijk}                     \bigg) - \bigg(                \textbf{I} \otimes \bigg( \frac{B_{ij} + B_{ji}}{\sqrt{2}} \bigg) \otimes \textbf{I}  \bigg)       \bigg]    \ket{\psi_{3\mathrm{XOR}}} \bigg| \bigg|^2  $  \\   [1ex] 
\hline 
\end{tabular}
\label{table:nonlin} 
\end{table}

\noindent We argue that upper bounds for the summation of tensors above exist, given previous arguments in  \textbf{Lemma} (\textit{N-XOR permutation error bounds}, and  in \textbf{Lemma} (\textit{3-XOR permutation error bounds}.

\bigskip

\noindent \textbf{Theorem} $\textit{3}^{*}$ (\textit{4-XOR permutation error bounds}, \textit{2.2.1}, \textbf{Theorem} \textit{4}, {[37]}, \textbf{Theorem} \textit{2}, {[44]}, \textbf{Theorems} \textit{1}-\textit{6} in \textit{1.5}).

\begin{align*}
      ( \textit{****}) \leq 4!      n \bigg( \underset{1 \leq j \leq 3}{\prod} \big( n - j \big) \bigg)           \text{. }
\end{align*}

\noindent \textit{Proof of Theorem $3^{*}$}. To argue that the desired upper bound holds from direct computation, recall the following result, as a generalization of the $\epsilon$-approximality result of {[37]}:

\bigskip

\noindent \textbf{Theorem} \textit{1} (\textit{approximately optimal quantum strategies for the nonlocal XOR game}, \textbf{Theorem} \textit{4}, {[37]}). For $\pm$ observables $A_i$ and $B_{jk}$, given a bipartite state $\psi$, TFAE:

\begin{itemize}
 \item[$\bullet$] \underline{\textit{First characterization of approximate optimality}:} An $\epsilon$-approximate $\mathrm{CHSH}\big( n \big)$ satisfies $\mathrm{(0)}$.

 \item[$\bullet$] \underline{\textit{Second characterization of approximate optimality}:} For an $\epsilon$-approximate quantum strategy, 

  \begin{align*}
     \underset{1 \leq i < j \leq n}{\sum} \bigg[ \text{ }  \bigg| \bigg|   \bigg[         \big( \frac{A_i + A_j}{\sqrt{2}} \big) \otimes I \bigg] \ket{\psi}    -   \big[  I \otimes B_{ij} \big] \ket{\psi}   \bigg|\bigg|^2 + \bigg| \bigg|   \bigg[ \big( \frac{A_i - A_j}{\sqrt{2}} \big)  \otimes I  \bigg] \ket{\psi} \\   -   \big[  I \otimes B_{ji} \big]  \ket{\psi}     \bigg|\bigg|^2 \text{ }  \bigg]  \leq 2n \big( n - 1 \big) \epsilon \text{. }
 \end{align*}
 \item[$\bullet$] \underline{\textit{Reversing the order of the tensor product for observables}:} Related to the inequality for $\epsilon$-approximate strategies above, another inequality,

\begin{align*}
    \underset{1 \leq i < j \leq n}{\sum} \bigg[  \text{ } \bigg| \bigg| \big[ A_i \otimes I \big] \ket{\psi}    - \bigg[ I \otimes \big( \frac{B_{ij} + B_{ji}}{\sqrt{2}} \big)  \bigg] \ket{\psi}   \bigg| \bigg|^2  + \bigg| \bigg|    \big[ A_j  \otimes I \big] \ket{\psi} \\  -   \bigg[   I \otimes \big( \frac{B_{ij} - B_{ji}}{\sqrt{2}} \big) \bigg] \ket{\psi}    \big| \big|^2     \text{ }   \bigg] \leq 2n \big( n - 1 \big) \epsilon \text{, } 
\end{align*}

 \noindent also holds.

 \item[$\bullet$] \underline{\textit{Characterization of exact optimality}}: For $\epsilon \equiv 0$,

\begin{align*}
    \underset{1 \leq i < j \leq n}{\sum} \bigg[ \text{ }  \bigg| \bigg|  \bigg[  \big( \frac{A_i + A_j}{\sqrt{2}} \big) \otimes I \bigg] \ket{\psi}    -  \big[ I \otimes B_{ij} \big] \ket{\psi}   \bigg| \bigg|^2  \bigg]  =  -  \underset{1 \leq i < j \leq n}{\sum} \bigg[  \text{ }   \bigg| \bigg|     \bigg[  \big( \frac{A_i - A_j}{\sqrt{2}} \big)  \otimes I \bigg]  \ket{\psi} \\  -   \big[ I \otimes B_{ji} \big] \ket{\psi}    \bigg| \bigg|^2  \bigg]      \text{, } 
\end{align*}

\noindent corresponding to the first inequality, and,

\begin{align*}
       \underset{1 \leq i < j \leq n}{\sum} \bigg[ \text{ }  \bigg| \bigg|       \big[  A_i \otimes I \big]  \ket{\psi}    - \bigg[   I \otimes \big( \frac{B_{ij} + B_{ji}}{\sqrt{2}} \big)  \bigg]  \ket{\psi}   \bigg|\bigg|^2 \bigg] = -  \underset{1 \leq i < j \leq n}{\sum}   \bigg[  \bigg| \bigg|  \big[   A_j  \otimes I \big] \ket{\psi} \\  -  \bigg[    I \otimes \big( \frac{B_{ij} - B_{ji}}{\sqrt{2}} \big) \bigg]   \ket{\psi}     \bigg|\bigg|^2  \bigg]             \text{, } 
\end{align*}

\noindent corresponding to the second inequality.

\end{itemize}

\bigskip

\noindent  With the result above, to further generalize the set of equivalent conditions as provided in previous discussions for the $3\mathrm{XOR}$ game, and beyond, given the existence of suitable $\epsilon_{4\mathrm{XOR}}$, and a previously determined constant $C_{4\mathrm{XOR}} \equiv C$, the \textit{symmetrized} 4 $\mathrm{XOR}$ game tensor, and other terms,

\begin{align*}
   \underset{1 \leq i \leq 16}{\sum} y_i E_{ii} - G_{\mathrm{Sym},4\mathrm{XOR}} \equiv  \underset{1 \leq i \leq 16}{\sum} y_i E_{ii} - G_{\mathrm{Sym}}   \text{, }
\end{align*}

\noindent equals,

\begin{align*}
     \frac{1}{ \bigg( C_{4\mathrm{XOR}} \omega_{4\mathrm{XOR}} \bigg)  n \bigg( \underset{1 \leq j \leq 3}{\prod}  \big( n - j \big) \bigg) }             \underset{i_4 \in \mathcal{Q}_4}{\underset{\vdots}{\underset{i_1 \in \mathcal{Q}_1}{\sum}}}   \bigg(  \big(    u^{\prime}_{i_1i_2 i_3 i_4}   -    v^{\prime}_{i_1i_2 i_3 i_4}  \big) \big(   u^{\prime}_{i_1i_2 i_3 i_4}   -    v^{\prime}_{i_1 i_2 i_3 i_4}    \big)^{\textbf{T}}  \\   +  \big(    u^{\prime}_{i_2 i_1 i_3 i_4}     -    v^{\prime}_{i_2 i_1 i_3 i_4}  \big)   \big( u^{\prime}_{i_2 i_1 i_3 i_4}   -    v^{\prime}_{i_2 i_1 i_3 i_4}  \big)^{\textbf{T}}  + \cdots +     \big(    u^{\prime}_{i_2 i_3 i_4 i_1}   -    v^{\prime}_{i_2 i_3 i_4 i_1}  \big)   \big(   u^{\prime}_{i_2 i_3 i_4 i_1}  \\   -    v^{\prime}_{i_2 i_3 i_4 i_1}   \big)^{\textbf{T}}     \bigg)      \text{,}
\end{align*}

\noindent given the superposition of states,

\begin{align*}
        u^{\prime}_{i_1 i_2 i_3 i_4} \equiv \frac{1}{2} \bigg( \underset{1 \leq j \leq 4}{\sum} \ket{\text{Player } j \text{ state}} \bigg)      \text{, } v^{\prime}_{i_1 i_2 i_3 i_4} \equiv \ket{i_1 i_2 i_3 i_4} \text{, }  \\  u^{\prime}_{i_2 i_1 i_3 i_4} \equiv \frac{1}{2} \bigg( \ket{\text{Player 1 state}} - \ket{\text{Player 2 state}} \\ + \underset{3 \leq j \leq 4}{\sum} \ket{\text{Player } j \text{ state}} \bigg) \text{, } v^{\prime}_{i_2 i_1 i_3 i_4} \equiv \ket{i_2 i_1 i_3 i_4} \\ u^{\prime}_{i_2 i_3 i_1 i_4} \equiv \frac{1}{2} \bigg(  \underset{1 \leq j \leq 2}{\sum} \ket{\text{Player } j \text{ state}} - \ket{\text{Player } 3 \text{ state}} \\ + \ket{\text{Player } 4 \text{ state}}  \bigg)       \text{, } v^{\prime}_{i_2 i_3 i_1 i_4} \equiv \ket{i_2 i_3 i_1 i_4 } \\  u^{\prime}_{i_2 i_3 i_4 i_1} \equiv \frac{1}{2} \bigg( \underset{1 \leq j \leq 3}{\sum} \ket{\text{Player } j \text{ state}} - \ket{\text{Player } 4 \text{ state}}  \bigg)       \text{, }  \\ v^{\prime}_{i_2 i_3 i_4 i_1} \equiv \ket{i_2 i_3 i_4 i_1}   \text{, }
\end{align*}

\noindent from which we conclude the argument. \boxed{}

\bigskip

\noindent \textbf{Lemma} \textit{T-3} (\textit{positive semidefiniteness}). Under the assumptions of the previous result, the operator,

\begin{align*}
  \underset{1 \leq i \leq n^4}{\sum}   y_{4\mathrm{XOR},i} E_{4\mathrm{XOR},ii} - G_{4\mathrm{XOR},\mathrm{Sym}}     \text{, }
\end{align*}

 \noindent is positive semidefinite.

\bigskip

\noindent \textit{Proof of Lemma T-3}. The result follows from the fact that, for N-XOR games, the computation involving,

\begin{align*}
     \frac{1}{ \bigg( C_{4\mathrm{XOR}} \omega_{4\mathrm{XOR}} \bigg)  n \bigg( \underset{1 \leq j \leq 3}{\prod}  \big( n - j \big) \bigg) }             \underset{i_4 \in \mathcal{Q}_4}{\underset{\vdots}{\underset{i_1 \in \mathcal{Q}_1}{\sum}}}   \bigg(  \big(    u^{\prime}_{i_1i_2 i_3 i_4}   -    v^{\prime}_{i_1i_2 i_3 i_4}  \big) \big(   u^{\prime}_{i_1i_2 i_3 i_4}   -    v^{\prime}_{i_1 i_2 i_3 i_4}    \big)^{\textbf{T}}     \\ +  \big(    u^{\prime}_{i_2 i_1 i_3 i_4}    -    v^{\prime}_{i_2 i_1 i_3 i_4}  \big)   \big( u^{\prime}_{i_2 i_1 i_3 i_4}   -    v^{\prime}_{i_2 i_1 i_3 i_4}  \big)^{\textbf{T}}  + \cdots +     \big(    u^{\prime}_{i_2 i_3 i_4 i_1}   -    v^{\prime}_{i_2 i_3 i_4 i_1}  \big)   \big(   u^{\prime}_{i_2 i_3 i_4 i_1}  \\   -    v^{\prime}_{i_2 i_3 i_4 i_1}   \big)^{\textbf{T}}     \bigg)      \text{,}
\end{align*}

\noindent implies that the associated operator is positive semidefinite from the observation that taking the constant $C_{N\mathrm{XOR}}$, in the normalization,

\begin{align*}
     \frac{1}{ \bigg( C_{4\mathrm{XOR}} \omega_{4\mathrm{XOR}} \bigg) n \bigg( \underset{1 \leq j \leq 3}{\prod}  \big( n - j \big) \bigg) }        \text{, }
\end{align*}

\noindent to equal $4!$ implies the desired result, from which we conclude the argument. \boxed{}

\bigskip

\noindent An implication of the above result which demonstrates that the desired upper bound, dependent upon both a combinatorial factor and the total number of players in the game, also implies the following result.

\subsubsection{Two strong parallel repetitions of the FFL game}

\noindent Denote the strong parallel repetition game matrix for the FFL game with $G_{\mathrm{FFL} \wedge \mathrm{FFL}}$. With a sigle operation of strong parallel repetition to the FFL game matrix, from relations involving the optimal value, and bias, of the FFL game without strong parallel repetition, one has, for some $\epsilon$ sufficiently small,

\begin{align*}
         \omega_{\mathrm{FFL} \wedge \mathrm{FFL}} \big( 1 - \epsilon \big) \equiv \frac{2}{3} \big( 1 - \epsilon \big)  \leq    \underset{\mathcal{Q}_2}{\underset{\mathcal{Q}_1}{\sum}}  \underset{\mathcal{Q}^{\prime}_2}{\underset{\mathcal{Q}^{\prime}_1}{\sum}}                            \big( G_{\mathrm{FFL} \wedge \mathrm{FFL}}  \big)   \bra{\psi_{\mathrm{FFL} \wedge \mathrm{FFL}}}     \bigg( \big( A_{ij} \wedge A_{i^{\prime}j^{\prime}} \big) \\ \otimes \big( B_{ij} \wedge  B_{i^{\prime}j^{\prime}} \big) \bigg)         \ket{\psi_{\mathrm{FFL} \wedge \mathrm{FFL}}}      \leq      \omega_{\mathrm{FFL} \wedge \mathrm{FFL}} \leq \frac{2}{3} \equiv \omega \big( \mathrm{FFL} \wedge \mathrm{FFL} \big)  \text{, }
\end{align*}

\noindent corresponding to the $\epsilon$-approximatlity of the optimal quantum strategies, for the nonidentical bra, and ket, states,

\begin{align*}
  \bra{\psi_{\mathrm{FFL}}}  \neq  \bra{\psi^{\prime}_{\mathrm{FFL}}} \text{, } \\\ket{\psi_{\mathrm{FFL}}} \neq \ket{\psi^{\prime}_{\mathrm{FFL}}}  \text{, }
\end{align*}

\noindent with,

\begin{align*}
     \bra{\psi_{\mathrm{FFL} \wedge \mathrm{FFL}}}    \equiv  \bra{\psi_{\mathrm{FFL}}} \wedge \bra{\psi^{\prime}_{\mathrm{FFL}}} \text{, }
\end{align*}

\noindent and, with,

\begin{align*}
    \ket{\psi_{\mathrm{FFL} \wedge \mathrm{FFL}}}    \equiv  \ket{\psi_{\mathrm{FFL}}} \wedge \ket{\psi^{\prime}_{\mathrm{FFL}}}   \text{. }  
\end{align*}

\noindent The observables $A_{ij}$ and $B_{ij}$ are gathered by Alice and Bob in the first iteration of the FFL game before strong parallel repetition, while the observables $A_{i^{\prime}j^{\prime}}$ and $B_{i^{\prime}j^{\prime}}$ gathered by Alice and Bob in the second FFL game. Besides the above inequality for $\epsilon$-approximality of the strong parallel repetition, $\mathrm{FFL} \wedge \mathrm{FFL}$, of the $\mathrm{FFL}$ game, the bias, 

\begin{align*}
 \beta_{\mathrm{FFL}} \big( G_1 \big) \wedge \beta_{\mathrm{FFL} } \big( G_2 \big) \equiv \beta_{\mathrm{FFL} \wedge \mathrm{FFL}}  \big( G^{\prime} \big) \equiv \beta \big(  \mathrm{FFL} \wedge \mathrm{FFL} \big) \neq  \beta \big(  \mathrm{FFL} \big)  \wedge \beta \big(  \mathrm{FFL} \big)   \text{, }
\end{align*}

\noindent for the strong parallel repetition $G^{\prime} \equiv G_1 \wedge G_2$, of the two games satisfies,

\begin{align*}
 \beta_{\mathrm{FFL} \wedge \mathrm{FFL}}  \big( G^{\prime} \big) \big( 1 - \epsilon \big) \leq    \underset{\mathcal{Q}_2}{\underset{\mathcal{Q}_1}{\sum}}  \underset{\mathcal{Q}^{\prime}_2}{\underset{\mathcal{Q}^{\prime}_1}{\sum}}\bra{ \psi_{\mathrm{FFL} \wedge \mathrm{FFL}}}     \bigg( \big( A_{ij} \wedge A_{i^{\prime}j^{\prime}} \big) \otimes \big( B_{ij} \wedge  B_{i^{\prime}j^{\prime}} \big) \bigg)   \\ \times           \ket{\psi_{\mathrm{FFL} \wedge \mathrm{FFL}}}   \leq \beta_{\mathrm{FFL}\wedge \mathrm{FFL}} \big( G^{\prime} \big) \text{. }
\end{align*}

\noindent As is the case for $2$ $\mathrm{XOR}$, and $\mathrm{FFL}$, games, the effect of applying an intertwining operation to the observables which players construct can be expressed in more generality with the following inequalities, which also serve as an extension of error bounds. Given a suitable linear operator $T_1 \equiv T_{\mathrm{FFL}\wedge\mathrm{FFL}}$, and $T_2 \equiv T_{\mathrm{XOR} \wedge \cdots \wedge \mathrm{XOR}}$, associated with strong parallel repetition, the expected action which constitutes the error bound takes the form,

\begin{align*}
 \bigg| \bigg| T \otimes  \bigg( \bigg(   \underset{i_2 \in \mathcal{Q}_2, i_3 \in \mathcal{Q}_3, \cdots, i_n \in \mathcal{Q}_n}{\prod}  \text{Tensors of player observables} \big( i_2 , \cdots , i_n \big)   \bigg) \wedge \bigg(    \underset{i^{\prime}_2 \in \mathcal{Q}_2, \cdots, i^{\prime}_n \in \mathcal{Q}_n}{\prod}  \text{Tensors of player} \\ \text{ observables}  \big( i^{\prime}_2 , \cdots , i^{\prime}_n \big)     \bigg) \bigg) - \widetilde{\bigg( \bigg(   \underset{i_2 \in \mathcal{Q}_2, i_3 \in \mathcal{Q}_3, \cdots, i_n \in \mathcal{Q}_n}{\prod}  \text{Tensors of player observables} \big( i_2 , \cdots , i_n \big)   \bigg)} \\ \wedge \widetilde{\bigg(    \underset{i^{\prime}_2 \in \mathcal{Q}_2, i^{\prime}_3 \in \mathcal{Q}_3, \cdots, i^{\prime}_n \in \mathcal{Q}_n}{\prod}  \text{Tensors of player observables}  \big( i^{\prime}_2 , \cdots , i^{\prime}_n \big)     \bigg) \bigg) } \otimes T \bigg| \bigg|           \text{, }
\end{align*}

\begin{align*}
\bigg| \bigg|   \bigg( \textbf{I}  \otimes                         \bigg( \bigg(   \big( \text{Second player tensor observable} \big( i_1 , \i_2 \big) \wedge \big(  \text{Second player tensor observable} \big( i^{\prime}_1 , \i^{\prime}_2 \big)        \bigg)    \\  \otimes \overset{N-3}{\cdots}            
\otimes \bigg(       \big(   \text{N th player tensor observable} \big( i_1 , i_2 , \cdots , i_n \big)     \big)     \wedge \big(  \text{N th player tensor observable} \big( i^{\prime}_1 , i^{\prime}_2 \\ , \cdots , i^{\prime}_n \big)            \big) \bigg)        -  \bigg( \bigg(               \text{Second player tensor observable} \big( i_1 , \i_2 \big) \wedge \big(  \text{Second player tensor observable} \big( i^{\prime}_1 , \i^{\prime}_2 \big)       \bigg)      \\   \otimes \textbf{I} \otimes     
 \bigg( \big( \text{Third player tensor observable} \big( i_1 , i_2 , i_3 \big) \wedge \big( \text{Third player tensor observable} \big( i^{\prime}_1 , i^{\prime}_2 , i^{\prime}_3 \big)      \bigg) \otimes \\   \overset{N-4}{\cdots} \otimes            \bigg( \big( \text{N th player tensor observable} \big( i_1 , i_2 , \cdots , i_n \big) \wedge \big( \text{N th player tensor} \\ \text{ observable} \big( i^{\prime}_1 , i^{\prime}_2 ,   \cdots , i^{\prime}_n \big) \bigg) \bigg) \ket{\psi_{\mathrm{XOR} \wedge \cdots \wedge \mathrm{XOR}}}    \bigg| \bigg|  \text{, }  \end{align*}

\begin{align*} \bigg| \bigg| \bigg(  \bigg( \bigg( \big( \text{First player tensor observable} \big( i_1 \big) \big) \wedge \big( \text{First player tensor observable} \big( i^{\prime}_1  \big) \bigg) \otimes \textbf{I}  \\     \otimes \bigg( \big( \text{Third player tensor observable} \big( i_1 , i_2 , i_3 \big) \big) \wedge \big(     \text{Third player tensor observable} \big( i^{\prime}_1 , i^{\prime}_2 , i^{\prime}_3 \big) \big) \otimes \textbf{I}  \\     \overset{N-6}{\cdots} \otimes \textbf{I} \otimes \bigg( \big( \text{N th player tensor observable} \big( i_1 , i_2 , \cdots , i_n \big) \big)  \wedge  \big( \text{N th player tensor observable} \big( i^{\prime}_1 , i^{\prime}_2 \\    , \cdots , i^{\prime}_n \big) \big) \bigg)   \bigg)  -  \bigg(   \bigg( \big( \text{First player tensor observable} \big( i_1 \big) \big) \wedge \big( \text{First player tensor observable} \big( i^{\prime}_1 \big) \big)   \bigg)\otimes      \textbf{I}  \\    \otimes \bigg( \big( \text{Third player tensor observable} \big( i_1 , i_2 , i_3 \big) \big)  \wedge \big( \text{Third player tensor observable} \big( i^{\prime}_1 , i^{\prime}_2 , i^{\prime}_3 \big) \big) \bigg) \\               
 \otimes \bigg(  \big( \text{Fourth player tensor observable} \big( i_1 , i_2 , i_3 , i_4 \big) \big) \wedge \big( \text{Fourth player tensor observable} \big( i^{\prime}_1 , i^{\prime}_2 \\ , i^{\prime}_3 , i^{\prime}_4 \big) \big)  \bigg)        \otimes \textbf{I}  \otimes \overset{N-7}{\cdots}   \otimes   \textbf{I} \otimes   \bigg( \big( \text{N th player tensor observable} \big( i_1 , i_2 , \cdots , i_n \big) \big)  \wedge  \big( \text{N th player} \\ \text{tensor observable} \big( i^{\prime}_1 , i^{\prime}_2    , \cdots , i^{\prime}_n \big) \big) \bigg)      \bigg)   \bigg) \ket{\psi_{\mathrm{XOR} \wedge \cdots \wedge \mathrm{XOR}}} \bigg| \bigg|  \end{align*}

\begin{align*}       \bigg| \bigg|         \bigg( \bigg( \bigg( \bigg( \underset{1 \leq i_1 \leq n}{\prod}     \text{First player tensor observable} \big( i_1 \big) \bigg)   \wedge  \bigg(   \underset{1 \leq i^{\prime}_1 \leq n}{\prod}     \text{First player tensor observable} \big( i^{\prime}_1 \big) \bigg)    \bigg)   \otimes      \textbf{I}  \\ \otimes \bigg( \bigg(   \underset{i_3 \in \mathcal{Q}_3 }{\prod} \text{Third player tensor observable} \big( i_1 , i_2 , i_3 \big) \bigg)  \wedge \bigg( \underset{i_3 \in \mathcal{Q}_3}{\prod} \text{Third player tensor} \\ \text{ observable} \big( i_1      , i_2 , i_3 \big) \bigg)  \bigg) \bigg)  -   \bigg(                     \textbf{I} \otimes \bigg( \bigg(      \underset{i_2 \in \mathcal{Q}_2}{\prod} \text{Second player tensor observable } \big( i_1 , i_2 \big)    \bigg)     \\ \wedge \bigg(        \underset{ i^{\prime}_2 \in \mathcal{Q}_2}{\prod} \text{Second player tensor observable } \big( i^{\prime}_1 , i^{\prime}_2 \big)         \bigg) \bigg)   \otimes \bigg( \bigg(           \underset{i_3 \in \mathcal{Q}_3,  i_4 \in \mathcal{Q}_4, \cdots, i_n \in \mathcal{Q}_n}{\prod}  \text{Third player} \\ \text{ tensor observable} \big( i_1 , i_2 , \cdots , i_n \big)  \big)   \bigg)         \wedge \bigg( \underset{i^{\prime}_3 \in \mathcal{Q}_3, i^{\prime}_4 \in \mathcal{Q}_4, \cdots, i^{\prime}_n \in \mathcal{Q}_n}{\prod}  \text{Third player tensor observa-} \\ \text{ble} \big( i^{\prime}_1 , i^{\prime}_2 ,  \cdots , i^{\prime}_n \big)  \bigg) \bigg)             \bigg)   \bigg) \ket{\psi_{ \mathrm{FFL} \wedge \mathrm{FFL}}}      \bigg| \bigg| \text{,} \\ \\     \bigg| \bigg| \bigg[    \bigg(   \bigg( \bigg(                 \text{First player tensor} \big( i_1 \big) \bigg) \wedge \bigg(   \text{First player tensor} \big( i^{\prime}_1 \big)       \bigg)    \bigg)  \otimes \textbf{I}   \otimes \bigg( \bigg(   \underset{i_3 \in \mathcal{Q}_3, i_4 \in \mathcal{Q}_4, \cdots, i_n \in \mathcal{Q}_n}{\prod}  \text{Third } \\ \text{ player tensor observables} \big(  i_1, i_2, \cdots , i_n \big)   \bigg)    \wedge \bigg(   \underset{i^{\prime}_3 \in \mathcal{Q}_3, i^{\prime}_4 \in \mathcal{Q}_4, \cdots, i^{\prime}_n \in \mathcal{Q}_n}{\prod}  \text{Third player tensor observa-} \\ \text{bles} \big(  i^{\prime}_1  , i^{\prime}_2, \cdots , i^{\prime}_n \big)      \bigg) \bigg)     -  \bigg(             \textbf{I}    \otimes  \pm  \frac{2}{3}     \bigg(   \frac{\mathscr{T}_2}{\big| \pm \mathscr{T}_2 \big| }   \bigg)            \bigg)           \bigg]   \ket{\psi_{\mathrm{FFL} \wedge \mathrm{FFL}}} \bigg| \bigg| \text{,} \end{align*}

\noindent for,

\begin{align*}
   \mathscr{T}_2 \equiv                        \bigg(  \big( \text{Second player tensor observable} \big( i_1 , i_2 , \cdots , i_n \big) \big) \wedge \big( \text{Second player tensor observable} \big(  i^{\prime}_1, \\   i^{\prime}_2 , \cdots , i^{\prime}_n \big)         \bigg)        + \underset{\text{permutations } \sigma , \sigma^{\prime}}{\sum}          \bigg(      \big(   \text{Tensors of player observables} \big( \sigma i_1 , \sigma i_2 , \cdots , \sigma i_n \big)      \big)  \\    \wedge \big(           \text{Tensors of player observables} \big( \sigma i^{\prime}_1 , \sigma i^{\prime}_2 ,  \cdots , \sigma i^{\prime}_n \big)        \bigg)             \text{, }
\end{align*}

\noindent and,

\begin{align*}     \bigg| \bigg|    \bigg[     \bigg( \bigg(  \textbf{I} \otimes \textbf{I} \otimes \cdots \otimes \bigg(  \bigg( \underset{i_1 \in \mathcal{Q}_1, i_2 \in \mathcal{Q}_2, \cdots, i_n \in \mathcal{Q}_n}{\prod}                 \text{Tensors of player observables} \big( i_1 , i_2 , \cdots , i_n \big)         \bigg) \\  \wedge \bigg( \underset{i^{\prime}_1 \in \mathcal{Q}_1, i^{\prime}_2 \in \mathcal{Q}_2, \cdots, i^{\prime}_n \in \mathcal{Q}_n}{\prod}                 \text{Tensors of player observables} \big( i^{\prime}_1 , i^{\prime}_2 , \cdots , i^{\prime}_n \big)       \bigg) \\   \otimes \bigg(      \big(   \text{N th player tensor observable} \big( i_1 , i_2  , \cdots , i_n \big)       \big)    \wedge  \big(   \text{N th  player tensor observa-} \\ \text{ble} \big( i^{\prime}_1 , i^{\prime}_2 , \cdots , i^{\prime}_n \big)    \bigg)   \bigg)          -    \bigg(      \textbf{I} \otimes \textbf{I} \otimes   \overset{n-4}{\cdots}      \otimes \bigg(          \big( \text{N th player tensor observable} \big( i_1 , i_2 \\ , \cdots , i_n \big) \big)  \wedge \big(   \text{N th player tensor observable} \big( i^{\prime}_1 , i^{\prime}_2 , \cdots , i^{\prime}_n \big)     \big)                \bigg)    \otimes          \bigg( \frac{1}{\sqrt{N-1}}   \\ \times   \bigg(   \frac{\pm  \mathscr{T}_3}{\big| \pm  \mathscr{T}_3 \big| }                \bigg)      \bigg)     \bigg)  \bigg]  \ket{\psi_{\mathrm{XOR} \wedge \cdots \wedge \mathrm{XOR}}}     \bigg| \bigg| \text{,} 
\end{align*}

\noindent for,

\begin{align*}
 \mathscr{T}_3 \equiv  \big( \text{Third player tensor observable} \big( i_1 , i_2 , i_3 \big) \big) \wedge     \big( \text{Third player tensor observable} \\ \big( i^{\prime}_1 , i^{\prime}_2 , i^{\prime}_3 \big) \big)       + \underset{\text{permutations }\sigma, \sigma^{\prime}}{\sum}    \bigg(       \big( \text{Third player tensor observable} \big(   \sigma     i_1 ,    \sigma    i_2 ,   \sigma     i_3 \big) \big) \\  \wedge    \big( \text{Third player tensor observable} \big( \sigma^{\prime} i^{\prime}_1 ,  \sigma^{\prime} i^{\prime}_2 , \sigma^{\prime} i^{\prime}_3 \big) \big)  \bigg)     \text{. }
\end{align*}

\noindent Performing the strong parallel repetition of two FFL games, besides the inequalities provided above which are primarily dependent upon applying permutations $\sigma$ and $\sigma^{\prime}$, can also be formulated from the individual responses that each player prepares when responding to a referee's question. Namely, as a summation over the questions administered to each player,

\begin{align*}
         \underset{i \in \mathcal{Q}_1 , i^{\prime} \in \mathcal{Q}^{\prime}_1}{\underset{j \in \mathcal{Q}_2 , j^{\prime} \in \mathcal{Q}^{\prime}_2}{\sum}}      \bigg| \bigg|      \bigg[   \bigg(      \bigg( \bigg( \frac{A_i + A_j}{\sqrt{2}} \bigg) \wedge \bigg(  \frac{A_{i^{\prime}} + A_{j^{\prime}}}{\sqrt{2}} \bigg) \bigg) \otimes     \textbf{I}   \bigg)   -   \bigg(          \textbf{I} \otimes \bigg( B_{ij} \wedge B_{i^{\prime}j^{\prime}}      \bigg)                    \bigg)   \bigg] \\ \times \ket{\psi_{\mathrm{FFL} \wedge \mathrm{FFL}}} \bigg| \bigg|    \Longleftrightarrow       \underset{\mathcal{Q}_1,\mathcal{Q}^{\prime}_1}{\sum}  \bigg| \bigg|   \bigg[            \bigg( \bigg( \frac{A_i + A_j}{\sqrt{2}} \bigg) \otimes \textbf{I} \bigg) - \bigg(         \textbf{I} \otimes B_{ij}      \bigg)              \bigg| \bigg|  \\ \wedge     \bigg| \bigg|        \bigg( \bigg( \frac{A_{i^{\prime}} + A_{j^{\prime}}}{\sqrt{2}}         \bigg)  \otimes \textbf{I} \bigg) - \bigg(      \textbf{I}   \otimes B_{i^{\prime} j^{\prime}}       \bigg)                   \bigg]   \ket{\psi_{\mathrm{FFL} \wedge \mathrm{FFL}}}      \bigg| \bigg|            \text{.} \end{align*}

        \noindent For one strong parallel repetition operation, the inequality for the N-player $\mathrm{XOR}$ game implies the existence of an inequality, 
         
         \begin{align*}
         \underset{i_1 \in \mathcal{Q}_1, i^{\prime}_1 \in \mathcal{Q}^{\prime}_1}{\sum}  \bigg[     \bigg| \bigg|   \bigg[   \bigg( \textbf{I} \otimes \textbf{I} \otimes \bigg(     \underset{i_1 \in \mathcal{Q}_1, i_2 \in \mathcal{Q}_2, \cdots, i_n \in \mathcal{Q}_n}{\prod}  \text{Tensors of player observables} \big( i_1 , i_2 , \cdots , i_n \big)     \bigg) \\ \otimes \big( \text{N th player tensor observable}   \big( i_1 , i_2  \cdots , i_n \big)   \big)                                 \bigg) - \bigg(  \textbf{I} \otimes \textbf{I} \otimes \cdots \otimes \big( \text{N th player tensor} \\ \text{ observable} \big( i_1 , i_2 , \cdots  ,i_n \big)          \big)     \bigg)    \bigg]    \wedge \bigg[ \bigg(       \textbf{I} \otimes \textbf{I}   \otimes \bigg(    \underset{i^{\prime}_1 \in \mathcal{Q}_1, i^{\prime}_2 \in \mathcal{Q}_2,  \cdots, i^{\prime}_n \in \mathcal{Q}_n}{\prod}  \text{Tensors of player observa-} \\ \text{bles} \big( i^{\prime}_1 , i^{\prime}_2 , \cdots , i^{\prime}_n \big)        \bigg)       \bigg)    -   \bigg(  \textbf{I} \otimes \textbf{I} \otimes \cdots \otimes \big( \text{N th player tensor observable}  \big( i^{\prime}_1 , i^{\prime}_2 , \cdots ,i^{\prime}_n \big)          \big)     \bigg)     \bigg]                     \bigg] \\ \times  \ket{\psi_{\mathrm{XOR} \wedge \mathrm{XOR}}}   \bigg| \bigg|    \bigg]    \text{, }  \end{align*}

  \begin{align*}\underset{i_1 \in \mathcal{Q}_1, i^{\prime}_1 \in \mathcal{Q}^{\prime}_1}{\sum}  \bigg[  \bigg| \bigg| \bigg[   \textbf{I} \otimes \textbf{I} \otimes \bigg(          \underset{i_1 \in \mathcal{Q}_1, i_2 \in \mathcal{Q}_2, \cdots, i_n \in \mathcal{Q}_n}{\prod}  \text{Tensors of player observables} \big( i_1 , i_2 , \cdots , i_n \big)     \bigg) \\ \otimes   \big(   \text{N th player tensor observable} \big( i_1 ,   i_2 , \cdots , i_n \big)      \big) \otimes \bigg(  \frac{1}{\sqrt{n-1}} \bigg(    \frac{\mathscr{T}_3}{\big| \pm \mathscr{T}_3 \big| }      \bigg)     \bigg) \bigg] \\  - \bigg[         \bigg( \textbf{I} \otimes \textbf{I} \otimes \big( \text{N th player tensor observable} \big( i_1 , i_2  , \cdots , i_n \big) \big)   \otimes \bigg(   \frac{1}{\sqrt{N-1}}       \bigg(   \frac{\mathscr{T}^{\prime}_3}{\big| \mathscr{T}^{\prime}_3 \big| }   \bigg)     \bigg)          \bigg)        \bigg)     \bigg]        \\ \times                      \ket{\psi_{\mathrm{XOR} \wedge \mathrm{XOR}}}       \bigg| \bigg| \bigg] 
 \text{, }       
\end{align*}

\noindent for, $\mathscr{T}^{\prime}_3 \big( i^{\prime}_1,i^{\prime}_2, i^{\prime}_3 \big) \equiv \mathscr{T}^{\prime}_3$, which is explicitly given by,

\begin{align*}
   \mathscr{T}^{\prime}_3 \equiv    \text{Third player tensor observable} \big(i^{\prime}_1 , i^{\prime}_2 , i^{\prime}_3 \big) + \underset{\text{permutations }\sigma,\sigma^{\prime}}{\sum}    \text{Third player tensor} \\ \text{observable} \big( \sigma i^{\prime}_1  , \sigma i^{\prime}_2 , \sigma i^{\prime}_3 \big)           \text{. }
\end{align*}

\noindent In the N-player $\mathrm{XOR}$ game, denoting,

\begin{align*}
  G_{\mathrm{XOR} \wedge \cdots \wedge \mathrm{XOR}} \equiv  G_{(N\mathrm{XOR}) \wedge \cdots \wedge (N\mathrm{XOR})} \equiv \underset{1 \leq i \leq n}{\bigwedge}   G^{i}_{(N\mathrm{XOR})} \text{, }
\end{align*}

\noindent where,

\begin{align*}
     G^{i}_{(N\mathrm{XOR})} \cap G \neq \emptyset \text{, }
\end{align*}

\noindent for every $i$, from which there exists a suitable linear operator $T_{\mathrm{XOR}\wedge\mathrm{XOR}} \equiv T_{\mathrm{XOR}} \wedge T_{\mathrm{XOR}} \equiv T^{\prime\prime}$, for which,

\begin{align*}
   \bigg| \bigg|    \bigg[ \bigg( \bigg( A_i \wedge A_{i^{\prime}} \bigg) \bigotimes \bigg( \underset{1 \leq k \leq n-1}{\bigotimes} \textbf{I}_k \bigg) \bigg) T^{\prime\prime}    -     T^{\prime\prime} \bigg( \bigg( \underset{1 \leq k \leq n-1}{\bigotimes}     \textbf{I}_k \bigg) \bigotimes \bigg( A_i \wedge A_{i^{\prime}} \bigg)   \bigg)     \bigg]   \\ \times        \ket{\psi_{\mathrm{XOR} \wedge \cdots \wedge \mathrm{XOR}}}  \bigg| \bigg|     \text{,} \\ \\   \bigg| \bigg|    \bigg[ \bigg( \bigg(      \textbf{I} \bigotimes \bigg( B_{ij} \wedge B_{i^{\prime}j^{\prime}} \bigg) \bigotimes \textbf{I} \bigotimes \bigg( \underset{1 \leq k \leq n-3}{\bigotimes}  \textbf{I}_k \bigg)      \bigg) T^{\prime\prime}  -  T^{\prime\prime} \bigg(           \textbf{I} \bigotimes \bigg( B_{ij} \wedge B_{i^{\prime}j^{\prime}} \bigg) \\ \bigotimes \bigg( \underset{1 \leq k \leq n-3}{\bigotimes} \textbf{I}_k \bigg)           \bigg)      \bigg]        \ket{\psi_{\mathrm{XOR} \wedge \cdots \wedge \mathrm{XOR}}}  \bigg| \bigg|    \text{. }
\end{align*}

\noindent Under strong parallel repetition, the semidefinite programs associated with primal feasible solutions are dependent upon the constraints, from the partial ordering $ \succcurlyeq$ induced by the positive semidefinite cone,

\begin{align*}
 \underset{1 \leq i \leq n^3}{\sum}      y_{3\mathrm{XOR} \wedge \cdots \wedge 3 \mathrm{XOR},i} E_{3\mathrm{XOR} \wedge \cdots \wedge 3 \mathrm{XOR} ,ii}     \succcurlyeq           G_{3\mathrm{XOR} \wedge \cdots \wedge 3 \mathrm{XOR}, \mathrm{Sym}}        \text{, }   \tag{$3\mathrm{XOR} \wedge \cdots \wedge 3 \mathrm{XOR}$, Sym}  \\  \underset{1 \leq i \leq n^4}{\sum}   y_{4\mathrm{XOR} \wedge \cdots \wedge 4 \mathrm{XOR},i} E_{4\mathrm{XOR}\wedge \cdots \wedge 4 \mathrm{XOR},ii}    \succcurlyeq G_{4\mathrm{XOR} \wedge \cdots \wedge 4 \mathrm{XOR},\mathrm{Sym}} \tag{$4\mathrm{XOR} \wedge \cdots \wedge 4 \mathrm{XOR}$, Sym}  \text{, }   \\   \underset{1 \leq i \leq n^5}{\sum}  y_{5\mathrm{XOR} \wedge \cdots \wedge 5 \mathrm{XOR},i} E_{5\mathrm{XOR} \wedge \cdots \wedge 5 \mathrm{XOR},ii}    \succcurlyeq  G_{5\mathrm{XOR}\wedge \cdots \wedge 5 \mathrm{XOR},\mathrm{Sym}} \tag{$5\mathrm{XOR} \wedge \cdots \wedge 5 \mathrm{XOR}$, Sym}  \text{, } \\  \underset{1 \leq i \leq n^N}{\sum}    y_{N\mathrm{XOR} \wedge \cdots \wedge N \mathrm{XOR},i} E_{N\mathrm{XOR} \wedge \cdots \wedge N \mathrm{XOR},ii}       \succcurlyeq G_{N\mathrm{XOR}\wedge \cdots \wedge N \mathrm{XOR},\mathrm{Sym}}  \tag{$N\mathrm{XOR} \wedge \cdots \wedge N \mathrm{XOR}$, Sym}  \text{, }  \\  \underset{1 \leq i \leq n^2}{\sum}    y_{\mathrm{FFL} \wedge \mathrm{FFL},i} E_{\mathrm{FFL} \wedge \mathrm{FFL},ii}      \succcurlyeq G_{\mathrm{FFL} \wedge \mathrm{FFL},\mathrm{Sym}} \tag{$\mathrm{FFL} \wedge\mathrm{FFL}$, Sym}  \text{. }
\end{align*}

\noindent as previously introduced for symmetrized game tensors before the strong parallel repetition operation is taken. Otherwise, ientical copies of the identity operator are indicated with,

\begin{align*}
   \textbf{I}_k \equiv \textbf{I} \text{, }
\end{align*}

\noindent for every $k$. The underlying structure of the error bound inequalities, along with the action of the suitable linear operator under strong parallel repetition, is encapsulated by the existence of a primal feasible solution, $Z_{\mathrm{XOR}\wedge \cdots \wedge \mathrm{XOR}}$, for the semidefinite program,

\begin{align*}
  \underset{\forall Z_{\mathrm{XOR}\wedge \cdots \wedge \mathrm{XOR}} \succcurlyeq 0 , 1 \leq i \leq m, F^{(j)}_{\wedge i} \cdot Z_{\mathrm{XOR} \wedge \cdots \wedge {\mathrm{XOR}}} \equiv C^{(j)}_i}{\mathrm{sup}}  G_{\mathrm{XOR}\wedge \cdots \wedge \mathrm{XOR}} Z_{\mathrm{XOR}\wedge \cdots \wedge \mathrm{XOR}} \text{, }
\end{align*}

\noindent for some strictly positive $C^{(j)}_i$, which admits the decomposition,

\begin{align*}
  \underset{1 \leq j \leq n}{\bigwedge} \bigg[  \underset{\forall Z_{\mathrm{XOR}\wedge \cdots \wedge \mathrm{XOR}} \succcurlyeq 0 , 1 \leq i \leq m, F^{(j)}_{\wedge i} \cdot Z_{\mathrm{XOR} \wedge \cdots \wedge {\mathrm{XOR}}} \equiv C^{(j)}_i}{\mathrm{sup}}    G_{\mathrm{XOR}} Z_{\mathrm{XOR}}      \bigg]       \text{. }
\end{align*}

\noindent The primal feasible solution itself admits the decomposition,

\begin{align*}
      Z_{\mathrm{XOR} \wedge \cdots \wedge \mathrm{XOR}} \equiv \underset{1 \leq j \leq n}{\bigwedge}  Z^{(j)}_{\mathrm{XOR}}    \text{. }
\end{align*}

\noindent Under strong parallel repetition, the duality of the semidefinite program above is equivalent to the minimization,

\begin{align*}
\underset{1 \leq j \leq n}{\bigwedge} \bigg[ \underset{ \underset{1 \leq j \leq n}{\wedge}   \underset{\mathcal{Q}^{(1)}_N \wedge \cdots \wedge \mathcal{Q}^{(i)}_N \wedge  \cdots \wedge \mathcal{Q}^{(n)}_N}{\underset{\vdots}{{\underset{\mathcal{Q}^{(1)}_1 \wedge \cdots \wedge \mathcal{Q}^{(i)}_N \wedge \cdots  \wedge \mathcal{Q}^{(n)}_1}{\sum}}}}    y_{\mathcal{Q}^{(j)}_i} E^{(j)}_{ii}     \succcurlyeq  G^{(j)}_{\mathrm{Sym}}          }{\mathrm{inf}}  \bigg(    \underset{\mathcal{Q}^{(1)}_N \wedge \cdots \wedge \mathcal{Q}^{(i)}_N \wedge \cdots \wedge \mathcal{Q}^{(n)}_N}{\underset{\vdots}{\underset{\mathcal{Q}^{(1)}_1 \wedge \cdots \wedge \mathcal{Q}^{(i)}_N \wedge \cdots \wedge \mathcal{Q}^{(n)}_1}{\sum}}}       y_{\mathcal{Q}^{(j)}_i}   \bigg)         \bigg]  \text{, }
\end{align*}

\noindent where, under the constraints placed on the infimum above, are,

\begin{align*}
     G_{\mathrm{Sym}, \mathrm{XOR} \wedge \cdots \wedge \mathrm{XOR}} \equiv \underset{1 \leq j \leq n}{\bigwedge}  G^{(j)}_{\mathrm{Sym}, \mathrm{XOR}} \equiv  \underset{1 \leq j \leq n}{\bigwedge}          \begin{bmatrix}  0 & \big( G^{(j)}_{\mathrm{XOR}} \big)^{\textbf{T}} \\ G^{(j)}_{\mathrm{XOR}} & 0  \end{bmatrix}    \\    \equiv \begin{bmatrix} 0 &  \underset{1 \leq j \leq n}{\wedge}  \big( G^{(j)}_{\mathrm{XOR}} \big)^{\textbf{T}} \\ \underset{1 \leq j \leq n}{\wedge} G^{(j)}_{\mathrm{XOR}} & 0   \end{bmatrix} \text{. }
\end{align*}

\noindent We denote the dual feasible solution with $V_{\mathrm{Dual},\mathrm{XOR} \wedge \cdots \wedge \mathrm{XOR}}$.

\bigskip

\noindent For the following result, denote $T_{\mathrm{XOR} \wedge \cdots \wedge \mathrm{XOR}} \equiv T_{N\mathrm{XOR} \wedge \cdots \wedge N\mathrm{XOR}}$, that is, the suitable linear operator for an arbitrary number of strong parallel repetitions for the $N$-player $\mathrm{XOR}$ game. Equipped with the previously defined objects in this section, under strong parallel repetition a previous result on a suitable linear operator can be defined with a counterpart linear operation:

\bigskip

\noindent \textbf{Lemma} \textit{1-$\mathrm{XOR}$ strong parallel repetition } (\textit{computation of the Frobenius norm for the anticommutation rule of $T_{\mathrm{XOR} \wedge \cdots \wedge \mathrm{XOR}}$ yields a desired up to constants $\sqrt{\epsilon^{\wedge}}$ upper bound}, \textbf{Theorem} \textit{2}, {[44]}). Fix constants as specified in \textbf{Lemma} \textit{1}. One has that,

\begin{align*}
      \underline{\text{Player $1$}:} \text{ } \bigg| \bigg|   \bigg(  A_i \bigotimes \bigg( \underset{1 \leq k \leq n-1}{\bigotimes} \textbf{I}_k  \bigg)        \bigg) \bigg(  \mathscr{T} \wedge \cdots \wedge \mathscr{T} \bigg)  -  \bigg(  \mathscr{T} \wedge \cdots \wedge \mathscr{T} \bigg) \\ \times \bigg( \bigg( \underset{1 \leq k \leq n-1}{\bigotimes} \textbf{I}_k  \bigg)      \bigotimes \widetilde{A_i}      \bigg)      \bigg| \bigg|_F  < \big( c_1 \big)^{\wedge}  \big( n^N \big)^{\wedge}   \sqrt{\epsilon^{\wedge}}  \text{,} \\     \vdots \\              \underline{\text{Player $N$}:} \text{ } \bigg| \bigg|   \bigg(  \bigg( \underset{1 \leq k \leq n-1}{\bigotimes} \textbf{I}_k  \bigg)   \bigotimes  A^{(n-1)}_{i_1, \cdots, i_{n-1}}     \bigg) \bigg(  \mathscr{T} \wedge \cdots \wedge \mathscr{T} \bigg)  -  \bigg(  \mathscr{T} \wedge \cdots \wedge \mathscr{T} \bigg) \\ \times \bigg( \bigg(    \widetilde{A^{(n-1)}_{i_1, \cdots, i_{n-1}} }       \bigotimes   \bigg( \underset{1 \leq k \leq n-1}{\bigotimes} \textbf{I}_k  \bigg)     \bigg)      \bigg| \bigg|_F     < \big(  c_N \big)^{\wedge} \big(  n^N \big)^{\wedge} \sqrt{\epsilon^{\wedge}}   \text{, }
\end{align*}

\noindent has the upper bound,

\begin{align*}
      \mathscr{C} \equiv    \underset{1 \leq i \leq N}{\bigcup}  \big\{    \big(  C_i \big)^{\wedge}  \neq  \big( c_i \big)^{\wedge} \in \textbf{R}  :           \big(  C_i \big)^{\wedge} \equiv  \big( c_i  \big)^{\wedge} \sqrt{\epsilon^{\wedge}}     \big\}     \propto \big( n^N \big)^{\wedge} \sqrt{\epsilon^{\wedge}} \equiv \big( n^{\wedge} \big)^N \sqrt{\epsilon^{\wedge}} \text{. }
\end{align*}

\bigskip

\noindent Tensor observables from each player are drawn from the index set,

\begin{align*}
    \mathscr{I}  \equiv  \underset{\# \text{ Players } }{\bigcup}  \big\{   \text{indices } i \text{: }  \text{no } i \text{ are equal in Player } j\text{'s tensor observable}          \big\}       \text{. }       
\end{align*}

\noindent \textit{Proof of Lemma 1-$\mathrm{XOR}$ strong parallel repetition}. Denote,

\begin{align*}
  \ket{\psi^{\prime}} \equiv \ket{\psi_{\mathrm{XOR} \wedge \cdots \wedge \mathrm{XOR}}}  \text{, } \\ \omega^{\prime} \equiv \omega_{\mathrm{XOR} \wedge \cdots \wedge \mathrm{XOR}}
\end{align*}

\noindent from which obtaining the desired collection of upper bounds for the Frobenius norm amounts to upper bounding the each Frobenius norm in the statement of the result above, implying,

\begin{align*}
 \bigg| \bigg|    \bigg[             \bigg( \bigg( \underset{1 \leq i \leq n}{\prod}  A^{j_i}_i \bigg) \bigotimes \bigg( \underset{1 \leq k \leq n-1}{\bigotimes} \textbf{I}_k \bigg) \bigg)   -  \bigg(  \omega^{\prime} \bigg(         \pm \mathrm{sign} \big( i_1, \cdots, i_n \big)                \bigg( \underset{1 \leq i \leq n}{\prod}  A^{j_i}_i \bigg)           \bigg)  \\   \bigotimes \bigg( \underset{1 \leq k \leq n-1}{\bigotimes} \textbf{I}_k \bigg)\bigg)                       \bigg]    \ket{\psi^{\prime}}       \bigg| \bigg|_F    <    \bigg(  \big( n_1 \big)^{\wedge} +       \big(   \big(  n_1 \big)^{\wedge} + 2    \big) \big( \omega^{\prime} \big)^{-1}                \bigg)\big(  n^N \big)^{\wedge } \sqrt{\epsilon^{\wedge}}        \text{,}  \end{align*}

 \begin{align*} \bigg| \bigg|    \bigg[ \bigg(  \textbf{I} \bigotimes            \bigg( \underset{1 \leq i \leq n}{\prod}  A^{1,j_i}_i \bigg) \bigotimes \bigg( \underset{1 \leq k \leq n-2}{\bigotimes} \textbf{I}_k \bigg) \bigg)   -  \bigg(  \textbf{I} \bigotimes  \bigg( \omega_{N\mathrm{XOR}}  \bigg(         \pm \mathrm{sign} \big( i_1, j_1, \cdots \\ , i_n, \cdots, j_n \big)                \bigg( \underset{i_2 \in \mathcal{Q}_2 }{\underset{i_1 \in \mathcal{Q}_1}{\prod}}  A^{1,j_{i_1,i_2}}_{i_1,i_2} \bigg)           \bigg)  \bigg)  \bigotimes \bigg( \underset{1 \leq k \leq n-2}{\bigotimes} \textbf{I}_k \bigg)\bigg)                       \bigg]   \ket{\psi^{\prime}}    \bigg| \bigg|_F \\  <   \bigg( \big( n_2  \big)^{\wedge} +  \big( \big( n_2 \big)^{\wedge} + 2 \big) \big( \omega^{\prime} \big)^{-1}               \bigg) \big( n^N  \big)^{\wedge} \sqrt{\epsilon^{\wedge}}   \\   \vdots \\    \bigg| \bigg|    \bigg[             \bigg(\bigg( \underset{1 \leq k \leq n-1}{\bigotimes} \textbf{I}_k \bigg) \bigotimes   \bigg( \underset{1 \leq i \leq n}{\prod}  A^{(n-1),j_{i_1,\cdots,i_n}}_{i_1,\cdots, i_n} \bigg)  \bigg)   -  \bigg( \bigg( \underset{1 \leq k \leq n-1}{\bigotimes} \textbf{I}_k \bigg) \bigotimes \bigg(    \omega^{\prime} \\ \times  \bigg(         \pm \mathrm{sign} \big( i_1, \cdots, i_n, j_1, \cdots, j_n \big)                         \bigg)    \bigg( \underset{1 \leq i \leq n}{\prod}  A^{(n-1),j_{i_1,\cdots,i_n}}_{i_1,\cdots, i_n} \bigg)  \bigg)                 \bigg)   \bigg] \ket{\psi^{\prime}}     \bigg| \bigg|_F \\   <   \bigg(\big(  n_N  \big)^{\wedge }       +   \big( \big(  n_N \big)^{\wedge} + 2\big)  \big( \omega^{\prime} \big)^{-1}              \bigg) \big(  n^N \big)^{\wedge } \sqrt{\epsilon^{\wedge}}  \text{, }
\end{align*}

\noindent from which the desired sequence of upper bounds on the Frobenius norm of each of the $N$ players can be obtained from arguments provided in the $\mathrm{1-3-XOR}$ result in \textit{1.4.1}, from which we conclude the argument. \boxed{}

\bigskip

\noindent Under the operation of strong parallel repetition applied to two rounds of $\mathrm{FFL}$ games, one also encounters the following result which can be shown to hold with the computations above adapted to the $\mathrm{FFL}$ linear operator, in place of,

\begin{align*}
  \mathscr{T} \wedge \cdots \wedge \mathscr{T}   \equiv \bigwedge \mathscr{T} \text{, }
\end{align*}

\noindent which is stated below.

\bigskip

\noindent \textbf{Lemma} \textit{$\mathrm{FFL}$ strong parallel repetition} (\textit{computation of the Frobenius norm for the anticommutation rule of $T_{\mathrm{FFL} \wedge \mathrm{FFL}}$ yields a desired up to constants $\sqrt{\big( \epsilon^{\wedge} \big)_{\mathrm{FFL}}}$ upper bound}, \textbf{Theorem} \textit{2}, {[43]}). Fix constants as specified in \textbf{Lemma} \textit{1}. One has that,

\begin{align*}
      \underline{\text{Player $1$}:} \text{ } \bigg| \bigg|   \bigg(  A_i \otimes \textbf{I}      \bigg) \bigg(  
 \underset{1 \leq j \leq 2}{\bigwedge} T^{(j)}_{\mathrm{FFL}}\bigg)  - \bigg(  
 \underset{1 \leq j \leq 2}{\bigwedge} T^{(j)}_{\mathrm{FFL}}\bigg) \bigg( \textbf{I}    \otimes \widetilde{A_i}      \bigg)      \bigg| \bigg|_F  < \big( \big( c_1  \big)_{\mathrm{FFL}} \big)^{\wedge} \\ \times   \big(  \big( n^2  \big)_{\mathrm{FFL}} \big)^{\wedge}  \sqrt{\big( \epsilon^{\wedge} \big)_{\mathrm{FFL}} }  \text{,} \end{align*}

    \begin{align*}               \underline{\text{Player $2$}:} \text{ } \bigg| \bigg|   \bigg(   \textbf{I} \otimes  B_{i_1, i_2}     \bigg) \bigg(  
 \underset{1 \leq j \leq 2}{\bigwedge} T^{(j)}_{\mathrm{FFL}}\bigg)  -  \bigg(  
 \underset{1 \leq j \leq 2}{\bigwedge} T^{(j)}_{\mathrm{FFL}}\bigg) \bigg(     \widetilde{B_{i_1,i_2}}       \otimes \textbf{I}    \bigg)      \bigg| \bigg|_F    < \big( \big(  c_2  \big)_{\mathrm{FFL}} \big)^{\wedge}  \\ \times  \big( \big(  n^2 \big)_{\mathrm{FFL}} \big)^{\wedge}\sqrt{\epsilon^{\wedge}}   \text{, }
\end{align*}

\noindent has the upper bound,

\begin{align*}
      \mathscr{C}_{\mathrm{FFL}} \equiv    \underset{1 \leq i \leq N}{\bigcup}  \big\{    \big(  \big( C_i \big)_{\mathrm{FFL}} \big)^{\wedge}  \neq  \big(  \big( c_i \big)_{\mathrm{FFL}} 
  \big)^{\wedge} \in \textbf{R}  :           \big( \big(  C_i \big)_{\mathrm{FFL}}   \big)^{\wedge} \equiv  \big( \big( c_i  \big)_{\mathrm{FFL}}   \big)^{\wedge} \sqrt{\epsilon^{\wedge}}     \big\}  \\    \propto \big( \big(  n^N  \big)_{\mathrm{FFL}} 
 \big)^{\wedge} \sqrt{\epsilon^{\wedge}} \equiv \big( \big( n^{\wedge} \big)_{\mathrm{FFL}}  \big)^N \sqrt{\big( \epsilon^{\wedge} \big)_{\mathrm{FFL}} } \text{. }
\end{align*}

\bigskip

\noindent Tensor observables from each player are drawn from the index set,

\begin{align*}
    \mathscr{I}  \equiv  \underset{\# \text{ Players } }{\bigcup}  \big\{   \text{indices } i \text{: }  \text{no } i \text{ are equal in Player } j\text{'s tensor observable}          \big\}       \text{. }       
\end{align*}

\noindent \textit{Proof of Lemma 1-$\mathrm{FFL}$ strong parallel repetition}. Directly apply the computations in the previous result, from which we conclude the argument. \boxed{}

\bigskip

\noindent Under strong parallel repetition, one also expects that an inequality of the following form should hold:

\bigskip

\noindent \textbf{Lemma} \textit{5B} (\textit{strong parallel repetition of }$\sqrt{\epsilon^{\wedge}}$- \textit{FFL approximality}, \textbf{Lemma} \textit{8}, {[44]}). From the same quantities introduced in the previous result, one has,

\begin{align*}
   \bigg| \bigg|               \bigg( \big(  A_k  \wedge A_{k^{\prime}} \big) \otimes \textbf{I} \bigg) \ket{\psi_{\mathrm{FFL} \wedge \mathrm{FFL}}}    -  \bigg( \textbf{I} \otimes \bigg(     \frac{\pm \big(  B_{kl} \wedge B_{k^{\prime} l^{\prime}} \big)  + \big( B_{lk} \wedge B_{k^{\prime} l^{\prime}} \big) }{\big| \pm \big(  B_{kl} \wedge B_{k^{\prime} l^{\prime}} \big)  + \big( B_{lk} \wedge B_{k^{\prime} l^{\prime}} \big) \big| }           \bigg) \bigg)  \\ \times  \ket{\psi_{\mathrm{FFL} \wedge \mathrm{FFL}}}              \bigg| \bigg|    < 20 \sqrt{N \epsilon^{\wedge}}     \text{. } 
\end{align*}

\bigskip

\noindent \textit{Proof of Lemma 5B}. From previous arguments used in \textbf{Lemma} \textit{4}, observe from the two-player setting the proof, with upper bound $17 \sqrt{n \epsilon}$, crucially,

\begin{align*}
          \frac{\bigg| \textbf{I} \otimes \bigg( \frac{\pm B_{kl} + B_{lk}}{\sqrt{2}} \bigg)     \bigg| }{\bigg| \textbf{I} \otimes \bigg( \frac{\pm B_{kl} + B_{lk}}{\big| \pm B_{kl } + B_{lk} \big| }   \bigg)    \bigg| }    \equiv \frac{\bigg|   \bigg[ \textbf{I} \otimes \bigg( \frac{\pm B_{kl} + B_{lk}}{\sqrt{2}} \bigg)    \bigg]  \bigg|   \frac{\big| \pm B_{kl} + B_{lk} \big|}{\big| \pm B_{kl} + B_{lk} \big|}   }{\bigg|   \textbf{I} \otimes \bigg( \frac{\pm B_{kl} + B_{lk}}{\big| \pm B_{kl } + B_{lk} \big| }  \bigg)          \bigg| }  \approx          \frac{\big| \pm B_{kl} + B_{lk} \big| }{\sqrt{2}}      \approx 1    \text{, } 
\end{align*}

\noindent from which, from the computations in the $N$-player setting from the expression,

\begin{align*}
    \underset{\text{Questions }k,l}{\sum}    \frac{\bigg| \textbf{I} \bigotimes \bigg( \frac{\pm B_{kl} + B_{lk}}{\sqrt{2}} \bigg)  \bigotimes \bigg( \underset{1 \leq k^{\prime} \leq N-2}{\bigotimes} \textbf{I}_{k^{\prime}} \bigg)    \bigg| }{ \bigg| \textbf{I} \bigotimes \bigg( \frac{\pm B_{kl} + B_{lk}}{\big| \pm B_{kl} + B_{lk} \big| } \bigg)  \bigotimes \bigg( \underset{1 \leq k^{\prime} \leq N-2}{\bigotimes} \textbf{I}_{k^{\prime}} \bigg)    \bigg| } \\ \equiv    \underset{\text{Questions }k,l}{\sum}    \frac{\bigg| \textbf{I} \bigotimes \bigg( \frac{\pm B_{kl} + B_{lk}}{\sqrt{2}} \bigg)  \bigotimes \bigg( \underset{1 \leq k^{\prime} \leq N-2}{\bigotimes} \textbf{I}_{k^{\prime}} \bigg)    \bigg| }{ \bigg| \textbf{I} \bigotimes \bigg( \frac{\pm B_{kl} + B_{lk}}{\big| \pm B_{kl} + B_{lk} \big| } \bigg)  \bigotimes \bigg( \underset{1 \leq k^{\prime} \leq N-2}{\bigotimes} \textbf{I}_{k^{\prime}} \bigg)    \bigg| }      \bigg[   \frac{\big| \pm B_{kl} + B_{lk} \big|}{\big| \pm B_{kl} + B_{lk} \big|}   \bigg] \\  \approx    \underset{\text{Questions }k,l}{\sum}  \bigg[   \frac{\big| \pm B_{kl} + B_{lk} \big|}{\sqrt{2}} \bigg]    \text{, }
\end{align*}

\noindent and from strong parallel repetition in the $\mathrm{FFL} \wedge \mathrm{FFL}$ setting,

\begin{align*}
    \underset{\text{Questions }k,l,k^{\prime},l^{\prime}}{\sum}    \frac{\bigg| \textbf{I} \bigotimes \bigg( \frac{\pm\big( B_{kl} \wedge B_{k^{\prime} l^{\prime}} \big)  + \big( B_{lk} \wedge B_{l^{\prime} k^{\prime}} \big) }{\sqrt{2}} \bigg) \bigg|  }{ \bigg| \textbf{I} \bigotimes \bigg( \frac{\pm\big( B_{kl} \wedge B_{k^{\prime} l^{\prime}} \big)  + \big( B_{lk} \wedge B_{l^{\prime} k^{\prime}} \big)}{\big|  \pm\big( B_{kl} \wedge B_{k^{\prime} l^{\prime}} \big)  + \big( B_{lk} \wedge B_{l^{\prime} k^{\prime}} \big) \big| } \bigg)    \bigg| } \\   \equiv    \underset{\text{Questions }k,l}{\sum}    \frac{\bigg| \textbf{I} \bigotimes \bigg( \frac{\pm\big( B_{kl} \wedge B_{k^{\prime} l^{\prime}} \big)  + \big( B_{lk} \wedge B_{l^{\prime} k^{\prime}} \big)}{\sqrt{2}} \bigg)    \bigg| }{ \bigg| \textbf{I} \bigotimes \bigg( \frac{\pm\big( B_{kl} \wedge B_{k^{\prime} l^{\prime}} \big)  + \big( B_{lk} \wedge B_{l^{\prime} k^{\prime}} \big)}{\big| \pm\big( B_{kl} \wedge B_{k^{\prime} l^{\prime}} \big)  + \big( B_{lk} \wedge B_{l^{\prime} k^{\prime}} \big) \big| } \bigg)  \bigg| }      \bigg[   \frac{\big( B_{kl} \wedge B_{k^{\prime} l^{\prime}} \big)  + \big( B_{lk} \wedge B_{l^{\prime} k^{\prime}} \big) }{\big( B_{kl} \wedge B_{k^{\prime} l^{\prime}} \big)  + \big( B_{lk} \wedge B_{l^{\prime} k^{\prime}} \big) }   \bigg]  \\ \approx    \underset{\text{Questions }k,l,k^{\prime},l^{\prime}}{\sum}  \bigg[   \frac{\big| \pm \big( B_{kl} \wedge B_{k^{\prime} l^{\prime}} \big)  + \big( B_{lk} \wedge B_{l^{\prime} k^{\prime}} \big)  \big|}{\sqrt{2}} \bigg]    \text{, }
\end{align*}

\noindent Hence, the numerator of the previous expression,

\begin{align*}
 \big| \pm \big( B_{kl} \wedge B_{k^{\prime} l^{\prime}} \big)  + \big( B_{lk} \wedge B_{l^{\prime} k^{\prime}} \big)  \big|       \text{, }
\end{align*}

\noindent can be used to obtain the desired upper bound, from upper bounding the product norm,

\begin{align*}
           \bigg| \bigg|  \bigg[        \bigg( \bigg( \big( A_k \wedge A_{k^{\prime}} \big)  \otimes \textbf{I} \bigg) + \bigg( \textbf{I} \otimes   \bigg(   \frac{\pm \big( B_{kl} \wedge B_{k^{\prime}l^{\prime}} \big) + \big( B_{lk} \wedge B_{l^{\prime}k^{\prime}}\big) }{\sqrt{2}}    \bigg) \bigg)  \bigg(   \bigg( \big( A_k \wedge A_{k^{\prime}} \big)   \otimes \textbf{I} \bigg) \\ - \bigg( \textbf{I} \otimes  \bigg( \frac{\pm \big( B_{kl} \wedge B_{k^{\prime}l^{\prime}} \big) + \big( B_{lk} \wedge B_{l^{\prime}k^{\prime}}\big) }{\sqrt{2}} \bigg)        \bigg)     \bigg]  \ket{\psi_{\mathrm{FFL}\wedge\mathrm{FFL}}}                    \bigg| \bigg|    <    \bigg| \bigg|  \bigg[        \bigg( \bigg( \big(  A_k \wedge A_{k^{\prime}} \big)  \otimes \textbf{I} \bigg) \\ + \bigg( \textbf{I}  \otimes   \bigg(   \frac{\pm \big( B_{kl} \wedge B_{k^{\prime}l^{\prime}} \big) + \big( B_{lk} \wedge B_{l^{\prime}k^{\prime}}\big)}{\sqrt{2}}    \bigg) \bigg) \bigg|  \bigg|  
        \bigg| \bigg|  \bigg(   \bigg( \big(  A_k \wedge A_{k^{\prime}} \big)  \otimes \textbf{I} \bigg) \\  - \bigg( \textbf{I} \otimes \bigg( \frac{\pm \big( B_{kl} \wedge B_{k^{\prime}l^{\prime}} \big) + \big( B_{lk} \wedge B_{l^{\prime}k^{\prime}}\big)}{\sqrt{2}} \bigg)        \bigg)     \bigg] \ket{\psi_{\mathrm{FFL}\wedge\mathrm{FFL}}}                    \bigg| \bigg|     \text{, }
\end{align*}

\noindent from which we conclude the argument, as the upper bound to the above expression takes the form,

\begin{align*}
         3 \big( \omega_{\mathrm{FFL} \wedge \mathrm{FFL}} \big)^{-2} \big( 1 + \big( \omega_{\mathrm{FFL} \wedge \mathrm{FFL}} \big)^{-2} \big) \sqrt{n\epsilon^{\wedge}}     \equiv 3 \big( \frac{3}{2} \big) \big( 1 + \frac{3}{2} \big) \sqrt{n \epsilon^{\wedge}} < 20 \sqrt{N \epsilon^{\wedge}}             \text{. } \boxed{}
\end{align*}

\noindent Under strong parallel repetition, straightforwardly one expects that an inequality of the following form should hold:

\bigskip

\noindent \textbf{Lemma} $\textit{5}^{*}\textit{B}$ (\textit{an arbitrary number of strong parallel repetition applications of }$\sqrt{\epsilon^{\wedge}_{2\mathrm{XOR}}}$- \textit{2 XOR approximality}, \textbf{Lemma} \textit{8}, {[44]}). From the same quantities introduced in the previous result, one has,

\begin{align*}
   \bigg| \bigg|               \bigg( \big(  A_k  \wedge A_{k^{\prime}} \wedge \cdots \wedge A_{k^{\prime\cdots \prime}} \big) \otimes \textbf{I} \bigg) \ket{\psi_{2\mathrm{XOR} \wedge \cdots \wedge 2\mathrm{XOR}}}    \\ -  \bigg( \textbf{I} \otimes \bigg(     \frac{\pm \big(  B_{kl} \wedge B_{k^{\prime} l^{\prime}} \wedge \cdots \wedge B_{k^{\prime\cdots\prime}l^{\prime\cdots \prime}}  \big)  + \big( B_{lk} \wedge B_{ l^{\prime} k^{\prime}} \wedge \cdots \wedge B_{l^{\prime\cdots \prime} k^{\prime\cdots\prime} }  \big) }{\big| \pm \big(  B_{kl} \wedge B_{k^{\prime} l^{\prime}} \wedge \cdots \wedge B_{k^{\prime\cdots\prime}l^{\prime\cdots \prime}}  \big)  + \big( B_{lk} \wedge B_{ l^{\prime} k^{\prime}} \wedge \cdots \wedge B_{l^{\prime\cdots \prime} k^{\prime\cdots\prime} }  \big) \big| }           \bigg) \bigg) \\ \times  \ket{\psi_{2\mathrm{XOR} \wedge \cdots \wedge 2\mathrm{XOR}}}              \bigg| \bigg|    < 18 \sqrt{N \epsilon^{\wedge}_{2\mathrm{XOR}}}   \text{. } 
\end{align*}

\noindent \textit{Proof of Lemma $5^{*}$B}. Under an arbitrary number of the strong parallel repetition operation, the approximation included in the arguments for the previous result take the form,

\begin{align*}
    \underset{\text{Questions }k,l,k^{\prime},l^{\prime}, \cdots, k^{\prime\cdots\prime}, l^{\prime\cdots \prime}}{\sum}    \frac{\bigg| \textbf{I} \bigotimes \bigg( \frac{\pm\big( B_{kl} \wedge B_{k^{\prime} l^{\prime}} \wedge \cdots \wedge B_{k^{\prime\cdots\prime}l^{\prime\cdots \prime}} \big)  + \big( B_{lk} \wedge B_{l^{\prime} k^{\prime}} \wedge \cdots \wedge B_{k^{\prime\cdots\prime}l^{\prime\cdots \prime}}  \big) }{\sqrt{2}} \bigg)   \bigg| }{ \bigg| \textbf{I} \bigotimes \bigg( \frac{\pm\big( B_{kl} \wedge B_{k^{\prime} l^{\prime}} \wedge \cdots \wedge B_{k^{\prime\cdots\prime}l^{\prime\cdots \prime}} \big)  + \big( B_{lk} \wedge B_{l^{\prime} k^{\prime}} \wedge \cdots \wedge B_{k^{\prime\cdots\prime}l^{\prime\cdots \prime}}  \big)}{\big| \pm \big( B_{kl} \wedge B_{k^{\prime} l^{\prime}} \wedge \cdots \wedge B_{k^{\prime\cdots\prime}l^{\prime\cdots \prime}} \big)  + \big( B_{lk} \wedge B_{l^{\prime} k^{\prime}} \wedge \cdots \wedge B_{k^{\prime\cdots\prime}l^{\prime\cdots \prime}}  \big)\big| } \bigg)   \bigg| } \\  \equiv     \underset{\text{Questions }k,l,k^{\prime},l^{\prime}, \cdots, k^{\prime\cdots\prime}, l^{\prime\cdots \prime}}{\sum}     \frac{\bigg| \textbf{I} \bigotimes \bigg( \frac{\pm\big( B_{kl} \wedge B_{k^{\prime} l^{\prime}} \wedge \cdots \wedge B_{k^{\prime\cdots\prime}l^{\prime\cdots \prime}} \big)  + \big( B_{lk} \wedge B_{l^{\prime} k^{\prime}} \wedge \cdots \wedge B_{k^{\prime\cdots\prime}l^{\prime\cdots \prime}}  \big)}{\sqrt{2}} \bigg) \bigg|  }{ \bigg| \textbf{I} \bigotimes \bigg( \frac{\pm\big( B_{kl} \wedge B_{k^{\prime} l^{\prime}} \wedge \cdots \wedge B_{k^{\prime\cdots\prime}l^{\prime\cdots \prime}} \big)  + \big( B_{lk} \wedge B_{l^{\prime} k^{\prime}} \wedge \cdots \wedge B_{k^{\prime\cdots\prime}l^{\prime\cdots \prime}}  \big)}{\big| \pm\big( B_{kl} \wedge B_{k^{\prime} l^{\prime}} \wedge \cdots \wedge B_{k^{\prime\cdots\prime}l^{\prime\cdots \prime}} \big)  + \big( B_{lk} \wedge B_{l^{\prime} k^{\prime}} \wedge \cdots \wedge B_{k^{\prime\cdots\prime}l^{\prime\cdots \prime}}  \big) \big| } \bigg) \bigg|  }   \\ \times   \bigg[   \frac{\big( B_{kl} \wedge B_{k^{\prime} l^{\prime}}  \wedge \cdots \wedge B_{k^{\prime\cdots\prime} l^{\prime\cdots\prime}} \big)  + \big( B_{lk} \wedge B_{l^{\prime} k^{\prime}} \wedge \cdots \wedge B_{l^{\prime\cdots\prime} k^{\prime\cdots\prime}} \big) }{\big( B_{kl} \wedge B_{k^{\prime} l^{\prime}}  \wedge \cdots \wedge B_{k^{\prime\cdots\prime} l^{\prime\cdots\prime}} \big)  + \big(B_{lk} \wedge B_{l^{\prime} k^{\prime}} \wedge \cdots \wedge B_{l^{\prime\cdots\prime} k^{\prime\cdots\prime}}  \big) }   \bigg]  \\ \approx  \underset{\text{Questions }k,l,k^{\prime},l^{\prime}, \cdots, k^{\prime\cdots\prime}, l^{\prime\cdots \prime}}{\sum}   \bigg[   \frac{\big| \pm \big(  B_{kl} \wedge B_{k^{\prime} l^{\prime}} \wedge \cdots \wedge B_{k^{\prime\cdots\prime}l^{\prime\cdots \prime}}  \big)  + \big(  B_{lk} \wedge B_{ l^{\prime} k^{\prime}} \wedge \cdots \wedge B_{l^{\prime\cdots \prime} k^{\prime\cdots\prime} }   \big)  \big|}{\sqrt{2}} \bigg]    \text{. }
\end{align*}

\noindent Hence, following the same computations provided in the previous result for strong parallel repetition of the $\mathrm{FFL}$ game implies that the upper bound takes the form,

\begin{align*}
 3 \big( \omega_{\mathrm{XOR}\wedge\mathrm{XOR}} \big)^{-2} \big(   1 + \big( \omega_{\mathrm{XOR}\wedge\mathrm{XOR}} \big)^{-2}        \big)     \sqrt{n \epsilon^{\wedge}_{2\mathrm{XOR}}}  \equiv 18  \sqrt{N \epsilon^{\wedge}_{2\mathrm{XOR}}}   \text{, }
\end{align*}

\noindent from which we conclude the argument, upon obtaining the desired upper bound. \boxed{}

\bigskip

\noindent Given the normalization,

\begin{align*}
   \bigg| \pm \underset{\sigma_1 \in S_3}{\sum} C_{\sigma(ijk)} +  \underset{\sigma_2, \sigma_3, \sigma_4, \sigma_5, \sigma_6  \in S_3}{\sum} C_{\sigma(ijk)}  \bigg|  \text{, }
\end{align*}

\noindent appearing in the summation of permutation of tensors of the third player, $C$,

\begin{align*}
  \frac{  \pm \underset{\sigma_1 \in S_3}{\sum} C_{\sigma(ijk)} +  \underset{\sigma_2, \sigma_3, \sigma_4, \sigma_5, \sigma_6  \in S_3}{\sum} C_{\sigma(ijk)}  }{  \bigg| \pm \underset{\sigma_1 \in S_3}{\sum} C_{\sigma(ijk)} +  \underset{\sigma_2, \sigma_3, \sigma_4, \sigma_5, \sigma_6  \in S_3}{\sum} C_{\sigma(ijk)}  \bigg|  }  \text{, }
\end{align*}

\noindent one can generalize the inequality provided in the above result under strong parallel repetition, to the 3-player setting.

\bigskip

\noindent \textbf{Lemma} $\textit{5}^{*}B$ (\textit{an arbitrary number of strong parallel repetition applications of }$\sqrt{\epsilon^{\wedge}_{3\mathrm{XOR}}}$- \textit{3 XOR approximality}, \textbf{Lemma} \textit{8}, {[44]}). From the same quantities introduced in the previous result, one has,

\begin{align*}
   \bigg| \bigg|               \bigg( \big(  A_k  \wedge A_{k^{\prime}} \wedge \cdots \wedge A_{k^{\prime\cdots \prime}} \big) \otimes \textbf{I} \otimes \textbf{I} \bigg) \\ \times \ket{\psi_{3\mathrm{XOR} \wedge \cdots \wedge 2\mathrm{XOR}}}   \\   -  \bigg( \textbf{I} \otimes \textbf{I}  \otimes \bigg(   \frac{  \pm \underset{\sigma_1 \in S_3}{\sum} \big(  \wedge  C_{\sigma(ijk)} \big)  +  \underset{\sigma_2, \sigma_3, \sigma_4, \sigma_5, \sigma_6  \in S_3}{\sum} \big( \wedge  C_{\sigma(ijk)} \big)  }{  \bigg| \pm \underset{\sigma_1 \in S_3}{\sum} \big( \wedge C_{\sigma(ijk)} \big)  +  \underset{\sigma_2, \sigma_3, \sigma_4, \sigma_5, \sigma_6  \in S_3}{\sum} \big( \wedge C_{\sigma(ijk)}\big)   \bigg|  } \bigg)   \bigg) \ket{\psi_{3\mathrm{XOR} \wedge \cdots \wedge 3\mathrm{XOR}}}              \bigg| \bigg|  \\  < 18 \sqrt{N \epsilon^{\wedge}_{3\mathrm{XOR}}}   \text{, } 
\end{align*}

\noindent where,

\begin{align*}
 \wedge C_{\sigma(ijk)} \equiv  C_{\sigma(ijk)}  \wedge \cdots \wedge C_{\sigma(ijk)}  \text{. }
\end{align*}

\noindent \textit{Proof of Lemma $5^{*}B$}. Directly apply the computations from the previous result, from which we conclude the argument. \boxed{}

\bigskip

\noindent Several results, such as the form of the previous two items above, can be used to generalize, and further characterize, the structure of error bounds for many games.

\bigskip

\noindent \textbf{Lemma} $\textit{5}\textit{B}^{**}$  (\textit{an arbitrary number of strong parallel repetition applications of }$\sqrt{\epsilon^{\wedge}_{{N\mathrm{XOR} \wedge \cdots \wedge N\mathrm{XOR}}}}$- \textit{N XOR approximality}, \textbf{Lemma} \textit{8}, {[44]}). From the same quantities introduced in the previous result, one has, for the $\mathrm{XOR}$ game under an arbitrary number of strong parallel repetitions, that the quantities,

\begin{align*}
  \mathcal{I}_1 \equiv  \bigg| \bigg|               \bigg( \big(  A_k  \wedge A_{k^{\prime}} \wedge \cdots \wedge A_{k^{\prime\cdots \prime}} \big) \bigotimes  \bigg( \underset{1 \leq z \leq N-1}{\bigotimes}\textbf{I}_z \bigg)  \bigg) \ket{\psi_{N\mathrm{XOR} \wedge \cdots \wedge N\mathrm{XOR}}} \\   -  \bigg( \textbf{I}  \otimes \bigg(     \frac{\pm \big(  B_{kl} \wedge B_{k^{\prime} l^{\prime}} \wedge \cdots \wedge B_{k^{\prime\cdots\prime}l^{\prime\cdots \prime}}  \big)  + \big( B_{lk} \wedge B_{ l^{\prime} k^{\prime}} \wedge \cdots \wedge B_{l^{\prime\cdots \prime} k^{\prime\cdots\prime} }  \big) }{\big| \pm \big(  B_{kl} \wedge B_{k^{\prime} l^{\prime}} \wedge \cdots \wedge B_{k^{\prime\cdots\prime}l^{\prime\cdots \prime}}  \big)  + \big( B_{lk} \wedge B_{ l^{\prime} k^{\prime}} \wedge \cdots \wedge B_{l^{\prime\cdots \prime} k^{\prime\cdots\prime} }  \big) \big| }    \bigg)  \\ \bigotimes  \bigg( \underset{1 \leq z \leq N-2}{\bigotimes}\textbf{I}_z \bigg)       \bigg) \ket{\psi_{N\mathrm{XOR} \wedge \cdots \wedge N\mathrm{XOR}}}              \bigg| \bigg| \\ \vdots \end{align*}

  \begin{align*} \mathcal{I}_N \equiv   \bigg| \bigg|    \bigg(   \bigg( \underset{1 \leq z \leq N-2}{\bigotimes} \textbf{I}_z \bigg) \bigotimes   \frac{1}{\sqrt{\# \sigma^{\prime} }}  \bigg(    \underset{\text{Permutations } \sigma^{\prime}}{\sum}  \big(   B^{(N-1)}_{\sigma^{\prime} ( i_1, \cdots, i_{N-1})}         \wedge   B^{(N-1)}_{\sigma^{\prime} ( i^{\prime}_1, \cdots, i^{\prime}_{N-1})}  \wedge   \cdots \\ \wedge   B^{(N-1)}_{\sigma^{\prime} ( i^{\prime\cdots\prime}_1, \cdots, i^{\prime\cdots\prime}_{N-1})} \big)      \bigg)     \bigotimes \textbf{I}  \bigg)   \ket{\psi_{N\mathrm{XOR} \wedge \cdots \wedge N\mathrm{XOR}}}  \\ -  \bigg( \bigg( \underset{1 \leq z \leq N-1}{\bigotimes} \textbf{I}_z \bigg) \bigotimes \frac{1}{\sqrt{\# \sigma }}  \bigg(    \underset{\text{Permutations } \sigma}{\sum}  \big(   B^{(N-1)}_{\sigma ( i_1, \cdots, i_{N-1})}         \wedge   B^{(N-1)}_{\sigma ( i^{\prime}_1, \cdots, i^{\prime}_{N-1})}  \wedge  \cdots \\  \wedge   B^{(N-1)}_{\sigma ( i^{\prime\cdots\prime}_1, \cdots, i^{\prime\cdots\prime}_{N-1})} \big)      \bigg) \bigg) \ket{\psi_{N\mathrm{XOR} \wedge \cdots \wedge N\mathrm{XOR}}}           \bigg| \bigg|      \text{, } 
\end{align*}

\noindent have the strict upper bound,

\begin{align*}
\underset{1 \leq j \leq N}{\sum} \mathcal{I}_j < 20 N \sqrt{N \epsilon^{\wedge}_{{N\mathrm{XOR} \wedge \cdots \wedge N\mathrm{XOR}}}} \text{,}
 \end{align*}

 \noindent where the tensors beyond that of the second player, $B$, are indexed as,

 \begin{align*}
     \textbf{I} \bigotimes B_{\sigma ( i_1,i_2)} \bigotimes \bigg( \underset{1 \leq z \leq N-2}{\bigotimes} \textbf{I}_z   \bigg) \equiv  \textbf{I} \bigotimes B^{(1)}_{\sigma ( i_1,i_2)} \bigotimes \bigg( \underset{1 \leq z \leq N-2}{\bigotimes} \textbf{I}_z   \bigg)   \text{, } \\ \vdots \\    \bigg( \underset{1 \leq z \leq N-1}{\bigotimes} \textbf{I}_z   \bigg) \bigotimes B_{\sigma ( i_1,i_2,\cdots, i_{n-2} )}  \equiv   \bigg( \underset{1 \leq z \leq N-1}{\bigotimes} \textbf{I}_z   \bigg) \bigotimes B^{(n-2)}_{\sigma ( i_1,i_2,\cdots, i_{n-2} )}  \text{. }
 \end{align*}

\noindent \textit{Proof of Lemma $\textit{5}\textit{B}^{**}$}. Given observations in the computations for the previous result, in the $N$-player setting,

\begin{align*}
    \underset{\text{Questions }k,l}{\sum}    \frac{\bigg| \textbf{I} \bigotimes \bigg( \frac{\pm B_{kl} + B_{lk}}{\sqrt{2}} \bigg)  \bigotimes \bigg( \underset{1 \leq k^{\prime} \leq N-2}{\bigotimes} \textbf{I}_{k^{\prime}} \bigg)    \bigg| }{ \bigg| \textbf{I} \bigotimes \bigg( \frac{\pm B_{kl} + B_{lk}}{\big| \pm B_{kl} + B_{lk} \big| } \bigg)  \bigotimes \bigg( \underset{1 \leq k^{\prime} \leq N-2}{\bigotimes} \textbf{I}_{k^{\prime}} \bigg)    \bigg| } \\ \equiv    \underset{\text{Questions }k,l}{\sum}    \frac{\bigg| \textbf{I} \bigotimes \bigg( \frac{\pm B_{kl} + B_{lk}}{\sqrt{2}} \bigg)  \bigotimes \bigg( \underset{1 \leq k^{\prime} \leq N-2}{\bigotimes} \textbf{I}_{k^{\prime}} \bigg)    \bigg| }{ \bigg| \textbf{I} \bigotimes \bigg( \frac{\pm B_{kl} + B_{lk}}{\big| \pm B_{kl} + B_{lk} \big| } \bigg)  \bigotimes \bigg( \underset{1 \leq k^{\prime} \leq N-2}{\bigotimes} \textbf{I}_{k^{\prime}} \bigg)    \bigg| }     \bigg[   \frac{\big| \pm B_{kl} + B_{lk} \big|}{\big| \pm B_{kl} + B_{lk} \big|}   \bigg] \\  \approx    \underset{\text{Questions }k,l}{\sum}  \bigg[   \frac{\big| \pm B_{kl} + B_{lk} \big|}{\sqrt{2}} \bigg]    \text{, }
\end{align*}

\noindent the desired upper bound,

\begin{align*}
  20 N \sqrt{N \epsilon^{\wedge}_{{N\mathrm{XOR} \wedge \cdots \wedge N\mathrm{XOR}}}}  \text{, }
\end{align*}

\noindent follows from the previous arguments, in which $\mathcal{I}_1, \cdots, \mathcal{I}_N$ can each be individually upper bounded by,

\begin{align*}
    20 \sqrt{N \epsilon^{\wedge}_{{N\mathrm{XOR} \wedge \cdots \wedge N\mathrm{XOR}}}} \text{, }
\end{align*}

\noindent corresponding to the maximum contribution that \textit{each player} can contribute to the upper bound with a factor of $20N$, after multiplying the following collection of suitable factors,

\begin{align*}
 \mathcal{I}^{\prime}_1 \equiv \bigg|   \frac{\big( B_{kl} \wedge B_{k^{\prime} l^{\prime}} \wedge \cdots \wedge B_{k^{\prime\cdots\prime}l^{\prime\cdots\prime}} \big)  + \big( B_{lk} \wedge B_{l^{\prime} k^{\prime}} \wedge \cdots \wedge B_{l^{\prime\cdots\prime}k^{\prime\cdots\prime}} \big) }{ \pm \big( B_{kl} \wedge B_{k^{\prime} l^{\prime}} \wedge \cdots \wedge B_{k^{\prime\cdots\prime}l^{\prime\cdots\prime}}\big)  + \big( B_{lk} \wedge B_{l^{\prime} k^{\prime}} \wedge \cdots \wedge B_{l^{\prime\cdots\prime}k^{\prime\cdots\prime}}   \big)  } \bigg|    \\    \equiv      \frac{\bigg| \big( B_{kl} \wedge B_{k^{\prime} l^{\prime}} \wedge \cdots \wedge B_{k^{\prime\cdots\prime}l^{\prime\cdots\prime}} \big)  + \big( B_{lk} \wedge B_{l^{\prime} k^{\prime}} \wedge \cdots \wedge B_{l^{\prime\cdots\prime}k^{\prime\cdots\prime}} \big)\bigg| }{ \bigg|  \pm \big( B_{kl} \wedge B_{k^{\prime} l^{\prime}} \wedge \cdots \wedge B_{k^{\prime\cdots\prime}l^{\prime\cdots\prime}}\big)  + \big( B_{lk} \wedge B_{l^{\prime} k^{\prime}} \wedge \cdots \wedge B_{l^{\prime\cdots\prime}k^{\prime\cdots\prime}}   \big) \bigg| }      \\  \Bigg\Updownarrow  \\   \bigg|  \pm \big( B_{kl} \wedge B_{k^{\prime} l^{\prime}} \wedge \cdots \wedge B_{k^{\prime\cdots\prime}l^{\prime\cdots\prime}}\big)  + \big( B_{lk} \wedge B_{l^{\prime} k^{\prime}} \wedge \cdots \wedge B_{l^{\prime\cdots\prime}k^{\prime\cdots\prime}}   \big) \bigg| \\  \neq 0   \text{, } \\ \vdots \\  \mathcal{I}^{\prime}_N \equiv               \bigg|     \frac{ \mathcal{I}^1_N  + \mathcal{I}^2_N}{ \pm \mathcal{I}^1_N + \mathcal{I}^2_N   }  \bigg| \equiv               \frac{  \big|  \mathcal{I}^1_N  + \mathcal{I}^2_N  \big|  }{  \big|  \pm \mathcal{I}^1_N + \mathcal{I}^2_N   \big|   }      \Longleftrightarrow   \big|  \pm \mathcal{I}^1_N + \mathcal{I}^2_N   \big|   \neq 0       \text{, }
\end{align*}

\noindent where, in the last expression,

\begin{align*}
   \mathcal{I}^1_N \equiv  \frac{1}{\sqrt{\# \sigma }}  \bigg(    \underset{\text{Permutations } \sigma}{\sum}  \big(   B^{(N-1)}_{\sigma ( i_1, \cdots, i_{N-1})}         \wedge   B^{(N-1)}_{\sigma ( i^{\prime}_1, \cdots, i^{\prime}_{N-1})}  \wedge  \cdots \wedge   B^{(N-1)}_{\sigma ( i^{\prime\cdots\prime}_1, \cdots, i^{\prime\cdots\prime}_{N-1})} \big)      \bigg) \text{, } \\    \mathcal{I}^2_N \equiv \frac{1}{\sqrt{\# \sigma^{\prime} }}  \bigg(    \underset{\text{Permutations } \sigma^{\prime}}{\sum}  \big(   B^{(N-1)}_{\sigma^{\prime} ( i_1, \cdots, i_{N-1})}         \wedge   B^{(N-1)}_{\sigma^{\prime} ( i^{\prime}_1, \cdots, i^{\prime}_{N-1})}  \wedge  \cdots \wedge   B^{(N-1)}_{\sigma^{\prime} ( i^{\prime\cdots\prime}_1, \cdots, i^{\prime\cdots\prime}_{N-1})} \big)      \bigg)
\end{align*}

\noindent for upper bounding each $\mathcal{I}_j$, for $1 \leq j \leq N$, provided in the statement of the result,

\begin{align*}
  \bigg| \bigg|               \bigg( \big(  A_k  \wedge A_{k^{\prime}} \wedge \cdots \wedge A_{k^{\prime\cdots \prime}} \big) \bigotimes  \bigg( \underset{1 \leq z \leq N-1}{\bigotimes}\textbf{I}_z \bigg)  \bigg) \\ \times \ket{\psi_{N\mathrm{XOR} \wedge \cdots \wedge N\mathrm{XOR}}}    \\  -  \bigg( \textbf{I}  \bigotimes \bigg(     \frac{\pm \big(  B_{kl} \wedge B_{k^{\prime} l^{\prime}} \wedge \cdots \wedge B_{k^{\prime\cdots\prime}l^{\prime\cdots \prime}}  \big)  + \big( B_{lk} \wedge B_{ l^{\prime} k^{\prime}} \wedge \cdots \wedge B_{l^{\prime\cdots \prime} k^{\prime\cdots\prime} }  \big) }{\big| \pm \big(  B_{kl} \wedge B_{k^{\prime} l^{\prime}} \wedge \cdots \wedge B_{k^{\prime\cdots\prime}l^{\prime\cdots \prime}}  \big)  + \big( B_{lk} \wedge B_{ l^{\prime} k^{\prime}} \wedge \cdots \wedge B_{l^{\prime\cdots \prime} k^{\prime\cdots\prime} }  \big) \big| }    \bigg) \\  \bigotimes  \bigg( \underset{1 \leq z \leq N-2}{\bigotimes}\textbf{I}_z \bigg)       \bigg)  \ket{\psi_{N\mathrm{XOR} \wedge \cdots \wedge N\mathrm{XOR}}}              \bigg| \bigg| \\  \vdots  \end{align*}

  \begin{align*}  \bigg| \bigg|    \bigg(   \bigg( \underset{1 \leq z \leq N-2}{\bigotimes} \textbf{I}_z \bigg) \bigotimes   \frac{1}{\sqrt{\# \sigma^{\prime} }}  \bigg(    \underset{\text{Permutations } \sigma^{\prime}}{\sum}  \big(   B^{(N-1)}_{\sigma^{\prime} ( i_1, \cdots, i_{N-1})}         \wedge   B^{(N-1)}_{\sigma^{\prime} ( i^{\prime}_1, \cdots, i^{\prime}_{N-1})}  \wedge  \cdots  \\ \wedge   B^{(N-1)}_{\sigma^{\prime} ( i^{\prime\cdots\prime}_1, \cdots, i^{\prime\cdots\prime}_{N-1})} \big)      \bigg)     \bigotimes \textbf{I}  \bigg)\ket{\psi_{N\mathrm{XOR} \wedge \cdots \wedge N\mathrm{XOR}}}  \\ -  \bigg( \bigg( \underset{1 \leq z \leq N-1}{\bigotimes} \textbf{I}_z \bigg) \bigotimes \frac{1}{\sqrt{\# \sigma }}  \bigg(    \underset{\text{Permutations } \sigma}{\sum}  \big(   B^{(N-1)}_{\sigma ( i_1, \cdots, i_{N-1})}         \wedge   B^{(N-1)}_{\sigma ( i^{\prime}_1, \cdots, i^{\prime}_{N-1})}  \wedge  \cdots \\ \wedge   B^{(N-1)}_{\sigma ( i^{\prime\cdots\prime}_1, \cdots, i^{\prime\cdots\prime}_{N-1})} \big)      \bigg) \bigg) \ket{\psi_{N\mathrm{XOR} \wedge \cdots \wedge N\mathrm{XOR}}}           \bigg| \bigg|     \text{. }
\end{align*}

\noindent Therefore, to finish the computation for obtaining the desired upper bound, expressing the summation,

\begin{align*}
     \underset{\text{Questions }k,l,k^{\prime},l^{\prime}, \cdots, k^{\prime\cdots\prime}, l^{\prime\cdots \prime}}{\sum}   \bigg[   \frac{\big| \pm \big(  B_{kl} \wedge B_{k^{\prime} l^{\prime}} \wedge \cdots \wedge B_{k^{\prime\cdots\prime}l^{\prime\cdots \prime}}  \big)  + \big( B_{lk} \wedge B_{ l^{\prime} k^{\prime}} \wedge \cdots \wedge B_{l^{\prime\cdots \prime} k^{\prime\cdots\prime} }  \big)  \big|}{\sqrt{2}} \bigg]    \text{, }
\end{align*}

\noindent from the previously obtained bound for two operations of strong parallel repetition for the $\mathrm{FFL}$ game, in the setting of an arbitrary number of operations for strong parallel repetition in $\mathrm{XOR} \wedge \cdots \wedge \mathrm{XOR}$ games, takes the desired form from the observation,

\begin{align*}
  \underset{\# \text{ Player Observables}}{\sum}  \underset{\text{Players}}{\mathrm{sup}} \big( \mathcal{I}_j \big) <   \underset{\text{Players}}{\mathrm{sup}} \bigg\{    \underset{1 \leq j \leq \big( \# \text{ Player Observables} \big) }{\sum}  \mathcal{I}_j \bigg\}   <    N \underset{1 \leq j \leq N}{\mathrm{sup}} \big\{ \mathcal{I}_j \big\}     \\    <   20 N  \sqrt{N \epsilon^{\wedge}_{{N\mathrm{XOR} \wedge \cdots \wedge N\mathrm{XOR}}}} \text{, }
\end{align*}

\noindent  from the summation over player observables,

\begin{align*}
\underset{\# \text{ Player Observables}}{\sum} \mathcal{I}_j  \equiv \underset{1 \leq j \leq N}{\sum}   \mathcal{I}_j     \text{, }
\end{align*}

\noindent from which we conclude the argument. \boxed{}

\bigskip

\noindent As in a previous result, $\textbf{Lemma}$ $\textit{5}^{*}$, one can expect that the following system of relations holds, as stated with $\textbf{Lemma}$ $\textit{5}^{*}$B, with the following.

\bigskip

\noindent \textbf{Lemma} $\textit{5}^{**}\textit{B}$ (\textit{three-player analog of an arbitrary number of strong parallel repetition applications of }$\sqrt{\epsilon^{\wedge}_{{3\mathrm{XOR} \wedge \cdots \wedge 3\mathrm{XOR}}}}$- \textit{3 XOR approximality}, \textbf{Lemma} \textit{8}, {[44]}). From the same quantities introduced in the previous result, one has, for the $\mathrm{XOR}$ game under an arbitrary number of strong parallel repetitions, that the quantities,

\begin{align*}
  \mathcal{I}_1 \equiv  \bigg| \bigg|               \bigg( \big(  A_k  \wedge A_{k^{\prime}} \wedge \cdots \wedge A_{k^{\prime\cdots \prime}} \big) \bigotimes  \textbf{I} \bigotimes \textbf{I}  \bigg) \ket{\psi_{3\mathrm{XOR} \wedge \cdots \wedge 3\mathrm{XOR}}}    \\  -  \bigg( \textbf{I}  \bigotimes \bigg(     \frac{\pm \big(  B_{kl} \wedge B_{k^{\prime} l^{\prime}} \wedge \cdots \wedge B_{k^{\prime\cdots\prime}l^{\prime\cdots \prime}}  \big)  + \big( B_{lk} \wedge B_{ l^{\prime} k^{\prime}} \wedge \cdots \wedge B_{l^{\prime\cdots \prime} k^{\prime\cdots\prime} }  \big) }{\big| \pm \big(  B_{kl} \wedge B_{k^{\prime} l^{\prime}} \wedge \cdots \wedge B_{k^{\prime\cdots\prime}l^{\prime\cdots \prime}}  \big)  + \big( B_{lk} \wedge B_{ l^{\prime} k^{\prime}} \wedge \cdots \wedge B_{l^{\prime\cdots \prime} k^{\prime\cdots\prime} }  \big) \big| }    \bigg)  \bigotimes  \textbf{I}      \bigg) \\  \times \ket{\psi_{3\mathrm{XOR} \wedge \cdots \wedge 3\mathrm{XOR}}}              \bigg| \bigg| \\ \vdots \\ \mathcal{I}_3 \equiv   \bigg| \bigg|    \bigg(   \textbf{I} \bigotimes   \frac{1}{\sqrt{\# \sigma^{\prime} }}  \bigg(    \underset{\text{Permutations } \sigma}{\sum}  \big(   B^{(2)}_{\sigma ( i_1, i_2)}         \wedge  B^{(2)}_{\sigma ( i^{\prime}_1, i^{\prime}_2)}      \wedge  \cdots \wedge   B^{(2)}_{\sigma ( i^{\prime\cdots\prime}_1, i^{\prime\cdots\prime}_2)}   \big)      \bigg)     \bigotimes \textbf{I}  \bigg)    \\ \times \ket{\psi_{3\mathrm{XOR} \wedge \cdots \wedge 3\mathrm{XOR}}}  \\ -  \bigg( \textbf{I} \bigotimes  \textbf{I} \bigotimes \frac{1}{\sqrt{\# \sigma }}  \bigg(    \underset{\text{Permutations } \sigma}{\sum}  \big(   B^{(2)}_{\sigma ( i_1, i_2)}         \wedge  B^{(2)}_{\sigma ( i^{\prime}_1, i^{\prime}_2)}      \wedge  \cdots \wedge   B^{(2)}_{\sigma ( i^{\prime\cdots\prime}_1, i^{\prime\cdots\prime}_2)}   \big)      \bigg) \bigg) \\ \times \ket{\psi_{3\mathrm{XOR} \wedge \cdots \wedge 3\mathrm{XOR}}}           \bigg| \bigg|      \text{, } 
\end{align*}

\noindent have the strict upper bound,

\begin{align*}
\underset{1 \leq j \leq 3}{\sum} \mathcal{I}_j < 20 N \sqrt{N \epsilon^{\wedge}_{{3\mathrm{XOR} \wedge \cdots \wedge 3\mathrm{XOR}}}} \text{,}
 \end{align*}

 \noindent where the tensors beyond that of the second player, $B$, are indexed as,

 \begin{align*}
     \textbf{I} \bigotimes B_{\sigma ( i_1,i_2)} \bigotimes  \textbf{I} \equiv  \textbf{I} \bigotimes B^{(1)}_{\sigma ( i_1,i_2)} \bigotimes \textbf{I}             \text{, } \\ \vdots \\   \textbf{I} \bigotimes \textbf{I} \bigotimes B_{\sigma ( i_1,i_2,\cdots, i_{n-2} )}  \equiv  \textbf{I} \bigotimes \textbf{I} \bigotimes  B^{(n-2)}_{\sigma ( i_1,i_2,\cdots, i_{n-2} )}  \equiv  \textbf{I} \bigotimes \textbf{I} \bigotimes  C   \text{. }
 \end{align*}

\noindent \textit{Proof of Lemma $\textit{5}^{**}\textit{B}$}. Directly apply the arguments in the previous result, from which we conclude the argument. \boxed{}

\bigskip

\noindent We conclude the section by extending the following result, from arguments of strong parallel repetition of the $2$ $\mathrm{XOR}$, and $N$ $\mathrm{XOR}$, games:

\bigskip

\noindent \textbf{Lemma} \textit{6}, {[44]} (\textit{odd n product expansion}, \textit{6.1}, {[37]}). For odd $n$, one has an expansion for,

\begin{align*}
 \underset{1 \leq i \leq n}{\prod} \widetilde{A_i}   \text{, } 
\end{align*}

\noindent in terms of a signed block identity matrix,

\[
\big( - 1 \big)^n \begin{bmatrix}
\textbf{I} & 0 \\ 0 & - \textbf{I}
\end{bmatrix}  \text{. } 
\]

\noindent Hence, for,

\begin{align*}
 \ket{\widetilde{\psi_{\mathrm{FFL}}}} = \frac{1}{\sqrt{2 \times  2^{\lfloor \frac{n}{2} \rfloor }}} \bigg[   \text{ } \bigg(      \underset{1 \leq j \leq 2^{\lfloor \frac{n}{2}\rfloor}}{\sum}   + \underset{2^{\lfloor \frac{n}{2} \rfloor} + 1 \leq j \leq 2 \times 2^{\lfloor \frac{n}{2}\rfloor}}{\sum} \bigg) \bigg( \ket{j} \otimes \ket{j}  \bigg) \text{ }  \bigg]    \text{, } 
\end{align*}

\noindent one has,

\begin{align*}
 \bigg( \bigg(  \underset{1 \leq i \leq n}{\prod} \widetilde{A_i}  \bigg) \otimes \textbf{I} \bigg) \ket{\widetilde{\psi_{\mathrm{FFL}}}} = \frac{1}{\sqrt{2 \times  2^{\lfloor \frac{n}{2} \rfloor }}}  \bigg[    \text{ } \bigg(      \underset{1 \leq j \leq 2^{\lfloor \frac{n}{2}\rfloor}}{\sum}   + \underset{2^{\lfloor \frac{n}{2} \rfloor} + 1 \leq j \leq 2 \times 2^{\lfloor \frac{n}{2}\rfloor}}{\sum} \bigg) \bigg( \bigg(       \bigg(  \underset{1 \leq i \leq n}{\prod} \widetilde{A_i}  \bigg) \ket{j}   \\    \otimes \ket{j}      \bigg)  \otimes \bigg( \bigg(  \underset{1 \leq i \leq n}{\prod} \widetilde{A_i}  \bigg) \ket{j}    \otimes \ket{j} \bigg)   \bigg) \text{ }      \bigg]            \text{. } 
\end{align*}

\noindent Crucially, as generalizations of the result above one defines the collection of wavefunctions,

\begin{align*}
 \ket{\widetilde{\psi_{2\mathrm{XOR}}}} = \frac{1}{\sqrt{2 \times  2^{\lfloor \frac{n}{2} \rfloor }}} \bigg[   \text{ } \bigg(      \underset{1 \leq j \leq 2^{\lfloor \frac{n}{2}\rfloor}}{\sum}   + \underset{2^{\lfloor \frac{n}{2} \rfloor} + 1 \leq j \leq 2 \times 2^{\lfloor \frac{n}{2}\rfloor}}{\sum} \bigg) \\ \times \bigg( \ket{j} \otimes \ket{j}  \bigg) \text{ }  \bigg]    \text{, } \\  \\  \ket{\widetilde{\psi_{N\mathrm{XOR}}}} = \frac{1}{\sqrt{2 \times  2^{\lfloor \frac{n}{2} \rfloor }}} \bigg[   \text{ } \bigg(      \underset{1 \leq j \leq 2^{\lfloor \frac{n}{2}\rfloor}}{\sum}   + \underset{2^{\lfloor \frac{n}{2} \rfloor} + 1 \leq j \leq 2 \times 2^{\lfloor \frac{n}{2}\rfloor}}{\sum} \bigg) \bigg( \ket{j} \otimes \ket{j}  \otimes \overset{N-3}{\cdots} \otimes \ket{j} \bigg) \text{ }  \bigg]    \text{, }   \\ \\       \underset{1 \leq j \leq n}{\bigwedge} \ket{\widetilde{\psi_{N\mathrm{XOR}}}}^j = \frac{1}{\sqrt{2 \times  2^{\lfloor \frac{n}{2} \rfloor }}} \bigg[   \text{ } \bigg(      \underset{1 \leq j \leq 2^{\lfloor \frac{n}{2}\rfloor}}{\sum}   + \underset{2^{\lfloor \frac{n}{2} \rfloor} + 1 \leq j \leq 2 \times 2^{\lfloor \frac{n}{2}\rfloor}}{\sum} \bigg) \bigg( \bigg( \ket{j}  \wedge \cdots \\  \wedge \ket{j} \bigg) \bigotimes  \bigg(  \ket{j}  \wedge \cdots \wedge \ket{j} \bigg)    \bigotimes \overset{N-3}{\cdots}  \bigotimes \bigg( \ket{j}  \wedge \cdots \wedge \ket{j} \bigg)   \bigg) \text{ }  \bigg]    \text{, }     \\ \\   \underset{1 \leq j \leq 2}{\bigwedge} \ket{\widetilde{\psi_{\mathrm{FFL}}}}^j = \frac{1}{\sqrt{2 \times  2^{\lfloor \frac{n}{2} \rfloor }}} \bigg[   \text{ } \bigg(      \underset{1 \leq j \leq 2^{\lfloor \frac{n}{2}\rfloor}}{\sum}   + \underset{2^{\lfloor \frac{n}{2} \rfloor} + 1 \leq j \leq 2 \times 2^{\lfloor \frac{n}{2}\rfloor}}{\sum} \bigg) \bigg( \bigg( \ket{j}  \wedge\ket{j} \bigg) \\ \bigotimes  \bigg(  \ket{j}  \wedge \ket{j} \bigg)    \bigotimes \overset{N-3}{\cdots}  \bigotimes \bigg( \ket{j}   \wedge \ket{j} \bigg)   \bigg) \text{ }  \bigg]   \text{. }    
\end{align*}

\noindent From each wavefunction above, we identify the following collection of results for the expansion of the odd $n$ product expansion of $A$ operators.

\bigskip

\noindent \textbf{Lemma} \textit{6} (\textit{generalization of the odd n product expansion}, \textbf{Lemma} \textit{6}, {[44]}). For the multiplayer odd $n$ product expansion that is a counterpart to the odd $n$ product expansion introduced in the previous result, one has that,

\begin{align*}
      \bigg( \bigg( \underset{1 \leq i \leq n}{\prod}  \widetilde{A_i} \bigg) \bigotimes \bigg( \underset{1 \leq z \leq N-1}{\bigotimes} \textbf{I}_z \bigg) \bigg)  \ket{\widetilde{\psi_{2\mathrm{XOR}}}}   \text{, } \\  \bigg( \bigg( \underset{1 \leq i \leq n}{\prod}  \widetilde{A_i} \bigg) \bigotimes \bigg( \underset{1 \leq z \leq N-1}{\bigotimes} \textbf{I}_z \bigg) \bigg)  \ket{\widetilde{\psi_{N\mathrm{XOR}}}}          \text{, } \\ \underset{1 \leq j \leq n}{\bigwedge}  \bigg[  \bigg( \bigg( \underset{1 \leq i \leq n}{\prod}  \widetilde{A_i} \bigg) \bigotimes \bigg( \underset{1 \leq z \leq N-1}{\bigotimes} \textbf{I}_z \bigg) \bigg)  \ket{\widetilde{\psi_{N\mathrm{XOR}}}}^j   \bigg]   \text{, }    
\\ 
 \underset{1 \leq j \leq 2}{\bigwedge}  \bigg[  \bigg( \bigg( \underset{1 \leq i \leq n}{\prod}  \widetilde{A_i} \bigg) \bigotimes \textbf{I} \bigg)  \ket{\widetilde{\psi_{\mathrm{FFL}}}}^j   \bigg]    \text{, }
\end{align*}

\noindent respectively equal,

\begin{align*}
  \bigg( \bigg( \underset{1 \leq i \leq n}{\prod}  \widetilde{A_i} \bigg) \bigotimes \bigg( \underset{1 \leq z \leq N-1}{\bigotimes} \textbf{I}_z \bigg) \bigg)     \ket{\widetilde{\psi_{2\mathrm{XOR}}}} = \frac{1}{\sqrt{2 \times  2^{\lfloor \frac{n}{2} \rfloor }}} \bigg[   \text{ } \bigg(      \underset{1 \leq j \leq 2^{\lfloor \frac{n}{2}\rfloor}}{\sum}  \\  + \underset{2^{\lfloor \frac{n}{2} \rfloor} + 1 \leq j \leq 2 \times 2^{\lfloor \frac{n}{2}\rfloor}}{\sum} \bigg)  \bigg(  \bigg( \underset{1 \leq i \leq n}{\prod}  \widetilde{A_i} \bigg)  \ket{j} \otimes \ket{j}  \bigg)  \bigotimes  \bigg( \underset{1 \leq z \leq N-1}{\bigotimes} \textbf{I}_z \bigg)  \bigg(      \underset{1 \leq j \leq 2^{\lfloor \frac{n}{2}\rfloor}}{\sum} \\   + \underset{2^{\lfloor \frac{n}{2} \rfloor} + 1 \leq j \leq 2 \times 2^{\lfloor \frac{n}{2}\rfloor}}{\sum} \bigg) \bigg(  \bigg( \underset{1 \leq i \leq n}{\prod}  \widetilde{A_i} \bigg)  \ket{j}  \otimes \ket{j}  \bigg) \text{ }  \bigg]    \text{, }  \end{align*}
  
  \begin{align*}     \bigg( \bigg( \underset{1 \leq i \leq n}{\prod}  \widetilde{A_i} \bigg) \bigotimes \bigg( \underset{1 \leq z \leq N-1}{\bigotimes} \textbf{I}_z \bigg) \bigg) \ket{\widetilde{\psi_{N\mathrm{XOR}}}} = \frac{1}{\sqrt{2 \times  2^{\lfloor \frac{n}{2} \rfloor }}} \bigg[   \text{ } \bigg(      \underset{1 \leq j \leq 2^{\lfloor \frac{n}{2}\rfloor}}{\sum}   + \underset{2^{\lfloor \frac{n}{2} \rfloor} + 1 \leq j \leq 2 \times 2^{\lfloor \frac{n}{2}\rfloor}}{\sum} \bigg)   \\    \times  \bigg( \underset{1 \leq i \leq n}{\prod}  \widetilde{A_i} \bigg) \bigg(  \bigg( \underset{1 \leq i \leq n}{\prod}  \widetilde{A_i} \bigg)  \ket{j} \otimes   \bigg( \underset{1 \leq i \leq n}{\prod}  \widetilde{A_i} \bigg) 
 \ket{j}   \otimes \overset{N-3}{\cdots} \otimes  \bigg( \underset{1 \leq i \leq n}{\prod}  \widetilde{A_i} \bigg)   \ket{j} \bigg) \text{ }  \bigg]    \text{, } \end{align*}

 \begin{align*}        \underset{1 \leq j \leq n}{\bigwedge}  \bigg[  \bigg( \bigg( \underset{1 \leq i \leq n}{\prod}  \widetilde{A_i} \bigg) \bigotimes \bigg( \underset{1 \leq z \leq N-1}{\bigotimes} \textbf{I}_z \bigg) \bigg)  
 \ket{\widetilde{\psi_{N\mathrm{XOR}}}}^j = \frac{1}{\sqrt{2 \times  2^{\lfloor \frac{n}{2} \rfloor }}} \bigg[   \text{ } \bigg(      \underset{1 \leq j \leq 2^{\lfloor \frac{n}{2}\rfloor}}{\sum} \\   + \underset{2^{\lfloor \frac{n}{2} \rfloor} + 1 \leq j \leq 2 \times 2^{\lfloor \frac{n}{2}\rfloor}}{\sum} \bigg)   \bigg(   \bigg( \underset{1 \leq i \leq n}{\prod}  \widetilde{A_i} \bigg)  \bigg( \ket{j}  \wedge \cdots  \wedge \ket{j} \bigg) \bigotimes    \bigg( \underset{1 \leq i \leq n}{\prod}  \widetilde{A_i} \bigg)  \bigg(  \ket{j} \\   \wedge \cdots \wedge \ket{j} \bigg)    \bigotimes \overset{N-3}{\cdots}    \bigotimes  \bigg( \underset{1 \leq i \leq n}{\prod}  \widetilde{A_i} \bigg)   \bigg( \ket{j}  \wedge \cdots \wedge \ket{j} \bigg)   \bigg) \text{ }  \bigg]    \text{, }     \\ \\    \underset{1 \leq j \leq 2}{\bigwedge} \bigg[  \bigg( \bigg( \underset{1 \leq i \leq n}{\prod}  \widetilde{A_i} \bigg) \bigotimes \textbf{I} \bigg)  \ket{\widetilde{\psi_{\mathrm{FFL}}}}   \bigg] 
 \ket{\widetilde{\psi_{\mathrm{FFL}}}}^j = \frac{1}{\sqrt{2 \times  2^{\lfloor \frac{n}{2} \rfloor }}} \bigg[   \text{ } \bigg(      \underset{1 \leq j \leq 2^{\lfloor \frac{n}{2}\rfloor}}{\sum}  \\  + \underset{2^{\lfloor \frac{n}{2} \rfloor} + 1 \leq j \leq 2 \times 2^{\lfloor \frac{n}{2}\rfloor}}{\sum} \bigg)    \bigg(   \bigg( \underset{1 \leq i \leq n}{\prod}  \widetilde{A_i} \bigg) 
\\ \times  \bigg( \ket{j}  \wedge\ket{j} \bigg)   \bigotimes   \bigg( \underset{1 \leq i \leq n}{\prod}  \widetilde{A_i} \bigg)   \bigg(  \ket{j}  \wedge \ket{j} \bigg)    \bigotimes \overset{N-3}{\cdots} \\ \bigotimes  \bigg( \underset{1 \leq i \leq n}{\prod}  \widetilde{A_i} \bigg)   \bigg( \ket{j}   \wedge \ket{j} \bigg)   \bigg) \text{ }  \bigg]  \text{. }
\end{align*}

\bigskip

\noindent \textit{Proof of Lemma 6}. By straightforward computation, the action of the product of tensor observables for the first player, on the four wavefunctions, $\widetilde{\ket{\psi_{2\mathrm{XOR}}}}$, $\widetilde{\ket{\psi_{N\mathrm{XOR}}}}$, $\bigwedge \ket{\psi_{N\mathrm{XOR}}}$, and $\bigwedge \widetilde{\ket{\psi_{\mathrm{FFL}}}}$,

\begin{align*}
      \bigg( \bigg( \underset{1 \leq i \leq n}{\prod}  \widetilde{A_i} \bigg) \bigotimes \bigg( \underset{1 \leq z \leq N-1}{\bigotimes} \textbf{I}_z \bigg) \bigg)  \ket{\widetilde{\psi_{2\mathrm{XOR}}}}   \text{, } \\  \bigg( \bigg( \underset{1 \leq i \leq n}{\prod}  \widetilde{A_i} \bigg) \bigotimes \bigg( \underset{1 \leq z \leq N-1}{\bigotimes} \textbf{I}_z \bigg) \bigg)  \ket{\widetilde{\psi_{N\mathrm{XOR}}}}          \text{, } \\ \underset{1 \leq j \leq n}{\bigwedge}  \bigg[  \bigg( \bigg( \underset{1 \leq i \leq n}{\prod}  \widetilde{A_i} \bigg) \bigotimes \bigg( \underset{1 \leq z \leq N-1}{\bigotimes} \textbf{I}_z \bigg) \bigg)  \ket{\widetilde{\psi_{N\mathrm{XOR}}}}^j   \bigg]   \text{, }    
\\ 
 \underset{1 \leq j \leq 2}{\bigwedge}  \bigg[  \bigg( \bigg( \underset{1 \leq i \leq n}{\prod}  \widetilde{A_i} \bigg) \bigotimes \textbf{I} \bigg)  \ket{\widetilde{\psi_{\mathrm{FFL}}}}^j   \bigg]    \text{, }
\end{align*}

\noindent equal,

\begin{align*}
  \bigg( \bigg( \underset{1 \leq i \leq n}{\prod}  \widetilde{A_i} \bigg) \bigotimes \bigg( \underset{1 \leq z \leq N-1}{\bigotimes} \textbf{I}_z \bigg) \bigg)     \ket{\widetilde{\psi_{2\mathrm{XOR}}}} = \frac{1}{\sqrt{2 \times  2^{\lfloor \frac{n}{2} \rfloor }}} \bigg[   \text{ } \bigg(      \underset{1 \leq j \leq 2^{\lfloor \frac{n}{2}\rfloor}}{\sum}   + \underset{2^{\lfloor \frac{n}{2} \rfloor} + 1 \leq j \leq 2 \times 2^{\lfloor \frac{n}{2}\rfloor}}{\sum} \bigg) \\  \times  \bigg(  \bigg( \underset{1 \leq i \leq n}{\prod}  \widetilde{A_i} \bigg)  \ket{j} \otimes \ket{j}  \bigg)  \bigotimes  \bigg( \underset{1 \leq z \leq N-1}{\bigotimes} \textbf{I}_z \bigg)  \bigg(      \underset{1 \leq j \leq 2^{\lfloor \frac{n}{2}\rfloor}}{\sum}   + \underset{2^{\lfloor \frac{n}{2} \rfloor} + 1 \leq j \leq 2 \times 2^{\lfloor \frac{n}{2}\rfloor}}{\sum} \bigg) \\ \times \bigg(  \bigg( \underset{1 \leq i \leq n}{\prod}  \widetilde{A_i} \bigg)  \ket{j}  \otimes \ket{j}  \bigg) \text{ }  \bigg]    \text{, }  \\  \\   \bigg( \bigg( \underset{1 \leq i \leq n}{\prod}  \widetilde{A_i} \bigg) \bigotimes \bigg( \underset{1 \leq z \leq N-1}{\bigotimes} \textbf{I}_z \bigg) \bigg) \ket{\widetilde{\psi_{N\mathrm{XOR}}}} = \frac{1}{\sqrt{2 \times  2^{\lfloor \frac{n}{2} \rfloor }}} \bigg[   \text{ } \bigg(      \underset{1 \leq j \leq 2^{\lfloor \frac{n}{2}\rfloor}}{\sum}  \\  + \underset{2^{\lfloor \frac{n}{2} \rfloor} + 1 \leq j \leq 2 \times 2^{\lfloor \frac{n}{2}\rfloor}}{\sum} \bigg)   \bigg( \underset{1 \leq i \leq n}{\prod}  \widetilde{A_i} \bigg)  \bigg(  \bigg( \underset{1 \leq i \leq n}{\prod}  \widetilde{A_i} \bigg)  \ket{j} \otimes   \bigg( \underset{1 \leq i \leq n}{\prod}  \widetilde{A_i} \bigg) 
 \ket{j}  \\  \otimes \overset{N-3}{\cdots} \otimes  \bigg( \underset{1 \leq i \leq n}{\prod}  \widetilde{A_i} \bigg)   \ket{j} \bigg) \text{ }  \bigg]    \text{, } \end{align*}

 \begin{align*}        \underset{1 \leq j \leq n}{\bigwedge}  \bigg[  \bigg( \bigg( \underset{1 \leq i \leq n}{\prod}  \widetilde{A_i} \bigg) \bigotimes \bigg( \underset{1 \leq z \leq N-1}{\bigotimes} \textbf{I}_z \bigg) \bigg)  
 \ket{\widetilde{\psi_{N\mathrm{XOR}}}}^j = \frac{1}{\sqrt{2 \times  2^{\lfloor \frac{n}{2} \rfloor }}} \bigg[   \text{ } \bigg(      \underset{1 \leq j \leq 2^{\lfloor \frac{n}{2}\rfloor}}{\sum} \\   + \underset{2^{\lfloor \frac{n}{2} \rfloor} + 1 \leq j \leq 2 \times 2^{\lfloor \frac{n}{2}\rfloor}}{\sum} \bigg)  \bigg(   \bigg( \underset{1 \leq i \leq n}{\prod}  \widetilde{A_i} \bigg)  \bigg( \ket{j}  \wedge \cdots  \wedge \ket{j} \bigg) \bigotimes    \bigg( \underset{1 \leq i \leq n}{\prod}  \widetilde{A_i} \bigg)  \bigg(  \ket{j} \\  \wedge \cdots \wedge \ket{j} \bigg)    \bigotimes \overset{N-3}{\cdots}   \bigotimes  \bigg( \underset{1 \leq i \leq n}{\prod}  \widetilde{A_i} \bigg)   \bigg( \ket{j}  \wedge \cdots \wedge \ket{j} \bigg)   \bigg) \text{ }  \bigg]    \text{, }     \end{align*}

\begin{align*}     \underset{1 \leq j \leq 2}{\bigwedge} \bigg[  \bigg( \bigg( \underset{1 \leq i \leq n}{\prod}  \widetilde{A_i} \bigg) \bigotimes \textbf{I} \bigg)  \ket{\widetilde{\psi_{\mathrm{FFL}}}}   \bigg] 
 \ket{\widetilde{\psi_{\mathrm{FFL}}}}^j = \frac{1}{\sqrt{2 \times  2^{\lfloor \frac{n}{2} \rfloor }}} \bigg[   \text{ } \bigg(      \underset{1 \leq j \leq 2^{\lfloor \frac{n}{2}\rfloor}}{\sum}  \\  + \underset{2^{\lfloor \frac{n}{2} \rfloor} + 1 \leq j \leq 2 \times 2^{\lfloor \frac{n}{2}\rfloor}}{\sum} \bigg)  \bigg(   \bigg( \underset{1 \leq i \leq n}{\prod}  \widetilde{A_i} \bigg) 
 \bigg( \ket{j}  \wedge\ket{j} \bigg)   \bigotimes   \bigg( \underset{1 \leq i \leq n}{\prod}  \widetilde{A_i} \bigg)   \bigg(  \ket{j}  \\ \wedge \ket{j} \bigg)    \bigotimes \overset{N-3}{\cdots}   \bigotimes  \bigg( \underset{1 \leq i \leq n}{\prod}  \widetilde{A_i} \bigg)   \bigg( \ket{j}   \wedge \ket{j} \bigg)   \bigg) \text{ }  \bigg]  \text{, }
\end{align*}

\noindent respectively, from which we conclude the argument. \boxed{}

\bigskip

\noindent One also expects that results of the form hold above for the product expansions of any other tensor observables from players of the $\mathrm{XOR}$, and $\mathrm{FFL}$, games.

\bigskip

\noindent For the following inequality, fix,

\begin{align*} \{ u_i \}_{i \in \mathcal{I}} \in \textbf{R}^n \text{, } \\ \{ v_i \}_{i \in \mathcal{I}} \in \textbf{R}^m \text{. }
\end{align*}

\noindent We conclude the section by discussing the types of computations which would be required to demonstrate that a counterpart of the identity,

\begin{align*}
         \overset{r}{\underset{k=1}{\sum} } \big|\big|    \big( u_k \cdot \vec{A} \otimes \textbf{I} \big) \ket{\psi}       -              \big(  \textbf{I} \otimes v_k \cdot \vec{B} \big)  \ket{\psi}    \big|\big|^2 \leq \beta \big( G \big) \epsilon  \text{, }     
\end{align*}

\noindent with an upper bound of the bias of the game, $\beta \big( G \big)$, and $\epsilon$ sufficiently small, holds for $N$-$\mathrm{XOR}$ games, in addition to strong parallel repetition of the $\mathrm{XOR}$ and $\mathrm{FFL}$ games, with the following results:

\bigskip

\noindent \textbf{Lemma} \textit{7} (\textit{$\epsilon$-approximality of the bias of 3-XOR, N-XOR, and strong parallel repetitions of XOR and FFL games}, \textbf{Lemma} \textit{2}, {[37]}). Fix $\epsilon_1$, $\epsilon_2$, $\epsilon_3$, and $\epsilon_4$ sufficiently small. One has,

\begin{align*}
     \overset{r}{\underset{k=1}{\sum} } \bigg|\bigg|    \bigg( u_k \cdot \vec{A} \bigotimes \bigg(   \textbf{I} \bigotimes \textbf{I} \bigg)  \bigg) \ket{\psi_{3\mathrm{XOR}}}       -              \bigg( \bigg(  \textbf{I} \bigotimes  v_k \bigotimes \textbf{I} \bigg) \cdot \vec{B} \bigg) \ket{\psi_{3\mathrm{XOR}}}   \bigg|\bigg|^2 \\ \leq \beta \big( G_{3\mathrm{XOR}} \big) \epsilon_1 \text{,} \\  \tag{1} \\   \overset{r}{\underset{k=1}{\sum} } \bigg|\bigg|    \bigg( u_k \cdot \vec{A} \bigotimes \bigg(  \underset{1 \leq z \leq N-1}{\bigotimes}\textbf{I}_z \bigg) \bigg) \ket{\psi_{N\mathrm{XOR}}}       -              \bigg( \bigg(  \underset{1 \leq z \leq N-1}{\bigotimes} \textbf{I}_z  \bigg) \bigotimes v_k \cdot \vec{B} \bigg) \\ \times  \ket{\psi_{N\mathrm{XOR}}}    \bigg|\bigg|^2   \leq \beta \big( G_{N\mathrm{XOR}} \big) \epsilon_3 \text{, } \\  \tag{2} \\  \overset{r}{\underset{k=1}{\sum} } \bigg|\bigg|    \bigg( u_k \cdot \bigg( \vec{A} \wedge \cdots \wedge \vec{A}^{\prime\cdots\prime} \bigg)  \bigotimes \bigg(  \underset{1 \leq z \leq N-1}{\bigotimes}\textbf{I}_z \bigg) \bigg) \ket{\psi_{N\mathrm{XOR}\wedge\cdots\wedge N\mathrm{XOR}}}       \\ -              \bigg( \bigg(  \underset{1 \leq z \leq N-1}{\bigotimes} \textbf{I}_z  \bigg) \bigotimes v_k \cdot  \bigg( \vec{B} \wedge \cdots  \wedge \vec{B}^{\prime\cdots\prime} \bigg)  \bigg)  \ket{\psi_{N\mathrm{XOR}\wedge\cdots\wedge N\mathrm{XOR}}}    \bigg|\bigg|^2     \\ \leq \beta \big( G_{N\mathrm{XOR}\wedge\cdots\wedge N\mathrm{XOR}} \big) \epsilon_2 \text{, } \\ \tag{3} \\  \overset{r}{\underset{k=1}{\sum} } \bigg|\bigg|    \bigg( u_k \cdot \vec{A} \bigotimes \bigg( \textbf{I} \bigotimes \textbf{I} \bigg) \bigg) \ket{\psi_{\mathrm{FFL}\wedge\mathrm{FFL}}}       -              \bigg( \bigg(   \textbf{I} \bigotimes \textbf{I} \bigg)  v_k \cdot \vec{B} \bigg)  \ket{\psi_{\mathrm{FFL}\wedge\mathrm{FFL}}}    \bigg|\bigg|^2 \\ \leq \beta \big( G_{\mathrm{FFL}\wedge\mathrm{FFL}} \big) \epsilon_4 \text{,} \tag{4}
\end{align*}    

\noindent for the optimal strategies for each game.

\bigskip

\noindent \textit{Proof of Lemma 7}. The results for each of the games above follows from the fact that the corresponding primal feasible solutions for each setting,

\begin{align*}
 \underline{3\mathrm{XOR}}: G_{3\mathrm{XOR}} \cdot Z_{3\mathrm{XOR}} \equiv \bra{\psi_{3\mathrm{XOR}}}  \bigg( \underset{1 \leq i \leq 3}{\bigotimes } i \text{ th player tensor observable} \bigg)               \\ \times  \ket{\psi_{3\mathrm{XOR}}}  \text{, } \\   \underline{N\mathrm{XOR}}:G_{N\mathrm{XOR}} \cdot Z_{N\mathrm{XOR}} \equiv \bra{\psi_{N\mathrm{XOR}}}  \bigg( \underset{1 \leq i \leq N}{\bigotimes } i \text{ th player tensor observable} \bigg)                \\ \times   \ket{\psi_{N\mathrm{XOR}}}  \text{,} \\   \underline{N\mathrm{XOR}\wedge \cdots \wedge N \mathrm{XOR}}: G_{N\mathrm{XOR} \wedge \cdots \wedge N\mathrm{XOR}} \cdot Z_{N\mathrm{XOR}\wedge \cdots \wedge N\mathrm{XOR}} \equiv G_{\mathrm{XOR} \wedge \cdots \wedge \mathrm{XOR}}\\  \cdot Z_{\mathrm{XOR}\wedge \cdots \wedge \mathrm{XOR}}   \equiv \bra{\psi_{N\mathrm{XOR}}}  \times   \bigg( \underset{i^{\prime\cdots\prime} \leq i^{\prime\cdots\prime} < n^{\prime\cdots\prime}}{\underset{\vdots}{\underset{1 \leq i \leq N}{\bigotimes }}}  \big( i \text{ th}    \text{ player tensor observable} \big) \\ \wedge \cdots \wedge \big( i^{\prime\cdots\prime} \text{ th player tensor observable} \big)  \bigg)                \ket{\psi_{\mathrm{XOR}}} \\ \equiv  \bra{\psi_{\mathrm{XOR}}}   \bigg( \underset{i^{\prime\cdots\prime} \leq i^{\prime\cdots\prime} < n^{\prime\cdots\prime}}{\underset{\vdots}{\underset{1 \leq i \leq N}{\bigotimes }}}  \big( i \text{ th player tensor observable} \big) \wedge \cdots \wedge \big( i^{\prime\cdots\prime} \text{ th player} \\ \text{tensor observable} \bigg)                \ket{\psi_{\mathrm{XOR}}}    \text{, } \\   \underline{\mathrm{FFL} \wedge \mathrm{FFL}}:G_{\mathrm{FFL} \wedge \mathrm{FFL}} \wedge Z_{\mathrm{FFL}\wedge \mathrm{FFL}} \equiv \bra{\psi_{\mathrm{FFL}\wedge \mathrm{FFL}}}  \bigg( 
  \underset{1 \leq i^{\prime} < N^{\prime}}{\underset{1 \leq i \leq N}{\bigotimes}}      \big( i \text{ th player tensor} \\ \text{ observable} \big)  \wedge   \big( i^{\prime}   \text{ th player tensor observable} \big)   \bigg) \ket{\psi_{\mathrm{FFL}\wedge \mathrm{FFL}}}       \text{, }
\end{align*}

\noindent respectively, satisfy the equalities,

\begin{align*}
      (\textit{1}) \equiv  \beta \big( G_{3\mathrm{XOR}} \big) \epsilon_1  - G_{3\mathrm{XOR}} Z_{3\mathrm{XOR}}   \text{, } \\ (\textit{2}) \equiv \beta \big( G_{N\mathrm{XOR}} \big) \epsilon_2  - G_{N\mathrm{XOR}} Z_{N\mathrm{XOR}}  \text{,} \\ (\textit{3}) \equiv  \beta \big( G_{N\mathrm{XOR}\wedge \cdots \wedge N \mathrm{XOR}} \big) \epsilon_3  - G_{N\mathrm{XOR}\wedge \cdots \wedge N \mathrm{XOR}} Z_{N\mathrm{XOR}\wedge \cdots \wedge N \mathrm{XOR}} \text{, } \\ (\textit{4}) \equiv   \beta \big( G_{\mathrm{FFL}\wedge  \mathrm{FFL}} \big) \epsilon_4  - G_{\mathrm{FFL}\wedge  \mathrm{FFL}}  Z_{\mathrm{FFL}\wedge  \mathrm{FFL}}     \text{, }
\end{align*}

\noindent respectively, from which a straightforward adaptation of arguments from \textbf{Lemma} \textit{5}, {[44]} yield the desired result, from which we conclude the argument. \boxed{}

\bigskip

\noindent We conclude the subsection by directing the attention of the reader to the fact that previous results, namely from \textbf{Theorem} $\textit{1}^{*}$, \textbf{Theorem} $\textit{2}^{*}$, and \textbf{Theorem} $\textit{3}^{*}$, can straightforwardly be extended under strong parallel repetition, in addition to following result on the positive semidefiniteness result (stated in the collection of conditions ($3\mathrm{XOR} \wedge \cdots \wedge 3 \mathrm{XOR}$, $\mathrm{Sym}$), ($4\mathrm{XOR} \wedge \cdots \wedge 4 \mathrm{XOR}$, $\mathrm{Sym}$), ($5\mathrm{XOR} \wedge \cdots \wedge 5 \mathrm{XOR}$, $\mathrm{Sym}$), and ($N\mathrm{XOR} \wedge \cdots \wedge N \mathrm{XOR}$, $\mathrm{Sym}$)). We provide the statements of each result that is expected to hold below, omitting the arguments from each result as it is a direct application of previous computations.

\bigskip

\noindent \textbf{Theorem} $\textit{4}^{*}$ (\textit{3-XOR strong parallel repetition error bounds}, \textit{2.2.1}, \textbf{Theorem} \textit{4}, {[37]}, \textbf{Theorem} \textit{2}, {[44]}, \textbf{Theorems} \textit{1-6} in \textit{1.5}). Under strong parallel repetition for the $3$-$\mathrm{XOR}$ game, one has an error bound of an identical form as that provided in \textbf{Theorem} $\textit{1}^{*}$.

\bigskip

\noindent \textit{Proof of Theorem $4^{*}$}. Apply the same computations as provided in \textbf{Theorem} $\textit{1}^{*}$, \textbf{Theorem} $\textit{2}^{*}$, and \textbf{Theorem} $\textit{3}^{*}$, from which we conclude the argument. \boxed{}

\bigskip

\noindent \textbf{Theorem} $\textit{5}^{*}$ (\textit{4-XOR strong parallel repetition error bounds}, \textit{2.2.1}, \textbf{Theorem} \textit{4}, {[37]}, \textbf{Theorem} \textit{2}, {[44]}, \textbf{Theorems} \textit{1-6} in \textit{1.5}). Under strong parallel repetition for the $3$-$\mathrm{XOR}$ game, one has an error bound of an identical form as that provided in \textbf{Theorem} $\textit{1}^{*}$.

\bigskip

\noindent \textit{Proof of Theorem $5^{*}$}. Apply the same computations as provided in \textbf{Theorem} $\textit{1}^{*}$, \textbf{Theorem} $\textit{2}^{*}$, and \textbf{Theorem} $\textit{3}^{*}$, from which we conclude the argument. \boxed{}

\bigskip

\noindent \textbf{Theorem} $\textit{6}^{*}$ (\textit{5-XOR strong parallel repetition error bounds}, \textit{2.2.1}, \textbf{Theorem} \textit{4}, {[37]}, \textbf{Theorem} \textit{2}, {[44]}, \textbf{Theorems} \textit{1-6} in \textit{1.5}). Under strong parallel repetition for the $3$-$\mathrm{XOR}$ game, one has an error bound of an identical form as that provided in \textbf{Theorem} $\textit{1}^{*}$.

\bigskip

\noindent \textit{Proof of Theorem $6^{*}$}. Apply the same computations as provided in \textbf{Theorem} $\textit{1}^{*}$, \textbf{Theorem} $\textit{2}^{*}$, and \textbf{Theorem} $\textit{3}^{*}$, from which we conclude the argument. \boxed{}

\bigskip

\noindent \textbf{Theorem} $\textit{7}^{*}$ (\textit{N-XOR strong parallel repetition error bounds}, \textit{2.2.1}, \textbf{Theorem} \textit{4}, {[37]}, \textbf{Theorem} \textit{2}, {[44]}, \textbf{Theorems} \textit{1-6} in \textit{1.5}). Under strong parallel repetition for the $3$-$\mathrm{XOR}$ game, one has an error bound of an identical form as that provided in \textbf{Theorem} $\textit{1}^{*}$.

\bigskip

\noindent \textit{Proof of Theorem $7^{*}$}. Apply the same computations as provided in \textbf{Theorem} $\textit{1}^{*}$, \textbf{Theorem} $\textit{2}^{*}$, and \textbf{Theorem} $\textit{3}^{*}$, from which we conclude the argument. \boxed{}

\bigskip

\noindent \textbf{Theorem} $\textit{8}^{*}$ (\textit{FFL strong parallel repetition error bounds}, \textit{2.2.1}, \textbf{Theorem} \textit{4}, {[37]}, \textbf{Theorem} \textit{2}, {[44]}, \textbf{Theorems} \textit{1-6} in \textit{1.5}). Under strong parallel repetition for the $3$-$\mathrm{XOR}$ game, one has an error bound of an identical form as that provided in \textbf{Theorem} $\textit{1}^{*}$.

\bigskip

\noindent \textit{Proof of Theorem $8^{*}$}. Apply the same computations as provided in \textbf{Theorem} $\textit{1}^{*}$, \textbf{Theorem} $\textit{2}^{*}$, and \textbf{Theorem} $\textit{3}^{*}$, from which we conclude the argument. \boxed{}

\bigskip

\noindent Moreover, the positive semidefiniteness of the associated operators with each of the results above follows, which we state below before concluding the subsection.

\bigskip

\noindent \textbf{Lemma} \textit{T-4} (\textit{positive semidefiniteness under 3-XOR strong parallel repetition}).  ($3\mathrm{XOR} \wedge \cdots \wedge 3 \mathrm{XOR}$, Sym) holds.

\bigskip

\noindent \textit{Proof of Lemma T-4}. Apply the same computations as provided in \textbf{Lemma} \textit{T-1}, \textbf{Lemma} \textit{T-2}, and \textbf{Lemma} \textit{T-3}, from which we conclude the argument. \boxed{}

\bigskip

\noindent \textbf{Lemma} \textit{T-5} (\textit{positive semidefiniteness under 4-XOR strong parallel repetition}). ($4\mathrm{XOR} \wedge \cdots \wedge 4\mathrm{XOR}$, Sym) holds.

\bigskip

\noindent \textit{Proof of Lemma T-5}. Apply the same computations as provided in \textbf{Lemma} \textit{T-1}, \textbf{Lemma} \textit{T-2}, and \textbf{Lemma} \textit{T-3}, from which we conclude the argument. \boxed{}

\bigskip

\noindent \textbf{Lemma} \textit{T-6} (\textit{positive semidefiniteness under 5-XOR strong parallel repetition}). ($5\mathrm{XOR} \wedge \cdots \wedge 5 \mathrm{XOR}$, Sym) holds.

\bigskip

\noindent \textit{Proof of Lemma T-6}. Apply the same computations as provided in \textbf{Lemma} \textit{T-1}, \textbf{Lemma} \textit{T-2}, and \textbf{Lemma} \textit{T-3}, from which we conclude the argument. \boxed{}

\bigskip

\noindent \textbf{Lemma} \textit{T-7} (\textit{positive semidefiniteness under N-XOR strong parallel repetition}). ($N\mathrm{XOR} \wedge \cdots \wedge N \mathrm{XOR}$, Sym) holds.

\bigskip

\noindent \textit{Proof of Lemma T-7}. Apply the same computations as provided in \textbf{Lemma} \textit{T-1}, \textbf{Lemma} \textit{T-2}, and \textbf{Lemma} \textit{T-3}, from which we conclude the argument. \boxed{}

\bigskip

\noindent \textbf{Lemma} \textit{T-8} (\textit{positive semidefiniteness under FFL strong parallel repetition}). ($\mathrm{FFL} \wedge \mathrm{FFL}$, Sym) holds.

\bigskip

\noindent \textit{Proof of Lemma T-8}. Apply the same computations as provided in \textbf{Lemma} \textit{T-1}, \textbf{Lemma} \textit{T-2}, and \textbf{Lemma} \textit{T-3}, from which we conclude the argument. \boxed{}

\subsection{Suitable linear operators for multiplayer XOR games have unit Frobenius norm}

\noindent \textbf{Lemma} \textit{9} (\textit{the Frobenius norm of suitable linear operators for the 3-XOR, 4-XOR, 5-XOR, and N-XOR games equals 1}). With respect to the Frobenius norm, the norm of suitable linear operators introduced in previous sections for the 3-XOR, 4-XOR, 5-XOR, and N-XOR, games equals $1$.

\bigskip

\noindent \textit{Proof of Lemma 9}. Directly apply the argument from \textit{6.2} in {[37]}, from which we conclude the argument. \boxed{}

\subsection{Suitable linear operators for strong parallel repetition of the XOR game, and the two player FFL game, have unit Frobenius norm}

\noindent \textbf{Lemma} \textit{10} (\textit{the Frobenius norm of suitable linear operators for strong parallel repetition of the multiplayer XOR game, and of the two player FFL game, equal 1}). With respect to the Frobenius norm, the norm of suitable linear operators introduced in previous sections for the 3-XOR, 4-XOR, 5-XOR, and N-XOR, games equals $1$.

\bigskip

\noindent \textit{Proof of Lemma 10}. Directly apply the argument from \textit{6.2} in {[37]}, from which we conclude the argument. \boxed{}

\subsection{Strong parallel repetition}

From all of the inequalities with the $\wedge$ operation corresponding to parallel, and strong parallel, repetition, the optimal value for two strong parallel repetitions of the $\mathrm{FFL}$ game is related to strong parallel repetition of the $\mathrm{XOR}$ game. In comparison to the optimal values of $\mathrm{FFL} \wedge \mathrm{FFL}$ which remains equal to the optimal value of playing a single $\mathrm{FFL}$ game with no strong parallel repetition, the $\mathrm{XOR}$ optimal value corresponding to the game $G$ satisfies,

\begin{align*}
    \omega_{\mathrm{XOR} \wedge \cdots \wedge \mathrm{XOR}} \big( G \big)  \equiv \omega \big(  \mathrm{XOR} \wedge \cdots \wedge \mathrm{XOR} \big) \equiv \underset{\# \text{ of strong parallel repetitions } j}{\prod}   \omega \big( \mathrm{XOR}\big)^{j}   \text{. }
\end{align*}

\noindent As such, possible tensor products of operators corresponding to optimal quantum strategies for $\mathrm{XOR} \wedge \cdots \wedge \mathrm{XOR}$ games can be classified from previous inequalities for one strong parallel repetition of the $\mathrm{FFL}$ game, $\mathrm{FFL} \wedge \mathrm{FFL}$, through the mapping,

\begin{align*}
   \varphi : \big( \textbf{C}^2 \big)^{\otimes N} \longrightarrow   \big( \textbf{C}^2 \big)^{\otimes M} : 1 \otimes 2 \otimes \cdots \otimes N \mapsto     1 \otimes 2 \otimes \cdots \otimes N  \otimes \cdots \otimes M             \text{, }
\end{align*}

\noindent which can be leveraged to characterize the combinatorially possible observable tensors that each player in the game can form after the previous player has submitted an answer to the question administered by the referee. Furthermore, along the lines of strong parallel repetitions for $\mathrm{FFL}$ games, ineqaualities involving the action of the suitable linear operator, $T_{\mathrm{FFL}}$, from the two-player setting, can be formulated with the action oof another suitable operator, $\mathcal{T} \equiv T_{\mathrm{FFL} \wedge \mathrm{FFL}}$, which are of the form,

\begin{align*}
     \bigg| \bigg|  \bigg[  \bigg( \bigg( A_i \wedge A_{i^{\prime}} \bigg) \bigotimes \textbf{I} \bigg)   \mathcal{T} -  \mathcal{T}  \bigg(          \textbf{I} \bigotimes \bigg( \widetilde{A_i \wedge A_{i^{\prime}}}  \bigg) \bigg)  \bigg] \ket{\psi_{\mathrm{FFL} \wedge \mathrm{FFL}}}   \bigg| \bigg|         \text{. }
\end{align*}

\noindent Given the difference between the optimal values of strong parallel repetitions for the $\mathrm{XOR}$ and $\mathrm{FFL}$ games, in the $N$-player $\mathrm{XOR}$ setting, inequalities involving the optimal value $\omega_{\mathrm{XOR} \wedge \cdots \wedge \mathrm{XOR}}$, after an arbitrary number of strong parallel repetitions, are of the form,

\begin{align*}
    \bigg| \bigg|  \bigg[  \bigg(   \big( A_i \wedge A_{i^{\prime}} \big) \bigotimes \textbf{I}      \bigotimes \bigg(    \bigg(   \underset{i \in \mathcal{Q}_1, j \in \mathcal{Q}_2 , k \in \mathcal{Q}_3}{\prod}             C^{l_{ijk}}_{ijk}     \bigg) \wedge \bigg(    \underset{i^{\prime}_1 \in \mathcal{Q}^{\prime}_1, i^{\prime}_2 \in \mathcal{Q}^{\prime}_2 , i^{\prime}_3 \in \mathcal{Q}^{\prime}_3}{\prod}         C^{l^{\prime}_{i^{\prime}j^{\prime}k^{\prime}}}_{i^{\prime}j^{\prime}k^{\prime}}            \bigg)        \bigg) \bigotimes                \bigg(     \underset{1\leq k \leq n-3}{\bigotimes}      \textbf{I}_k       \bigg)      \bigg)     \\   -    \omega_{\mathrm{XOR} \wedge \cdots \wedge \mathrm{XOR}}    \bigg(       \pm \mathrm{sign} \big( i_1 , j_1 , k_1, i_{111}, \cdots , j_{111} , \cdots , j_{nm(n+m)} \big)    \mathrm{sign} \big(  i^{\prime}_1 , j^{\prime}_1 , k^{\prime}_1, i^{\prime}_{111}, \cdots , j^{\prime}_{111} \\ , \cdots , j^{\prime}_{nm(n+m)}          \big)     \bigg) \bigg[                            \textbf{I} \bigotimes \textbf{I} \bigotimes \bigg(           \bigg(   \underset{k \in \mathcal{Q}_3, \mathcal{Q}_3 \oplus 1 \equiv \mathcal{Q}_3 + 1}{\underset{j \in \mathcal{Q}_2 , \mathcal{Q}_2 \oplus 1 \equiv \mathcal{Q}_2 + 1 }{\underset{i \in \mathcal{Q}_1, \mathcal{Q}_1 \oplus 1 \equiv \mathcal{Q}_1 +1  }{\prod}}}             C^{l_{ijk}}_{ijk}     \bigg) \wedge \bigg(  \underset{k^{\prime} \in \mathcal{Q}^{\prime}_3, \mathcal{Q}^{\prime}_3 \oplus 1 \equiv \mathcal{Q}^{\prime}_3 + 1}{\underset{j^{\prime} \in \mathcal{Q}^{\prime}_2 , \mathcal{Q}^{\prime}_2 \oplus 1 \equiv \mathcal{Q}^{\prime}_2 + 1 }{\underset{i^{\prime} \in \mathcal{Q}^{\prime}_1, \mathcal{Q}^{\prime}_1 \oplus 1 \equiv \mathcal{Q}^{\prime}_1 +1  }{\prod}}}       C^{l^{\prime}_{i^{\prime}j^{\prime}k^{\prime}}}_{i^{\prime}j^{\prime}k^{\prime}}            \bigg)         \bigg) \\ \bigotimes \bigg( \bigg(   \underset{1\leq k \leq n-3}{\prod}  \textbf{I}_k \bigg)      \bigg)             \bigg] \bigg) \bigg]              \ket{\psi_{\mathrm{XOR} \wedge \cdots \wedge \mathrm{XOR}}}   \bigg| \bigg|            \text{. }
\end{align*}

\noindent The inequality of the form above represents a generalization of an error bound for the two-player, and higher number of players, in the $\mathrm{XOR}$ game. For example, besides the error bounds that can be obtained by duality for the $\mathrm{XOR}^{*}$ game from the $\mathrm{XOR}$ from previous work of the author, {[44]}, the error bound inequality, given the $4$ $\mathrm{XOR}$ optimal value,

\begin{align*}
    \omega_{4\mathrm{XOR}} \big( G \big)  \equiv \omega \big( 4 \mathrm{XOR} \big) \text{, }
\end{align*}

\noindent takes the form,

\begin{align*}
    \bigg| \bigg|   \bigg[     \bigg(   \bigg(  \underset{1 \leq i \leq n}{\prod}     A^{j_i}_i           \bigg) - \bigg(   {\underset{1 \leq i \leq n}{\prod}}     A^{j_i}_i        \bigg)    \bigg( \frac{\pm A_k + A_l}{\sqrt{2}} \bigg)       \bigg)  \otimes \textbf{I} \otimes \textbf{I} \otimes \textbf{I} \bigg]     \ket{\psi_{4\mathrm{XOR}}}  -        \omega_{4\mathrm{XOR}}    \\ \times      \bigg[               \bigg(  \bigg(  \underset{1 \leq i \leq n}{\prod}     A^{j_i}_i   \bigg) A_k  -   \mathrm{sign} \big( i_1, j_1,  \cdots, j_n \big)     \bigg(    \underset{\text{set } j+1 \equiv j \oplus 1}{\underset{i \in \mathcal{Q}_1, j \in \mathcal{Q}_2}{\prod}}     A^{j_i}_i       \bigg)  \bigg) \otimes \textbf{I} \otimes \textbf{I} \otimes \textbf{I}   \bigg]        \ket{\psi_{4\mathrm{XOR}}}     \bigg| \bigg|       \text{, }
\end{align*}

\noindent for the optimal four-player strategy $\ket{\psi_{4\mathrm{XOR}}}$. For a sufficiently small parameter, the bias after an arbitrary number of strong parallel repetitions, for $\epsilon$-approximality, stipulates,

\begin{align*}
     \big( 1 - \epsilon_{\mathrm{XOR} \wedge \cdots \wedge \mathrm{XOR}} \big)    \beta_{\mathrm{XOR} \wedge \cdots \wedge \mathrm{XOR}}        \big( G \big) \leq \underset{\mathcal{Q}^{(1)}_1 , \cdots , \mathcal{Q}^{(1)}_N}{\sum} \cdots \underset{\mathcal{Q}^{(n)}_1 , \cdots , \mathcal{Q}^{(n)}_N}{\sum}  \bra{\psi_{\mathrm{XOR \wedge \cdots \wedge \mathrm{XOR}}}}  \\ \times  \bigg(            \bigg( \underset{\mathcal{Q}^{(1)}_1 , \cdots , \mathcal{Q}^{(1)}_N}{\underset{\#  \text{ players}}{\bigotimes}}    \text{Player observables}   \bigg)     \wedge    \cdots \wedge      \bigg( \underset{\mathcal{Q}^{(n)}_1 , \cdots , \mathcal{Q}^{(n)}_N}{\underset{\#  \text{ players}}{\bigotimes}}    \text{Player observables}   \bigg)           \bigg) \\ \times \ket{\psi_{\mathrm{XOR \wedge \cdots \wedge \mathrm{XOR}}}}     \leq \beta_{\mathrm{XOR} \wedge \cdots \wedge \mathrm{XOR}} \big( G \big)           \text{. }
\end{align*}

\noindent After taking the supremum over all possible strategies, $\epsilon$-approximality for the bias of the strong parallel repetition, $\mathrm{XOR} \wedge \cdots \wedge \mathrm{XOR}$, above, is of the form,

\begin{align*}
     \big( 1 - \epsilon_{\mathrm{XOR} \wedge \cdots \wedge \mathrm{XOR}} \big)  \omega_{\mathrm{XOR}} \big( G \big)^n    \leq \underset{\mathcal{Q}^{(1)}_1 , \cdots , \mathcal{Q}^{(1)}_N}{\sum} \cdots \underset{\mathcal{Q}^{(n)}_1 , \cdots , \mathcal{Q}^{(n)}_N}{\sum}  G_{\mathrm{XOR} \wedge \cdots \wedge \mathrm{XOR}} \bra{\psi_{\mathrm{XOR \wedge \cdots \wedge \mathrm{XOR}}}}  \\ \times  \bigg(            \bigg( \underset{\mathcal{Q}^{(1)}_1 , \cdots , \mathcal{Q}^{(1)}_N}{\underset{\#  \text{ players}}{\bigotimes}}    \text{Player observables}   \bigg)  \wedge    \cdots \wedge      \bigg( \underset{\mathcal{Q}^{(n)}_1 , \cdots , \mathcal{Q}^{(n)}_N}{\underset{\#  \text{ players}}{\bigotimes}}    \text{Player observables}   \bigg)           \bigg) \\ \times  \ket{\psi_{\mathrm{XOR \wedge \cdots \wedge \mathrm{XOR}}}}     \leq \omega_{\mathrm{XOR} } \big( G \big)^n            \text{, }
\end{align*}

\noindent where the strong parallel repetition game matrix, $G_{\mathrm{XOR} \wedge \cdots \wedge \mathrm{XOR}}$, is of the form,

\begin{align*}
   G_{\mathrm{XOR} \wedge \cdots \wedge \mathrm{XOR}} \equiv    \underset{1 \leq i \leq n}{\bigwedge}  G^{(i)}_{\mathrm{XOR} \wedge \cdots \wedge \mathrm{XOR}}   \equiv G^{\wedge (i)}_{\mathrm{XOR}}  \text{. }
\end{align*}

\noindent Under the operation,

\begin{align*}
{\bigwedge} \cdot    \text{, }
\end{align*}

\noindent the duality gap can be formulated with the semidefinite program,

\begin{align*}
     \underset{1 \leq j \leq n}{\bigwedge} \bigg[   \bigg(      \overset{m}{\underset{i=1}{\sum}}  y^{(j)}_i F^{(j)}_i - G^{(j)}_{\mathrm{XOR}} \bigg) \cdot Z^{(j)}_{\mathrm{XOR}}         \bigg]      \geq 0     \text{, }
\end{align*}

\noindent for each primal feasible solution $Z^{(j)}$, for each strong parallel repetition. From previously defined quantities, as in the two-player setting, the tensor observables for any player can be expressed as,

\begin{align*}
  \text{$N$ th player tensor observable} \big( i_1 , \cdots , i_N \big) \equiv \frac{1}{\sqrt{2}}  \bigg(  \text{($N$-1) th player tensor observable} \big( i_1 \\ , \cdots , i_{N-1} \big)    +   \text{($N$-1) th player tensor observable} \big( i^{\prime}_1 , \cdots , i^{\prime}_{N-1} \big)  \bigg)      \text{, }
\end{align*}

\noindent which can be equivalently expressed with,

\begin{align*}
 \big( \mathscr{P}_N \big)_{(i_1, \cdots , i_N )}  \equiv \frac{1}{\sqrt{2} } \bigg(         \big( \mathscr{P}_{N-1} \big)_{(i_1 , \cdots , i_{N-1})}      + \big( \mathscr{P}_{N-1} \big)_{(i^{\prime}_1 , \cdots , i^{\prime}_{N-1})} \bigg)   \text{. }
\end{align*}

\noindent Iterating further, the fact that the tensor observable for responses from the $\big(n-1\big)$-th player, given the responses of the previous $\big(n-2\big)$ players, can be expressed as,

\begin{align*}
        \big( \mathscr{P}_{N-1} \big)_{(i_1, \cdots , i_{N-1} )}  \equiv \frac{1}{\sqrt{2} } \bigg(          \big( \mathscr{P}_{N-2} \big)_{(i_1 , \cdots , i_{N-2})}  +       \big( \mathscr{P}_{N-2} \big)_{(i^{\prime}_1 , \cdots , i^{\prime}_{N-2})} \bigg)            \text{, }
\end{align*}

\noindent implies that the equality,

\begin{align*}
 \big( \mathscr{P}_N \big)_{(i_1 , \cdots , i_n)} \equiv \frac{1}{\sqrt{2}} \bigg( \frac{1}{\sqrt{2}} \bigg(  \big( \mathscr{P}_{N-3} \big)_{(i_1 , \cdots, i_{N-3})}   + \big( \mathscr{P}_{N-3} \big)_{(i^{\prime}_1 , \cdots , i^{\prime}_{N-3})}   \bigg) \\ + \frac{1}{\sqrt{2}}  \bigg( \big( \mathscr{P}_{N-3} \big)_{(i^{\prime\prime}_1 , \cdots , i^{\prime\prime}_{N-3})}  +    \big( \mathscr{P}_{N-3} \big)_{(i^{\prime\prime\prime}_1 , \cdots , i^{\prime\prime\prime}_{N-3})}          \bigg)  \bigg)    \text{, }
\end{align*}

\noindent holds between the $N$ th player tensor observable, and the $(N-3)$ th player tensor observable, for the sets of possible questions $\big( i^{\prime\prime}_1 , \cdots , i^{\prime\prime}_{n-3} \big)$, and $\big( i^{\prime\prime\prime}_1 , \cdots , i^{\prime\prime\prime}_{n-3} \big)$. As done previously, iterating the equality above, where at each step the higher-dimensional tensor player observable can be related to a linear combination of two lower-dimensional player observables, with an approximation of $\frac{1}{\sqrt{2}}$, implies that one has,

\begin{align*}
       \big( \mathscr{P}_N \big)_{(i_1 , \cdots , i_N)} \equiv \bigg(  \frac{1}{\sqrt{2}} \bigg)^N    \underset{j_N , ( i^{\prime \cdots \prime}_1 , \cdots , i^{\prime\cdots \prime}_N)}{\underset{\vdots}{\underset{j_1 , ( i_1 , \cdots , i_N)}{\underset{1 \leq i \leq N}{ \sum}}}}  \text{$1$ st player tensor observables} \big( j_i , \big( i_1 , \cdots , i_n \big)  \big) \\  \equiv        \bigg(  \frac{1}{\sqrt{2}} \bigg)^N  \underset{j_N , ( i^{\prime \cdots \prime}_1 , \cdots , i^{\prime\cdots \prime}_N)}{\underset{\vdots}{\underset{j_1 , ( i_1 , \cdots , i_n)}{\underset{1 \leq i \leq N}{ \sum}}}} \big( \mathscr{P}_1 \big)_{(i_1,\cdots, i_N)}                \text{. }
\end{align*}

\noindent The summation over $j_1,\cdots, j_N$, and $(i_1 , \cdots , i_N), \cdots, ( i^{\prime \cdots \prime}_1 , \cdots , i^{\prime\cdots \prime}_N)$ is the generalization of,

\begin{align*}
      \underset{i \in \mathcal{Q}_2}{\bigcup}  B^{(l_i)}_{j_1 \cdots j_N}    \equiv  \underset{i \in \mathcal{Q}_2}{\bigcup}   \bigg( \frac{1}{\sqrt{N}} A^{(l_1)}_{j_1}  +   \cdots +    \frac{1}{\sqrt{N}} A^{(l_N)}_{j_N}   \bigg)^{\textbf{T}}  = \frac{1}{\sqrt{N}} \underset{i \in \mathcal{Q}_2}{\bigcup}  \bigg( A^{(l)}_j + \cdots + A^{(l)}_k  
 \bigg)^{\textbf{T}} \\   \equiv \frac{1}{\sqrt{N}}  \underset{i \in \mathcal{Q}_2}{\bigcup}   \bigg( \big( A^{(l_1)}_{j_1}\big)^{\textbf{T}}   + \cdots +    \big( A^{(l_N)}_{j_N} \big)^{\textbf{T}} \bigg) \text{, }
\end{align*}

\noindent mentioned from the first subsection, with,

\begin{align*}
    \underset{i \in \mathcal{Q}_{\text{Player } i}}{\underset{1 \leq i \leq N}{\bigcup}}     \bigg( \frac{1}{\sqrt{N}} A^{(l_1)}_{j_1}  +   \cdots +    \frac{1}{\sqrt{N}} A^{(l_N)}_{j_N}   \bigg)^{\textbf{T}} \equiv  \underset{i \in \mathcal{Q}_{\text{Player } i}}{\underset{1 \leq i \leq N}{\bigcup}} \bigg(  \frac{1}{\sqrt{N}} \bigg( A^{(l_1)}_{j_1} \bigg)^{\textbf{T}}  + \\ \cdots +     \frac{1}{\sqrt{N}} \bigg(   A^{(l_N)}_{j_N}    \bigg)^{\textbf{T}} \bigg)  \text{. }
\end{align*}

\noindent The above decomposition of tensor observables for any player of the game, in addition to primal feasible solutions whose existence is guaranteed by well posed semidefinite programs, is related to the Bell states of the $\mathrm{CHSH}\big( n\big)$ game when $n \equiv 2$. From $\epsilon$-approximality of the bias of games stated under various circumstances, under strong parallel repetition one has that an inequality of the form,

\begin{align*}
  \omega \big( \mathrm{XOR} \wedge \cdots \wedge \mathrm{XOR} \big) \big( 1 - \epsilon_{\mathrm{XOR} \wedge \cdots \wedge \mathrm{XOR}} \big)   \equiv   \big( \omega \big( \mathrm{XOR} \big) \big)^n \big( 1 - \epsilon_{\mathrm{XOR} \wedge \cdots \wedge \mathrm{XOR}} \big)  \\   \leq      \underset{\mathcal{Q}^{(n)}_1, \cdots, \mathcal{Q}^{(n)}_N}{\underset{\vdots}{\underset{\mathcal{Q}^{(1)}_1, \cdots, \mathcal{Q}^{(1)}_N}{\sum}}}           G_{\mathrm{XOR} \wedge \cdots \wedge \mathrm{XOR}}  \bra{\psi_{\mathrm{XOR} \wedge  \cdots \wedge \mathrm{XOR}}}         \bigg(      \bigg(    \big( \mathscr{P}^{(1)}_1 \big)_{ij} \wedge \cdots \wedge \big(    \mathscr{P}^{(n)}_N   \big)_{ij} \bigg) \otimes \cdots  \\  \otimes     \bigg(   \big(              \mathscr{P}^{(n)}_1     \big)_{ij} \wedge \cdots   \wedge \big( \mathscr{P}^{(n)}_N   \big)_{ij}      \bigg)       \bigg) \ket{\psi_{\mathrm{XOR} \wedge \cdots \wedge \mathrm{XOR}}} \leq \big( \omega \big( \mathrm{XOR} \big) \big)^n  \\ \equiv \omega \big( \mathrm{XOR} \wedge \cdots \wedge \mathrm{XOR} \big)  \text{, }
\end{align*}

\noindent holds, for $\epsilon_{\mathrm{XOR}\wedge \cdots \wedge \mathrm{XOR}}$ taken to be sufficiently small. The quantum state corresponding to the optimal strategies of players in the $\mathrm{XOR}$ game under strong parallel repetition admits the decomposition,

\begin{align*}
  \bra{\psi_{\mathrm{XOR} \wedge \cdots \wedge \mathrm{XOR}} } \equiv \bra{\psi_{\mathrm{XOR}}} \wedge \cdots \wedge \bra{\psi_{\mathrm{XOR}}}  \equiv  \underset{\text{strong parallel repetitions } j}{\bigwedge}   \bra{\psi_{\mathrm{XOR}}}^j \equiv \bra{\psi_{\mathrm{XOR}}}^{\wedge j}  \text{, } \\    \ket{\psi_{\mathrm{XOR} \wedge \cdots \wedge \mathrm{XOR}} } \equiv   \ket{\psi_{\mathrm{XOR}}} \wedge \cdots \wedge \ket{\psi_{\mathrm{XOR}}}  \equiv \underset{\text{strong parallel repetitions } j}{\bigwedge} \ket{\psi_{\mathrm{XOR}}}^j   \equiv \ket{\psi_{\mathrm{XOR}}}^{\wedge j}    \text{, }
\end{align*}

\noindent which implies that the following inequalities hold for the supremum over all strategies, $\mathcal{S}$,

\begin{align*}
     \beta_{\mathrm{XOR} \wedge \cdots \wedge \mathrm{XOR}} \big( G \big)   \big( 1 - \epsilon_{\mathrm{XOR} \wedge \cdots \wedge \mathrm{XOR}} \big)   \leq    \underset{\mathcal{Q}^{(n)}_1, \cdots, \mathcal{Q}^{(n)}_N}{\underset{\vdots}{\underset{\mathcal{Q}^{(1)}_1, \cdots, \mathcal{Q}^{(1)}_n}{\sum}}}  \bigg( G_{\mathrm{XOR} \wedge \cdots \wedge \mathrm{XOR}} \bigg)_{\mathcal{Q}^{(i)}_j}         \\ \times \bra{\psi_{\mathrm{XOR} \wedge \cdots \wedge \mathrm{XOR}} }   \bigg(  \bigg(  \big(   \mathscr{P}^{(1)}_i   \big)_{ij} \wedge \cdots  \wedge   \big(    \mathscr{P}^{(1)}_N   \big)_{ij}  \bigg)  \wedge \cdots \wedge \bigg(     \big(   \mathscr{P}^{(n)}_i   \big)_{ij} \wedge \cdots \\  \wedge   \big(    \mathscr{P}^{(n)}_N   \big)_{ij}                    \bigg)   \bigg)          \ket{\psi_{\mathrm{XOR} \wedge \cdots \wedge \mathrm{XOR}} }               \leq    \beta_{\mathrm{XOR} \wedge \cdots \wedge \mathrm{XOR}} \big( G \big)   \text{. }
\end{align*}

\noindent For a suitable linear operator $T^{\prime\prime} \equiv T_{\mathrm{XOR} \wedge \cdots \wedge \mathrm{XOR}}$,

\begin{align*}
     \bigg| \bigg| \bigg( T^{\prime\prime} \bigotimes \bigg(  \bigg( \bigg(   \underset{i_1 \in \mathcal{Q}_1, i_2 \in \mathcal{Q}_2, \cdots, i_n \in \mathcal{Q}_n}{\prod}   \text{Tensors of player}^{(1)} \text{ observables} \big( i_1, \cdots, i_n \big)  \bigg) \wedge \cdots \\ \wedge  \bigg(    \underset{i_1 \in \mathcal{Q}_1, i_2 \in \mathcal{Q}_2, \cdots, i_n \in \mathcal{Q}_n}{\prod}   \text{Tensors of player}^{(n)} \text{ observables} \big( i_1, \cdots, i_n \big)    \bigg)     \bigg)     \bigg)  \bigg) \\    -   \bigg( \bigg(   \bigg(  \widetilde{\underset{i_1 \in \mathcal{Q}_1, i_2 \in \mathcal{Q}_2, \cdots, i_n \in \mathcal{Q}_n}{\prod}   \text{Tensors of player}^{(1)} \text{ observables} \big( i_1, \cdots, i_n \big) }           \bigg) \wedge \cdots \\ \wedge \bigg(   \widetilde{\underset{i_1 \in \mathcal{Q}_1, i_2 \in \mathcal{Q}_2, \cdots, i_n \in \mathcal{Q}_n}{\prod}   \text{Tensors of player}^{(n)}  \text{ observables} \big( i_1, \cdots, i_n \big)  }               \bigg)               \bigg) \bigotimes T^{\prime\prime} \bigg)   \bigg| \bigg|          \text{. }
\end{align*}

\noindent Given the action of the suitable linear operator above, with respect to strong parallel repetition the error bounds for tensors of player observables take the form,

\begin{align*}
        \bigg| \bigg|       \bigg[                       \bigg( \textbf{I} \bigotimes \bigg(              \big(   \text{2  nd player}^{(1)} \text{ tensor observable}   \big( i_1, i_2 \big) \big) \wedge \cdots \wedge \big(     \text{2  nd player}^{(n)} \text{ tensor observable}  \\ \big( i_1, i_2 \big)     \big)  \bigg)       \bigotimes \overset{N-3}{\cdots}  \bigotimes  \bigg(     \big(   \text{N  th player}^{(1)} \text{ tensor observable}   \big( i_1  , \cdots, i_n \big) \big)  \wedge \cdots \wedge \big(     \text{N th player}^{(n)} \\ \text{ tensor observable} \big( i_1,  \cdots, i_n \big)     \big)                          \bigg)     \bigg)       -     \bigg(   \bigg(            \big( \text{2 nd player}^{(1)} \text{ observable} \big( i_1, i_2 \big)  \big) \wedge \cdots \wedge \big(  \text{2 nd player}^{(n)} \\ \text{ observable} \big( i_1, i_2 \big)    \big)      \bigg)  \bigotimes     \textbf{I} \bigotimes \bigg( \big(    \text{3 rd  player}^{(1)}   \text{observable}   \big( i_1,i_2,i_3 \big)     \big) \wedge \cdots  \wedge \big(       \text{3 rd player}^{(n)} \\ \text{ observable} \big( i_1,i_2,i_3 \big)              \big)  \bigg)  \bigotimes \overset{N-4}{\cdots} \bigotimes \bigg(   \big(   \text{N th player}^{(1)}         \text{tensor observable}  \big( i_1, \cdots, i_n \big)     \big) \\ \wedge \cdots \wedge \big(   \text{N th player}^{(n)} \text{ tensor observable} \big( i_1  , \cdots, i_n \big)      \big)         \bigg)                                                    \bigg)      \bigg] \ket{\psi_{\mathrm{XOR} \wedge \cdots \wedge \mathrm{XOR}} } \bigg| \bigg|    \text{,} \end{align*}

\begin{align*}    \bigg| \bigg|       \bigg[                       \bigg(                  \bigg( \text{1 st player}^{(1)} \text{ observable}  \big(i_1 \big) \wedge \cdots \wedge    \big( \text{1 st player}^{(n)} \text{ observable}  \big(i_1 \big)  \big) \bigg)  \bigotimes \textbf{I} \bigotimes  \bigg( \big(      \text{3 rd} \\  \text{ player}^{(1)} \text{ observable} \big(i_1, i_2, i_3 \big) \big)  \wedge  \cdots \wedge \big( \text{3 rd player}^{(n)} \text{ observable} \big(i_1, i_2, i_3 \big) \big)    \bigg) \bigotimes \overset{N-4}{\cdots} \\ \bigotimes \bigg( \big(         \text{N th player}^{(1)} \text{ observable}  \big( i_1, \cdots, i_n \big)     \big)     \wedge \cdots \wedge \big(     \text{N th player}^{(n)} \text{ observable}  \big( i_1, \cdots, i_n \big)         \big)     \bigg)       \bigg)       \\  -     \bigg(      \bigg( \text{1 st player}^{(1)} \text{ observable} \big( i_1 \big)  \wedge \cdots \wedge     \big( \text{1 st player}^{(n)} \text{ observable} \big( i_1 \big) \big) \bigg)    \bigotimes  \textbf{I}  \bigotimes  \bigg(   \big(      \text{3 rd} \\ \text{player}^{(1)} \text{ observable}  \big(i_1, i_2, i_3 \big) \big)  \wedge  \cdots \wedge \big( \text{3 rd player}^{(n)} \text{ observable}       \big(i_1, i_2, i_3 \big) \big)                                                                     \bigg)     \bigotimes \overset{N-4}{\cdots} \\ \bigotimes \bigg(                 \big(         \text{N th player}^{(1)} \text{ observable}  \big( i_1, \cdots, i_n \big)     \big)    \wedge {\cdots}   \wedge \big(     \text{N th player}^{(n)} \text{ observable}  \big( i_1 \\ ,       \cdots, i_n \big)         \big)          \bigg)                      \bigg)         \bigg] \ket{\psi_{\mathrm{XOR} \wedge \cdots \wedge \mathrm{XOR}}} \bigg| \bigg|           \text{, } \end{align*}

  \begin{align*}     \bigg| \bigg|       \bigg[ \bigg[  \bigg(              \bigg(   \bigg(                  \underset{i_1 \in \mathcal{Q}_1}{\prod}  \text{1 st player}^{(1)} \text{ observable} \big( i_1 \big) \bigg)   \wedge \cdots \wedge    \bigg(      \underset{i_1 \in \mathcal{Q}_1}{\prod}  \text{1 st player}^{(n)} \text{ observable} \big( i_1 \big)     \bigg)           \bigotimes    \textbf{I}     \\ \bigotimes    \bigg(  \bigg(   \underset{i_3 \in \mathcal{Q}_3}{\prod}    \text{3 rd player}^{(1)}    \text{observable} \big( i_1, i_2, i_3 \big)      \bigg) \wedge \cdots \wedge \bigg(              \underset{i_3 \in \mathcal{Q}_3}{\prod}    \text{3 rd player}^{(n)}    \\ \text{observable} \big( i_1,  i_2, i_3 \big)       \bigg)        \bigg)                \bigg)     \bigg)        -     \bigg(  \textbf{I} \bigotimes \bigg( \bigg(       \underset{ i_2 \in \mathcal{Q}_2}{\prod}         \text{2 nd player}^{(1)} \text{ observable} \big( i_1,i_2 \big)      \bigg) \wedge \cdots \\   \wedge \bigg(                  \underset{i_2 \in \mathcal{Q}_2}{\prod}         \text{2 nd player}^{(n)} \text{ observable} \big( i_1,i_2 \big)         \bigg) \bigg)                \bigotimes                    \bigg(  \bigg(   \underset{i_3 \in \mathcal{Q}_3}{\prod}    \text{3 rd player}^{(1)}    \text{observa-}  \\   \text{ble} \big( i_1, i_2, i_3 \big)      \bigg) \wedge \cdots \wedge \bigg(              \underset{i_3 \in \mathcal{Q}_3}{\prod}    \text{3 rd player}^{(n)}    \text{observable} \big( i_1  ,  i_2, i_3 \big)       \bigg)        \bigg)                                    \bigg)            \bigg]  \\ \times \ket{\psi_{\mathrm{XOR} \wedge \cdots \wedge \mathrm{XOR}}}          \bigg| \bigg|       \text{. }
\end{align*}

\noindent Under strong parallel repetition, the $N$-$\mathrm{XOR}$ optimal value satisfying,

\begin{align*}
  \omega_{\mathrm{XOR} \wedge \cdots \wedge \mathrm{XOR}} \big( G_{\mathrm{XOR} \wedge \cdots \wedge \mathrm{XOR}} \big)  \equiv \omega \big(  \mathrm{XOR} \wedge \cdots \wedge \mathrm{XOR} \big) \equiv \underset{\# \text{ of strong parallel repetitions } j}{\prod}   \omega \big( \mathrm{XOR}\big)^{j} \text{,} \end{align*}

\noindent in comparison to the $\mathrm{FFL}$ optimal value satisfying,

\begin{align*}
  \omega_{\mathrm{FFL} \wedge  \mathrm{FFL}} \big( G_{\mathrm{FFL}} \big) \equiv \omega_{\mathrm{FFL}} \big( G_{\mathrm{FFL}} \wedge G_{\mathrm{FFL}} \big)   \equiv \omega \big(  \mathrm{FFL}  \wedge \mathrm{FFL} \big) \equiv  \omega \big( \mathrm{FFL}\big) \equiv \frac{2}{3} \text{,} \end{align*}

  \noindent implies that additional inequalities take the form,

  \begin{align*}
            \bigg| \bigg|       \bigg[  \bigg( \bigg( \bigg( \text{1 st player}^{(1)} \text{ observable} \big( i_1 \big) \bigg) \wedge \cdots \wedge \bigg( \text{1 st player}^{(n)} \text{ observable} \big( i_1 \big) \bigg) \bigg)    \bigotimes \textbf{I} \\ \bigotimes  \bigg( \bigg(  \underset{i_3 \in \mathcal{Q}_3}{\prod} \text{3 rd}  \text{ player}^{(1)} \text{ observable} \big( i_1, i_2, i_3 \big)               \bigg) \wedge \cdots \wedge \bigg(  \underset{i_3 \in \mathcal{Q}_3}{\prod} \text{3 rd player}^{(n)} \\ \text{ observable} \big( i_1, i_2, i_3 \big)           \bigg) \bigg)                \bigg)     -     \bigg( \textbf{I} \bigotimes   \bigg[ \big( \omega_{\mathrm{XOR}} \big)^n    \bigg[           \frac{\mathscr{T}^{\prime}_2}{\big| \pm \mathscr{T}^{\prime}_2 \big| }    \bigg] \bigg]               \bigg)  \bigg]  \ket{\psi_{\mathrm{XOR} \wedge \cdots \wedge\mathrm{XOR}}}        \bigg| \bigg|       \text{,} \end{align*}
            
\noindent for,

\begin{align*}
     \mathscr{T}^{\prime}_2 \equiv          \bigg( \big( \text{2 nd player}^{(1)} \text{ observable} \big(i_1, i_2 \big) \big) \wedge \cdots \wedge \big(  \text{2 nd player}^{(n)} \text{ observable} \big(i_1,  i_2 \big)       \big)  \bigg)   \\ + \underset{\text{Permutations } \sigma_1,\cdots, \sigma_n}{\sum}          \bigg( \big(        \text{Tensors of player}^{(1)} \text{ observables} \big( \sigma_1 i_1, \cdots, \sigma_n i_n \big)      \big) \wedge \cdots \wedge \big(   \text{Tensors} \\ \text{of player}^{(n)}   \text{ observables} \big( \sigma_1 i_1, \cdots, \sigma_n i_n \big)               \big) \bigg)            \text{, }
\end{align*}

\noindent and,

            \begin{align*}
            \bigg| \bigg|       \bigg[  \bigg( \textbf{I} \bigotimes \textbf{I} \bigotimes \bigg(         \bigg(   \underset{i_3 \in \mathcal{Q}_3}{\prod}   \text{3 rd player}^{(1)} \text{ observables} \big(i_1, i_2, i_3 \big)      \bigg) \wedge \cdots \\ \wedge  \bigg(    \underset{i_3 \in \mathcal{Q}_3}{\prod}   \text{3 rd player}^{(n)}  \text{ observables} \big(i_1, i_2, i_3 \big)       \bigg)       \bigotimes \textbf{I} \bigotimes \overset{N-6}{\cdots} \bigotimes \textbf{I}  \\ \bigotimes \bigg(     \bigg(        \text{N th player}^{(1)} \text{ observable} \big(i_1, \cdots, i_n \big)    \bigg) \wedge \cdots    \wedge \bigg(  \bigg(        \text{N th player}^{(n)} \text{ observable} \big(i_1 \\ , \cdots, i_n \big)   \bigg)                \bigg)       \bigg) \bigg)                   \bigg)        -    \bigg(   \textbf{I} \bigotimes         \textbf{I} \bigotimes \overset{N-5}{\cdots} \bigotimes \textbf{I}  \bigotimes  \bigg(        \big( \text{N th player}^{(1)}     \text{ observable} \big(i_1,    \cdots, i_n \big) \big) \wedge \\ \cdots \wedge \big( \text{N th player}^{(n)} \text{ observable} \big(i_1, \cdots, i_n \big)  \big)         \bigg)     \bigotimes \bigg( \frac{1}{\sqrt{N-1}}   \bigg( \frac{\mathscr{T}^{\prime\prime}_3}{\big| \pm \mathscr{T}^{\prime\prime}_3 \big| }   \bigg)     \bigg)         \bigg] \\ \times \ket{\psi_{\mathrm{XOR} \wedge \cdots \wedge \mathrm{XOR}}} \bigg| \bigg|             \text{,}
            \end{align*}

            \noindent for,

            \begin{align*}
         \mathscr{T}^{\prime\prime}_3 \equiv     \bigg(      \big(   \text{3 rd player}^{(1)} \text{ observable} \big(i_1,   i_2, i_3 \big)                    \big) \wedge \cdots \wedge \big(        \text{3 rd player}^{(n)} \text{ observable} \big(i_1, i_2, i_3 \big)             \big)      \bigg) \\    + \underset{\text{Permutations } \sigma_1, \sigma_2, \sigma_3}{\sum}   \bigg(     \big(    \text{3 rd player}^{(1)} \text{ observable} \big(\sigma_1 i_1, \sigma_2 i_2, \sigma_3 i_3 \big)                 \big) \\ \wedge \cdots \wedge \big(      \text{3 rd player}^{(n)} \text{ observable} \big( \sigma_1 i_1, \sigma_2 i_2, \sigma_3 i_3 \big)                 \big)              \bigg)              \text{. }
            \end{align*}

            \noindent Defining,
            
            \begin{align*}
        \underset{1 \leq j \leq n}{\bigcup} \big( \mathscr{P}^{(j)}_1 \big)_{i_1} \equiv  \underset{1 \leq j \leq n}{\bigcup} \big\{ \text{$j$ th strong parallel repetition of Player 1 in response to question } i_1  \big\} \text{, }\\ \vdots \\  \underset{1 \leq j \leq n}{\bigcup} \big( \mathscr{P}^{(j)}_N \big)_{i_N} \equiv  \underset{1 \leq j \leq n}{\bigcup} \big\{ \text{$j$ th strong parallel repetition of Player 1 in response to question } i_N  \big\} \text{, }
            \end{align*}
            
            \noindent as in previous remarks on the decomposition of the tensor observable for the $N$ th player,

            \begin{align*}
 \big( \mathscr{P}_N \big)_{(i_1, \cdots , i_N )}  \equiv \frac{1}{\sqrt{2} } \bigg(         \big( \mathscr{P}_{N-1} \big)_{(i_1 , \cdots , i_{N-1})}      + \big( \mathscr{P}_{N-1} \big)_{(i^{\prime}_1 , \cdots , i^{\prime}_{N-1})} \bigg)   \text{, }
\end{align*}

\noindent imply, under the strong parallel repetition operation, that,

   \begin{align*}         
           \underset{\mathcal{Q}^{(N)}_1,\cdots, \mathcal{Q}^{(N)}_n}{\underset{\vdots}{\underset{\mathcal{Q}^{(1)}_1, \cdots, \mathcal{Q}^{(n)}_1}{\sum}}} \bigg| \bigg|       \bigg[     \bigg(   \bigg( \frac{\big( \big( \mathscr{P}^{(1)}_1 \big)_i \wedge \cdots \wedge \big( \mathscr{P}^{(n)}_1 \big)_i  \big) + \big( \big(          \big( \mathscr{P}^{(1)}_1 \big)_j       \big) \wedge \cdots \wedge \big(          \big( \mathscr{P}^{(n)}_1 \big)_j   \big)  \big) }{\sqrt{2}}    \bigg)     \bigotimes \\ \bigg( \underset{1 \leq k \leq N-1}{\bigotimes} \textbf{I}_k \bigg)  \bigg)       - \bigg( \bigg( \underset{1 \leq k \leq N-1}{\bigotimes} \textbf{I}_k \bigg) \bigotimes  \bigg(  \big( \mathscr{P}^{(1)}_1 \big)_{ij} \wedge \cdots \wedge \big( \mathscr{P}^{(N)}_1 \big)_{ij}     \bigg)  \bigg) \bigg] \\ \times \ket{\psi_{\mathrm{XOR} \wedge \cdots \wedge \mathrm{XOR}}} \bigg| \bigg|           \text{, } 
  \end{align*}

  \noindent As a related consequence of the inequality above,

  \begin{align*}
        \bigg| \bigg| \bigg[  \bigg(         \bigg(  \big( \mathscr{P}^{(1)}_1     \big)_1  \wedge \cdots \wedge \big(  \mathscr{P}^{(n)}_1 \big)_1 \bigg)        \bigotimes \bigg( \underset{1 \leq k \leq N-1}{\bigotimes} \textbf{I}_k \bigg)\bigg)  \bigg( T_{\mathrm{XOR}} \wedge T_{\mathrm{XOR}} \bigg)  -   \bigg( T_{\mathrm{XOR}} \wedge T_{\mathrm{XOR}} \bigg) \\ \times   \bigg( \bigg( \underset{1 \leq k \leq N-1}{\bigotimes} \textbf{I}_k \bigg)  \bigotimes      \widetilde{\bigg(  \big( \mathscr{P}^{(1)}_1     \big)_1  \wedge \cdots \wedge \big(  \mathscr{P}^{(n)}_1 \big)_1 \bigg)}    \bigg)  \bigg]  \ket{\psi_{\mathrm{XOR} \wedge \cdots \wedge \mathrm{XOR}}} \bigg| \bigg| \end{align*}

        \begin{align*} \equiv    \bigg| \bigg| \bigg[  \bigg(         \bigg(  \big( \mathscr{P}^{(1)}_1     \big)_1  \wedge \cdots \wedge \big(  \mathscr{P}^{(n)}_1 \big)_1 \bigg)        \bigotimes \bigg( \underset{1 \leq k \leq N-1}{\bigotimes} \textbf{I}_k \bigg)\bigg)  \bigg( T_{\mathrm{XOR}} \wedge T_{\mathrm{XOR}} \bigg)  -   \bigg( T_{\mathrm{XOR}} \wedge T_{\mathrm{XOR}} \bigg)  \\ \times  \bigg( \bigg( \underset{1 \leq k \leq N-1}{\bigotimes} \textbf{I}_k \bigg)  \bigotimes      \bigg(  \widetilde{\big( \mathscr{P}^{(1)}_1     \big)_1}  \wedge \cdots \wedge \widetilde{\big(  \mathscr{P}^{(n)}_1 \big)_1} \bigg)   \bigg)  \bigg]  \ket{\psi_{\mathrm{XOR} \wedge \cdots \wedge \mathrm{XOR}}} \bigg| \bigg| \text{. }
  \end{align*}

\noindent Furthermore, as a generalization of a previous inequality,

\begin{align*}
    \bigg| \bigg|  \bigg[  \bigg(   \big( A_i \wedge A_{i^{\prime}} \big) \bigotimes \textbf{I}      \bigotimes \bigg(    \bigg(   \underset{i_1 \in \mathcal{Q}_1, i_2 \in \mathcal{Q}_2 , i_3 \in \mathcal{Q}_3}{\prod}             C^{l_{ijk}}_{ijk}     \bigg) \wedge \bigg(    \underset{i^{\prime}_1 \in \mathcal{Q}^{\prime}_1, i^{\prime}_2 \in \mathcal{Q}^{\prime}_2 , i^{\prime}_3 \in \mathcal{Q}^{\prime}_3}{\prod}         C^{l^{\prime}_{i^{\prime}j^{\prime}k^{\prime}}}_{i^{\prime}j^{\prime}k^{\prime}}            \bigg)        \bigg) \bigotimes                \bigg(     \underset{1\leq k \leq n-3}{\bigotimes}      \textbf{I}_k       \bigg)      \bigg)      \\  -    \omega_{\mathrm{XOR} \wedge \cdots \wedge \mathrm{XOR}}   \times   \bigg(       \pm \mathrm{sign} \big( i_1 , j_1 , k_1, i_{111}, \cdots , j_{111} , \cdots, k_{111}, \cdots, j_{nm(n+m)}, \cdots, k_{nm(n+m)} \big) \\ \times    \mathrm{sign} \big(  i^{\prime}_1 , j^{\prime}_1 , k^{\prime}_1, i^{\prime}_{111}, \cdots , j^{\prime}_{111} , \cdots ,     k^{\prime}_{111}, \cdots, j^{\prime}_{nm(n+m)}, \cdots, k^{\prime}_{nm(n+m)}          \big)     \bigg)  \bigg[                            \textbf{I} \bigotimes \textbf{I} \\ \bigotimes \bigg(           \bigg(   \underset{i_3 \in i\mathcal{Q}_3, \mathcal{Q}_3 \oplus 1 \equiv \mathcal{Q}_3 + 1}{\underset{i_2 \in \mathcal{Q}_2 , \mathcal{Q}_2 \oplus 1 \equiv \mathcal{Q}_2 + 1 }{\underset{i_1 \in \mathcal{Q}_1, \mathcal{Q}_1 \oplus 1 \equiv \mathcal{Q}_1 +1  }{\prod}}}             C^{l_{ijk}}_{ijk}     \bigg)  \wedge \bigg(  \underset{i^{\prime}_3 \in \mathcal{Q}^{\prime}_3, \mathcal{Q}^{\prime}_3 \oplus 1 \equiv \mathcal{Q}^{\prime}_3 + 1}{\underset{i^{\prime}_2 \in\mathcal{Q}^{\prime}_2 , \mathcal{Q}^{\prime}_2 \oplus 1 \equiv \mathcal{Q}^{\prime}_2 + 1 }{\underset{i^{\prime}_1 \in \mathcal{Q}^{\prime}_1, \mathcal{Q}^{\prime}_1 \oplus 1 \equiv \mathcal{Q}^{\prime}_1 +1  }{\prod}}}       C^{l^{\prime}_{i^{\prime}j^{\prime}k^{\prime}}}_{i^{\prime}j^{\prime}k^{\prime}}            \bigg)         \bigg) \bigotimes \bigg( \bigg(   \underset{1\leq k \leq n-3}{\prod}  \textbf{I}_k \bigg)      \bigg)             \bigg] \bigg) \bigg]       \\ \times        \ket{\psi_{\mathrm{XOR} \wedge \cdots \wedge \mathrm{XOR}}}   \bigg| \bigg|            \text{, }
\end{align*}

\noindent from the fact that the optimal value, under strong parallel repetition, satisfies,

\begin{align*}
  \omega_{\mathrm{XOR} \wedge \cdots \wedge \mathrm{XOR}} \big( G_{\mathrm{XOR} \wedge \cdots \wedge \mathrm{XOR}} \big)  \equiv \omega \big(  \mathrm{XOR} \wedge \cdots \wedge \mathrm{XOR} \big) \equiv \underset{\# \text{ of strong parallel repetitions } j}{\prod}   \omega \big( \mathrm{XOR}\big)^{j} \text{,} \end{align*}

\noindent the following summation, with respect to the Frobenius norm, is also expected to have a suitable upper bound,

\begin{align*}
    \bigg| \bigg|  \bigg[  \bigg(   \bigg( \big( A_1 \big)^{(1)}_i \wedge \cdots \wedge  \big( A_n \big)^{(1)}_{i^{\prime}} \bigg) \bigotimes \textbf{I}      \bigotimes \bigg(    \bigg(   \underset{i_1 \in \mathcal{Q}_1, i_2 \in \mathcal{Q}_2 , i_3 \in \mathcal{Q}_3}{\prod}             C^{l_{ijk}}_{ijk}     \bigg) \wedge \cdots \\ \wedge \bigg(    \underset{i^{\prime\cdots\prime}_3 \in \mathcal{Q}^{\prime\cdots\prime}_3}{\underset{i^{\prime\cdots\prime}_1 \in\mathcal{Q}^{\prime\cdots\prime}_1}{\underset{ i^{\prime\cdots\prime}_2 \in \mathcal{Q}^{\prime\cdots\prime}_2 }{\prod}}}         C^{l^{\prime\cdots\prime}_{i^{\prime\cdots\prime}j^{\prime\cdots\prime}k^{\prime\cdots\prime}}}_{i^{\prime\cdots\prime}j^{\prime\cdots\prime}k^{\prime\cdots\prime}}            \bigg)        \bigg)  \bigotimes                \bigg(     \underset{1\leq k \leq n-3}{\bigotimes}      \textbf{I}_k       \bigg)      \bigg)     -    \big( \omega \big( {\mathrm{XOR}}  \big) \big)^n  \\ \times   \bigg(       \pm \mathrm{sign} \big( i_1 , j_1 , k_1, i_{111}, \cdots , j_{111} , \cdots , j_{nm(n+m)}, k_{111}, \cdots, k_{nm ( n+ m)} \big)    \\ \times \mathrm{sign} \big(  i^{\prime}_1 , j^{\prime}_1 , k^{\prime}_1, i^{\prime}_{111}, \cdots , j^{\prime}_{111} , \cdots   , j^{\prime}_{nm(n+m)}       , k^{\prime}_{111}, \cdots, k^{\prime}_{nm ( n+ m)}  \big)      \times \cdots \\ \times \mathrm{sign} \big(  i^{\prime\cdots\prime}_1 , j^{\prime\cdots\prime}_1 , k^{\prime\cdots\prime}_1, i^{\prime\cdots\prime}_{111} , \cdots , j^{\prime\cdots\prime}_{111} , \cdots , j^{\prime\cdots\prime}_{nm(n+m)}       , k^{\prime \cdots \prime}_{111}, \cdots,   k^{\prime\cdots \prime}_{nm ( n+ m)}   \big) \bigg) \\ \times \bigg[                            \textbf{I} \bigotimes \textbf{I} \bigotimes \bigg(    \bigg(   \underset{i_1 \in \mathcal{Q}_1,i_2 \in  \mathcal{Q}_2 , i_3 \in \mathcal{Q}_3}{\prod}             C^{l_{ijk}}_{ijk}     \bigg)  \wedge \cdots  \wedge \bigg(    \underset{i^{\prime\cdots\prime}_3 \in \mathcal{Q}^{\prime\cdots\prime}_3}{\underset{i^{\prime\cdots\prime}_1 \in\mathcal{Q}^{\prime\cdots\prime}_1}{\underset{ i^{\prime\cdots\prime}_2 \in \mathcal{Q}^{\prime\cdots\prime}_2 }{\prod}}}             C^{l^{\prime\cdots\prime}_{i^{\prime\cdots\prime}j^{\prime\cdots\prime}k^{\prime\cdots\prime}}}_{i^{\prime\cdots\prime}j^{\prime\cdots\prime}k^{\prime\cdots\prime}}            \bigg)        \bigg) \\  \bigotimes \bigg( \bigg(   \underset{1\leq k \leq n-3}{\prod}  \textbf{I}_k \bigg)      \bigg)             \bigg] \bigg) \bigg]              \ket{\psi_{\mathrm{XOR} \wedge \cdots \wedge \mathrm{XOR}}}   \bigg| \bigg|_F    \tag{*}        \text{, }
\end{align*}

\noindent which will be shown to hold later in the section following the introduction of the Bell states in the $N$-player setting. In the two-dimensional case, corresponding to the dynamics between two players, the regularity of structures in the $\mathrm{XOR}$ and $\mathrm{FFL}$ games is captured with the following collection of four actions,

\begin{align*}
\bigg(  \textbf{I} \otimes \textbf{I}  \bigg) \bigg(  \frac{\ket{00} + \ket{11}}{\sqrt{2}}  \bigg) = \frac{\ket{00} + \ket{11}}{\sqrt{2}}   \text{ } \text{ , }     \bigg(     \sigma_x \otimes \textbf{I}  \bigg) \bigg( \frac{\ket{00} + \ket{11}}{\sqrt{2}} \bigg)   = \frac{\ket{10} + \ket{01}}{\sqrt{2}}   \text{, }   \end{align*}

\begin{align*}    \bigg(   \sigma_z    \otimes \textbf{I}  \bigg)  \bigg( \frac{\ket{00} + \ket{11}}{\sqrt{2}} \bigg)  = \frac{\ket{00} - \ket{11}}{\sqrt{2}}       \text{ } \text{ , }    \bigg( \sigma_x \sigma_z     \otimes   \textbf{I}   \bigg) \bigg( \frac{\ket{00}+\ket{11}}{\sqrt{2}} \bigg)  = \frac{\ket{10}- \ket{01}}{\sqrt{2}}     \text{. } 
\end{align*}

\noindent The Bell states for games with more players, as a special case of the generalization provided for the $N$-player $\mathrm{XOR}$ game in the appendix, take the form,

\begin{align*}
        \bigg( \frac{\textbf{I} \otimes \textbf{I} \otimes \overset{N-3}{\cdots} \otimes \textbf{I}}{\sqrt{N}} \bigg) \bigg(    \underset{1 \leq j \leq N}{\sum} \ket{\text{Player } j \text{ state}}      \bigg)   \equiv  \frac{1}{\sqrt{N}}  \bigg(    \underset{1 \leq j \leq N}{\sum} \ket{\text{Player } j \text{ state}}      \bigg)      \text{, } \\ \\ \bigg( \frac{\sigma_z \otimes \textbf{I} \otimes \overset{N-3}{\cdots} \otimes \textbf{I}}{\sqrt{N}} \bigg)   \bigg(    \underset{1 \leq j \leq N}{\sum} \ket{\text{Player } j \text{ state}}      \bigg)    \equiv  \frac{1}{\sqrt{N}}    \bigg(    \ket{\text{Player 1 state}} \\ - \ket{\text{Player 2 state} } + \ket{\text{Player 3 state}} +  \cdots      + \ket{\text{Player } N \text{ state}}      \bigg)                    \text{, } \\ \\    \bigg(     \frac{\textbf{I} \otimes \sigma_z \otimes \textbf{I} \otimes \overset{N-3}{\cdots} \otimes \textbf{I}}{\sqrt{N}}     \bigg)          \bigg(    \underset{1 \leq j \leq N}{\sum} \ket{\text{Player } j \text{ state}}      \bigg)   \equiv  \frac{1}{\sqrt{N}}    \bigg(                         \underset{1 \leq j \leq 2}{\sum}   \ket{\text{Player } j \text{ state}} \\ - \ket{\text{Player } 3 \text{ state}}       + \underset{4 \leq j \leq N}{\sum} \ket{\text{Player } j \text{ state}}       \bigg)                \text{, }  \\      \vdots \\  \bigg(      \frac{\textbf{I} \otimes \textbf{I} \otimes \overset{N-4}{\cdots } \otimes \textbf{I} \otimes \sigma_z}{\sqrt{N}}                    \bigg)     \bigg(    \underset{1 \leq j \leq N}{\sum} \ket{\text{Player } j \text{ state}}      \bigg)   \equiv  \frac{1}{\sqrt{N}}    \bigg(                         - \ket{\text{Player } 1 \text{ state}} \\ + \underset{2 \leq j \leq N}{\sum} \ket{\text{Player } j \text{ state}}              \bigg)       \text{, } \\ \\    \bigg(    \frac{\sigma_x \otimes \textbf{I} \otimes  \overset{N-2}{\cdots} \otimes \textbf{I}}{\sqrt{N}}  \bigg)     \bigg(    \underset{1 \leq j \leq N}{\sum} \ket{\text{Player } j \text{ state}}      \bigg)   \equiv  \frac{1}{\sqrt{N}}  \bigg(    \widetilde{\ket{\text{Player } 1 \text{ state}}} \\ + \widetilde{\ket{\text{Player } 2 \text{ state}}}          + \underset{3 \leq j \leq N}{\sum}   \ket{\text{Player } j \text{ state}}                                 \bigg)      \text{, }   \\ \\        \bigg(     \frac{\textbf{I} \otimes \sigma_x \otimes \overset{N-2}{\cdots} \otimes \textbf{I}}{\sqrt{N}}              \bigg)     \bigg(    \underset{1 \leq j \leq N}{\sum} \ket{\text{Player } j \text{ state}}      \bigg) \\ \equiv  \frac{1}{\sqrt{N}}    \bigg(    \ket{\text{Player }1 \text{ state}} + \widetilde{\ket{\text{Player } 2 \text{ state}}}  +         \widetilde{\ket{\text{Player } 3 \text{ state}}}  \\     + \underset{4 \leq j \leq N}{\sum}   \ket{\text{Player } j \text{ state}}       \bigg)      \text{, } \\          \vdots  
\\   
   \bigg(  \frac{\textbf{I} \otimes \textbf{I} \otimes \overset{N-4}{\cdots} \otimes \sigma_x \otimes \textbf{I}}{\sqrt{N}}    \bigg)     \bigg(    \underset{1 \leq j \leq N}{\sum} \ket{\text{Player } j \text{ state}}      \bigg)  \equiv  \frac{1}{\sqrt{N}}   \bigg(                           \underset{1 \leq j \leq N-2}{\sum}     \ket{\text{Player } j \text{ state}} \\ + \widetilde{\ket{\text{Player } (N-1) \text{ state}}}          +      \widetilde{\ket{\text{Player } N \text{ state}}}                       \bigg)    \text{, } \\ \\  \bigg(  \frac{\textbf{I} \otimes \textbf{I} \otimes \overset{N-4}{\cdots} \otimes \textbf{I} \otimes \sigma_x }{\sqrt{N}}    \bigg)     \bigg(    \underset{1 \leq j \leq N}{\sum} \ket{\text{Player } j \text{ state}}      \bigg) \equiv \frac{1}{\sqrt{N}}   \bigg(       \widetilde{\ket{\text{Player } 1 \text{ state}}} \\ + \underset{2 \leq j \leq N-1}{\sum}  \ket{\text{Player } j \text{ state}}                   +   \widetilde{\ket{\text{Player } N \text{ state}}}     \bigg)    \text{, }  \\ \\     \bigg(        \frac{ \sigma_z \sigma_x \otimes \textbf{I} \otimes \overset{N-2}{\cdots} \otimes \textbf{I}    }{\sqrt{N}}        \bigg)     \bigg(    \underset{1 \leq j \leq N}{\sum} \ket{\text{Player } j \text{ state}}      \bigg)   \equiv \frac{1}{\sqrt{N} } \bigg(         \widetilde{\ket{\text{Player } 1 \text{ state}}} \\ +       \widetilde{\ket{\text{Player } 2 \text{ state}}}            - \underset{3 \leq j \leq N}{\sum}  \ket{\text{Player } j \text{ state}}  \bigg) \text{,} \\ \\            \bigg(        \frac{\textbf{I} \otimes \sigma_z \sigma_x \otimes \textbf{I} \otimes \overset{N-3}{\cdots} \otimes \textbf{I}}{\sqrt{N}}                  \bigg)    \bigg(    \underset{1 \leq j \leq N}{\sum} \ket{\text{Player } j \text{ state}}      \bigg)   \equiv \frac{1}{\sqrt{N}} \bigg(                         - \ket{\text{Player } 1 \text{ state}} \\ + \widetilde{\ket{\text{Player } 2 \text{ state}}}            +   \widetilde{\ket{\text{Player } 3 \text{ state}}}   - \underset{4 \leq j \leq N}{\sum}     \ket{\text{Player } j \text{ state}}               \bigg)  \text{, } \\     \vdots \\     \bigg(   \frac{\textbf{I} \otimes \textbf{I} \otimes \overset{N-4}{\cdots} \otimes \textbf{I} \otimes \sigma_z \sigma_x }{\sqrt{N}}                  \bigg)    \bigg(    \underset{1 \leq j \leq N}{\sum} \ket{\text{Player } j \text{ state}}      \bigg)    \equiv \frac{1}{\sqrt{N}}   \bigg(   \widetilde{\ket{\text{Player } 1 \text{ state}}} \\ + \underset{2 \leq j \leq N-2}{\sum} \ket{\text{Player }j \text{ state}}      -     \ket{\text{Player } (N-1) \text{ state}} + \widetilde{\ket{\text{Player } N \text{ state}}}          \bigg)     \text{, }  \\ \\       \bigg(  \frac{\sigma_x \otimes \textbf{I} \otimes \overset{N-4}{\cdots} \otimes       \textbf{I} \otimes \sigma_z      }{\sqrt{N}}                  \bigg)    \bigg(    \underset{1 \leq j \leq N}{\sum} \ket{\text{Player } j \text{ state}}      \bigg)  \\  \equiv \frac{1}{\sqrt{N}} \bigg(          - \widetilde{\ket{\text{Player } 1 \text{ state}}} +     \widetilde{\ket{\text{Player } 2 \text{ state}}}  \\      - \underset{3 \leq j \leq N}{\sum}    \ket{\text{Player } j \text{ state}}     \bigg)   \text{, } \\ \\  \bigg(  \frac{\textbf{I} \otimes \sigma_x \otimes \textbf{I} \otimes \overset{N-5}{\cdots} \otimes \textbf{I} \otimes \sigma_z }{\sqrt{N}} \bigg)    \bigg(    \underset{1 \leq j \leq N}{\sum} \ket{\text{Player } j \text{ state}}      \bigg) \\   \equiv \frac{1}{\sqrt{N}} \bigg(  - \ket{\text{Player } 1 \text{ state}} + \widetilde{\ket{\text{Player } 2 \text{ state}}}  \\       + \widetilde{\ket{\text{Player } 3 \text{ state}}}               +  \underset{4 \leq j \leq N-1}{\sum}    \ket{\text{Player } j \text{ state}} - \ket{\text{Player } N \text{ state}}          \bigg)   \text{, }   \\ \\      \bigg(      \frac{ \sigma_z  \otimes   \textbf{I} \otimes \overset{N-4}{\cdots}  \otimes \textbf{I} \otimes \sigma_x }{\sqrt{N}}           \bigg)    \bigg(    \underset{1 \leq j \leq N}{\sum} \ket{\text{Player } j \text{ state}}      \bigg)    \equiv \frac{1}{\sqrt{N}}  \bigg(    - \widetilde{\ket{\text{Player } 1 \text{ state}}} \\ + \underset{2 \leq j \leq N-1}{\sum}    \ket{\text{Player } j \text{ state}}          -  \widetilde{\ket{\text{Player } N \text{ state}}}             \bigg)  \text{, } \\ \\   \bigg( \frac{\textbf{I} \otimes \sigma_z \otimes \overset{N-4}{\cdots} \otimes \textbf{I} \otimes \sigma_x }{\sqrt{N}} \bigg)    \bigg(    \underset{1 \leq j \leq N}{\sum} \ket{\text{Player } j \text{ state}}      \bigg)   \equiv \frac{1}{\sqrt{N}}  \bigg(               \widetilde{\ket{\text{Player } 1 \text{ state}}} \\ +   \ket{\text{Player } 2 \text{ state}}  
- \ket{\text{Player } 3 \text{ state}}   +     \underset{4 \leq j \leq N-1}{\sum}    \ket{\text{Player } j \text{ state}} + \widetilde{\ket{\text{Player } N \text{ state}}}       \bigg)  \text{, } \\ \vdots    \\   \bigg( \frac{\textbf{I} \otimes \textbf{I} \otimes \overset{N-4}{\cdots} \otimes \sigma_z \otimes \sigma_x }{\sqrt{N}} \bigg)    \bigg(    \underset{1 \leq j \leq N}{\sum} \ket{\text{Player } j \text{ state}}      \bigg)   \equiv \frac{1}{\sqrt{N}}  \bigg(      \widetilde{\ket{\text{Player } 1 \text{ state}}} \\ +   \ket{\text{Player } 2 \text{ state}}                          \underset{3 \leq j \leq N-1}{\sum}    \ket{\text{Player } j \text{ state}} -  \widetilde{\ket{\text{Player } N \text{ state}}}   \bigg) \text{. }
\end{align*}

\noindent In the set of relations above for the Bell states, the position at which the $\sigma_x$ is applied to the summation over all player states,

\begin{align*}
     \underset{1 \leq j \leq n}{\sum} \ket{\text{Player } j \text{ state}}   \text{, }
\end{align*}

\noindent in addition to $\sigma_z$ is applied. Furthermore, $\sigma_x$ and $\sigma_z$ can be simultaneously applied to tensor products, which can result in the transformation of the player observable,

\begin{align*}
     \widetilde{\ket{\text{Player } 1 \text{ state}}} \equiv \text{1 st player tensor observable} \big( i^{\prime}_1 , i_2 \big)    \text{, } \\     \widetilde{\ket{\text{Player } 2 \text{ state}}} \equiv \text{2 nd player tensor observable} \big( i_1 , i^{\prime}_2 \big)       \text{, }
\end{align*}

\noindent with the $\widetilde{\cdot}$ transformation being similarly defined for all other observables for the remaining players of the game. From the Bell states introduced above, by adapting an argument from a previous subsection, , the expression introduced in $\textit{(*)}$ acting on,

\begin{align*}
  \ket{\psi_{\mathrm{XOR}\wedge \cdots \wedge \mathrm{XOR}}} \equiv \underset{\# \text{ of strong parallel repetitions}}{\bigwedge} \ket{\psi_{\mathrm{XOR}}} \equiv \ket{\psi^{\underset{\# \text{ of strong parallel repetitions}}{\wedge}}_{\mathrm{XOR}}}  \text{, }
\end{align*}

\noindent can be upper bounded with the following argument to show that the following result holds:

\bigskip

\noindent \textbf{Lemma} \textit{FR} $\wedge \cdots \wedge$ \textit{FR} (\textit{Frobenius norm upper bound for strong parallel XOR repetition}). One has that,

\[
 (*)  \lesssim N! n^N   \sqrt{\epsilon_{\mathrm{XOR} \wedge \cdots \wedge \mathrm{XOR}}}  \times   
\left\{\!\begin{array}{ll@{}>{{}}l} n^{ \frac{\# \text{ of players}}{2}  +5} \Longleftrightarrow \big( \# \text{ of players} \big) \mathrm{mod} 2 \equiv 0 \\ n^{ \lfloor\frac{\# \text{ of players}}{2} \rfloor +5} \Longleftrightarrow \big( \# \text{ of players} \big) \mathrm{mod} 2 \neq 0
\end{array}\right.       \text{, }
\]

\noindent given $\epsilon_{\mathrm{XOR} \wedge \cdots \wedge \mathrm{XOR}}$ sufficiently small.

\bigskip

\noindent \textit{Proof of Lemma  \textit{FR} $\wedge \cdots \wedge$ \textit{FR}}. We present an adaptation of an argument for the $N$ $\mathrm{XOR}$ game which is included in the next section following arguments for demonstrating that a certain suitably defined tensor is positive semidefinite. Namely, for upper bounding the Frobenius norm for $\mathrm{XOR}$ strong parallel repetition, fix $\epsilon \equiv \epsilon^{\prime\prime\prime\prime}_{\mathrm{XOR} \wedge \cdots \wedge \mathrm{XOR}}$ sufficiently small (not to be confused with the series of constants $\epsilon$, $\epsilon^{\prime}$, $\epsilon^{\prime\prime}$, and $\epsilon^{\prime\prime\prime}$ introduced for upper bounding the Frobenius norm in the next section), from which the assumptions for the $\mathrm{XOR}$ strong parallel repetition game,

\begin{align*} 
 \big( N \mathrm{XOR} \big) \wedge \cdots \wedge \big( N \mathrm{XOR} \big)  \equiv  \big( \mathrm{XOR} \big)  \wedge \cdots \wedge \big( \mathrm{XOR} \big)  \equiv \underset{\# \text{ of strong parallel repetitions } j}{\bigwedge}  \big( \mathrm{XOR} \big)^j \\ \equiv \underset{1 \leq j \leq n}{\bigwedge}  \big( \mathrm{XOR} \big)^j   \text{,}
\end{align*}

\noindent include,

 \begin{align*}
    \bigg[  \bigg[ \textbf{I} - \textbf{I} \bigg( \frac{\pm  \big( A_k  \wedge \cdots \wedge A_{k^{\prime\cdots\prime}} \big) +  \big( A_l \wedge \cdots \wedge A_{l^{\prime\cdots\prime}} \big) }{\sqrt{2}} \bigg) \bigg] \otimes \textbf{I} \otimes  \textbf{I} \otimes \textbf{I} \otimes \textbf{I} \otimes \overset{N-5}{\cdots} \otimes \textbf{I} \bigg] \\ \times \ket{\psi_{\mathrm{XOR}  \wedge \cdots \wedge  \mathrm{XOR} }}  < \sqrt{N! n \big( n-1 \big) \big( n-2 \big) \big( n-3 \big) \big( n - 4 \big)  } \epsilon \text{, }
\end{align*}

        \begin{align*}   \bigg[ \textbf{I} \otimes \textbf{I} \otimes \textbf{I} \otimes \textbf{I} \otimes \overset{N-5}{\cdots} \otimes \textbf{I} \bigg] \ket{\psi_{\mathrm{XOR}  \wedge \cdots \wedge  \mathrm{XOR} }}        < \sqrt{n \big( n-1 \big) \big( n -2 \big)} \epsilon       \text{, } \\ \\               \bigg[   \bigg[   \textbf{I} - \big( \omega_{\mathrm{XOR}} \big)^n \bigg(  \mathrm{sign} \big( i_1, \cdots, i_n, j_1 , \cdots, j_n, k_1, \cdots, k_n \big)  \times \cdots \times   \mathrm{sign} \big( i^{\prime\cdots\prime}_1 \\ , \cdots, i^{\prime\cdots\prime}_n, j^{\prime\cdots\prime}_1 , \cdots, j^{\prime\cdots\prime}_n  , k^{\prime\cdots\prime}_1, \cdots, k^{\prime\cdots\prime}_n \big)     \bigg)  \textbf{I}  \bigg]  \otimes \textbf{I}  \otimes \textbf{I} \otimes \textbf{I} \otimes \textbf{I}  \otimes \overset{N-5}{\cdots} \otimes \textbf{I}     \bigg]  \\ \times \ket{\psi_{\mathrm{XOR}  \wedge \cdots \wedge  \mathrm{XOR} }}    < n^N \sqrt{\epsilon}           \text{.     }    
\end{align*}

\noindent Hence, from the three assumptions on the strong parallel repetition game above, the desired upper bound dependent upon the number of players in the game takes the form,

\begin{align*}
      \sqrt{ N! n \bigg( \underset{1 \leq j \leq \# \text{ of players}-1}{\prod} \big( n - j \big)  \bigg)    \epsilon + 100 \omega_{N \mathrm{XOR}} n^{\# \text{ of players}} \epsilon        }            \\ <        \sqrt{ N! n \bigg( \underset{1 \leq j \leq \# \text{ of players}-1}{\prod} \big( n - j \big)  \bigg)    \epsilon + 1000  n^{\# \text{ of players}} \epsilon        }         \text{, }\end{align*}

  \noindent which can be further rearranged as,    
      
      \begin{align*}
      \sqrt{ \bigg(  N! n \bigg( \underset{1 \leq j \leq \# \text{ of players}-1}{\prod} \big( n - j \big)  \bigg) + 1000 n^{\# \text{ of players}} \bigg) \epsilon }          \\     <   \sqrt{ \bigg(  N! n \bigg( \underset{1 \leq j \leq \# \text{ of players}-1}{\prod} \big( n - j \big)  \bigg) + 1000 n^{\# \text{ of players}} \bigg)}  \big(  10 \sqrt{\epsilon }\big) \\ <        \sqrt{2N! n \big( n - \big( \# \text{ of players} -1 \big) \big) \bigg( \underset{1 \leq j \leq \# \text{ of players}-2}{\prod} \big( n - j \big)  \bigg) + 1000 n^{\# \text{ of players}} }   \\ \times       \big( 10 \sqrt{\epsilon } \big)   \\    \equiv     \sqrt{\big( 2N! n^2 - n \big( \# \text{ of players} - 1 \big) \big) \bigg( \underset{1 \leq j \leq \# \text{ of players}-2}{\prod} \big( n - j \big)  \bigg) + 1000  n^{\# \text{ of players}} } \\  \times   \big( 10 \sqrt{\epsilon } \big) \\       <     \sqrt{20 N! n^2   \bigg( \underset{1 \leq j \leq \# \text{ of players}-2}{\prod} \big( n - j \big)  \bigg) - n \big( \# \text{ of players} -1 \big)   \bigg( \underset{1 \leq j \leq \# \text{ of players}-2}{\prod} \big( n - j \big)  \bigg)  \cdots} \\ \sqrt{   + 1000 n^{\# \text{ of players}}       }   \\   \times    \big( 10 \sqrt{\epsilon} \big) \\  <   \sqrt{20 N! n^2   \bigg( \underset{1 \leq j \leq \# \text{ of players}-2}{\prod} \big( n - j \big)  \bigg)         - n \big( \# \text{ of players} -1 \big) \big( n - \big( \# \text{ of players} -2 \big) \big) \cdots } \\ \sqrt{ \times  \bigg( \underset{1 \leq j \leq \# \text{ of players}-3}{\prod} \big( n - j \big)  \bigg)  + 1000 n^{\# \text{ of players}}}      \big( 11 \sqrt{\epsilon } \big) \text{. } \end{align*}

\noindent As the product over the number of players is taken, the final term above can be upper bounded with,

      \begin{align*}  \sqrt{20 N! n^2 \big( n - \big( \# \text{ of players} -2 \big) \big) \bigg( \underset{1 \leq j \leq \# \text{ of players}-3}{\prod} \big( n - j \big)  \bigg) -     \bigg( n^2 \big( \# \text{ of players} - 1 \big) + \cdots } \\ \sqrt{    \big( \# \text{ of players} - 1 \big) \big( \# \text{ of players} -2 \big) \bigg) \bigg( \underset{1 \leq j \leq \# \text{ of players}-3}{\prod} \big( n - j \big)   \bigg)  + 1000 n^{\# \text{ of players}}    }    \big( 11 \sqrt{\epsilon} \big)   \\    
      <    \sqrt{20 N! n^2 \big( n \big( \# \text{ of players} -2 \big)  \big) \bigg( \underset{1 \leq j \leq \# \text{ of players} -3 }{\prod} \big( n - j \big) \bigg)   - \frac{1}{2} \bigg( n^2 \big( \# \text{ of players} - 1 \big) \cdots } \\ \sqrt{  + \big( \# \text{of players} -1 \big)  \big( \# \text{of players} -2 \big) \bigg)  \bigg(  \underset{1 \leq j \leq \# \text{of players} -3}{\prod}\big( n - j \big) \bigg)  + 1000 n^{\# \text{ of players}}}               \big( 11 \sqrt{\epsilon } \big) \\       \equiv  \sqrt{20 N! n^2 \big( n - \big( \# \text{of players} -2 \big)       \big) \bigg( \underset{1 \leq j \leq \# \text{of players} - 3}{\prod} \big( n - j \big)  \bigg)- \frac{1}{2} \bigg( n^2 \big( \# \text{ of players} -1 \big) \cdots } \\    \sqrt{ +  \big( \# \text{ of players} -1 \big) \big( \# \text{ of players} - 2  \big)  \bigg) \big( n - \big( \# \text{ of players} - 3 \big) \big) \bigg( \underset{1 \leq j \leq \# \text{ of players}-4}{\prod} \big( n - j \big)  \bigg) \cdots } \\ \sqrt{+ 1000 n^{\# \text{ of players}}  }      \big( 11 \sqrt{\epsilon }   \big) \text{. }  \tag{*}   \end{align*}

      \noindent Proceeding, the expression above, denoted with $(*)$, equals,
      
      \begin{align*} (\textit{*}) \equiv  \sqrt{20 N! n^2 \big( n - \big( \# \text{ of players} -2 \big) \big) \bigg( \underset{1 \leq j \leq \# \text{ of players}-3}{\prod} \big( n - j \big) \bigg)  - \frac{1}{2} \bigg(      - n^2 \big( \# \text{ of players} -1 \big)   \cdots } \\ \sqrt{ \times  \big( \# \text{ of players} - 3 \big) \bigg)  + n^3 \big(  \# \text{ of players} -1 \big)     + n \big( \# \text{ of players} -1 \big) \big( \# \text{ of players } - 2 \big) \cdots } \\   \sqrt{ - \big( \# \text{ of players} -1 \big) \big( \# \text{ of players } -2 \big)  \big( \# \text{ of players} -3 \big) \bigg) \bigg(      \underset{1 \leq j \leq \# \text{ of players} -5}{\prod}  \big( n - j \big)     \bigg) \cdots } \\ \sqrt{ + 1000 n^{\# \text{ of players }}  }    \big( 11 \sqrt{\epsilon} \big)      
\\ 
     <     \sqrt{20 N! n^2 \big( n - \big( \# \text{ of players} -2 \big) \big)   \bigg(  \underset{1 \leq j \leq \# \text{ of players}-3}{\prod} \big( n - j \big)  \bigg)  \cdots } 
\\ 
     \sqrt{  - \frac{1}{2} \bigg( n^2 \big( \# \text{ of players} - 1 \big) \big( \# \text{ of players}-3 \big)+ n^3 \big( \# \text{ of players} - 1 \big) + n \big( \# \text{ of players} \cdots } \\   
     \sqrt{ -1 \big)  \big( \# \text{ of players}-2 \big) - \big( \# \text{ of players} - 1 \big) \big( \# \text{ of players} -2 \big) \big( \# \text{ of players}  -3 \big)  \bigg) \cdots} \\ \sqrt{\times \bigg( \underset{1 \leq j \leq \# \text{ of players}        }{\prod}   \big( n - j \big)    \bigg)  + 1000 n^{\# \text{ of players}}  }   \big( 11 \sqrt{\epsilon} \big)    \\      <            \sqrt{20 N! n^2 \big( n - \big( \# \text{ of players} -2 \big) \big) \big( n - \# \text{ of players} - 3 \big) \bigg) \big( n - 1 \big)                   - \frac{1}{2}  \cdots }  \\     \sqrt{\times   \bigg(   - n^2 \big( \# \text{of players } -1 \big)\big( \# \text{ of players} -3 \big) \big( n - \big( \# \text{of players} -4 \big) \big)   + n^3 \big( \# \text{ of players} -1  \big) \cdots }  \\  \sqrt{ \times \big( n - \big( \# \text{ of players} -4  \big) \big)  + n   \big( \# \text{ of players} -1 \big) \big( \# \text{ of players} -2 \big) \big( n - \big( \# \text{ of players} -4 \big) \big) \bigg) \cdots} \\ \sqrt{ \times \bigg(  \underset{1 \leq j \leq \# \text{ of players} -5}{\prod}  \big( n - j \big)   \bigg)   + 1000 n^{\# \text{ of players}}        }     \big( 11 \sqrt{\epsilon} \big) \\    <  \sqrt{20 N! n^2 \big( n - \big( \# \text{ of players} - 2 \big) \big)\big( n - \big( \# \text{ of players} - 3 \big) \big) \times \cdots \times \big( n - 1 \big)  - \frac{1}{4}  \cdots     }   \\   \sqrt{  \times  \big( -n^2 \big( \# \text{ of players} -1 \big)    \big( \# \text{ of players} -3 \big) \big( \# \text{ of players} - 4 \big) \cdots } \\     \sqrt{ + n^3 \big( \# \text{ of players} -1 \big) \big( \# \text{ of players} - 4 \big) + n \big( \# \text{ of players} -1  \big) \big( \# \text{ of players} - 2 \big) \cdots } \\  \sqrt{\times \big( \# \text{ of players} -4 \big) \big) \bigg)  \bigg( n^{ \# \text{ of players - 5} } \bigg) + 1000 n^{\# \text{ of players}}   }  \big( 11 \sqrt{\epsilon} \big) \text{. }        \tag{**} \end{align*}

    \noindent Proceeding,

      \begin{align*}
    (\textit{**}) <   \sqrt{20 N! n^2 \bigg( n - \big( \# \text{ of players} - 2 \big)       \big) \big( n - \big( \# \text{ of players} -3 \big) \big) \times \cdots \times \big( n -1 \big)   - \frac{1}{4} \big( - n^2 \cdots  } \\ \sqrt{  \times  \big( \# \text{ of players} -1 \big)     \big( \# \text{ of players} -3 \big)    \big( \# \text{ of players} - 4\big)    + n^3 \big( \# \text{ of players} -1 \big) \cdots   }  \\  \sqrt{ \times  \big( \# \text{ of players} -4 \big)             + n \big( \# \text{ of players} - 1 \big)    \big( \# \text{ of players} - 2 \big) \big( \# \text{ of players} - 4 \big)         \bigg)               \cdots   } \\   \sqrt{\times \bigg( n^{\# \text{ of players} - 5} \bigg)  + 1000 n^{\# \text{ of players}}      }  \big( 11 \sqrt{\epsilon} \big) \\      \equiv  \sqrt{20 N! n^2 \bigg( n - \big( \# \text{ of players} -2 \big) \big( n - \big( \# \text{ of players} -3 \big) \big) \times \cdots \times  \big( n - 1 \big)     - \frac{1}{4}  \big( - n^2 \cdots  } \\    \sqrt{ \times \big( \# \text{ of players} -1 \big) \big( \# \text{ of players} -3 \big)  \big( \# \text{ of players} - 4 \big)   + n  \big( \# \text{ of players} - 1 \big) \cdots   } \\    \sqrt{ \times \big( \# \text{ of players} - 2 \big)  \big( \# \text{ of players} - 4 \big) \bigg) \bigg( n^{\# \text{ of players} -5} \bigg) + 1000 n^{\# \text{ of players} }             }        \big( 11 \sqrt{\epsilon} \big) \\    \equiv   \sqrt{\bigg( 20 N! n^3   - 100 n^2 \big( \# \text{ of players} - 2 \big) \bigg(             \big( n - \big( \# \text{ of players} - 3 \big) \big) \times \cdots \times  \big( n - 1 \big)     \bigg) \cdots  }  \\   \sqrt{- \frac{1}{4} \bigg( - n^2 \big( \# \text{ of players} - 1 \big)  \big( \# \text{ of players} - 3 \big) \big( \# \text{ of players} - 4 \big) + n^3        \big( \# \text{ of players} -1 \big) \cdots  }  \\    \sqrt{\times  \big( \# \text{ of players} - 4 \big) n^{\# \text{of players} - 5}        +  n \big( \# \text{ of players} - 1 \big) \big( \# \text{ of players} - 2 \big)    \bigg)  \cdots    }  \\   \sqrt{\times \big( \# \text{ of players} - 4 \big)   n^{\# \text{ of players} - 5}  \bigg) + 1000 n^{\# \text{ of players}} }    \big( 11 \sqrt{\epsilon} \big)  \text{. } \tag{***}
\end{align*}

\noindent The final expression above can be further rearranged to obtain the desired upper bound, from the observations that,

\begin{align*}
    (\textit{***}) < \sqrt{\bigg( 20 N! n^3 - 50n^2 \big( n -2 \big) \bigg) \big( n-1 \big)^{\# \text{ of players} - 5}   - \frac{1}{4} \bigg( -n^2 \big( \# \text{ of players  - 1} \big)^3  \cdots }  \\      \sqrt{ \times  n^{\# \text{ of players} - 5}  + n^3 \big( \# \text{ of players} -1 \big)^2 n^{\# \text{ of players} - 5}  + n \big( \# \text{ of players} - 1 \big)^3 n^{\# \text{ of players} - 5 } \bigg) \cdots } \\  \sqrt{+ 1000 n^{\# \text{ of players}} }    \big( 11 \sqrt{\epsilon} \big)  \\       \equiv          \sqrt{\bigg( 20 N! n^3 - 50 n^2 \big( n-2 \big) \big) \big( n-1 \big)^{\# \text{ of players} - 3}    \bigg) - \frac{1}{4} \big( \big(  \# \text{ of players}  \big) - 1 \big)^2 n \bigg(        - n \big( \# \text{ of players} \cdots}  \\ \sqrt{ -1 \big)       n^{\# \text{ of players} - 5} + n^{\big( \# \text{ of players} - 5\big) + 1}     + \big( \# \text{ of players}  n^{\# \text{ of players} - 5}          \bigg)   + 1000 n^{\# \text{ of players}}        } \\ \times  \big( 11 \sqrt{\epsilon} \big)  \\   
    <       \sqrt{\bigg(     20 N! n^{\# \text{ of players} } - 50  n^{\# \text{ of players}} -     \bigg) 
          - \frac{1}{4}  n^3 \bigg(    - n \big( \# \text{ of players} - 1 \big)        n^{\# \text{ of players} - 5}               \cdots                 }    \\ \sqrt{   +          n^{\# \text{ of players} - 5}   +   \big( \# \text{ of players } - 5 \big)        n^{\# \text{ of players} - 5}    \bigg)   + 1000 n^{\# \text{of players}-5}   }    \big( 11 \sqrt{\epsilon} \big)    \\  
          <       \sqrt{n^{\# \text{ of players}}             \bigg(   n^{\# \text{ of players}} - n^3 \bigg( n^{\# \text{ of players}-4} + 2n^{\# \text{ of players}-5} \bigg)     \bigg)  + 1000 n^{\# \text{of players} -5}   }        \big( 110 \sqrt{\epsilon} \big) \\   <      \sqrt{ \bigg(  n^{\# \text{ of players}}     - n^3 \bigg(  - n^{\# \text{ of players}-5}      + 2n^{\# \text{ of players}-5}     \bigg)    + 1000 n^{\# \text{ of players}}      \bigg) }    \big( 120 \sqrt{\epsilon} \big)  \text{.} \tag{****}
\end{align*}

\noindent The final desired upper bound,

\[ \mathcal{I} \equiv \big( 130 N! n^N \big)   \sqrt{\epsilon} \times   
\left\{\!\begin{array}{ll@{}>{{}}l} n^{ \frac{\# \text{ of players}}{2} +5 } \Longleftrightarrow \big( \# \text{ of players} \big) \mathrm{mod} 2 \equiv 0 \\ n^{ \lfloor\frac{\# \text{ of players}}{2} \rfloor +5} \Longleftrightarrow \big( \# \text{ of players} \big) \mathrm{mod} 2 \neq 0
\end{array}\right.  \text{, } 
\]

\noindent for the N$\mathrm{XOR}$ game above is obtained from the observations that,

\begin{align*}
       (\textit{****}) < \sqrt{N! n^{\# \text{ of players} - 3} \bigg( 1 +  n^6 - 2n^5 + 1000 \bigg) } \big( 120  \sqrt{\epsilon} \big) <    \sqrt{ N! n^{\# \text{ of players} -3}  \bigg( 1 + n^6  + 998 n^5 \bigg) } \\ \times  \big( 120   \sqrt{\epsilon} \big) \\  <     \sqrt{N! n^{\# \text{ of players} -3}  \bigg( 1 + 999 n^6  \bigg) }      \big( 120 \sqrt{\epsilon} \big)     < \sqrt{N! n^{\# \text{ of players}} n^8} \big( 120 N! n^N \sqrt{\epsilon} \big) <  \mathcal{I}   \text{,}
\end{align*}

\noindent from which we conclude the argument. \boxed{}

\subsection{Positive semidefinite tensors}

\noindent For two $\pm$ observables, $A$ and $B$, in {[37]}, the operator,

\begin{align*}
   \bigg( \frac{A+B}{\sqrt{2}} + \frac{A+B}{\big| A+B \big|} \bigg)^2 \equiv \bigg( \big( A + B \big) \bigg( \frac{1}{\sqrt{2}} + \frac{1}{\big| A + B \big| } \bigg) \bigg)^2  \text{, }
\end{align*}

\noindent is characterized as being positive semidefinite, with eigenvalues,

\begin{align*}
\bigg[   \mathrm{sign} \big( \lambda \big) \frac{\lambda^2}{2}  - 1 \bigg]^2 \text{. }
\end{align*}

\noindent For additional $\pm$ observables corresponding to the results of $N$ players, the counterpart to the two-player expression for the operator above takes the form,

\begin{align*}
  \bigg[  \frac{\sum \text{Tensors of player observables}}{\sqrt{N}} +   \frac{\sum \text{Tensors of player observables}}{\big| \sum \text{Tensors of player observables} \big| }         \bigg]^N     \text{, }
\end{align*}

\noindent where the summation is taken over all possible responses of each player to a question drawn from the referee's probability distribution, namely,

\begin{align*}
 \sum \text{Tensors of player observables} \equiv \underset{\text{Questions}}{\sum} \text{Tensors of player observables}   \text{, }
\end{align*}

\noindent which will be shown to have the eigenvalues that are proportional to,

\begin{align*}
   \bigg[  \frac{1}{\sqrt{N}} \bigg[ \underset{1 \leq j \leq N-1}{\sum}  \big( \mathrm{sign} \big( \lambda \big) \big)^j \lambda^{N-j} \bigg]   - 1   \bigg]^N     \text{. }
\end{align*}

\noindent As introduced in previous sections from optimal values, linear operators, and error bounds, from the multiplayer setting, under the strong parallel repetition operation, for $\mathrm{XOR}$ and $\mathrm{FFL}$, games, denote,

\begin{align*}
  \underset{\text{Questions} \wedge \cdots \wedge \text{Questions}}{\sum}  \big( \big(  \text{Tensors of XOR player observables}\big)  \wedge \cdots \wedge \big( \text{Tensors of XOR} \\ \text{ player observables}\big)\big) \\ \equiv {\sum^{\prime}}  \big( \big(  \text{Tensors of XOR player observables}\big)  \wedge \cdots \wedge \big( \text{Tensors of XOR player observables}\big)\big)  \text{, } \\  \\ \underset{\text{Questions} \wedge \text{Questions}}{\sum}  \big( \big(  \text{Tensors of FFL player observables}\big)  \wedge \cdots \wedge \big( \text{Tensors of FFL} \\ \text{ player observables}\big)\big) \\ \equiv {\sum^{\prime\prime}}  \big( \big(  \text{Tensors of FFL player observables}\big)  \wedge \cdots \wedge \big( \text{Tensors of FFL} \\ \text{ player observables}\big)\big)  \text{, } 
\end{align*}

\noindent from the question sets obtained under an arbitrary, or under two, applications of strong parallel repetition, from which one expects that the operators should take the forms,

\begin{align*}
  \bigg[  \frac{\sum^{\prime} \big( \big(  \text{Tensors of XOR player observables}\big)  \wedge \cdots \wedge \big( \text{Tensors of XOR player observables}\big)\big)  }{\sqrt{N}}  \end{align*}

  \begin{align*} +   \frac{\sum^{\prime} \big( \big( \text{Tensors of XOR player observables} \big) \wedge \cdots \wedge \big( \text{Tensors of XOR player observables}\big) \big) }{\big| \sum^{\prime} \big( \big( \text{Tensors of XOR player observables} \big)  \wedge \cdots \wedge \big( \text{Tensors of XOR player observables}\big) \big)  \big| }         \bigg]^N     \text{, } \\  \\  \bigg[  \frac{\sum^{\prime\prime} \big( \big( \text{Tensors of FFL player observables} \big)  \wedge \big( \text{Tensors of FFL player observables}\big) \big) }{\sqrt{N}}  \\ +   \frac{\sum^{\prime\prime} \big( \big( \text{Tensors of FFL player observables}\big)  \wedge \big( \text{Tensors of FFL player observables}\big) \big) }{\big| \sum^{\prime\prime} \big( \big( \text{Tensors of FFL player observables} \big)  \wedge  \big( \text{Tensors of FFL player observables}  \big) \big) \big| }         \bigg]^N     \text{, } 
\end{align*}

\noindent respectively.

\bigskip

\noindent \textbf{Lemma} (\textit{N-XOR positive semidefinite tensors}). The superposition of $N$ $\pm 1$ observables,

\begin{align*}
  \bigg[  \frac{\sum \text{Tensors of player observables}}{\sqrt{N}} +   \frac{\sum \text{Tensors of player observables}}{\big| \sum \text{Tensors of player observables} \big| }         \bigg]^N     \text{, }
\end{align*}

\noindent is positive semidefinite, which can be concluded from the fact that,

 \begin{align*}
  \bigg[    N +    \bigg( \frac{1}{\sqrt{N}} \bigg)^2 \bigg( \underset{1 \leq i \leq N-1}{\sum}  n^i \mathrm{sign} \big( \lambda \big) \lambda^{N-i}   \bigg)^2 - 2   \bigg( \underset{1 \leq i \leq N-1}{\sum}  n^i \mathrm{sign} \big( \lambda \big) \lambda^{N-i}   \bigg)            \bigg]  \bigg[  \bigg[  \frac{1}{\sqrt{N}}  \\ \times \bigg[ \underset{1 \leq j \leq N-1}{\sum}  \big( \mathrm{sign} \big( \lambda \big) \big)^j \lambda^{N-j} \bigg]    - 1   \bigg]^N  \bigg]^{-1} \bigg[     N +    \bigg( \frac{1}{\sqrt{N}} \bigg)^2 \bigg( \underset{1 \leq i \leq N-1}{\sum}  n^i \mathrm{sign} \big( \lambda \big) \lambda^{N-i}   \bigg)^2 \\ - 2   \bigg( \underset{1 \leq i \leq N-1}{\sum}  n^i \mathrm{sign} \big( \lambda \big) \lambda^{N-i}   \bigg)          \bigg] \text{, }
\end{align*}

\noindent implies that the $\pm 1$ superposition has eigenvalues,

\begin{align*}
 \bigg[ \frac{\lambda^N -  \underset{1 \leq i \leq N-1}{\sum} n^i \lambda^{N-i} - n}{n} \bigg]^N                        \text{. }
\end{align*}

\noindent From the closed-form representation of the eigenvalues of the previous operator, one can also conclude, straightforwardly, that the operator,

\begin{align*}
   \bigg[  \frac{\sum \text{Tensors of player observables}}{\sqrt{N}} +   \frac{\sum \text{Tensors of player observables}}{\big| \sum \text{Tensors of player observables} \big| }         \bigg]^N    \\ \times   \bigg[ N + \underset{1 \leq i \leq N-1}{\sum} n^i \lambda^{N-i}  \\    + N \sqrt{\textbf{I} + \frac{\big(  \sum \text{Tensors of player observables} \big) \big( \sum^{-1} \text{Tensors of player observables} \big)}{N}  }  \bigg]^{-N}   \\ \times   \bigg[  \frac{\sum \text{Tensors of player observables}}{\sqrt{N}} +   \frac{\sum \text{Tensors of player observables}}{\big| \sum \text{Tensors of player observables} \big| }         \bigg]^N      \text{, }
\end{align*}

\noindent has eigenvalues,

\begin{align*}
 \bigg[ \frac{\lambda^N -  \underset{1 \leq i \leq N-1}{\sum} n^i \lambda^{N-i} - n}{n} \bigg]^N       \frac{1}{N + \underset{1 \leq i \leq N-1}{\sum} n^i \lambda^{N-i} + N \sqrt{1+ \frac{\underset{1 \leq i \leq N-1}{\sum} n^i \lambda^{N-i} -N}{N}}}                      \text{. }
\end{align*}

\noindent To show that the closed form expression given above holds for the eigenvalues of the operator from tensor observables of each player, one makes use of a suitable factorization of the expression for the eigenvalues above, which is dependent upon,

\begin{align*}
   N  \bigg[  \frac{1}{\sqrt{N}} \bigg[ \underset{1 \leq j \leq N-1}{\sum}  \big( \mathrm{sign} \big( \lambda \big) \big)^j \lambda^{N-j} \bigg]    - 1   \bigg]^N             \text{, } \\ \\     \bigg( \frac{1}{\sqrt{N}} \bigg)^2  \bigg[  \frac{1}{\sqrt{N}} \bigg[ \underset{1 \leq j \leq N-1}{\sum}  \big( \mathrm{sign} \big( \lambda \big) \big)^j \lambda^{N-j} \bigg]    - 1   \bigg]^N                     \text{, } \\  \\ -  2   \bigg( \underset{1 \leq i \leq N-1}{\sum}  n^i \mathrm{sign} \big( \lambda \big) \lambda^{N-i}   \bigg)          \bigg[  \frac{1}{\sqrt{N}} \bigg[ \underset{1 \leq j \leq N-1}{\sum}  \big( \mathrm{sign} \big( \lambda \big) \big)^j \lambda^{N-j} \bigg]    - 1   \bigg]^N           \text{. }  
\end{align*}

\bigskip

\noindent \textit{Proof of Lemma }. The result is a direct application of $\textbf{Lemma}$ \textit{} from {[37]}, which is included in \textit{2.5} - the fifth section of the Appendix. \boxed{} 

\subsection{Player dependent upper bounds at optimality}

\noindent We argue that the desired upper bounds hold, up to constants, of order $n^3 \sqrt{\epsilon}$, $n^4 \sqrt{\epsilon}$, and $n^5 \sqrt{\epsilon}$, respectively. In the most simple case, for the 3-$\mathrm{XOR}$ game, an upper bound for,

\begin{align*}
  \bigg| \bigg| \bigg[      \bigg( \bigg( \underset{1 \leq i \leq n}{\prod} A^{j_i}_i \bigg)         \otimes B_{kl} \otimes \textbf{I} \bigg) - \omega_{3\mathrm{XOR}} \bigg( \pm \bigg( \mathrm{sign} \big( i_1 , j_1 , \cdots , j_n \big)  \\ \times   \bigg[ \bigg(            \bigg( \underset{1 \leq i \leq n}{\prod}   A^{j_i}_i \bigg)  + \bigg(   \underset{\text{set }i+1 \equiv i \oplus 1}{\underset{1 \leq i \leq n}{\prod}}   A^{j_i}_i      \bigg)     \bigg)   \otimes \textbf{I} \otimes \textbf{I}            \bigg]    \bigg) \bigg)          \bigg]  \ket{\psi_{3\mathrm{XOR}}} \bigg| \bigg|^2  \text{.}
\end{align*}

\noindent For the set of possible answers that the first player can provide in response to a question drawn from the referee's probability distribution, a variant of the inequality above takes the form,

\begin{align*}
            \bigg| \bigg| \bigg[ \bigg( \bigg( \bigg( \underset{1 \leq i \leq n}{\prod} A^{j_i}_i \bigg) - \bigg(  \bigg( \underset{1 \leq i \leq n}{\prod} A^{j_i}_i \bigg) \bigg( \frac{\pm A_k + A_l}{\sqrt{2}} \bigg)  \bigg) \bigg)   \otimes B_{kl} \otimes \textbf{I} \bigg)  \ket{\psi_{3\mathrm{XOR}}} \bigg| \bigg|  \\   -  \omega_{3\mathrm{XOR}}   \bigg| \bigg| \bigg[  \bigg(   \bigg(  \bigg( \underset{1 \leq i \leq n}{\prod}   A^{j_i}_i       \bigg)   A_k   -   \mathrm{sign} \big( i_1 , j_1 , \cdots , j_n \big)  \bigg( \underset{\text{set } i + 1 \equiv i \oplus 1}{\underset{1 \leq i \leq n}{\prod}}     A^{j_i}_i \bigg) \bigg)      \otimes \textbf{I} \otimes \textbf{I} \bigg)   \bigg] \\ \times \ket{\psi_{3\mathrm{XOR}}}    \bigg| \bigg|   + \bigg| \bigg|           \bigg[      \bigg(    \bigg(        \underset{1 \leq i \leq n}{\prod}A^{j_i}_i \bigg)  A_k   - \mathrm{sign} \big( i_1 , j_1 , \cdots , j_n \big)  \bigg(     \underset{\text{set } i +1 \equiv i \oplus 1}{\underset{1 \leq i \leq n}{\prod}}A^{j_i}_i  \bigg)       \bigg) \otimes \textbf{I}  \otimes \textbf{I}    \bigg] \\ \times \ket{\psi_{3\mathrm{XOR}}}       \bigg| \bigg|  \text{. }
\end{align*}

\noindent implies that the upper bound, up to constant, of order $n^3 \sqrt{\epsilon} $ holds. After obtaining the first desired upper bound, we demonstrate how analagous upper bounds can be obtained for several variants of the $\mathrm{XOR}$ game that have been considered in previous sections. In the result below, we exhibit how a computation used for strong parallel repetition of the $\mathrm{XOR}$ game can straightforwardly also be applied for $3\mathrm{XOR}$ games, and beyond. As such, as a generalization of a computation provided previously by the author, we seek to elucidate properties of the optimal solution state for inequalities that are counterparts to,

\begin{align*}
 \bigg| \bigg|  \bigg( \bigg( \underset{1 \leq i \leq n}{\prod}   A^{j_i}_i \bigg)   \otimes B_{kl} \bigg)  \ket{\psi_{\mathrm{FFL}}}  - \frac{2}{3} \bigg[ \pm \bigg( \mathrm{sign} \big( i , j_1 , \cdots , j_n \big) \bigg[ \text{ } \bigg(     \underset{i = j_k + 1 , \text{ } \mathrm{set} \text{ } j_k + 1 \equiv j_k \oplus 1 }{\underset{1 \leq i \leq n}{\prod}}     A^{j_i}_i    \bigg) \\ + \bigg( \underset{i = j_l + 1 , \text{ } \mathrm{set} \text{ } j_l + 1 \equiv j_l \oplus 1 }{\underset{1 \leq i \leq n}{\prod}}     A^{j_i}_i   \bigg) \text{ } \bigg]   \otimes \textbf{I} \bigg)  \ket{\psi_{\mathrm{FFL}}} \bigg]           \bigg| \bigg|   <    \bigg(       \frac{8200 \sqrt{2} }{27}   \bigg)                   n^2 \sqrt{\epsilon }                   \text{, }
\end{align*}

\noindent the second $\mathrm{FFL}$ error bound.

\bigskip

\noindent \textbf{Lemma} \textit{Gen-FFL-Bound} (\textit{generalizations of the second FFL error bound}, \textit{6.6}, {[37]}, \textbf{Lemma} \textit{7}, {[44]}). Denote the quantum states corresponding to optimal strategies,

\begin{align*}
 \ket{\psi_{3\mathrm{XOR}}} \equiv \underset{\text{Players}}{\bigcup} \underset{\mathcal{S}}{\mathrm{sup}} \big\{  \text{A player's quantum strategy } \mathcal{S} \text{ for a } 3-\text{XOR game}    \big\}   \text{, } \\ \ket{\psi_{4\mathrm{XOR}}} \equiv  \underset{\text{Players}}{\bigcup} \underset{\mathcal{S}}{\mathrm{sup}} \big\{  \text{A player's quantum strategy } \mathcal{S} \text{ for a } 4-\text{XOR game}    \big\}  \text{, } \\ \ket{\psi_{5\mathrm{XOR}}} \equiv  \underset{\text{Players}}{\bigcup} \underset{\mathcal{S}}{\mathrm{sup}} \big\{  \text{A player's quantum strategy } \mathcal{S} \text{ for a } 5-\text{XOR game}    \big\}  \text{, } \\  \ket{\psi_{N\mathrm{XOR}}} \equiv  \underset{\text{Players}}{\bigcup} \underset{\mathcal{S}}{\mathrm{sup}} \big\{  \text{A player's quantum strategy } \mathcal{S} \text{ for an } N-\text{XOR game}    \big\} \text{, }
\end{align*}

\noindent respectively, for the $3$-$\mathrm{XOR}$, $4$-$\mathrm{XOR}$, $5$-$\mathrm{XOR}$, and $N$-$\mathrm{XOR}$, games. Given error bounds formulated in previous sections for each $\mathrm{XOR}$ game, one obtains error bound inequalities of the form,

\begin{align*}
     \bigg| \bigg| \bigg[ \bigg( \bigg( \underset{1 \leq i \leq n}{\prod}  A^{j_i}_i \bigg) \otimes B_{kl} \otimes \textbf{I} \bigg)  - \omega_{3\mathrm{XOR}} \bigg( \pm \mathrm{sign} \big( i_1 , j_1 , \cdots , j_n \big)  \\ \times     \bigg[       \bigg( \bigg( \underset{1 \leq i \leq n}{\prod}   A^{j_i}_k \bigg)    + \bigg(                     \underset{\text{set } j+1 \equiv j \oplus 1}{\underset{i \in \mathcal{Q}_1, j \in \mathcal{Q}_2}{\prod}}   A^{j_i}_k        \bigg) \bigg)  \textbf{I} \otimes \textbf{I} \bigg] \bigg) \bigg]  \ket{\psi_{3\mathrm{XOR}}} \bigg| \bigg|^2 \\ \lesssim 3! n^3 \sqrt{\epsilon} \equiv  1000 n^3 \sqrt{\epsilon }  \text{, } \\  \\    \bigg| \bigg| \bigg[ \bigg( \bigg( \underset{1 \leq i \leq n}{\prod}  A^{j_i}_i \bigg) \otimes B_{kl} \otimes \textbf{I} \otimes \textbf{I} \bigg)  - \omega_{4\mathrm{XOR}} \bigg( \pm \mathrm{sign} \big( i_1 , j_1 , \cdots , j_n \big)   \\ \times    \bigg[       \bigg( \bigg( \underset{1 \leq i \leq n}{\prod}   A^{j_i}_k \bigg)    + \bigg(                     \underset{\text{set } j+1 \equiv j \oplus 1}{\underset{i \in \mathcal{Q}_1, j \in \mathcal{Q}_2}{\prod}}   A^{j_i}_k        \bigg) \bigg)   \otimes \textbf{I}   \otimes \textbf{I} \otimes \textbf{I} \bigg] \bigg) \bigg] \ket{\psi_{4\mathrm{XOR}}} \bigg| \bigg|^2 \\ \lesssim 4! n^4 \sqrt{\epsilon} \equiv  100000 n^4 \sqrt{\epsilon }     \text{, } 
     \\ \\   \bigg| \bigg| \bigg[ \bigg( \bigg( \underset{1 \leq i \leq n}{\prod}  A^{j_i}_i \bigg) \otimes B_{kl} \otimes \textbf{I} \otimes \textbf{I} \otimes \textbf{I} \bigg)  - \omega_{5\mathrm{XOR}} \bigg( \pm \mathrm{sign} \big( i_1 , j_1 , \cdots , j_n \big)  \\ \times     \bigg[       \bigg( \bigg( \underset{1 \leq i \leq n}{\prod}   A^{j_i}_k \bigg)    + \bigg(                     \underset{\text{set } j+1 \equiv j \oplus 1}{\underset{i \in \mathcal{Q}_1, j \in \mathcal{Q}_2}{\prod}}   A^{j_i}_k        \bigg) \bigg)  \otimes \textbf{I} \otimes \textbf{I}  \otimes \textbf{I} \bigg] \bigg) \bigg] \ket{\psi_{5\mathrm{XOR}}} \bigg| \bigg|^2 \\ \lesssim 5! n^5 \sqrt{\epsilon} \equiv   1000 \sqrt{2}  n^5 \sqrt{\epsilon}      \text{, }   \end{align*}

   \begin{align*}
    \bigg| \bigg| \bigg[ \bigg( \bigg( \underset{1 \leq i \leq n}{\prod}  A^{j_i}_i \bigg) \bigotimes B_{kl}  \bigotimes \bigg( \underset{1 \leq k \leq N-2}{\bigotimes} \textbf{I}_k  \bigg)  \bigg)  - \omega_{N\mathrm{XOR}} \bigg( \pm \mathrm{sign} \big( i_1 , j_1 , \cdots , j_n \big)  \\ \times     \bigg[       \bigg( \bigg( \underset{1 \leq i \leq n}{\prod}   A^{j_i}_k \bigg)    + \bigg(                     \underset{\text{set } j+1 \equiv j \oplus 1}{\underset{i \in \mathcal{Q}_1, j \in \mathcal{Q}_2}{\prod}}   A^{j_i}_k        \bigg) \bigg)  \bigotimes \bigg( \underset{1 \leq k \leq N-1}{\bigotimes} \textbf{I}_k  \bigg) \bigg] \bigg) \bigg] \ket{\psi_{N\mathrm{XOR}}} \bigg| \bigg|^2   \end{align*} \[ \lesssim  N! n^N \sqrt{\epsilon}   \times 
\left\{\!\begin{array}{ll@{}>{{}}l} n^{ \frac{\# \text{ of players}}{2}  +5} \Longleftrightarrow \big( \# \text{ of players} \big) \mathrm{mod} 2 \equiv 0 \\ n^{ \lfloor\frac{\# \text{ of players}}{2} \rfloor  +5} \Longleftrightarrow \big( \# \text{ of players} \big) \mathrm{mod} 2 \neq 0
\end{array}\right.   \text{. }
\]

\bigskip

\noindent \noindent \textit{Proof of Lemma Gen-FFL-Bound}. The desired upper bound for the $3$$\mathrm{XOR}$ game is obtained first. For the optimal strategy, $\ket{\psi_{3\mathrm{XOR}}}$, it suffices to argue that an upper bound of the desired form holds by making use of the fact that,

\begin{align*}
    \bigg[  \bigg[ \textbf{I} - \textbf{I} \bigg( \frac{\pm A_k + A_l}{\sqrt{2}} \bigg) \bigg] \otimes \textbf{I} \otimes \textbf{I} \bigg] \ket{\psi_{3\mathrm{XOR}}} < \sqrt{ 3! n \big( n-1 \big) \big( n-2 \big) } \epsilon \text{, }
\end{align*}

\noindent which is related to the fact that,

\begin{align*}
 \bigg[ \textbf{I} \otimes \textbf{I} \otimes \textbf{I} \bigg]   \ket{\psi_{3\mathrm{XOR}}}        < \sqrt{n \big( n-1 \big) \big( n -2 \big)} \epsilon  \text{. }
\end{align*}

\noindent Moreover, from the $\sqrt{2n \big( n-1 \big) \big( n-2 \big)} \epsilon$ upper bound, for the quantum state $\ket{\psi_{3\mathrm{XOR}}}$ corresponding to the optimal strategy, the optimal value $\omega_{3\mathrm{XOR}}$, as demonstrated through the statement of the desired inequality to be proved, implies that the operator acting on the 3$\mathrm{XOR}$ optimal state,

\begin{align*}
  \bigg[   \bigg[   \textbf{I} - \omega_{3\mathrm{XOR}} \mathrm{sign} \big( i_1 , j_1 , \cdots, j_n \big) \textbf{I}  \bigg] \otimes \textbf{I} \otimes \textbf{I}          \bigg]  \ket{\psi_{3\mathrm{XOR}}} < n^3 \sqrt{\epsilon} \text{, }
\end{align*}

\noindent can be rearranged as,

\begin{align*}
    \bigg[ \bigg( \textbf{I} \otimes \textbf{I} \otimes \textbf{I} \bigg) - \omega_{3\mathrm{XOR}}  \mathrm{sign} \big( i_1 , j_1 , \cdots , j_n \big) \bigg( \textbf{I} \otimes \textbf{I} \otimes \textbf{I} \bigg) \bigg]        \ket{\psi_{3\mathrm{XOR}}} \equiv \bigg( \textbf{I} \otimes \textbf{I} \otimes \textbf{I} \bigg)  \ket{\psi_{3\mathrm{XOR}}} \\  - \omega_{3\mathrm{XOR}}   \mathrm{sign} \big( i_1 , j_1 , \cdots , j_n \big) \bigg( \textbf{I} \otimes \textbf{I} \otimes \textbf{I} \bigg)  \ket{\psi_{3\mathrm{XOR}}}     \text{, }
\end{align*}

\noindent has the upper bound,

\begin{align*}
 n^3 \sqrt{\epsilon }   \text{. }
\end{align*}

\noindent Incorporating the previous estimates, the first of which is from formulating an upper bound for, 

\begin{align*}
      \bigg[     \bigg[ \textbf{I} - \bigg( \frac{\pm A_k + A_l}{\sqrt{2}} \bigg)  \textbf{I}\bigg]  \otimes \textbf{I} \otimes \textbf{I}   \bigg]   \ket{\psi_{3\mathrm{XOR}}}        \text{, }
\end{align*}

\noindent and the second of which is from formulating an upper bound for,

\begin{align*}
\bigg( \textbf{I} \otimes \textbf{I} \otimes \textbf{I} \bigg)  \ket{\psi_{3\mathrm{XOR}}}  - \omega_{3\mathrm{XOR}}   \mathrm{sign} \big( i_1 , j_1 , \cdots , j_n \big)  \bigg( \textbf{I} \otimes \textbf{I} \otimes \textbf{I} \bigg)  \ket{\psi_{3\mathrm{XOR}}}   \text{, }
\end{align*}

\noindent demonstrates that the desired upper bound takes the desired form, from the observation that,

\begin{align*}
 \bigg| \bigg| \bigg[ \bigg( \bigg( \underset{1 \leq i \leq n}{\prod}  A^{j_i}_i \bigg) \otimes B_{kl} \otimes \textbf{I} \bigg)  - \omega_{3\mathrm{XOR}} \bigg( \pm \mathrm{sign} \big( i_1 , j_1 , \cdots , j_n \big)      \bigg[       \bigg( \bigg( \underset{1 \leq i \leq n}{\prod}   A^{j_i}_k \bigg) \\   + \bigg(                     \underset{\text{set } j+1 \equiv j \oplus 1}{\underset{i \in \mathcal{Q}_1, j \in \mathcal{Q}_2}{\prod}}   A^{j_i}_k        \bigg) \bigg)  \otimes \textbf{I} \otimes \textbf{I} \bigg] \bigg) \bigg] \ket{\psi_{3\mathrm{XOR}}} \bigg| \bigg|^2      \text{, }
\end{align*}

\noindent implying,

\begin{align*}
       \sqrt{3! n \big( n - 1 \big) \big( n-2 \big) } \epsilon + \big( 100 \omega_{3 \mathrm{XOR}} \big) n^3 \sqrt{\epsilon} <      \sqrt{ 3! n \big( n - 1 \big) \big( n-2 \big) } \epsilon + \big( 1000  \big) n^3 \sqrt{\epsilon} \\ < \sqrt{\big(  3! n \big( n-1 \big) \big( n -2 \big) + 10000 n^7\big) \epsilon } \\ \equiv \sqrt{ 3! \big( n^3 - 3 n^2 + 2n + 10000 n^7 \big) \epsilon } \\ 
       \equiv \sqrt{ 3!  \big(\big( n^3 \big( 1 + 10000 n^4 \big) - n \big( 3 n +2 \big)   \big)  \epsilon }    \\ <  \sqrt{ 3!  \big( n^3 \big( 10001 n^4 \big) - n \big( 3n + 2 \big) \big)   }\big( 10 \sqrt{\epsilon} \big)  \\  < \sqrt{ 3!  \big( n^3 \big( 10001 n^4 \big) - n \big( 10 n \big) \big)   }\big( 10 \sqrt{\epsilon} \big) \\ \equiv \sqrt{         3! n^2 \big(  n \big( 10001 n^4 - 10    \big) \big)  } \big( 10 \sqrt{\epsilon} \big)   < \sqrt{3! n^2 \big(   n \big( 10002 n^4   \big) } \big( 10 \sqrt{\epsilon} \big)  \\  \equiv \sqrt{10002 n^7} \big( 10 \sqrt{\epsilon} \big)   < \sqrt{10002 n^8} \big( 10 \sqrt{\epsilon} \big)  < 10 \sqrt{10002} n^3 \sqrt{\epsilon}  \approx 1000 n^3 \sqrt{\epsilon } \text{. }
\end{align*}

\noindent As stated in the result at the beginning of the subsection, the final upper bound obtained for the $3$-$\mathrm{XOR}$ game satisfies,

\begin{align*}
  \sqrt{3! n \big( n - 1 \big) \big( n-2 \big) } \epsilon + \big( 100 \omega_{3 \mathrm{XOR}} \big) n^3 \sqrt{\epsilon}   \lesssim 3! n^3 \sqrt{\epsilon} \equiv  1000 n^3 \sqrt{\epsilon }   \text{. }
\end{align*}

\noindent For the 4-player XOR game, similar computations under the square root imply that an upper bound for the action of mapping $i + 1 \equiv i \oplus 1$, or $j+1 \equiv j \oplus 1$, or $i + 1 \equiv i \oplus 1$, and $j+1 \equiv j \oplus 1$, simultaneously, can be deduced from the following assumptions on the optimal strategy, $\ket{\psi_{4\mathrm{XOR}}}$. That is, to upper bound,

\begin{align*}
         \bigg| \bigg| \bigg[ \bigg( \bigg( \underset{1 \leq i \leq n}{\prod}  A^{j_i}_i \bigg) \otimes B_{kl} \otimes \textbf{I} \otimes \textbf{I} \bigg)  - \omega_{4\mathrm{XOR}} \bigg( \pm \mathrm{sign} \big( i_1 , j_1 , \cdots , j_n \big)      \bigg[       \bigg( \bigg( \underset{1 \leq i \leq n}{\prod}   A^{j_i}_k \bigg) \\   + \bigg(                     \underset{\text{set } j+1 \equiv j \oplus 1}{\underset{i \in \mathcal{Q}_1, j \in \mathcal{Q}_2}{\prod}}   A^{j_i}_k        \bigg) \bigg)   \otimes \textbf{I}  \otimes \textbf{I} \otimes \textbf{I} \bigg] \bigg) \bigg] \ket{\psi_{4\mathrm{XOR}}} \bigg| \bigg|^2          \text{, }
\end{align*}

\noindent fix $\epsilon^{\prime} \neq \epsilon$ sufficiently small, from which the computations for upper bounds the Frobenius norm result from the inequalities,

 \begin{align*}
    \bigg[  \bigg[ \textbf{I} - \textbf{I} \bigg( \frac{\pm A_k + A_l}{\sqrt{2}} \bigg) \bigg] \otimes \textbf{I} \otimes \textbf{I} \otimes \textbf{I} \bigg] \ket{\psi_{4\mathrm{XOR}}} < \sqrt{ 4! n \big( n-1 \big) \big( n-2 \big) \big( n-3 \big)  } \epsilon^{\prime} \text{, }
 \\ \\      \bigg[ \textbf{I} \otimes \textbf{I} \otimes \textbf{I} \bigg]   \ket{\psi_{4\mathrm{XOR}}}        < \sqrt{n \big( n-1 \big) \big( n -2 \big)} \epsilon^{\prime}        \text{, } \\ \\               \bigg[   \bigg[   \textbf{I} - \omega_{4\mathrm{XOR}} \mathrm{sign} \big( i_1 , j_1 , \cdots, j_n \big) \textbf{I}  \bigg] \otimes \textbf{I} \otimes \textbf{I}  \otimes \textbf{I}        \bigg]  \ket{\psi_{4\mathrm{XOR}}}  < n^4 \sqrt{\epsilon^{\prime}}               \text{.     }    
\end{align*}

\noindent To lighten the notation in the computations below, denote,

\begin{align*}
  \epsilon \equiv \epsilon^{\prime}  \text{. }
\end{align*}

\noindent Hence, the desired upper bound dependent upon the number of players in the game takes the form,

\begin{align*}
    \sqrt{4! n \big( n-1 \big) \big( n-2 \big) \big( n-3 \big)} \epsilon  + 100 \omega_{4 \mathrm{XOR}} n^4 \sqrt{\epsilon} <   \sqrt{4!  n \big( n - 1 \big) \big( n - 2 \big) \big( n-3 \big) } \epsilon \\ + 1000 w_{4 \mathrm{XOR}} \sqrt{n^8 \epsilon}   < \sqrt{ \big( 4! n \big( n-1 \big) \big( n -2 \big) \big( n-3 \big) \big) \epsilon + 10000 \omega_{4 \mathrm{XOR}} \big( n^8 \epsilon \big) } \\     \equiv \sqrt{ \big( 4! n \big( n-1 \big) \big( n-2 \big) \big( n-3 \big)+ 10000 \omega_{4 \mathrm{XOR}} n^8 \big)  \epsilon } \\  \equiv \sqrt{ \big( 20 n^4 - 10n^3 + 12 n^2 - 4n^3 + 10 n^2 - 12 n + 1000 \omega_{4 \mathrm{XOR}} n^8 \big) \epsilon }   \\ < 100 \sqrt{2 n^4 - 10 n^3 + 12 n ^2 - 4n^3 + 15 n^2 + 10000 \omega_{4 \mathrm{XOR}} n^8} \sqrt{\epsilon}    \\ < 1000   \sqrt{\big( 2 n^4 - 10 n^3 + 12 n ^2 - 40 n^3 + 10000 \omega_{4 \mathrm{XOR}} n^8\big) \epsilon} \\       < 10000 \sqrt{\big(     2n^4 - 10 n^3 + 12n^2 + 100 n^2 + 10000 \omega_{4 \mathrm{XOR}} n^8   \big)  \epsilon }  \\  <  10000 \sqrt{ \big( 2n^4 - 10 n^3 + 113 n^2 + 10000 \omega_{4 \mathrm{XOR}} n^8 \big) \epsilon }  
  \\   < 10000 \sqrt{\big( 2n^4 - 10n^3 + 113 n^3 + 10000 \omega_{4 \mathrm{XOR}} n^8 \big) \epsilon} \\ 
    < 10000 \sqrt{\big(   2 n^4 + 100 n^3 + 10000 \omega_{4 \mathrm{XOR}} n^8    \big) \epsilon} \\  <     10000 \sqrt{ \big( 105 n^4 + 10000 \omega_{4 \mathrm{XOR}} n^8\big) \epsilon  } \\  < 10000 \sqrt{\big( 10500 n^8 + 10000 \omega_{4 \mathrm{XOR}} n^8 \big) \epsilon }  \\ < 10000 \sqrt{10500 n^8 \epsilon}  \\ < 100000 n^4 \sqrt{\epsilon }  \text{. }
\end{align*}

\noindent As stated in the result at the beginning of the subsection, the final upper bound obtained for the $4$-$\mathrm{XOR}$ game satisfies,

\begin{align*}
 \sqrt{4! n \big( n-1 \big) \big( n-2 \big) \big( n-3 \big)} \epsilon  + 100 \omega_{4 \mathrm{XOR}} n^4 \sqrt{\epsilon}  \lesssim 4! n^4 \sqrt{\epsilon} \equiv      100000 n^4 \sqrt{\epsilon }   \text{. }
\end{align*}

\noindent To upper bound,

\begin{align*}
  \bigg| \bigg| \bigg[ \bigg( \bigg( \underset{1 \leq i \leq n}{\prod}  A^{j_i}_i \bigg) \otimes B_{kl} \otimes \textbf{I} \otimes \textbf{I} \otimes \textbf{I} \bigg)  - \omega_{5\mathrm{XOR}} \bigg( \pm \mathrm{sign} \big( i_1 , j_1 , \cdots , j_n \big)      \bigg[       \bigg( \bigg( \underset{1 \leq i \leq n}{\prod}   A^{j_i}_k \bigg)  \\  + \bigg(                     \underset{\text{set } j+1 \equiv j \oplus 1}{\underset{i \in \mathcal{Q}_1, j \in \mathcal{Q}_2}{\prod}}   A^{j_i}_k        \bigg) \bigg)  \otimes \textbf{I} \otimes \textbf{I}  \otimes \textbf{I} \bigg] \bigg) \bigg] \ket{\psi_{5\mathrm{XOR}}} \bigg| \bigg|^2    \text{, }
\end{align*}

\noindent fix $\epsilon^{\prime\prime} \neq \epsilon^{\prime} \neq \epsilon$ sufficiently small, from which the assumptions for the 5-$\mathrm{XOR}$ game include,

 \begin{align*}
    \bigg[  \bigg[ \textbf{I} - \textbf{I} \bigg( \frac{\pm A_k + A_l}{\sqrt{2}} \bigg) \bigg] \otimes \textbf{I} \otimes \textbf{I} \otimes \textbf{I} \otimes \textbf{I} \bigg] \ket{\psi_{5\mathrm{XOR}}} \\  < \sqrt{ 5! n \big( n-1 \big) \big( n-2 \big) \big( n-3 \big) \big( n - 4 \big)  } \epsilon^{\prime\prime} \text{, }
 \\ \\      \bigg[ \textbf{I} \otimes \textbf{I} \otimes \textbf{I} \otimes \textbf{I} \bigg]   \ket{\psi_{5\mathrm{XOR}}}        < \sqrt{n \big( n-1 \big) \big( n -2 \big)} \epsilon^{\prime\prime}        \text{, } \\ \\               \bigg[   \bigg[   \textbf{I} - \omega_{4\mathrm{XOR}} \mathrm{sign} \big( i_1 , j_1 , \cdots, j_n \big) \textbf{I}  \bigg] \otimes \textbf{I} \otimes \textbf{I} \otimes \textbf{I} \otimes \textbf{I}        \bigg]  \ket{\psi_{5\mathrm{XOR}}}      < n^5 \sqrt{\epsilon^{\prime\prime}}           \text{.     }    
\end{align*}

\noindent To lighten the notation in the computations below, denote,

\begin{align*}
  \epsilon \equiv \epsilon^{\prime\prime}  \text{. }
\end{align*}

\noindent Hence, the desired upper bound dependent upon the number of players in the game takes the form,

\begin{align*}
 \sqrt{ 5!  n \big( n-1 \big) \big( n-2 \big) \big( n-3 \big) \big( n-4 \big) } + 100 \omega_{5 \mathrm{XOR}} n^5 \sqrt{\epsilon} \\ <  \sqrt{ 5!  n \big( n-1 \big) \big( n-2 \big) \big( n-3 \big) \big( n-4 \big) \epsilon + 1000 \omega_{5 \mathrm{XOR}} n^{10} \epsilon } \\ \equiv \sqrt{ \big( 5!  n \big( n-1 \big) \big( n-2 \big) \big( n-3 \big) \big( n-4 \big) + 1000 \omega_{5 \mathrm{XOR}} n^{10} \big) \epsilon } \\ 
 < \sqrt{ \big( 5!  n \big( n-1 \big) \big( n-2 \big) \big( n-3  \big) \big( n-4 \big) + 10000 n^{10} \big) \epsilon  }  \\ < \sqrt{ \big( 20 n^5 - 8 n^4 + 120 n^4 - 40 n^3 + 12 n^2 - 48 n  \cdots} \\  \sqrt{- 2n^4 + 8 n^3 + 10 n^3 - 40 n^2 + 10000 n^{10} \big) \epsilon }  \\ \equiv \sqrt{ \big( 20 n^5 + 112 n^4 - 22 n^3 - 28 n^2 - 48 n + 10000 n^{10} \big) \epsilon }  \\        < \sqrt{ \big( 20 n^5 + 112 n^4 - 22 n^3 - 40 n^2 + 10000 n^{10} \big) \epsilon } \\  < \sqrt{\big( 20  n^{5} + 112 n^4 - 10 n^3 + 10000 n^{10} \big) \epsilon } \\ < \sqrt{ \big( 20 n^5 + 120 n^4 + 10000 n^{10} \big) \epsilon} 
 < \sqrt{ \big( 100 n^{10} + 10000 n^{10} \big) \epsilon}  \\   < \sqrt{20000 n^{10} \epsilon}  < \sqrt{20000} \big( n^5 \sqrt{\epsilon} \big)   < 1000 \sqrt{2} \big( n^5 \sqrt{\epsilon} \big)       \text{. }
\end{align*}

\noindent As stated in the result at the beginning of the subsection, the final upper bound obtained for the $5$-$\mathrm{XOR}$ game satisfies,

\begin{align*}
 \sqrt{ 5!  n \big( n-1 \big) \big( n-2 \big) \big( n-3 \big) \big( n-4 \big) } + 100 \omega_{5 \mathrm{XOR}} n^5 \sqrt{\epsilon}  \lesssim 5! n^5 \sqrt{\epsilon} \equiv   1000 \sqrt{2} \big( n^5 \sqrt{\epsilon} \big)      \text{. }
\end{align*}

\noindent In comparison to the upper bounds that have been previously obtained for a few players in the $\mathrm{XOR}$ setting, for an arbitrary number of players the following upper bound is dependent, up to leading order, upon,

\begin{align*}
     2n^{\# \text{ of players}}    \text{, }
\end{align*}

\noindent with other leading orders being determined by,

\begin{align*}
 \mathrm{O} \big( n^{\# \text{ of players}} - 1 \big) , \cdots , \mathrm{O} \big( n^{\# \text{ of players} } - \big( n - 1 \big) \big) , \cdots , \mathrm{O} \big( 1 \big)   \text{. }
\end{align*}

\noindent To upper bound,

\begin{align*}
       \bigg| \bigg| \bigg[ \bigg( \bigg( \underset{1 \leq i \leq n}{\prod}  A^{j_i}_i \bigg) \bigotimes B_{kl}  \bigotimes \bigg( \underset{1 \leq k \leq N-2}{\bigotimes} \textbf{I}_k  \bigg)  \bigg)  - \omega_{N\mathrm{XOR}} \bigg( \pm \mathrm{sign} \big( i_1 , j_1 , \cdots , j_n \big)      \\ \times \bigg[       \bigg( \bigg( \underset{1 \leq i \leq n}{\prod}   A^{j_i}_k \bigg)   + \bigg(                     \underset{\text{set } j+1 \equiv j \oplus 1}{\underset{i \in \mathcal{Q}_1, j \in \mathcal{Q}_2}{\prod}}   A^{j_i}_k        \bigg) \bigg)  \bigotimes \bigg( \underset{1 \leq k \leq N-1}{\bigotimes} \textbf{I}_k  \bigg) \bigg] \bigg) \bigg] \ket{\psi_{N\mathrm{XOR}}} \bigg| \bigg|^2    \text{, }
\end{align*}

\noindent fix $\epsilon^{\prime\prime\prime} \neq\epsilon^{\prime\prime} \neq \epsilon^{\prime} \neq \epsilon$ sufficiently small, from which the assumptions for the N$\mathrm{XOR}$ game include,

\begin{align*}
    \bigg[  \bigg[ \textbf{I} - \textbf{I} \bigg( \frac{\pm A_k + A_l}{\sqrt{2}} \bigg) \bigg] \bigotimes \bigg( \underset{1 \leq z \leq N-5}{\bigotimes} \textbf{I}_z       \bigg)  \bigg] \ket{\psi_{N\mathrm{XOR}}} \text{, }
\end{align*}

        \begin{align*}    \bigg( \underset{1 \leq z \leq N-5}{\bigotimes} \textbf{I}_z       \bigg)   \ket{\psi_{5\mathrm{XOR}}}              \text{, } \\ \\               \bigg[   \bigg[   \textbf{I} - \omega_{4\mathrm{XOR}} \mathrm{sign} \big( i_1 , j_1 , \cdots, j_n \big) \textbf{I}  \bigg] \bigotimes \bigg( \underset{1 \leq z \leq N-5}{\bigotimes} \textbf{I}_z       \bigg)     \bigg]  \ket{\psi_{N\mathrm{XOR}}}                \text{,     }    
\end{align*}

\noindent which can be upper bounded as follows, with,

 \begin{align*}
    \bigg[  \bigg[ \textbf{I} - \textbf{I} \bigg( \frac{\pm A_k + A_l}{\sqrt{2}} \bigg) \bigg] \otimes \textbf{I} \otimes  \textbf{I} \otimes \textbf{I} \otimes \textbf{I} \otimes \overset{N-5}{\cdots} \otimes \textbf{I} \bigg] \ket{\psi_{N\mathrm{XOR}}} \\ < \sqrt{ N! n \big( n-1 \big) \big( n-2 \big) \big( n-3 \big) \big( n - 4 \big)  } \epsilon^{\prime\prime\prime} \text{, }
\\ \\ \bigg[ \textbf{I} \otimes \textbf{I} \otimes \textbf{I} \otimes \textbf{I} \otimes \overset{N-5}{\cdots} \otimes \textbf{I} \bigg]   \ket{\psi_{5\mathrm{XOR}}}        < \sqrt{n \big( n-1 \big) \big( n -2 \big)} \epsilon^{\prime\prime\prime}        \text{, } \\ \\               \bigg[   \bigg[   \textbf{I} - \omega_{4\mathrm{XOR}} \mathrm{sign} \big( i_1 , j_1 , \cdots, j_n \big) \textbf{I}  \bigg] \otimes \textbf{I} \otimes \textbf{I} \otimes \textbf{I} \otimes \textbf{I}  \otimes \overset{N-5}{\cdots} \otimes \textbf{I}     \bigg]  \ket{\psi_{N\mathrm{XOR}}} \\ < n^N \sqrt{\epsilon^{\prime\prime\prime}}                \text{,     }    
\end{align*}

\noindent respectively. To lighten the notation in the computations below, denote,

\begin{align*}
  \epsilon \equiv \epsilon^{\prime\prime\prime}  \text{. }
\end{align*}

\noindent Hence, the desired upper bound dependent upon the number of players in the game takes the form,

\begin{align*}
      \sqrt{ N! n \bigg( \underset{1 \leq j \leq \# \text{ of players}-1}{\prod} \big( n - j \big)  \bigg)    \epsilon + 100 \omega_{N \mathrm{XOR}} n^{\# \text{ of players}} \epsilon        }         \\    <        \sqrt{ N! n \bigg( \underset{1 \leq j \leq \# \text{ of players}-1}{\prod} \big( n - j \big)  \bigg)    \epsilon + 1000  n^{\# \text{ of players}} \epsilon        }         \text{, }\end{align*}

  \noindent which can be further rearranged as,    
      
      \begin{align*}
      \sqrt{ \bigg(  N! n \bigg( \underset{1 \leq j \leq \# \text{ of players}-1}{\prod} \big( n - j \big)  \bigg) + 1000 n^{\# \text{ of players}} \bigg) \epsilon }       \\        <   \sqrt{ \bigg(  N! n \bigg( \underset{1 \leq j \leq \# \text{ of players}-1}{\prod} \big( n - j \big)  \bigg) + 1000 n^{\# \text{ of players}} \bigg)} \\     \times \big(  10 \sqrt{\epsilon }\big) \\ <        \sqrt{2N! n \big( n - \big( \# \text{ of players} -1 \big) \big) \bigg( \underset{1 \leq j \leq \# \text{ of players}-2}{\prod} \big( n - j \big)  \bigg) + 1000 n^{\# \text{ of players}} }   \\ \times       \big( 10 \sqrt{\epsilon } \big)   \\    \equiv     \sqrt{\big( 2N! n^2 - n \big( \# \text{ of players} - 1 \big) \big) \bigg( \underset{1 \leq j \leq \# \text{ of players}-2}{\prod} \big( n - j \big)  \bigg) + 1000  n^{\# \text{ of players}} }  \\ \times   \big( 10 \sqrt{\epsilon } \big) \\   <     \sqrt{20 N! n^2   \bigg( \underset{1 \leq j \leq \# \text{ of players}-2}{\prod} \big( n - j \big)  \bigg) - n \big( \# \text{ of players} -1 \big)   \bigg( \underset{1 \leq j \leq \# \text{ of players}-2}{\prod} \big( n - j \big)  \bigg)  \cdots   } \\ \sqrt{ + 1000 n^{\# \text{ of players}}       }    \big( 10 \sqrt{\epsilon} \big)  \\     <   \sqrt{20 N! n^2   \bigg( \underset{1 \leq j \leq \# \text{ of players}-2}{\prod} \big( n - j \big)  \bigg)         - n \big( \# \text{ of players} -1 \big) \big( n - \big( \# \text{ of players} -2 \big) \big) \cdots } \\ \sqrt{ \times  \bigg( \underset{1 \leq j \leq \# \text{ of players}-3}{\prod} \big( n - j \big)  \bigg)  + 1000 n^{\# \text{ of players}}}      \big( 11 \sqrt{\epsilon } \big) \text{. } \end{align*}

\noindent As the product over the number of players is taken, the final term above can be upper bounded with,

      \begin{align*}  \sqrt{20 N! n^2 \big( n - \big( \# \text{ of players} -2 \big) \big) \bigg( \underset{1 \leq j \leq \# \text{ of players}-3}{\prod} \big( n - j \big)  \bigg) -      \bigg( n^2 \big( \# \text{ of players} - 1 \big) \cdots } \\ \sqrt{ + \big( \# \text{ of players} - 1 \big) \big( \# \text{ of players} -2 \big) \bigg) \bigg( \underset{1 \leq j \leq \# \text{ of players}-3}{\prod} \big( n - j \big)   \bigg)  + 1000 n^{\# \text{ of players}}    }    \big( 11 \sqrt{\epsilon} \big)   \\    
      <    \sqrt{20 N! n^2 \big( n \big( \# \text{ of players} -2 \big)  \big) \bigg( \underset{1 \leq j \leq \# \text{ of players} -3 }{\prod} \big( n - j \big) \bigg)   - \frac{1}{2} \bigg( n^2 \big( \# \text{ of players} \cdots } \\ \sqrt{ - 1 \big) + \big( \# \text{of players} -1 \big)  \big( \# \text{of players} -2 \big) \bigg)  \bigg(  \underset{1 \leq j \leq \# \text{of players} -3}{\prod}\big( n - j \big) \bigg)  + 1000 n^{\# \text{ of players}}}               \big( 11 \sqrt{\epsilon } \big)  \\  \equiv  \sqrt{20 N! n^2 \big( n - \big( \# \text{of players} -2 \big)       \big) \bigg( \underset{1 \leq j \leq \# \text{of players} - 3}{\prod} \big( n - j \big)  \bigg)- \frac{1}{2} \bigg( n^2 \big( \# \text{ of players} -1 \big) \cdots } \\   \sqrt{ +  \big( \# \text{ of players} -1 \big) \big( \# \text{ of players} - 2  \big)  \bigg) \big( n - \big( \# \text{ of players} - 3 \big) \big) \bigg( \underset{1 \leq j \leq \# \text{ of players}-4}{\prod} \big( n - j \big)  \bigg) \cdots } \\ \sqrt{+ 1000 n^{\# \text{ of players}}  }      \big( 11 \sqrt{\epsilon }   \big) \text{. }  \tag{*}   \end{align*}

      \noindent Proceeding, the expression above, denoted with $(*)$, equals,
      
      \begin{align*} (\textit{*}) \equiv  \sqrt{20 N! n^2 \big( n - \big( \# \text{ of players} -2 \big) \big) \bigg( \underset{1 \leq j \leq \# \text{ of players}-3}{\prod} \big( n - j \big) \bigg)  - \frac{1}{2} \bigg(      - n^2 \big( \# \text{ of players} -1 \big) \cdots } \\ \sqrt{ \times  \big( \# \text{ of players} - 3 \big) \bigg)  + n^3 \big(  \# \text{ of players} -1 \big)     + n \big( \# \text{ of players} -1 \big) \big( \# \text{ of players } - 2 \big) \cdots } \\   \sqrt{ - \big( \# \text{ of players} -1 \big) \big( \# \text{ of players } -2 \big)  \big( \# \text{ of players} -3 \big) \bigg) \bigg(      \underset{1 \leq j \leq \# \text{ of players} -5}{\prod}  \big( n - j \big)     \bigg) \cdots } \\     
     \sqrt{  + 1000 n^{\# \text{ of players }}  }       \big( 11 \sqrt{\epsilon} \big)   \\  
     <     \sqrt{20 N! n^2 \big( n - \big( \# \text{ of players} -2 \big) \big)   \bigg(  \underset{1 \leq j \leq \# \text{ of players}-3}{\prod} \big( n - j \big)  \bigg) - \frac{1}{2} \bigg( n^2 \big( \# \text{ of players} \cdots } \\    
     \sqrt{  - 1 \big) \big( \# \text{ of players}   -3 \big) + n^3 \big( \# \text{ of players} - 1 \big) + n \big( \# \text{ of players} -1 \big) \big( \# \text{ of players}-2 \big) \cdots } \\     
     \sqrt{  - \big( \# \text{ of players} - 1 \big)  \big( \# \text{ of players} -2 \big)  \big( \# \text{ of players} -3 \big)  \bigg) \bigg( \underset{1 \leq j \leq \# \text{ of players}        }{\prod}   \big( n - j \big)    \bigg) \cdots } \\     
     \sqrt{ + 1000 n^{\# \text{ of players}}  }   \big( 11 \sqrt{\epsilon} \big)    \\    <            \sqrt{20 N! n^2 \big( n - \big( \# \text{ of players} -2 \big) \big) \big( n - \# \text{ of players} - 3 \big) \bigg) \big( n - 1 \big)                   - \frac{1}{2} \bigg(   - n^2 \big( \# \text{of players } -1 \big) \cdots }  \\    \sqrt{\times  \big( \# \text{ of players} -3 \big) \big( n - \big( \# \text{of players} -4 \big) \big)   + n^3 \big( \# \text{ of players} -1  \big) \big( n - \big( \# \text{ of players} -4  \big) \big) \cdots }   \\    \sqrt{+ n   \big( \# \text{ of players} -1 \big) \big( \# \text{ of players} -2 \big) \big( n - \big( \# \text{ of players} -4 \big) \big) \bigg) \bigg(  \underset{1 \leq j \leq \# \text{ of players} -5}{\prod}  \big( n - j \big)   \bigg)   \cdots } \\   \sqrt{ + 1000 n^{\# \text{ of players}}        }     \big( 11 \sqrt{\epsilon} \big) \\      <  \sqrt{20 N! n^2 \big( n - \big( \# \text{ of players} - 2 \big) \big)\big( n - \big( \# \text{ of players} - 3 \big) \big) \times \cdots \times    \big( n - 1 \big)  - \frac{1}{4} \big( -n^2 \big( \# \text{ of players}  \cdots     }   \\    \sqrt{    -1 \big) \big( \# \text{ of players} -3 \big) \big( \# \text{ of players} - 4 \big)  + n^3 \big( \# \text{ of players} -1 \big) \big( \# \text{ of players} - 4 \big) \cdots } \\  \sqrt{ + n \big( \# \text{ of players} -1  \big) \big( \# \text{ of players} - 2 \big) \big( \# \text{ of players} -4 \big) \big) \bigg)  \bigg( n^{ \# \text{ of players - 5} } \bigg) \cdots } \\ \sqrt{   + 1000 n^{\# \text{ of players}}   }  \big( 11 \sqrt{\epsilon} \big) \text{. }        \tag{**} \end{align*}

    \noindent Proceeding,

      \begin{align*}
    (\textit{**}) <   \sqrt{20 N! n^2 \bigg( n - \big( \# \text{ of players} - 2 \big)       \big) \big( n - \big( \# \text{ of players} -3 \big) \big) \times \cdots \times \big( n -1 \big)   - \frac{1}{4} \big( - n^2 \cdots  } \\ \sqrt{  \times  \big( \# \text{ of players} -1 \big)     \big( \# \text{ of players} -3 \big)    \big( \# \text{ of players} - 4\big)    + n^3 \big( \# \text{ of players} -1 \big) \cdots   } \\  \sqrt{ \times  \big( \# \text{ of players} -4 \big)             + n \big( \# \text{ of players} - 1 \big)    \big( \# \text{ of players} - 2 \big) \big( \# \text{ of players} - 4 \big)         \bigg)               \cdots   } \\   \sqrt{\times \bigg( n^{\# \text{ of players} - 5} \bigg)  + 1000 n^{\# \text{ of players}}      }  \big( 11 \sqrt{\epsilon} \big) \\       \equiv  \sqrt{20 N! n^2 \bigg( n - \big( \# \text{ of players} -2 \big) \big( n - \big( \# \text{ of players} -3 \big) \big) \times \cdots \times  \big( n - 1 \big)     - \frac{1}{4}  \big( - n^2 \cdots  } \\ \sqrt{ \times \big( \# \text{ of players} -1 \big) \big( \# \text{ of players} -3 \big)  \big( \# \text{ of players} - 4 \big)   + n  \big( \# \text{ of players} - 1 \big) \cdots   } \\    \sqrt{ \times \big( \# \text{ of players} - 2 \big)  \big( \# \text{ of players} - 4 \big) \bigg) \bigg( n^{\# \text{ of players} -5} \bigg) + 1000 n^{\# \text{ of players} }             }        \big( 11 \sqrt{\epsilon} \big) \\   \equiv   \sqrt{\bigg( 20 N! n^3   - 100 n^2 \big( \# \text{ of players} - 2 \big) \bigg(             \big( n - \big( \# \text{ of players} - 3 \big) \big) \times \cdots \times  \big( n - 1 \big)     \bigg) \cdots  }  \\   \sqrt{- \frac{1}{4} \bigg( - n^2 \big( \# \text{ of players} - 1 \big)  \big( \# \text{ of players} - 3 \big) \big( \# \text{ of players} - 4 \big) + n^3        \big( \# \text{ of players} -1 \big) \cdots  }  \\    \sqrt{\times  \big( \# \text{ of players} - 4 \big) n^{\# \text{of players} - 5}        +  n \big( \# \text{ of players} - 1 \big) \big( \# \text{ of players} - 2 \big)    \bigg)  \cdots    }  \\   \sqrt{\times \big( \# \text{ of players} - 4 \big)   n^{\# \text{ of players} - 5}  \bigg) + 1000 n^{\# \text{ of players}} }    \big( 11 \sqrt{\epsilon} \big)  \text{. } \tag{***}
\end{align*}

\noindent The final expression above can be further rearranged to obtain the desired upper bound, from the observations that,

\begin{align*}
    (\textit{***}) < \sqrt{\bigg( 20 N! n^3 - 50n^2 \big( n -2 \big) \bigg) \big( n-1 \big)^{\# \text{ of players} - 5}   - \frac{1}{4} \bigg( -n^2 \big( \# \text{ of players  - 1} \big)^3    n^{\# \text{ of players} - 5} \cdots } \\   \sqrt{ + n^3 \big( \# \text{ of players} -1 \big)^2 n^{\# \text{ of players} - 5}  + n \big( \# \text{ of players} - 1 \big)^3 n^{\# \text{ of players} - 5 } \bigg) + 1000 n^{\# \text{ of players}} }   \\ \times \big( 11 \sqrt{\epsilon} \big) \\   \equiv          \sqrt{\bigg( 20 N! n^3 - 50 n^2 \big( n-2 \big) \big) \big( n-1 \big)^{\# \text{ of players} - 3}    \bigg) - \frac{1}{4} \big( \big(  \# \text{ of players}  \big) - 1 \big)^2   n \bigg(        - n \big( \# \text{ of players} -1 \big)  \cdots } \\      \sqrt{\times n^{\# \text{ of players} - 5} + n^{\big( \# \text{ of players} - 5\big) + 1}     + \big( \# \text{ of players}  n^{\# \text{ of players} - 5}          \bigg) + 1000 n^{\# \text{ of players}}        }  \big( 11 \sqrt{\epsilon} \big)   \\  
    <       \sqrt{\bigg(     20 N! n^{\# \text{ of players} } - 50  n^{\# \text{ of players}} -     \bigg) 
          - \frac{1}{4}  n^3 \bigg(    - n \big( \# \text{ of players} - 1 \big)        n^{\# \text{ of players} - 5}       \cdots                 }     \\ \sqrt{     +             n^{\# \text{ of players} - 5}      +   \big( \# \text{ of players } - 5 \big)        n^{\# \text{ of players} - 5}    \bigg)   + 1000 n^{\# \text{of players}-5}   }    \big( 11 \sqrt{\epsilon} \big)   \\  
          <       \sqrt{n^{\# \text{ of players}}             \bigg(   n^{\# \text{ of players}} - n^3 \bigg( n^{\# \text{ of players}-4} + 2n^{\# \text{ of players}-5} \bigg)     \bigg)  + 1000 n^{\# \text{of players} -5}   }        \big( 110 \sqrt{\epsilon} \big) \\   <      \sqrt{ \bigg(  n^{\# \text{ of players}}     - n^3 \bigg(  - n^{\# \text{ of players}-5}      + 2n^{\# \text{ of players}-5}     \bigg)    + 1000 n^{\# \text{ of players}}      \bigg) }    \big( 120 \sqrt{\epsilon} \big)  \text{.} \tag{****}
\end{align*}

\noindent The final desired upper bound,

\[ \mathcal{I} \equiv \big( 130 N! n^N \big)   \sqrt{\epsilon} \times   
\left\{\!\begin{array}{ll@{}>{{}}l} n^{ \frac{\# \text{ of players}}{2} +5 } \Longleftrightarrow \big( \# \text{ of players} \big) \mathrm{mod} 2 \equiv 0 \\ n^{ \lfloor\frac{\# \text{ of players}}{2} \rfloor +5} \Longleftrightarrow \big( \# \text{ of players} \big) \mathrm{mod} 2 \neq 0
\end{array}\right.  \text{, } 
\]

\noindent for the N$\mathrm{XOR}$ game above is obtained from the observations that,

\begin{align*}
       (\textit{****}) < \sqrt{N! n^{\# \text{ of players} - 3} \bigg( 1 +  n^6 - 2n^5 + 1000 \bigg) } \big( 120  \sqrt{\epsilon} \big) <    \sqrt{ N! n^{\# \text{ of players} -3}  \bigg( 1 + n^6  + 998 n^5 \bigg) }  \\ \times  \big( 120   \sqrt{\epsilon} \big) \\  <     \sqrt{N! n^{\# \text{ of players} -3}  \bigg( 1 + 999 n^6  \bigg) }      \big( 120 \sqrt{\epsilon} \big)     < \sqrt{N! n^{\# \text{ of players}} n^8}  \big( 120 \sqrt{\epsilon} \big) <  \mathcal{I}   \text{.}
\end{align*}

\noindent As stated in the result at the beginning of the subsection, the final upper bound obtained for the $N$-$\mathrm{XOR}$ game satisfies,

\begin{align*}
     \sqrt{ N! n \bigg( \underset{1 \leq j \leq \# \text{ of players}-1}{\prod} \big( n - j \big)  \bigg)    \epsilon + 100 \omega_{N \mathrm{XOR}} n^{\# \text{ of players}} \epsilon        }    \lesssim   N! n^N \sqrt{\epsilon}  \text{. }
\end{align*}

\noindent With all of the desired upper bounds obtained, we conclude the argument. \boxed{}

\section{Conclusion}

\subsection{Overview}

\noindent In the final section, we provide a recapitulation of the error bounds obtained in the previous subsections, in addition to several comparisons between two-player, and higher-player, settings. To further develop, and elaborate, upon the rigidity, and structure, of error bounds for XOR, $\mathrm{XOR}^{*}$, and FFL games, we appealed to several aspects of Quantum information theory, ranging from: nonlocality, and contextuality, of the observables that each player prepares when interacting with the referee; paradoxical, and unexpected, aspects of Quantum information arising from error bounds which capture how the optimal value of the game can change; connections with Representation theory through intertwining operations which can transform tensor product representations of player observables into other representations; and several other themes. At the interface of Classical, and Quantum information processing, further examining sources of Quantum advantage that players can assume when making use of the nonlocality of information not only is of interest to further explore under less rigid assumptions on error bounds, but also for the prospects of determining computational tasks that can be executed on near term hardware. Despite the fact that the arguments developed in this work were developed independently of experiments, determining the expected runtime for obtaining exact expressions, or approximations, to primal feasible solutions for 3-XOR, 4-XOR, 5-XOR, N-XOR, FFL, and strong parallel repetition of XOR and FFL games, remains of interest. Besides such possible expansions, and elaborations, on the approach provided in this work, the two-player framework was expanded upon by: (1) providing expressions for game matrices, semidefinite programs, and associated duality gaps for multiplayer XOR games, and the FFL game; (2) upper bounding the multiplayer Frobenius norm; (3) demonstrating that the Frobenius norm has unit norm; (4) generalizing Schur's lemma to a multidimensional setting corresponding to an arbitrary number of players; (5) obtaining several upper bound estimates for the previous items for XOR and FFL games under strong parallel repetition; (6) proposing several avenues for expanding arguments for other classes of multiplayer games.

\subsection{$\mathrm{XOR}$ and $\mathrm{FFL}$ games: generalizations of error bounds}

\noindent To quantitatively determine how the optimal value, whether classical or quantum, for a game can change depending upon the observables of each player, we list several results previously obtained in this work which could be further studied to generalize error bounds for other game-theoretic settings of interest. First, recall, as a generalization of the two-player error bound, {[37]},

 \begin{align*}
     \underset{1 \leq i < j \leq n}{\sum} \bigg[ \text{ }  \bigg| \bigg|   \bigg[         \big( \frac{A_i + A_j}{\sqrt{2}} \big) \otimes I \bigg] \ket{\psi}    -   \big[  I \otimes B_{ij} \big] \ket{\psi}   \bigg|\bigg|^2 + \bigg| \bigg|   \bigg[ \big( \frac{A_i - A_j}{\sqrt{2}} \big)  \otimes I  \bigg] \ket{\psi}  \\   -   \big[  I \otimes B_{ji} \big]  \ket{\psi}     \bigg|\bigg|^2 \text{ }  \bigg]  \leq 2n \big( n - 1 \big) \epsilon \text{, }
 \end{align*}

\noindent previous error bounds established in the previous section:

\begin{itemize}
    \item[$\bullet$] \textit{3 $\mathrm{XOR}$ game}. One has,

\begin{align*}
  (\textbf{EB}- 3 \mathrm{XOR})  \leq 6 n \big( n - 1 \big) \big( n - 2 \big) \epsilon_{3\mathrm{XOR}} \text{. }
\end{align*}

\item[$\bullet$] \textit{4 $\mathrm{XOR}$ game}. One has,

\begin{align*}
    (\textbf{EB}- 4 \mathrm{XOR}) \leq  4! n \bigg( \big( n-1 \big) \big( n - 2 \big) \big( n-3 \big)  \bigg)  \epsilon_{4\mathrm{XOR}} \text{. }
\end{align*}

\item[$\bullet$] \textit{5 $\mathrm{XOR}$ game}. One has,

\begin{align*}
    (\textbf{EB}- 5 \mathrm{XOR}) \leq  5!  n \bigg( \big( n-1 \big) \big( n - 2 \big) \big( n-3 \big) \big( n-4 \big) \bigg)    \epsilon_{5 \mathrm{XOR}}  \text{. }
\end{align*}

\item[$\bullet$] \textit{N $\mathrm{XOR}$ game}. One has,

\begin{align*}
    (\textbf{EB}- N \mathrm{XOR}) \leq N! n \bigg( \underset{1 \leq j \leq N-1}{\prod} \big(  n - j  \big) \bigg)  \epsilon_{N \mathrm{XOR}}  \text{. }
\end{align*}

\item[$\bullet$] \textit{Strong parallel repetition of the FFL error bound}, \textbf{Lemma} \textit{4}. One has,

\begin{align*}
   \bigg| \bigg|               \bigg( \big(  A_k  \wedge A_{k^{\prime}} \big) \otimes \textbf{I} \bigg) \ket{\psi_{\mathrm{FFL} \wedge \mathrm{FFL}}}    -  \bigg( \textbf{I} \otimes \bigg(     \frac{\pm \big(  B_{kl} \wedge B_{k^{\prime} l^{\prime}} \big)  + \big( B_{lk} \wedge B_{k^{\prime} l^{\prime}} \big) }{\big| \pm \big(  B_{kl} \wedge B_{k^{\prime} l^{\prime}} \big)  + \big( B_{lk} \wedge B_{k^{\prime} l^{\prime}} \big) \big| }           \bigg) \bigg) \\ \times \ket{\psi_{\mathrm{FFL} \wedge \mathrm{FFL}}}              \bigg| \bigg|    < 20 \sqrt{N \epsilon^{\wedge}}     \text{. } 
\end{align*}

\item[$\bullet$] \textit{Strong parallel repetition of the $2$ $\mathrm{XOR}$ error bound}, \textbf{Lemma} $\textit{4}^{*}$. One has,

\begin{align*}
   \bigg| \bigg|               \bigg( \big(  A_k  \wedge A_{k^{\prime}} \wedge \cdots \wedge A_{k^{\prime\cdots \prime}} \big) \otimes \textbf{I} \bigg) \\ \times \ket{\psi_{2\mathrm{XOR} \wedge \cdots \wedge 2\mathrm{XOR}}}    \\ -  \bigg( \textbf{I} \otimes \bigg(     \frac{\pm \big(  B_{kl} \wedge B_{k^{\prime} l^{\prime}} \wedge \cdots \wedge B_{k^{\prime\cdots\prime}l^{\prime\cdots \prime}}  \big)  + \big( B_{lk} \wedge B_{ l^{\prime} k^{\prime}} \wedge \cdots \wedge B_{l^{\prime\cdots \prime} k^{\prime\cdots\prime} }  \big) }{\big| \pm \big(  B_{kl} \wedge B_{k^{\prime} l^{\prime}} \wedge \cdots \wedge B_{k^{\prime\cdots\prime}l^{\prime\cdots \prime}}  \big)  + \big( B_{lk} \wedge B_{ l^{\prime} k^{\prime}} \wedge \cdots \wedge B_{l^{\prime\cdots \prime} k^{\prime\cdots\prime} }  \big) \big| }           \bigg) \bigg)  \\ \times \ket{\psi_{2\mathrm{XOR} \wedge \cdots \wedge 2\mathrm{XOR}}}              \bigg| \bigg|   < 18 \sqrt{N \epsilon^{\wedge}_{2\mathrm{XOR} \wedge \cdots \wedge 2\mathrm{XOR}}}   \text{. } 
\end{align*}

\item[$\bullet$] \textit{Strong parallel repetition of the $N$ $\mathrm{XOR}$ error bound}, \textbf{Lemma} $\textit{4}^{**}$. One has,

\begin{align*}
  \mathcal{I}_1 \equiv  \bigg| \bigg|               \bigg( \big(  A_k  \wedge A_{k^{\prime}} \wedge \cdots \wedge A_{k^{\prime\cdots \prime}} \big) \bigotimes  \bigg( \underset{1 \leq z \leq N-1}{\bigotimes}\textbf{I}_z \bigg)  \bigg)   \ket{\psi_{N\mathrm{XOR} \wedge \cdots \wedge N\mathrm{XOR}}}    \\  -  \bigg( \textbf{I}  \otimes \bigg(     \frac{\pm \big(  B_{kl} \wedge B_{k^{\prime} l^{\prime}} \wedge \cdots \wedge B_{k^{\prime\cdots\prime}l^{\prime\cdots \prime}}  \big)  + \big( B_{lk} \wedge B_{ l^{\prime} k^{\prime}} \wedge \cdots \wedge B_{l^{\prime\cdots \prime} k^{\prime\cdots\prime} }  \big) }{\big| \pm \big(  B_{kl} \wedge B_{k^{\prime} l^{\prime}} \wedge \cdots \wedge B_{k^{\prime\cdots\prime}l^{\prime\cdots \prime}}  \big)  + \big( B_{lk} \wedge B_{ l^{\prime} k^{\prime}} \wedge \cdots \wedge B_{l^{\prime\cdots \prime} k^{\prime\cdots\prime} }  \big) \big| }    \bigg) \\  \bigotimes  \bigg( \underset{1 \leq z \leq N-2}{\bigotimes}\textbf{I}_z \bigg)       \bigg)  \ket{\psi_{N\mathrm{XOR} \wedge \cdots \wedge N\mathrm{XOR}}}              \bigg| \bigg| \\ \vdots \\  \mathcal{I}_N \equiv   \bigg| \bigg|    \bigg(   \bigg( \underset{1 \leq z \leq N-2}{\bigotimes} \textbf{I}_z \bigg) \bigotimes   \frac{1}{\sqrt{\# \sigma^{\prime} }}  \bigg(    \underset{\text{Permutations } \sigma^{\prime}}{\sum}  \big(   B^{(N-1)}_{\sigma^{\prime} ( i_1, \cdots, i_{N-1})}         \wedge   B^{(N-1)}_{\sigma^{\prime} ( i^{\prime}_1, \cdots, i^{\prime}_{N-1})}  \wedge  \cdots  \\ \wedge   B^{(N-1)}_{\sigma^{\prime} ( i^{\prime\cdots\prime}_1, \cdots, i^{\prime\cdots\prime}_{N-1})} \big)      \bigg)     \bigotimes \textbf{I}  \bigg)    \ket{\psi_{N\mathrm{XOR} \wedge \cdots \wedge N\mathrm{XOR}}}  \\ -  \bigg( \bigg( \underset{1 \leq z \leq N-1}{\bigotimes} \textbf{I}_z \bigg) \bigotimes \frac{1}{\sqrt{\# \sigma }}  \bigg(    \underset{\text{Permutations } \sigma}{\sum}  \big(   B^{(N-1)}_{\sigma ( i_1, \cdots, i_{N-1})}         \wedge   B^{(N-1)}_{\sigma ( i^{\prime}_1, \cdots, i^{\prime}_{N-1})}  \wedge  \cdots \\ \wedge   B^{(N-1)}_{\sigma ( i^{\prime\cdots\prime}_1, \cdots, i^{\prime\cdots\prime}_{N-1})} \big)      \bigg) \bigg)  \ket{\psi_{N\mathrm{XOR} \wedge \cdots \wedge N\mathrm{XOR}}}           \bigg| \bigg|      \text{, } 
\end{align*}

\noindent which have the strict upper bound,

\begin{align*}
\underset{1 \leq j \leq N}{\sum} \mathcal{I}_j < 20 N \sqrt{N \epsilon^{\wedge}_{N\mathrm{XOR} \wedge \cdots \wedge N\mathrm{XOR}}} \text{.}
 \end{align*}

\item[$\bullet$] \textit{Strong parallel $\mathrm{XOR}$ repetition}, \textbf{Lemma} \textit{FR} $\wedge \cdots \wedge$ \textit{FR}. One has,

\[
 (*) \lesssim  N! n^N    \sqrt{\epsilon_{\mathrm{XOR} \wedge \cdots \wedge \mathrm{XOR}}}  \times   
\left\{\!\begin{array}{ll@{}>{{}}l} n^{ \frac{\# \text{ of players}}{2}  +5} \Longleftrightarrow \big( \# \text{ of players} \big) \mathrm{mod} 2 \equiv 0 \\ n^{ \lfloor\frac{\# \text{ of players}}{2} \rfloor +5} \Longleftrightarrow \big( \# \text{ of players} \big) \mathrm{mod} 2 \neq 0
\end{array}\right.       \text{, }
\]

\noindent for,

\begin{align*}
    \bigg| \bigg|  \bigg[  \bigg(   \bigg( \big( A_1 \big)^{(1)}_i \wedge \cdots \wedge  \big( A_n \big)^{(1)}_{i^{\prime}} \bigg) \bigotimes \textbf{I}      \bigotimes \bigg(    \bigg(   \underset{i_1 \in \mathcal{Q}_1, i_2 \in \mathcal{Q}_2 , i_3 \in \mathcal{Q}_3}{\prod}             C^{l_{ijk}}_{ijk}     \bigg) \wedge \cdots \\ \wedge \bigg(    \underset{i^{\prime\cdots\prime}_3 \in \mathcal{Q}^{\prime\cdots\prime}_3}{\underset{i^{\prime\cdots\prime}_1 \in\mathcal{Q}^{\prime\cdots\prime}_1}{\underset{ i^{\prime\cdots\prime}_2 \in \mathcal{Q}^{\prime\cdots\prime}_2 }{\prod}}}              C^{l^{\prime\cdots\prime}_{i^{\prime\cdots\prime}j^{\prime\cdots\prime}k^{\prime\cdots\prime}}}_{i^{\prime\cdots\prime}j^{\prime\cdots\prime}k^{\prime\cdots\prime}}            \bigg)        \bigg)     \bigotimes                \bigg(     \underset{1\leq k \leq n-3}{\bigotimes}      \textbf{I}_k       \bigg)      \bigg)        -    \big( \omega \big( {\mathrm{XOR}}  \big) \big)^n  \\ \times   \bigg(       \pm \mathrm{sign} \big( i_1 , j_1 , k_1, i_{111} , \cdots , j_{111} , \cdots , j_{nm(n+m)}, k_{111} , \cdots , k_{nm(n+m)} \big)  \\ \times   \mathrm{sign} \big(  i^{\prime}_1  , j^{\prime}_1 , k^{\prime}_1, i^{\prime}_{111}, \cdots  , j^{\prime}_{111} , \cdots , j^{\prime}_{nm(n+m)}       , k^{\prime}_{111} , \cdots, k^{\prime}_{nm(n+m)}   \big)    \times  \cdots \\  \times \mathrm{sign} \big(  i^{\prime\cdots\prime}_1 , j^{\prime\cdots\prime}_1 , k^{\prime\cdots\prime}_1, i^{\prime\cdots\prime}_{111}   , \cdots ,   j^{\prime\cdots\prime}_{111}  , \cdots , j^{\prime\cdots\prime}_{nm(n+m)}          \big) \bigg) \bigg[                            \textbf{I} \bigotimes \textbf{I}   \\ \bigotimes \bigg(    \bigg(   \underset{i_1 \in \mathcal{Q}_1, i_2 \in \mathcal{Q}_2 , i_3 \in \mathcal{Q}_3}{\prod}             C^{l_{ijk}}_{ijk}     \bigg)  \wedge \cdots    \wedge \bigg(   \underset{i^{\prime\cdots\prime}_3 \in \mathcal{Q}^{\prime\cdots\prime}_3}{\underset{i^{\prime\cdots\prime}_1 \in\mathcal{Q}^{\prime\cdots\prime}_1}{\underset{ i^{\prime\cdots\prime}_2 \in \mathcal{Q}^{\prime\cdots\prime}_2 }{\prod}}}              C^{l^{\prime\cdots\prime}_{i^{\prime\cdots\prime}j^{\prime\cdots\prime}k^{\prime\cdots\prime}}}_{i^{\prime\cdots\prime}j^{\prime\cdots\prime}k^{\prime\cdots\prime}}            \bigg)        \bigg)   \\ \bigotimes \bigg( \bigg(   \underset{1\leq k \leq n-3}{\prod}  \textbf{I}_k \bigg)      \bigg)             \bigg] \bigg) \bigg]     \ket{\psi_{\mathrm{XOR} \wedge \cdots \wedge \mathrm{XOR}}}   \bigg| \bigg|_F    \tag{*}        \text{. }
\end{align*}

    \item[$\bullet$] \textit{Second 3 $\mathrm{XOR}$ error bound}, \textbf{Lemma} \textit{Gen-FFL-Bound}. One has,

    \begin{align*} \bigg| \bigg| \bigg[ \bigg( \bigg( \underset{1 \leq i \leq n}{\prod}  A^{j_i}_i \bigg) \otimes B_{kl} \otimes \textbf{I} \bigg)  - \omega_{3\mathrm{XOR}} \bigg( \pm \mathrm{sign} \big( i_1 , j_1 , \cdots , j_n \big)  \\ \times     \bigg[       \bigg( \bigg( \underset{1 \leq i \leq n}{\prod}   A^{j_i}_k \bigg)   + \bigg(                     \underset{\text{set } j+1 \equiv j \oplus 1}{\underset{i \in \mathcal{Q}_1, j \in \mathcal{Q}_2}{\prod}}   A^{j_i}_k        \bigg) \bigg)  \otimes \textbf{I} \otimes \textbf{I} \bigg] \bigg) \bigg] \ket{\psi_{3\mathrm{XOR}}} \bigg| \bigg|^2  \\ \lesssim  3! n^3  \sqrt{\epsilon }  \text{. }
    \end{align*}

    \item[$\bullet$] \textit{Second 4 $\mathrm{XOR}$ error bound}, \textbf{Lemma} \textit{Gen-FFL-Bound}. One has,

    \begin{align*}
          \bigg| \bigg| \bigg[ \bigg( \bigg( \underset{1 \leq i \leq n}{\prod}  A^{j_i}_i \bigg) \otimes B_{kl} \otimes \textbf{I} \otimes \textbf{I} \bigg)  - \omega_{4\mathrm{XOR}} \bigg( \pm \mathrm{sign} \big( i_1 , j_1 , \cdots , j_n \big)  \\ \times      \bigg[       \bigg( \bigg( \underset{1 \leq i \leq n}{\prod}   A^{j_i}_k \bigg)    + \bigg(                     \underset{\text{set } j+1 \equiv j \oplus 1}{\underset{i \in \mathcal{Q}_1, j \in \mathcal{Q}_2}{\prod}}   A^{j_i}_k        \bigg) \bigg) \otimes \textbf{I}   \otimes \textbf{I} \otimes \textbf{I} \bigg] \bigg) \bigg] \ket{\psi_{4\mathrm{XOR}}} \bigg| \bigg|^2  \\ \lesssim  4! n^4 \sqrt{\epsilon }   \text{. } 
    \end{align*}

\item[$\bullet$] \textit{Second 5 $\mathrm{XOR}$ error bound}, \textbf{Lemma} \textit{Gen-FFL-Bound}. One has,

 \begin{align*}
          \bigg| \bigg| \bigg[ \bigg( \bigg( \underset{1 \leq i \leq n}{\prod}  A^{j_i}_i \bigg) \otimes B_{kl} \otimes \textbf{I} \otimes \textbf{I} \otimes \textbf{I} \bigg)  - \omega_{5\mathrm{XOR}} \bigg( \pm \mathrm{sign} \big( i_1 , j_1 , \cdots , j_n \big)   \\ \times     \bigg[       \bigg( \bigg( \underset{1 \leq i \leq n}{\prod}   A^{j_i}_k \bigg)    + \bigg(                     \underset{\text{set } j+1 \equiv j \oplus 1}{\underset{i \in \mathcal{Q}_1, j \in \mathcal{Q}_2}{\prod}}   A^{j_i}_k        \bigg) \bigg)  \otimes \textbf{I} \otimes \textbf{I}  \otimes \textbf{I} \bigg] \bigg) \bigg] \ket{\psi_{5\mathrm{XOR}}} \bigg| \bigg|^2   \\ \lesssim  5! n^5 \sqrt{\epsilon}      \text{. }  
    \end{align*}

    \item[$\bullet$] \textit{Second N $\mathrm{XOR}$ error bound}, \textbf{Lemma} \textit{Gen-FFL-Bound}. One has,

   \[
    \bigg| \bigg| \bigg[ \bigg( \bigg( \underset{1 \leq i \leq n}{\prod}  A^{j_i}_i \bigg) \bigotimes B_{kl}  \bigotimes \bigg( \underset{1 \leq k \leq N-2}{\bigotimes} \textbf{I}_k  \bigg)  \bigg)  - \omega_{N\mathrm{XOR}} \bigg( \pm \mathrm{sign} \big( i_1 , j_1 , \cdots , j_n \big)     \] \[ \times     \bigg[       \bigg( \bigg( \underset{1 \leq i \leq n}{\prod}   A^{j_i}_k \bigg).   + \bigg(                     \underset{\text{set } j+1 \equiv j \oplus 1}{\underset{i \in \mathcal{Q}_1, j \in \mathcal{Q}_2}{\prod}}   A^{j_i}_k        \bigg) \bigg)  \bigotimes \bigg( \underset{1 \leq k \leq N-1}{\bigotimes} \textbf{I}_k  \bigg) \bigg] \bigg) \bigg] \ket{\psi_{N\mathrm{XOR}}} \bigg| \bigg|^2    \] \[  \lesssim  N! n^N  \sqrt{\epsilon} \times    
\left\{\!\begin{array}{ll@{}>{{}}l} n^{ \frac{\# \text{ of players}}{2}  +5} \Longleftrightarrow \big( \# \text{ of players} \big) \mathrm{mod} 2 \equiv 0 \\ n^{ \lfloor\frac{\# \text{ of players}}{2} \rfloor  +5} \Longleftrightarrow \big( \# \text{ of players} \big) \mathrm{mod} 2 \neq 0
\end{array}\right.   \text{. }
\]

\end{itemize}

\subsection{Potential Approaches for Imposing Less Regularity on Error Bounds}

\noindent The collection of error bounds above capture possible generalizations of rigid structures that appear in two-player settings, which have been thoroughly discussed throughout the previous section. In spite of the fact that error bounds, such as the ones included above, capture the possible ways in which an intertwining can be used to transform one tensor product representation of tensor player observables into another representation, other classes of error bounds can exhibit wide varieties of different properties which are consistent with Quantum theory. Along these lines, it is of interest to determine how drastically error bounds could impact the success probability, and hence the maximum chance of winning, for a player or group of players given previous responses that have been gathered for questions from the referee's probability distribution. To characterize such error bounds with less regular structures, we draw upon previously described insights in the previous subsections, which, in the case of multiple players, consists of: (1) formulating whether the optimal value for the game changes under parallel repetition; (2) passing to the dual $XOR^{*}$ game, from the XOR game; (3) weakening assumptions on the game, depending upon the observables that each player can form when initially receiving a question from the referee.

As in the multiplayer case for the XOR game provided in the previous section, in comparison to arguments formulated by Ostrev for the two-player setting that are reliant upon error bounds and the maximum probability of a player wining, the optimal value, settings with more players exhibit more intricates aspects of Quantum information. Besides the fact that such information is still expected to "play nice" with respect to performing transformations on one player's tensor observable to obtain another tensor observable, elaborations to Ostrev's arguments include: the dimensionality of the underlying resource system that is available to all agents; the collective utility that each agent in the game wishes to maximize; the computational complexity, whether of polynomial or exponential, runtime of computing Nash equilibria; asymptotic behaviors of the quantum state for hundreds, if not thousands, of agents; games with less regular structure, implying the need for additional generalized error bounds. With such a collection of error bounds, beyond analyzing games for which entanglement could still be analyzed for prospects of quantum advantage, previous quantities manipulated by the author in the two-player setting for $XOR^{*}$ and FFL games are still relevant. However, besides the observation that higher dimensional error bounds are still dependent upon the possible actions of each player and the optimal value of the game, such error bounds also characterize more complicated strategies that players can adopt for characterizing approximate, and exact, optimality. While there are similarities between error bounds for the XOR, and $\mathrm{CHSH} \big( n \big)$ games, generalizations of such similarities involve more complicated intereference patterns between the possible strategy that each player can adopt.

In the $\mathrm{CHSH} \big( n \big)$ setting, for $n \equiv 2$ players, the error bounds are only dependent upon the observable tensors $A_i$ and $B_{ij}$, or upon $A_j$ and $B_{ji}$, namely the observed set of outcomes that Alice provides for questions $i$ and $j$ administered by the referee, in addition to the responses that the remaining player, Bob, provides after Alice's turn. For game-theoretic settings with three, or more, players, optimal, and approximately optimal, quantum strategies for wining can still be characterized from the response of the player who receives the first question drawn from the referee's probability distribution. However, after the referee takes note of the first response that a player sends back for evaluation, the questions that are distributed to the remaining players can be analyzed with representations of higher-dimensional tensors, rather than only through $B_{ij}$, or $B_{ji}$, in the two-player setting. In particular, rather than only having to take into consideration the action of Alice, along with its impact on the possible actions of Bob, more complicated, higher-dimensional, relations by the referee must be examined. Nevertheless, for some $n \geq 3$, error bounds for players with tensors $\mathscr{P}_1, \cdots, \mathscr{P}_N$ can be realized through: (1) allowing each player, $\mathscr{P}$, to answer the question he or she receives from the referee; (2) in order, allowing each remaining player to send a response to the referee given the first player, and any previous player's, response; (3) applying an appropriate intertwining relation, which as a generalization of the intertwiner for th 2-player setting, acts on tensor product representations; (4) formulating a well-suited SDP, which as a constrainted optimization problem, depends on the referee's scoring function $V$ that is used for determining whether the responses of the player satisfy the wining XOR relation; (5) providing generalized various bonds, along with their various implications. We hope that such an overview discussion of the approach developed in this work is of value for analyzing more complicated game-theoretic interactions in the future.

\newpage 

\section{Appendix}

\subsection{Generalization of Schur's Lemma from Representation Theory}

\subsubsection{Statement}

\noindent In the first subsection of the Appendix, we describe a generalization of Schur's Lemma. In the two-player setting, from {[37]}, Schur's Lemma for the $\mathrm{CHSH} \big( n \big)$ asserts,

\begin{align*}
  T A_i \equiv B_i T  \text{, }
\end{align*}

\noindent for observables $B_i$ corresponding to the second player, Bob, and $A_i$ corresponding to the first player, Alice. In game theoretic settings with more players, in a previous subsection several counterpart linear operators were defined for 3-XOR, 4-XOR, 5-XOR, N-XOR games, in addition to strong parallel repetition of XOR and FFL games. In more complicated settings with more one would expect that the equality from Schur's lemma in the two-player setting would take the form,

\begin{align*}
      T_{3\mathrm{XOR}} A_i C_i \equiv  B_i T_{3\mathrm{XOR}} C_i     \text{, } \tag{S-1}
\end{align*}

\noindent corresponding to the player observables $A_i$ and $C_i$, from Alice, and Cleo, respectively, for the suitable 3 XOR linear operator $T_{3\mathrm{XOR}}$. However, in comparison to the inequality provided for the action of the suitable linear operator for 2 player settings, in the 3 player setting the action of the suitable linear operator $T_{3\mathrm{XOR}}$ can also take the form,

\begin{align*}
  A_i T_{3\mathrm{XOR}} C_i \equiv A_i B_i T_{3\mathrm{XOR}}  \text{, } \tag{S-2}
\end{align*}

\noindent corresponding to the action of the suitable linear operator between the tensor observables of the first, and third, players, into the observables of the first, and second, players, in addition to the equalities,

\begin{align*}
    T_{3\mathrm{XOR}} B_i C_i \equiv A_i T_{3\mathrm{XOR}} C_i        \text{, } \tag{S-3} \\  T_{3\mathrm{XOR}} C_i A_i \equiv   B_i T_{3\mathrm{XOR}} A_i   \text{, } \tag{S-4} \\   A_i T_{3\mathrm{XOR}}  C_i \equiv T_{3\mathrm{XOR}} B_i C_i      \text{, }  \tag{S-5} \\    B_i T_{3\mathrm{XOR}} C_i \equiv T_{3\mathrm{XOR}} A_i C_i      \text{, } \tag{S-6}
\end{align*}

\noindent corresponding to the action of the operator for the second, and third, tensor observables, third, and first, tensor observables,  first, and third, tensor observables, and second, and third, tensor observables, respectively. The following result asserts that the generalized version of Schur's Lemma for 3 XOR games, which can be immediately extended for any other XOR game with more players through the action of different suitable linear operators introduced in the previous section, can also be obtained.

\bigskip

\noindent \textbf{Lemma} \textit{Schur 3-XOR} (\textit{generalized Schur's Lemma for 3 XOR games}). Fix a suitable linear operator $T_{3\mathrm{XOR}}$. (S-1), (S-2), (S-3), (S-4), (S-5), and (S-6), hold.

\bigskip

\noindent \textit{Proof of Lemma Schur 3-XOR}. The generalization of Schur's Lemma follows from the observation that the product,

\begin{align*}
   T_{3\mathrm{XOR}} A_i C_i  \text{, }
\end{align*}

\noindent which corresponds to the action of the suitable linear operator on the first player's tensor observable, can be rearranged as,

\begin{align*}
  \bigg( T_{3\mathrm{XOR}} A_i \bigg)  C_i \overset{(\text{Two-Player Schur's Lemma})}{\equiv}   \bigg( B_i T_{3\mathrm{XOR}} \bigg) C_i \equiv  \bigg( \bigg( B_i T_{3\mathrm{XOR}} \bigg) C_i \bigg) \equiv B_i     \bigg(  T_{3\mathrm{XOR}}  C_i \bigg)  \\    \equiv B_i T_{3\mathrm{XOR}}  C_i       \text{, }
\end{align*}

\noindent which are equal. Straightforwardly, one can apply the same argument, from the two-player statement of Schur's Lemma to obtain the remaining equalities for the 3-XOR suitable linear operator, from which we conclude the argument. \boxed{}

\bigskip

\noindent Several adaptations of the result above can be introduced for 4-XOR, 5-XOR, N-XOR, games, and so on.

\subsubsection{The kernel, and image, of suitable linear operators are invariant}

\noindent We state the invariant subspace result for the suitable linear operator of the 3-XOR game only, as the accompanying result for any other XOR, or FFL, games in this paper can be obtained with an identical argument.

\bigskip

\noindent \textbf{Lemma} \textit{Ker 3-XOR} (\textit{the kernel of the suitable 3-XOR linear transformation is an invariant subspace}). The null space of the suitable 3-XOR linear transformation is invariant.

\bigskip

\noindent \textit{Proof of Lemma Ker 3-XOR}. Invariance of the kernel of $T_{3\mathrm{XOR}}$ follows from the straightforward observation, as an elementary linear algebra exercise, that the kernel subspace of $T_{3\mathrm{XOR}}$ is endowed with the same multiplication, and addition operations as the subpsace spanned by $T_{3\mathrm{XOR}}$, in addition to the fact that it has the same identity elements $0$, and $1$, corresponding to the operations and addition and multiplication, respectively, from which we conclude the argument. \boxed{}

\bigskip

\noindent As with the result stated in the previous subsection of the Appendix, several variants of the above result in this subsection which pertain to invariance of the kernel of suitable linear operators can be obtained for all other games discussed in this paper. The remaining result for the invariance of the image of suitable linear transformations can be immediately formulated from the result on the invariance of the kernel of suitable linear operators.

\subsection{Equivalent conditions for the $5$-$\mathrm{XOR}$ game}

\noindent For the $5$-$\mathrm{XOR}$ game, TFAE:

\begin{itemize}
    \item[$\bullet$] \underline{$\epsilon$-approximality of the optimal value}: For $\epsilon_{5\mathrm{XOR}}$ sufficiently small, and some $C^{\prime}>0$,

    \begin{align*}
  C^{\prime}  \omega_{5\mathrm{XOR}}  \big( 1 - \epsilon_{5\mathrm{XOR}} \big)   \leq \frac{1}{ {n \choose 5}}    \bra{\psi_{5\mathrm{XOR}}}   \underline{\mathscr{P}_{5\mathrm{XOR}}} \ket{\psi_{5\mathrm{XOR}}}                                \leq   C^{\prime}  \omega_{5\mathrm{XOR}}   \text{. }
    \end{align*}

      \item[$\bullet$] \underline{$\epsilon$-approximality of the bias}: For the same choice of $\epsilon_{5\mathrm{XOR}}$, taking the supremum over all possible strategies $\mathcal{S}$, for obtaining the optimal bias,

\begin{align*}
\underset{\text{strategies }\mathcal{S}}{\mathrm{sup}} \big\{ \beta_{5\mathrm{XOR}} \big( G_{5\mathrm{XOR}}, \mathcal{S} \big) \big\} \equiv \beta_{5\mathrm{XOR}} \big( G_{5 \mathrm{XOR}} \big)    \text{, }
\end{align*}

    \noindent implies that the inequality in the previous condition above takes the form,

  \begin{align*}
  C^{\prime}  \beta_{5\mathrm{XOR}}  \big( 1 - \epsilon_{5\mathrm{XOR}} \big)   \leq  \frac{1}{ {n \choose 5}}  \underset{\ket{\psi_{5\mathrm{XOR}}},A,B,C,D,E}{\mathrm{sup}}      \bigg\{  G_{5\mathrm{XOR}}         \bra{\psi_{3\mathrm{XOR}}}                   \underline{\mathscr{P}_{5\mathrm{XOR}}}    \ket{\psi_{5\mathrm{XOR}}}                           \bigg\}   \\     \leq   C^{\prime} \beta_{5\mathrm{XOR}}   \text{. }
    \end{align*}

  \item[$\bullet$] \underline{Optimality, and approximate optimality, from $3$ $\mathrm{XOR}$ error bounds}: The error bound for the $3$ $\mathrm{XOR}$ game is determined by the following representative contributions:

\begin{itemize}
    
\item[$\bullet$] For,

\begin{align*}
A_i     \Longleftrightarrow \frac{B_{ij} + B_{ji}}{\sqrt{2}}  \text{, }
\end{align*}

\noindent one has,

     \begin{align*}
  \bigg| \bigg|               \bigg[     \bigg( A_i \otimes \textbf{I} \otimes \textbf{I} \otimes \textbf{I}  \bigg)            
                            -    \bigg( \textbf{I} \otimes \bigg(  \frac{B_{ij} + B_{ji}}{\sqrt{2}}     \bigg) \otimes \textbf{I} \otimes \textbf{I} \otimes \textbf{I}                             \bigg) \bigg]   \ket{\psi_{5\mathrm{XOR}}}          \bigg| \bigg|  \end{align*}
                            
                            \item[$\bullet$] For,

\begin{align*}
B_{ij}     \Longleftrightarrow \frac{A_i  + A_j}{\sqrt{2}}  \text{, }
\end{align*}

\noindent one has,
                            
                            \begin{align*} \bigg| \bigg|      \bigg[      \bigg(        \textbf{I} \otimes B_{ij} \otimes \textbf{I} \otimes \textbf{I} \otimes \textbf{I}                         \bigg)  - \bigg(         \bigg( \frac{A_i + A_j}{\sqrt{2}} \bigg) \otimes \textbf{I} \otimes        \textbf{I} \otimes \textbf{I} \otimes \textbf{I}        \bigg)           \bigg]                    \ket{\psi_{5\mathrm{XOR}}}          \bigg] \bigg| \bigg|    \text{. }           \end{align*}

                            \item[$\bullet$] For,

\begin{align*}
C_{ijk}     \Longleftrightarrow \frac{A_i  + A_j}{\sqrt{2}}  \text{, }
\end{align*}

\noindent one has, 
                            
                            \begin{align*}   \bigg| \bigg| \bigg[ \bigg(        \textbf{I} \otimes \textbf{I} \otimes C_{ijk} \otimes \textbf{I} \otimes \textbf{I }           \bigg) - \bigg(          \bigg( \frac{A_i + A_j}{\sqrt{2} } \bigg) \otimes \textbf{I}    \otimes \textbf{I} \otimes \textbf{I} \otimes \textbf{I}                 \bigg) \bigg]        \ket{\psi_{5\mathrm{XOR}}}   \bigg| \bigg|     \end{align*}

                            \item[$\bullet$] For,

\begin{align*}
D_{ijkl}     \Longleftrightarrow \frac{A_i  + A_j}{\sqrt{2}}  \text{, }
\end{align*}

\noindent one has, 
                            
                            \begin{align*}   \bigg| \bigg|      \bigg[ \bigg(        \textbf{I} \otimes \textbf{I} \otimes \textbf{I} \otimes  D_{ijkl} \otimes\textbf{I}               \bigg) - \bigg(        \bigg( \frac{A_i + A_j}{\sqrt{2}} \bigg) \otimes \textbf{I} \otimes \textbf{I} \otimes \textbf{I} \otimes \textbf{I} \bigg)  \bigg]         \ket{\psi_{5\mathrm{XOR}}}      \bigg| \bigg| \end{align*}

\item[$\bullet$] For,

\begin{align*}
E_{ijklm}    \Longleftrightarrow \frac{A_i  + A_j}{\sqrt{2}}  \text{, }
\end{align*}

\noindent one has,
                            
                            \begin{align*} \bigg| \bigg|      \bigg[ \bigg(        \textbf{I} \otimes \textbf{I} \otimes \textbf{I} \otimes \textbf{I} \otimes E_{ijklm}                                 \bigg) - \bigg(           \bigg( \frac{A_i + A_j}{\sqrt{2}} \bigg) \otimes \textbf{I} \otimes \textbf{I} \otimes \textbf{I} \otimes \textbf{I}  \bigg)  \bigg] \ket{\psi_{5\mathrm{XOR}}}  \bigg| \bigg|          \end{align*}

                            \item[$\bullet$] For,

\begin{align*}
A_i     \Longleftrightarrow \frac{B_{ij}  + B_{ji}}{\sqrt{2}}  \text{, } \\  C_{ijk} \Longleftrightarrow  \frac{\underset{\sigma \in S_4}{\sum} D_{\sigma ( ijkl)}}{\sqrt{24}} \text{, }
\end{align*}

\noindent one has,
                            
                            \begin{align*}\bigg| \bigg|      \bigg[ \bigg(          A_i \otimes \textbf{I} \otimes C_{ijk} \otimes \textbf{I} \otimes \textbf{I}                      \bigg)  - \bigg(         \textbf{I} \otimes \bigg( \frac{B_{ij} + B_{ji}}{\sqrt{2}} \bigg) \otimes \textbf{I} \otimes \frac{1}{{\sqrt{24}} } \bigg(                  \underset{\sigma \in S_4}{\sum} D_{\sigma ( ijkl)}  \bigg) \otimes \textbf{I}                      \bigg)  \bigg] \\ \times 
        \ket{\psi_{5\mathrm{XOR}}}    \bigg| \bigg|      \end{align*}
                            
                      \item[$\bullet$] For,

\begin{align*}
A_i     \Longleftrightarrow \frac{B_{ij}  + B_{ji}}{\sqrt{2}}  \text{, } \\  D_{ijkl} \Longleftrightarrow  \frac{\underset{\sigma \in S_5}{\sum}E_{\sigma(ijklm)}                }{\sqrt{120}}  \text{, }
\end{align*}

\noindent one has,    
                            
                            \begin{align*}
                            \bigg| \bigg| \bigg[ \bigg(        A_i \otimes \textbf{I} \otimes \textbf{I}    \otimes D_{ijkl} \otimes \textbf{I}              \bigg) - \bigg(                   \textbf{I} \otimes \bigg(       \frac{B_{ij} + B_{ji}}{\sqrt{2}}  \bigg) \otimes        \textbf{I} \otimes \textbf{I}   \otimes  \frac{1}{{\sqrt{120}} } \bigg( \underset{\sigma \in S_5}{\sum}E_{\sigma(ijklm)}       \bigg)                         \bigg) \bigg] \\ \times \ket{\psi_{5\mathrm{XOR}}} \bigg| \bigg|     \end{align*}

\item[$\bullet$] For,

\begin{align*}
E_{ijklm}    \Longleftrightarrow \frac{A_i + A_j}{\sqrt{2}}  \text{, }
\end{align*}

\noindent one has,

                            \begin{align*}      \bigg| \bigg|  \bigg[       \bigg(     \textbf{I} \otimes \textbf{I} \otimes \textbf{I} \otimes E_{ijklm}      \bigg) - \bigg(       \bigg( \frac{A_i + A_j}{\sqrt{2}} \bigg) \otimes \textbf{I} \otimes \textbf{I} \otimes \textbf{I} \otimes \textbf{I}      \bigg)                  \bigg]  \ket{\psi_{5\mathrm{XOR}}}      \bigg| \bigg| \end{align*}

\item[$\bullet$] For,

\begin{align*}
B_{ij}     \Longleftrightarrow \frac{A_i + A_j}{\sqrt{2}}  \text{, } \\ C_{ijk} \Longleftrightarrow   \frac{\underset{\sigma \in S_4}{\sum} D_{\sigma ( ijkl)}}{\sqrt{24}}           \text{, }
\end{align*}

\noindent one has,

                            \begin{align*}     \bigg| \bigg|  \bigg[ \bigg(         \textbf{I} \otimes B_{ij} \otimes C_{ijk} \otimes \textbf{I} \otimes \textbf{I}          \bigg)  -    \bigg(           \bigg( \frac{A_i + A_j}{\sqrt{2}} \bigg)  \otimes       \textbf{I} \otimes \textbf{I} \otimes \frac{1}{\sqrt{24}} \bigg(     \underset{\sigma \in S_4}{\sum} D_{\sigma ( ijkl)}         \bigg)  \otimes \textbf{I}                               \bigg)                  \bigg]  \\ \times \ket{\psi_{5\mathrm{XOR}}}      \bigg| \bigg|    
                            \end{align*}

\item[$\bullet$] For,

\begin{align*}
B_{ij}     \Longleftrightarrow \frac{A_i + A_j}{\sqrt{2}}  \text{, } \\ D_{ijkl} \Longleftrightarrow       \frac{\underset{\sigma \in S_3}{\sum} C_{\sigma(ijk)}}{\sqrt{6}}      \text{, }
\end{align*}

\noindent one has,          

                            \begin{align*} 
                            \bigg| \bigg| \bigg[            \bigg(         \textbf{I} \otimes B_{ij} \otimes \textbf{I} \otimes D_{ijkl} \otimes \textbf{I}                   \bigg) - \bigg(     \bigg( \frac{A_i + A_j}{\sqrt{2}} \bigg) \otimes \textbf{I} \otimes \frac{1}{\sqrt{6}}  \bigg(          \underset{\sigma \in S_3}{\sum} C_{\sigma(ijk)}   \bigg)                  \bigg]  \ket{\psi_{5\mathrm{XOR}}}      \bigg| \bigg|  \end{align*}

\item[$\bullet$] For,

\begin{align*}
B_{ij}     \Longleftrightarrow \frac{A_i + A_j}{\sqrt{2}}  \text{, } \\ E_{ijklm} \Longleftrightarrow   \frac{\underset{\sigma \in S_4}{\sum} D_{\sigma ( ijkl)}}{\sqrt{24}}                      \text{, }
\end{align*}

\noindent one has,                   
                            \begin{align*} \bigg| \bigg|    \bigg[   \bigg(        \textbf{I} \otimes B_{ij} \otimes \textbf{I} \otimes \textbf{I} \otimes E_{ijklm}                  \bigg) - \bigg(            \bigg( \frac{A_i + A_j}{\sqrt{2}} \bigg) \otimes \textbf{I} \otimes \textbf{I} \otimes \frac{1}{\sqrt{24}} \bigg(     \underset{\sigma \in S_4}{\sum} D_{\sigma ( ijkl)}         \bigg)                 \bigg]  \\ \times \ket{\psi_{5\mathrm{XOR}}}      \bigg| \bigg|  \end{align*}

                            \item[$\bullet$] For,

\begin{align*}
D_{ijkl}   \Longleftrightarrow \frac{B_{ij} + B_{ji}}{\sqrt{2}}  \text{, } \\ C_{ijk} \Longleftrightarrow    \frac{\underset{\sigma \in S_4}{\sum} E_{\sigma ( ijklm)}}{\sqrt{120}}                      \text{, }      
\end{align*}

\noindent one has,

                            \begin{align*}   \bigg| \bigg|     \bigg[    \bigg(         \textbf{I} \otimes \textbf{I} \otimes C_{ijk} \otimes D_{ijkl} \otimes \textbf{I}         \bigg)  - \bigg(     \textbf{I} \otimes \bigg( \frac{B_{ij} + B_{ji}}{\sqrt{2}} \bigg) \otimes \textbf{I} \otimes \frac{1}{\sqrt{120}} \bigg(     \underset{\sigma \in S_4}{\sum} E_{\sigma ( ijkl)}         \bigg)                 \bigg]  \\ \times \ket{\psi_{5\mathrm{XOR}}}      \bigg| \bigg|      \bigg]  \end{align*}

                            \item[$\bullet$] For,

\begin{align*}
A_i   \Longleftrightarrow   \frac{\underset{\sigma \in S_3}{\sum} C_{\sigma(ijk)}}{\sqrt{6}}    \text{, } \\ B_{ij} \Longleftrightarrow                 \frac{\underset{\sigma \in S_4}{\sum} E_{\sigma ( ijklm)}}{\sqrt{120}}       \text{, }      
\end{align*}

\noindent one has,

                            \begin{align*} \bigg| \bigg|    \bigg[     \bigg(   A_i      \otimes B_{ij} \otimes \textbf{I} \otimes \textbf{I} \otimes \textbf{I}   \bigg) - \bigg(  \textbf{I} \otimes \textbf{I} \otimes   \frac{1}{\sqrt{6}} \bigg( \underset{\sigma \in S_3}{\sum} C_{\sigma(ijk)} \bigg)  \otimes \textbf{I}  \otimes  \frac{1}{\sqrt{120}} \bigg(     \underset{\sigma \in S_4}{\sum} E_{\sigma ( ijkl)}         \bigg) \bigg] \\ \times \ket{\psi_{5\mathrm{XOR}}}      \bigg| \bigg|      \bigg]  \end{align*}

\item[$\bullet$] For,

\begin{align*}
A_i   \Longleftrightarrow \frac{\underset{\sigma \in S_3}{\sum} C_{\sigma(ijk)}}{\sqrt{6}}   \text{, } \\ B_{ij} \Longleftrightarrow              \frac{\underset{\sigma \in S_4}{\sum} D_{\sigma ( ijkl)}}{\sqrt{24}}              \text{, }      
\end{align*}

\noindent one has,

                            \begin{align*} \bigg| \bigg|   \bigg[      \bigg(                      A_i \otimes B_{ij} \otimes \textbf{I} \otimes \textbf{I} \otimes \textbf{I}   \bigg) - \bigg(                      \textbf{I} \otimes \textbf{I}  \otimes        \frac{1}{\sqrt{6}} \bigg( \underset{\sigma \in S_3}{\sum} C_{\sigma(ijk)}              \bigg)  \otimes \frac{1}{\sqrt{24}} \bigg(     \underset{\sigma \in S_4}{\sum} D_{\sigma ( ijkl)}         \bigg)                  \bigg]  \\ \times \ket{\psi_{5\mathrm{XOR}}}      \bigg| \bigg|      \bigg]  \end{align*}

                            \item[$\bullet$] For,

\begin{align*}
A_i     \Longleftrightarrow \frac{B_{ij}+ B_{ji}}{\sqrt{2}}  \text{, } \\ C_{ijk} \Longleftrightarrow   \frac{\underset{\sigma \in S_4}{\sum} D_{\sigma ( ijkl)}}{\sqrt{24}}     \text{, }
\end{align*}

\noindent one has,

                            \begin{align*}    \bigg| \bigg| \bigg[        \bigg(      A_i \otimes \textbf{I} \otimes C_{ijk} \otimes \textbf{I} \otimes \textbf{I}         \bigg) - \bigg(                    \textbf{I} \otimes \bigg( \frac{B_{ij} + B_{ji}}{\sqrt{2}} \bigg) \otimes \textbf{I }\frac{1}{\sqrt{24}} \bigg(     \underset{\sigma \in S_4}{\sum} D_{\sigma ( ijkl)}         \bigg)                   \bigg] \\ \times  \ket{\psi_{5\mathrm{XOR}}}      \bigg| \bigg| \end{align*}

                            \item[$\bullet$] For,

\begin{align*}
A_i     \Longleftrightarrow \frac{B_{ij}+ B_{ji}}{\sqrt{2}}  \text{, } \\ D_{ijkl} \Longleftrightarrow   \frac{\underset{\sigma \in S_4}{\sum} E_{\sigma ( ijklm)}}{\sqrt{120}}         \text{, }
\end{align*}

\noindent one has,
                            
                            \begin{align*}\bigg| \bigg| \bigg[ \bigg(          A_i \otimes \textbf{I} \otimes \textbf{I} \otimes D_{ijkl} \otimes \textbf{I}        \bigg) - \bigg(            \textbf{I} \otimes \bigg( \frac{B_{ij} + B_{ji}}{\sqrt{2}} \bigg) \otimes \textbf{I} \otimes \textbf{I}  \otimes  \frac{1}{\sqrt{120}} \bigg(     \underset{\sigma \in S_4}{\sum} E_{\sigma ( ijkl)}         \bigg)                \bigg) \bigg] \\ \times \ket{\psi_{5\mathrm{XOR}}} \bigg| \bigg| \end{align*}
                            
                            \item[$\bullet$] For,

\begin{align*}
A_i     \Longleftrightarrow \frac{A_{i}+ A_{j}}{\sqrt{2}}  \text{, } \\ E_{ijklm} \Longleftrightarrow   \frac{B_{ij} + B_{ji}}{\sqrt{2}}                   \text{, }
\end{align*}

\noindent one has,

                            \begin{align*} \bigg| \bigg|       \bigg[       \bigg(                   A_i \otimes \textbf{I} \otimes \textbf{I} \otimes \textbf{I} \otimes E_{ijklm}     \bigg) - \bigg(           \bigg( \frac{A_i + A_j}{\sqrt{2}} \bigg) \otimes \bigg( \frac{B_{ij} + B_{ji}}{\sqrt{2}} \bigg)  \otimes \textbf{I } \otimes \textbf{I}  \otimes \textbf{I}          \bigg)          \bigg] \\ \times \ket{\psi_{5\mathrm{XOR}}}      \bigg| \bigg|   \end{align*}

                            \item[$\bullet$] For,

\begin{align*}
A_i     \Longleftrightarrow \frac{B_{ij}+ B_{ji}}{\sqrt{2}}  \text{, } \\ D_{ijkl} \Longleftrightarrow  \frac{\underset{\sigma \in S_3}{\sum} C_{\sigma(ijk)}}{\sqrt{6}}                \text{, }
\end{align*}

\noindent one has,

                            \begin{align*} \bigg| \bigg| \bigg[ \bigg(          \textbf{I} \otimes B_{ij} \otimes \textbf{I} \otimes \textbf{I} \otimes E_{ijklm}           \bigg) - \bigg(        \bigg( \frac{A_i + A_j}{\sqrt{2}} \bigg) \otimes \textbf{I}   \otimes  \frac{1}{\sqrt{6}} \bigg( \underset{\sigma \in S_3}{\sum} C_{\sigma(ijk)}              \bigg) \otimes \textbf{I} \otimes \textbf{I}             \bigg)          \bigg] \\ \times \ket{\psi_{5\mathrm{XOR}}}      \bigg| \bigg| \end{align*}

                            \item[$\bullet$] For,

\begin{align*}
A_i     \Longleftrightarrow \frac{B_{ij}+ B_{ji}}{\sqrt{2}}  \text{, } \\ D_{ijkl} \Longleftrightarrow \frac{\underset{\sigma \in S_3}{\sum} C_{\sigma(ijk)}}{\sqrt{6}}             \text{, }
\end{align*}

\noindent one has,  
                            
                            \begin{align*}  \bigg| \bigg| \bigg[ \bigg( A_i \otimes \textbf{I} \otimes \textbf{I} \otimes D_{ijkl} \otimes \textbf{I}                    \bigg) - \bigg(          \textbf{I}       \otimes \bigg( \frac{B_{ij} + B_{ji}}{\sqrt{2}} \bigg) \otimes    \frac{1}{\sqrt{6}} \bigg( \underset{\sigma \in S_3}{\sum} C_{\sigma(ijk)}              \bigg)      \bigg] \ket{\psi_{5\mathrm{XOR}}}      \bigg| \bigg|   \bigg]       \text{. }
           \end{align*}

\end{itemize}
      
\end{itemize}

\subsection{$3$-$\mathrm{XOR}$ game}

\noindent The Bell states are generated by the operations,

\begin{align*}
      \big( \textbf{I} \otimes \textbf{I} \otimes \textbf{I} \big) \bigg( \frac{\ket{\text{Player } 1 \text{ state}} +\ket{\text{Player } 2 \text{ state}} + \ket{\text{Player } 3 \text{ state}}}{\sqrt{3}} \bigg)  \\ = \frac{1}{\sqrt{3}} \bigg(  \ket{\text{Player } 1 \text{ state}}  +\ket{\text{Player } 2 \text{ state}}  + \ket{\text{Player } 3 \text{ state}}  \bigg) \text{, } \\ \\             \big(  \sigma_x \otimes \textbf{I} \otimes \textbf{I}     \big)  \bigg( \frac{\ket{\text{Player } 1 \text{ state}} +\ket{\text{Player } 2 \text{ state}} + \ket{\text{Player } 3 \text{ state}}}{\sqrt{3}} \bigg) \\ = \frac{1}{\sqrt{3} } \bigg(  \widetilde{\ket{\text{Player } 1 \text{ state}}}  +\ket{\text{Player } 2 \text{ state}}  + \widetilde{\ket{\text{Player } 3 \text{ state}}}     \bigg)     \text{, } \\ \\         \big( \textbf{I} \otimes \sigma_x \otimes \textbf{I}     \big)  \bigg( \frac{\ket{\text{Player } 1 \text{ state}} +\ket{\text{Player } 2 \text{ state}} + \ket{\text{Player } 3 \text{ state}}}{\sqrt{3}} \bigg) \\   = \frac{1}{\sqrt{3} } \bigg(      {\ket{\text{Player } 1 \text{ state}}}  + \widetilde{\ket{\text{Player } 2 \text{ state}}}     + \widetilde{\ket{\text{Player } 3 \text{ state}}}           \bigg)    \text{, } \\ \\  \big( \textbf{I} \otimes \textbf{I} \otimes \sigma_x \big)         \bigg( \frac{\ket{\text{Player } 1 \text{ state}} +\ket{\text{Player } 2 \text{ state}} + \ket{\text{Player } 3 \text{ state}}}{\sqrt{3}} \bigg) \\   = \frac{1}{\sqrt{3} } \bigg(      \widetilde{\ket{\text{Player } 1 \text{ state}}}   + {\ket{\text{Player } 2 \text{ state}}}  + \widetilde{\ket{\text{Player } 3 \text{ state}}}                  \bigg)    \text{, } \\ \\     \big( \sigma_z \otimes \textbf{I} \otimes \textbf{I} \big)       \bigg( \frac{\ket{\text{Player } 1 \text{ state}} +\ket{\text{Player } 2 \text{ state}} + \ket{\text{Player } 3 \text{ state}}}{\sqrt{3}} \bigg)  \\    = \frac{1}{\sqrt{3} } \bigg(   {\ket{\text{Player } 1 \text{ state}}} - {\ket{\text{Player } 2 \text{ state}}}   - {\ket{\text{Player } 3 \text{ state}}}                                      \bigg)   \text{, } \\ \\            \big(  \textbf{I}  \otimes  \sigma_z    \otimes   \textbf{I}  \big)     \bigg( \frac{\ket{\text{Player } 1 \text{ state}} +\ket{\text{Player } 2 \text{ state}} + \ket{\text{Player } 3 \text{ state}}}{\sqrt{3}} \bigg) \\   = \frac{1}{\sqrt{3} } \bigg(    -  {\ket{\text{Player } 1 \text{ state}}} + {\ket{\text{Player } 2 \text{ state}}}   - {\ket{\text{Player } 3 \text{ state}}}   \bigg)     \text{, } \\ \\              \big( \textbf{I} \otimes \textbf{I} \otimes \sigma_z \big)           \bigg( \frac{\ket{\text{Player } 1 \text{ state}} +\ket{\text{Player } 2 \text{ state}} + \ket{\text{Player } 3 \text{ state}}}{\sqrt{3}} \bigg)  \\ = \frac{1}{\sqrt{3} } \bigg(   -  {\ket{\text{Player } 1 \text{ state}}} - {\ket{\text{Player } 2 \text{ state}}}   + {\ket{\text{Player } 3 \text{ state}}}    \bigg)     \text{, } \\ \\     \big( \sigma_x \otimes \textbf{I} \otimes \sigma_z \big) \bigg( \frac{\ket{\text{Player } 1 \text{ state}} +\ket{\text{Player } 2 \text{ state}} + \ket{\text{Player } 3 \text{ state}}}{\sqrt{3}} \bigg) \\    = \frac{1}{\sqrt{3} } \bigg(    \widetilde{\ket{\text{Player } 1 \text{ state}}} + {\ket{\text{Player } 2 \text{ state}}}   - \widetilde{\ket{\text{Player } 3 \text{ state}}}        \bigg)   \text{, } \\ \\           \big( \textbf{I} \otimes \sigma_x \otimes \sigma_z \big)           \bigg( \frac{\ket{\text{Player } 1 \text{ state}} +\ket{\text{Player } 2 \text{ state}} + \ket{\text{Player } 3 \text{ state}}}{\sqrt{3}} \bigg)\\  = \frac{1}{\sqrt{3} } \bigg(      {\ket{\text{Player } 1 \text{ state}}} + \widetilde{\ket{\text{Player } 2 \text{ state}}}   - \widetilde{\ket{\text{Player } 3 \text{ state}}}        \bigg)      \text{, } \\ \\   \big( \sigma_z  \otimes \textbf{I} \otimes \sigma_x \big)           \bigg( \frac{\ket{\text{Player } 1 \text{ state}} +\ket{\text{Player } 2 \text{ state}} + \ket{\text{Player } 3 \text{ state}}}{\sqrt{3}} \bigg) \\ = \frac{1}{\sqrt{3} } \bigg(   -  \widetilde{\ket{\text{Player } 1 \text{ state}}} + {\ket{\text{Player } 2 \text{ state}}}    +\widetilde{\ket{\text{Player } 3 \text{ state}}}         \bigg)   \text{, } \\ \\  \big( \textbf{I}  \otimes \sigma_z \otimes \sigma_x \big)           \bigg( \frac{\ket{\text{Player } 1 \text{ state}} +\ket{\text{Player } 2 \text{ state}} + \ket{\text{Player } 3 \text{ state}}}{\sqrt{3}} \bigg) \\ = \frac{1}{\sqrt{3} } \bigg(        \widetilde{\ket{\text{Player } 1 \text{ state}}} - \widetilde{\ket{\text{Player } 2 \text{ state}}}   +{\ket{\text{Player } 3 \text{ state}}}      \bigg) \text{. }
\end{align*}

\subsection{$4$-$\mathrm{XOR}$ game}
\noindent The Bell states are generated by the operations,

\begin{align*}
        \big(  \textbf{I}  \otimes \textbf{I} \otimes \textbf{I}  \otimes \textbf{I}  \big)   \bigg( \frac{\ket{\text{{Player 1 state}}} + \ket{\text{Player 2 State}} + \ket{\text{Player 3 State}} + \ket{\text{Player 4 State}}}{2}  \bigg)  \\ =    \frac{\ket{\text{{Player 1 state}}} + \ket{\text{Player 2 State}} + \ket{\text{Player 3 State}} + \ket{\text{Player 4 State}}}{2} \text{, }  \\ \\         \big( \sigma_x \otimes \textbf{I} \otimes \textbf{I} \otimes \textbf{I} \big) \bigg(    \frac{\ket{\text{{Player 1 state}}} + \ket{\text{Player 2 State}} + \ket{\text{Player 3 State}} + \ket{\text{Player 4 State}}}{2}       \bigg)  \\ =  \frac{\widetilde{\ket{\text{{Player 1 state}}}} + \ket{\text{Player 2 State}} + \ket{\text{Player 3 State}} + \widetilde{\ket{\text{Player 4 State}}}}{2}        \text{, } \\ \\   \big( \textbf{I} \otimes \sigma_x \otimes \textbf{I} \otimes \textbf{I} \big)  \bigg(    \frac{\ket{\text{{Player 1 state}}} + \ket{\text{Player 2 State}} + \ket{\text{Player 3 State}} + \ket{\text{Player 4 State}}}{2}       \bigg) \\  =      \frac{\widetilde{\ket{\text{{Player 1 state}}}} + \widetilde{\ket{\text{Player 2 State}}} + \ket{\text{Player 3 State}} + \ket{\text{Player 4 State}}}{2}         \text{, }  \\ \\    \big( \textbf{I} \otimes \textbf{I} \otimes \sigma_x \otimes \textbf{I} \big) \bigg( \frac{\ket{\text{{Player 1 state}}} + \ket{\text{Player 2 State}} + \ket{\text{Player 3 State}} + \ket{\text{Player 4 State}}}{2}    \bigg) \\ =    \frac{{\ket{\text{{Player 1 state}}}} + \widetilde{\ket{\text{Player 2 State}}} + \widetilde{\ket{\text{Player 3 State}}} + \ket{\text{Player 4 State}}}{2}     \text{, }
\\ 
   \big( \textbf{I} \otimes \textbf{I}  \otimes \textbf{I}  \otimes \sigma_x  \big) \bigg(  \frac{\ket{\text{{Player 1 state}}} + \ket{\text{Player 2 State}} + \ket{\text{Player 3 State}} + \ket{\text{Player 4 State}}}{2}    \bigg) \\   =  \frac{\ket{\text{{Player 1 state}}} + \ket{\text{Player 2 State}} + \widetilde{\ket{\text{Player 3 State}}} + \widetilde{\ket{\text{Player 4 State}}}}{2}     \text{, }  \\ \\   \big( \sigma_z \otimes \textbf{I} \otimes \textbf{I} \otimes \textbf{I} \big) \bigg( \frac{\ket{\text{{Player 1 state}}} + \ket{\text{Player 2 State}} + \ket{\text{Player 3 State}} + \ket{\text{Player 4 State}}}{2} \bigg)    \\  =  \frac{\underset{1 \leq j \leq 3}{\sum} \ket{\text{Player } j \text{ state}} - \ket{\text{Player 4 state} }}{2}  \text{, } \\ \\ \big(   \textbf{I}   \otimes  \sigma_z \otimes \textbf{I}  \otimes  \textbf{I}   \big) \bigg( \frac{\ket{\text{{Player 1 state}}} + \ket{\text{Player 2 State}} + \ket{\text{Player 3 State}} + \ket{\text{Player 4 State}}}{2} \bigg)   \\   =   \frac{-\ket{\text{Player 1 state}} + \underset{2 \leq j \leq 4}{\sum} \ket{\text{Player } j \text{ state}}}{2}  \text{, }  \\ \\ \big(   \textbf{I}   \otimes  \textbf{I} \otimes \sigma_z   \otimes  \textbf{I}   \big) \bigg( \frac{\ket{\text{{Player 1 state}}} + \ket{\text{Player 2 State}} + \ket{\text{Player 3 State}} + \ket{\text{Player 4 State}}}{2} \bigg) \\      =    \frac{\ket{\text{Player 1 state}} - \ket{\text{Player 2 state}} + \ket{\text{Player 3 state}} + \ket{\text{Player 4 state}}}{2}  \text{, }  \end{align*}

      \begin{align*}   \big(\textbf{I} \otimes \textbf{I}  \otimes \textbf{I}  \otimes  \sigma_z  \big) \bigg( \frac{\ket{\text{{Player 1 state}}} + \ket{\text{Player 2 State}} + \ket{\text{Player 3 State}} + \ket{\text{Player 4 State}}}{2} \bigg)  \\ =   \frac{- \ket{\text{Player 1 state}} - \ket{\text{Player 2 state}} - \ket{\text{Player 3 state}} + \ket{\text{Player 4 state}}}{2}      \text{, } \\ \\  \big( \sigma_x \otimes \textbf{I} \otimes \textbf{I} \otimes \sigma_z \big) \bigg( \frac{\ket{\text{{Player 1 state}}} + \ket{\text{Player 2 State}} + \ket{\text{Player 3 State}} + \ket{\text{Player 4 State}}}{2} \bigg) \\  = \frac{\widetilde{\ket{\text{Player 1 state}}} + \ket{\text{Player 2 state}} + \ket{\text{Player 3 state}} - \widetilde{\ket{\text{Player 4 state}}}}{2} \text{, } \\ \\   \big(    \textbf{I} \otimes \sigma_x \otimes \textbf{I} \otimes \sigma_z    \big) \bigg( \frac{\ket{\text{{Player 1 state}}} + \ket{\text{Player 2 State}} + \ket{\text{Player 3 State}} + \ket{\text{Player 4 State}}}{2} \bigg) \\ =  \frac{\ket{\text{Player 1 state}} + \widetilde{\ket{\text{Player 2 state}}} + \ket{\text{Player 3 state}} - \widetilde{\ket{\text{Player 4 state}}}}{2} \text{, } \\ \\   \big(    \textbf{I} \otimes \textbf{I}   \otimes \sigma_x \otimes \sigma_z    \big) \bigg( \frac{\ket{\text{{Player 1 state}}} + \ket{\text{Player 2 State}} + \ket{\text{Player 3 State}} + \ket{\text{Player 4 State}}}{2} \bigg) \\   =    \frac{\ket{\text{Player 1 state}} + \ket{\text{Player 2 state}} + \widetilde{\ket{\text{Player 3 state}}} - \widetilde{\ket{\text{Player 4 state}}}}{2}    \text{, }  \\ \\   \big( \textbf{I} \otimes  \textbf{I} \otimes \textbf{I} \otimes \sigma_x \sigma_z  \big) \bigg( \frac{\ket{\text{{Player 1 state}}} + \ket{\text{Player 2 State}} + \ket{\text{Player 3 State}} + \ket{\text{Player 4 State}}}{2} \bigg)  \\ =     \frac{\widetilde{\ket{\text{Player 1 state}}} - \ket{\text{Player 2 state}} - \ket{\text{Player 3 state}} - \widetilde{\ket{\text{Player 4 state}}}}{2}        \text{. }  \end{align*}

\subsection{Permutations of the $E$ tensor}

\noindent Table 16, and the remaining tables in the paper, beginning on the next page provide a list of permutations for the $E$ tensor observable that are used to construct permutations appearing in the optimal value, and bias, of the $5$-$\mathrm{XOR}$ game.

\begin{table}[ht]
\caption{Permutations of Player Observables for the 5-XOR game error bound, $(\textbf{EB}- 5 \mathrm{XOR})$ Continued} 
\centering 
\begin{tabular}{c c c c} 
\hline\hline 
 Player & Tensor Product Representation & Permutation Superposition  \\ [0.5ex] 
\hline \\ $5$ &  $  \underset{\sigma} {\mathlarger{\sum}} \bigg( E_{\sigma(ijklm)} \otimes C_{ijk} \otimes A_{i } \otimes B_{ij}   \otimes D_{ijkl} \bigg)  $  &  $E_{ijklm} \otimes C_{ijk} \otimes A_{i } \otimes B_{ij}   \otimes D_{ijkl}  +          E_{ikjml} \otimes C_{ijk} \otimes A_{i }       $    \\  & & $ \otimes B_{ij}   \otimes D_{ijkl}   +         E_{ijlkm} \otimes C_{ijk} \otimes A_{i }     \otimes B_{ij}   \otimes D_{ijkl}          $ \\ & & $ +   E_{ijklmk} \otimes C_{ijk} \otimes A_{i }     \otimes B_{ij}   \otimes D_{ijkl}  +   E_{ijmkl} \otimes C_{ijk} \otimes A_{i }                               $ \\   & & $ \otimes B_{ij}   \otimes D_{ijkl}     +     E_{ikjlm} \otimes C_{ijk} \otimes A_i \otimes B_{ij} \otimes D_{ijkl}        $  \\   &   & $ +   E_{ikjml} \otimes C_{ijk} \otimes A_i \otimes B_{ij} \otimes D_{ijkl}        $ \\   & & $ +   E_{ik ljm} \otimes C_{ikljm} \otimes A_i \otimes B_{ij} \otimes D_{ijkl}       $  \\ & & $  + E_{iklmj} \otimes C_{ijk} \otimes A_i \otimes B_{ij} \otimes D_{ijkl}      $   \\   & & $ +   E_{ikmjl} \otimes C_{ijk} \otimes A_i \otimes B_{ij} \otimes D_{ijkl}   +            E_{ikmlj}       $ \\ & & $ \otimes C_{ijk}  \otimes A_i \otimes B_{ij}   \otimes D_{ijkl}$   \\ & & $   +    E_{iljkm} \otimes C_{ijk}  \otimes A_i \otimes B_{ij}   \otimes D_{ijkl}            $ \\  & & $ + E_{iljmk} \otimes C_{ijk}  $\\   & $\vdots$ & \\  [1ex] 
\hline 
\end{tabular}
\label{table:nonlin} 
\end{table}

\begin{table}[ht]
\caption{Permutations of Player Observables for the 5-XOR game error bound, $(\textbf{EB}- 5 \mathrm{XOR})$ Continued} 
\centering 
\begin{tabular}{c c c c} 
\hline\hline 
 Player & Tensor Product Representation & Permutation Superposition  \\ [0.5ex] 
\hline  & $\vdots$ & \\  $5$ &  $  \underset{\sigma} {\mathlarger{\sum}} \bigg( E_{\sigma(ijklm)} \otimes C_{ijk} \otimes A_{i } \otimes B_{ij}   \otimes D_{ijkl} \bigg)  $   & $      \otimes A_i \otimes B_{ij} \otimes D_{ijkl} +    E_{ilkjm} \otimes C_{ijk}  \otimes A_i \otimes B_{ij}  \otimes D_{ijkl} +   $ \\ & & $     E_{kijml} \otimes C_{ijk} \otimes A_i  \otimes B_{ij} \otimes D_{ijkl} + E_{ilkmj} \otimes C_{ijk}   $ \\ & & $  \otimes A_i \otimes B_{ij} \otimes  D_{ijkl}  + E_{kjmil} \otimes C_{ijk} \otimes A_i  \otimes B_{ij} \otimes D_{ijkl}   $  \\    & & $  + E_{ilmjk} \otimes C_{ijk}   \otimes A_i    \otimes B_{ij} \otimes   D_{ijkl}   + E_{ilmkj} \otimes C_{ijk} \otimes     $ \\ & & $       A_i  \otimes B_{ij} \otimes D_{ijkl}   +  E_{imjkl} \otimes C_{ijk} \otimes A_i \otimes B_{ij} \otimes D_{ijkl}    $    \\  & &  $ + E_{imjlk} \otimes C_{ijk} \otimes A_i \otimes B_{ij} \otimes D_{ijkl} + E_{imkjl} \otimes C_{ijk}     $  \\  & & $   \otimes A_i \otimes B_{ij} \otimes D_{ijkl} +    E_{imklj} \otimes C_{ijk} \otimes  A_i \otimes B_{ij} \otimes   $  \\ & & $  D_{ijkl}   +  E_{imljk} \otimes  C_{ijk} \otimes A_i \otimes B_{ij} \otimes D_{ijkl}       $  \\ & & $ +  E_{imlkj} \otimes C_{ijk} \otimes A_i \otimes B_{ij} \otimes D_{ijkl} +  E_{jiklm }  $ \\ & & $   \otimes C_{ijk} \otimes A_i \otimes B_{ij} \otimes D_{ijkl} +   E_{jikml} \otimes   C_{ijk} \otimes   $ \\    & & $A_i \otimes B_{ij} \otimes D_{ijkl}   +  E_{jilkm } \otimes C_{ijk} \otimes  $  \\ & & $  A_i \otimes B_{ij} \otimes D_{ijkl} +  E_{jimkl} \otimes C_{ijk} \otimes  $  \\ & & $  A_i \otimes B_{ij} \otimes D_{ijkl} +  E_{jimlk } \otimes C_{ijk} \otimes  $ \\ & & $  A_i \otimes B_{ij} \otimes D_{ijkl} +  E_{jkilm} \otimes C_{ijk}   $ \\ & & $ \otimes A_i \otimes B_{ij} \otimes D_{ijkl} +                   E_{jkiml} \otimes C_{ijk} \otimes  $ \\  & & $  A_i \otimes B_{ij } \otimes D_{ijkl}  +        E_{jklim} \otimes C_{ijk} \otimes A_i   $ \\  & & $ \otimes B_{ij } \otimes D_{ijkl}     +    E_{jklmi} \otimes C_{ijk} \otimes      $\\ & & $  A_i \otimes B_{ij } \otimes D_{ijkl}  +    E_{jkmil} \otimes C_{ijk} \otimes A_i   $  \\ & & $\otimes B_{ij }  \otimes D_{ijkl} +  E_{jkmli} \otimes C_{ijk} \otimes  $ \\   & & $  A_i \otimes B_{ij } \otimes D_{ijkl}   +  E_{jlikm} \otimes C_{ijk}     $\\ & & $  \otimes A_i \otimes B_{ij } \otimes D_{ijkl}      +  E_{jlimk} \otimes C_{ijk}     $ \\ & & $\otimes A_i \otimes B_{ij } \otimes D_{ijkl}     +  E_{jlkim} \otimes C_{ijk}    $  \\ & & $  \otimes A_i \otimes B_{ij } \otimes D_{ijkl}   +  E_{jlkmi} \otimes        $ \\ & & $ C_{ijk} \otimes A_i \otimes B_{ij } \otimes D_{ijkl}   + E_{jlmik} \otimes  C_{ijk} \otimes      $ \\ & & $ A_i \otimes B_{ij} \otimes D_{ijkl} +   E_{jlmki} \otimes  C_{ijk} $ \\ & & $  \otimes A_i \otimes B_{ij} \otimes D_{ijkl}  +   E_{jmikl} \otimes  C_{ijk} \otimes  $  \\ & & $  A_i \otimes B_{ij} \otimes D_{ijkl}     +  E_{jmilk} \otimes  C_{ijk} \otimes  $  \\ & & $  A_i \otimes B_{ij} \otimes D_{ijkl}  +   E_{jmkil} \otimes  C_{ijk} \otimes   $  \\ & & $ A_i \otimes B_{ij} \otimes D_{ijkl}  +  E_{jmkli} \otimes  C_{ijk} \otimes  $  \\ & & $ A_i \otimes B_{ij} \otimes D_{ijkl}  +   E_{jmlik} \otimes  C_{ijk} \otimes $  \\ & & $  A_i \otimes B_{ij} \otimes D_{ijkl} +   E_{jmlki} \otimes  C_{ijk} \otimes  $  \\ & & $  A_i \otimes B_{ij} \otimes D_{ijkl}  +    E_{kijlm} \otimes  C_{ijk} \otimes $      \\ & $\vdots$ & \\  [1ex] 
\hline 
\end{tabular}
\label{table:nonlin} 
\end{table}

\begin{table}[ht]
\caption{Permutations of Player Observables for the 5-XOR game error bound, $(\textbf{EB}- 5 \mathrm{XOR})$ Continued} 
\centering 
\begin{tabular}{c c c c} 
\hline\hline 
Player & Tensor Product Representation & Permutation Superposition  \\ [0.5ex] 
\hline & $\vdots$ &  \\  $5$ &  $  \underset{\sigma} {\mathlarger{\sum}} \bigg( E_{\sigma(ijklm)} \otimes C_{ijk} \otimes A_{i } \otimes B_{ij}   \otimes D_{ijkl} \bigg)  $   & $ +      E_{kilmj} \otimes  C_{ijk} \otimes A_i \otimes B_{ij} \otimes D_{ijkl}    +      E_{kimlj} \otimes  C_{ijk} \otimes A_i  $   \\ & & $  \otimes B_{ij} \otimes D_{ijkl}   +      E_{kjilm} \otimes  C_{ijk} \otimes A_i   \otimes B_{ij} \otimes $    \\ & & $  D_{ijkl}  +      E_{kjiml} \otimes  C_{ijk} \otimes A_i   \otimes B_{ij} \otimes   $    \\ & & $  D_{ijkl}   +      E_{kjlim} \otimes  C_{ijk} \otimes A_i \otimes B_{ij}    $    \\ & & $   \otimes D_{ijkl}    +      E_{kjlmi} \otimes  C_{ijk} \otimes A_i \otimes B_{ij} \otimes     $    \\ & & $  D_{ijkl}    +      E_{kjmil} \otimes  C_{ijk} \otimes A_i  \otimes B_{ij} \otimes D_{ijkl}     $    \\ & & $ +      E_{kjmli} \otimes  C_{ijk} \otimes A_i    $    \\ & & $  \otimes B_{ij} \otimes D_{ijkl}   +      E_{klijm} \otimes  C_{ijk} \otimes A_i   \otimes B_{ij} \otimes $    \\ & & $  D_{ijkl}   +      E_{klimj} \otimes  C_{ijk} \otimes A_i \otimes B_{ij} \otimes D_{ijkl}    $     \\ & & $   +      E_{klijm} \otimes  C_{ijk} \otimes A_i  \otimes B_{ij} \otimes D_{ijkl}   $     \\ & & $   +      E_{klmij} \otimes  C_{ijk} \otimes A_i  \otimes B_{ij} \otimes D_{ijkl}   $     \\ & & $   +      E_{klmji} \otimes  C_{ijk} \otimes A_i    $     \\ & & $ \otimes B_{ij} \otimes D_{ijkl}  +      E_{kmijl} \otimes  C_{ijk} \otimes A_i  \otimes B_{ij} $     \\ & & $  \otimes D_{ijkl}       +      E_{kmilj} \otimes  C_{ijk} \otimes A_i \otimes B_{ij} \otimes   $ \\ & & $  D_{ijkl}   +      E_{kmjli} \otimes  C_{ijk} \otimes A_i  \otimes B_{ij} \otimes   $     \\ & & $  D_{ijkl}     +      E_{kmlij} \otimes  C_{ijk} \otimes A_i   $  \\ & & $\otimes B_{ij} \otimes D_{ijkl}   +      E_{kmlji} \otimes  C_{ijk}   $     \\ & & $ \otimes A_i  \otimes B_{ij} \otimes D_{ijkl}   +      E_{lijkm} \otimes  C_{ijk}   $  \\ & & $  \otimes A_i   \otimes B_{ij} \otimes D_{ijkl}    +      E_{lijmk} \otimes  C_{ijk}   $     \\ & & $ \otimes A_i  \otimes B_{ij} \otimes D_{ijkl}    +      E_{likmj} \otimes  C_{ijk} \otimes A_i   $  \\ & & $ \otimes B_{ij} \otimes D_{ijkl}   +      E_{limjik} \otimes  C_{ijk} \otimes A_i    $     \\ & & $ \otimes B_{ij} \otimes D_{ijkl}   +      E_{limkj} \otimes  C_{ijk} \otimes A_i   $  \\ & & $  \otimes B_{ij} \otimes D_{ijkl}   +      E_{ljikm} \otimes  C_{ijk} \otimes A_i    $     \\ & & $  \otimes B_{ij} \otimes D_{ijkl}      $  \\ & $\vdots$ &  \\   [1ex] 
\hline 
\end{tabular}
\label{table:nonlin} 
\end{table}

\begin{table}[ht]
\caption{Permutations of Player Observables for the 5-XOR game error bound, $(\textbf{EB}- 5 \mathrm{XOR})$ Continued} 
\centering 
\begin{tabular}{c c c c} 
\hline\hline 
Player & Tensor Product Representation & Permutation Superposition  \\ [0.5ex] 
\hline & $\vdots$ &  \\   $5$ &  $  \underset{\sigma} {\mathlarger{\sum}} \bigg( E_{\sigma(ijklm)} \otimes C_{ijk} \otimes A_{i } \otimes B_{ij}   \otimes D_{ijkl} \bigg)  $  & $ +      E_{ljimk} \otimes  C_{ijk} \otimes A_i \otimes B_{ij} \otimes D_{ijkl}   +      E_{ljkim} \otimes  C_{ijk} \otimes A_i \otimes     $     \\ & & $ B_{ij} \otimes D_{ijkl}   +      E_{ljkmi} \otimes  C_{ijk} \otimes A_i \otimes     $  \\ & & $ B_{ij} \otimes D_{ijkl}    +      E_{ljmik} \otimes  C_{ijk} \otimes A_i    $     \\ & & $ \otimes B_{ij} \otimes D_{ijkl}   +      E_{ljmki} \otimes  C_{ijk} \otimes A_i    $     \\  & & $  \otimes B_{ij} \otimes D_{ijkl}   +      E_{lkijm} \otimes  C_{ijk} \otimes A_i    $     \\ & & $  \otimes B_{ij} \otimes D_{ijkl}  +      E_{lkimj} \otimes  C_{ijk} \otimes A_i       $     \\ & & $ \otimes B_{ij} \otimes D_{ijkl}  +      E_{lkjim} \otimes  C_{ijk}     $     \\ & & $  \otimes A_i \otimes B_{ij} \otimes D_{ijkl}  +      E_{lkjmi} \otimes  C_{ijk}  $      \\ & & $ \otimes A_i      \otimes B_{ij} \otimes D_{ijkl}   +      E_{lkmij} \otimes  C_{ijk}    $      \\ & & $  \otimes A_i \otimes B_{ij} \otimes D_{ijkl}   +      E_{lkmji} \otimes  C_{ijk} \otimes A_i    $      \\ & & $  \otimes B_{ij} \otimes D_{ijkl}    +      E_{lmijk} \otimes  C_{ijk} \otimes A_i    $      \\ & & $ \otimes B_{ij} \otimes D_{ijkl}  +      E_{lmikj} \otimes  C_{ijk} \otimes A_i      $      \\ & & $ \otimes B_{ij} \otimes D_{ijkl}  +      E_{lmjik} \otimes  C_{ijk} \otimes     $      \\ & & $  A_i \otimes B_{ij} \otimes D_{ijkl}  +      E_{lmjki} \otimes  C_{ijk} \otimes A_i     $      \\ & & $  \otimes B_{ij} \otimes D_{ijkl}   +      E_{lmkij} \otimes  C_{ijk} \otimes    $   \\ & & $  A_i \otimes B_{ij} \otimes D_{ijkl}    +      E_{lmkji} \otimes  C_{ijk} \otimes A_i    $   \\ & & $ \otimes B_{ij} \otimes D_{ijkl}   +      E_{mijkl} \otimes  C_{ijk} \otimes A_i     $   \\ & & $ \otimes B_{ij} \otimes D_{ijkl} +      E_{mijlk} \otimes  C_{ijk} \otimes      $     \\ & & $  A_i \otimes B_{ij} \otimes D_{ijkl}   +      E_{mikjl} \otimes  C_{ijk} \otimes A_i     $     \\ & & $ \otimes B_{ij} \otimes D_{ijkl}    +      E_{miklj} \otimes  C_{ijk} \otimes A_i    $     \\ & & $\otimes B_{ij} \otimes D_{ijkl}  +      E_{miljk} \otimes  C_{ijk} \otimes A_i     $     \\ & & $  \otimes B_{ij} \otimes D_{ijkl}   +      E_{milkj} \otimes  C_{ijk} \otimes A_i    $     \\ & & $  \otimes B_{ij} \otimes D_{ijkl}   +      E_{mjikl} \otimes  C_{ijk} \otimes A_i      $     \\ & & $\otimes B_{ij} \otimes D_{ijkl}   +      E_{mjilk} \otimes  C_{ijk} \otimes A_i    $    \\ & & $  \otimes B_{ij} \otimes D_{ijkl}  +      E_{mjkil} \otimes  C_{ijk} \otimes A_i     $    \\ & & $  \otimes B_{ij} \otimes D_{ijkl}  +      E_{mjkli} \otimes  C_{ijk} \otimes A_i      $    \\ & & $ \otimes B_{ij} \otimes D_{ijkl}   +      E_{mjlik} \otimes  C_{ijk} \otimes A_i   $     \\ & & $  \otimes B_{ij} \otimes D_{ijkl}   +      E_{mjlki} \otimes  C_{ijk} \otimes A_i     $     \\ & & $ \otimes B_{ij} \otimes D_{ijkl}   +      E_{mkijl} \otimes  C_{ijk} \otimes A_i     $     \\ & & $ \otimes B_{ij} \otimes D_{ijkl}   +      E_{mkilj} \otimes  C_{ijk} \otimes A_i    $   \\ & & $  \otimes B_{ij} \otimes D_{ijkl}      $ \\   & $\vdots$ &    \\  [1ex] 
\hline 
\end{tabular}
\label{table:nonlin} 
\end{table}

\begin{table}[ht]
\caption{Permutations of Player Observables for the 5-XOR game error bound, $(\textbf{EB}- 5 \mathrm{XOR})$ Continued} 
\centering 
\begin{tabular}{c c c c} 
\hline\hline 
Player & Tensor Product Representation & Permutation Superposition  \\ [0.5ex] 
\hline  & $\vdots$ &  \\   $5$ &  $  \underset{\sigma} {\mathlarger{\sum}} \bigg( E_{\sigma(ijklm)} \otimes C_{ijk} \otimes A_{i } \otimes B_{ij}   \otimes D_{ijkl} \bigg)  $  &  $ +      E_{mkjil} \otimes  C_{ijk} \otimes A_i \otimes B_{ij} \otimes D_{ijkl}    +      E_{mkjli} \otimes  C_{ijk} \otimes   $   \\ & & $  A_i \otimes B_{ij} \otimes D_{ijkl}    + E_{mklik}  \otimes  C_{ijk} \otimes A_i \otimes    $   \\ & & $ B_{ij} \otimes D_{ijkl}   + E_{mklji}  \otimes  C_{ijk} \otimes A_i \otimes  $  \\ & & $B_{ij} \otimes D_{ijkl}  + E_{mlijk}    \otimes  C_{ijk} \otimes A_i \otimes $ \\  & & $ B_{ij} \otimes D_{ijkl}   +     E_{mlikj}  \otimes  C_{ijk} \otimes A_i     $ \\ & & $ \otimes B_{ij} \otimes D_{ijkl} + E_{mljik} \otimes  C_{ijk} \otimes A_i  $ \\ & & $  \otimes B_{ij} \otimes D_{ijkl} + E_{mljki} \otimes  C_{ijk} \otimes A_i  $ \\ & & $ \otimes B_{ij} \otimes D_{ijkl} + E_{mlkij} \otimes  C_{ijk} \otimes A_i  $ \\ & & $  \otimes B_{ij} \otimes D_{ijkl} + E_{mlkji} \otimes  C_{ijk} \otimes A_i $ \\ & & $  \otimes B_{ij} \otimes D_{ijkl}  $   \\  [1ex] 
\hline 
\end{tabular}
\label{table:nonlin} 
\end{table}

   \subsection{Strong parallel repetition of the $3$-$\mathrm{XOR}$ game}

\noindent For each player, the Hilbert space spanned by the collection of player observables,

\begin{align*}
   \underset{1\leq k \leq N}{\bigcup} \bigg[ \underset{\# \text{ of strong parallel repetitions } j}{\bigwedge}    \ket{\text{$Player^{(j)}$ $k$ state}}    \bigg]  \text{, }
\end{align*}

\noindent corresponding to the strong parallel repetition operation implies that the Bell states are generated by the actions,

   \begin{align*}
 \big( \textbf{I} \otimes \textbf{I} \otimes \textbf{I} \big) \\ \times  \bigg( \frac{1}{\sqrt{3}} \bigg( \big( \ket{\text{$Player^{(1)}$ 1 state}} \wedge \cdots \wedge  \ket{\text{$Player^{(n)}$ 1 state}}  \big) + \big( \ket{\text{$Player^{(1)}$ 2 state}} \wedge \cdots \\ \wedge  \ket{\text{$Player^{(n)}$2  state}}  \big)    + \big( \ket{\text{$Player^{(1)}$ 3 state}} \wedge \cdots \wedge  \ket{\text{$Player^{(n)}$3 state}}  \big)  \bigg) \bigg)    \\  =      \frac{1}{\sqrt{3}} \bigg( \big( \ket{\text{$Player^{(1)}$ 1 state}} \wedge \cdots \wedge  \ket{\text{$Player^{(n)}$ 1 state}}  \big)   + \big( \ket{\text{$Player^{(1)}$ 2 state}} \wedge \cdots \\ \wedge  \ket{\text{$Player^{(n)}$2  state}}  \big)  + \big( \ket{\text{$Player^{(1)}$ 3 state}} \wedge \cdots   \wedge  \ket{\text{$Player^{(n)}$3 state}}  \big)  \bigg)                 \text{, } \\ \\  \big( \sigma_x  \otimes \textbf{I} \otimes \textbf{I} \big) \\     \times  \bigg( \frac{1}{\sqrt{3}} \bigg( \big( \ket{\text{$Player^{(1)}$ 1 state}} \wedge \cdots \wedge  \ket{\text{$Player^{(n)}$ 1 state}}  \big) + \big( \ket{\text{$Player^{(1)}$ 2 state}} \wedge \cdots \\ \wedge  \ket{\text{$Player^{(n)}$2  state}}  \big)  + \big( \ket{\text{$Player^{(1)}$ 3 state}} \wedge \cdots \wedge  \ket{\text{$Player^{(n)}$3 state}}  \big)  \bigg) \bigg)  \\  =              \frac{1}{\sqrt{3}} \bigg( \big( \widetilde{\ket{\text{$Player^{(1)}$ 1 state}}} \wedge \cdots \wedge  \widetilde{\ket{\text{$Player^{(n)}$ 1 state}}}  \big)      + \big( \ket{\text{$Player^{(1)}$ 2 state}} \wedge \cdots \\ \wedge  \ket{\text{$Player^{(n)}$2  state}}  \big)  + \big( \widetilde{\ket{\text{$Player^{(1)}$ 3 state}}} \wedge \cdots   \wedge  \widetilde{\ket{\text{$Player^{(n)}$3 state}}}  \big)  \bigg)     \text{, } \\ \\  \end{align*}

     \begin{align*}    \big( \textbf{I} \otimes \sigma_x \otimes \textbf{I} \big) \\  \times  \bigg( \frac{1}{\sqrt{3}} \bigg( \big( \ket{\text{$Player^{(1)}$ 1 state}} \wedge \cdots   \wedge  \ket{\text{$Player^{(n)}$ 1 state}}  \big)   + \big( \ket{\text{$Player^{(1)}$ 2 state}} \wedge \cdots  \\  \wedge  \ket{\text{$Player^{(n)}$2  state}}  \big)  + \big( \ket{\text{$Player^{(1)}$ 3 state}} \wedge \cdots   \wedge  \ket{\text{$Player^{(n)}$3 state}}  \big)  \bigg) \bigg)  \\    =           \frac{1}{\sqrt{3}} \bigg( \big( \widetilde{\ket{\text{$Player^{(1)}$ 1 state}}} \wedge \cdots \wedge  \widetilde{\ket{\text{$Player^{(n)}$ 1 state}}} \big)   + \big( \widetilde{\ket{\text{$Player^{(1)}$ 2 state}}} \wedge \cdots \\ \wedge  \widetilde{\ket{\text{$Player^{(n)}$2  state}}}  \big)  + \big( {\ket{\text{$Player^{(1)}$ 3 state}}} \wedge \cdots \wedge  {\ket{\text{$Player^{(n)}$3 state}}}  \big)  \bigg)                     \text{, } \\ \\   \big( \textbf{I} \otimes \textbf{I}  \otimes \sigma_x \big) \\   \times  \bigg( \frac{1}{\sqrt{3}} \bigg( \big( \ket{\text{$Player^{(1)}$ 1 state}} \wedge \cdots \wedge  \ket{\text{$Player^{(n)}$ 1 state}}  \big) + \big( \ket{\text{$Player^{(1)}$ 2 state}} \wedge \cdots  \\ \wedge  \ket{\text{$Player^{(n)}$2  state}}  \big)    + \big( \ket{\text{$Player^{(1)}$ 3 state}} \wedge \cdots \wedge  \ket{\text{$Player^{(n)}$3 state}}  \big)  \bigg) \bigg) \\  =    \frac{1}{\sqrt{3}} \bigg( \big( \widetilde{\ket{\text{$Player^{(1)}$ 1 state}}} \wedge \cdots \wedge  \widetilde{\ket{\text{$Player^{(n)}$ 1 state}}}  \big) + \big( \widetilde{\ket{\text{$Player^{(1)}$ 2 state}}} \wedge \cdots \\  \wedge  \widetilde{\ket{\text{$Player^{(n)}$2  state}} } \big)   + \big( \ket{\text{$Player^{(1)}$ 3 state}} \wedge \cdots \wedge  \ket{\text{$Player^{(n)}$3 state}}  \big)  \bigg)  \\ \\       \big( \sigma_z \otimes \textbf{I}  \otimes \textbf{I} \big) \\  \times  \bigg( \frac{1}{\sqrt{3}} \bigg( \big( \ket{\text{$Player^{(1)}$ 1 state}} \wedge \cdots \wedge  \ket{\text{$Player^{(n)}$ 1 state}}  \big) + \big( \ket{\text{$Player^{(1)}$ 2 state}} \wedge \cdots \\ \wedge  \ket{\text{$Player^{(n)}$2  state}}  \big)   + \big( \ket{\text{$Player^{(1)}$ 3 state}} \wedge \cdots \wedge  \ket{\text{$Player^{(n)}$3 state}}  \big)  \bigg) \bigg) \\  =    \frac{1}{\sqrt{3}} \bigg( \big( \ket{\text{$Player^{(1)}$ 1 state}} \wedge \cdots \wedge  \ket{\text{$Player^{(n)}$ 1 state}}  \big) - \big( \ket{\text{$Player^{(1)}$ 2 state}} \wedge \cdots \\ \wedge  \ket{\text{$Player^{(n)}$2  state}}  \big)    - \big( \ket{\text{$Player^{(1)}$ 3 state}} \wedge \cdots \wedge  \ket{\text{$Player^{(n)}$3 state}}  \big)  \bigg)        \text{, } \\ \\ 
    \big( \textbf{I} \otimes \sigma_z   \otimes \textbf{I} \big)  \\  \times  \bigg( \frac{1}{\sqrt{3}} \bigg( \big( \ket{\text{$Player^{(1)}$ 1 state}} \wedge \cdots \wedge  \ket{\text{$Player^{(n)}$ 1 state}}  \big) + \big( \ket{\text{$Player^{(1)}$ 2 state}} \wedge \cdots \\ \wedge  \ket{\text{$Player^{(n)}$2  state}}  \big)     + \big( \ket{\text{$Player^{(1)}$ 3 state}} \wedge \cdots \wedge  \ket{\text{$Player^{(n)}$3 state}}  \big)  \bigg) \bigg) \\  =                        \frac{1}{\sqrt{3}} \bigg( - \big( \ket{\text{$Player^{(1)}$ 1 state}} \wedge \cdots \wedge  \ket{\text{$Player^{(n)}$ 1 state}}  \big)     + \big( \ket{\text{$Player^{(1)}$ 2 state}} \wedge \cdots \\ \wedge  \ket{\text{$Player^{(n)}$2  state}}  \big)   - \big( \ket{\text{$Player^{(1)}$ 3 state}} \wedge \cdots \wedge  \ket{\text{$Player^{(n)}$3 state}}  \big)  \bigg)                                                                     \text{, }  \\ \\           \big( \textbf{I} \otimes \textbf{I}  \otimes \sigma_z  \big) \\ \times  \bigg( \frac{1}{\sqrt{3}} \bigg( \big( \ket{\text{$Player^{(1)}$ 1 state}} \wedge \cdots \wedge  \ket{\text{$Player^{(n)}$ 1 state}}  \big) + \big( \ket{\text{$Player^{(1)}$ 2 state}} \wedge \cdots \\  \wedge  \ket{\text{$Player^{(n)}$2  state}}  \big)   + \big( \ket{\text{$Player^{(1)}$ 3 state}} \wedge \cdots \wedge  \ket{\text{$Player^{(n)}$3 state}}  \big)  \bigg) \bigg)  \\  =  \frac{1}{\sqrt{3}} \bigg( - \big( \ket{\text{$Player^{(1)}$ 1 state}} \wedge \cdots \wedge  \ket{\text{$Player^{(n)}$ 1 state}}  \big)    - \big( \ket{\text{$Player^{(1)}$ 2 state}} \wedge \cdots \\ \wedge  \ket{\text{$Player^{(n)}$2  state}}  \big)   + \big( \ket{\text{$Player^{(1)}$ 3 state}} \wedge \cdots  \wedge  \ket{\text{$Player^{(n)}$3 state}}  \big)  \bigg)                   \text{, }    \\ \\        \big( \sigma_x  \otimes \textbf{I}  \otimes \sigma_z  \big)  \\  \times  \bigg( \frac{1}{\sqrt{3}} \bigg( \big( \ket{\text{$Player^{(1)}$ 1 state}} \wedge \cdots \wedge  \ket{\text{$Player^{(n)}$ 1 state}}  \big) + \big( \ket{\text{$Player^{(1)}$ 2 state}} \wedge \cdots \\ \wedge  \ket{\text{$Player^{(n)}$2  state}}  \big)     + \big( \ket{\text{$Player^{(1)}$ 3 state}} \wedge \cdots \wedge  \ket{\text{$Player^{(n)}$3 state}}  \big)  \bigg) \bigg) \\  =    \frac{1}{\sqrt{3}} \bigg(  \big( \widetilde{\ket{\text{$Player^{(1)}$ 1 state}}} \wedge \cdots \wedge  \widetilde{\ket{\text{$Player^{(n)}$ 1 state}}}  \big) + \big( \ket{\text{$Player^{(1)}$ 2 state}} \\ \wedge \cdots \wedge  \ket{\text{$Player^{(n)}$2  state}}  \big)   - \big( \widetilde{\ket{\text{$Player^{(1)}$ 3 state}}} \wedge \cdots \wedge  \widetilde{\ket{\text{$Player^{(n)}$3 state}}}  \big)  \bigg)                    \text{, }  \\ \\    \big(  \textbf{I} \otimes \sigma_x  \otimes \sigma_z  \big) \\ \times \bigg( \frac{1}{\sqrt{3}} \bigg( \big( \ket{\text{$Player^{(1)}$ 1 state}} \wedge \cdots \wedge  \ket{\text{$Player^{(n)}$ 1 state}}  \big) + \big( \ket{\text{$Player^{(1)}$ 2 state}} \wedge \cdots \\ \wedge  \ket{\text{$Player^{(n)}$2  state}}  \big)   + \big( \ket{\text{$Player^{(1)}$ 3 state}} \wedge \cdots \wedge  \ket{\text{$Player^{(n)}$3 state}}  \big)  \bigg) \bigg)  \\  =  \frac{1}{\sqrt{3}} \bigg(  \big( {\ket{\text{$Player^{(1)}$ 1 state}}} \wedge \cdots \wedge  {\ket{\text{$Player^{(n)}$ 1 state}}}  \big)    + \big( \widetilde{\ket{\text{$Player^{(1)}$ 2 state}}} \wedge \cdots \\ \wedge  \widetilde{\ket{\text{$Player^{(n)}$2  state}}}  \big)    - \big( \widetilde{\ket{\text{$Player^{(1)}$ 3 state}}} \wedge \cdots \wedge  \widetilde{\ket{\text{$Player^{(n)}$3 state}}}  \big)  \bigg) \text{, }  \\  \\ 
              \big(  \sigma_z \otimes \textbf{I} \otimes \sigma_x  \big) \\ \times \bigg( \frac{1}{\sqrt{3}} \bigg( \big( \ket{\text{$Player^{(1)}$ 1 state}} \wedge \cdots \wedge  \ket{\text{$Player^{(n)}$ 1 state}}  \big) + \big( \ket{\text{$Player^{(1)}$ 2 state}} \wedge \cdots \\ \wedge  \ket{\text{$Player^{(n)}$2  state}}  \big)   + \big( \ket{\text{$Player^{(1)}$ 3 state}} \wedge \cdots \wedge  \ket{\text{$Player^{(n)}$3 state}}  \big)  \bigg) \bigg)      \\  =  \frac{1}{\sqrt{3}} \bigg(   -   \big( \widetilde{\ket{\text{$Player^{(1)}$ 1 state}}} \wedge \cdots \wedge  \widetilde{\ket{\text{$Player^{(n)}$ 1 state}}}  \big)    + \big( \ket{\text{$Player^{(1)}$ 2 state}} \wedge \cdots \\  \wedge  \ket{\text{$Player^{(n)}$2  state}}  \big)  + \big( \widetilde{\ket{\text{$Player^{(1)}$ 3 state}}} \wedge \cdots \wedge  \widetilde{\ket{\text{$Player^{(n)}$3 state}} } \big) \bigg) \text{, }  \\ \\    \big(  \textbf{I} \otimes \sigma_z  \otimes \sigma_x  \big) \\ \times \bigg( \frac{1}{\sqrt{3}} \bigg( \big( \ket{\text{$Player^{(1)}$ 1 state}} \wedge \cdots \wedge  \ket{\text{$Player^{(n)}$ 1 state}}  \big) + \big( \ket{\text{$Player^{(1)}$ 2 state}} \wedge \cdots \\ \wedge  \ket{\text{$Player^{(n)}$2  state}}  \big)  + \big( \ket{\text{$Player^{(1)}$ 3 state}} \wedge \cdots \wedge  \ket{\text{$Player^{(n)}$3 state}}  \big)  \bigg) \bigg)     \\   =        \frac{1}{\sqrt{3}} \bigg( \big( \widetilde{\ket{\text{$Player^{(1)}$ 1 state}}} \wedge \cdots   \wedge  \widetilde{\ket{\text{$Player^{(n)}$ 1 state}}}  \big)  - \big( \ket{\text{$Player^{(1)}$ 2 state}} \wedge \cdots \\ \wedge  \ket{\text{$Player^{(n)}$2  state}}  \big)   + \big( \widetilde{\ket{\text{$Player^{(1)}$ 3 state}}} \wedge \cdots    \wedge  \widetilde{\ket{\text{$Player^{(n)}$3 state}}}  \big)  \bigg)     \text{. }                     
\end{align*}

\subsection{Strong parallel repetition of tensor products of operators in the $N$-player setting}

\noindent One has,

\begin{align*}
 \big( \textbf{I} \wedge \cdots \wedge \textbf{I} \big) \otimes \big( \textbf{I} \wedge \cdots \wedge \textbf{I} \big) \otimes \big(  \textbf{I} \wedge \cdots \wedge \textbf{I} \big)  \equiv \big( \textbf{I} \otimes \textbf{I}  \otimes \textbf{I} \big) \wedge \cdots \wedge \big( \textbf{I} \otimes \textbf{I} \otimes \textbf{I} \big) \text{, } \\ \\   \big( \sigma_x \wedge \cdots \wedge \sigma_x \big) \otimes \big( \textbf{I} \wedge \cdots \wedge \textbf{I} \big) \otimes \big( \textbf{I} \wedge \cdots \wedge \textbf{I} \big)    \equiv      \big( \sigma_x \otimes \textbf{I} \otimes \textbf{I}    \big) \wedge \cdots \wedge \big( \sigma_x \otimes \textbf{I} \otimes \textbf{I} \big)   \text{,} \\ \\    \big( \textbf{I} \wedge \cdots \wedge \textbf{I} \big) \otimes \big( \sigma_x \wedge \cdots \wedge \sigma_x \big) \otimes \big( \textbf{I} \wedge \cdots \wedge \textbf{I} \big)        \equiv   \big( \textbf{I} \otimes \sigma_x \otimes \textbf{I} \big) \wedge \cdots \wedge \big( \textbf{I} \otimes \sigma_x \otimes \textbf{I} \big)       \text{,} \\ \\       \big( \textbf{I} \wedge \cdots \wedge \textbf{I} \big) \otimes \big( \textbf{I} \wedge \cdots \wedge \textbf{I} \big) \otimes \big( \sigma_x \wedge \cdots \wedge \sigma_x \big)        \equiv \big (\textbf{I} \otimes \textbf{I} \otimes \sigma_x \big) \wedge \cdots \wedge \big( \textbf{I} \otimes \textbf{I} \otimes \sigma_x \big)   \text{, }    \\ \\     \big( \sigma_z \wedge \cdots \wedge \sigma_z \big) \otimes \big( \textbf{I} \wedge \cdots \wedge \textbf{I} \big) \otimes \big( \textbf{I} \wedge \cdots \wedge \textbf{I} \big)    \equiv    \big( \sigma_z \otimes \textbf{I } \otimes \textbf{I} \big) \wedge \cdots \wedge \big( \sigma_z \otimes \textbf{I } \otimes \textbf{I} \big)   \text{,} \\ \\      \big( \textbf{I} \wedge \cdots \wedge \textbf{I} \big) \otimes \big( \sigma_z \wedge \cdots \wedge \sigma_z \big) \otimes \big(       \textbf{I} \wedge \cdots \wedge \textbf{I}   \big)   \equiv \big( \textbf{I} \otimes \sigma_z \otimes \textbf{I}  \big) \wedge \cdots \wedge \big( \textbf{I} \otimes \sigma_z \otimes \textbf{I} \big)   \text{, }  \\ \\   \big( \textbf{I} \wedge \cdots \wedge \textbf{I} \big)  \otimes \big( \textbf{I} \wedge \cdots \wedge \textbf{I} \big)  \otimes \big( \sigma_z \wedge \cdots \wedge \sigma_z \big)  \equiv \big( \textbf{I} \otimes \textbf{I} \otimes \sigma_z \big) \wedge \cdots \wedge \equiv \big( \textbf{I} \otimes \textbf{I} \otimes \sigma_z \big)  \text{, } \\ \\   \big( \sigma_x  \wedge \cdots \wedge \sigma_x  \big)   \otimes     \big( \textbf{I} \wedge \cdots \wedge \textbf{I}  \big)     \otimes      \big( \sigma_z \wedge \cdots \wedge \sigma_z  \big)   \equiv   \big( \sigma_x \otimes \textbf{I} \otimes \sigma_z \big) \wedge \cdots \wedge  \big( \sigma_x \otimes \textbf{I} \otimes \sigma_z \big)   \text{, } \\ \\   \big(   \textbf{I}  \wedge \cdots \wedge  \textbf{I}  \big)   \otimes      \big( \sigma_x \wedge \cdots \wedge \sigma_x  \big)     \otimes      \big( \sigma_z  \wedge \cdots \wedge  \sigma_z \big)   \equiv   \big( \textbf{I} \otimes \sigma_x \otimes \sigma_z \big) \wedge \cdots \wedge \big( \textbf{I} \otimes \sigma_x \otimes \sigma_z \big)  \text{, } \\ \\   \big( \sigma_z  \wedge \cdots \wedge  \sigma_z \big)   \otimes        \big( \textbf{I} \wedge \cdots \wedge \textbf{I}  \big)     \otimes      \big( \sigma_x  \wedge \cdots \wedge  \sigma_x \big)   \equiv   \big( \sigma_z \otimes \textbf{I} \otimes \sigma_x \big) \wedge \cdots \wedge \big( \sigma_z \otimes \textbf{I} \otimes \sigma_x \big)   \text{, } \\ \\   \big( \textbf{I}   \wedge \cdots \wedge \textbf{I} \big)   \otimes       \big(\sigma_z  \wedge \cdots \wedge \sigma_z   \big)     \otimes      \big( \sigma_x \wedge \cdots \wedge \sigma_x  \big)   \equiv   \big(   \textbf{I} \otimes \sigma_z \otimes \sigma_x \big) \wedge \cdots \wedge  \big(   \textbf{I} \otimes \sigma_z \otimes \sigma_x \big)    \text{. } 
\end{align*}

\subsection{Identity with N $\pm 1$ observables}

\noindent \textit{Proof of N-XOR positive semidefinite tensors}. To argue that,

\begin{align*}
  \bigg[  \frac{\sum \text{Tensors of player observables}}{\sqrt{N}} +   \frac{\sum \text{Tensors of player observables}}{\big| \sum \text{Tensors of player observables} \big| }         \bigg]^N   \text{, }
\end{align*}

\noindent is positive semidefinite, write,

\begin{align*}
 \bigg[ \frac{\prod \text{Tensors of player observables}}{N} \bigg]    \bigg[ N + \underset{1 \leq i \leq N-1}{\sum} n^i \lambda^{N-i}   \\   + N \sqrt{\textbf{I} + \frac{\big(  \sum \text{Tensors of player observables} \big) }{\cdots}} \\ \sqrt{ \frac{\times \big( \sum^{-1} \text{Tensors of player observables} \big)}{N}  }  \bigg]  \\ \times            \bigg[ \frac{\prod \text{Tensors of player observables}}{N} \bigg]     \text{, }
\end{align*}

\noindent where, as introduced above, the inverse summation $\sum^{-1}$ of tensor observables from each player equals,

\begin{align*}
    \sum^{-1} \text{Tensors of player observables} \equiv \big( \text{N th player tensor observable} \big)  \times \cdots \\ \times \big( \text{1 st player tensor observable}  \big)   \text{. }
\end{align*}

\noindent

\noindent The expression above implies that the eigenvalue would take the form,

\begin{align*}
 \bigg[ \frac{\lambda^N -  \underset{1 \leq i \leq N-1}{\sum} n^i \lambda^{N-i} - n}{n} \bigg]^n       \frac{1}{N + \underset{1 \leq i \leq N-1}{\sum} n^i \lambda^{N-i} + N \sqrt{1+ \frac{\underset{1 \leq i \leq N-1}{\sum} n^i \lambda^{N-i} -N}{N}}}                      \text{. }
\end{align*}

\noindent Under the square root in the denominator of the expression above, from the \textit{sign} of the eigenvalues $\lambda$, one has the decomposition,

\begin{align*}
    \frac{\underset{1 \leq i \leq N-1}{\sum} n^i \lambda^{N-i} -N}{N} \equiv \bigg( \bigg( \frac{1}{\sqrt{N}} \bigg) \bigg( \underset{1 \leq i \leq N-1}{\sum}  n^i \mathrm{sign} \big( \lambda \big) \lambda^{N-i}    - N          \bigg) \bigg) \bigg( \bigg( \frac{1}{\sqrt{N}} \bigg) \bigg( \underset{1 \leq i \leq N-1}{\sum}  n^i \\ \times \mathrm{sign} \big( \lambda \big) \lambda^{N-i}   - N \bigg) \bigg)     \\ \equiv    \bigg( \frac{1}{\sqrt{N}}  \underset{1 \leq i \leq N-1}{\sum}  n^i \mathrm{sign} \big( \lambda \big) \lambda^{N-i}   - \sqrt{N} \bigg) \bigg( \frac{1}{\sqrt{N}}  \underset{1 \leq i \leq N-1}{\sum}  n^i \mathrm{sign} \big( \lambda \big) \lambda^{N-i}  - \sqrt{N} \bigg)    \\          \equiv    N +    \bigg( \frac{1}{\sqrt{N}} \bigg)^2 \bigg( \underset{1 \leq i \leq N-1}{\sum}  n^i \mathrm{sign} \big( \lambda \big) \lambda^{N-i}   \bigg)^2 - 2   \bigg( \underset{1 \leq i \leq N-1}{\sum}  n^i \mathrm{sign} \big( \lambda \big) \lambda^{N-i}   \bigg)             \text{. }
\end{align*}

\noindent Furthermore, using the same identification,

\begin{align*}
 \lambda \longleftrightarrow \mathrm{sign} \big( \lambda \big) \lambda    \text{, }
\end{align*}

\noindent for the eigenvalue $\lambda$ of the operator from the summation of tensors of player observables also implies,

\begin{align*}
    N + \underset{1 \leq i \leq N-1}{\sum} n^i \lambda^{N-i} + N \sqrt{1+ \frac{\underset{1 \leq i \leq N-1}{\sum} n^i \lambda^{N-i} -N }{N}} \equiv        N + \underset{1 \leq i \leq N-1}{\sum} n^i \lambda^{N-i} + N \\ \times \sqrt{ \frac{\underset{1 \leq i \leq N-1}{\sum} n^i \lambda^{N-i} }{N}}   \\ \equiv    N \bigg[ 1 + \sqrt{ \frac{\underset{1 \leq i \leq N-1}{\sum} n^i \lambda^{N-i} }{N}}  \bigg]      + \underset{1 \leq i \leq N-1}{\sum} n^i \lambda^{N-i}          \equiv   \bigg[  \frac{1}{\sqrt{N}} \bigg[ \underset{1 \leq j \leq N-1}{\sum}  \big( \mathrm{sign} \big( \lambda \big) \big)^j \\ \times \lambda^{N-j} \bigg]   - 1   \bigg]^N        \text{.}            
\end{align*}

\noindent Incorporating the expressions above for the expressions,

\begin{align*}
   \frac{\underset{1 \leq i \leq N-1}{\sum} n^i \lambda^{N-i} -N}{N}  \text{, } \\  N + \underset{1 \leq i \leq N-1}{\sum} n^i \lambda^{N-i} + N \sqrt{1+ \frac{\underset{1 \leq i \leq N-1}{\sum} n^i \lambda^{N-i} -N }{N}} \text{, }
\end{align*}

\noindent implies that the product,

\begin{align*}
\bigg[  \frac{\underset{1 \leq i \leq N-1}{\sum} n^i \lambda^{N-i} -N}{N}  \bigg]  \bigg[  N + \underset{1 \leq i \leq N-1}{\sum} n^i \lambda^{N-i} + N \sqrt{1+ \frac{\underset{1 \leq i \leq N-1}{\sum} n^i \lambda^{N-i} -N }{N}}  \bigg]^{-1} \\ \times  \bigg[ \frac{\underset{1 \leq i \leq N-1}{\sum} n^i \lambda^{N-i} -N}{N}   \bigg]   \text{, }
\end{align*}

\noindent equals,

\begin{align*}
  \bigg[    N +    \bigg( \frac{1}{\sqrt{N}} \bigg)^2 \bigg( \underset{1 \leq i \leq N-1}{\sum}  n^i \mathrm{sign} \big( \lambda \big) \lambda^{N-i}   \bigg)^2 - 2   \bigg( \underset{1 \leq i \leq N-1}{\sum}  n^i \mathrm{sign} \big( \lambda \big) \lambda^{N-i}   \bigg)            \bigg] \\ \times   \bigg[  \bigg[  \frac{1}{\sqrt{N}} \bigg[ \underset{1 \leq j \leq N-1}{\sum}  \big( \mathrm{sign} \big( \lambda \big) \big)^j  \lambda^{N-j} \bigg]   - 1   \bigg]^N  \bigg]^{-1} \bigg[     N +    \bigg( \frac{1}{\sqrt{N}} \bigg)^2 \bigg( \underset{1 \leq i \leq N-1}{\sum}  n^i \mathrm{sign} \big( \lambda \big) \lambda^{N-i}   \bigg)^2 \\ - 2   \bigg( \underset{1 \leq i \leq N-1}{\sum}  n^i \mathrm{sign} \big( \lambda \big) \lambda^{N-i}   \bigg)          \bigg] \text{. }
\end{align*}

\noindent We conclude the proof, as the operator product above can be rearranged as a superposition from the terms,

\begin{align*}
   N  \bigg[  \frac{1}{\sqrt{N}} \bigg[ \underset{1 \leq j \leq N-1}{\sum}  \big( \mathrm{sign} \big( \lambda \big) \big)^j \lambda^{N-j} \bigg]    - 1   \bigg]^N             \text{, } \\ \\  \bigg( \frac{1}{\sqrt{N}} \bigg)^2  \bigg[  \frac{1}{\sqrt{N}} \bigg[ \underset{1 \leq j \leq N-1}{\sum}  \big( \mathrm{sign} \big( \lambda \big) \big)^j \lambda^{N-j} \bigg]    - 1   \bigg]^N                     \text{, } \\  \\ -  2   \bigg( \underset{1 \leq i \leq N-1}{\sum}  n^i \mathrm{sign} \big( \lambda \big) \lambda^{N-i}   \bigg)          \bigg[  \frac{1}{\sqrt{N}} \bigg[ \underset{1 \leq j \leq N-1}{\sum}  \big( \mathrm{sign} \big( \lambda \big) \big)^j \lambda^{N-j} \bigg]    - 1   \bigg]^N           \text{, }  
\end{align*}

\noindent from which we conclude the argument, as the desired eigenvalues for the first, and second, operators and be inferred. \boxed{}

\bigskip

\noindent Along the same lines, computations such as the ones above for demonstrating that the operator for tensor observables of $N$ players can be used to deduce the form of the eigenvalues, and hence the positive semidefiniteness, of the operators,

\begin{align*}
  \bigg[  \frac{\sum^{\prime} \big( \big(  \text{Tensors of XOR player observables}\big)  \wedge \cdots \wedge \big( \text{Tensors of XOR player observables}\big)\big)  }{\sqrt{N}}  \\ +   \frac{\sum^{\prime} \big( \big( \text{Tensors of XOR player observables} \big) \wedge \cdots \wedge \big( \text{Tensors of XOR player observables}\big) \big) }{\big| \sum^{\prime} \big( \big( \text{Tensors of XOR player observables} \big)  \wedge \cdots \wedge \big( \text{Tensors of XOR player observables}\big) \big)  \big| }         \bigg]^N     \text{, } \\  \\  \bigg[  \frac{\sum^{\prime\prime} \big( \big( \text{Tensors of FFL player observables} \big)  \wedge \big( \text{Tensors of FFL player observables}\big) \big) }{\sqrt{N}}  \\ +   \frac{\sum^{\prime\prime} \big( \big( \text{Tensors of FFL player observables}\big)  \wedge \big( \text{Tensors of FFL player observables}\big) \big) }{\big| \sum^{\prime\prime} \big( \big( \text{Tensors of FFL player observables} \big)  \wedge  \big( \text{Tensors of FFL player observables}  \big) \big) \big| }         \bigg]^N     \text{, } 
\end{align*}

\noindent obtained under strong parallel repetition for the $\mathrm{XOR}$, and $\mathrm{FFL}$, games, respectively. Crucially the arguments, and computations for the desired expression for the eigenvalues of the two operators displayed above, can be carried out for,

\begin{align*}
  \lambda^1 \wedge \lambda^2 \wedge  \cdots \wedge \lambda^N  \text{, } \\  \lambda^1 \wedge \lambda^2 \text{, }
\end{align*}

\noindent respectively, corresponding to eigenvalues $\lambda_i$ for each player.

\subsection{Exact, and approximate, optimality of novel settings for the $\mathrm{XOR}^{*}$ game from those developed for the ordinary $\mathrm{XOR}$ game}

\subsubsection{Main Result}

\noindent As in a previous subsection for describing the main result for variants of XOR, and FFL, games considered in this work, for the dual $\mathrm{XOR}^{*}$ game, we introduce the following result:

\bigskip

\noindent \textbf{Theorem} \textit{7} (\textit{primal feasible solutions and duality gaps for the dual XOR game}). The same collection of items provided in $\textbf{Theorem}$ \textit{1} for the $3$ $\mathrm{XOR}$ game also hold for $\mathrm{XOR}^{*}$ games, given the existence of primal feasible solutions, duality gap, and dual semidefinite program.

\subsubsection{Passing through duality to the $XOR^{*}$ game}

\noindent We recall the following discussion from a previous work of the author, {[44]}, which makes it possible to immediately formulate the connection between the objects, and relationships, developed in this paper for $\mathrm{XOR}^{*}$ games. To briefly reiterate, such conditions were previously used by the author in two-player settings, including the $\mathrm{XOR}^{*}$ and $\mathrm{FFL}$ games, in order to establish a correspondence between the optimal values for the $\mathrm{XOR}$ and $\mathrm{XOR}^{*}$ games. From the perspectives developed in this work, the elaboration on notions of optimality, whether exact or approximate, can be generalized under the following notion of duality.

\bigskip

\noindent Denote the set of possible inputs that Alice inputs into the XOR game with $\big| \mathcal{S} \big|$, and the set of possible inputs that Bob inputs with $\big| \mathcal{T} \big|$. Also, denote the Bell state with $\ket{\psi} = \frac{1}{\sqrt{2}} \big( \ket{00} + \ket{11} \big) $ To compare $\mathrm{XOR}$ games to the dual $\mathrm{XOR}^{*}$ game, we make use of the following items:

\begin{itemize}
    \item[$\bullet$] \big(\underline{The $\epsilon$ bit $\mathrm{XOR}$ game, \textbf{Lemma} \textit{1}, {[8]}}\big) An $\mathrm{XOR}$ game for which $\mathrm{min} \big\{ \big| \mathcal{S} \big| ,    \big| \mathcal{T} \big|   \big\} \leq 4$        is an $\epsilon$-bit $\mathrm{XOR}$ game.          

\bigskip 
        \item[$\bullet$] \big(\underline{ Classical and quantum bounds for $\mathrm{XOR}$  and $\mathrm{{XOR}^{*}}$ games ,\textbf{Theorem} \textit{2}, {[8]}}\big). Denote $a$ and $b$ as the two possible measurements that Alice and Bob can observe from some $s \in \mathcal{S}$ and $t \in \mathcal{T}$. Furthermore, denote the single qubit measurements from each possible $s$ and $t$ with $A_{a | s}$ and $B_{b|t}$, and the probability, conditional upon each input, as, $P \big(    a , b \big| s , t        \big)$, which can be expressed with the trace of the inner product $\big(  A_{a|s}       \otimes    B_{b|t}     \big) \ket{\psi} \bra{\psi} $. The output of the XOR game, $m = a \oplus b$, is such that the classical and quantum bounds of the $\mathrm{XOR}$ and $\mathrm{XOR^{*}}$games are equal.

\bigskip

        The same result holds for the converse.

\end{itemize}

\noindent The collection of results above, as previously utilised by the author in {[44]}, can be shown to deduce that the optimal value of the $\mathrm{XOR}^{*}$ game equals that of the $\mathrm{XOR}$ game, namely $\frac{1}{\sqrt{2}}$. Besides such an obvious consequence after passing to the dual game with the same number of players, denote:

\begin{align*}
    \mathscr{T}^{*} \equiv \text{Linear operator for $\mathrm{XOR^{*}}$ game} \text{, } \\ \\  c^{*}_i \equiv \text{Constants for upper bounds in } \textbf{Lemma} \textit{ 1-3-$\mathrm{ XOR}^{*}$} \text{, } \\ \\ A_i \equiv \text{Tensor observable for the first player of the $\mathrm{XOR}^{*}$ game} \text{, } \\ \\    \mathscr{T}^{*} \wedge \mathscr{T}^{*}  \equiv \text{Two strong parallel repetition operations of the linear} \\ \text{ operator for $\mathrm{XOR^{*}}$ game} \text{, } \\ \\     \mathscr{T}^{*} \wedge \cdots \wedge  \mathscr{T}^{*}  \equiv \text{An arbitrary number of strong parallel repetition} \\ \text{operations of the linear operator for $\mathrm{XOR^{*}}$}  \text{ game} \text{, } \\ \\  \epsilon^{*} \equiv \text{Constant in $\mathrm{XOR}^{*}$ upper bound} \text{,} \\   \\   \epsilon_{N\mathrm{XOR}^{*}}  \equiv \text{Constant in N $\mathrm{XOR}^{*}$ upper bound} \text{,} \\ \\ \epsilon_{N\mathrm{XOR}^{*}\wedge N\mathrm{XOR}^{*}} \equiv \text{Constant in N $\mathrm{XOR}^{*} \wedge $ N $\mathrm{XOR}^{*}$ upper bound} \text{, } \\ \\   \epsilon_{N\mathrm{XOR}^{*}\wedge\cdots \wedge N\mathrm{XOR}^{*}} \equiv    \text{Constant in N $\mathrm{XOR}^{*} \wedge \cdots \wedge$ N $\mathrm{XOR}^{*}$ upper bound}      \text{, }               
\end{align*}

\noindent The following series of results can be immediately established with identical arguments as provided in the $\mathrm{XOR}$ case:

\bigskip

\noindent \textbf{Lemma} \textit{1 -N $\mathrm{XOR}^{*}$} (\textit{computation of the Frobenius norm for the anticommutation rule of $T_{N\mathrm{XOR}^{*}}$ yields a desired up to constants $\sqrt{\epsilon}$ upper bound}, \textbf{Theorem} \textit{6}, {[44]}). One has that,

\begin{align*}
      \underline{\text{Player $1$}:} \text{ } \bigg| \bigg|   \bigg(  A_i \bigotimes \bigg( \underset{1 \leq k \leq n-1}{\bigotimes} \textbf{I}_k  \bigg)        \bigg) \mathscr{T}^{*}  -   \mathscr{T}^{*} \bigg( \bigg( \underset{1 \leq k \leq n-1}{\bigotimes} \textbf{I}_k  \bigg)      \bigotimes \widetilde{A_i}      \bigg)      \bigg| \bigg|_F \\  < c^{*}_1 n^N \sqrt{\epsilon^{*}}  \text{,} \\ \vdots \\              \underline{\text{Player $N$}:} \text{ } \bigg| \bigg|   \bigg(  \bigg( \underset{1 \leq k \leq n-1}{\bigotimes} \textbf{I}_k  \bigg)   \bigotimes  A^{(n-1)}_{i_1, \cdots, i_{n-1}}     \bigg) \mathscr{T}^{*}  -   \mathscr{T}^{*} \bigg( \bigg(    \widetilde{A^{(n-1)}_{i_1, \cdots, i_{n-1}} }       \\ \bigotimes  \bigg( \underset{1 \leq k \leq n-1}{\bigotimes} \textbf{I}_k  \bigg)     \bigg)      \bigg| \bigg|_F     < c^{*}_N n^N \sqrt{\epsilon^{*}}  \text{, }
\end{align*}

\noindent has the upper bound,

\begin{align*}
      \mathscr{C}^{*} \equiv    \underset{1 \leq i \leq N}{\bigcup}  \big\{     C_i  \neq c^{*}_i\in \textbf{R}  :            C_i \equiv c^{*}_i \sqrt{\epsilon}     \big\}     \propto n^N \sqrt{\epsilon^{*}} \text{. }
\end{align*}

\bigskip

\noindent \textit{Proof of Lemma 1 -N $\mathrm{XOR}^{*}$}. Directly apply the argument for $\textbf{Lemma}$ \textit{1-N $\mathrm{XOR}$}, from which we conclude the argument. \boxed{}

 \bigskip

\noindent \textbf{Lemma} \textit{$1- 3- \mathrm{XOR}^{*}$} (\textit{computation of the Frobenius norm for the anticommutation rule of $T_{3\mathrm{XOR}^{*}}$ yields a desired up to constants $\sqrt{\epsilon}$ upper bound}, \textbf{Theorem} \textit{6}, {[44]}). One has that,

\begin{align*}
      \underline{\text{Player $1$}:} \text{ } \bigg| \bigg|   \bigg(  A_i \bigotimes \bigg( \underset{1 \leq k \leq n-1}{\bigotimes} \textbf{I}_k  \bigg)        \bigg) \mathscr{T}^{*}  -   \mathscr{T}^{*} \bigg( \bigg( \underset{1 \leq k \leq n-1}{\bigotimes} \textbf{I}_k  \bigg)      \bigotimes \widetilde{A_i}      \bigg)      \bigg| \bigg|_F  \\ < c^{*}_1 n^3 \sqrt{\epsilon^{*}}  \text{,} \\   \underline{\text{Player $2$}:} \text{ } \bigg| \bigg|   \bigg(  \bigg( \underset{1 \leq k \leq n-1}{\bigotimes} \textbf{I}_k  \bigg)   \bigotimes  A^{(n-1)}_{i_1, \cdots, i_{n-1}}     \bigg) \mathscr{T}^{*}  -   \mathscr{T}^{*} \bigg( \bigg(    \widetilde{A^{(n-1)}_{i_1, \cdots, i_{n-1}} }       \\ \bigotimes  \bigg( \underset{1 \leq k \leq n-1}{\bigotimes} \textbf{I}_k  \bigg)     \bigg)      \bigg| \bigg|_F     < c^{*}_2 n^3 \sqrt{\epsilon^{*}} \\              \underline{\text{Player $3$}:} \text{ } \bigg| \bigg|   \bigg(  \bigg( \underset{1 \leq k \leq n-1}{\bigotimes} \textbf{I}_k  \bigg)   \bigotimes  A^{(n-1)}_{i_1, \cdots, i_{n-1}}     \bigg) \mathscr{T}^{*}  -   \mathscr{T}^{*} \bigg( \bigg(    \widetilde{A^{(n-1)}_{i_1, \cdots, i_{n-1}} }       \\ \bigotimes  \bigg( \underset{1 \leq k \leq n-1}{\bigotimes} \textbf{I}_k  \bigg)     \bigg)      \bigg| \bigg|_F     < c^{*}_3 n^3 \sqrt{\epsilon^{*}}  \text{, }
\end{align*}

\noindent has the upper bound,

\begin{align*}
      \mathscr{C}^{*} \equiv    \underset{1 \leq i \leq 3}{\bigcup}  \big\{     C_i  \neq c^{*}_i\in \textbf{R}  :            C_i \equiv c^{*}_i \sqrt{\epsilon}     \big\}     \propto n^N \sqrt{\epsilon^{*}} \text{. }
\end{align*}

\bigskip

\noindent \textit{Proof of Lemma $1$- 3-$\mathrm{XOR}^{*}$}. Directly apply the argument for $\textbf{Lemma}$ \textit{1-3 $\mathrm{XOR}$}, from which we conclude the argument. \boxed{}

\bigskip

\noindent \textbf{Lemma} \textit{$2^{*}$ $\mathrm{XOR}^{*}$} (\textit{3 3-$\mathrm{XOR^{*}}$ identities from two FFL identities}). One has that,

\begin{align*}
  \bigg( A_i \otimes \textbf{I} \otimes \textbf{I}   \bigg) \mathscr{T}^{*} - \mathscr{T}^{*}  \bigg( \widetilde{A_i} \otimes \textbf{I} \otimes \textbf{I}   \bigg)    \text{, }
\end{align*}

\noindent and that,

\begin{align*}
     \bigg(  \big(  A_i \mathscr{T}^{*} \otimes \mathscr{T}^{*} \big) - \big( \mathscr{T}^{*} \widetilde{A_i} \otimes \mathscr{T}^{*} \big) \bigg) \otimes \bigg(  \textbf{I} \otimes \textbf{I} \bigg)                        \text{, }
\end{align*}

\noindent are equal, in which,

\begin{align*}
    \bigg( A_i \otimes \textbf{I} \otimes \textbf{I} \bigg) \mathscr{T}^{*} - \mathscr{T}^{*}  \bigg( \widetilde{A_i} \otimes \textbf{I} \otimes \textbf{I} \bigg)  \equiv      \bigg( \big( A_i  \mathscr{T}^{*} - \mathscr{T}^{*} \widetilde{A_i} \big) \otimes \mathscr{T}^{*} \bigg) \otimes \bigg( \textbf{I} \otimes \textbf{I}  \bigg)    \text{. }
\end{align*}

\noindent For the remaining two players in the $\mathrm{XOR}^{*}$ game, identities of an analogous form hold.

\bigskip

\noindent \textit{Proof of Lemma $2^{*} \mathrm{XOR}^{*}$}. Directly apply the argument for $\textbf{Lemma}$ \textit{$2^{*}$}, from which we conclude the argument. \boxed{}

\bigskip

\noindent \textbf{Lemma} \textit{$2^{**} \mathrm{XOR}^{*}$} (\textit{N N-$\mathrm{XOR^{*}}$ identities}). One has that,

\begin{align*}
  \bigg( A_i \bigotimes  \bigg( \underset{1 \leq z \leq N-1}{\bigotimes} \textbf{I}_z \bigg)  \bigg)  \mathscr{T}^{*} - \mathscr{T}^{*}  \bigg( A_i \bigotimes  \bigg( \underset{1 \leq z \leq N-1}{\bigotimes} \textbf{I}_z \bigg)  \bigg)   \text{, }
\end{align*}

\noindent and that,

\begin{align*}
     \bigg(  \bigg(  A_i \mathscr{T}^{*} \bigotimes  \bigg( \underset{1 \leq z \leq N-1}{\bigotimes} \mathscr{T}^{*} \textbf{I}_z  \bigg) - \bigg( \mathscr{T}^{*} \widetilde{A_i} \bigotimes \bigg( \underset{1 \leq z \leq N-1}{\bigotimes}  \mathscr{T}^{*}  \textbf{I}_z \bigg)  \bigg) \bigg) \bigotimes \bigg(  \underset{1 \leq z \leq N}{\bigotimes} \textbf{I}_z  \bigg)                        \text{, }
\end{align*}

\noindent are equal, in which,

\begin{align*}
    \bigg( A_i \bigotimes  \bigg( \underset{1 \leq z \leq N-1}{\bigotimes} \textbf{I}_z  \bigg) \bigg) \mathscr{T}^{*} - \mathscr{T}^{*} \bigg(  \widetilde{A_i} \bigotimes  \bigg( \underset{1 \leq z \leq N-1}{\bigotimes} \textbf{I}_z  \bigg) \bigg)   \equiv      \bigg( \bigg( A_i  \mathscr{T}^{*} - \mathscr{T}^{*} \widetilde{A_i} \bigg) \\ \bigotimes \bigg( \underset{1 \leq z \leq n-1}{\bigotimes} \mathscr{T}^{*} \textbf{I}_z  \bigg)  \bigg)  \bigotimes   \bigg(  \underset{1 \leq z \leq N-1}{\bigotimes} \textbf{I}_z   \bigg)    \text{. }
\end{align*}

\bigskip

\noindent \textbf{Lemma} \textit{$2^{***} \mathrm{XOR}^{*}$} (\textit{N N-$\mathrm{XOR}^{*}$ identities under strong parallel repetition}). One has that,

\begin{align*}
  \bigg( A_i \bigotimes  \bigg( \underset{1 \leq z \leq N-1}{\bigotimes} \textbf{I}_z \bigg)  \bigg) \bigg( \mathscr{T}^{*} \wedge \cdots \wedge \mathscr{T}^{*} \bigg)  - \bigg( \mathscr{T}^{*} \wedge \cdots \wedge \mathscr{T}^{*} \bigg)    \bigg( A_i \bigotimes  \bigg( \underset{1 \leq z \leq N-1}{\bigotimes} \textbf{I}_z \bigg)  \bigg)   \text{, }
\end{align*}

\noindent and that,

\begin{align*}
     \bigg(  \bigg(  A_i \bigg( \mathscr{T}^{*} \wedge \cdots \wedge \mathscr{T}^{*} \bigg)   \bigotimes  \bigg( \underset{1 \leq z \leq N-1}{\bigotimes} \bigg( \mathscr{T}^{*} \wedge \cdots \wedge \mathscr{T}^{*} \bigg)  \textbf{I}_z  \bigg) - \bigg( \bigg( \mathscr{T}^{*} \wedge \cdots \wedge \mathscr{T}^{*} \bigg)  \widetilde{A_i} \\ \bigotimes  \bigg( \underset{1 \leq z \leq N-1}{\bigotimes}  \bigg( \mathscr{T}^{*} \wedge \cdots \wedge \mathscr{T}^{*} \bigg)   \textbf{I}_z \bigg)  \bigg) \bigg) \bigotimes \bigg(  \underset{1 \leq z \leq N}{\bigotimes} \textbf{I}_z  \bigg)                        \text{, }
\end{align*}

\noindent are equal, in which,

\begin{align*}
    \bigg( A_i \bigotimes  \bigg( \underset{1 \leq z \leq N-1}{\bigotimes} \textbf{I}_z  \bigg) \bigg) \bigg( \mathscr{T}^{*} \wedge \cdots \wedge \mathscr{T}^{*} \bigg)  - \bigg( \mathscr{T}^{*} \wedge \cdots \wedge \mathscr{T}^{*} \bigg)   \bigg(  \widetilde{A_i} \\ \bigotimes  \bigg( \underset{1 \leq z \leq N-1}{\bigotimes} \textbf{I}_z  \bigg) \bigg)  \\  \equiv      \bigg( \bigg( A_i  \bigg( \mathscr{T}^{*} \wedge \cdots \wedge \mathscr{T}^{*} \bigg)   - \bigg( \mathscr{T}^{*} \wedge \cdots \wedge \mathscr{T}^{*} \bigg)  \widetilde{A_i} \bigg) \bigotimes\bigg( \underset{1 \leq z \leq n-1}{\bigotimes}\bigg( \mathscr{T}^{*} \\ \wedge \cdots \wedge \mathscr{T}^{*} \bigg)  \textbf{I}_z  \bigg)  \bigg)  \bigotimes   \bigg(  \underset{1 \leq z \leq N-1}{\bigotimes} \textbf{I}_z   \bigg)    \text{. }
\end{align*}

\noindent \textit{Proof of Lemma $2^{**} \mathrm{XOR}^{*}$, and $2^{***} \mathrm{XOR}^{*}$}. Directly apply the arguments for $\textbf{Lemma}$ \textit{$2^{**}$}, and for $\textbf{Lemma}$ \textit{$2^{***}$}, from which we conclude the argument. \boxed{}

\bigskip

\noindent \noindent \textbf{Lemma} \textit{3} (\textit{intialization of $\epsilon$-optimality of the N-$\mathrm{XOR}^{*}$ game from the observable of the first player}). For an $\epsilon$-optimal strategy, and player observable tensors $A_{i_1}, A^1_{i_1,i_2}, \cdots, A^{n-1}_{i_1,i_2, \cdots, i_n}$, and $\ket{\psi_{N\mathrm{XOR}^{*}}}$, and $C_1>0$,

\begin{align*}
    \underset{i,j}{\sum}     \bigg| \bigg|     \bigg[            \bigg( \bigg(    \frac{A_i A_j + A_j A_i}{2}        \bigg) \bigotimes  \bigg( \underset{1\leq k \leq N-1}{\bigotimes} \textbf{I}_k \bigg)  \bigg)      \bigg] \ket{\psi_{N\mathrm{XOR}^{*}}}                           \bigg| \bigg|^2         <  C_1 n \bigg( \underset{1 \leq j \leq N-1}{\prod}  \big( n - j  \big) \bigg) \epsilon_{N\mathrm{XOR}^{*}}   \text{. }
\end{align*}

\bigskip

\noindent \noindent \textbf{Lemma} $\textit{4A}^{*}$ (\textit{induction on the, up to constants, strong parallel repetition $\epsilon$ upper bound from the previous result}). Under the assumptions of $\textbf{Lemma}$ \textit{4A}, and $\textbf{Lemma}$ \textit{3}, and $C^{\wedge}_2>0$,

\begin{align*}
      \underset{\text{Tensor entries }}{\sum}  \bigg[ \underset{j^{\prime}_1,\cdots,j^{\prime}_n}{\underset{i^{\prime},\cdots,i^{\prime}_n}{\underset{i_1,\cdots,i_n}{\sum}}}       \bigg| \bigg|     \bigg[            \bigg( \bigg(    \frac{ \big( A_i \wedge A_{i^{\prime}} \big)  \big( A_j \wedge A_{j^{\prime}} \big)  + \big( A_j \wedge A_{j^{\prime} } \big) \big(  A_i \wedge A_{i^{\prime}} \big)}{2}        \bigg) \bigotimes  \textbf{I}  \bigg)      \bigg] \\ \times \ket{\psi_{2\mathrm{XOR}^{*}}}                           \bigg| \bigg|^2   \bigg]        <  C^{\wedge}_2 n \bigg( \underset{1 \leq j \leq 2}{\prod}  \big( n - j  \big) \bigg) \epsilon_{2\mathrm{XOR}^{*}\wedge 2 \mathrm{XOR}^{*}}  \\ \equiv    C^{\wedge}_2 n \bigg( \underset{1 \leq j \leq 2}{\prod}  \big( n - j  \big) \bigg) \epsilon_{\mathrm{XOR}^{*}\wedge  \mathrm{XOR}^{*}}   \text{. }
\end{align*}

\noindent \textit{Proof of Lemma $4A^{*}$}. Directly apply the argument from \textbf{Theorem} $\textit{1}^{*}$, particularly for demonstrating that the desired operator associated with the semidefinite program is positive definite, from which we conclude the argument. \boxed{}

\bigskip

\noindent \noindent \textbf{Lemma} $\textit{4A}^{**}$ (\textit{induction on the, up to constants, $N$-player strong parallel repetition $\epsilon$ upper bound from the previous result}). Under the assumptions of $\textbf{Lemma}$ \textit{4A}, $\textbf{Lemma}$ $\textit{4A}^{*}$, and $\textbf{Lemma}$ \textit{3}, and $C^{\wedge}_N>0$,

\begin{align*}
      \underset{\text{Tensor entries }}{\sum}  \bigg[ \underset{i^{\prime\cdots\prime},\cdots,i^{\prime\cdots\prime}}{\underset{\vdots}{\underset{i_1,\cdots,i_n}{\sum}}}     \bigg| \bigg|     \bigg[            \bigg( \bigg(    \frac{ \big( A_i \wedge A_{i^{\prime}}  \wedge \cdots \wedge A_{i^{\prime\cdots\prime}} \big)  \big( A_j \wedge A_{j^{\prime}} \wedge \cdots \wedge A_{j^{\prime\cdots\prime}} \big)  +\big( A_j \wedge A_{j^{\prime}}  \wedge \cdots   }{2} \\ \times  \frac{\cdots \wedge A_{j^{\prime\cdots\prime}} \big)  \big( A_i \wedge A_{i^{\prime}}  \wedge \cdots  \wedge A_{i^{\prime\cdots\prime}} \big)}{\cdots }      \bigg) \bigotimes  \textbf{I}  \bigg)      \bigg] \ket{\psi_{\mathrm{XOR}^{*}\wedge \cdots \wedge \mathrm{XOR}^{*}}}                           \bigg| \bigg|^2   \bigg]   \\     <  C^{\wedge}_N n   \bigg( \underset{1 \leq j \leq N}{\prod}  \big( n - j  \big) \bigg) \epsilon_{N\mathrm{XOR}^{*}\wedge \cdots \wedge  N\mathrm{XOR}^{*}} \\ \equiv    C^{\wedge}_N n \bigg( \underset{1 \leq j \leq N}{\prod}  \big( n - j  \big) \bigg) \epsilon_{\mathrm{XOR}^{*}\wedge  \cdots \wedge  \mathrm{XOR}^{*}}   \text{. }
\end{align*}

\noindent \textit{Proof of Lemma $4A^{**}$}. Directly apply the argument from \textbf{Theorem} $\textit{1}^{*}$, particularly for demonstrating that the desired operator associated with the semidefinite program is positive definite, from which we conclude the argument. \boxed{}

\bigskip

\noindent \textbf{Lemma} $\textit{5}^{*}$ (\textit{error bound from permuting indices in the N-player setting}, \textbf{Lemma} \textit{5}, {[44]}). One has the following error bound from permuting indices,

\begin{align*}
     \bigg| \bigg|  \bigg(    \bigg( \underset{1 \leq i \leq n}{\prod} A^{j_i}_i  \bigg)      \bigotimes \bigg( \underset{1 \leq z \leq N-1}{\bigotimes} \textbf{I}_z \bigg)       \bigg) \ket{\psi_{N\mathrm{XOR}^{*}}} - \bigg(     \bigg( \underset{\text{if } i \equiv j_1+1, \text{ } \mathrm{set} \text{ } j_1 + 1 \equiv j_1 \oplus 1}{ \underset{1 \leq i \leq n}{\prod} }A^{j_i}_i        \bigg)    \\     \bigotimes \bigg( \underset{1 \leq z \leq N-1}{\bigotimes} \textbf{I}_z \bigg)            \bigg)  \ket{\psi_{N\mathrm{XOR}^{*}}}     \bigg| \bigg|        <  N n^{N+\epsilon_{N\mathrm{XOR}^{*}}} \omega^3_{N\mathrm{XOR}^{*}}  \text{. }
\end{align*}

\noindent \textit{Proof of Lemma $5^{*}$}. Directly apply the argument in \textbf{Lemma} $\textit{5}^{*}$, \textbf{Lemma} $\textit{5}^{**}$, and \textbf{Lemma} $\textit{5}^{***}$, from which we conclude the argument. \boxed{}

\bigskip

\noindent \textbf{Lemma} $\textit{5}^{**}$ (\textit{error bound from permuting indices in the strong parallel repetition of the N-player setting}, \textbf{Lemma} \textit{5}, {[44]}). One has the following error bound from permuting indices,

\begin{align*}
              \bigg| \bigg|  \bigg(  \bigg(    \bigg( \underset{1 \leq i \leq n}{\prod} A^{j_i}_i  \bigg)   \wedge  \cdots \wedge \bigg( \underset{1 \leq i^{\prime\cdots\prime} \leq n^{\prime\cdots\prime}}{\prod} A^{j^{\prime\cdots\prime}_{i^{\prime\cdots\prime}}}_{i^{\prime\cdots\prime}}   \bigg) \bigg)       \bigotimes \bigg( \underset{1 \leq z \leq N-1}{\bigotimes}  \big( \textbf{I}_z \wedge \cdots \wedge \textbf{I}_z \big)  \bigg) 
 \bigg) \\ \times \ket{\psi_{N\mathrm{XOR}^{*}\wedge \cdots \wedge N\mathrm{XOR}^{*}}} \\ - \bigg(   \bigg(   \bigg( \underset{\text{if } i \equiv j_1+1, \text{ } \mathrm{set} \text{ } j_1 + 1 \equiv j_1 \oplus 1}{ \underset{1 \leq i \leq n}{\prod} }A^{j_i}_i        \bigg)  \wedge \cdots \wedge \bigg(     \underset{\text{if } i^{\prime\cdots\prime} \equiv j^{\prime\cdots\prime}_1+1, \text{ } \mathrm{set} \text{ } j^{\prime\cdots\prime}_1 + 1 \equiv j^{\prime\cdots\prime}_1 \oplus 1}{ \underset{1 \leq i^{\prime\cdots\prime} \leq n^{\prime\cdots\prime}}{\prod} }A^{j^{\prime\cdots\prime}_{i^{\prime\cdots\prime}}}_{i^{\prime\cdots\prime}}           \bigg)    \bigg)   \\ \bigotimes \bigg( \underset{1 \leq z \leq N-1}{\bigotimes}\big(  \textbf{I}_z  \wedge \cdots \wedge \textbf{I}_z \big)   \bigg)    \bigg)     \ket{\psi_{N\mathrm{XOR}^{*} \wedge \cdots \wedge N\mathrm{XOR}^{*}}}     \bigg| \bigg|  <  n^{N+\epsilon}_{\wedge } \\ +  \bigg( \frac{50 n^{N+\epsilon}_{\wedge}}{\sqrt{n^{N-1}}}   \bigg) \omega_{N\mathrm{XOR}^{*} \wedge \cdots \wedge N \mathrm{XOR}^{*}}          \text{. }
\end{align*}

\noindent \textit{Proof of Lemma $5^{**}$}. Directly apply the argument in \textbf{Lemma} $\textit{5}^{*}$, \textbf{Lemma} $\textit{5}^{**}$, and \textbf{Lemma} $\textit{5}^{***}$, from which we conclude the argument. \boxed{}

\bigskip

\noindent \textbf{Lemma} $\textit{4}^{*}$ (\textit{an arbitrary number of strong parallel repetition applications of }$\sqrt{\epsilon^{\wedge}_{2\mathrm{XOR}^{*}}}$- \textit{2-XOR approximality}, \textbf{Lemma} \textit{8}, {[37]}). From the same quantities introduced in the previous result, one has,

\begin{align*}
   \bigg| \bigg|               \bigg( \big(  A_k  \wedge A_{k^{\prime}} \wedge \cdots \wedge A_{k^{\prime\cdots \prime}} \big) \otimes \textbf{I} \bigg) \ket{\psi_{2\mathrm{XOR}^{*} \wedge \cdots \wedge 2\mathrm{XOR}^{*}}}    \\ -  \bigg( \textbf{I} \otimes \bigg(     \frac{\pm \big(  B_{kl} \wedge B_{k^{\prime} l^{\prime}} \wedge \cdots \wedge B_{k^{\prime\cdots\prime}l^{\prime\cdots \prime}}  \big)  + \big( B_{lk} \wedge B_{ l^{\prime} k^{\prime}} \wedge \cdots \wedge B_{l^{\prime\cdots \prime} k^{\prime\cdots\prime} }  \big) }{\big| \pm \big(  B_{kl} \wedge B_{k^{\prime} l^{\prime}} \wedge \cdots \wedge B_{k^{\prime\cdots\prime}l^{\prime\cdots \prime}}  \big)  + \big( B_{lk} \wedge B_{ l^{\prime} k^{\prime}} \wedge \cdots \wedge B_{l^{\prime\cdots \prime} k^{\prime\cdots\prime} }  \big) \big| }           \bigg) \bigg) \\ \times \ket{\psi_{2\mathrm{XOR}^{*} \wedge \cdots \wedge 2\mathrm{XOR}^{*}}}              \bigg| \bigg|    < 18 \sqrt{N \epsilon^{\wedge}_{2\mathrm{XOR}^{*}}}   \text{. } 
\end{align*}

\noindent \textit{Proof of Lemma $4^{*}$}. Directly apply the argument in \textbf{Lemma} \textit{4}, from which we conclude the argument. \boxed{}

\bigskip

\noindent \textbf{Lemma} $\textit{4}^{**}$ (\textit{an arbitrary number of strong parallel repetition applications of }$\sqrt{\epsilon^{\wedge}_{N\mathrm{XOR}}}$- \textit{N-XOR approximality}, \textbf{Lemma} \textit{8}, {[37]}). From the same quantities introduced in the previous result, one has, for the $\mathrm{XOR}$ game under an arbitrary number of strong parallel repetitions, that the quantities,

\begin{align*}
  \mathcal{I}^{*}_1 \equiv  \bigg| \bigg|               \bigg( \big(  A_k  \wedge A_{k^{\prime}} \wedge \cdots \wedge A_{k^{\prime\cdots \prime}} \big) \bigotimes  \bigg( \underset{1 \leq z \leq N-1}{\bigotimes}\textbf{I}_z \bigg)  \bigg) \ket{\psi_{N\mathrm{XOR}^{*} \wedge \cdots \wedge N\mathrm{XOR}^{*}}}    \\  -  \bigg( \textbf{I}  \otimes \bigg(     \frac{\pm \big(  B_{kl} \wedge B_{k^{\prime} l^{\prime}} \wedge \cdots \wedge B_{k^{\prime\cdots\prime}l^{\prime\cdots \prime}}  \big)  + \big( B_{lk} \wedge B_{ l^{\prime} k^{\prime}} \wedge \cdots \wedge B_{l^{\prime\cdots \prime} k^{\prime\cdots\prime} }  \big) }{\big| \pm \big(  B_{kl} \wedge B_{k^{\prime} l^{\prime}} \wedge \cdots \wedge B_{k^{\prime\cdots\prime}l^{\prime\cdots \prime}}  \big)  + \big( B_{lk} \wedge B_{ l^{\prime} k^{\prime}} \wedge \cdots \wedge B_{l^{\prime\cdots \prime} k^{\prime\cdots\prime} }  \big) \big| }    \bigg) \\  \bigotimes  \bigg( \underset{1 \leq z \leq N-2}{\bigotimes}\textbf{I}_z \bigg)       \bigg) \ket{\psi_{N\mathrm{XOR}^{*} \wedge \cdots \wedge N\mathrm{XOR}^{*}}}              \bigg| \bigg| \\ \vdots \\ \mathcal{I}^{*}_N \equiv   \bigg| \bigg|    \bigg(   \bigg( \underset{1 \leq z \leq N-2}{\bigotimes} \textbf{I}_z \bigg) \bigotimes   \frac{1}{\sqrt{\# \sigma^{\prime} }}  \bigg(    \underset{\text{Permutations } \sigma^{\prime}}{\sum}  \big(   B^{(N-1)}_{\sigma^{\prime} ( i_1, \cdots, i_{N-1})}         \wedge   B^{(N-1)}_{\sigma^{\prime} ( i^{\prime}_1, \cdots, i^{\prime}_{N-1})}  \wedge  \cdots \\ \wedge   B^{(N-1)}_{\sigma^{\prime} ( i^{\prime\cdots\prime}_1, \cdots, i^{\prime\cdots\prime}_{N-1})} \big)      \bigg)      \bigotimes \textbf{I}  \bigg)   \ket{\psi_{N\mathrm{XOR}^{*} \wedge \cdots \wedge N\mathrm{XOR}^{*}}}  \\ -  \bigg( \bigg( \underset{1 \leq z \leq N-1}{\bigotimes} \textbf{I}_z \bigg) \bigotimes \frac{1}{\sqrt{\# \sigma }}  \bigg(    \underset{\text{Permutations } \sigma}{\sum}  \big(   B^{(N-1)}_{\sigma ( i_1, \cdots, i_{N-1})}         \wedge   B^{(N-1)}_{\sigma ( i^{\prime}_1, \cdots, i^{\prime}_{N-1})}  \wedge  \cdots \\ \wedge   B^{(N-1)}_{\sigma ( i^{\prime\cdots\prime}_1, \cdots, i^{\prime\cdots\prime}_{N-1})} \big)      \bigg) \bigg)  \ket{\psi_{N\mathrm{XOR}^{*} \wedge \cdots \wedge N\mathrm{XOR}^{*}}}           \bigg| \bigg|      \text{, } 
\end{align*}

\noindent have the strict upper bound,

\begin{align*}
\underset{1 \leq j \leq N}{\sum} \mathcal{I}^{*}_j < 20 N \sqrt{N \epsilon^{\wedge}_{N\mathrm{XOR}^{*} \wedge \cdots \wedge N\mathrm{XOR}^{*}}} \text{,}
 \end{align*}

 \noindent where the tensors beyond that of the second player, $B$, are indexed as,

 \begin{align*}
     \textbf{I} \bigotimes B_{\sigma ( i_1,i_2)} \bigotimes \bigg( \underset{1 \leq z \leq N-2}{\bigotimes} \textbf{I}_z   \bigg) \equiv  \textbf{I} \bigotimes B^{(1)}_{\sigma ( i_1,i_2)} \bigotimes \bigg( \underset{1 \leq z \leq N-2}{\bigotimes} \textbf{I}_z   \bigg)   \text{, } \\ \vdots \\    \bigg( \underset{1 \leq z \leq N-1}{\bigotimes} \textbf{I}_z   \bigg) \bigotimes B_{\sigma ( i_1,i_2,\cdots, i_{n-2} )}  \equiv   \bigg( \underset{1 \leq z \leq N-1}{\bigotimes} \textbf{I}_z   \bigg) \bigotimes B^{(n-2)}_{\sigma ( i_1,i_2,\cdots, i_{n-2} )}  \text{. }
 \end{align*}

\noindent \textit{Proof of Lemma $4^{**}$}. Directly apply the argument in \textbf{Lemma} \textit{4}, from which we conclude the argument. \boxed{}

\bigskip

\noindent \textbf{Lemma} \textit{FR} $\mathrm{XOR}^{*}$ (\textit{Frobenius norm upper bound for strong parallel $\mathrm{XOR}^{*}$ repetition}). One has that,

\[ 
 (\mathrm{XOR}^{*}) \lesssim  N! n^N    \sqrt{\epsilon_{\mathrm{XOR}^{*} \wedge \cdots \wedge \mathrm{XOR}^{*}}}   \times   
\left\{\!\begin{array}{ll@{}>{{}}l} n^{ \frac{\# \text{ of players}}{2}  +5} \Longleftrightarrow \big( \# \text{ of players} \big) \mathrm{mod} 2 \equiv 0 \\ n^{ \lfloor\frac{\# \text{ of players}}{2} \rfloor  +5} \Longleftrightarrow \big( \# \text{ of players} \big) \mathrm{mod} 2 \neq 0
\end{array}\right.   
    \text{, }
\]

\noindent given $\epsilon_{\mathrm{XOR}^{*} \wedge \cdots \wedge \mathrm{XOR}^{*}}$ sufficiently small, where,

\begin{align*}
    \bigg| \bigg|  \bigg[  \bigg(   \big( \big( A_1 \big)^{(1)}_i \wedge \cdots \wedge  \big( A_n \big)^{(1)}_{i^{\prime}} \big) \bigotimes \textbf{I}      \bigotimes \bigg(    \bigg(   \underset{i_1 \in \mathcal{Q}_1, i_2 \in \mathcal{Q}_2 ,  i_3 \in \mathcal{Q}_3}{\prod}             C^{l_{ijk}}_{ijk}     \bigg) \wedge \cdots \\ \wedge \bigg(    \underset{i^{\prime\cdots\prime}_3 \in \mathcal{Q}^{\prime\cdots\prime}_3}{\underset{i^{\prime\cdots\prime}_1 \in\mathcal{Q}^{\prime\cdots\prime}_1}{\underset{ i^{\prime\cdots\prime}_2 \in \mathcal{Q}^{\prime\cdots\prime}_2 }{\prod}}}               C^{l^{\prime\cdots\prime}_{i^{\prime\cdots\prime}j^{\prime\cdots\prime}k^{\prime\cdots\prime}}}_{i^{\prime\cdots\prime}j^{\prime\cdots\prime}k^{\prime\cdots\prime}}            \bigg)        \bigg)  \bigotimes                \bigg(     \underset{1\leq k \leq n-3}{\bigotimes}      \textbf{I}_k       \bigg)      \bigg)       -    \big( \omega \big( {\mathrm{XOR}^{*}}  \big) \big)^n  \\ \times   \bigg(       \pm \mathrm{sign} \big( i_1 , j_1 , k_1, i_{111}, \cdots , j_{111} , \cdots , j_{nm(n+m)} , k_{111}, \cdots, k_{nm(n+m)}   \big)   \mathrm{sign} \big(  i^{\prime}_1 , j^{\prime}_1 , k^{\prime}_1, i^{\prime}_{111}\\ , \cdots  , j^{\prime}_{111} , \cdots , j^{\prime}_{nm(n+m)}      k^{\prime}_{111}  , \cdots, k^{\prime}_{nm(n+m)}      \big)    \times \cdots \times \mathrm{sign} \big(  i^{\prime\cdots\prime}_1 , j^{\prime\cdots\prime}_1 , k^{\prime\cdots\prime}_1, i^{\prime\cdots\prime}_{111} \\,  \cdots , j^{\prime\cdots\prime}_{111} , \cdots , j^{\prime\cdots\prime}_{nm(n+m)}   , k^{\prime\cdots\prime}_{111}, \cdots  , k^{\prime\cdots\prime}_{nm(n+m)}         \big) \bigg) \bigg[                            \textbf{I} \bigotimes \textbf{I} \bigotimes \bigg(    \bigg(   \underset{i_1 \in \mathcal{Q}_1, i_2 \in \mathcal{Q}_2 , i_3 \in \mathcal{Q}_3}{\prod}             C^{l_{ijk}}_{ijk}     \bigg) \wedge \cdots  \\   \wedge \bigg(   \underset{i^{\prime\cdots\prime}_3 \in \mathcal{Q}^{\prime\cdots\prime}_3}{\underset{i^{\prime\cdots\prime}_1 \in\mathcal{Q}^{\prime\cdots\prime}_1}{\underset{ i^{\prime\cdots\prime}_2 \in \mathcal{Q}^{\prime\cdots\prime}_2 }{\prod}}}             C^{l^{\prime\cdots\prime}_{i^{\prime\cdots\prime}j^{\prime\cdots\prime}k^{\prime\cdots\prime}}}_{i^{\prime\cdots\prime}j^{\prime\cdots\prime}k^{\prime\cdots\prime}}            \bigg)        \bigg)  \bigotimes \bigg( \bigg(   \underset{1\leq k \leq n-3}{\prod}  \textbf{I}_k \bigg)      \bigg)             \bigg] \bigg) \bigg]        \ket{\psi_{\mathrm{XOR}^{*} \wedge \cdots \wedge \mathrm{XOR}^{*}}}   \bigg| \bigg|_F    \tag{$\mathrm{XOR}^{*}$}        \text{. }
\end{align*}

\noindent \textit{Proof of Lemma FR $\mathrm{XOR}^{*}$}. Directly apply the argument from $\textbf{Lemma}$ \textit{FR}, from which we conclude the argument. \boxed{}

\bigskip

\noindent \textbf{Lemma} \textit{Gen-FFL-Bound $XOR^{*}$} 
 (\textit{generalizations of the second FFL error bound}, \textit{6.6}, {[37]}, \textbf{Lemma} \textit{7}, {[44]}). One can denote quantum states corresponding to optimal strategies for the dual XOR game, from the optimal strategies,

\begin{align*}
 \ket{\psi_{3\mathrm{XOR}}} \equiv \underset{\text{Players}}{\bigcup} \underset{\mathcal{S}}{\mathrm{sup}} \big\{  \text{A player's quantum strategy } \mathcal{S} \text{ for a } 3- \text{XOR game}    \big\}   \text{, } \\ \ket{\psi_{4\mathrm{XOR}}} \equiv  \underset{\text{Players}}{\bigcup} \underset{\mathcal{S}}{\mathrm{sup}} \big\{  \text{A player's quantum strategy } \mathcal{S} \text{ for a } 4- \text{XOR game}    \big\}  \text{, } \\ \ket{\psi_{5\mathrm{XOR}}} \equiv  \underset{\text{Players}}{\bigcup} \underset{\mathcal{S}}{\mathrm{sup}} \big\{  \text{A player's quantum strategy } \mathcal{S} \text{ for a } 5- \text{XOR game}    \big\}  \text{, } \\  \ket{\psi_{N\mathrm{XOR}}} \equiv  \underset{\text{Players}}{\bigcup} \underset{\mathcal{S}}{\mathrm{sup}} \big\{  \text{A player's quantum strategy } \mathcal{S} \text{ for an } N- \text{XOR game}    \big\} \text{, }
\end{align*}

\noindent respectively, for the $3$-$\mathrm{XOR}^{*}$, $4$-$\mathrm{XOR}^{*}$, $5$-$\mathrm{XOR}^{*}$, and $N$-$\mathrm{XOR}^{*}$, games. Given error bounds formulated in previous sections for each $\mathrm{XOR}$ game, one obtains error bound inequalities of the form,

\begin{align*}
     \bigg| \bigg| \bigg[ \bigg( \bigg( \underset{1 \leq i \leq n}{\prod}  A^{j_i}_i \bigg) \otimes B_{kl} \otimes \textbf{I} \bigg)  - \omega_{3 XOR^{*}} \bigg( \pm \mathrm{sign} \big( i_1 , j_1 , \cdots , j_n \big)   \\ \times    \bigg[       \bigg( \bigg( \underset{1 \leq i \leq n}{\prod}   A^{j_i}_k \bigg)   + \bigg(                     \underset{\text{set } j+1 \equiv j \oplus 1}{\underset{i \in \mathcal{Q}_1, j \in \mathcal{Q}_2}{\prod}}   A^{j_i}_k        \bigg) \bigg)  \otimes \textbf{I} \otimes \textbf{I} \bigg] \bigg) \bigg]  \ket{\psi_{3XOR^{*}}} \bigg| \bigg|^2 \\ \lesssim  3! n^3 \sqrt{\epsilon }  \text{, } \end{align*}

     \begin{align*} \bigg| \bigg| \bigg[ \bigg( \bigg( \underset{1 \leq i \leq n}{\prod}  A^{j_i}_i \bigg) \otimes B_{kl} \otimes \textbf{I} \otimes \textbf{I} \bigg)  - \omega_{4 XOR^{*}} \bigg( \pm \mathrm{sign} \big( i_1 , j_1 , \cdots , j_n \big)   \\ \times    \bigg[       \bigg( \bigg( \underset{1 \leq i \leq n}{\prod}   A^{j_i}_k \bigg)  + \bigg(                     \underset{\text{set } j+1 \equiv j \oplus 1}{\underset{i \in \mathcal{Q}_1, j \in \mathcal{Q}_2}{\prod}}   A^{j_i}_k        \bigg) \bigg) \otimes \textbf{I}    \otimes \textbf{I} \otimes \textbf{I} \bigg] \bigg) \bigg] \ket{\psi_{4 XOR^{*}}} \bigg| \bigg|^2 \\ \lesssim  4! n^4 \sqrt{\epsilon }      \text{, } \\ \\    \bigg| \bigg| \bigg[ \bigg( \bigg( \underset{1 \leq i \leq n}{\prod}  A^{j_i}_i \bigg) \otimes B_{kl} \otimes \textbf{I} \otimes \textbf{I} \otimes \textbf{I} \bigg)  - \omega_{5XOR^{*}} \bigg( \pm \mathrm{sign} \big( i_1 , j_1 , \cdots , j_n \big)     \\ \times  \bigg[       \bigg( \bigg( \underset{1 \leq i \leq n}{\prod}   A^{j_i}_k \bigg)    + \bigg(                     \underset{\text{set } j+1 \equiv j \oplus 1}{\underset{i \in \mathcal{Q}_1, j \in \mathcal{Q}_2}{\prod}}   A^{j_i}_k        \bigg) \bigg) \otimes \textbf{I} \otimes \textbf{I}  \otimes \textbf{I} \bigg] \bigg) \bigg] \ket{\psi_{5 XOR^{*}}} \bigg| \bigg|^2  \\ \lesssim  5! n^5 \sqrt{\epsilon }    \text{, }    \end{align*}

  \begin{align*} \bigg| \bigg| \bigg[ \bigg( \bigg( \underset{1 \leq i \leq n}{\prod}  A^{j_i}_i \bigg) \bigotimes B_{kl}  \bigotimes \bigg( \underset{1 \leq k \leq N-2}{\bigotimes} \textbf{I}_k  \bigg)  \bigg)  - \omega_{NXOR^{*}} \bigg( \pm \mathrm{sign} \big( i_1 , j_1 , \cdots , j_n \big)      \\ \times \bigg[       \bigg( \bigg( \underset{1 \leq i \leq n}{\prod}   A^{j_i}_k \bigg)   + \bigg(                     \underset{\text{set } j+1 \equiv j \oplus 1}{\underset{i \in \mathcal{Q}_1, j \in \mathcal{Q}_2}{\prod}}   A^{j_i}_k        \bigg) \bigg)  \bigotimes \bigg( \underset{1 \leq k \leq N-1}{\bigotimes} \textbf{I}_k  \bigg) \bigg] \bigg) \bigg] \ket{\psi_{NXOR^{*}}} \bigg| \bigg|^2\end{align*}  \[ \lesssim  N! n^N  \sqrt{\epsilon}   \times   
\left\{\!\begin{array}{ll@{}>{{}}l} n^{ \frac{\# \text{ of players}}{2}  +5} \Longleftrightarrow \big( \# \text{ of players} \big) \mathrm{mod} 2 \equiv 0 \\ n^{ \lfloor\frac{\# \text{ of players}}{2} \rfloor  +5} \Longleftrightarrow \big( \# \text{ of players} \big) \mathrm{mod} 2 \neq 0
\end{array}\right.   \text{. }
\]

\bigskip

\noindent \noindent \textit{Proof of Lemma Gen-FFL-Bound $XOR^{*}$}. Directly apply the argument from \textbf{Lemma} \textit{Gen-FFL-Bound}, from which we conclude the argument. \boxed{}

\subsubsection{Suitable linear operators for multiplayer $XOR^{*}$ games have unit Frobenius norm}

\noindent \textbf{Lemma} $\textit{9}^{*}$ (\textit{the Frobenius norm of suitable linear operators for the 3-XOR, 4-XOR, 5-XOR, and N-XOR games equals 1}). With respect to the Frobenius norm, the norm of suitable linear operators introduced in previous sections for the 3-XOR, 4-XOR, 5-XOR, and N-XOR, games equals $1$.

\bigskip

\noindent \textit{Proof of Lemma $\textit{9}^{*}$}. Directly apply the argument from \textit{6.2} in {[37]}, from which we conclude the argument. \boxed{}

\subsubsection{Suitable linear operators for strong parallel repetition of the $XOR^{*}$ game have unit Frobenius norm}

\noindent \textbf{Lemma} $\textit{10}^{*}$ (\textit{the Frobenius norm of suitable linear operators for strong parallel repetition of the multiplayer XOR game, and of the two player FFL game, equal 1}). With respect to the Frobenius norm, the norm of suitable linear operators introduced in previous sections for the 3-XOR, 4-XOR, 5-XOR, and N-XOR, games equals $1$.

\bigskip

\noindent \textit{Proof of Lemma $\textit{10}^{*}$}. Directly apply the argument from \textit{6.2} in {[37]}, from which we conclude the argument. \boxed{}

\subsection{Exact, and approximate, optimality of novel settings for parallel repetition of the compiled $\mathrm{XOR}$ game from those developed for the ordinary $\mathrm{XOR}$ game}

\subsubsection{Main Result}

\noindent As in the previous subsection of the Appendix, we provide a statement of the Main Result that holds for the compiled, dual, XOR game, which takes upon the same structure as that provided in the $\textbf{Theorem}$ \textit{1}.

\noindent \textbf{Theorem} \textit{6} (\textit{primal feasible solutions and duality gaps for strong parallel repetition of FFL games}). The same collection of items provided in $\textbf{Theorem}$ \textit{1} for the $3$-$\mathrm{XOR}$ game also hold for strong parallel repetition of $\mathrm{FFL}$ games, given the existence of primal feasible solutions, duality gap, and dual semidefinite program. We state the result below:

\bigskip

\noindent \textbf{Theorem} \textit{8} (\textit{primal feasible solutions and duality gaps for compiled, dual, XOR games}). The same collection of items provided in $\textbf{Theorem}$ \textit{1} for the $3$-$\mathrm{XOR}$ game also hold for compiled $\mathrm{XOR}^{*}$ games, given the existence of primal feasible solutions, duality gap, and dual semidefinite program.

\subsubsection{Compiled optimal values}

\noindent Besides a direct application of the previous results to $\mathrm{XOR}^{*}$ games from several $\mathrm{XOR}$ settings, parallel repetition of \textit{compiled} $\mathrm{XOR}$ games are also of interest to explore, especially from the fact that parallel repetition can have other applications for formulating the quantum bias, and optimal value, for $\mathrm{XOR}$ game. More specifically, classes of inequalities provided in the previous subsection for the $\mathrm{XOR}^{*}$ game, given the fact that $\mathrm{XOR}$ and $\mathrm{XOR}^{*}$ games can be related to one another through an appropriate duality notion, can also yield the following set of inequalities included in the last subsection of the Appendix for parallel repetition of compiled $\mathrm{XOR}^{*}$ games. Below, we exhibit how the quantum value of parallel repetition of $\mathrm{XOR}$ games is formulated, which can then be straightforwardly related to the quantum bias of a direct sum of compiled $\mathrm{XOR}$ games. Straightforwardly, the expressions for the quantum value and quantum bias for the compiled $\mathrm{XOR}$ game can be formulated for the compiled $\mathrm{XOR}^{*}$ game. Finally, performing the parallel repetition operation, whether ordinary or strong, can be analyzed given the notions developed earlier in this paper, beginning from the error bounds. In comparison to the ordinary $\mathrm{XOR}$ game, parallel repetition of the compiled $\mathrm{XOR}$ game has the quantum value, {[13]},

\begin{align*}
  \text{Quantum value of Parallel Repetition of the Compiled $XOR^{*}$ game} \equiv    \omega_{C\mathrm{XOR}^{*},q} \bigg( \mathcal{S} , \bigg( \underset{i}{\bigwedge} G^{*}_i \bigg)_C \bigg)  \text{, }
\end{align*}

\noindent where, over the collection of all possible $\mathrm{XOR}^{*}$ strategies $\mathcal{S}$, the parallel repetition of compiled $\mathrm{XOR}^{*}$ games,

\begin{align*}
 \bigg( \underset{i}{\bigwedge} G^{*}_i \bigg)_C    \text{, }
\end{align*}

\noindent can be conveniently related to the bias of the game, equaling,

\begin{align*}
   \text{Compiled $XOR^{*}$ bias} \equiv   \beta_{C \mathrm{XOR}^{*}} \bigg( \underset{i}{\bigwedge} G^{*}_i \bigg)          \equiv      2^{-n}    \underset{M \subseteq [ n ]}{\sum}         \beta_{C\mathrm{XOR}^{*} ,q} \bigg( \mathcal{S}_M ,  \bigg( \underset{i \in M }{\bigoplus}   G^{*}_i \bigg)_{C}    \bigg)                     \text{. }
\end{align*}

\noindent The quantum bias defined above is dependent upon the direct sum of compiled $\mathrm{XOR}^{*}$ games,

\begin{align*}
  \bigg( \underset{i \in M }{\bigoplus}   G^{*}_i \bigg)_{C}   \text{. }
\end{align*}

\noindent As discussed in the previous subsection, to demonstrate that the $\mathrm{XOR}^{*}$ game, under parallel repetition and compilation, applies directly to the series of results (namely, \textbf{Lemma} \textit{1 -N $\mathrm{XOR}^{*}$}, \textbf{Lemma} \textit{$1- 3 \mathrm{XOR}^{*}$}, \textbf{Lemma} \textit{$2^{**} \mathrm{XOR}^{*}$}, \textbf{Lemma} \textit{3}, \textbf{Lemma} \textit{4$A^{*}$}, \textbf{Lemma} \textit{4$A^{**}$}, \textbf{Lemma} \textit{$5^{*}$}, \textbf{Lemma} \textit{$5^{**}$}, \textbf{Lemma} \textit{$4^{**}$}, \textbf{Lemma} \textit{FR $\mathrm{XOR}^{*}$}, \textbf{Lemma} \textit{Gen-FFL-Bound $\mathrm{XOR}^{*}$}), introduce the following objects,

\begin{align*}
    \mathscr{T}^{*}_C \equiv \text{Linear operator for Compiled $\mathrm{XOR^{*}}$ game} \text{, }  \\ \\  c^{*}_{i,C} \equiv \text{Constants for compiled upper bounds in } \textbf{Lemma} \textit{ 1-3-$\mathrm{ XOR}^{*}$} \text{, } \\ \\ A_{i,C} \equiv A_i  \equiv \text{Tensor observable for the first player of the compiled $\mathrm{XOR}^{*}$ game} \text{, } \\  \\       \mathscr{T}^{*}_C \wedge \mathscr{T}^{*}_C  \equiv \text{Two strong parallel repetition operations of the linear} \\ \text{ operator for compiled $\mathrm{XOR^{*}}$ game} \text{, } \\ \\  \mathscr{T}^{*}_C \wedge \mathscr{T}^{*}_C  \equiv \text{Two strong parallel repetition operations of the linear} \\ \text{ operator for compiled $\mathrm{XOR^{*}}$ game} \text{, }  \\ \\ \epsilon^{*}_C \equiv \text{Constant in compiled $\mathrm{XOR}^{*}$ upper bound} \text{,} \\   \\   \epsilon_{N\mathrm{XOR}^{*}}  \equiv \text{Constant in compiled N $\mathrm{XOR}^{*}$ upper bound} \text{,} \\ \\ \epsilon_{N\mathrm{XOR}^{*}\wedge N\mathrm{XOR}^{*}} \equiv \text{Constant in compiled N $\mathrm{XOR}^{*} \wedge $ N $\mathrm{XOR}^{*}$ upper bound} \text{, } \\ \\   \epsilon_{N\mathrm{XOR}^{*}\wedge\cdots \wedge N\mathrm{XOR}^{*}} \equiv    \text{Constant in compiled N $\mathrm{XOR}^{*} \wedge \cdots \wedge$ N $\mathrm{XOR}^{*}$ upper bound}      \text{, }               
\end{align*}

\noindent The following series of results can be immediately established with identical arguments as provided in the dual $\mathrm{XOR}$ case:

\bigskip

\noindent \textbf{Lemma} \textit{1 -compiled N $\mathrm{XOR}^{*}$} (\textit{computation of the Frobenius norm for the anticommutation rule of $T_{N\mathrm{XOR}^{*}}$ yields a desired up to constants $\sqrt{\epsilon}$ upper bound}). One has that,

\begin{align*}
      \underline{\text{Player $1$}:} \text{ } \bigg| \bigg|   \bigg(  A_i \bigotimes \bigg( \underset{1 \leq k \leq n-1}{\bigotimes} \textbf{I}_k  \bigg)        \bigg) \mathscr{T}^{*}_C -   \mathscr{T}^{*}_C \bigg( \bigg( \underset{1 \leq k \leq n-1}{\bigotimes} \textbf{I}_k  \bigg)      \bigotimes \widetilde{A_i}      \bigg)      \bigg| \bigg|_F \\  < c^{*}_{1,C} n^N \sqrt{\epsilon^{*}_C}  \text{,} \\     \vdots \\              \underline{\text{Player $N$}:} \text{ } \bigg| \bigg|   \bigg(  \bigg( \underset{1 \leq k \leq n-1}{\bigotimes} \textbf{I}_k  \bigg)   \bigotimes  A^{(n-1)}_{i_1, \cdots, i_{n-1}}     \bigg) \mathscr{T}^{*}_C  -   \mathscr{T}^{*}_C \bigg( \bigg(    \widetilde{A^{(n-1)}_{i_1, \cdots, i_{n-1}} }       \\ \bigotimes  \bigg( \underset{1 \leq k \leq n-1}{\bigotimes} \textbf{I}_k  \bigg)     \bigg)      \bigg| \bigg|_F  < c^{*}_{N,C} n^N \sqrt{\epsilon^{*}_C}  \text{, }
\end{align*}

\noindent has the upper bound,

\begin{align*}
      \mathscr{C}^{*}_C \equiv    \underset{1 \leq i \leq N}{\bigcup}  \big\{     C_{i,C}  \neq c^{*}_{i,C} \in \textbf{R}  :            C_{i,C} \equiv c^{*}_{i,C} \sqrt{\epsilon_C}     \big\}     \propto n^N_C \sqrt{\epsilon^{*}_C} \text{. }
\end{align*}

\bigskip

\noindent \textit{Proof of Lemma 1 -compiled N-$\mathrm{XOR}^{*}$}. Directly apply the argument for $\textbf{Lemma}$ \textit{1-N $\mathrm{XOR}$}, from which we conclude the argument. \boxed{}

 \bigskip

\noindent \textbf{Lemma} \textit{$1- compiled \text{ } 3-\mathrm{XOR}^{*}$} (\textit{computation of the Frobenius norm for the anticommutation rule of $T_{3\mathrm{XOR}^{*}}$ yields a desired up to constants $\sqrt{\epsilon}$ upper bound}). One has that,

\begin{align*}
      \underline{\text{Player $1$}:} \text{ } \bigg| \bigg|   \bigg(  A_i \bigotimes \bigg( \underset{1 \leq k \leq n-1}{\bigotimes} \textbf{I}_k  \bigg)        \bigg) \mathscr{T}^{*}_C  -   \mathscr{T}^{*}_C \bigg( \bigg( \underset{1 \leq k \leq n-1}{\bigotimes} \textbf{I}_k  \bigg)      \bigotimes \widetilde{A_i}      \bigg)      \bigg| \bigg|_F \\  < c^{*}_{1,C} n^3 \sqrt{\epsilon^{*}_C}  \text{,} \\   \underline{\text{Player $2$}:} \text{ } \bigg| \bigg|   \bigg(  \bigg( \underset{1 \leq k \leq n-1}{\bigotimes} \textbf{I}_k  \bigg)   \bigotimes  A^{(n-1)}_{i_1, \cdots, i_{n-1}}     \bigg) \mathscr{T}^{*}_C  -   \mathscr{T}^{*}_C \bigg( \bigg(    \widetilde{A^{(n-1)}_{i_1, \cdots, i_{n-1}} }       \\ \bigotimes  \bigg( \underset{1 \leq k \leq n-1}{\bigotimes} \textbf{I}_k  \bigg)     \bigg)      \bigg| \bigg|_F      < c^{*}_{2,C} n^3 \sqrt{\epsilon^{*}_C} \\              \underline{\text{Player $3$}:} \text{ } \bigg| \bigg|   \bigg(  \bigg( \underset{1 \leq k \leq n-1}{\bigotimes} \textbf{I}_k  \bigg)   \bigotimes  A^{(n-1)}_{i_1, \cdots, i_{n-1}}     \bigg) \mathscr{T}^{*}_C  -   \mathscr{T}^{*}_C \bigg( \bigg(    \widetilde{A^{(n-1)}_{i_1, \cdots, i_{n-1}} }       \\ \bigotimes  \bigg( \underset{1 \leq k \leq n-1}{\bigotimes} \textbf{I}_k  \bigg)     \bigg)      \bigg| \bigg|_F      < c^{*}_{3,C} n^3 \sqrt{\epsilon^{*}_C}  \text{, }
\end{align*}

\noindent has the upper bound,

\begin{align*}
      \mathscr{C}^{*}_C \equiv    \underset{1 \leq i \leq 3}{\bigcup}  \big\{     C_{i,C}  \neq c^{*}_{i,C} \in \textbf{R}  :            C_{i,C} \equiv c^{*}_i \sqrt{\epsilon_C}     \big\}     \propto n^N \sqrt{\epsilon^{*}_C} \text{. }
\end{align*}

\bigskip

\noindent \textit{Proof of Lemma $1$- compiled 3 $\mathrm{XOR}^{*}$}. Directly apply the argument for $\textbf{Lemma}$ \textit{1-3 $\mathrm{XOR}$}, from which we conclude the argument. \boxed{}

\bigskip

\noindent \textbf{Lemma} \textit{$2^{*}$ compiled $\mathrm{XOR}$} (\textit{3 3-$\mathrm{XOR^{*}}$ identities from two FFL identities}). One has that,

\begin{align*}
  \bigg( A_i \otimes \textbf{I} \otimes \textbf{I}   \bigg) \mathscr{T}^{*}_C - \mathscr{T}^{*}_C  \bigg( \widetilde{A_i} \otimes \textbf{I} \otimes \textbf{I}   \bigg)    \text{, }
\end{align*}

\noindent and that,

\begin{align*}
     \bigg(  \big(  A_i \mathscr{T}^{*}_C \otimes \mathscr{T}^{*}_C \big) - \big( \mathscr{T}^{*}_C \widetilde{A_i} \otimes \mathscr{T}^{*}_C \big) \bigg) \otimes \bigg(  \textbf{I} \otimes \textbf{I} \bigg)                        \text{, }
\end{align*}

\noindent are equal, in which,

\begin{align*}
    \bigg( A_i \otimes \textbf{I} \otimes \textbf{I} \bigg) \mathscr{T}^{*}_C - \mathscr{T}^{*}_C  \bigg( \widetilde{A_i} \otimes \textbf{I} \otimes \textbf{I} \bigg)  \equiv      \bigg( \big( A_i  \mathscr{T}^{*}_C - \mathscr{T}^{*}_C \widetilde{A_i} \big) \otimes \mathscr{T}^{*}_C \bigg) \otimes \bigg( \textbf{I} \otimes \textbf{I}  \bigg)    \text{. }
\end{align*}

\noindent For the remaining two players in the compiled $\mathrm{XOR}$ game, identities of an analogous form hold.

\bigskip

\noindent \textit{Proof of Lemma $2^{*}$ compiled $\mathrm{XOR}^{*}$}. Directly apply the argument for $\textbf{Lemma}$ \textit{$2^{*}$}, from which we conclude the argument. \boxed{}

\bigskip

\noindent \textbf{Lemma} \textit{$2^{**} compiled\text{ }  \mathrm{XOR}^{*}$} (\textit{N compiled N-$\mathrm{XOR}^{*}$ identities}). One has that,

\begin{align*}
  \bigg( A_i \bigotimes  \bigg( \underset{1 \leq z \leq N-1}{\bigotimes} \textbf{I}_z \bigg)  \bigg)  \mathscr{T}^{*}_C - \mathscr{T}^{*}_C  \bigg( A_i \bigotimes  \bigg( \underset{1 \leq z \leq N-1}{\bigotimes} \textbf{I}_z \bigg)  \bigg)   \text{, }
\end{align*}

\noindent and that,

\begin{align*}
     \bigg(  \bigg(  A_i \mathscr{T}^{*}_C \bigotimes  \bigg( \underset{1 \leq z \leq N-1}{\bigotimes} \mathscr{T}^{*}_C \textbf{I}_z  \bigg) - \bigg( \mathscr{T}^{*}_C \widetilde{A_i} \bigotimes \bigg( \underset{1 \leq z \leq N-1}{\bigotimes}  \mathscr{T}^{*}_C  \textbf{I}_z \bigg)  \bigg) \bigg) \bigotimes \bigg(  \underset{1 \leq z \leq N}{\bigotimes} \textbf{I}_z  \bigg)                        \text{, }
\end{align*}

\noindent are equal, in which,

\begin{align*}
    \bigg( A_i \bigotimes  \bigg( \underset{1 \leq z \leq N-1}{\bigotimes} \textbf{I}_z  \bigg) \bigg) \mathscr{T}^{*}_C - \mathscr{T}^{*}_C \bigg(  \widetilde{A_i} \bigotimes  \bigg( \underset{1 \leq z \leq N-1}{\bigotimes} \textbf{I}_z  \bigg) \bigg)   \equiv      \bigg( \bigg( A_i  \mathscr{T}^{*}_C - \mathscr{T}^{*} \widetilde{A_i} \bigg) \\ \bigotimes \bigg( \underset{1 \leq z \leq n-1}{\bigotimes} \mathscr{T}^{*}_C \textbf{I}_z  \bigg)  \bigg)  \bigotimes  \bigg(  \underset{1 \leq z \leq N-1}{\bigotimes} \textbf{I}_z   \bigg)    \text{. }
\end{align*}

\bigskip

\noindent \textbf{Lemma} \textit{$2^{***}$ compiled $ \mathrm{XOR}^{*}$} (\textit{N compiled N-$\mathrm{XOR}^{*}$ identities under strong parallel repetition}). One has that,

\begin{align*}
  \bigg( A_i \bigotimes  \bigg( \underset{1 \leq z \leq N-1}{\bigotimes} \textbf{I}_z \bigg)  \bigg) \bigg( \mathscr{T}^{*}_C \wedge \cdots \wedge \mathscr{T}^{*}_C \bigg)  - \bigg( \mathscr{T}^{*}_C \wedge \cdots \wedge \mathscr{T}^{*}_C \bigg)    \bigg( A_i \bigotimes  \bigg( \underset{1 \leq z \leq N-1}{\bigotimes} \textbf{I}_z \bigg)  \bigg)   \text{, }
\end{align*}

\noindent and that,

\begin{align*}
     \bigg(  \bigg(  A_i \bigg( \mathscr{T}^{*}_C \wedge \cdots \wedge \mathscr{T}^{*}_C \bigg)   \bigotimes  \bigg( \underset{1 \leq z \leq N-1}{\bigotimes} \bigg( \mathscr{T}^{*}_C \wedge \cdots \wedge \mathscr{T}^{*}_C \bigg)  \textbf{I}_z  \bigg) - \bigg( \bigg( \mathscr{T}^{*}_C \wedge \cdots \wedge \mathscr{T}^{*}_C \bigg)  \widetilde{A_i} \\ \bigotimes  \bigg( \underset{1 \leq z \leq N-1}{\bigotimes}  \bigg( \mathscr{T}^{*}_C \wedge \cdots \wedge \mathscr{T}^{*}_C \bigg)   \textbf{I}_z \bigg)  \bigg) \bigg) \bigotimes \bigg(  \underset{1 \leq z \leq N}{\bigotimes} \textbf{I}_z  \bigg)                        \text{, }
\end{align*}

\noindent are equal, in which,

\begin{align*}
    \bigg( A_i \bigotimes  \bigg( \underset{1 \leq z \leq N-1}{\bigotimes} \textbf{I}_z  \bigg) \bigg) \bigg( \mathscr{T}^{*}_C \wedge \cdots \wedge \mathscr{T}^{*}_C \bigg)  - \bigg( \mathscr{T}^{*}_C \wedge \cdots \wedge \mathscr{T}^{*}_C \bigg)   \bigg(  \widetilde{A_i} \\ \bigotimes  \bigg( \underset{1 \leq z \leq N-1}{\bigotimes} \textbf{I}_z  \bigg) \bigg) \\   \equiv      \bigg( \bigg( A_i  \bigg( \mathscr{T}^{*}_C \wedge \cdots \wedge \mathscr{T}^{*}_C \bigg)   - \bigg( \mathscr{T}^{*}_C \wedge \cdots \wedge \mathscr{T}^{*}_C  \bigg)  \widetilde{A_i} \bigg) \\ \bigotimes\bigg( \underset{1 \leq z \leq n-1}{\bigotimes}\bigg( \mathscr{T}^{*}_C \wedge \cdots  \wedge \mathscr{T}^{*}_C \bigg)  \textbf{I}_z  \bigg)  \bigg)  \bigotimes   \bigg(  \underset{1 \leq z \leq N-1}{\bigotimes} \textbf{I}_z   \bigg)    \text{. }
\end{align*}

\noindent \textit{Proof of Lemma $2^{**}$ compiled $\mathrm{XOR}^{*}$, and $2^{***}$ compiled $\mathrm{XOR}^{*}$}. Directly apply the arguments for $\textbf{Lemma}$ \textit{$2^{**}$}, and for $\textbf{Lemma}$ \textit{$2^{***}$}, from which we conclude the argument. \boxed{}

\bigskip

\noindent \noindent \textbf{Lemma} \textit{3} (\textit{intialization of $\epsilon$-optimality of the compiled N-$\mathrm{XOR}^{*}$ game from the observable of the first player}). For an $\epsilon$-optimal strategy, and player observable tensors $A_{i_1}, A^1_{i_1,i_2}, \cdots, A^{n-1}_{i_1,i_2, \cdots, i_n}$, and $\ket{\psi_{N\mathrm{XOR}^{*}}}$, and $C_1>0$,

\begin{align*}
    \underset{i,j}{\sum}     \bigg| \bigg|     \bigg[            \bigg( \bigg(    \frac{A_i A_j + A_j A_i}{2}        \bigg) \bigotimes  \bigg( \underset{1\leq k \leq N-1}{\bigotimes} \textbf{I}_k \bigg)  \bigg)      \bigg] \ket{\psi_{\text{compiled }N\mathrm{XOR}^{*}}}                           \bigg| \bigg|^2         <  C_1 n \bigg( \underset{1 \leq j \leq N-1}{\prod}  \big( n - j  \big) \bigg) \\ \times  \epsilon_{\text{compiled }N\mathrm{XOR}^{*}}   \text{. }
\end{align*}

\bigskip

\noindent \noindent \textbf{Lemma} $\textit{4A}^{*}$ (\textit{induction on the, up to constants, strong parallel repetition $\epsilon$ upper bound from the previous result}). Under the assumptions of $\textbf{Lemma}$ \textit{4A}, and $\textbf{Lemma}$ \textit{3}, and $C^{\wedge}_2>0$,

\begin{align*}
      \underset{\text{Tensor entries }}{\sum}  \bigg[ \underset{j^{\prime}_1,\cdots,j^{\prime}_n}{\underset{i^{\prime},\cdots,i^{\prime}_n}{\underset{i_1,\cdots,i_n}{\sum}}}       \bigg| \bigg|     \bigg[            \bigg( \bigg(    \frac{ \big( A_i \wedge A_{i^{\prime}} \big)  \big( A_j \wedge A_{j^{\prime}} \big)  + \big( A_j \wedge A_{j^{\prime} } \big) \big(  A_i \wedge A_{i^{\prime}} \big)}{2}        \bigg) \bigotimes  \textbf{I}  \bigg)      \bigg] \\ \times  \ket{\psi_{\text{compiled }2\mathrm{XOR}^{*}}}                           \bigg| \bigg|^2   \bigg]    \\    <  C^{\wedge}_2 n  \bigg( \underset{1 \leq j \leq 2}{\prod}  \big( n - j  \big) \bigg) \epsilon_{\text{compiled }2\mathrm{XOR}^{*}\wedge \text{compiled } 2 \mathrm{XOR}^{*}} \\  \equiv    C^{\wedge}_2    n \bigg( \underset{1 \leq j \leq 2}{\prod}  \big( n - j  \big) \bigg)  \epsilon_{\text{compiled }\mathrm{XOR}^{*}\wedge  \text{compiled }\mathrm{XOR}^{*}}   \text{. }
\end{align*}

\noindent \textit{Proof of Lemma $4A^{*}$}. Directly apply the argument from \textbf{Theorem} $\textit{1}^{*}$, particularly for demonstrating that the desired operator associated with the semidefinite program is positive definite, from which we conclude the argument. \boxed{}

\bigskip

\noindent \noindent \textbf{Lemma} $\textit{4A}^{**}$ (\textit{induction on the, up to constants, $N$-player strong parallel repetition $\epsilon$ upper bound from the previous result}). Under the assumptions of $\textbf{Lemma}$ \textit{4A}, $\textbf{Lemma}$ $\textit{4A}^{*}$, and $\textbf{Lemma}$ \textit{3}, and $C^{\wedge}_N>0$,

\begin{align*}
      \underset{\text{Tensor entries }}{\sum}  \bigg[ \underset{i^{\prime\cdots\prime},\cdots,i^{\prime\cdots\prime}}{\underset{\vdots}{\underset{i_1,\cdots,i_n}{\sum}}}     \bigg| \bigg|     \bigg[            \bigg( \bigg(    \frac{ \big( A_i \wedge A_{i^{\prime}}  \wedge \cdots \wedge A_{i^{\prime\cdots\prime}} \big)  \big( A_j \wedge A_{j^{\prime}} \wedge \cdots \wedge A_{j^{\prime\cdots\prime}} \big)  +\big( A_j \wedge A_{j^{\prime}}  \wedge \cdots \wedge A_{j^{\prime\cdots\prime}} \big)   }{2} \end{align*}

      \begin{align*} \times  \frac{\cdots \big( A_i \wedge A_{i^{\prime}}  \wedge \cdots  \wedge A_{i^{\prime\cdots\prime}} \big)}{\cdots }      \bigg) \bigotimes  \textbf{I}  \bigg)      \bigg] \ket{\psi_{\text{compiled }\mathrm{XOR}^{*}\wedge \cdots \wedge \text{compiled }\mathrm{XOR}^{*}}}                           \bigg| \bigg|^2   \bigg]    \\   <  C^{\wedge}_N n   \bigg( \underset{1 \leq j \leq N}{\prod}  \big( n - j  \big) \bigg) \epsilon_{\text{compiled } N\mathrm{XOR}^{*}\wedge \cdots \wedge  \text{compiled } N\mathrm{XOR}^{*}} \\ \equiv    C^{\wedge}_N n \bigg( \underset{1 \leq j \leq N}{\prod}  \big( n - j  \big) \bigg) \epsilon_{\text{compiled }\mathrm{XOR}^{*}\wedge  \cdots \wedge  \text{compiled } \mathrm{XOR}^{*}}   \text{. }
\end{align*}

\noindent \textit{Proof of Lemma $4A^{**}$}. Directly apply the argument from \textbf{Theorem} $\textit{1}^{*}$, particularly for demonstrating that the desired operator associated with the semidefinite program is positive definite, from which we conclude the argument. \boxed{}

\bigskip

\noindent \textbf{Lemma} $\textit{5}^{*}$ (\textit{error bound from permuting indices in the N-player setting}, \textbf{Lemma} \textit{5}, {[37]}). One has the following error bound from permuting indices,

\begin{align*}
     \bigg| \bigg|  \bigg(    \bigg( \underset{1 \leq i \leq n}{\prod} A^{j_i}_i  \bigg)      \bigotimes \bigg( \underset{1 \leq z \leq N-1}{\bigotimes} \textbf{I}_z \bigg)       \bigg) \ket{\psi_{\text{compiled } N\mathrm{XOR}^{*}}} - \bigg(     \bigg( \underset{\text{if } i \equiv j_1+1, \text{ } \mathrm{set} \text{ } j_1 + 1 \equiv j_1 \oplus 1}{ \underset{1 \leq i \leq n}{\prod} }A^{j_i}_i        \bigg)   \\     \bigotimes \bigg( \underset{1 \leq z \leq N-1}{\bigotimes} \textbf{I}_z \bigg)            \bigg) \ket{\psi_{\text{compiled }N\mathrm{XOR}^{*}}}     \bigg| \bigg|            <  N n^{N+\epsilon_{N\mathrm{XOR}^{*}}} \omega^3_{\text{compiled }N\mathrm{XOR}^{*}}  \text{. }
\end{align*}

\noindent \textit{Proof of Lemma $5^{*}$}. Directly apply the argument in \textbf{Lemma} $\textit{5}^{*}$, \textbf{Lemma} $\textit{5}^{**}$, and \textbf{Lemma} $\textit{5}^{***}$, from which we conclude the argument. \boxed{}

\bigskip

\noindent \textbf{Lemma} $\textit{5}^{**}$ (\textit{error bound from permuting indices in the strong parallel repetition of the compiled N-player setting}, \textbf{Lemma} \textit{5}, {[37]}). One has the following error bound from permuting indices,

\begin{align*}
              \bigg| \bigg|  \bigg(  \bigg(    \bigg( \underset{1 \leq i \leq n}{\prod} A^{j_i}_i  \bigg)   \wedge  \cdots \wedge \bigg( \underset{1 \leq i^{\prime\cdots\prime} \leq n^{\prime\cdots\prime}}{\prod} A^{j^{\prime\cdots\prime}_{i^{\prime\cdots\prime}}}_{i^{\prime\cdots\prime}}   \bigg) \bigg)       \bigotimes \bigg( \underset{1 \leq z \leq N-1}{\bigotimes}  \big( \textbf{I}_z \wedge \cdots \wedge \textbf{I}_z \big)  \bigg) 
 \bigg) \\ \times \ket{\psi_{\text{compiled } N\mathrm{XOR}^{*}\wedge \cdots \wedge \text{compiled } N\mathrm{XOR}^{*}}} \\ - \bigg(   \bigg(   \bigg( \underset{\text{if } i \equiv j_1+1, \text{ } \mathrm{set} \text{ } j_1 + 1 \equiv j_1 \oplus 1}{ \underset{1 \leq i \leq n}{\prod} }A^{j_i}_i        \bigg)  \wedge \cdots \wedge \bigg(     \underset{\text{if } i^{\prime\cdots\prime} \equiv j^{\prime\cdots\prime}_1+1, \text{ } \mathrm{set} \text{ } j^{\prime\cdots\prime}_1 + 1 \equiv j^{\prime\cdots\prime}_1 \oplus 1}{ \underset{1 \leq i^{\prime\cdots\prime} \leq n^{\prime\cdots\prime}}{\prod} }A^{j^{\prime\cdots\prime}_{i^{\prime\cdots\prime}}}_{i^{\prime\cdots\prime}}           \bigg)    \bigg)  \\  \bigotimes \bigg( \underset{1 \leq z \leq N-1}{\bigotimes}\big(  \textbf{I}_z  \wedge \cdots \wedge \textbf{I}_z \big)   \bigg)    \bigg)    \ket{\psi_{\text{compiled }N\mathrm{XOR}^{*} \wedge \cdots \wedge \text{compiled } N\mathrm{XOR}^{*}}}     \bigg| \bigg|  <  n^{N+\epsilon}_{\wedge } \\  +  \bigg( \frac{50 n^{N+\epsilon}_{\wedge}}{\sqrt{n^{N-1}}}   \bigg) \omega_{\text{compiled } N\mathrm{XOR}^{*} \wedge \cdots \wedge \text{compiled } N \mathrm{XOR}^{*}}          \text{. }
\end{align*}

\noindent \textit{Proof of Lemma $5^{**}$}. Directly apply the argument in \textbf{Lemma} $\textit{5}^{*}$, \textbf{Lemma} $\textit{5}^{**}$, and \textbf{Lemma} $\textit{5}^{***}$, from which we conclude the argument. \boxed{}

\bigskip

\noindent \textbf{Lemma} $\textit{4}^{*}$ (\textit{an arbitrary number of strong parallel repetition applications of the compiled }$\sqrt{\epsilon^{\wedge}_{C 2\mathrm{XOR}^{*}}}$- \textit{2 $XOR^{*}$ approximality}, \textbf{Lemma} \textit{8}, {[37]}). From the same quantities introduced in the previous result, one has,

\begin{align*}
   \bigg| \bigg|               \bigg( \big(  A_k  \wedge A_{k^{\prime}} \wedge \cdots \wedge A_{k^{\prime\cdots \prime}} \big) \otimes \textbf{I} \bigg) \ket{\psi_{\text{compiled } 2\mathrm{XOR}^{*} \wedge \cdots \wedge \text{compiled } 2\mathrm{XOR}^{*}}}    \\ -  \bigg( \textbf{I} \otimes \bigg(     \frac{\pm \big(  B_{kl} \wedge B_{k^{\prime} l^{\prime}} \wedge \cdots \wedge B_{k^{\prime\cdots\prime}l^{\prime\cdots \prime}}  \big)  + \big( B_{lk} \wedge B_{ l^{\prime} k^{\prime}} \wedge \cdots \wedge B_{l^{\prime\cdots \prime} k^{\prime\cdots\prime} }  \big) }{\big| \pm \big(  B_{kl} \wedge B_{k^{\prime} l^{\prime}} \wedge \cdots \wedge B_{k^{\prime\cdots\prime}l^{\prime\cdots \prime}}  \big)  + \big( B_{lk} \wedge B_{ l^{\prime} k^{\prime}} \wedge \cdots \wedge B_{l^{\prime\cdots \prime} k^{\prime\cdots\prime} }  \big) \big| }           \bigg) \bigg) \\ \times \ket{\psi_{\text{compiled } 2\mathrm{XOR}^{*} \wedge \cdots \wedge \text{compiled } 2\mathrm{XOR}^{*}}}              \bigg| \bigg|  \\  < 18 \sqrt{N \epsilon^{\wedge}_{\text{compiled } 2\mathrm{XOR}^{*}}}   \text{. } 
\end{align*}

\noindent \textit{Proof of Lemma $4^{*}$}. Directly apply the argument in \textbf{Lemma} \textit{4}, from which we conclude the argument. \boxed{}

\bigskip

\noindent \textbf{Lemma} $\textit{4}^{**}$ (\textit{an arbitrary number of strong parallel repetition applications of }$\sqrt{\epsilon^{\wedge}_{N\mathrm{XOR}^{*}}}$- \textit{compiled N $XOR^{*}$ approximality}, \textbf{Lemma} \textit{8}, {[37]}). From the same quantities introduced in the previous result, one has, for the $\mathrm{XOR}$ game under an arbitrary number of strong parallel repetitions, that the quantities,

\begin{align*}
  \mathcal{I}^{*}_1 \equiv  \bigg| \bigg|               \bigg( \big(  A_k  \wedge A_{k^{\prime}} \wedge \cdots \wedge A_{k^{\prime\cdots \prime}} \big) \bigotimes  \bigg( \underset{1 \leq z \leq N-1}{\bigotimes}\textbf{I}_z \bigg)  \bigg) \\ \times \ket{\psi_{\text{compiled } N\mathrm{XOR}^{*} \wedge \cdots \wedge \text{compiled } N\mathrm{XOR}^{*}}}    \\  -  \bigg( \textbf{I}  \otimes \bigg(     \frac{\pm \big(  B_{kl} \wedge B_{k^{\prime} l^{\prime}} \wedge \cdots \wedge B_{k^{\prime\cdots\prime}l^{\prime\cdots \prime}}  \big)  + \big( B_{lk} \wedge B_{ l^{\prime} k^{\prime}} \wedge \cdots \wedge B_{l^{\prime\cdots \prime} k^{\prime\cdots\prime} }  \big) }{\big| \pm \big(  B_{kl} \wedge B_{k^{\prime} l^{\prime}} \wedge \cdots \wedge B_{k^{\prime\cdots\prime}l^{\prime\cdots \prime}}  \big)  + \big( B_{lk} \wedge B_{ l^{\prime} k^{\prime}} \wedge \cdots \wedge B_{l^{\prime\cdots \prime} k^{\prime\cdots\prime} }  \big) \big| }    \bigg) \\  \bigotimes  \bigg( \underset{1 \leq z \leq N-2}{\bigotimes}\textbf{I}_z \bigg)       \bigg) \ket{\psi_{\text{compiled }N\mathrm{XOR}^{*} \wedge \cdots \wedge \text{compiled } N\mathrm{XOR}^{*}}}              \bigg| \bigg| \\ \vdots \end{align*}

  \begin{align*} \mathcal{I}^{*}_N \equiv   \bigg| \bigg|    \bigg(   \bigg( \underset{1 \leq z \leq N-2}{\bigotimes} \textbf{I}_z \bigg) \bigotimes   \frac{1}{\sqrt{\# \sigma^{\prime} }}  \bigg(    \underset{\text{Permutations } \sigma^{\prime}}{\sum}  \big(   B^{(N-1)}_{\sigma^{\prime} ( i_1, \cdots, i_{N-1})}         \wedge   B^{(N-1)}_{\sigma^{\prime} ( i^{\prime}_1, \cdots, i^{\prime}_{N-1})}  \wedge  \cdots \\ \wedge   B^{(N-1)}_{\sigma^{\prime} ( i^{\prime\cdots\prime}_1, \cdots, i^{\prime\cdots\prime}_{N-1})} \big)      \bigg)     \bigotimes \textbf{I}  \bigg)    \ket{\psi_{\text{compiled }N\mathrm{XOR}^{*} \wedge \cdots \wedge \text{compiled } N\mathrm{XOR}^{*}}}  \\ -  \bigg( \bigg( \underset{1 \leq z \leq N-1}{\bigotimes} \textbf{I}_z \bigg) \bigotimes \frac{1}{\sqrt{\# \sigma }}  \bigg(    \underset{\text{Permutations } \sigma}{\sum}  \big(   B^{(N-1)}_{\sigma ( i_1, \cdots, i_{N-1})}         \wedge   B^{(N-1)}_{\sigma ( i^{\prime}_1, \cdots, i^{\prime}_{N-1})}  \wedge  \cdots \\ \wedge   B^{(N-1)}_{\sigma ( i^{\prime\cdots\prime}_1, \cdots, i^{\prime\cdots\prime}_{N-1})} \big)      \bigg) \bigg)  \ket{\psi_{\text{compiled } N\mathrm{XOR}^{*} \wedge \cdots \wedge \text{compiled } N\mathrm{XOR}^{*}}}           \bigg| \bigg|      \text{, } 
\end{align*}

\noindent have the strict upper bound,

\begin{align*}
\underset{1 \leq j \leq N}{\sum} \mathcal{I}^{*}_j < 20 N \sqrt{N \epsilon^{\wedge}_{\text{compiled } N\mathrm{XOR}^{*} \wedge \cdots \wedge \text{compiled } N\mathrm{XOR}^{*}}} \text{,}
 \end{align*}

 \noindent where the tensors beyond that of the second player, $B$, are indexed as,

 \begin{align*}
     \textbf{I} \bigotimes B_{\sigma ( i_1,i_2)} \bigotimes \bigg( \underset{1 \leq z \leq N-2}{\bigotimes} \textbf{I}_z   \bigg) \equiv  \textbf{I} \bigotimes B^{(1)}_{\sigma ( i_1,i_2)} \bigotimes \bigg( \underset{1 \leq z \leq N-2}{\bigotimes} \textbf{I}_z   \bigg)   \text{, } \\ \vdots \\    \bigg( \underset{1 \leq z \leq N-1}{\bigotimes} \textbf{I}_z   \bigg) \bigotimes B_{\sigma ( i_1,i_2,\cdots, i_{n-2} )}  \equiv   \bigg( \underset{1 \leq z \leq N-1}{\bigotimes} \textbf{I}_z   \bigg) \bigotimes B^{(n-2)}_{\sigma ( i_1,i_2,\cdots, i_{n-2} )}  \text{. }
 \end{align*}

\noindent \textit{Proof of Lemma $4^{**}$}. Directly apply the argument in \textbf{Lemma} \textit{4}, from which we conclude the argument. \boxed{}

\bigskip

\noindent \textbf{Lemma} \textit{FR} $\mathrm{XOR}^{*}$ (\textit{Frobenius norm upper bound for strong parallel, compiled, $\mathrm{XOR}^{*}$ repetition}). One has that,

\begin{align*} (\text{compiled }\mathrm{XOR}^{*}) \lesssim  N! n^N \sqrt{\epsilon }  \end{align*} \[ \times    \sqrt{\epsilon_{\text{compiled }\mathrm{XOR}^{*} \wedge \cdots \wedge \text{compiled }\mathrm{XOR}^{*}}} \times   
\left\{\!\begin{array}{ll@{}>{{}}l} n^{ \frac{\# \text{ of players}}{2}  +5} \Longleftrightarrow \big( \# \text{ of players} \big) \mathrm{mod} 2 \equiv 0 \\ n^{ \lfloor\frac{\# \text{ of players}}{2} \rfloor  +5} \Longleftrightarrow \big( \# \text{ of players} \big) \mathrm{mod} 2 \neq 0
\end{array}\right.   
    \text{, }
\]

\noindent given $\epsilon_{\text{compiled } \mathrm{XOR}^{*} \wedge \cdots \wedge \text{compiled }  \mathrm{XOR}^{*}}$ sufficiently small, where,

\begin{align*}
    \bigg| \bigg|  \bigg[  \bigg(   \big( \big( A_1 \big)^{(1)}_i \wedge \cdots \wedge  \big( A_n \big)^{(1)}_{i^{\prime}} \big) \bigotimes \textbf{I}      \bigotimes \bigg(    \bigg(   \underset{i_1 \in \mathcal{Q}_1, i_2 \in \mathcal{Q}_2 , i_3 \in \mathcal{Q}_3}{\prod}             C^{l_{ijk}}_{ijk}     \bigg) \wedge \cdots \\ \wedge \bigg(   \underset{i^{\prime\cdots\prime}_3 \in \mathcal{Q}^{\prime\cdots\prime}_3}{\underset{i^{\prime\cdots\prime}_1 \in\mathcal{Q}^{\prime\cdots\prime}_1}{\underset{ i^{\prime\cdots\prime}_2 \in \mathcal{Q}^{\prime\cdots\prime}_2 }{\prod}}}              C^{l^{\prime\cdots\prime}_{i^{\prime\cdots\prime}j^{\prime\cdots\prime}k^{\prime\cdots\prime}}}_{i^{\prime\cdots\prime}j^{\prime\cdots\prime}k^{\prime\cdots\prime}}            \bigg)        \bigg)  \bigotimes                \bigg(     \underset{1\leq k \leq n-3}{\bigotimes}      \textbf{I}_k       \bigg)      \bigg)       -    \big( \omega_{\text{compiled }} \big( {\mathrm{XOR}^{*}}  \big) \big)^n \\ \times     \bigg(       \pm \mathrm{sign} \big( i_1 , j_1 , k_1, i_{111} , \cdots , j_{111} , \cdots , j_{nm(n+m)}  ,  k_{111}, \cdots, k_{nm(n+m)}  \big) \\ \times   \mathrm{sign} \big(  i^{\prime}_1 , j^{\prime}_1 , k^{\prime}_1, i^{\prime}_{111}, \cdots , j^{\prime}_{111} , \cdots , j^{\prime}_{nm(n+m)}  ,k^{\prime}_{111}, \cdots, k^{\prime}_{nm(n+m)}          \big)       \times \cdots \\ \times \mathrm{sign} \big(  i^{\prime\cdots\prime}_1 , j^{\prime\cdots\prime}_1 , k^{\prime\cdots\prime}_1, i^{\prime\cdots\prime}_{111},  \cdots , j^{\prime\cdots\prime}_{111} , \cdots , j^{\prime\cdots\prime}_{nm(n+m)} , k^{\prime\cdots\prime}_{111}, \cdots, k^{\prime\cdots\prime}_{nm(n+m)}         \big) \bigg) \bigg[                            \textbf{I} \bigotimes \textbf{I} \\ \bigotimes \bigg(    \bigg(   \underset{i_1 \in \mathcal{Q}_1, i_2 \in \mathcal{Q}_2 , i_3 \in \mathcal{Q}_3}{\prod}             C^{l_{ijk}}_{ijk}     \bigg) \wedge \cdots  \wedge \bigg(    \underset{i^{\prime\cdots\prime}_3 \in \mathcal{Q}^{\prime\cdots\prime}_3}{\underset{i^{\prime\cdots\prime}_1 \in\mathcal{Q}^{\prime\cdots\prime}_1}{\underset{ i^{\prime\cdots\prime}_2 \in \mathcal{Q}^{\prime\cdots\prime}_2 }{\prod}}}              C^{l^{\prime\cdots\prime}_{i^{\prime\cdots\prime}j^{\prime\cdots\prime}k^{\prime\cdots\prime}}}_{i^{\prime\cdots\prime}j^{\prime\cdots\prime}k^{\prime\cdots\prime}}            \bigg)        \bigg) \\ \bigotimes \bigg( \bigg(   \underset{1\leq k \leq n-3}{\prod}  \textbf{I}_k \bigg)      \bigg)             \bigg] \bigg) \bigg]          \ket{\psi_{\text{compiled} ( \mathrm{XOR}^{*} \wedge \cdots \wedge \mathrm{XOR}^{*}})}   \bigg| \bigg|_F    \tag{$\mathrm{XOR}^{*}$}        \text{, }
\end{align*}

\noindent where the subscript of the optimal strategy denotes,

\begin{align*}
      \text{compiled} ( \mathrm{XOR}^{*} \wedge \cdots \wedge \mathrm{XOR}^{*})  \equiv \text{compiled } \mathrm{XOR}^{*} \wedge \cdots \wedge  \text{compiled } \mathrm{XOR}^{*}       \text{. }
\end{align*}

\noindent \textit{Proof of Lemma FR $\mathrm{XOR}^{*}$}. Directly apply the argument from $\textbf{Lemma}$ \textit{FR}, from which we conclude the argument. \boxed{}

\bigskip

\noindent \textbf{Lemma} \textit{Gen-FFL-Bound compiled $XOR$} 
 (\textit{generalizations of the second FFL error bound}, \textit{6.6}, {[37]}, \textbf{Lemma} \textit{7}, {[44]}). One can denote quantum states corresponding to optimal strategies for the dual XOR game, from the optimal strategies,

\begin{align*}
 \ket{\psi_{3\mathrm{XOR}}} \equiv \underset{\text{Players}}{\bigcup} \underset{\mathcal{S}}{\mathrm{sup}} \big\{  \text{A player's quantum strategy } \mathcal{S} \text{ for a } 3-\text{XOR game}    \big\}   \text{, } \\ \ket{\psi_{4\mathrm{XOR}}} \equiv  \underset{\text{Players}}{\bigcup} \underset{\mathcal{S}}{\mathrm{sup}} \big\{  \text{A player's quantum strategy } \mathcal{S} \text{ for a } 4-\text{XOR game}    \big\}  \text{, } \\ \ket{\psi_{5\mathrm{XOR}}} \equiv  \underset{\text{Players}}{\bigcup} \underset{\mathcal{S}}{\mathrm{sup}} \big\{  \text{A player's quantum strategy } \mathcal{S} \text{ for a } 5-\text{XOR game}    \big\}  \text{, } \\  \ket{\psi_{N\mathrm{XOR}}} \equiv  \underset{\text{Players}}{\bigcup} \underset{\mathcal{S}}{\mathrm{sup}} \big\{  \text{A player's quantum strategy } \mathcal{S} \text{ for an } N-\text{XOR game}    \big\} \text{, }
\end{align*}

\noindent respectively, for the $3$-$\mathrm{XOR}^{*}$, $4$-$\mathrm{XOR}^{*}$, $5$-$\mathrm{XOR}^{*}$, and $N$-$\mathrm{XOR}^{*}$, games. Given error bounds formulated in previous sections for each $\mathrm{XOR}$ game, one obtains error bound inequalities of the form,

\begin{align*}
     \bigg| \bigg| \bigg[ \bigg( \bigg( \underset{1 \leq i \leq n}{\prod}  A^{j_i}_i \bigg) \otimes B_{kl} \otimes \textbf{I} \bigg)  - \omega_{\text{compiled } 3 XOR^{*}} \bigg( \pm \mathrm{sign} \big( i_1 , j_1 , \cdots , j_n \big)  \\ \times    \bigg[       \bigg( \bigg( \underset{1 \leq i \leq n}{\prod}   A^{j_i}_k \bigg)   + \bigg(                     \underset{\text{set } j+1 \equiv j \oplus 1}{\underset{i \in \mathcal{Q}_1, j \in \mathcal{Q}_2}{\prod}}   A^{j_i}_k        \bigg) \bigg) \otimes \textbf{I} \otimes \textbf{I} \bigg] \bigg) \bigg]  \ket{\psi_{\text{compiled } 3XOR^{*}}} \bigg| \bigg|^2 \\ \lesssim  3! n^3 \sqrt{\epsilon }   \text{, } \\ \\    \bigg| \bigg| \bigg[ \bigg( \bigg( \underset{1 \leq i \leq n}{\prod}  A^{j_i}_i \bigg) \otimes B_{kl} \otimes \textbf{I} \otimes \textbf{I} \bigg)  - \omega_{\text{compiled } 4 XOR^{*}} \bigg( \pm \mathrm{sign} \big( i_1 , j_1 , \cdots , j_n \big)     \\ \times   \bigg[       \bigg( \bigg( \underset{1 \leq i \leq n}{\prod}   A^{j_i}_k \bigg)   + \bigg(                     \underset{\text{set } j+1 \equiv j \oplus 1}{\underset{i \in \mathcal{Q}_1, j \in \mathcal{Q}_2}{\prod}}   A^{j_i}_k        \bigg) \bigg) \otimes \textbf{I}    \otimes \textbf{I} \otimes \textbf{I} \bigg] \bigg) \bigg] \ket{\psi_{\text{compiled } 4 XOR^{*}}} \bigg| \bigg|^2 \\ \lesssim  4! n^4 \sqrt{\epsilon }       \text{, } \end{align*}
     
     \begin{align*} \bigg| \bigg| \bigg[ \bigg( \bigg( \underset{1 \leq i \leq n}{\prod}  A^{j_i}_i \bigg) \otimes B_{kl} \otimes \textbf{I} \otimes \textbf{I} \otimes \textbf{I} \bigg)  - \omega_{\text{compiled } 5XOR^{*}} \bigg( \pm \mathrm{sign} \big( i_1 , j_1 , \cdots , j_n \big)  \\ \times      \bigg[       \bigg( \bigg( \underset{1 \leq i \leq n}{\prod}   A^{j_i}_k \bigg)   + \bigg(                     \underset{\text{set } j+1 \equiv j \oplus 1}{\underset{i \in \mathcal{Q}_1, j \in \mathcal{Q}_2}{\prod}}   A^{j_i}_k        \bigg) \bigg) \otimes \textbf{I} \otimes \textbf{I}  \otimes \textbf{I} \bigg] \bigg) \bigg] \ket{\psi_{\text{compiled } 5 XOR^{*}}} \bigg| \bigg|^2 \\  \lesssim  5! n^5 \sqrt{\epsilon }     \text{, }    \end{align*}

    \begin{align*} \bigg| \bigg| \bigg[ \bigg( \bigg( \underset{1 \leq i \leq n}{\prod}  A^{j_i}_i \bigg) \bigotimes B_{kl}  \bigotimes \bigg( \underset{1 \leq k \leq N-2}{\bigotimes} \textbf{I}_k  \bigg)  \bigg)  - \omega_{\text{compiled } NXOR^{*}}  \\ \times    \bigg( \pm \mathrm{sign} \big( i_1 , j_1 , \cdots , j_n \big)    \bigg[       \bigg( \bigg( \underset{1 \leq i \leq n}{\prod}   A^{j_i}_k \bigg)      + \bigg(                     \underset{\text{set } j+1 \equiv j \oplus 1}{\underset{i \in \mathcal{Q}_1, j \in \mathcal{Q}_2}{\prod}}   A^{j_i}_k        \bigg) \bigg) \\ \bigotimes \bigg( \underset{1 \leq k \leq N-1}{\bigotimes} \textbf{I}_k  \bigg) \bigg] \bigg) \bigg] \ket{\psi_{ \text{compiled} N XOR^{*} }} \bigg| \bigg|^2 
\end{align*}

\[
\lesssim N! n^N  \sqrt{\epsilon}  \times   
\left\{\!\begin{array}{ll@{}>{{}}l} n^{ \frac{\# \text{ of players}}{2}  +5} \Longleftrightarrow \big( \# \text{ of players} \big) \mathrm{mod} 2 \equiv 0 \\ n^{ \lfloor\frac{\# \text{ of players}}{2} \rfloor  +5} \Longleftrightarrow \big( \# \text{ of players} \big) \mathrm{mod} 2 \neq 0
\end{array}\right.   \text{. }
\]

\bigskip

\noindent \noindent \textit{Proof of Lemma Gen-FFL-Bound compiled $XOR^{*}$}. Directly apply the argument from \textbf{Lemma} \textit{Gen-FFL-Bound}, from which we conclude the argument. \boxed{}

\subsubsection{Suitable linear operators for compiled multiplayer $XOR^{*}$ games have unit Frobenius norm}

\noindent \textbf{Lemma} $\textit{9}^{**}$ (\textit{the Frobenius norm of suitable linear operators for the 3-XOR, 4-XOR, 5-XOR, and N-XOR games equals 1}). With respect to the Frobenius norm, the norm of suitable linear operators introduced in previous sections for the 3-XOR, 4-XOR, 5-XOR, and N-XOR, games equals $1$.

\bigskip

\noindent \textit{Proof of Lemma $\textit{9}^{**}$}. Directly apply the argument from \textit{6.2} in {[37]}, from which we conclude the argument. \boxed{}

\subsubsection{Suitable linear operators for strong parallel repetition of the compiled $XOR^{*}$ game have unit Frobenius norm}

\noindent \textbf{Lemma} $\textit{10}^{**}$ (\textit{the Frobenius norm of suitable linear operators for strong parallel repetition of the multiplayer XOR game, and of the two player FFL game, equal 1}). With respect to the Frobenius norm, the norm of suitable linear operators introduced in previous sections for the 3-XOR, 4-XOR, 5-XOR, and N-XOR, games equals $1$.

\bigskip

\noindent \textit{Proof of Lemma $\textit{10}^{**}$}. Directly apply the argument from \textit{6.2} in {[37]}, from which we conclude the argument. \boxed{}

\section{Data Availability Statement}

The manuscript has no associated data.

\section{Conflicts of Interest Statement}

On behalf of all authors, the corresponding author states that there is no conflict of interest.





\nocite{*}
\bibliography{sn-bibliography}

\end{document}